\documentclass[12pt,preprint]{aastex}
\newcommand{\cmfgen}{{\sc cmfgen}}
\newcommand{\fastwind}{{\sc fastwind}}
\newcommand{\tlusty}{{\sc tlusty}}

\newcommand{\kms}{\hbox{km$\,$s$^{-1}$}}

\slugcomment{Accepted to the ApJ}

\shorttitle{\cmfgen\ vs \fastwind\ }

\shortauthors{Massey et al.}

\begin{document}

\title{A Bake-Off Between \cmfgen\ and \fastwind:\\ Modeling the Physical Properties of SMC and LMC O-type Stars}

\author{Philip Massey\altaffilmark{1},  Kathryn F. Neugent\altaffilmark{1}, D. John Hillier\altaffilmark{2}, and Joachim Puls\altaffilmark{3}}

\altaffiltext{1}{Lowell Observatory, 1400 W Mars Hill Road, Flagstaff, AZ 86001; phil.massey@lowell.edu; kneugent@lowell.edu}

\altaffiltext{2}{Department of Physics and Astronomy, University of Pittsburgh, Pittsburgh, PA 15260; hillier@pitt.edu}

\altaffiltext{3}{Universit\"{a}tssternwarte M\"{u}nchen, Scheinerstr.\ 1, 81679 M\"{u}nchen, Germany; uh101aw@usm.uni-muenchen.de}

\begin{abstract}

The model atmosphere programs \fastwind\  and \cmfgen\ are both elegantly designed to perform non-LTE analyses of the spectra of hot massive stars, and include sphericity and mass-loss.  The two codes differ primarily in their approach towards line blanketing, with \cmfgen\ treating all of the lines in the co-moving frame and \fastwind\  taking an approximate approach which speeds up execution times considerably.  Although both have been extensively used to model the spectra of O-type stars, no studies have used the codes to independently model the same spectra of the same stars and compare the derived physical properties.  We perform this task on ten O-type stars in the Magellanic Clouds.  For the late-type O supergiants, both \cmfgen\ and \fastwind\  have trouble fitting
some of the He I lines, and we discuss causes and cures.   We find that there is no difference in the average effective temperatures found by the two codes for the stars in our sample, although the dispersion is large, due primarily to the various difficulties each code has with He I.  The surface gravities determined using \fastwind\  are systematically lower by 0.12~dex compared to \cmfgen, a result we attribute to the better treatment of electron scattering by \cmfgen.  This has implications for the interpretation of the origin of the so-called mass discrepancy, as the masses derived by \fastwind\  are on average lower than inferred from stellar evolutionary models, while those found by \cmfgen\ are in better agreement.

\end{abstract}


\section{Introduction}

Modeling the stellar atmospheres of O-type stars is fraught with physical complexities that don't arise when modeling most other types of stars.  For instance, the assumption of local thermodynamic equilibrium (LTE), in which the temperature of the radiation field and that of the gas is taken to be the same,  is a not a good approximation for O-type stars, as their atmospheres are so hot that the degree of ionization is high.  This lowers the opacity considerably, making the atmospheres highly transparent.   Thus, the temperatures of the radiation field and that of the surrounding gas can be quite different, requiring non-LTE  calculations to obtain good fits, as first shown theoretically by Mihalas \& Auer (1970) and confirmed observationally by Conti (1973).  Additionally, these 
stars are losing mass due to radiatively driven winds.  Emission from the wind contaminates even the optical hydrogen and helium lines (along with those of some metals), and can produce strong P Cygni profiles in the resonance lines found in the ultraviolet (see, for example, Kudritzki \& Puls 2000).   The low wind densities also necessitate a non-LTE treatment.  
Rather than assuming static plane parallel geometry, modern O star model atmospheres adopt a spherical geometry and include the radial outflow of material, in addition to non-LTE.  Finally, it is crucial that O star stellar atmosphere models be {\it fully blanketed}; i.e., they must include the effects of thousands of overlapping metal lines, which occur at the (unobservable) short wavelengths where most of the flux of the star is produced (e.g., Hillier \& Miller 1998).

There are two fully line-blanketed non-LTE models in common use\footnote{TLUSTY (Hubeny \& Lanz 1995) is a very versatile, 
general-purpose static plane parallel stellar atmosphere code  that has often been used for the analysis of OB stars.  It also includes full
non-LTE line blanketing. It should be fine for OB stars that have weak stellar winds. For that matter, it should be noted that \cmfgen\ and \fastwind\  are not truly ``hydrodynamic", in that they don't {\it solve} for the wind structure, but rather adopt a wind law.}  for the analysis of the spectra of O stars, \cmfgen\footnote{the  CoMoving Frame GENeral} (Hiller \& Miller 1998; Hillier \& Lanz 2001) and \fastwind\footnote{Fast Analysis of STellar atmospheres with WINDs}  (Santolaya-Rey et al.\ 1997, Repolust et al.\ 2004, Puls et al.\ 2005).   The two codes differ primarily in their approach to line blanketing. \cmfgen\ is ``exact"
in that it treats all lines in the co-moving frame.  In order to achieve much shorter execution times, \fastwind\  makes an ``approximate but realistic" treatment of line blanketing (by using suitable averages for background line opacities and emissivities), and performs the detailed calculations only for the specific elements that will be used to fit the models (usually just H and He for O-type stars, but with the recent addition of N; see Rivero Gonz\'{a}lez et al.\ 2011, 2012a, 2012b).  On the other hand, \cmfgen, in addition to using H and He lines,  can utilize metal lines in both the UV and optical to provide additional constraints on key physical parameters.  Its ability to model the full spectrum (including the UV) has led to its success in modeling Wolf-Rayet spectra (Hillier 2003a).  More technical details of the codes are discussed in the Appendix.

A point made by Puls et al.\ (2005) is that far fewer stars have been analyzed with \cmfgen\ than with 
\fastwind\  due to the large differences in computational time.  In addition, \cmfgen\  modeling usually involves fitting more features than just H and He, and therefore is more effort is usually required.Although great effort has gone into 
making \cmfgen\ as efficient in its computations as possible (such as the use of ``super levels", which decrease the number of levels whose atomic populations much be explicitly solved; see Hillier \& Miller 1998 and Hillier \& Lanz 2001), 
the large number of lines treated results in much longer execution times compared to \fastwind.
(We estimate below, in Section~\ref{Sec-modeling}, that $\sim$450 hours of computer time are needed to completely model an O-type star with \cmfgen, vs.\ 1.5 hours with \fastwind\.)   As far as we are aware, only five stars have been 
analyzed in the optical using both codes, the ones listed 
in Table 4 of Massey et al.\ (2009)\footnote{Some Galactic O stars were analyzed with \fastwind\  in the optical and \cmfgen\ in the L-band by Najarro et al.\ (2011); see their Table A.2.  In addition, four Galactic (Cyg OB2) stars were analyzed
using \fastwind\ and a modified version of \cmfgen, as described in a conference proceeding by Herrero et al.\ (2003).}. The \cmfgen\ analyses of four of these
were done by Bouret et al.\ (2003) using data from Walborn et al.\ (2000) that may have been affected by a
reduction issue as discussed by Massey et al.\ (2005, 2009); in any event, two of these stars are likely composite
spectra (binaries) (Massey et al.\ 2009), and a third has reported radial velocity variations (Mokiem et al.\ 2006). 
Thus, the sample is inadequately small for a meaningful comparison between the results obtained with the two codes.  
In addition, to the best of our knowledge no stars have been analyzed using both \fastwind\ and \cmfgen\ {\it using the same
data}\footnote{The arguable exception is the study briefly described by Herrero et al.\ (2003).}.

Nevertheless, some interesting contrasts are known between the results of the two codes.   For instance, \cmfgen\ does not always do a good job of matching the He~I singlet lines, such as He~I $\lambda 4388$, producing too strong a line.
 Najarro et al.\ (2006) have attributed this
to a problem with the atomic data for the Fe IV lines that overlap the He~I resonance lines around 584~\AA.
While the wavelength of the transitions is known, there is considerable uncertainty in the oscillator strengths of the lines, as evidenced by the different values obtained in the calculations of Bob Kurucz, and Becker \& Butler (1995); see Najarro et al.\ (2006) for details.  \cmfgen\ treats the Fe IV atom using super-levels, but the individual transitions are included at their correct wavelength in the atomic model. From the analysis of He I singlet and triplet lines in O supergiants the effect appears to be real, but is smaller than would be predicted using the original Kurucz oscillator strengths. In the present calculation, the relevant oscillator strengths were reduced by a factor of 10 from the original Kurucz values.

In contrast, Massey et al.\ (2009) found that \fastwind\  does an adequate job with the He~I singlets, which they attribute to
the program being less sensitive to the uncertainty in the Fe IV oscillator strengths due to its approximate treatment of line blanketing.  On the other hand, \fastwind\  produces a He~I triplet $\lambda 4471$ model line that is often too weak for mid and late O-type giants
and supergiants for unknown reasons (Repolust et al.\ 2004).
As a result, fitting the He~I $\lambda 4471$ line has generally been eschewed in this spectral-type/luminosity regime 
by \fastwind\  aficionados (Herrero et al.\ 2000, 2002; Repolust et al.\ 2004; Massey et al.\ 2004, 2005, 2009).  This is in contrast to \cmfgen, where it is
clear from the literature that it does a very good job in matching the He I $\lambda 4471$ profiles for such stars: see, for example, the excellent fits for the O7 Iaf+ star AzV 83 and OC7.5((f)) star AzV 69, as shown in Figures 3 and 15, respectively, of Hillier et al.\ (2003a).   The He I $\lambda 4471$ \fastwind\  problem remains unresolved, although we find below we can largely solve this by increasing the adopted microturbulence velocity for late-type supergiants.

There have been numerous comparisons (both published and unpublished) between the output of \fastwind\  and \cmfgen\ for a given set of input parameters (see, for example, Rivero Gonz\'{a}lez et al.\ 2012b), but this is quite different than comparing the physical
parameters derived by modeling the same star with the two codes, particularly given the above discussion, as one relies more heavily on the He I singlets with \fastwind\  and the He I triplets with \cmfgen.

Therefore, we decided it would be useful to conduct the experiment of modeling the same data with \cmfgen\ and \fastwind\  independently.  We selected ten Magellanic Cloud O-type stars that have excellent, high signal-to-noise
optical data, and had been previously well modeled by an earlier version of
\fastwind\  (Massey et al.\ 2004, 2005, 2009, hereafter, Papers I, II, and III).  We remodeled these stars both with \cmfgen\ and the latest version of \fastwind\  in an effort to look for systematic differences in the resulting parameters.  For consistency the second author performed all of the modeling.

\section{Modeling Efforts}
\label{Sec-modeling}

We list in Table~\ref{tab:stars} the stars whose spectra were modeled for the present study.  The sample covers 
very early O-types (O3.5) to the latest O-types (O9.7), and includes both dwarfs and supergiants.  No giants were included to ensure both early and late O stars of extreme luminosity classes were included with a finite number of stars.  

Details of the ground-based optical spectroscopic data can be found in the papers referenced in Table~\ref{tab:stars}; here we will repeat that the data were all of high signal-to-noise, typically several hundred per 1-2~\AA\ spectral resolution element.  All stars were observed twice, once in the blue and once in the red, where the latter provides primarily the H$\alpha$ profile used in constraining the mass-loss rate.    The optical data for the sample of stars used here from Papers I and II were obtained with the CTIO 4-m Blanco telescope using the RC spectrograph and had a spectral resolution of 1.4~\AA\ in the blue (3750-4900~\AA) and 2.8~\AA\ in the red (5400-7800~\AA).  All of the optical data for the stars used here from Paper III were obtained with the Magellan Clay 6.5-m telescope and the Boller \& Chivens Spectrograph and had resolutions of 2.4~\AA\ in the blue (3500-5040~\AA) and 2.4~\AA\ in the red (5315-6950~\AA).

When modeling a star, we varied primarily three parameters to obtain a good fit to the optical Balmer and He lines: (1)  the effective temperature $T_{\rm eff}$,  (2) the logarithm of the effective surface gravity $\log g_{\rm eff}$ (where $g$ is expressed in cgs units),
and (3) the mass-loss rate, $\dot{M}$ (expressed in $M_\odot$ yr$^{-1}$).    To a first approximation, the line depths of the He~I and He~II lines are most sensitive to the
effective temperature, the wings of the Balmer lines (we relied heavily on H$\gamma$) to the surface gravity, and the H$\alpha$ profile to the mass-loss rate.   If neither the He~I nor He~II model lines were as strong as observed, we increased the He/H number ratio from the canonical 0.10 value; the maximum value we needed was 0.25.  For each star we kept a number of properties fixed, as shown in Table~\ref{tab:fixed}.  For instance, we adopted the terminal velocity $v_\infty$ of the stellar wind as measured from {\it HST} UV spectra as described in Papers I-III, as well as the parameter $\beta$ that characterizes the stellar wind law (Castor \& Lamers 1979,
Santolaya-Rey et al.\ 1997), 
although we were prepared to change the latter if we could not obtain a good fit to the H$\alpha$ profile.
Similarly, we adopted the values for the absolute visual magnitude ($M_V$) for each star from Papers I-III; this is needed to
compute the stellar radius for each set of input values, as well as computing the bolometric luminosity of the star. For the SMC, we fixed the metal abundances relative to the sun $Z/Z_\odot$ to 0.2 and for the LMC, 0.5.  (As noted later, we did also vary the CNO abundances in our \cmfgen\ fits to see if we could obtain better agreement with the optical lines.)  

Our modeling efforts began in mid-2009, and during the course of that time the \fastwind\  code was revised from version 9.2 to 10.1 (mid 2010);  all the \fastwind\  work was repeated using the latest version (10.1).  Differences were minor, as detailed below.  The \cmfgen\ code also received several improvements (including a fix to a small problem found as part of our modeling having to do with how certain input values were read); in the end, all final models come from the 7 April 2011 release.

The execution of \cmfgen\ and \fastwind\  both require rather complicated input files and a series of steps, as several different executable programs must be run in sequence (each with its own input) in order to produce synthetic spectra corresponding to a particular set of physical properties.   To simplify the process, we wrote a more user-friendly interface ``wrapper" for each program that simplified both the input and the running of the various executables. 
Of the two, \cmfgen\ tends to be fussier, requiring the user to fiddle with parameters as the model is running to ensure convergence within a reasonable time-period.  All models were run by K. F. N., and, once she was experienced, she found that the first \cmfgen\ model typically required about 50 hours of execution to finish with good convergence.  Subsequent models with small changes required about 30 hours to run, while 40 hours were needed for large changes.  Typically between 10-15 models were needed to obtain satisfactory fits with \cmfgen; i.e., the total execution time for each star (irrespective of examining the fits) was about 365-540 hours, or $\sim$ 450 hours on average.
The models were run on a MacPro with two 3 GHz quad-core Intel Xeon processors and 32 GB of memory.
(The eight CPUs were used singly for most of the computations, but allowed eight models to run simultaneously.)
By contrast, \fastwind\  required about 10-15 minutes per model, and fewer models, usually 5-10, or about 1.5 hours on average.   
\cmfgen\ required more models to be run as sometimes the convergence wasn't good, 
or additional diagnostics indicated a problem, such
as the x-ray luminosities being unrealistic requiring a manual adjustment of the filling factors.  We should explicitly note though that in some ways this comparison is not fair, in that we wanted to assure the x-ray luminosities were reasonable in order to make a valid comparison to the UV spectrum, a region which \fastwind\  doesn't attempt to model. 

We note that the number of models needed to fit the spectrum of an O-type star if mass-loss is included is 
considerably greater than those computed with, for instance, \tlusty.  When using plane-parallel models there are essentially only two key parameters, $\log g_{\rm eff}$ and $T_{\rm eff}$, although with the OB stars considered here, the helium abundance would also have to be adjusted. 
With spherical models, the stellar radius,  the mass-loss rate, the clumping stratification, and the velocity law are also potentially free parameters. In practice, the mass-loss rates were allowed to vary along with $\log g_{\rm eff}$ and $T_{\rm eff}$ and (as needed) the He abundance (with the radius
then fixed to provide the correct $M_V$); this accounts for the large number of models needed.  

In order to compare the theoretical spectrum to the observed spectrum, we used the built-in plotting routines for \cmfgen\ (which make calls to {\sc pgplot}\footnote{http://www.astro.caltech.edu/$\sim$tjp/pgplot}), and wrote our own code in {\sc idl} for the \fastwind\  output.
For comparing the model spectra with the observed spectra, we adjusted for the radial velocity $v_{\rm rad}$ of the star, and convolved the model profiles to match the projected rotational broadening $v \sin{i}$.  (Some amount of the latter quantity is due to instrumental broadening and possibly macroturbulence, as discussed below.)   The values we adopted are listed in Table~\ref{tab:fixed}, and mostly agree with what we list in Papers I-III.
(A detailed comparison is given in Sec.~\ref{Sec:diffs} below.)  In a few cases there is a significantly different radial velocity adopted for the red and blue spectra.  As explained in Papers I-III, the wavelength calibration was often based upon a single calibration exposure taken during the day, and thus these radial velocities are only approximately correct.   (These are the observed radial velocities of the spectra; the conversion to the heliocentric values
are given in a footnote to the table.)

Other than the treatment of blanketing  and the exact methodology used for treating non-LTE, there are two other differences between the codes that we discuss here.  The first concerns the chemical abundances. Although we scaled
to the SMC and LMC abundances from ``solar" abundances, the two codes did not assume identical  values for ``solar". 
\cmfgen\ adopts abundances relative to the solar abundance ratios tabulated by Cox (2000). These are based upon the older work of Anders \& Grevesse (1989) and Grevesse \& Noels (1993) as opposed to the newer abundances of Asplund et al.\ (2009).  Similarly we used the older values with \fastwind, although slightly different ones; i.e., from Grevesse \& Sauval (1998).  We note that the Asplund et al.\ (2009) values are also included in the \fastwind\  distribution.
The differences are relatively minor, by about 10\% in iron and somewhat more for oxygen and nitrogen. 
 At the same time, it should be noted that our runs of \cmfgen\ only explicitly included 
H, He, C, N, O, Si, P, S, and Fe, while \fastwind\  included all of the elements to Zn (i.e., atomic numbers 1-30), albeit in an approximate sense; only H and He were exactly computed.

The second concerns the values used for microturbulence.  
Microturbulence  has three effects.  First, it can affect the non-LTE occupation numbers, due to a different profile function and thus opacities and emissivities.  Secondly, because of this, it has a small influence on the atmospheric structure, particularly because of the radiation pressure.  Thirdly, it affects the emergent profiles, most strongly the metal lines, but also the He I profiles.
Both \fastwind\  and \cmfgen\ allow for one value to be used in constructing the stellar atmosphere, and another to be used
in computing the emergent line profiles.  In calculating the non-LTE occupation numbers and atmospheric structure, we adopted 10 km~s$^{-1}$ regardless of luminosity class with \fastwind\ , and 10 km~s$^{-1}$ for dwarfs but 20 km~s$^{-1}$ for supergiants with the \cmfgen\ models, in keeping with the advice of the creators of the code.   In computing the line profiles, we adopted 10 km s$^{-1}$ for all of the \cmfgen\ models.  However, for \fastwind\  we found we obtained more consistent results with the He I triplets and singlets if we used no microturbulence (0 km s$^{-1}$) for the hotter stars ($T_{\rm eff}>$36,000 K) and a value of 10 km s$^{-1}$ for the cooler stars in computing the line profiles.  The choice of microturbulence had no effect on the hydrogen lines.  We will find below that 15-20 km s$^{-1}$ gives better results for the late O-type supergiants with both codes. 

For both the \cmfgen\ and \fastwind\  models we assumed a stellar wind clumping law with a volume filling factor $f$ of 0.1 at $v_\infty$,  as argued by Owocki \& Cohen (2008).   This differs from our earlier  \fastwind\  calculations in Papers I-III that were run with an earlier version of the code with clumping turned off.  The effect should be that the currently derived mass-loss rates with \fastwind\  should be $\sim 3.3\times$ smaller than those found in Papers I-III, as unclumped models should overestimate the rates by $1/\sqrt{f}$
(e.g., Owocki \& Cohen 2008).  (The new mass-loss rates we find are a bit higher than this, as described below in Section~\ref{Sec:diffs}.)  Note though that we used different stratification laws for the clumping model.  By default, \cmfgen\ uses an exponential clumping law, resulting in a considerable part of the lower wind remaining unclumped.  By contrast, the clumping law we used with \fastwind\  involves constant clumping. Since most wind lines are formed in the lower wind, where \cmfgen\ is less clumped than \fastwind, one expects that the mass-loss rates derived from \cmfgen\ will be slightly higher that those derived from our \fastwind\  fits. 

Our fitting procedure was to obtain the ``best fits" (as judged by eye) to the H, He~I, and He~II lines by varying $T_{\rm eff}$, $\log g$, and $\dot{M}$ as described above.  The He/H ratio was varied (increased) only if absolutely necessary to make the He lines sufficiently strong. It was immediately apparent that the differences between the two codes described above were correct, that while \fastwind\  failed to produce a sufficiently strong He~I $\lambda 4471$ for mid-to-late type supergiants (and presumably giants), \cmfgen\ had no such difficulty. At the same time \cmfgen\ could not produce reliable singlet He~I lines (such as He~I $\lambda 4388$) as expected for the later O-types.  Once all of the modeling was done, we then varied the CNO abundances with \cmfgen\ to see if we could improve the fits for these elements.  We did not find any need to vary $T_{\rm eff}$, $\log g$, $\dot{M}$, or the He/H ratio from what we had found in these additional models.   We discuss the fits of the metal lines separately in Section~\ref{Sec-metals}.

\section{Results}

\begin{figure}
\epsscale{0.3}
\plotone{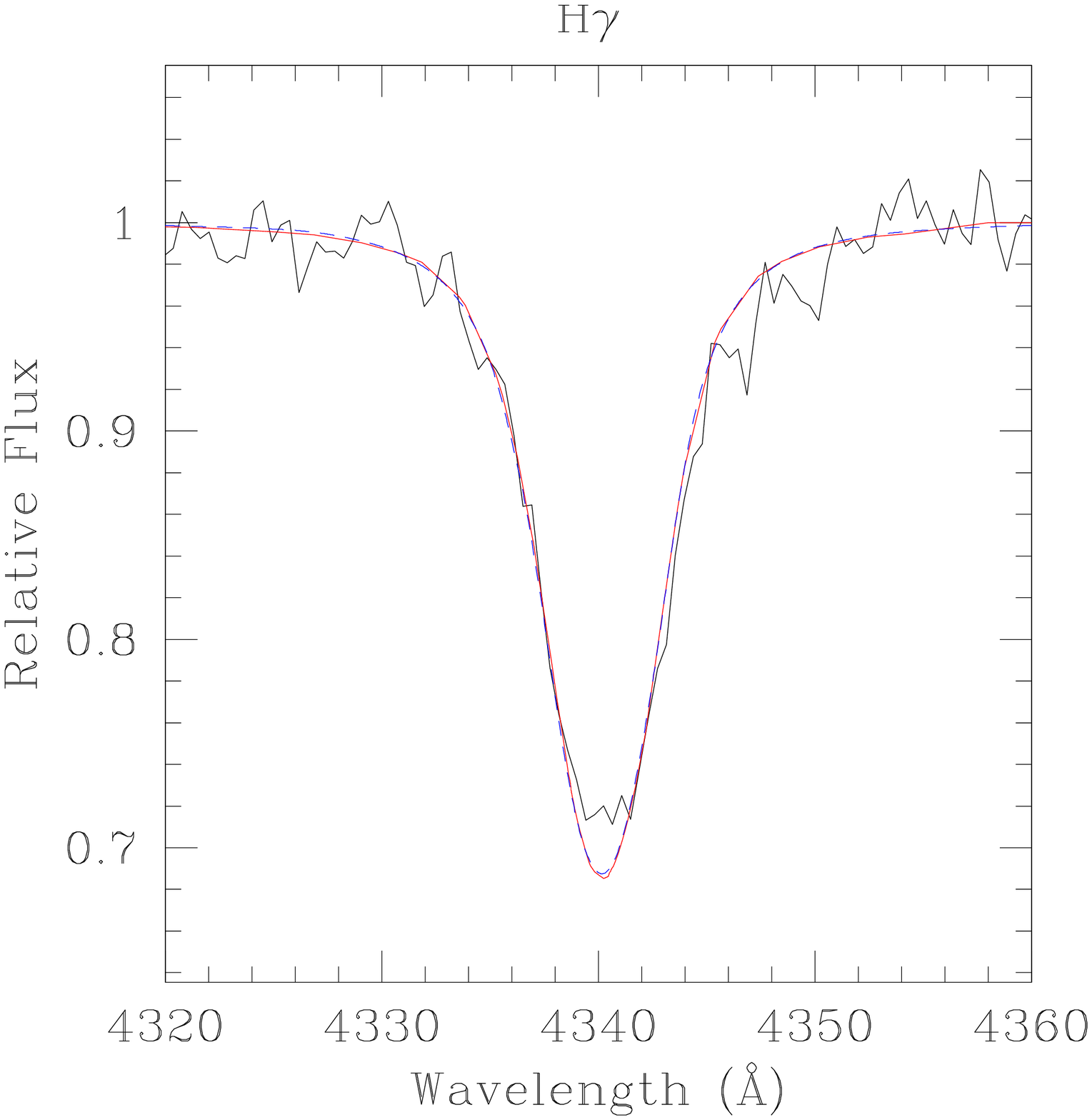}
\plotone{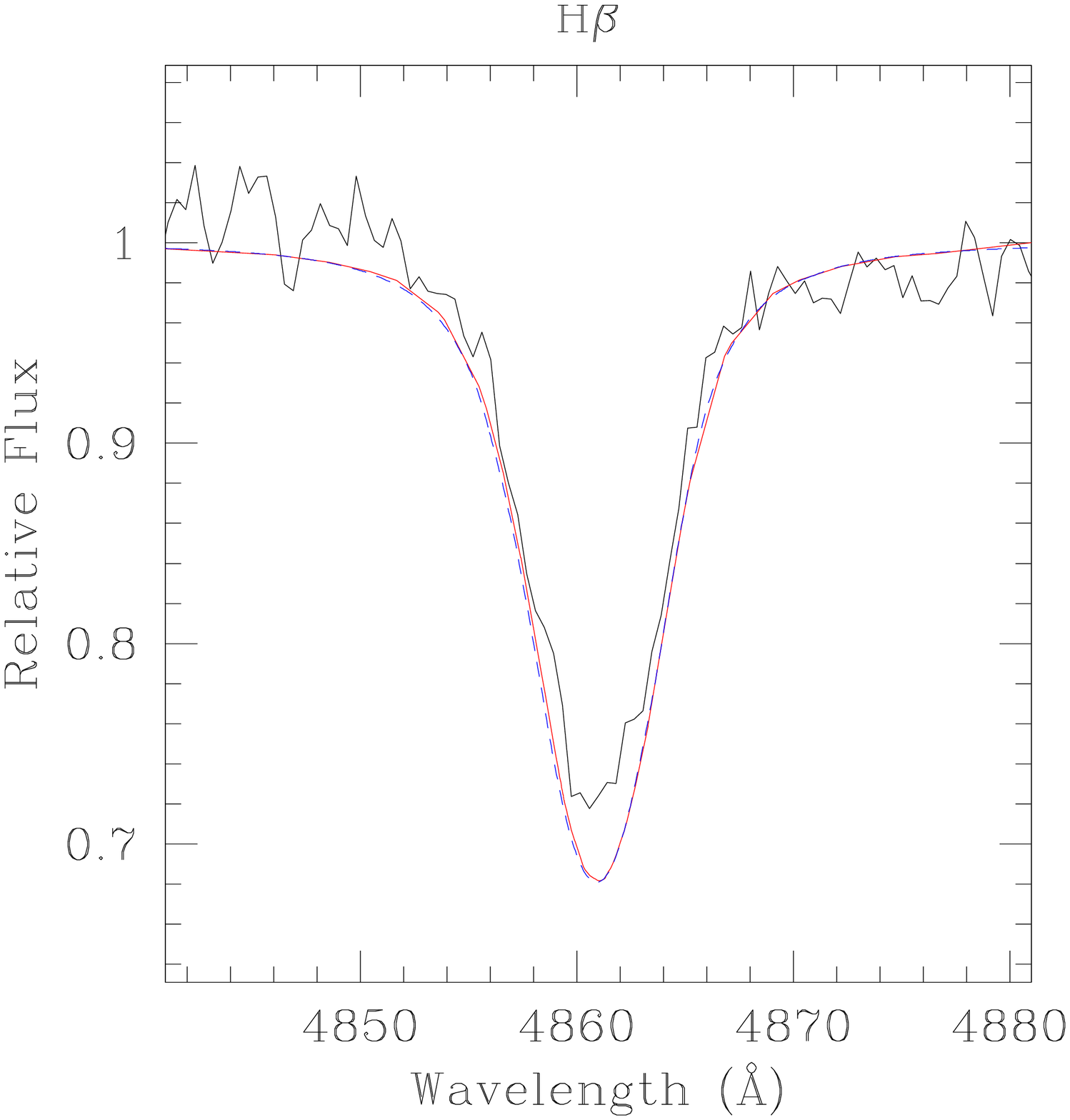}
\plotone{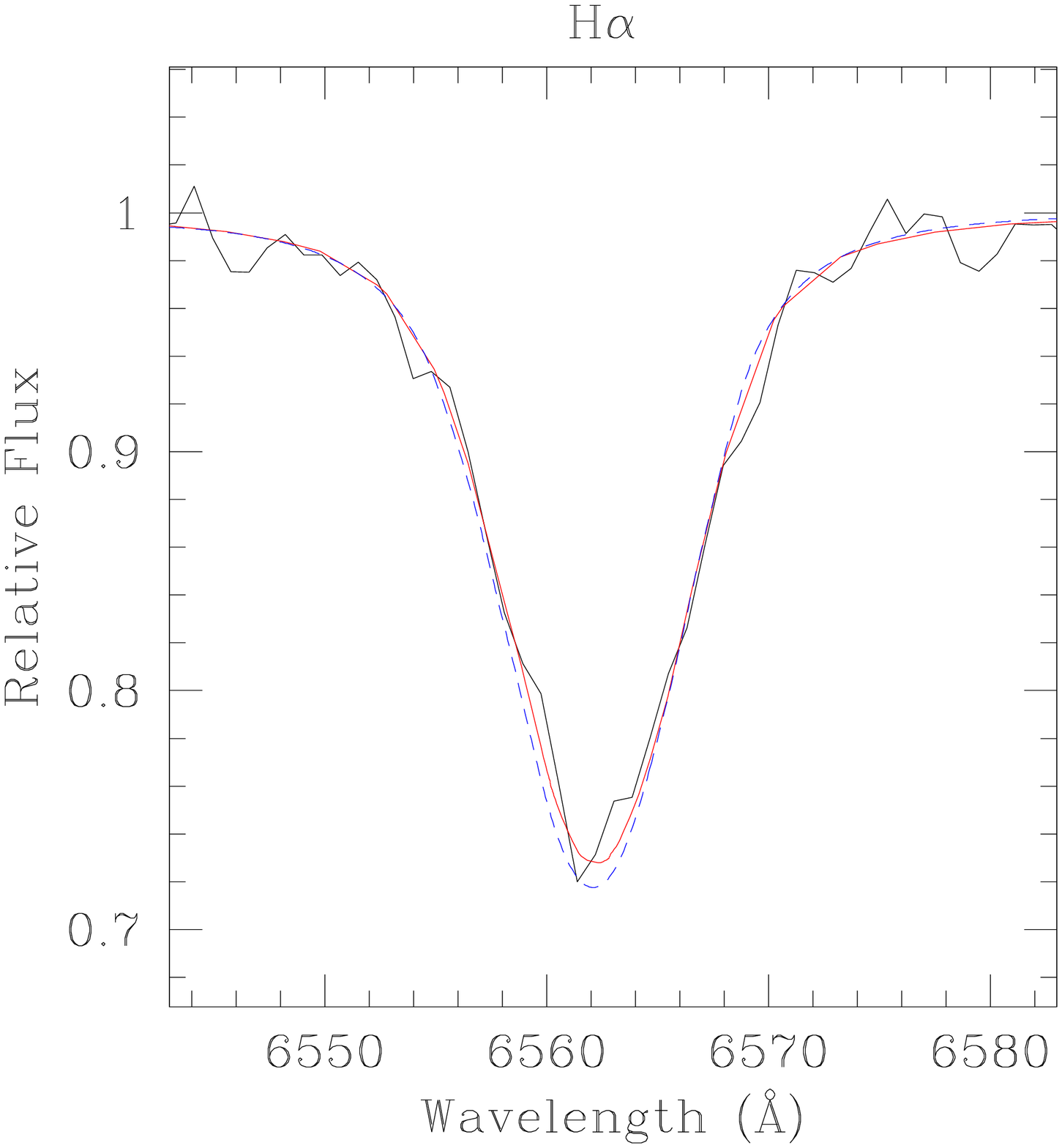}
\plotone{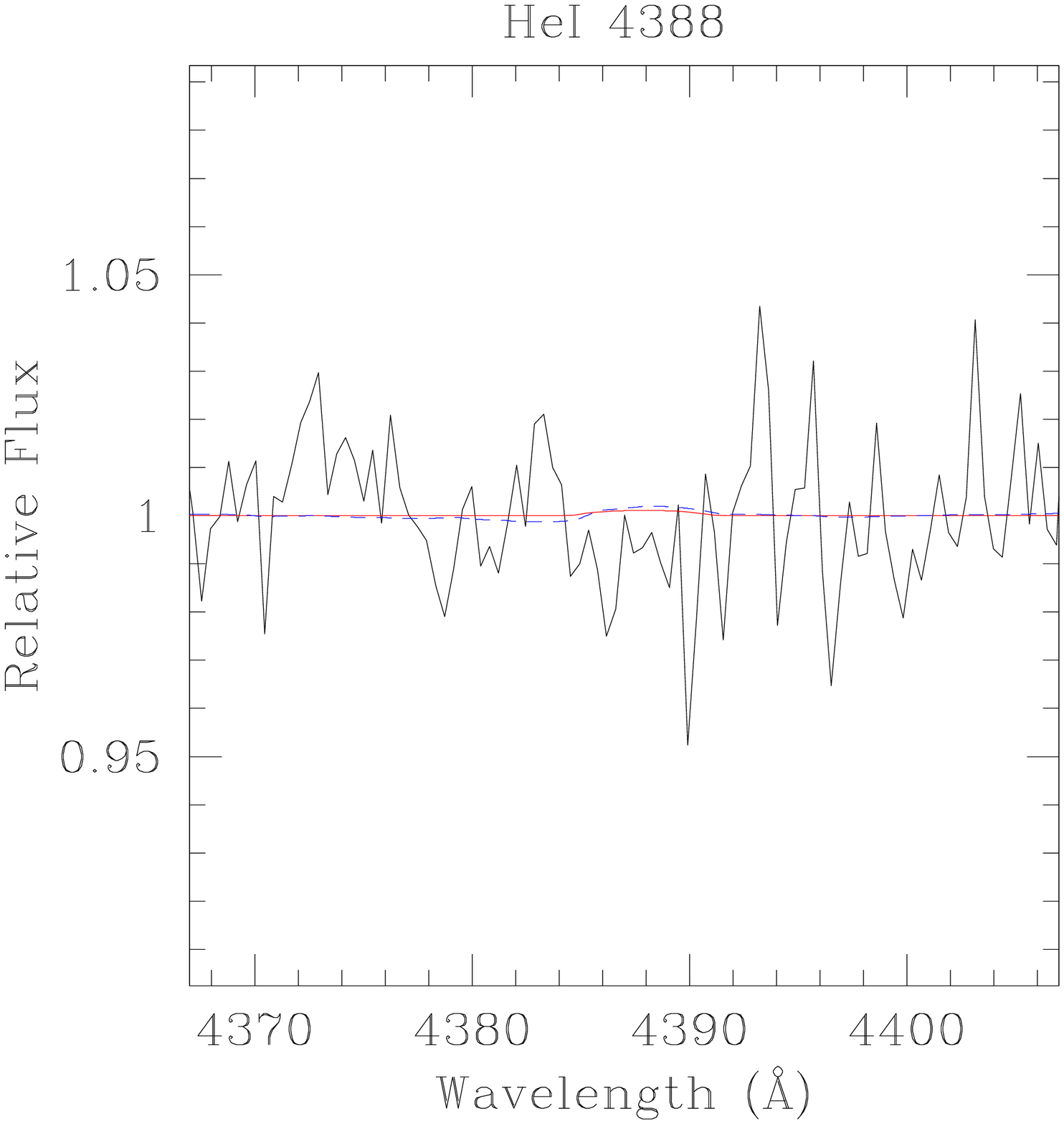}
\plotone{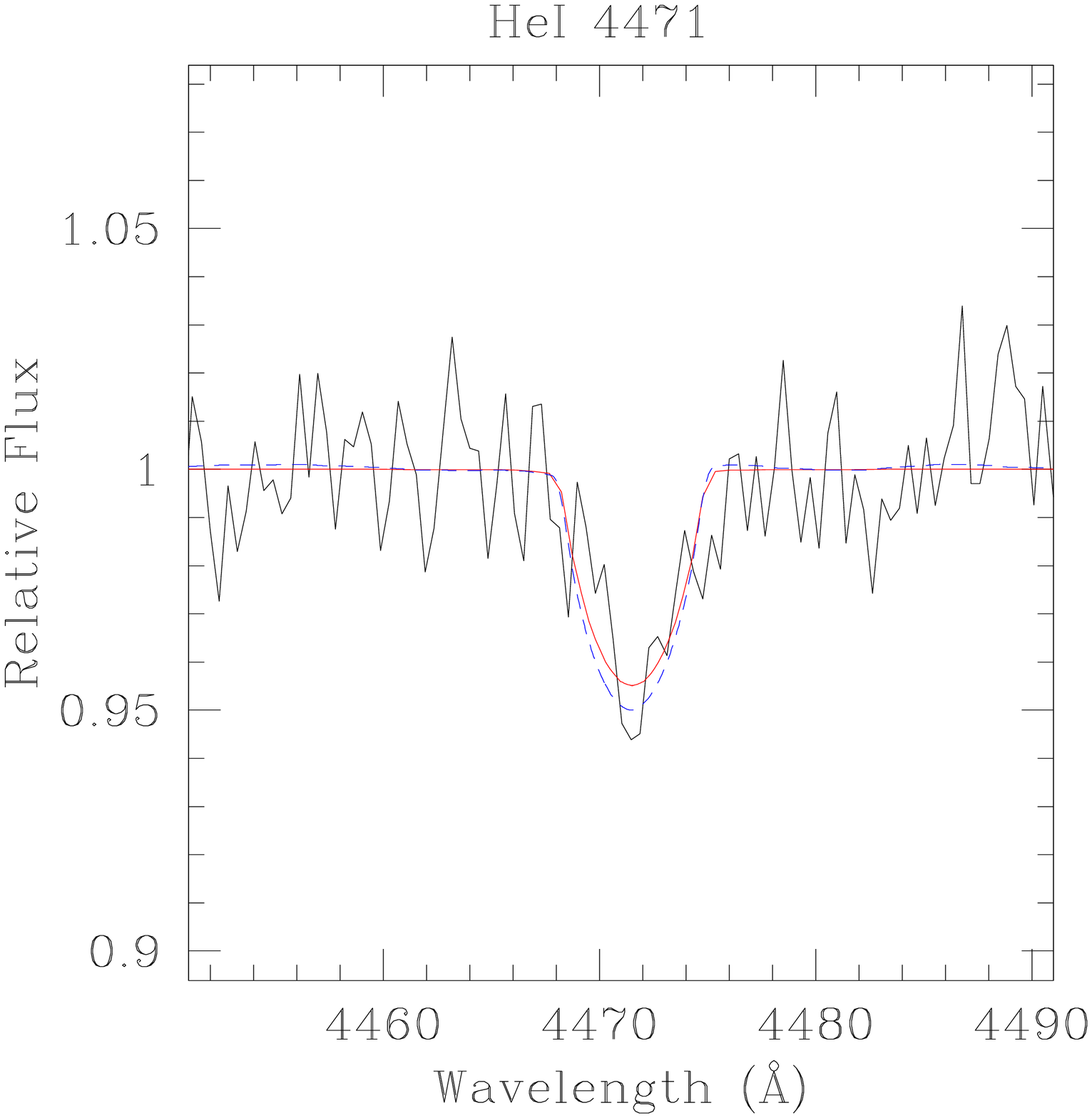}
\plotone{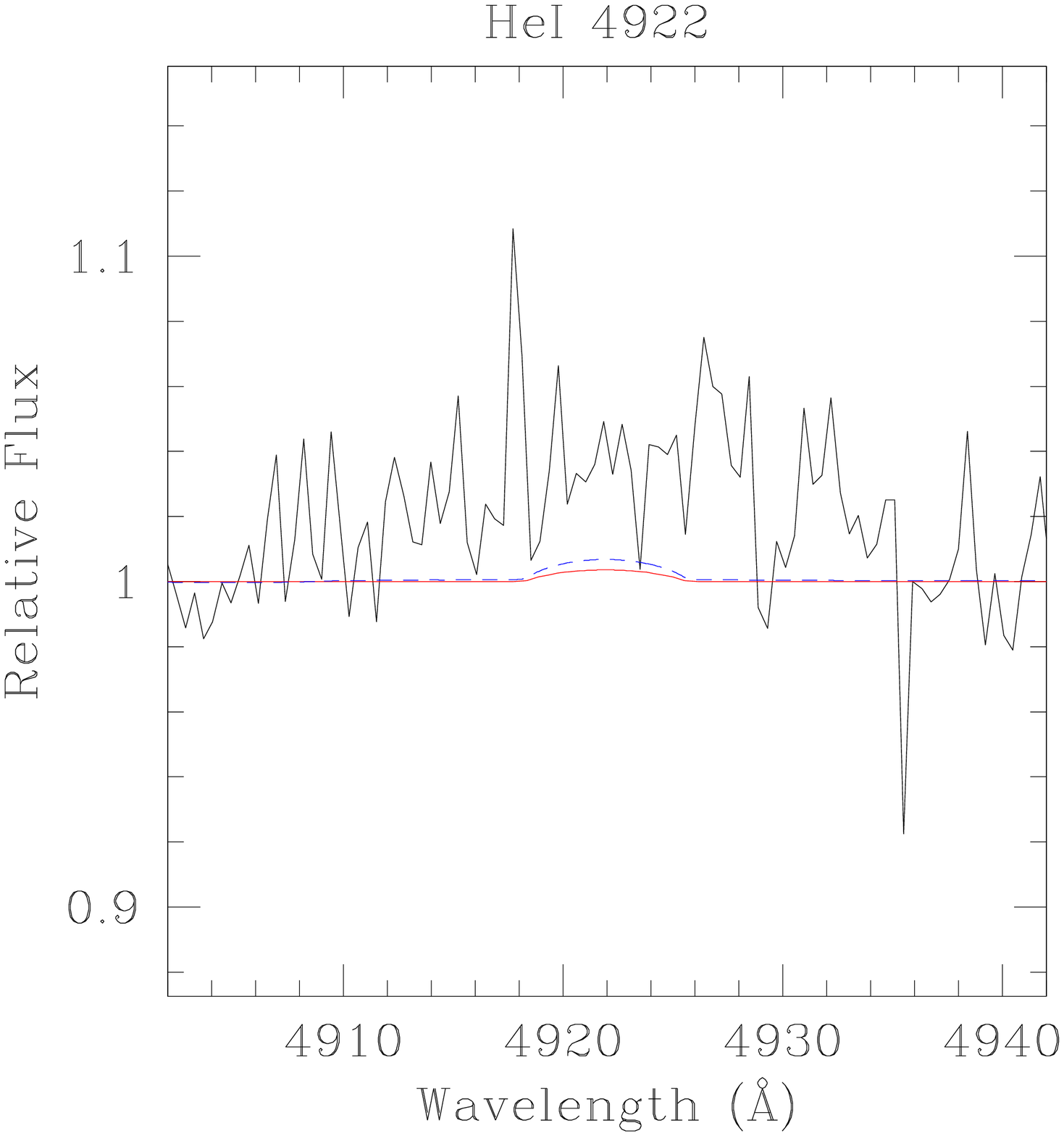}
\plotone{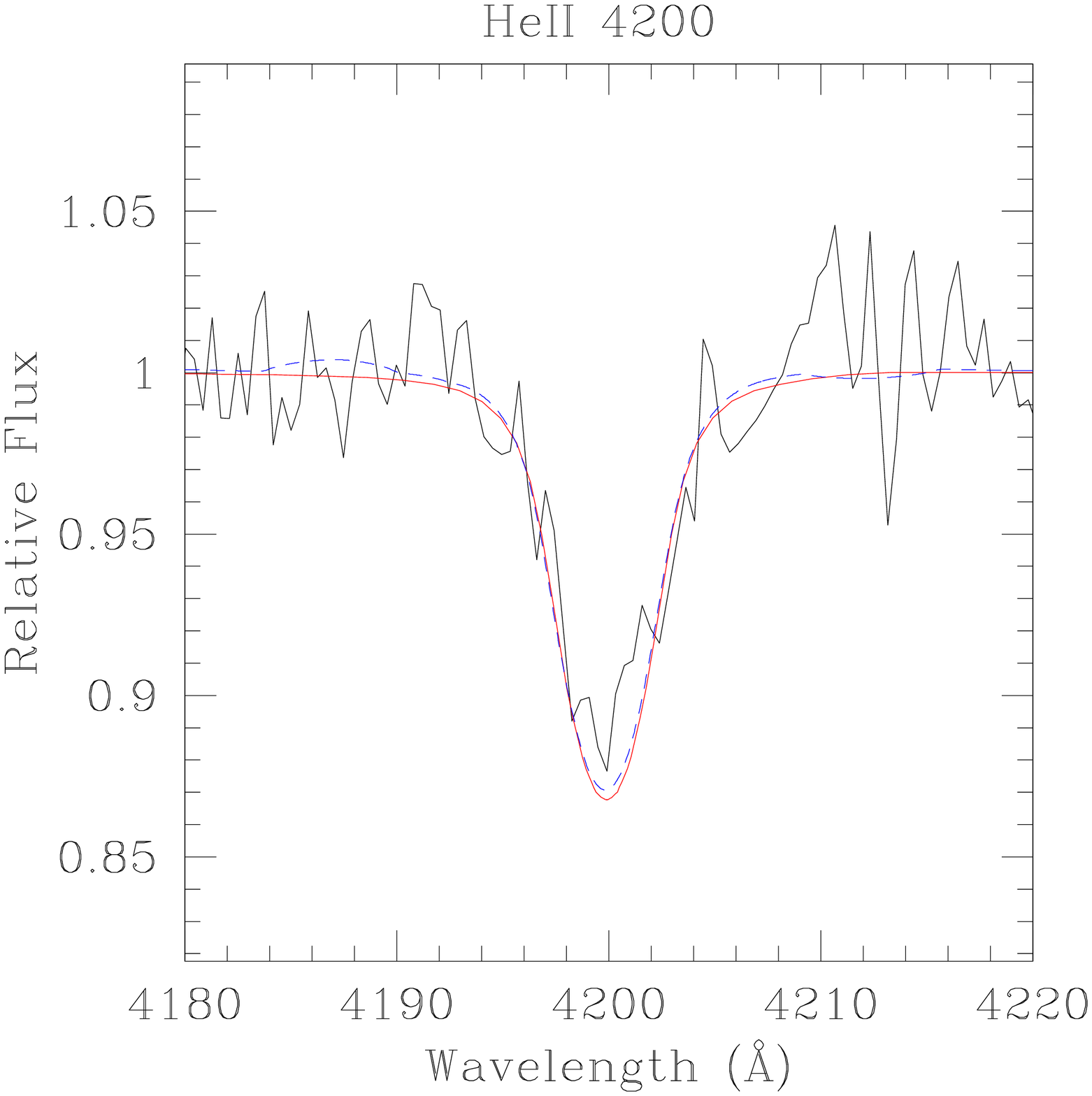}
\plotone{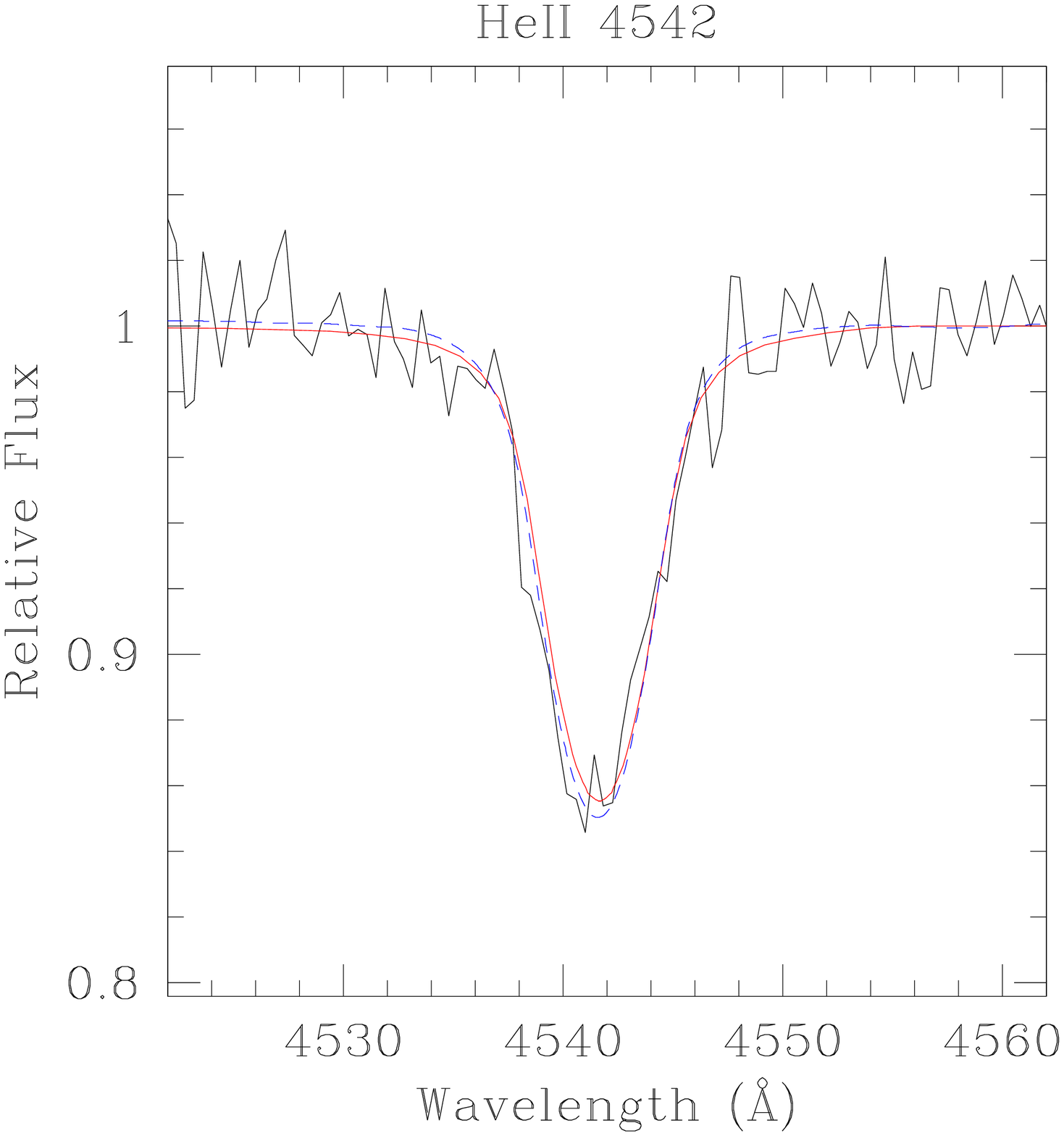}
\plotone{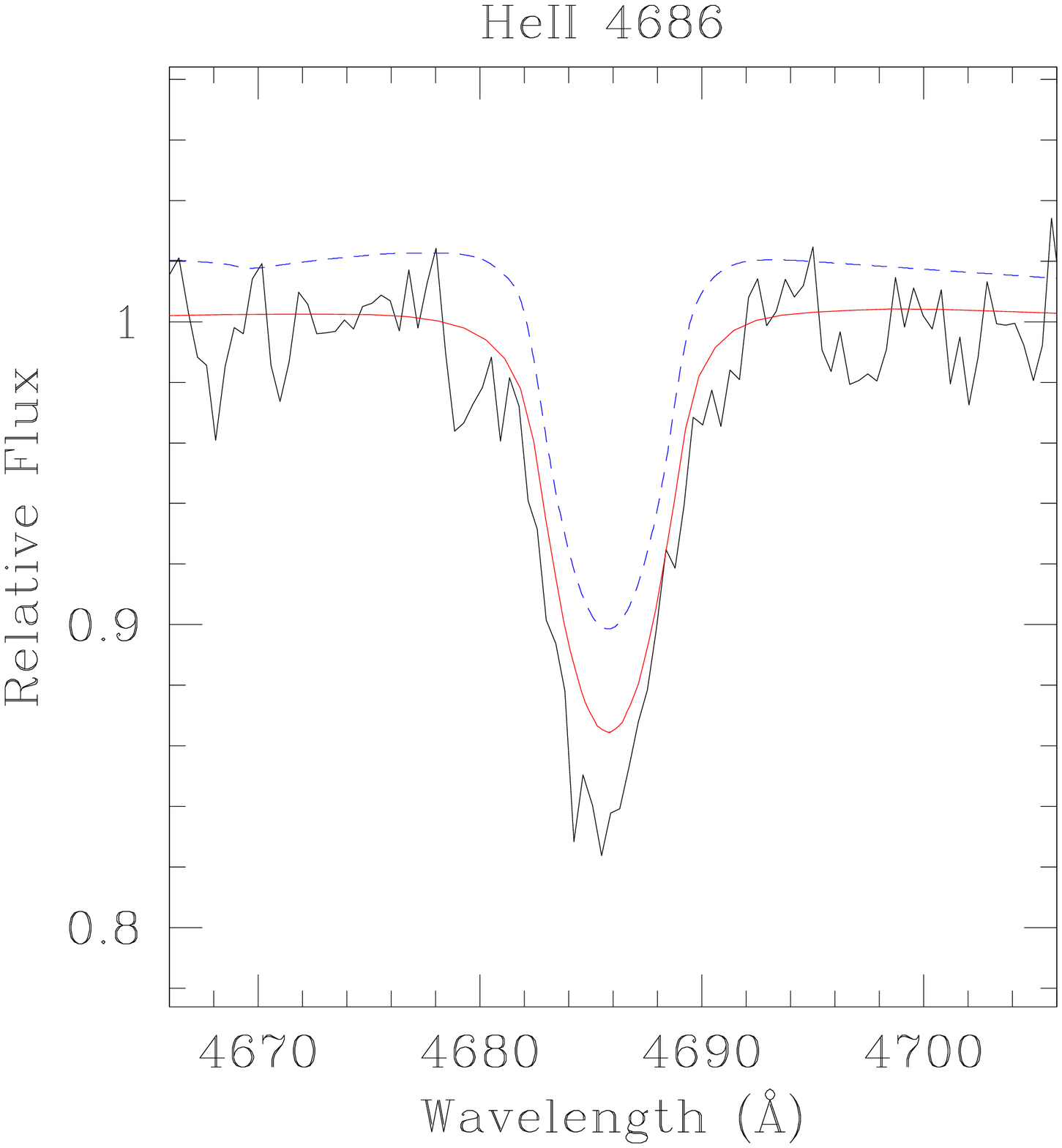}
\caption{\label{fig:AzV177} Model fits for AzV 177, an O4 V((f)) star in the SMC.  Black shows the observed spectrum, the red line shows the \fastwind\ fit, and the dashed blue line shows the \cmfgen\ fit. }
\end{figure}

\begin{figure}
\epsscale{0.3}
\plotone{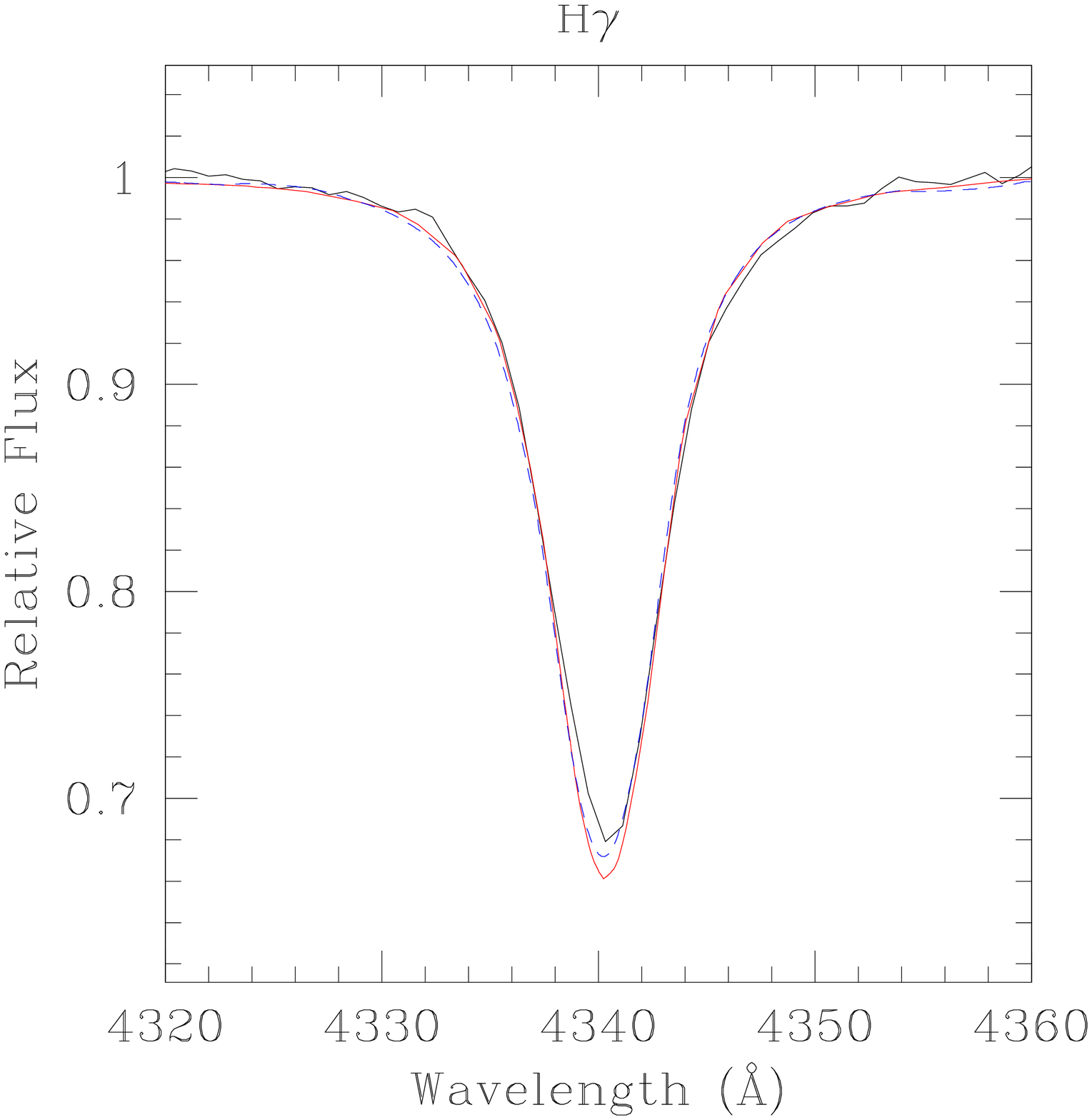}
\plotone{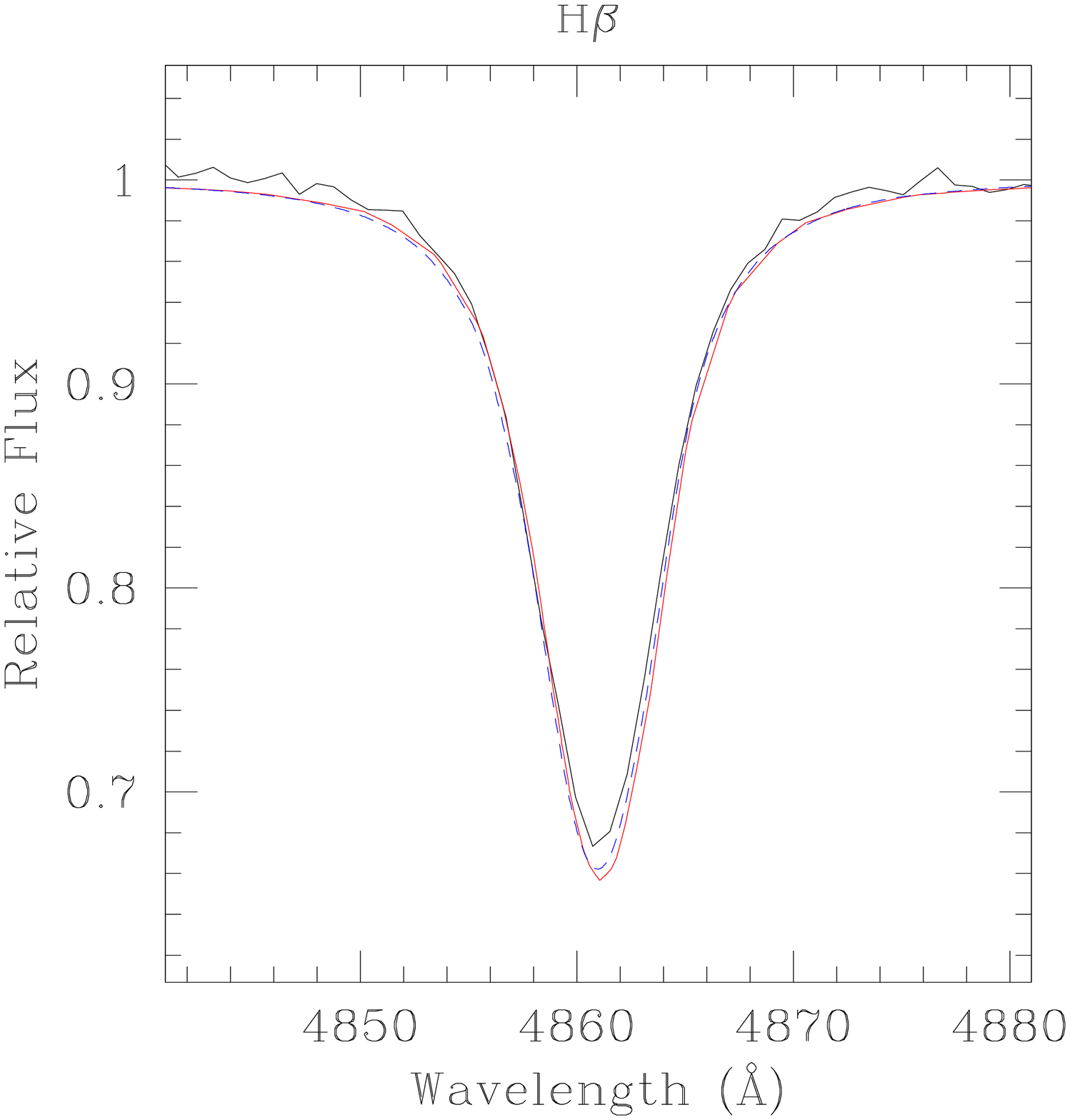}
\plotone{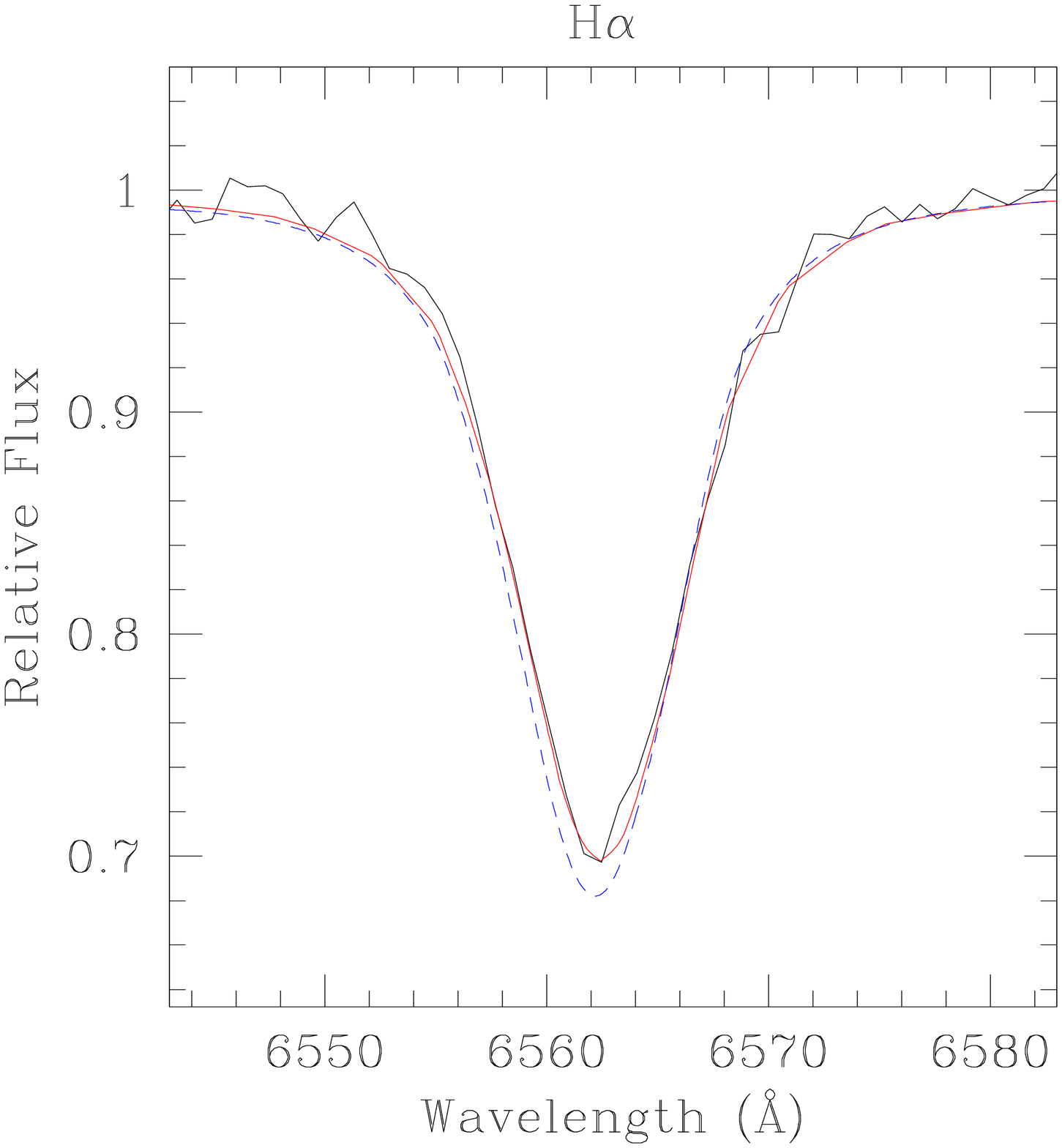}
\plotone{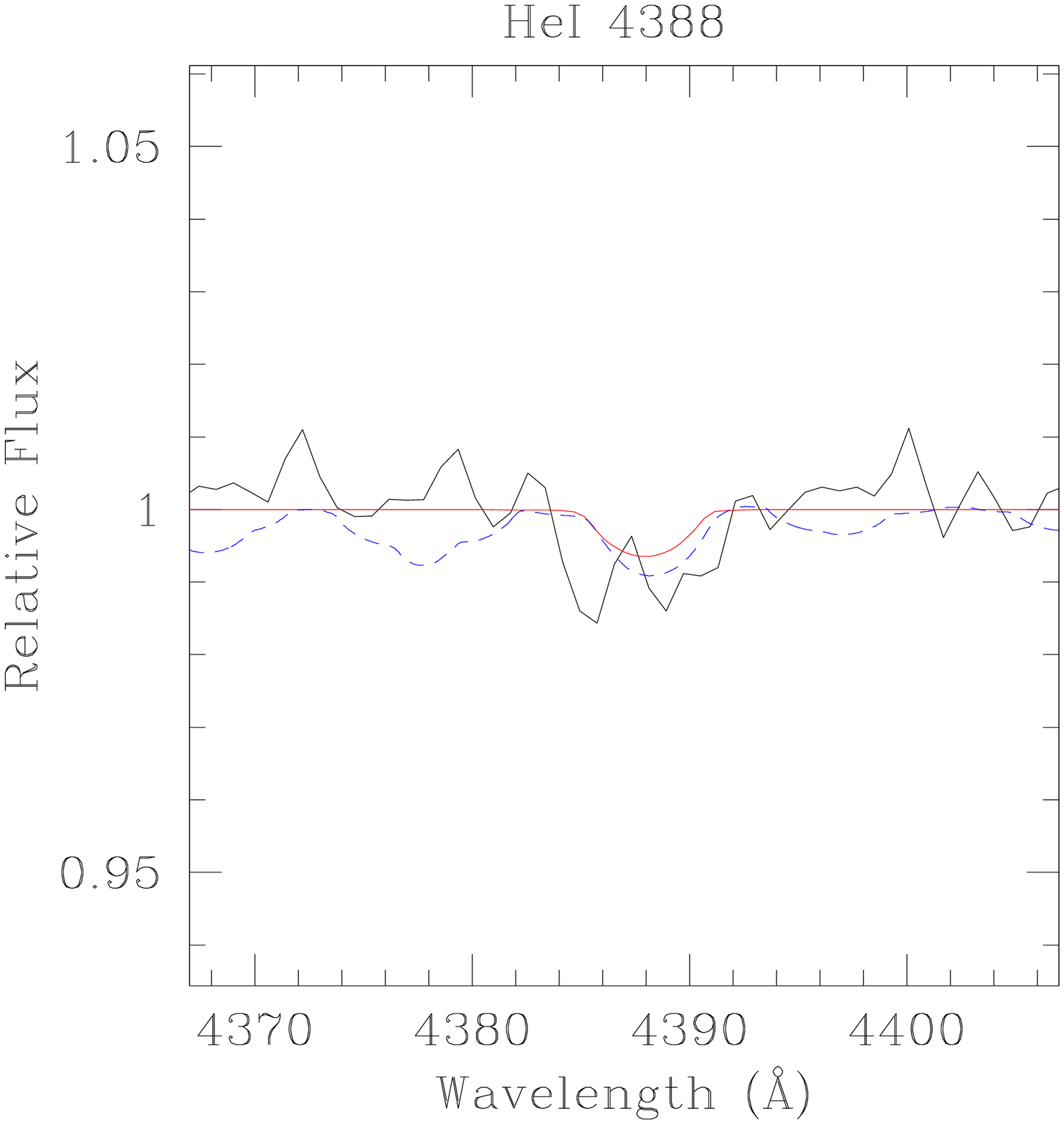}
\plotone{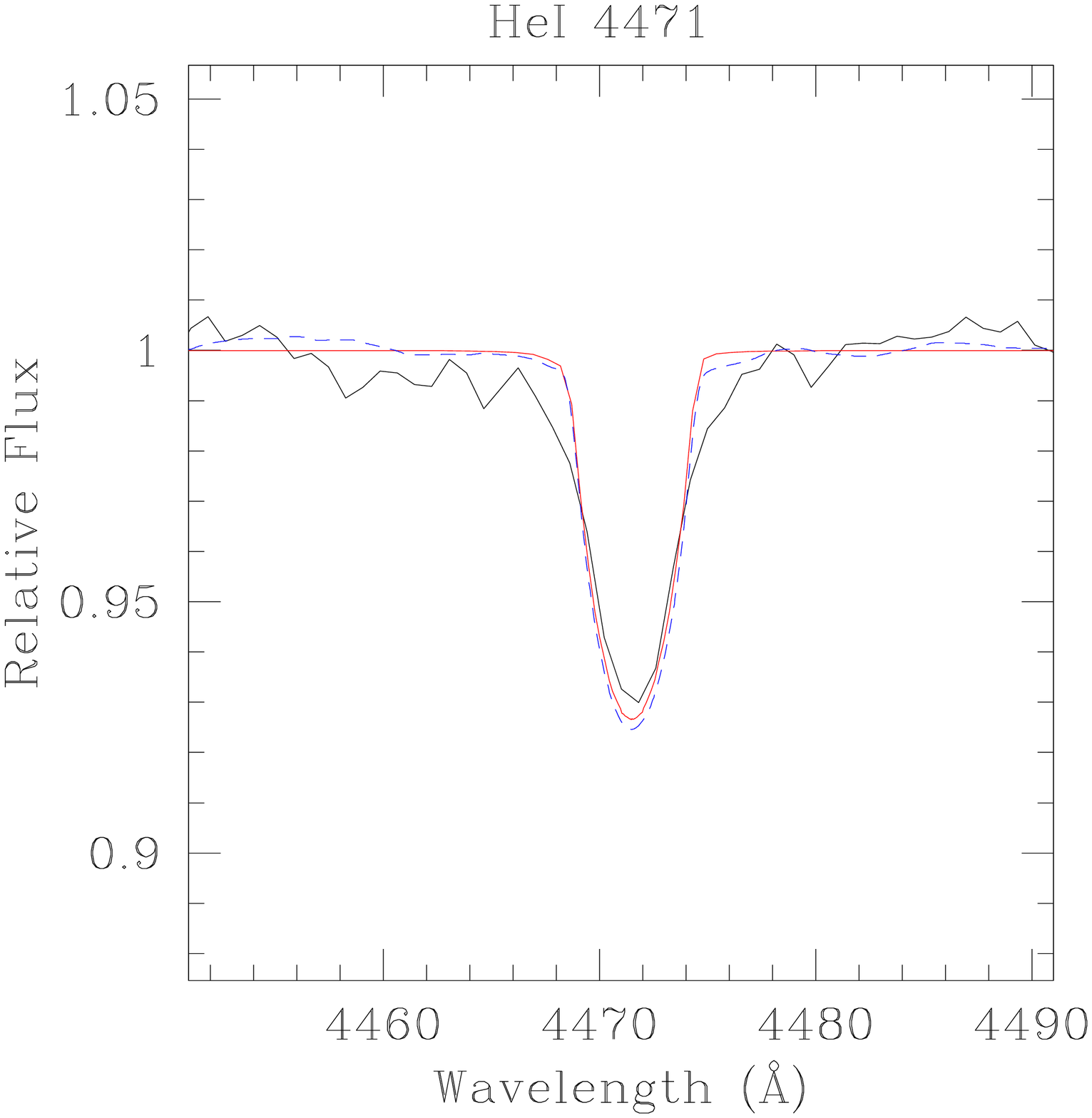}
\plotone{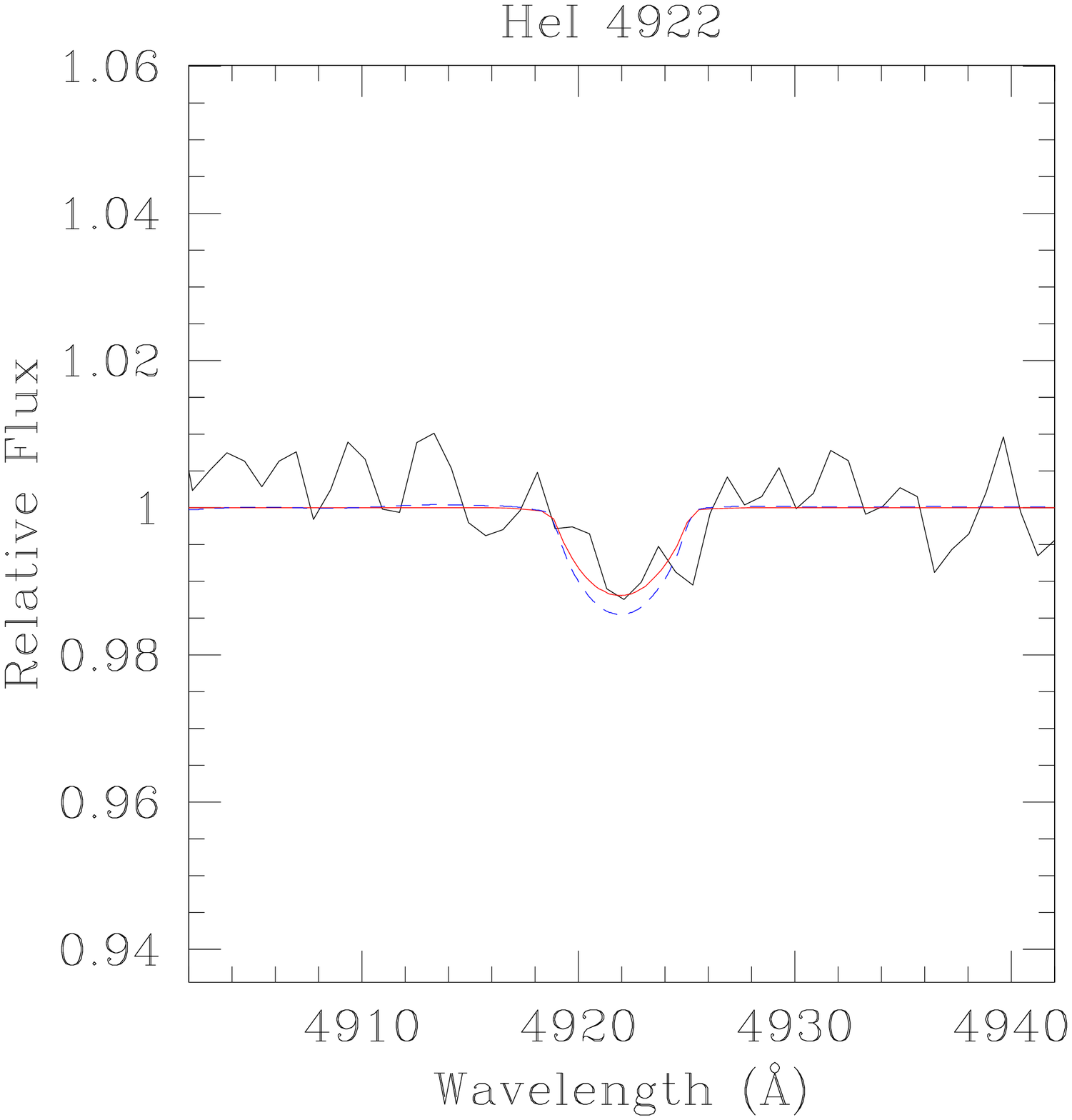}
\plotone{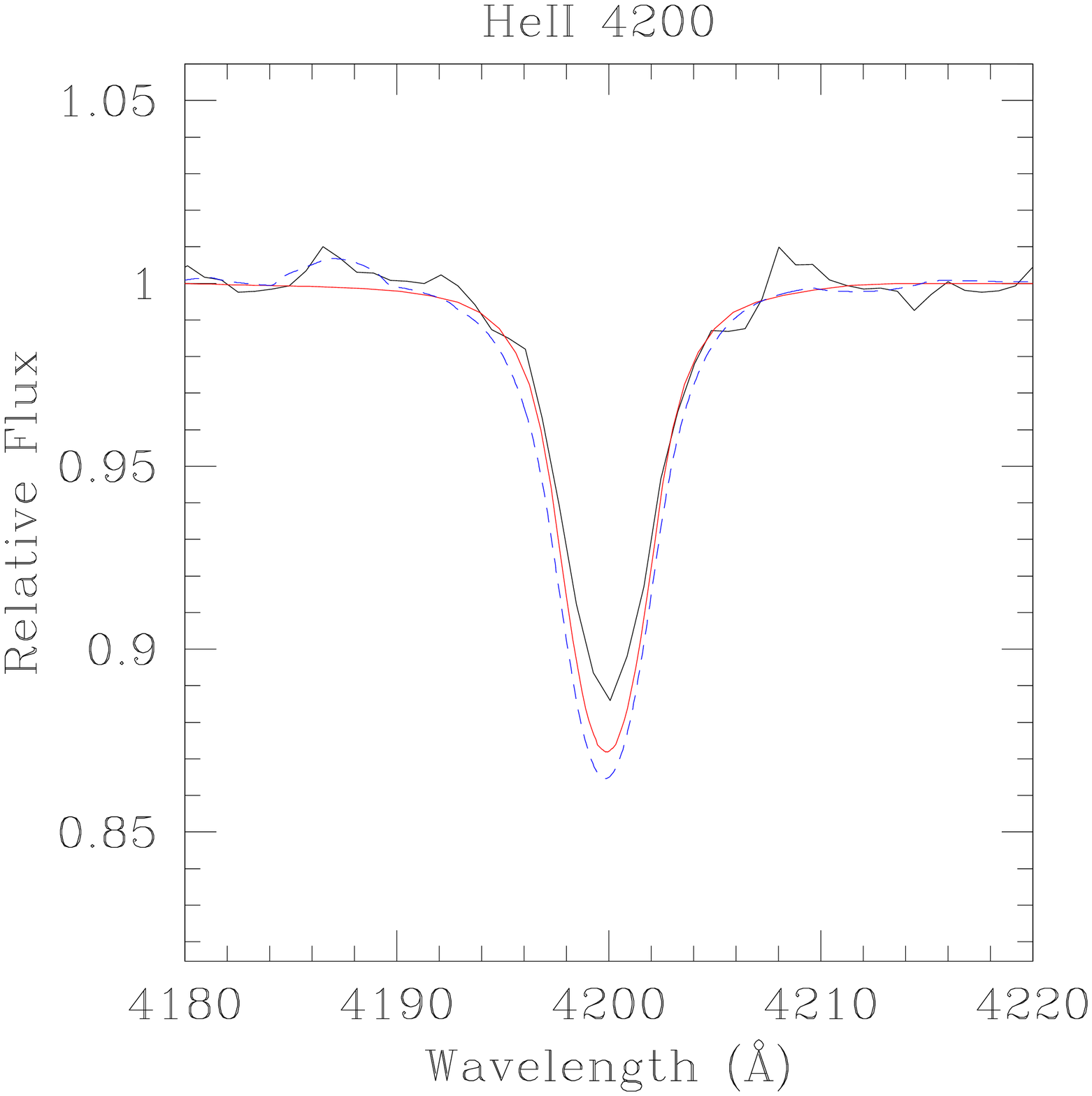}
\plotone{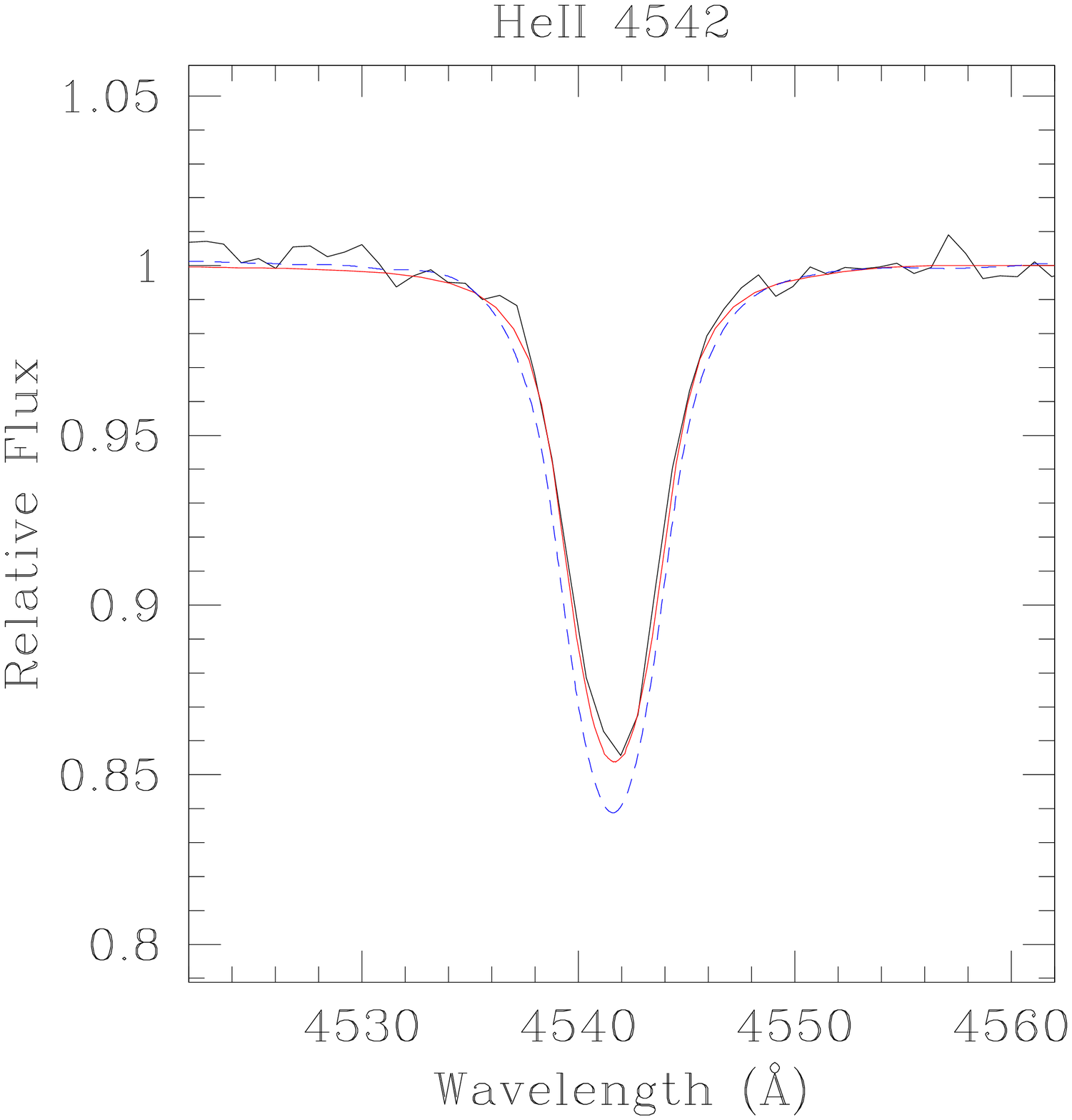}
\plotone{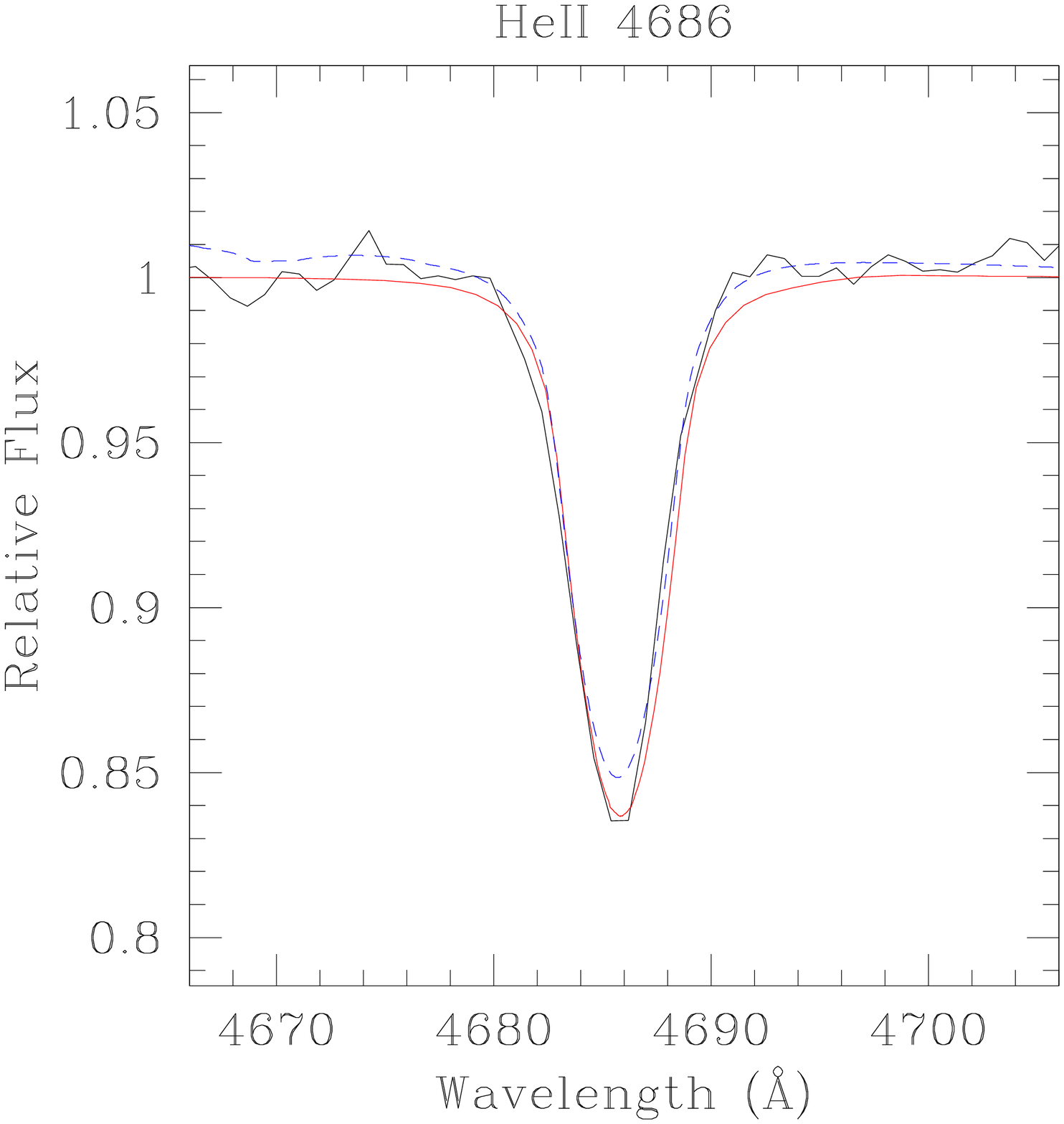}
\caption{\label{fig:AzV388} Model fits for AzV 388, an O5.5 V((f)) star in the SMC.  Black shows the observed spectrum, the red line shows the \fastwind\ fit, and the dashed blue line shows the \cmfgen\ fit. }
\end{figure}

\begin{figure}
\epsscale{0.3}
\plotone{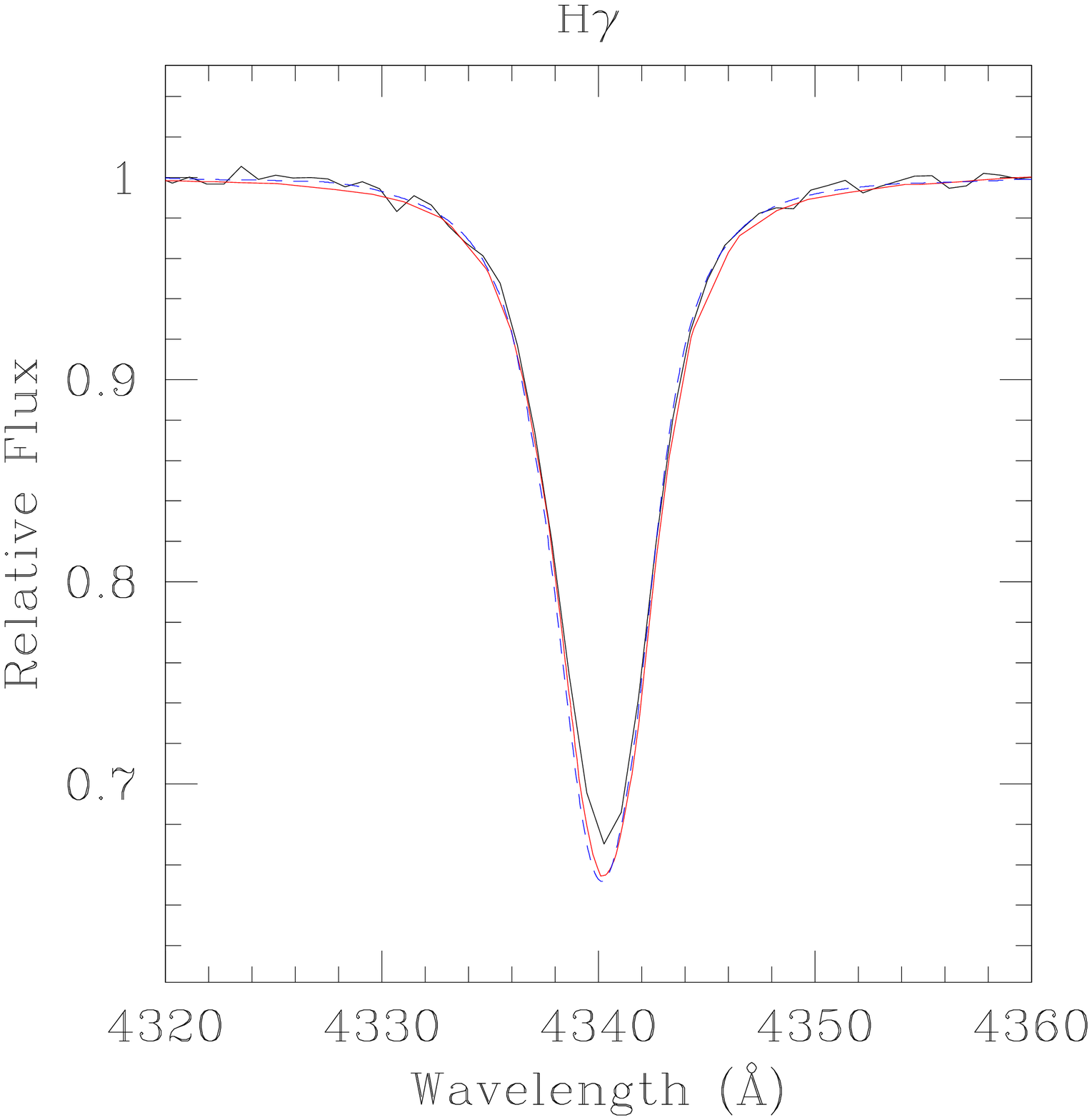}
\plotone{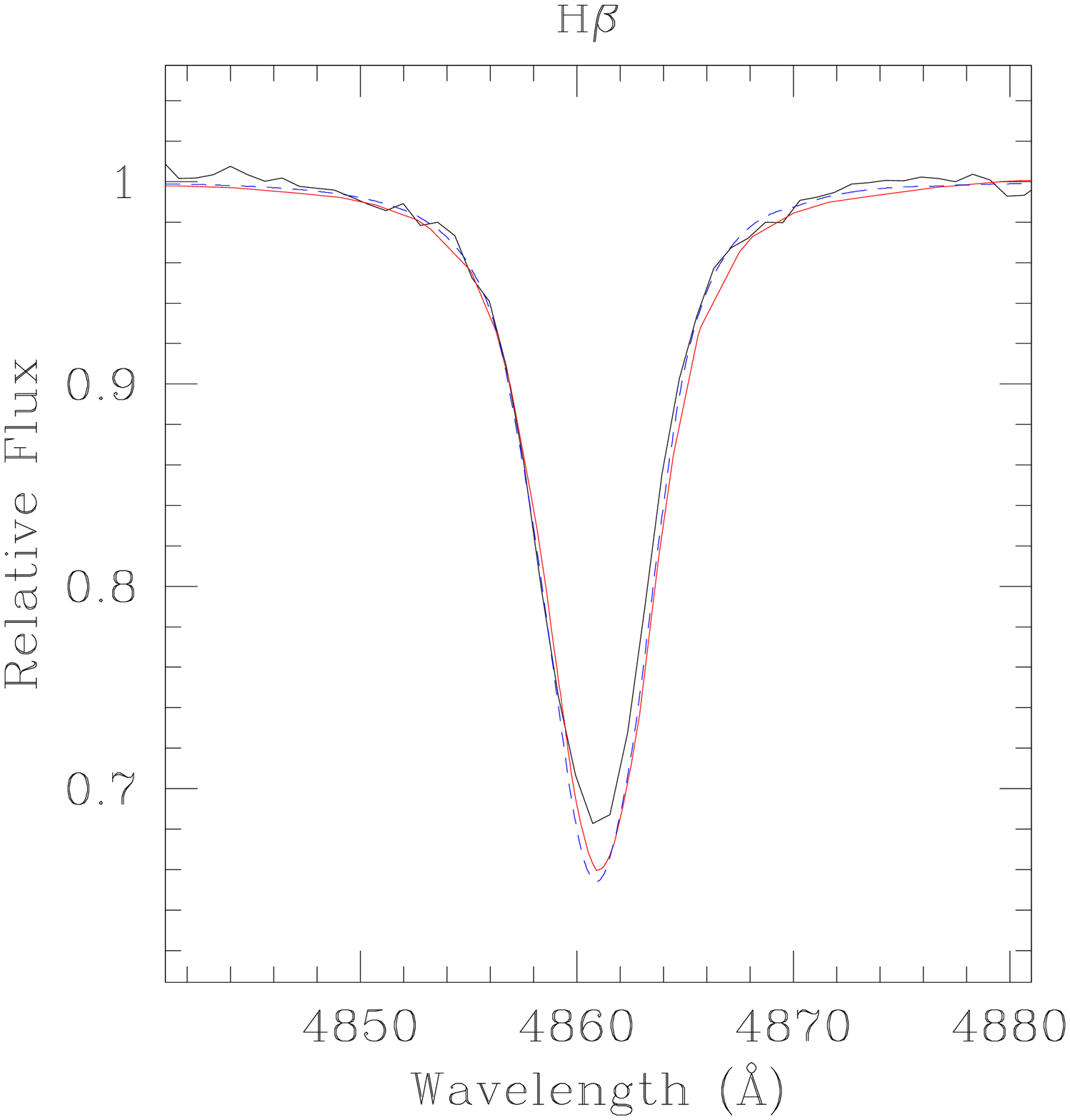}
\plotone{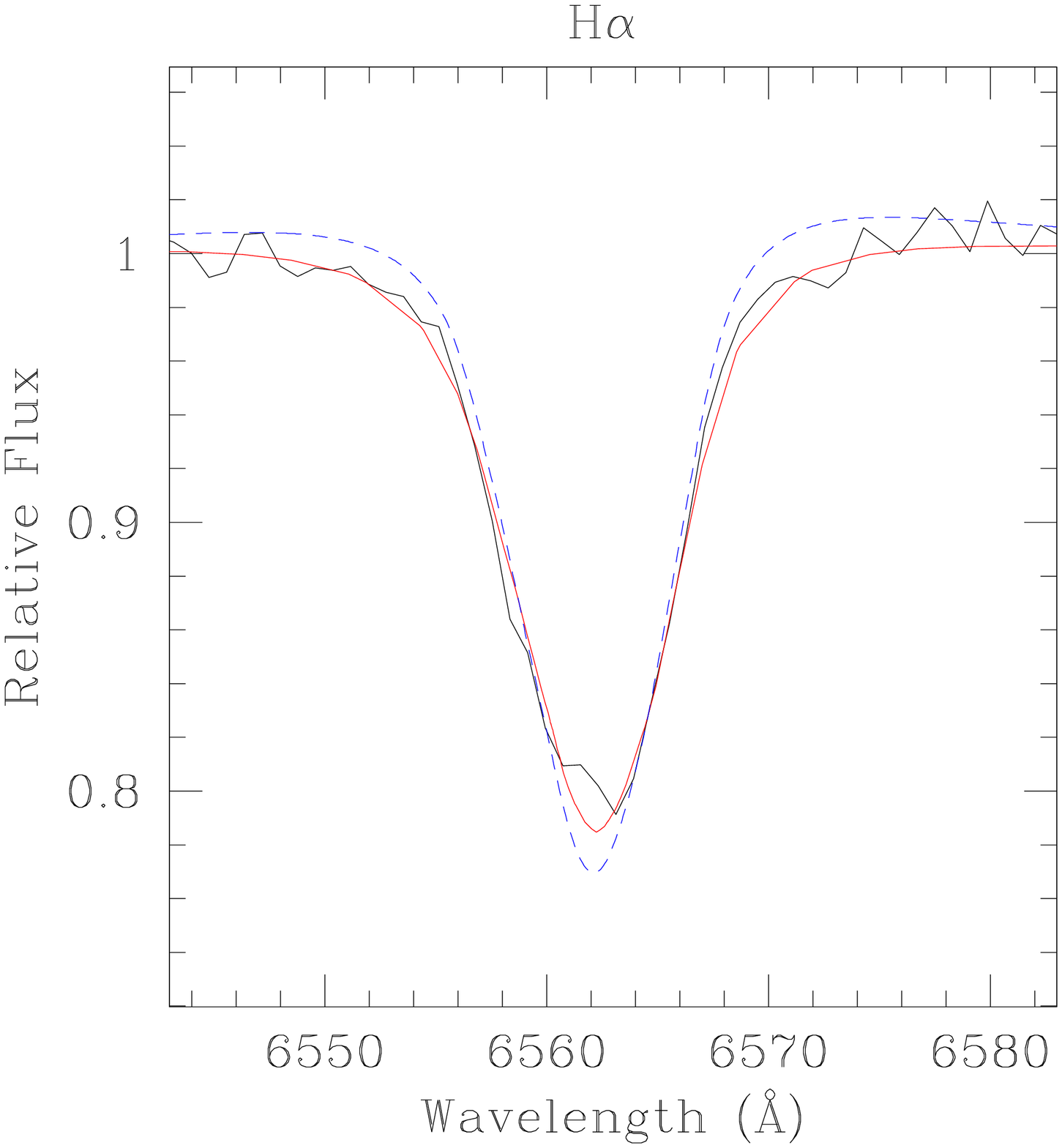}
\plotone{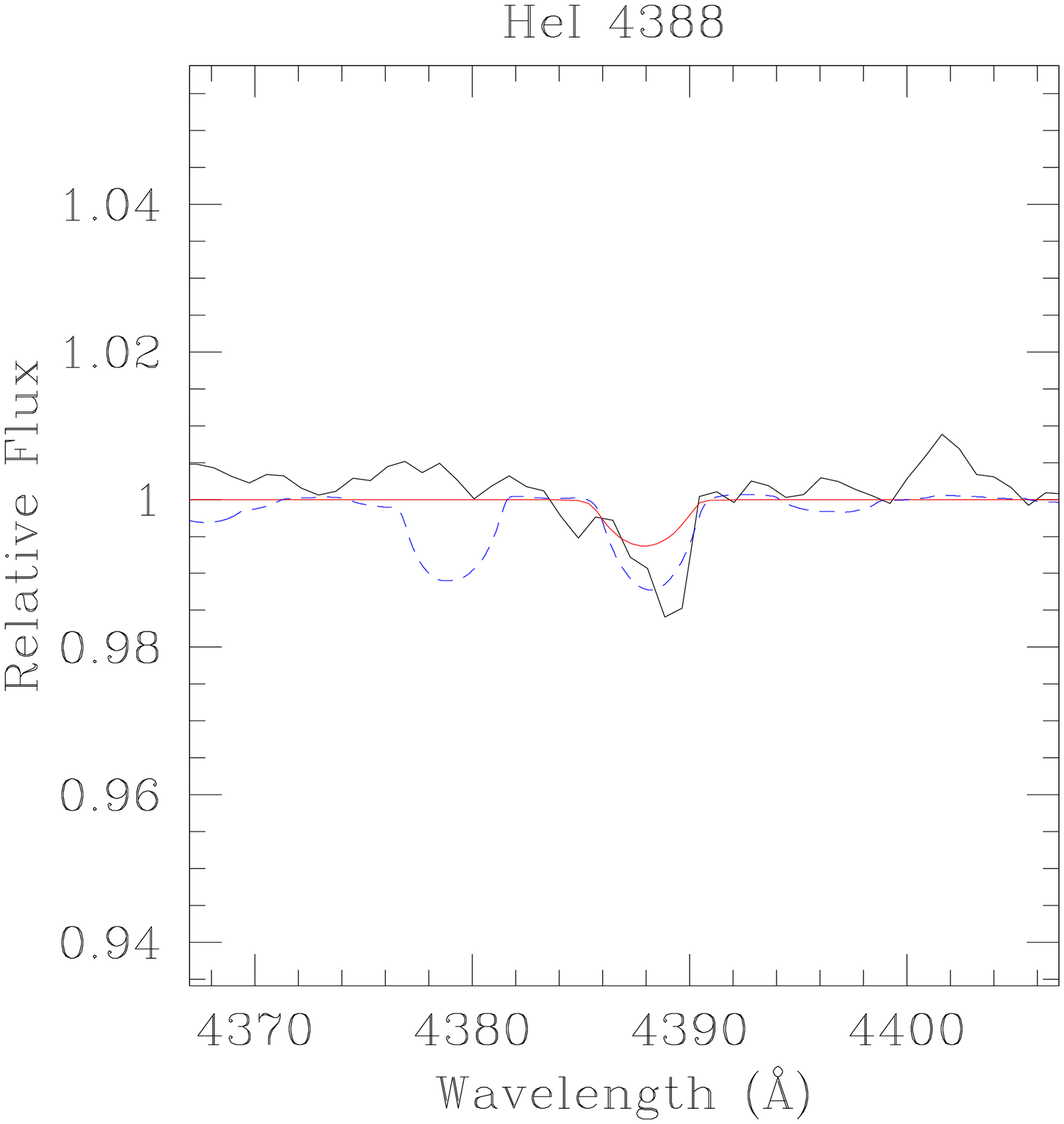}
\plotone{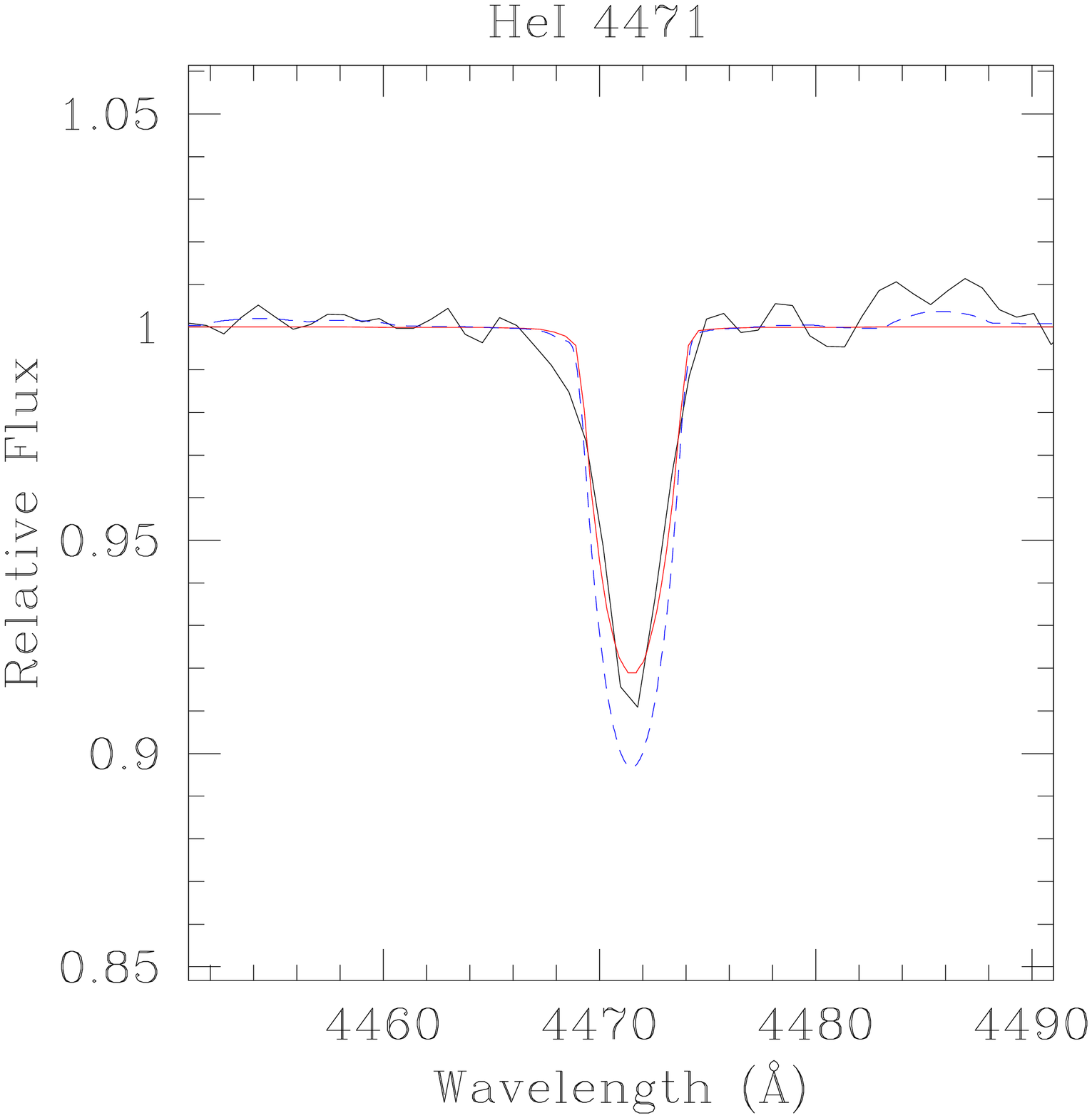}
\plotone{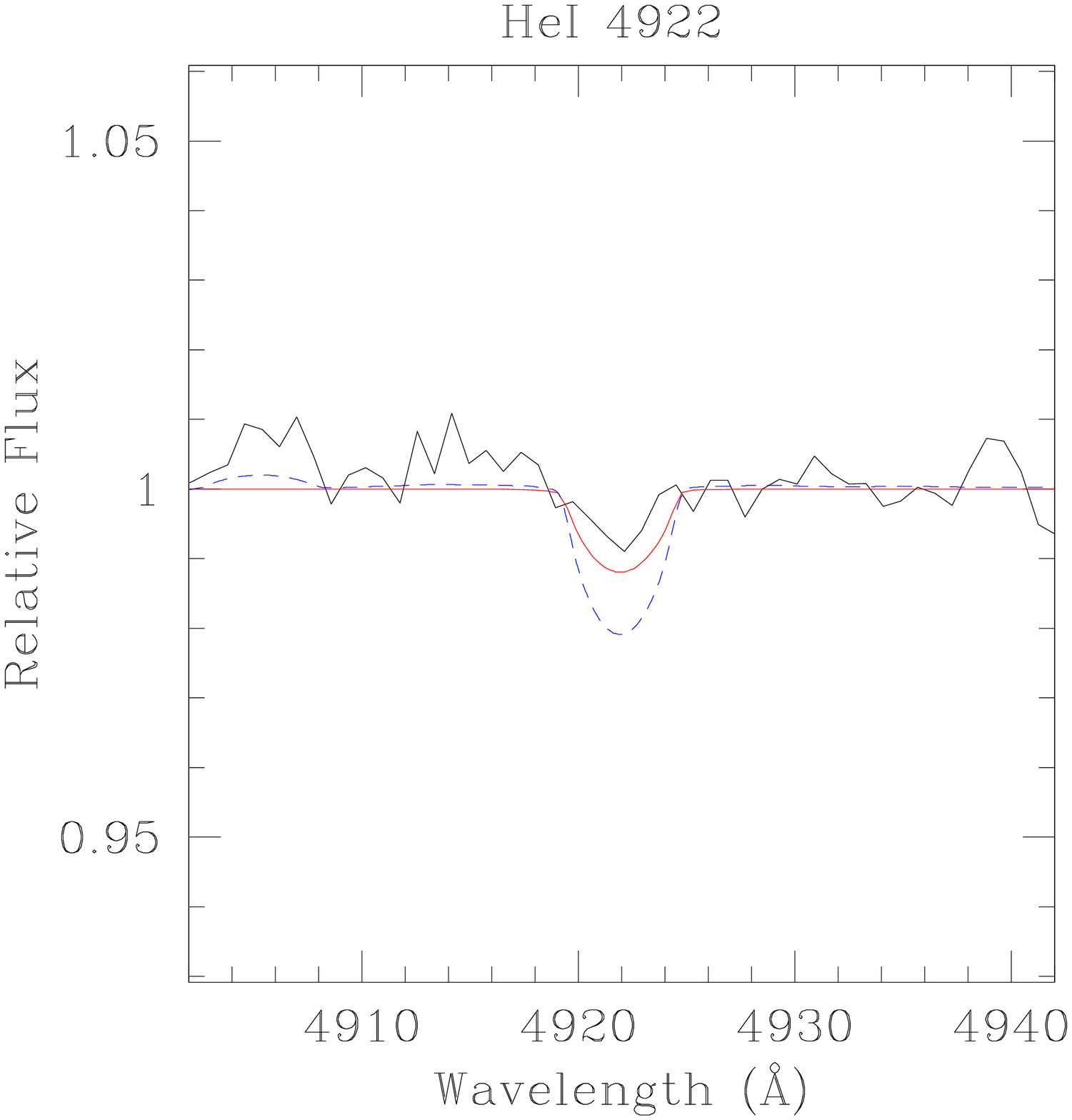}
\plotone{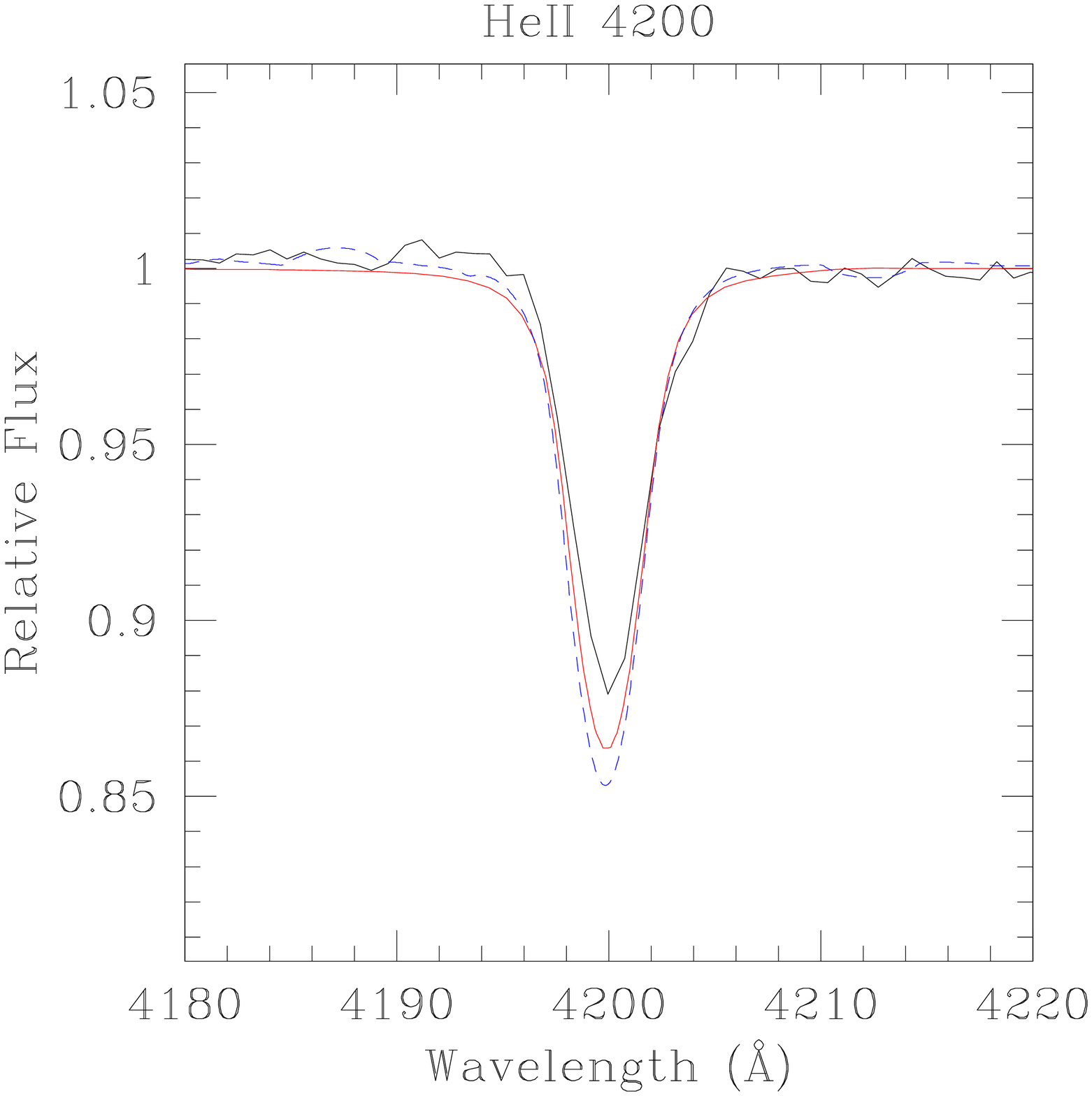}
\plotone{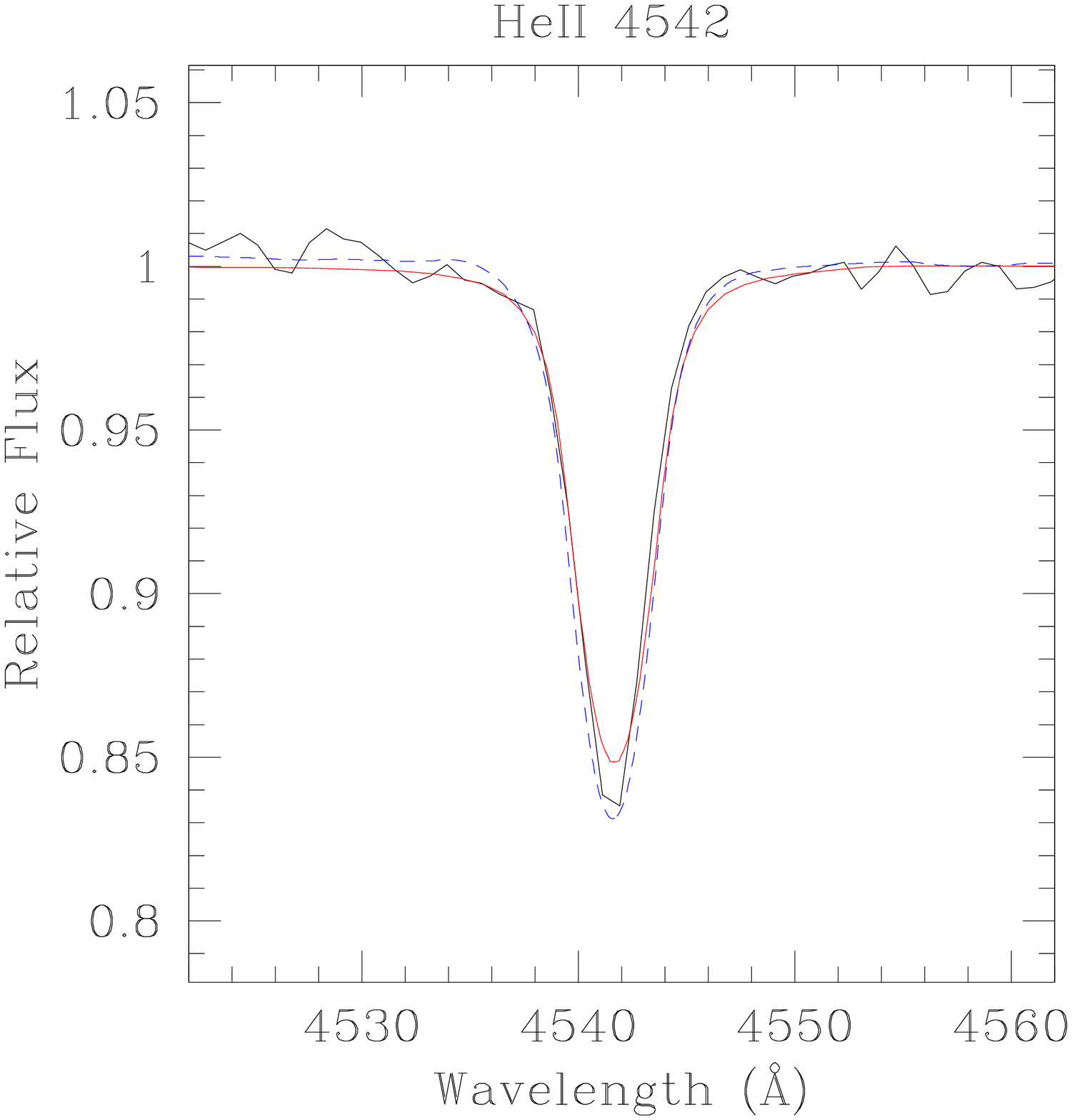}
\plotone{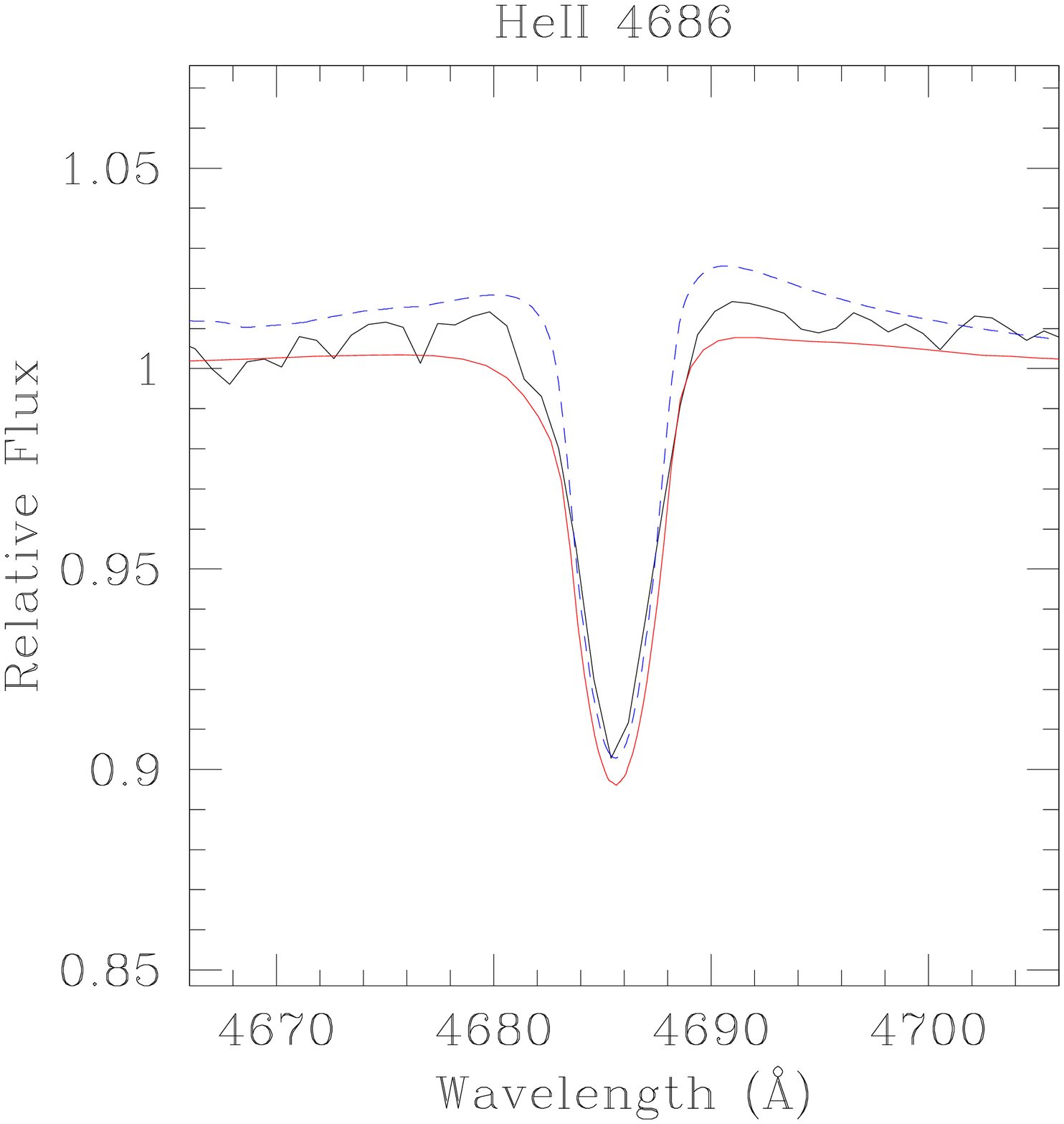}
\caption{\label{fig:AzV75} Model fits for AzV 75, an O5.5 I(f) star in the SMC.  Black shows the observed spectrum, the red line shows the \fastwind\ fit, and the dashed blue line shows the \cmfgen\ fit. }
\end{figure}

\begin{figure}
\epsscale{0.3}
\plotone{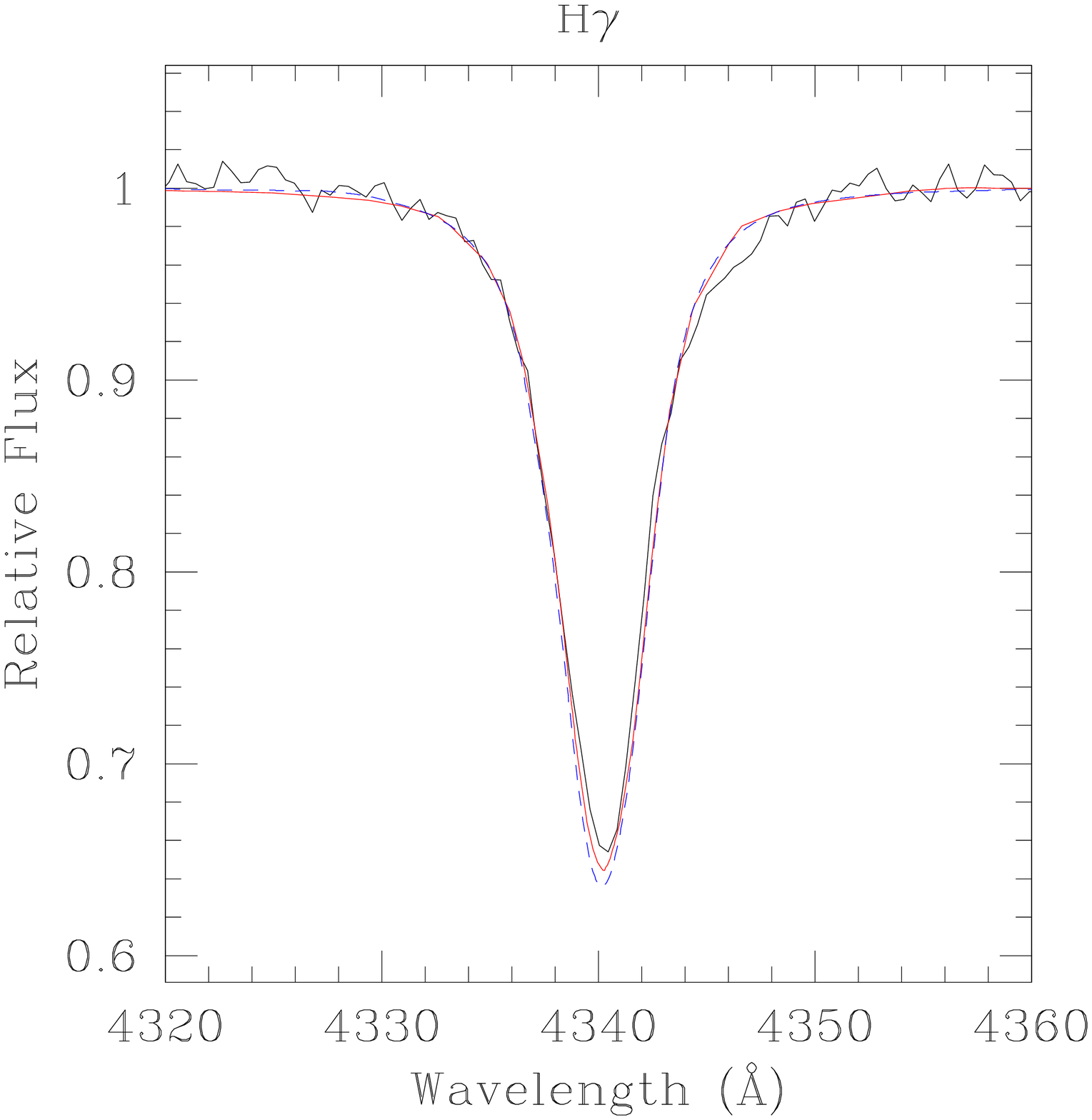}
\plotone{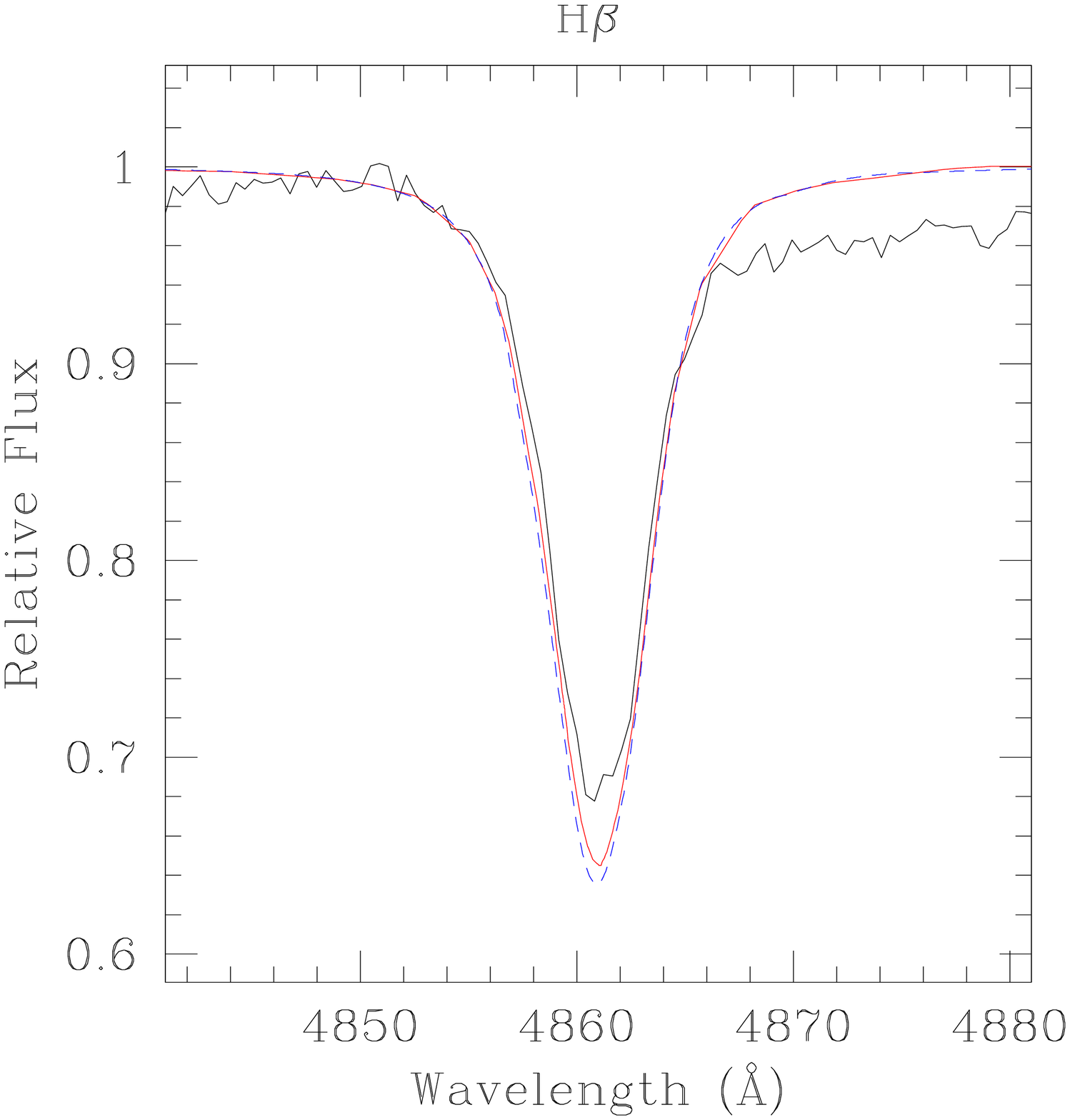}
\plotone{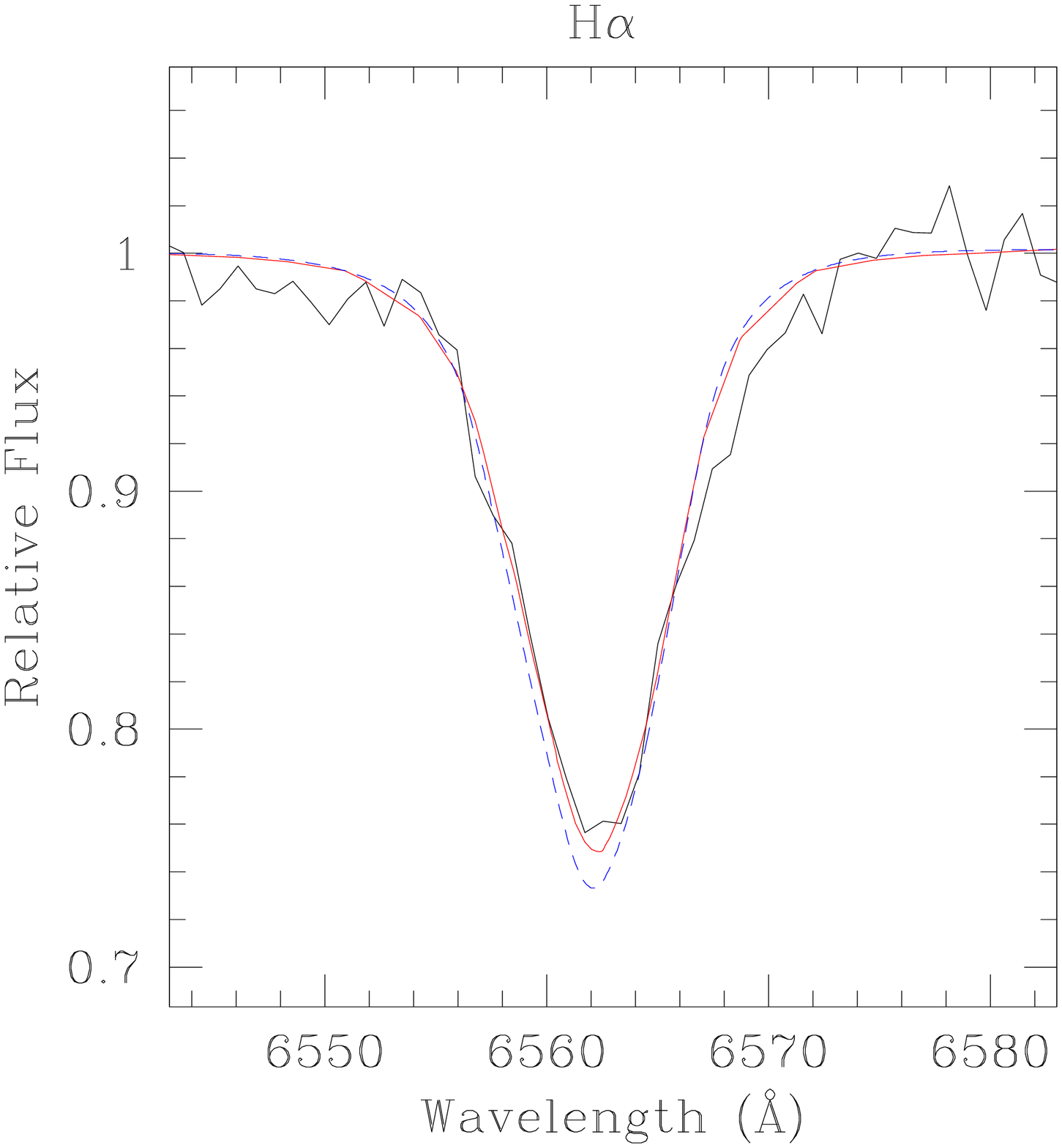}
\plotone{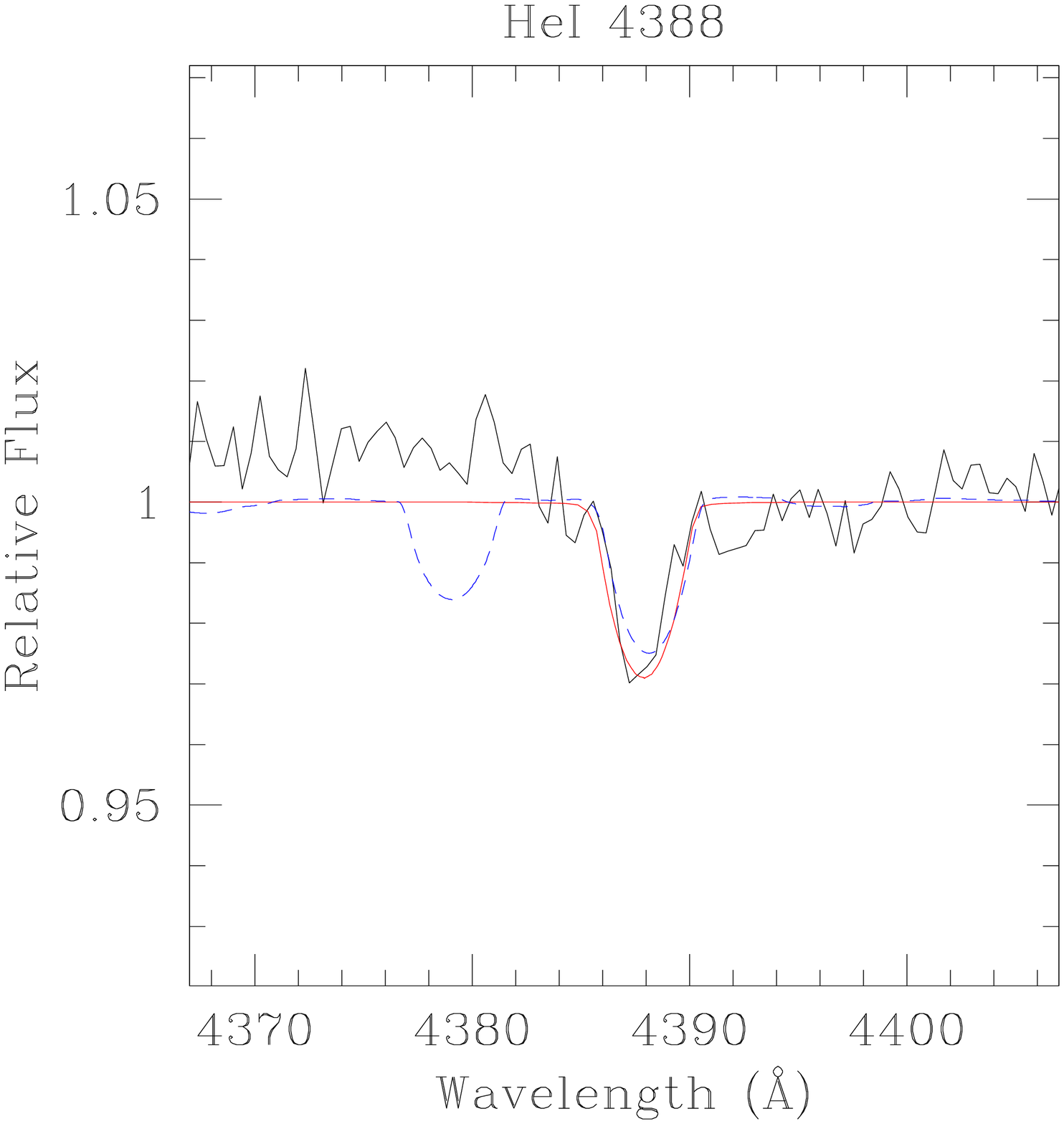}
\plotone{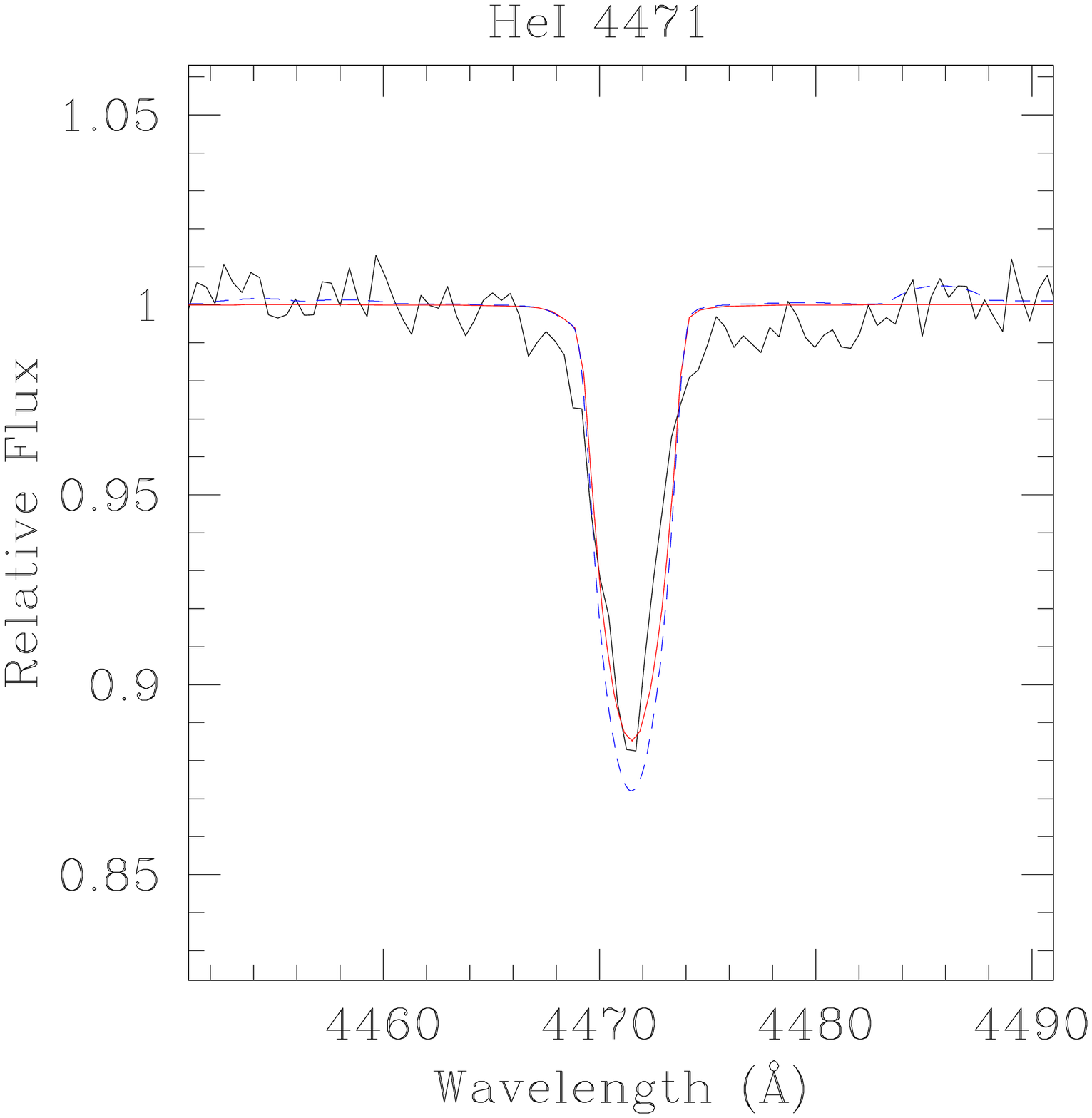}
\plotone{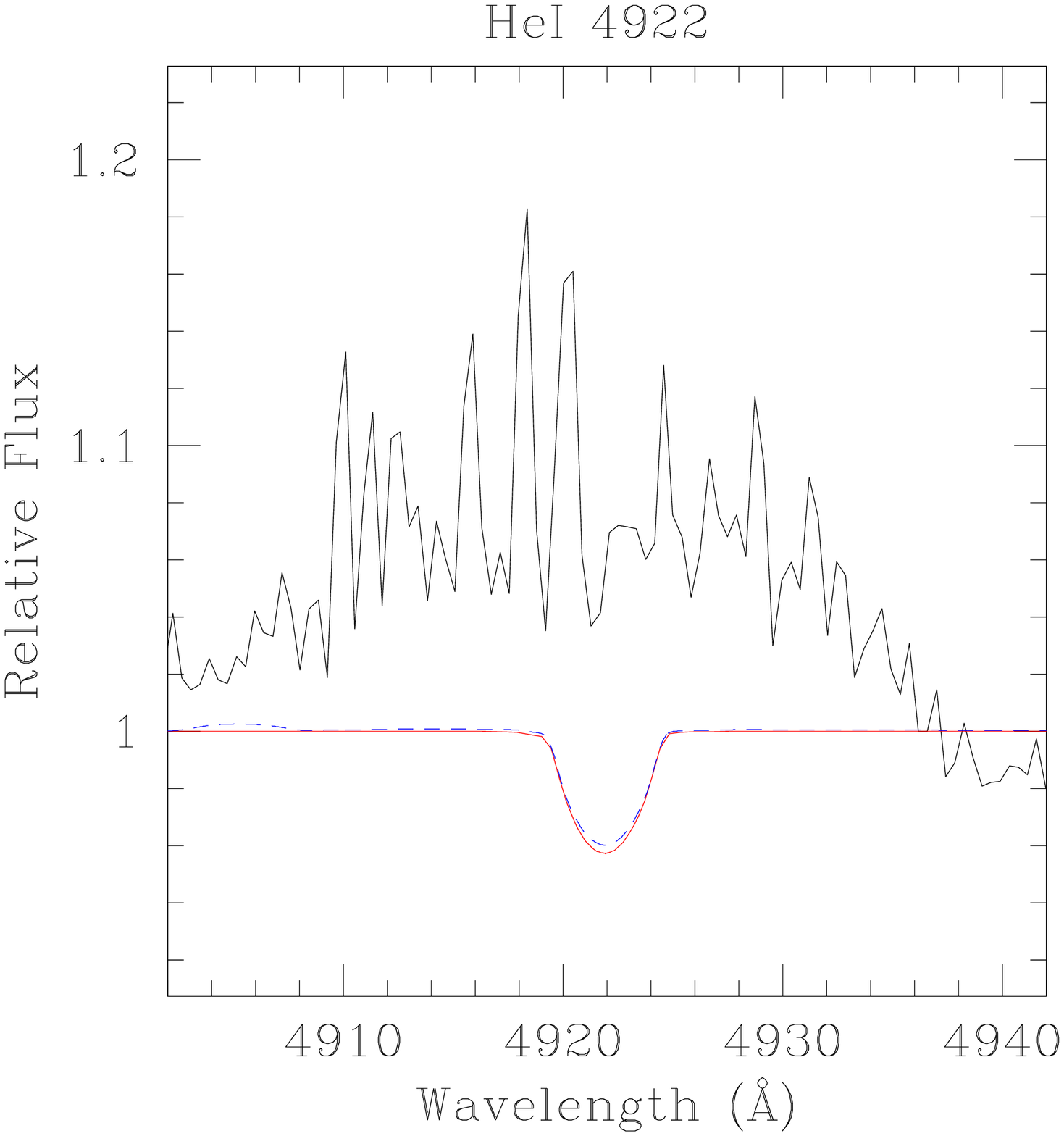}
\plotone{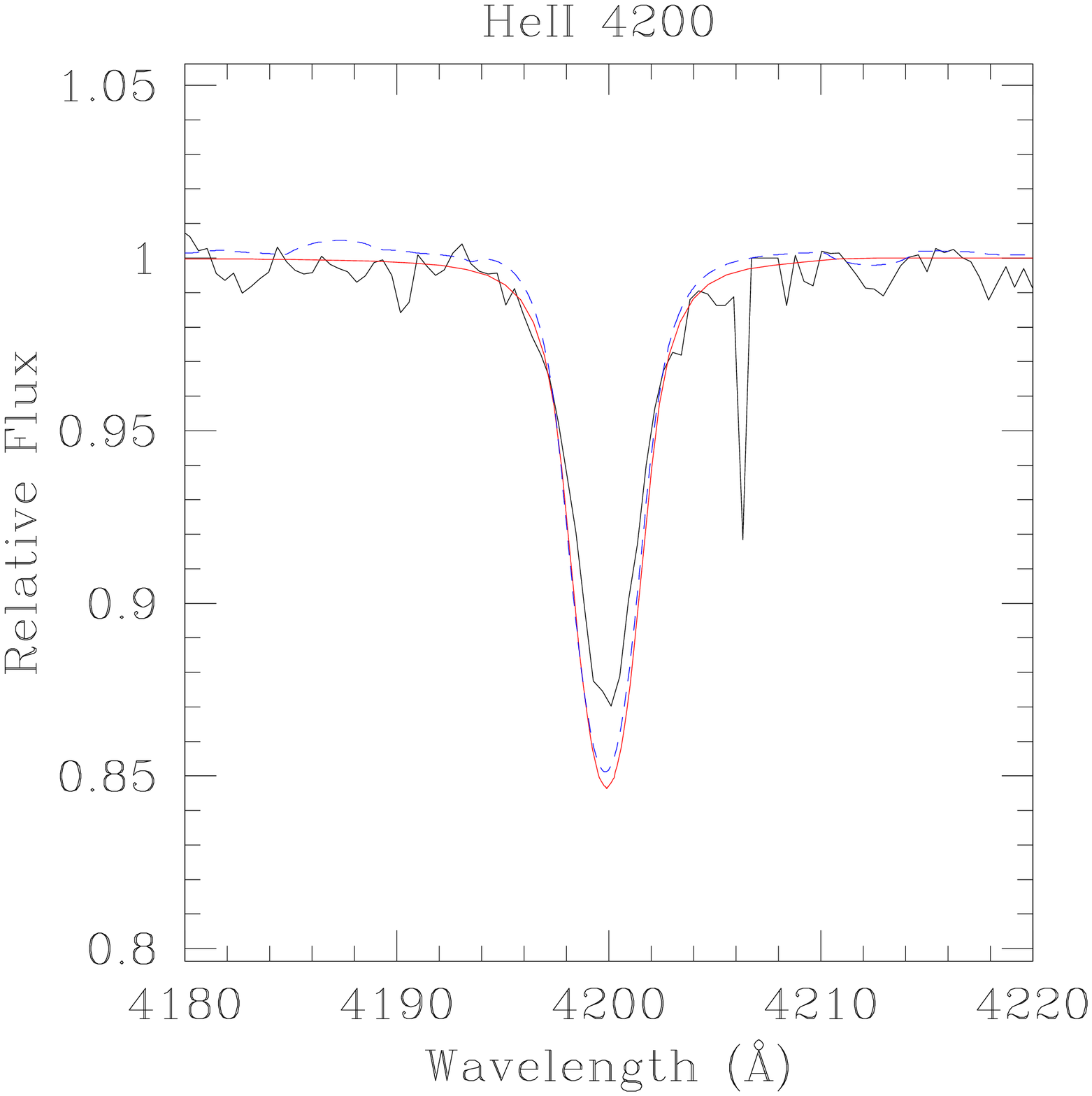}
\plotone{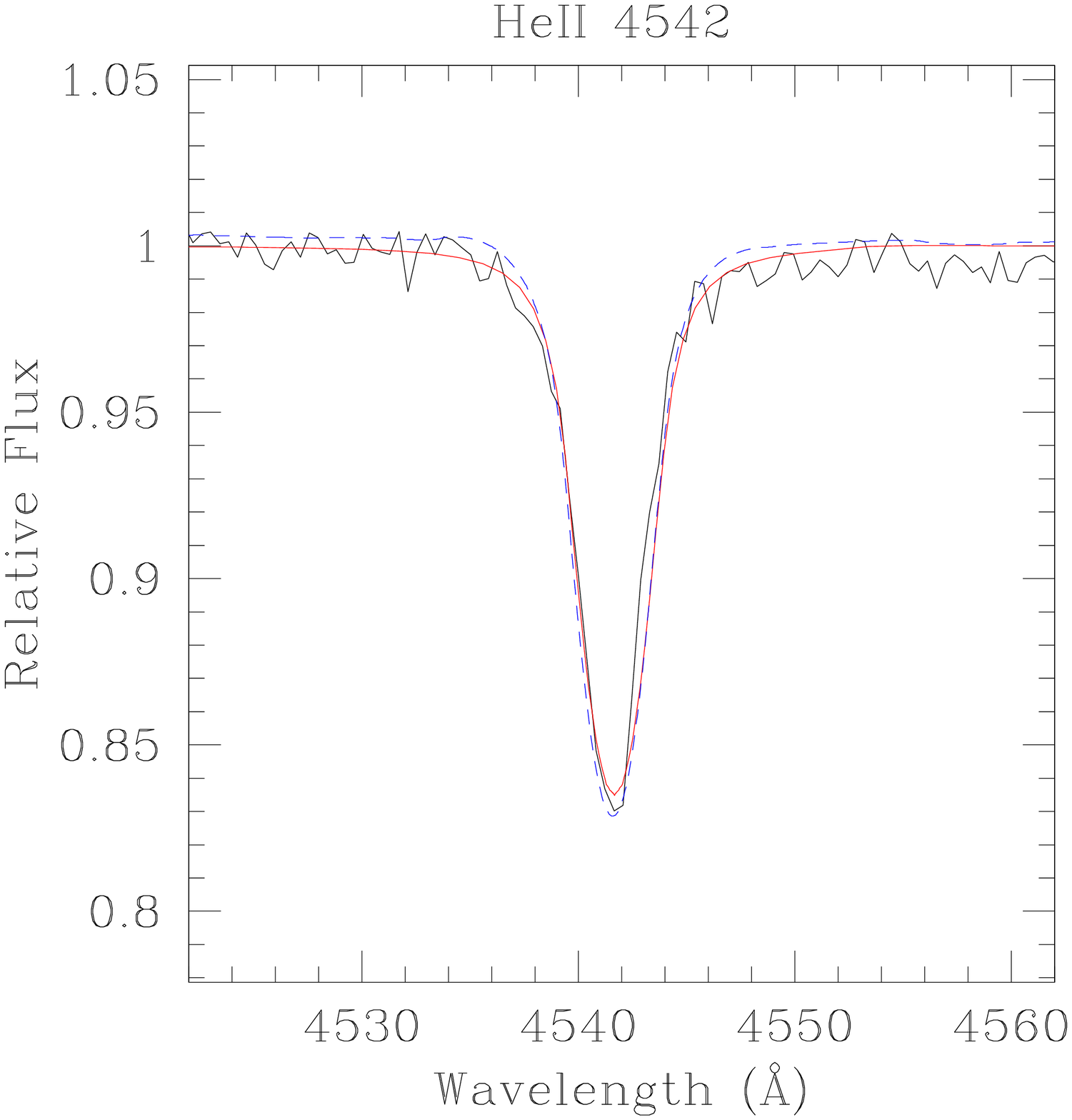}
\plotone{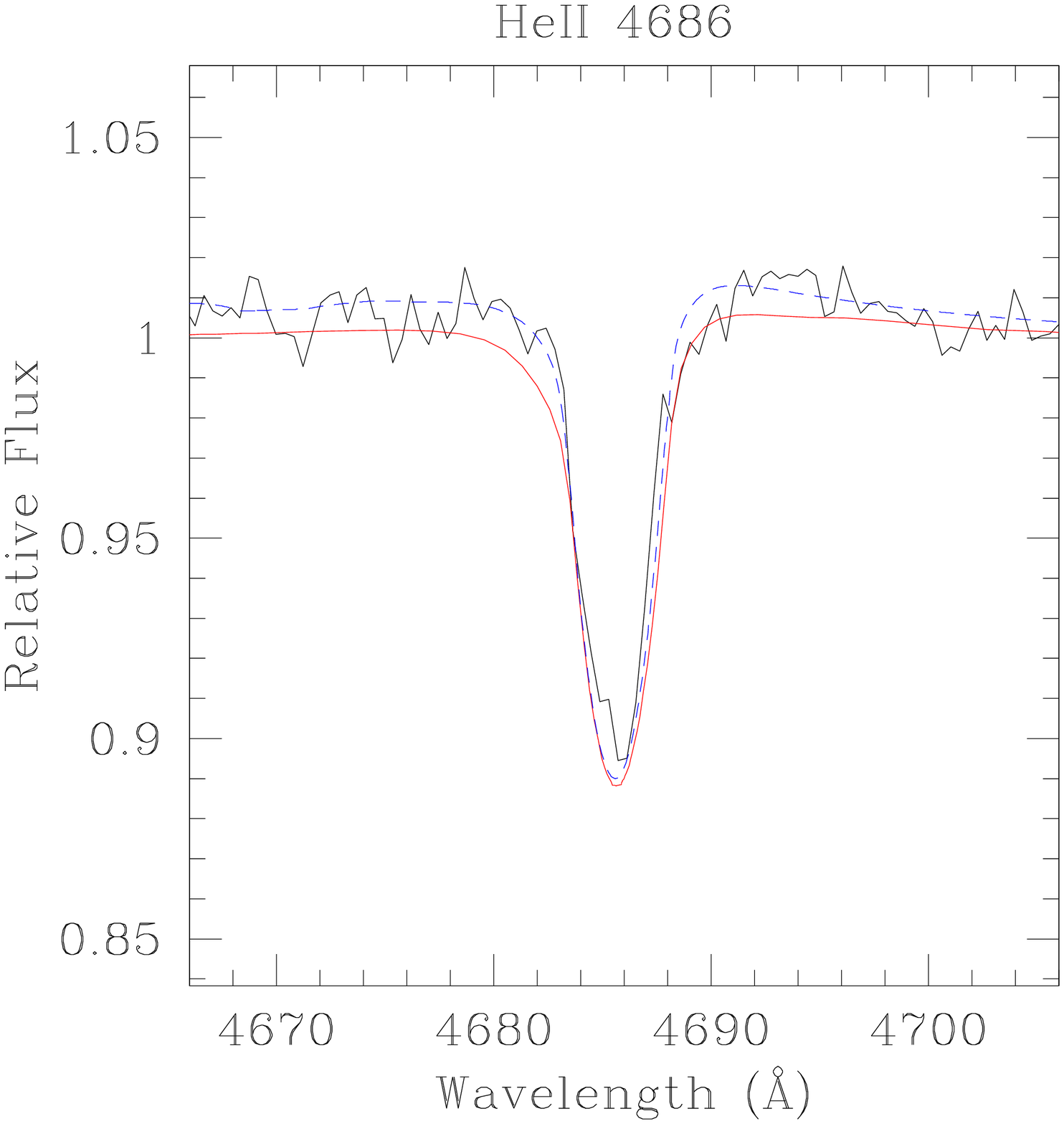}
\caption{\label{fig:AzV26} Model fits for AzV 26, an O6 I(f) star in the SMC.  Black shows the observed spectrum, the red line shows the \fastwind\ fit, and the dashed blue line shows the \cmfgen\ fit. }
\end{figure}

\begin{figure}
\epsscale{0.3}
\plotone{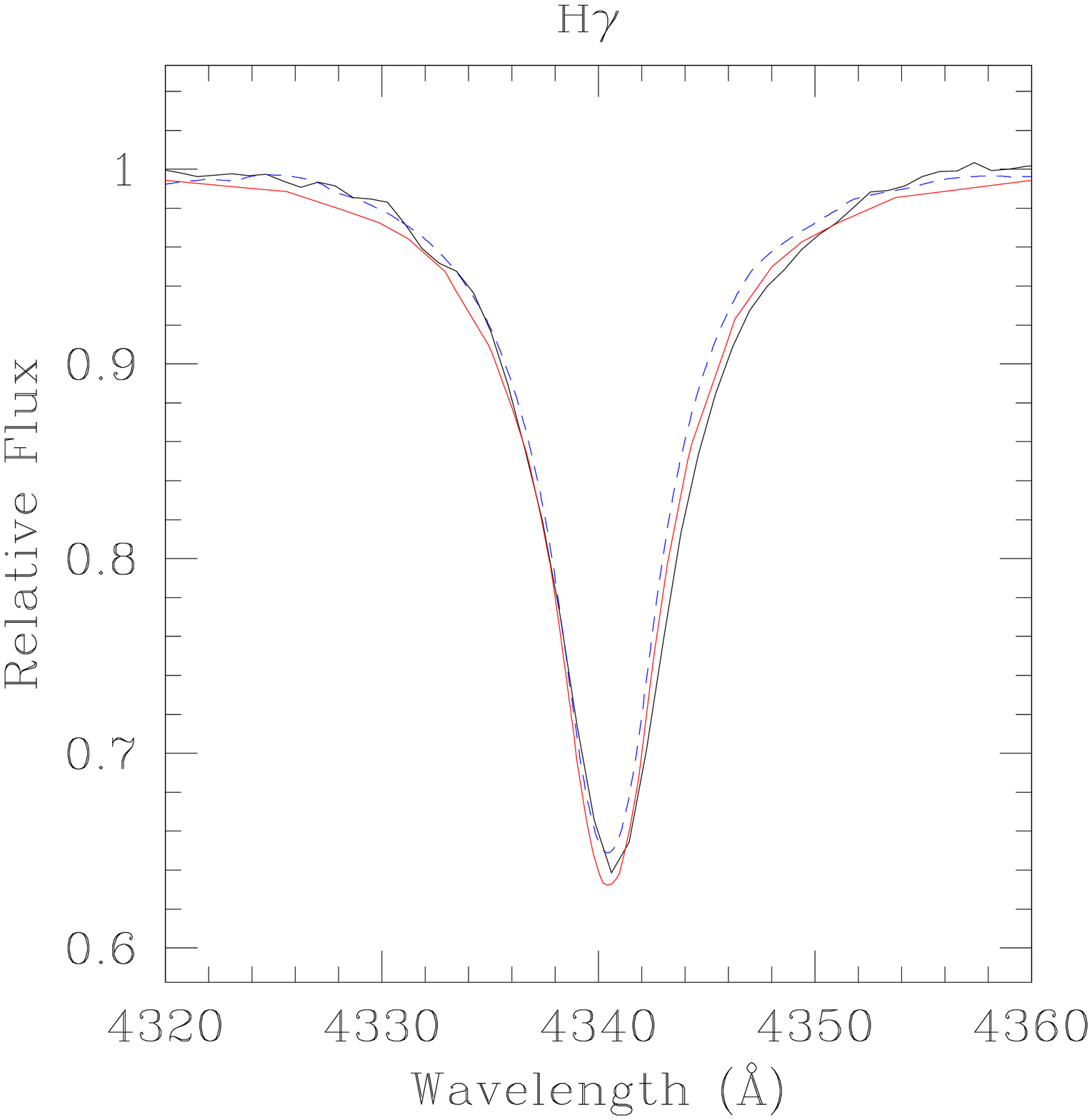}
\plotone{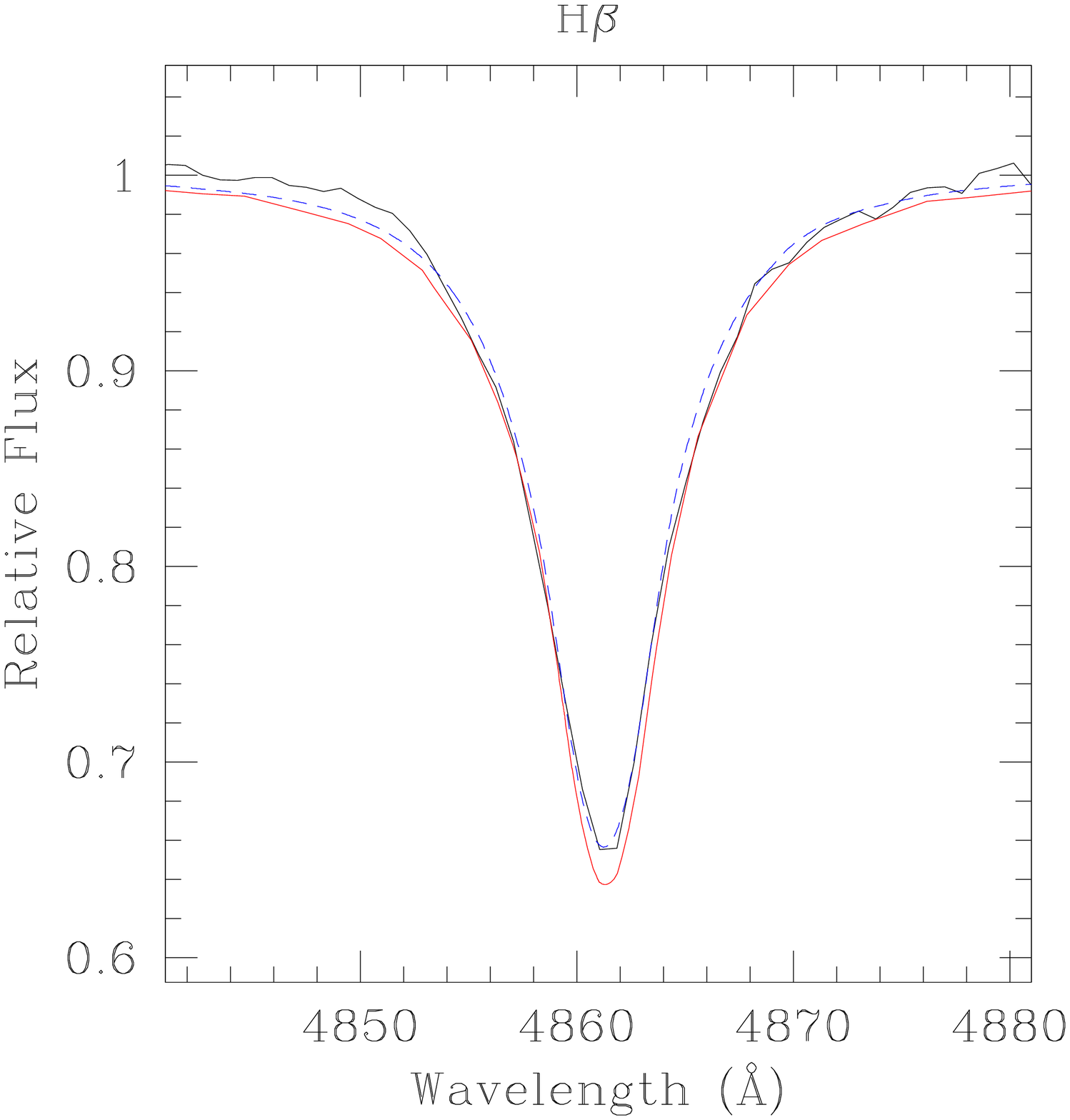}
\plotone{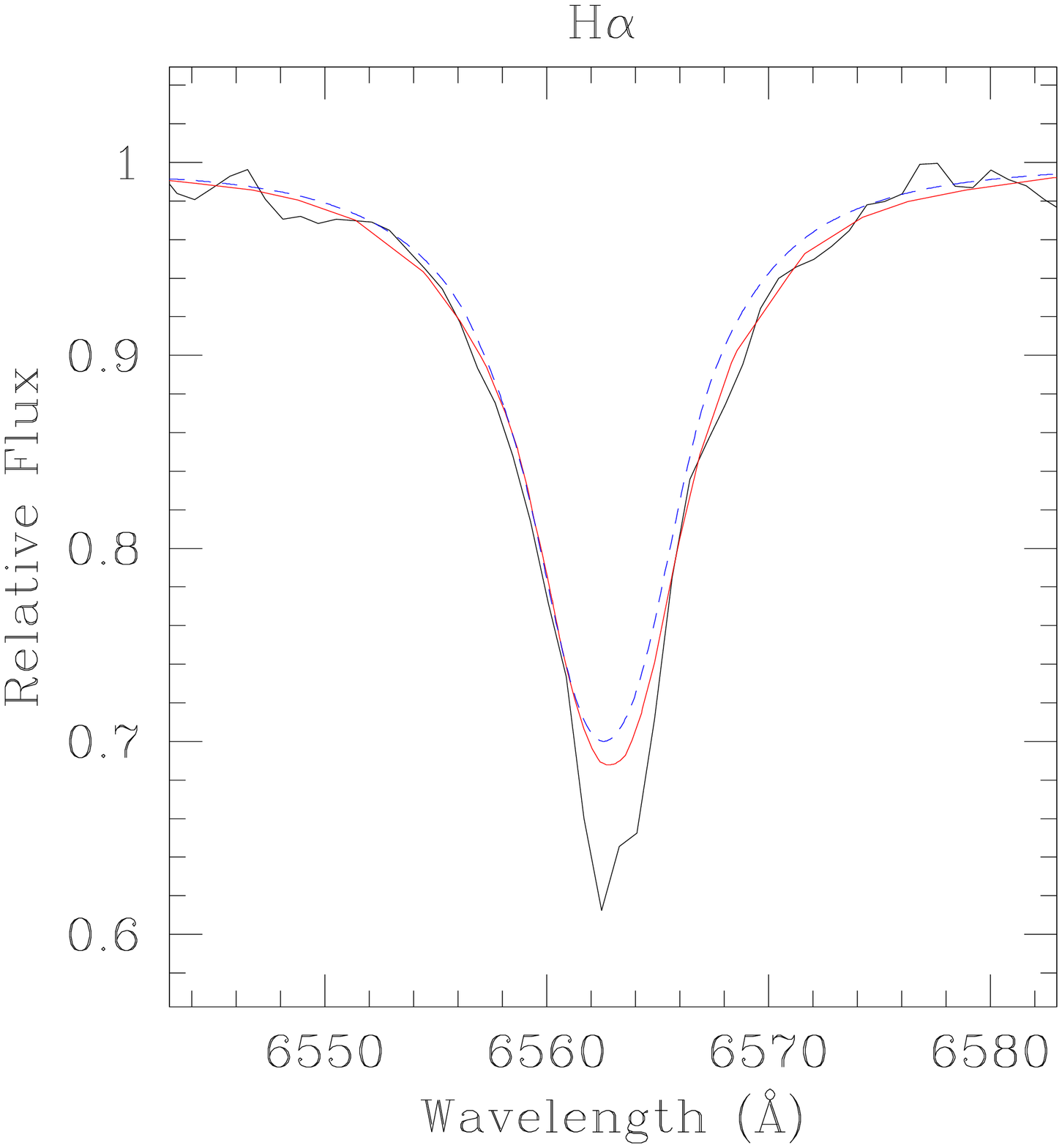}
\plotone{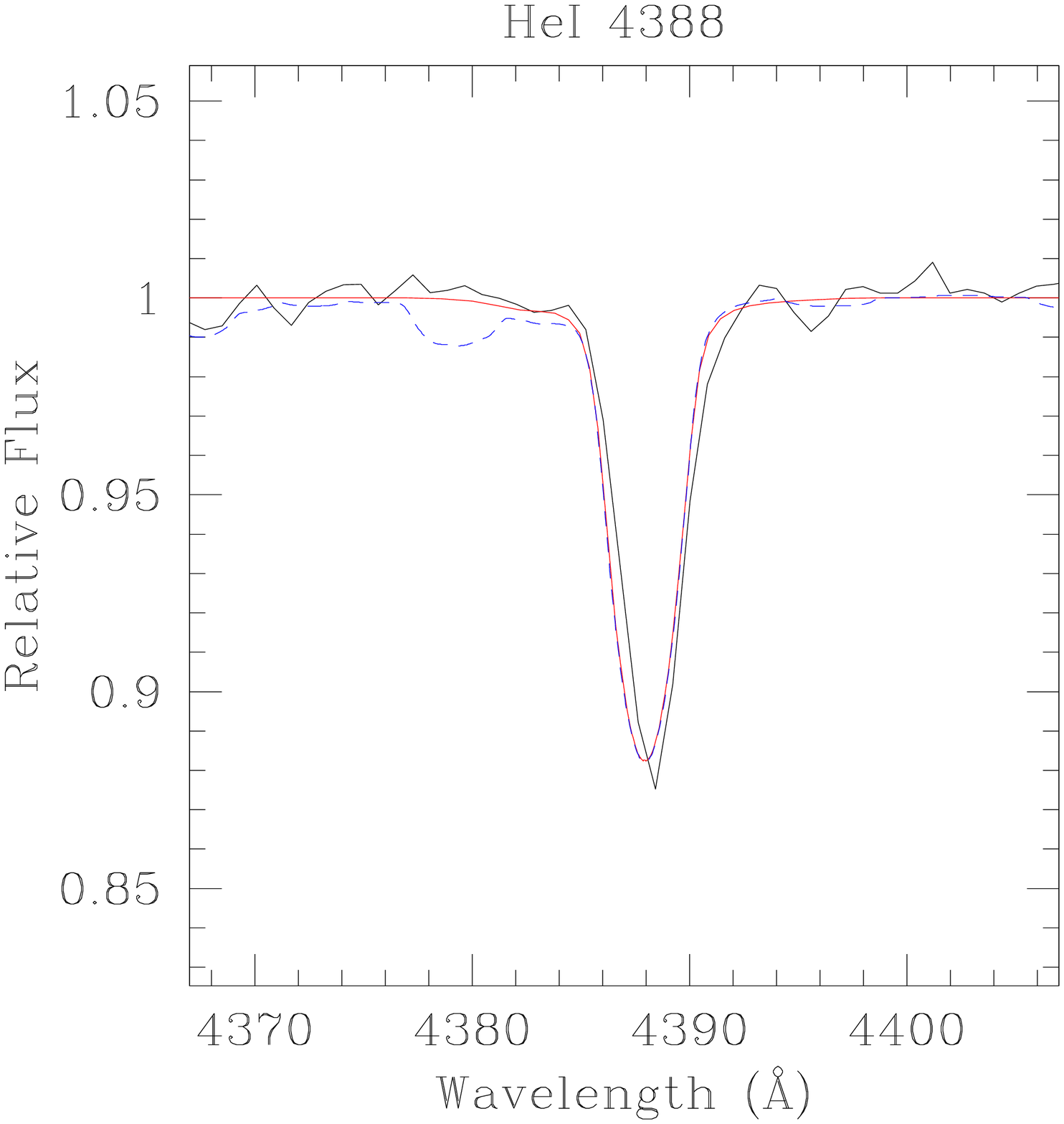}
\plotone{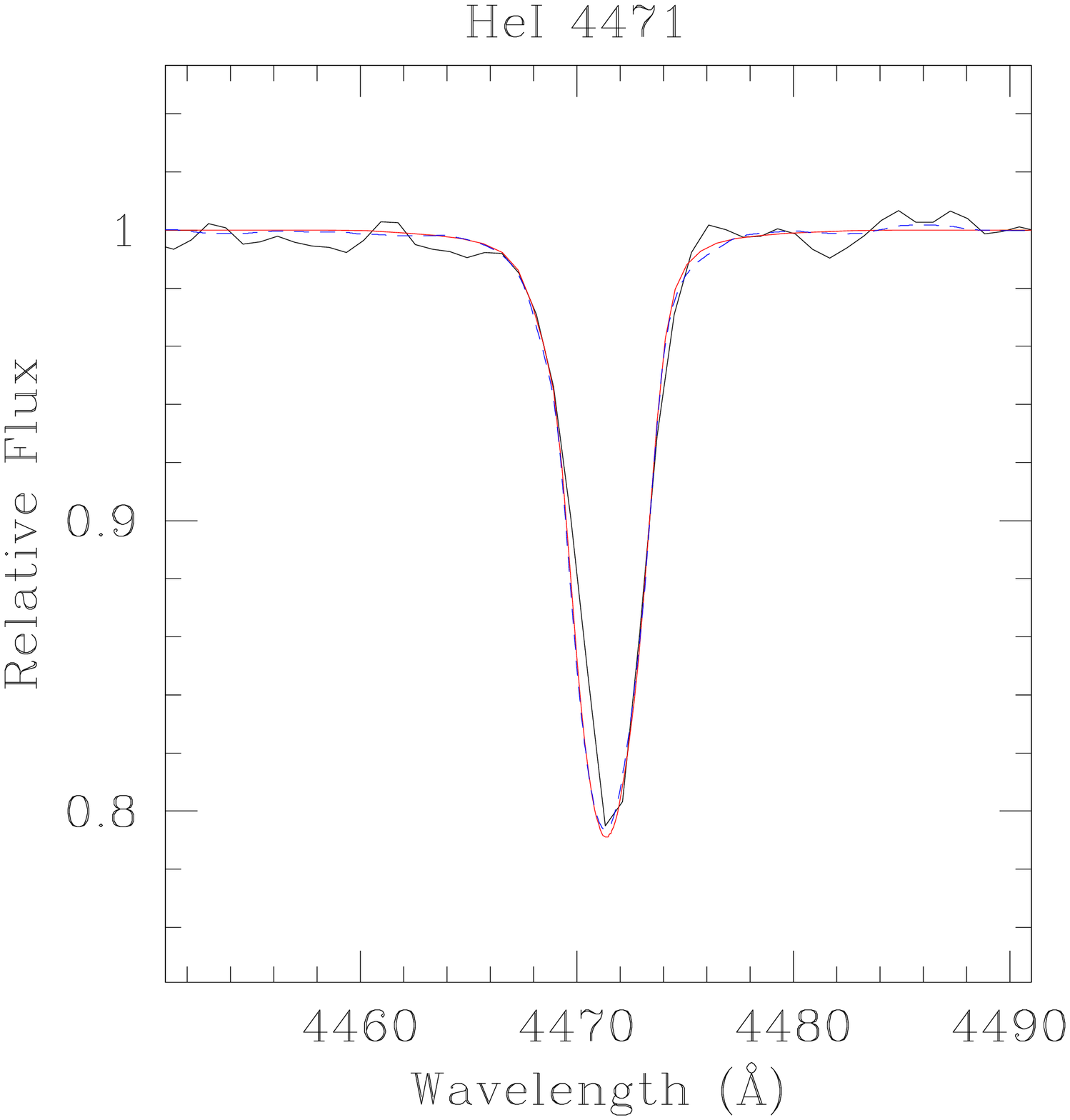}
\plotone{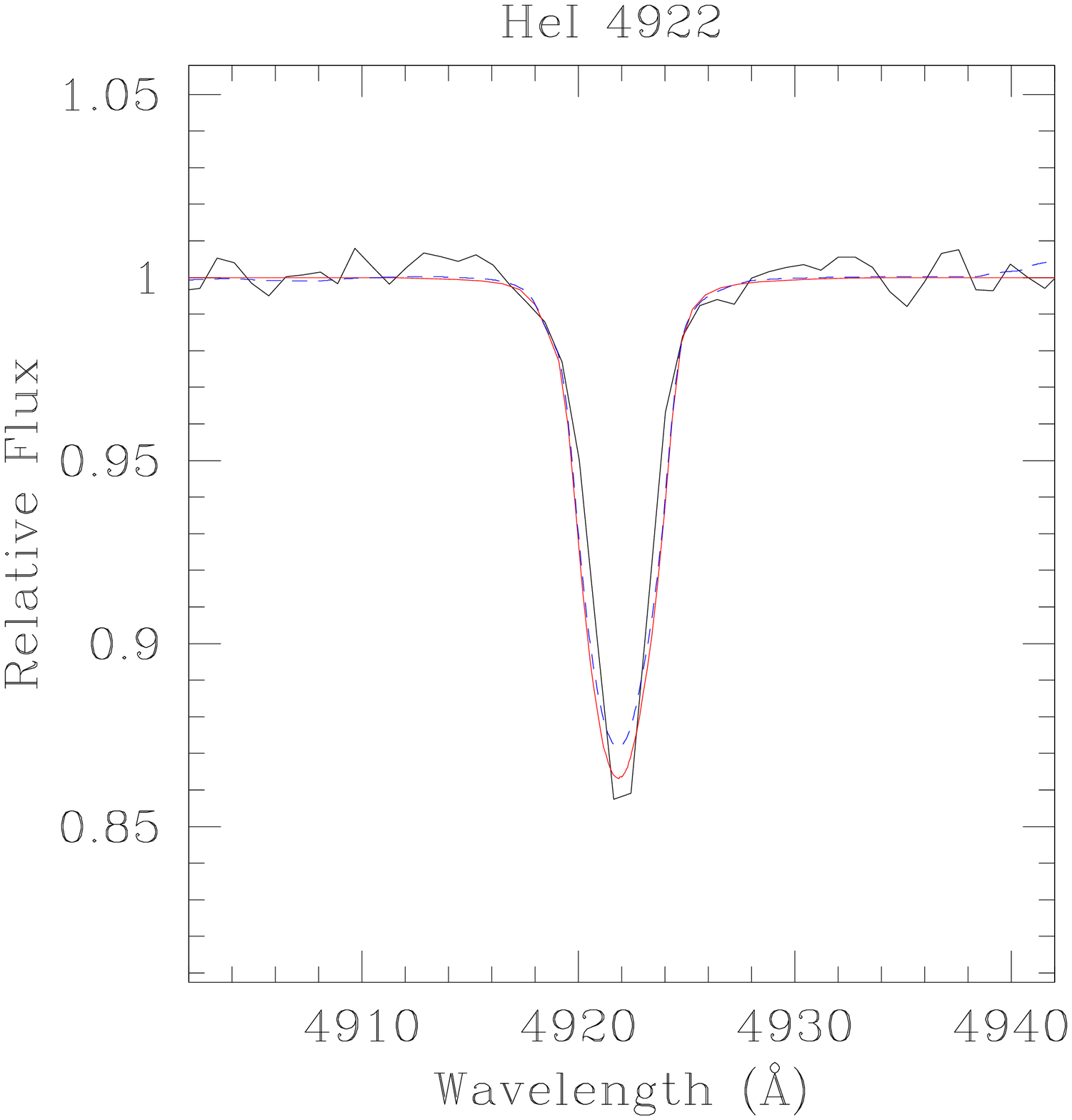}
\plotone{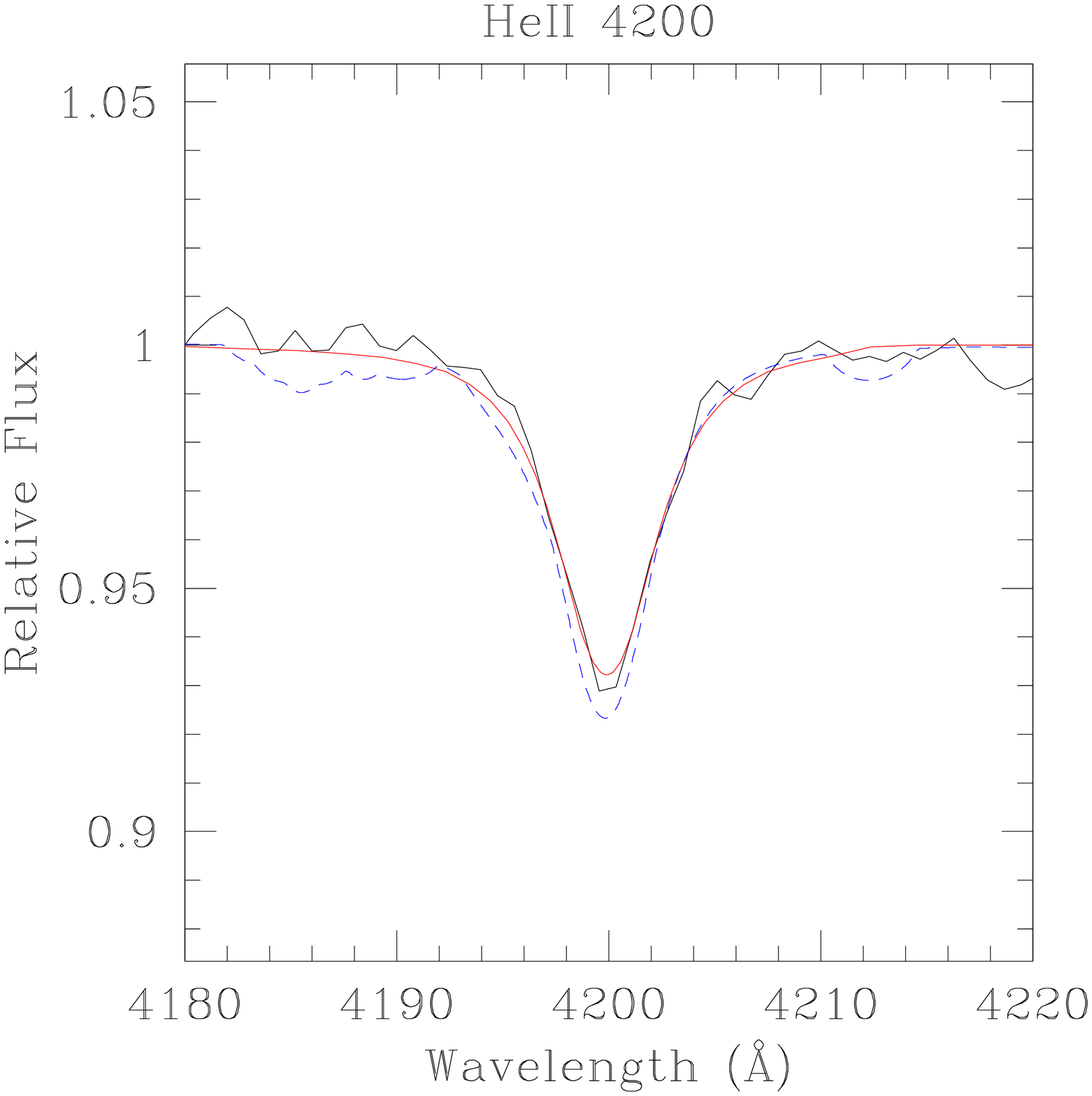}
\plotone{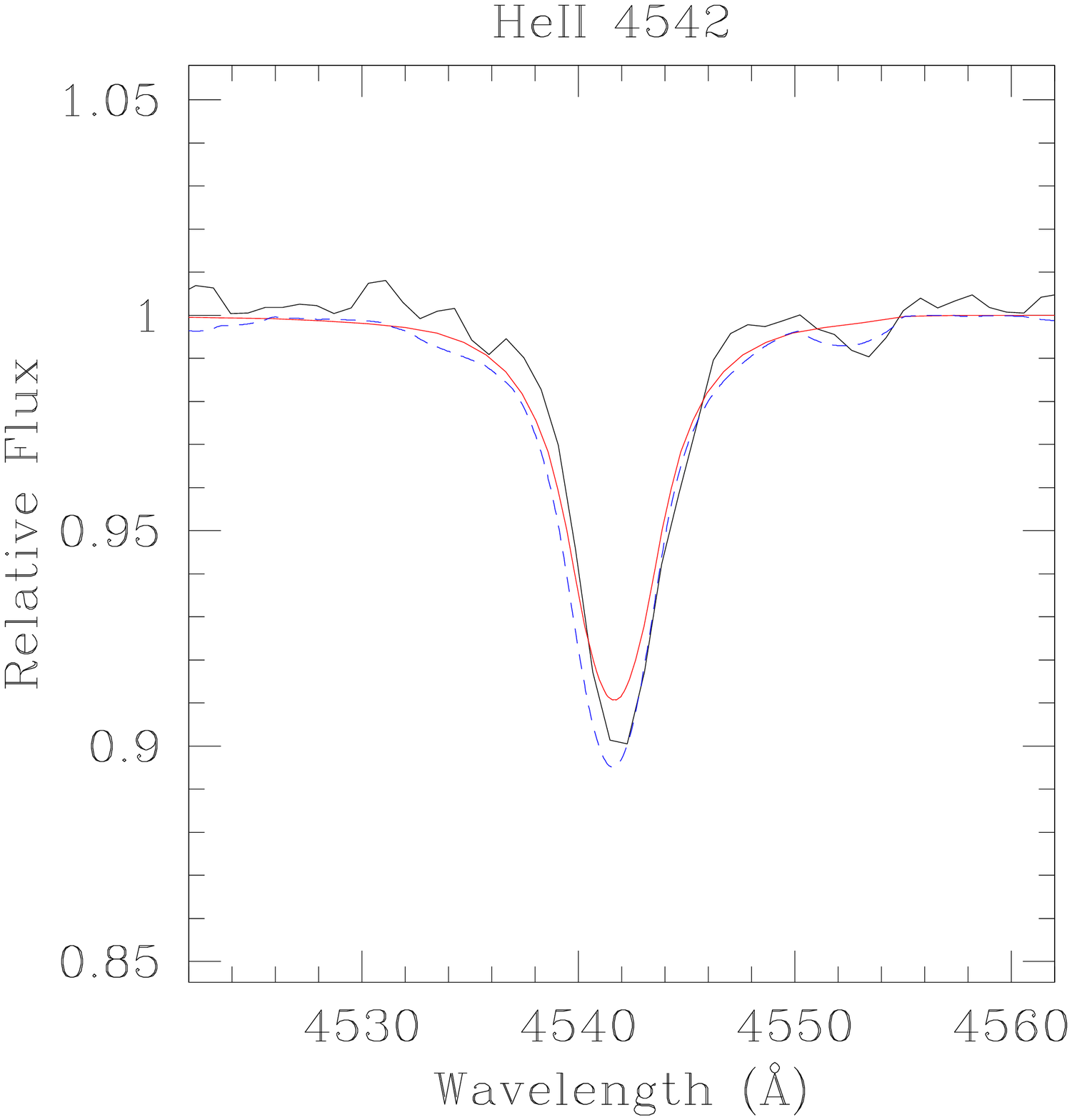}
\plotone{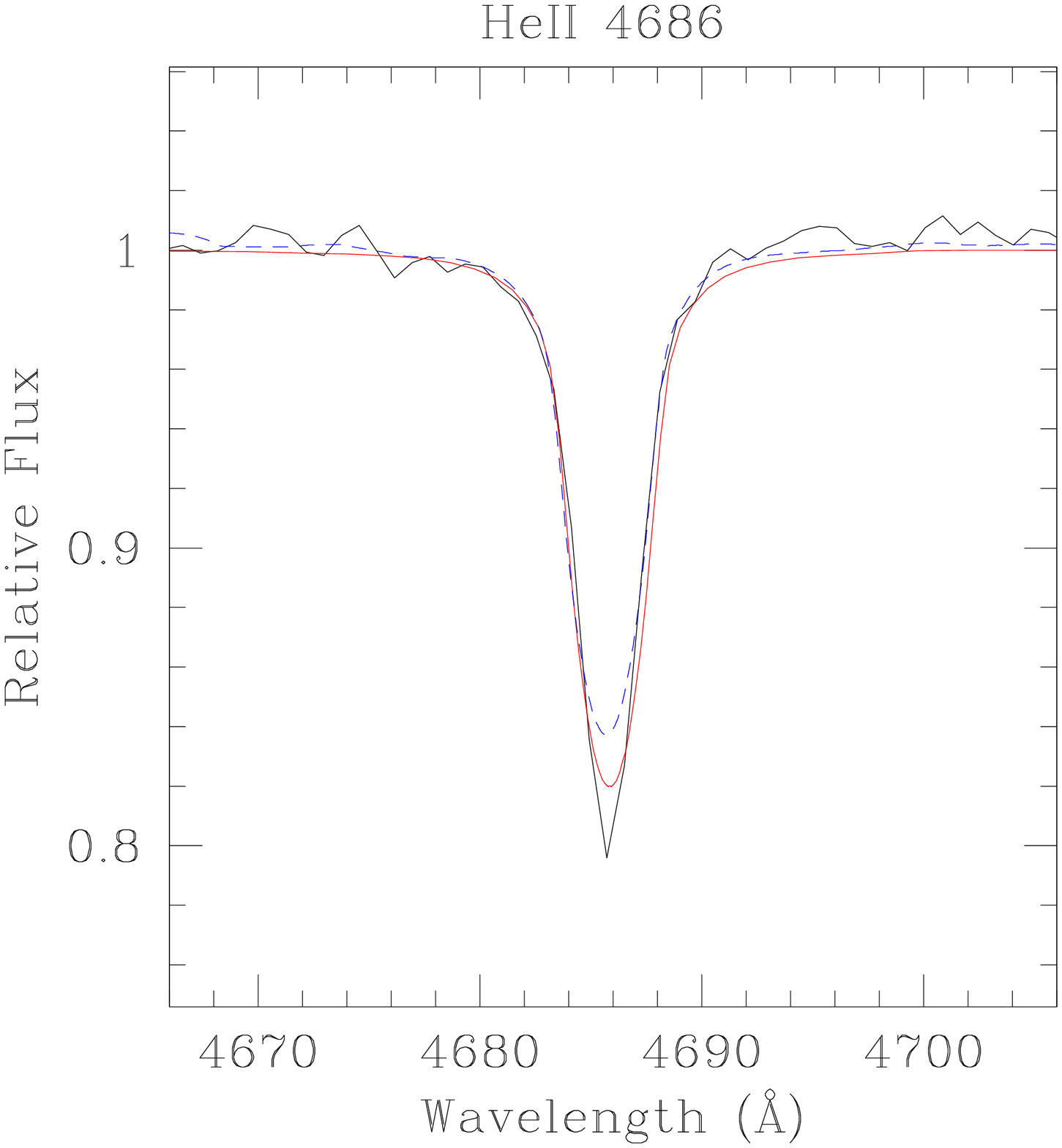}
\caption{\label{fig:NGC346-682} Model fits for NGC346-682, an O8 V star in the SMC.  Black shows the observed spectrum, the red line shows the \fastwind\ fit, and the dashed blue line shows the \cmfgen\ fit. }
\end{figure}

\begin{figure}
\epsscale{0.3}
\plotone{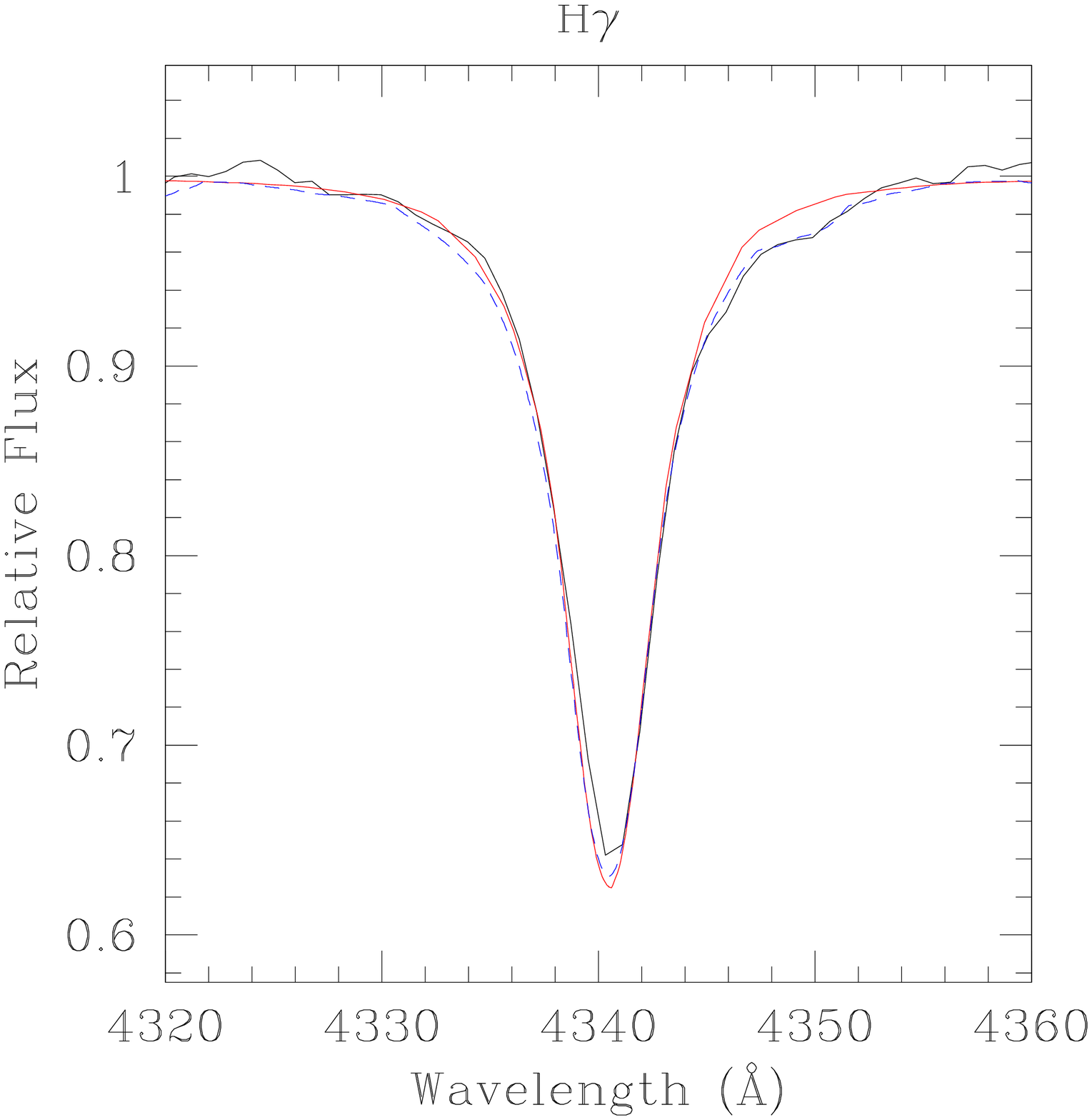}
\plotone{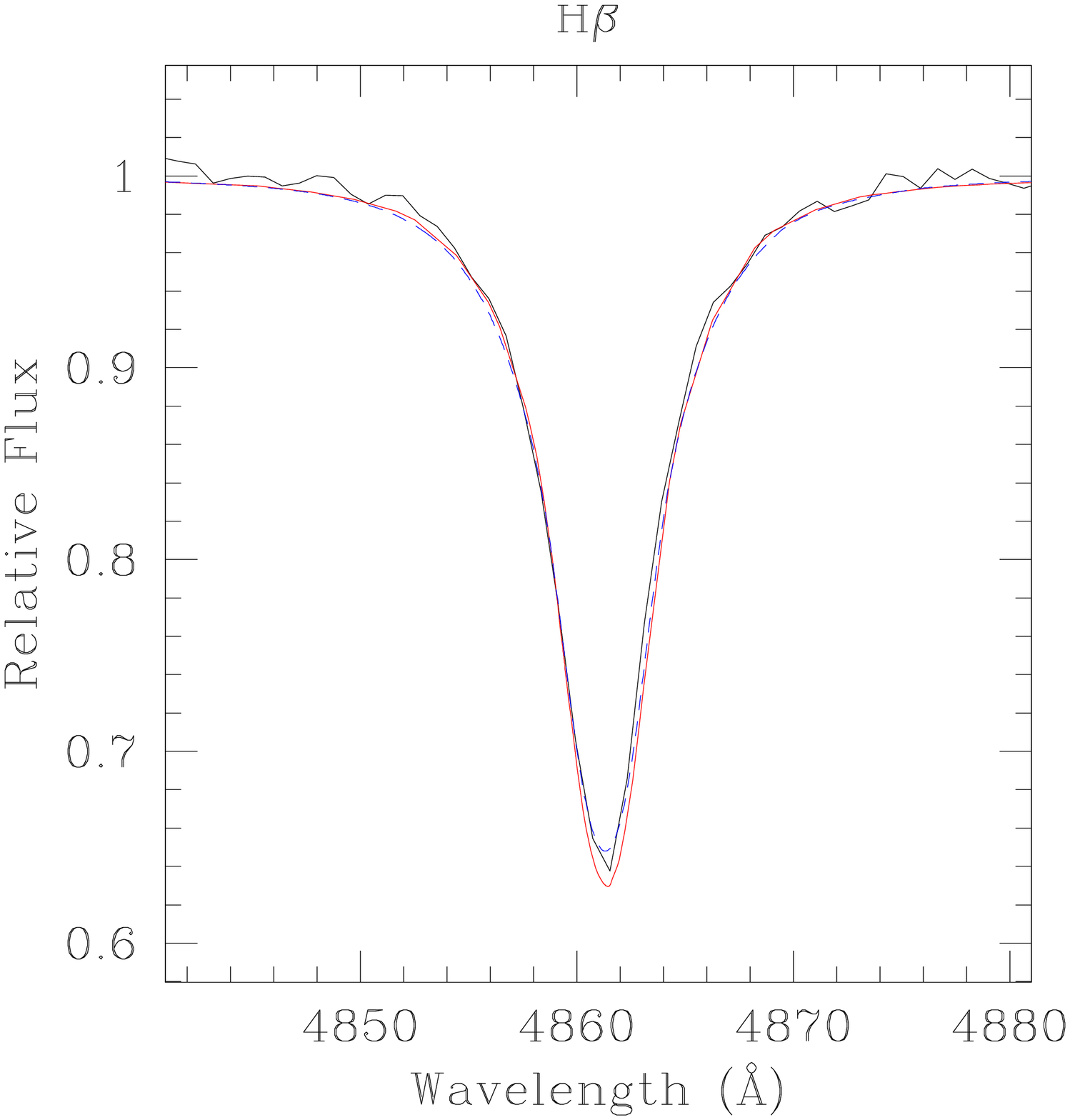}
\plotone{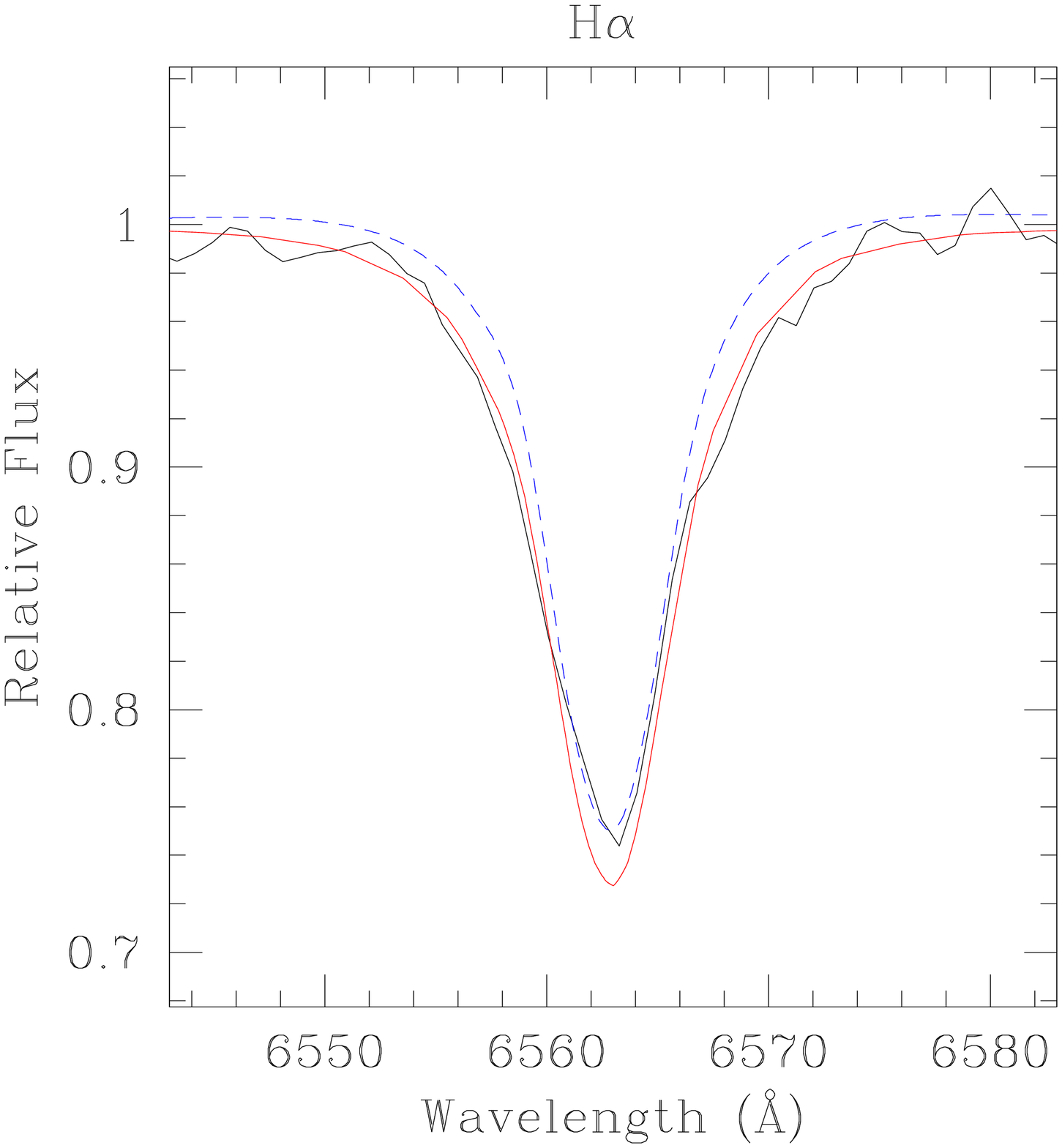}
\plotone{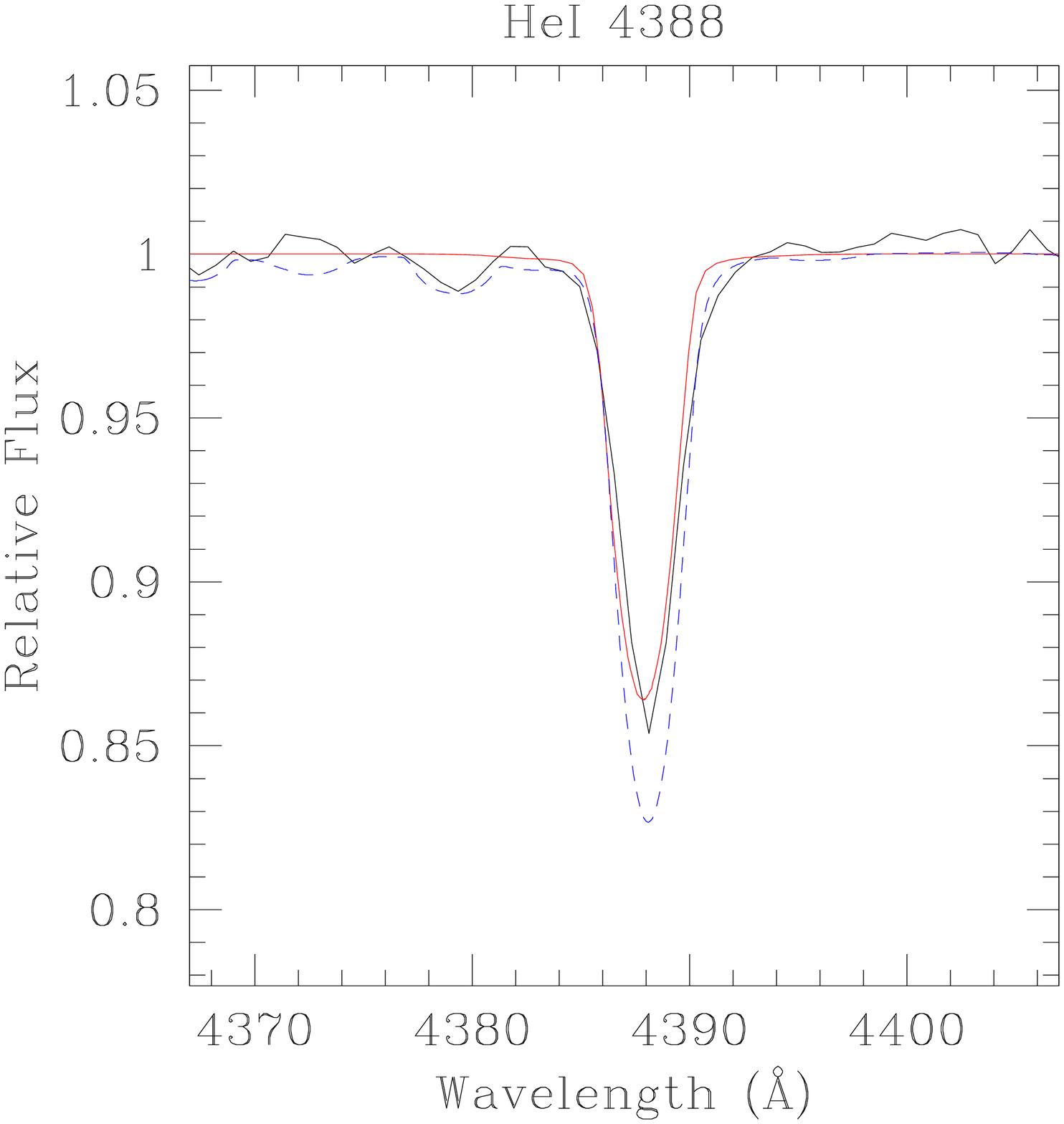}
\plotone{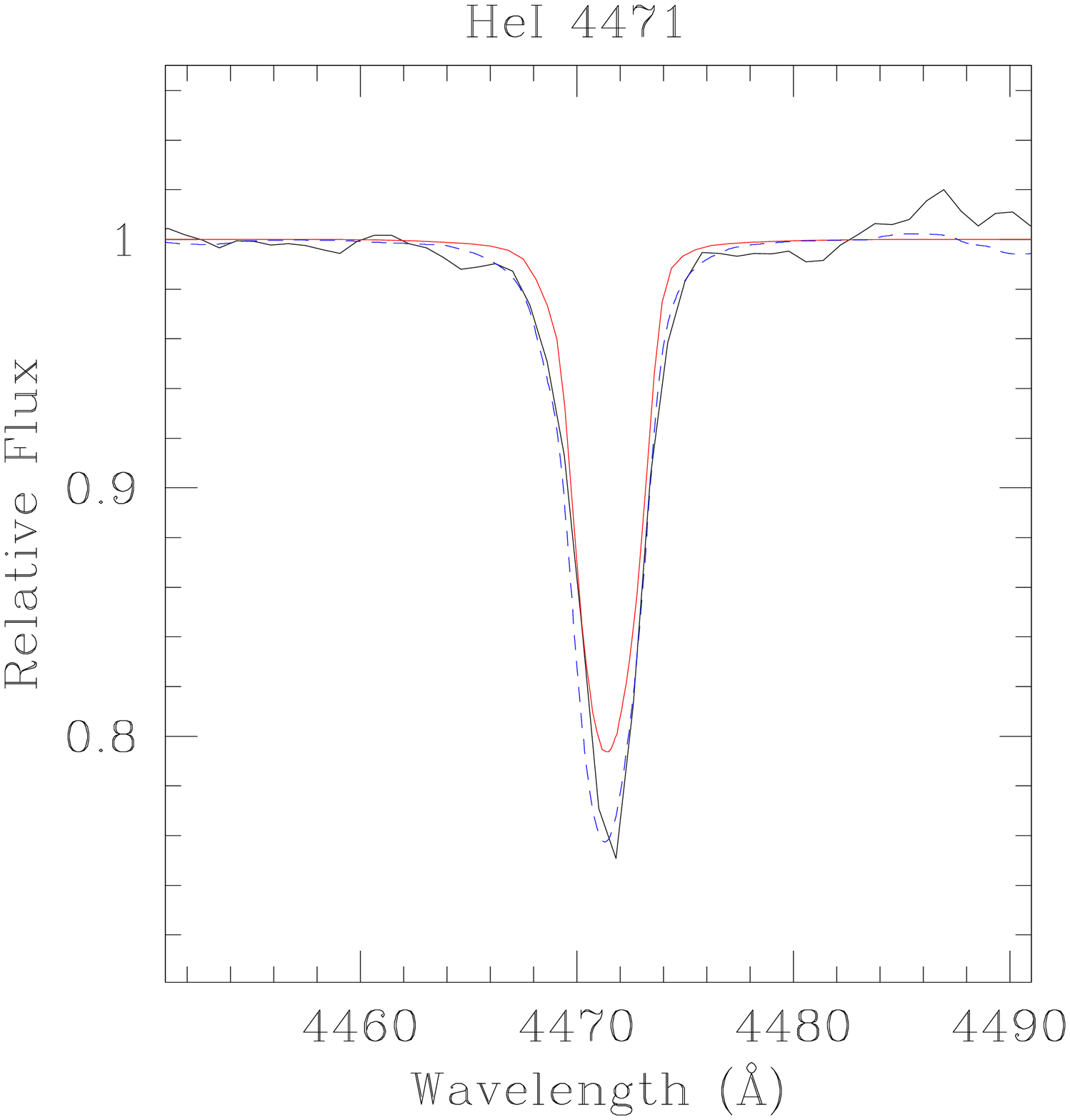}
\plotone{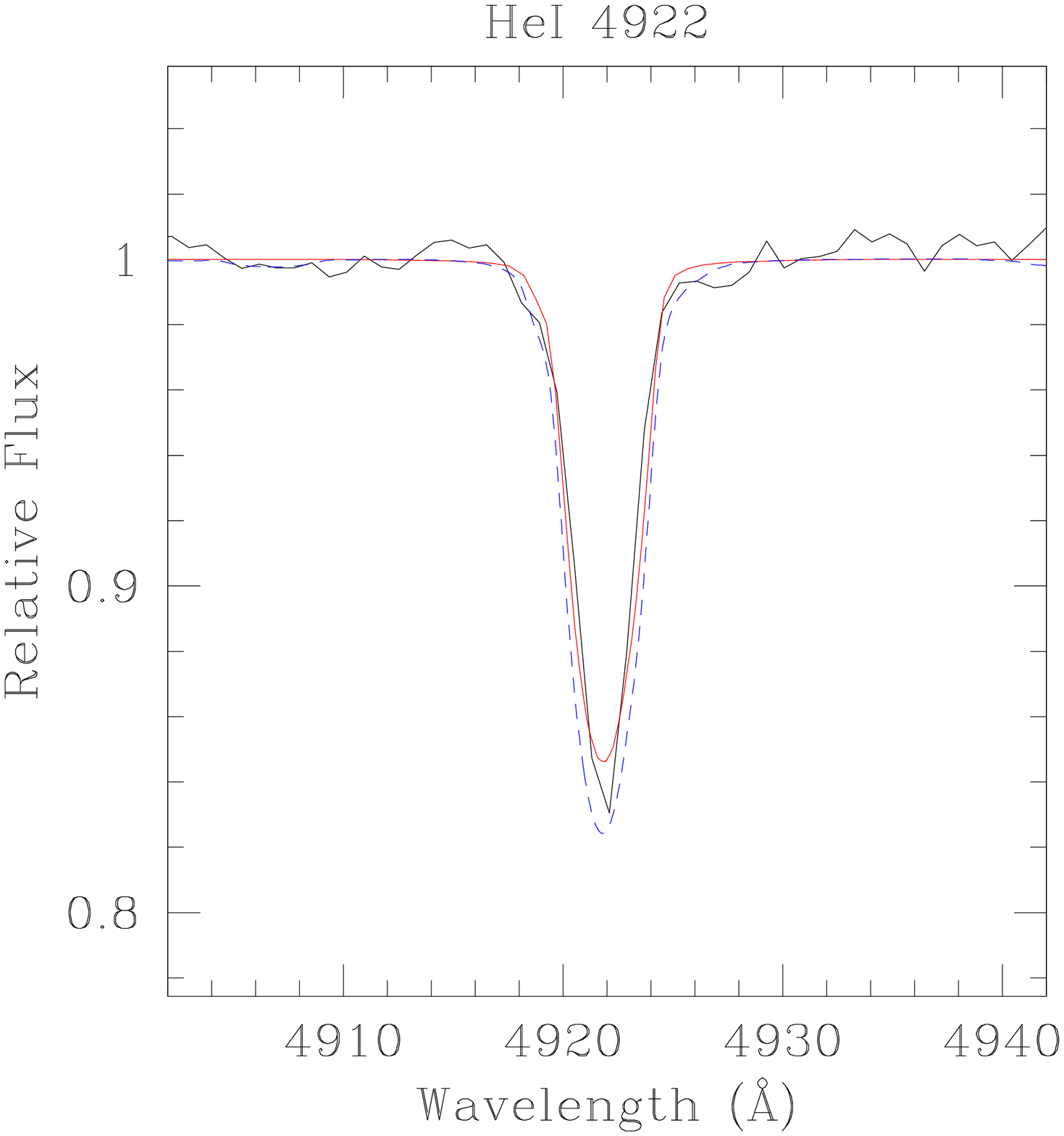}
\plotone{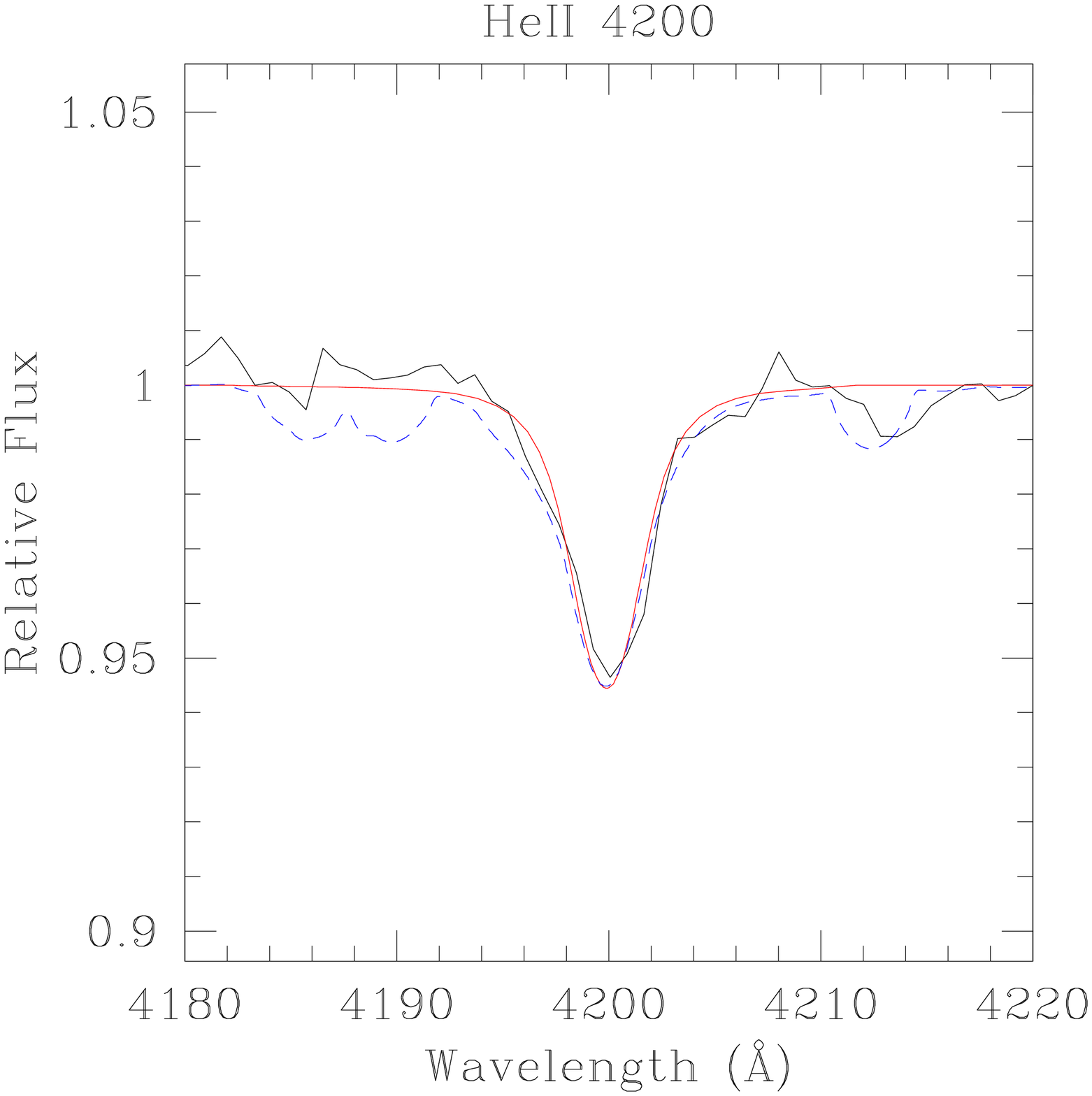}
\plotone{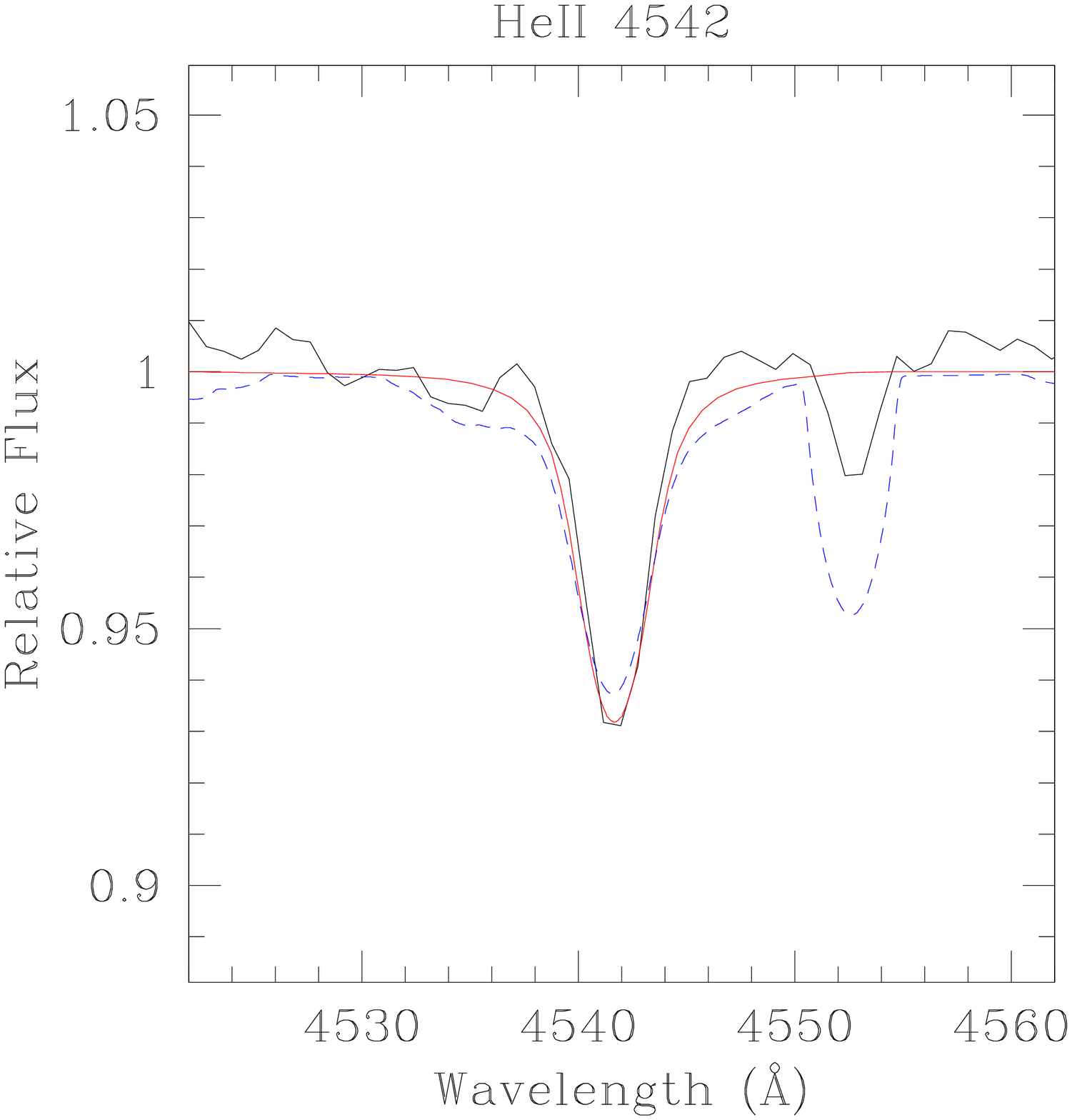}
\plotone{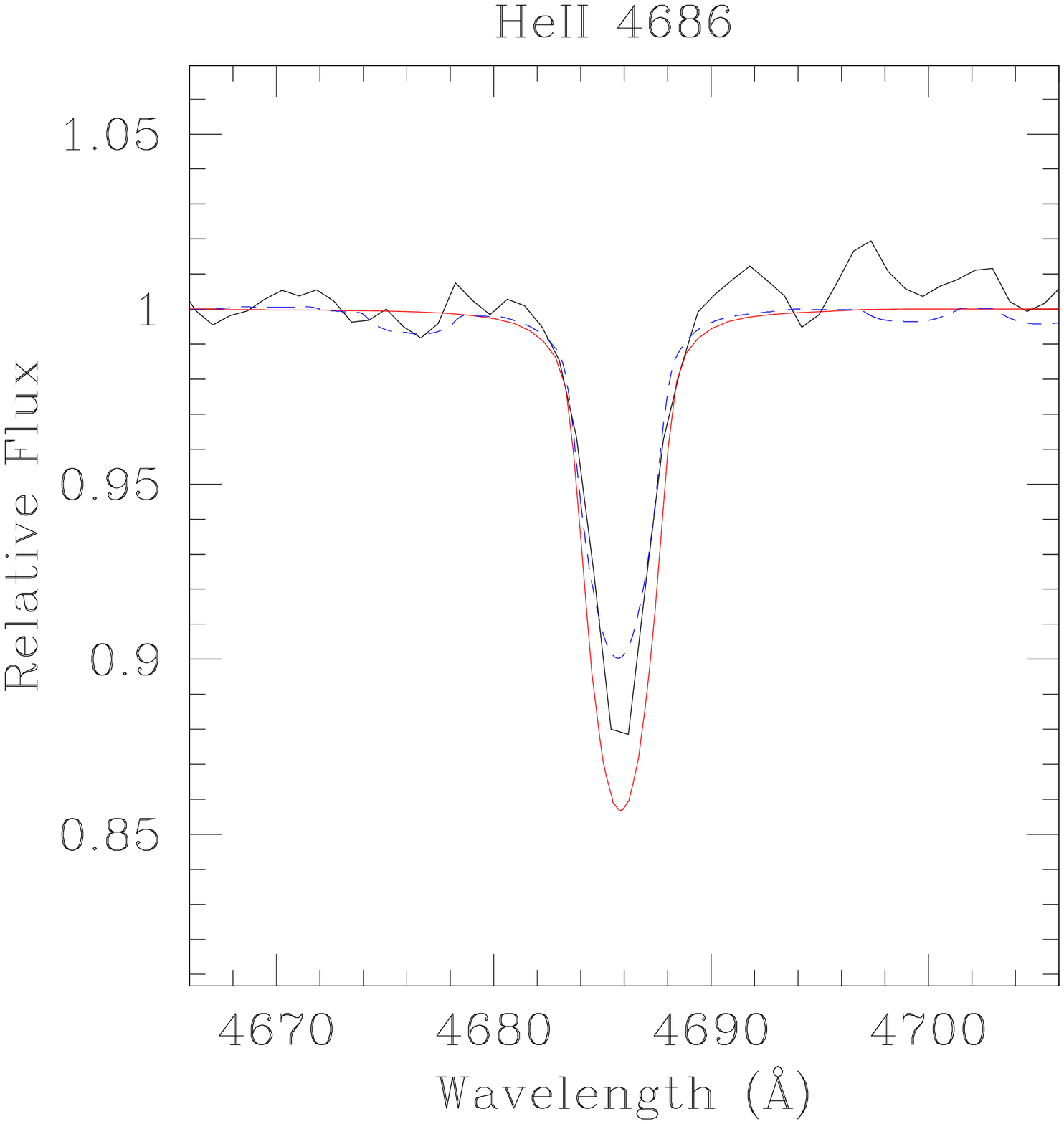}
\caption{\label{fig:AzV223} Model fits for AzV 223, an O9.5 II star in the SMC.  Black shows the observed spectrum, the red line shows the \fastwind\ fit, and the dashed blue line shows the \cmfgen\ fit. }
\end{figure}

\begin{figure}
\epsscale{0.3}
\plotone{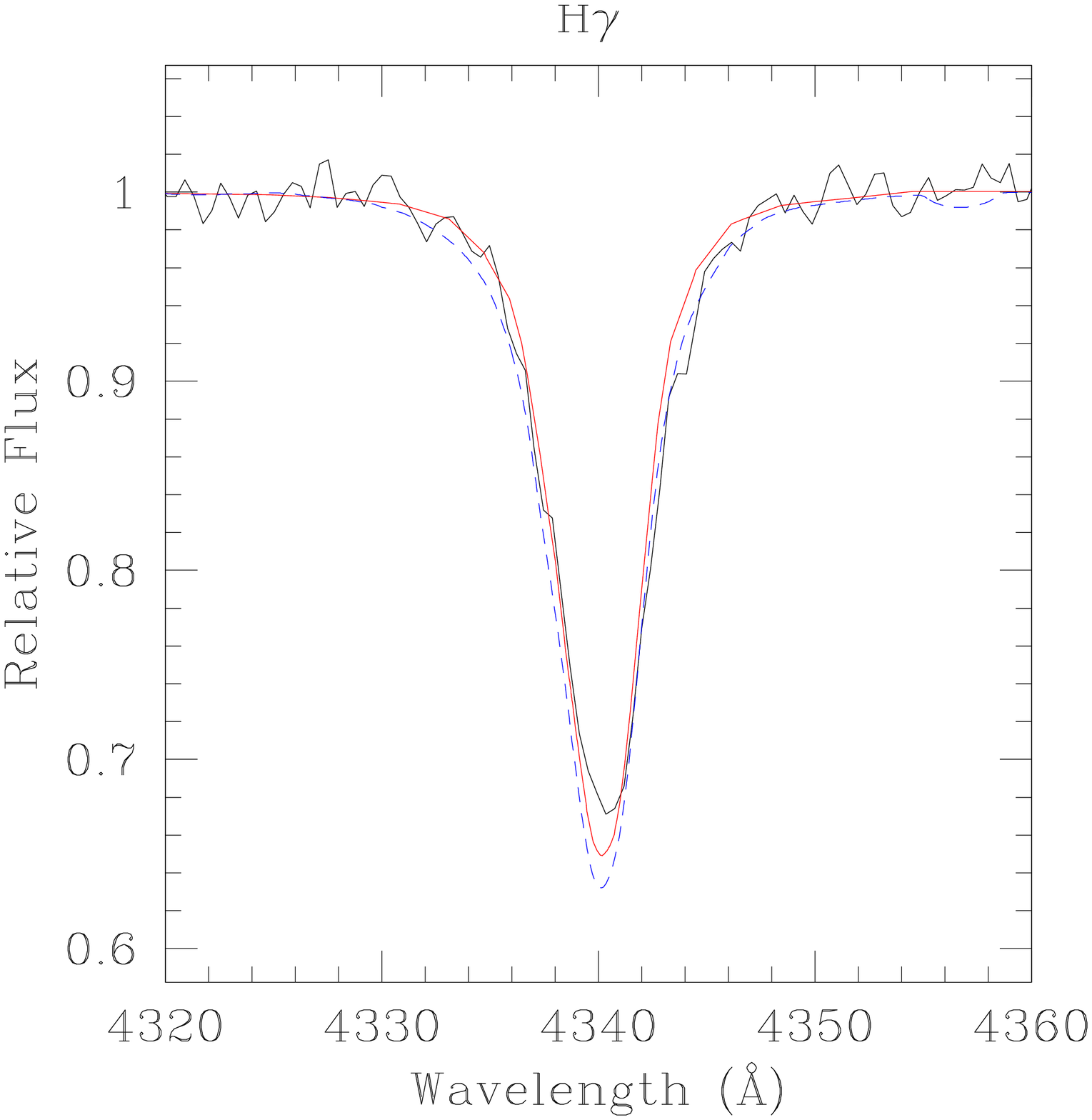}
\plotone{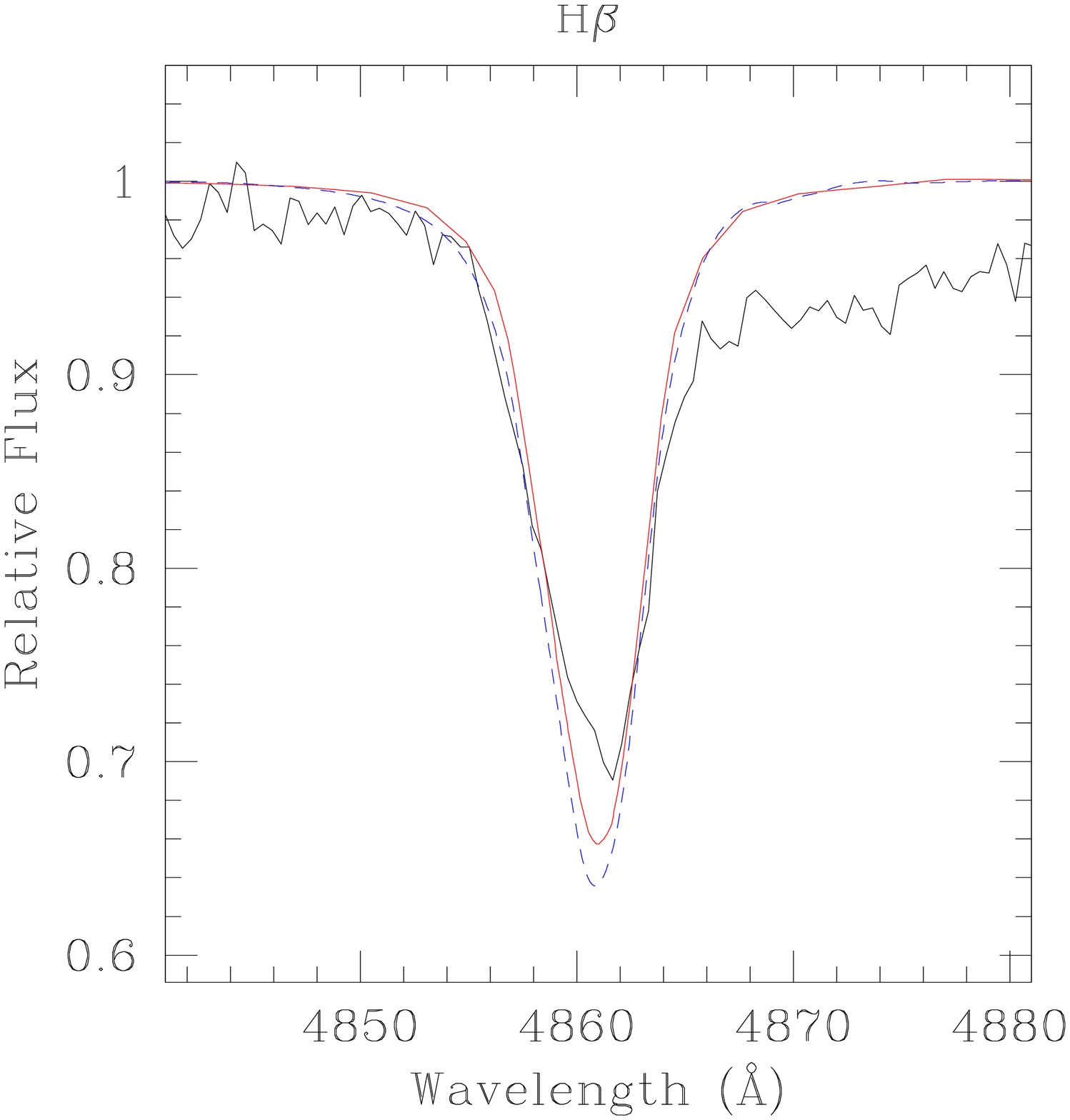}
\plotone{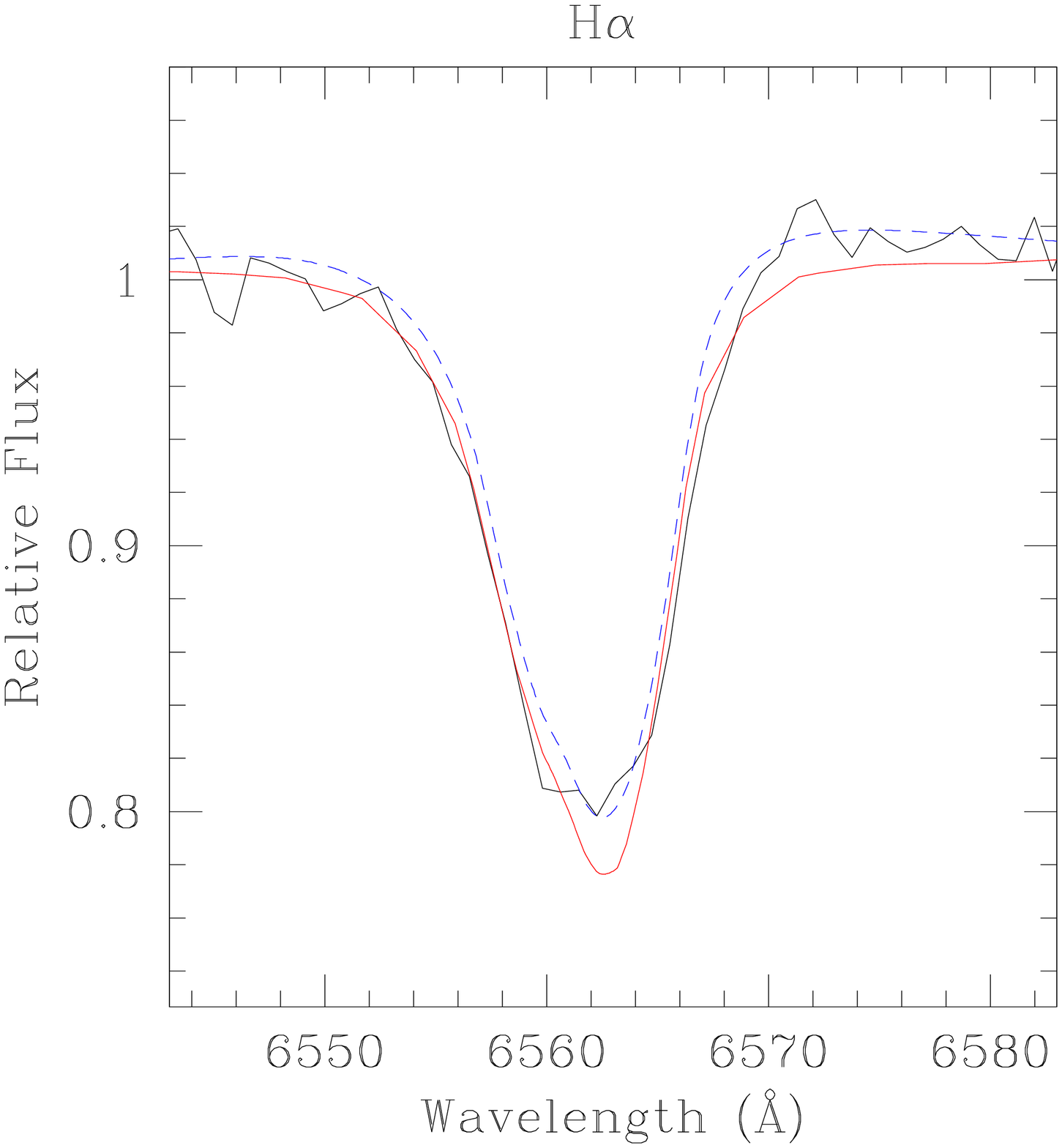}
\plotone{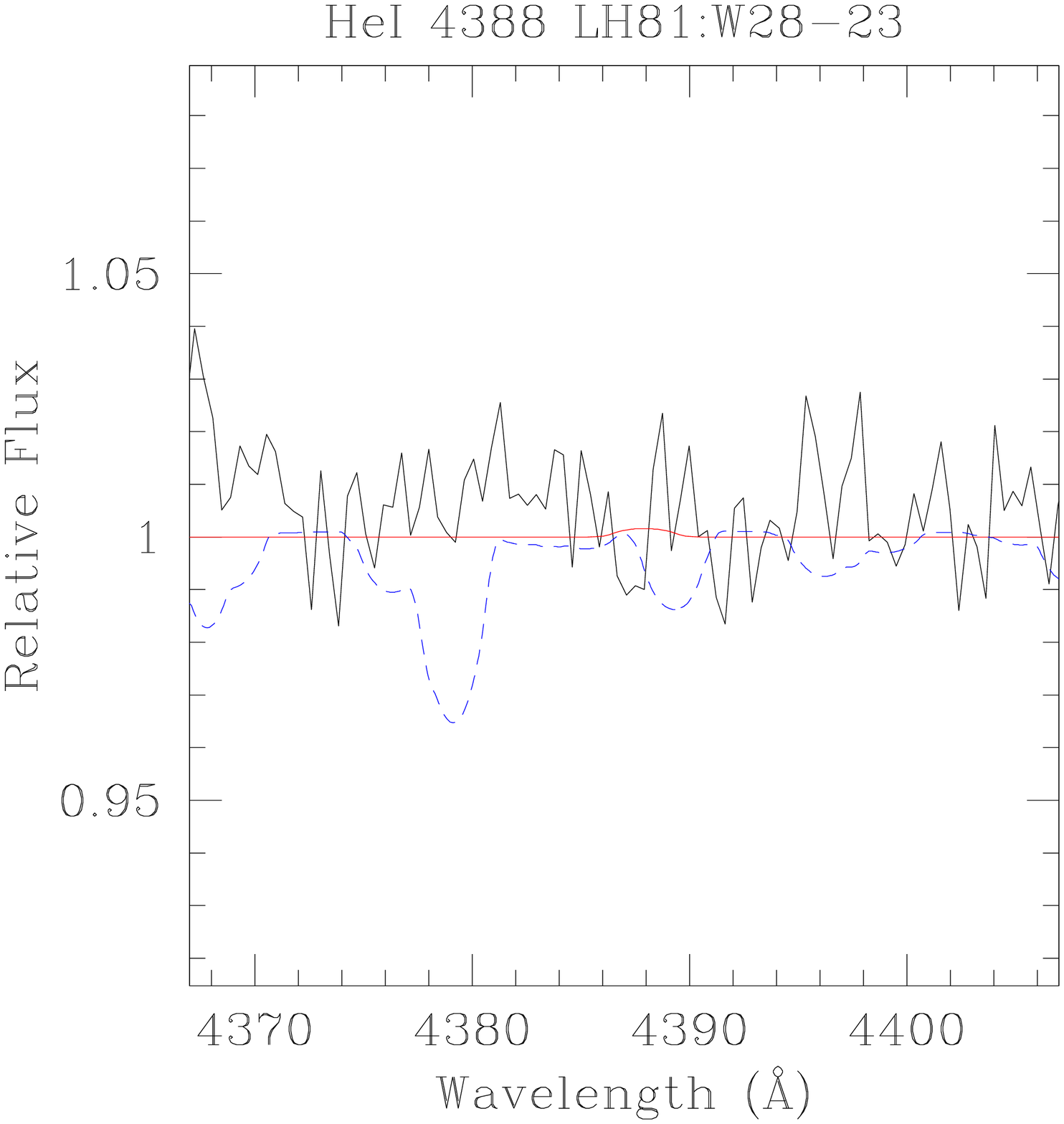}
\plotone{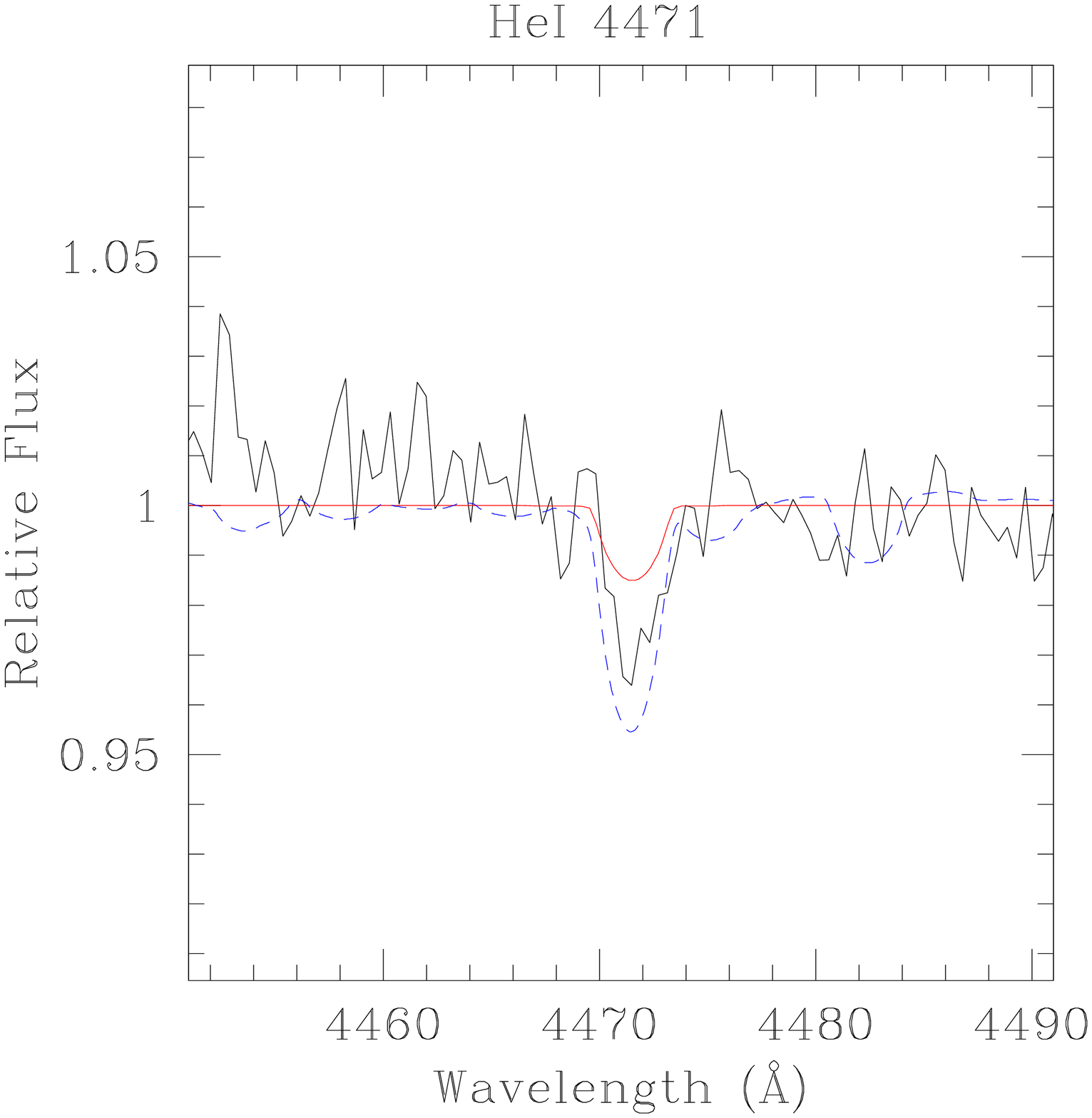}
\plotone{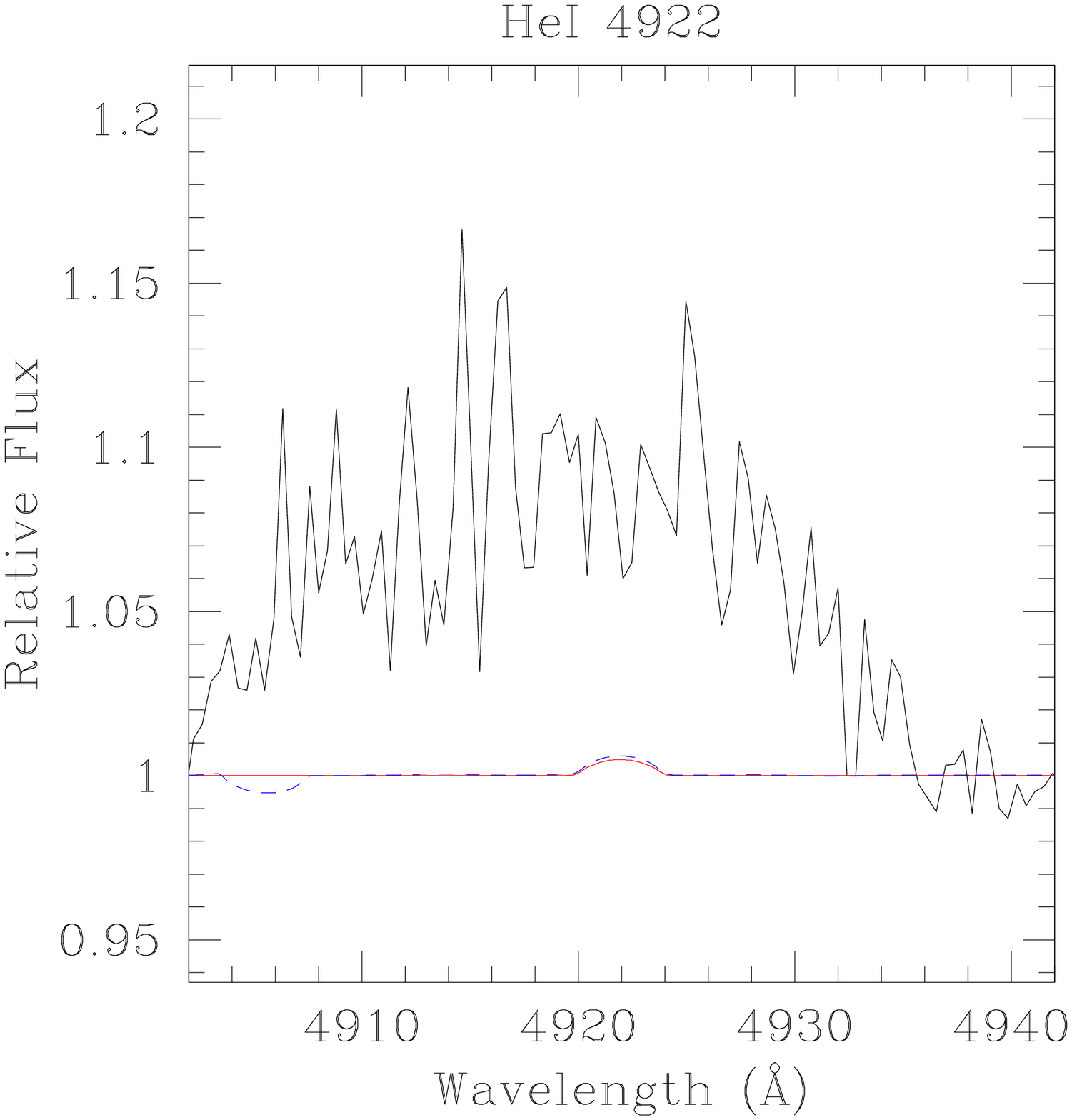}
\plotone{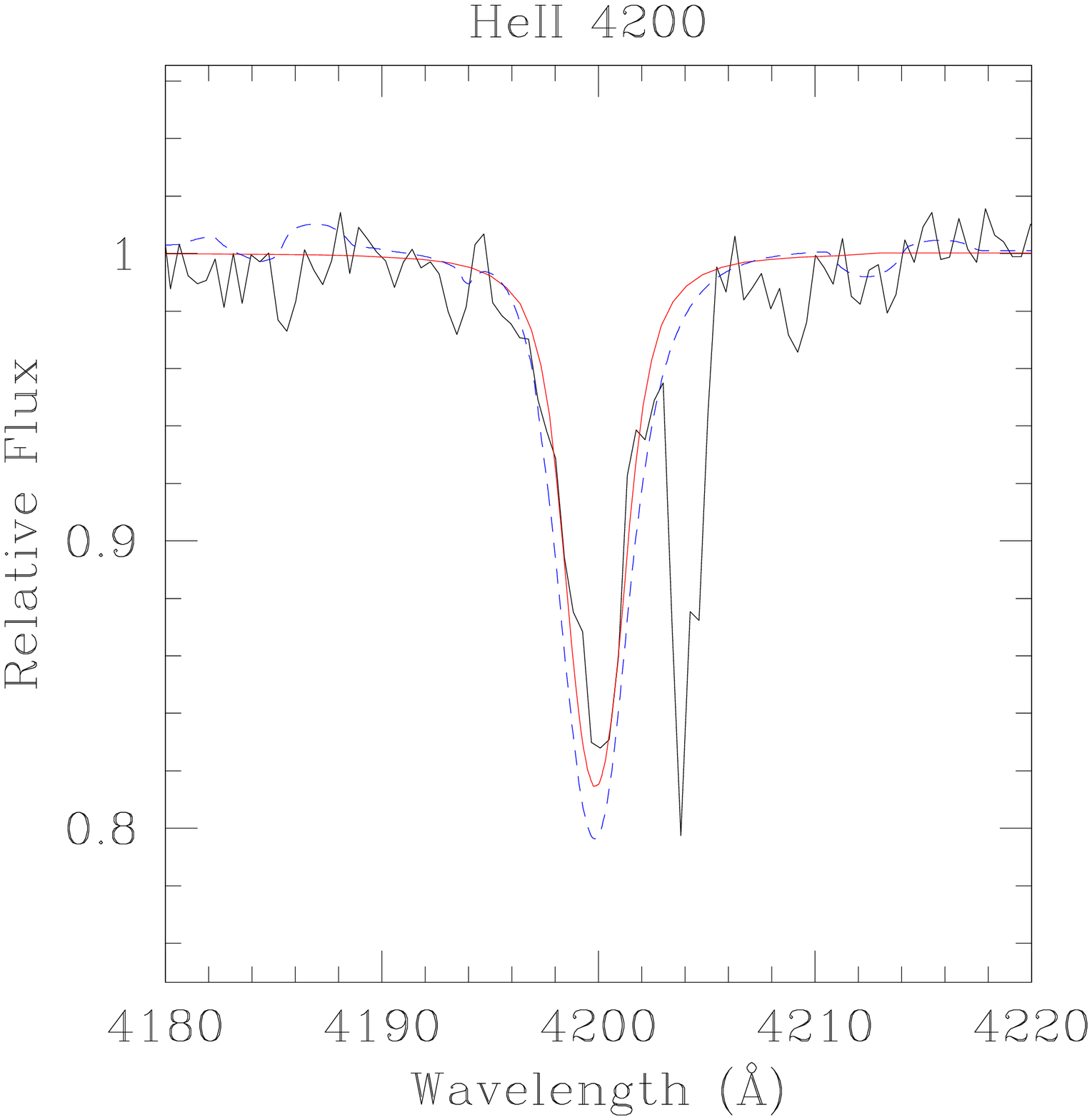}
\plotone{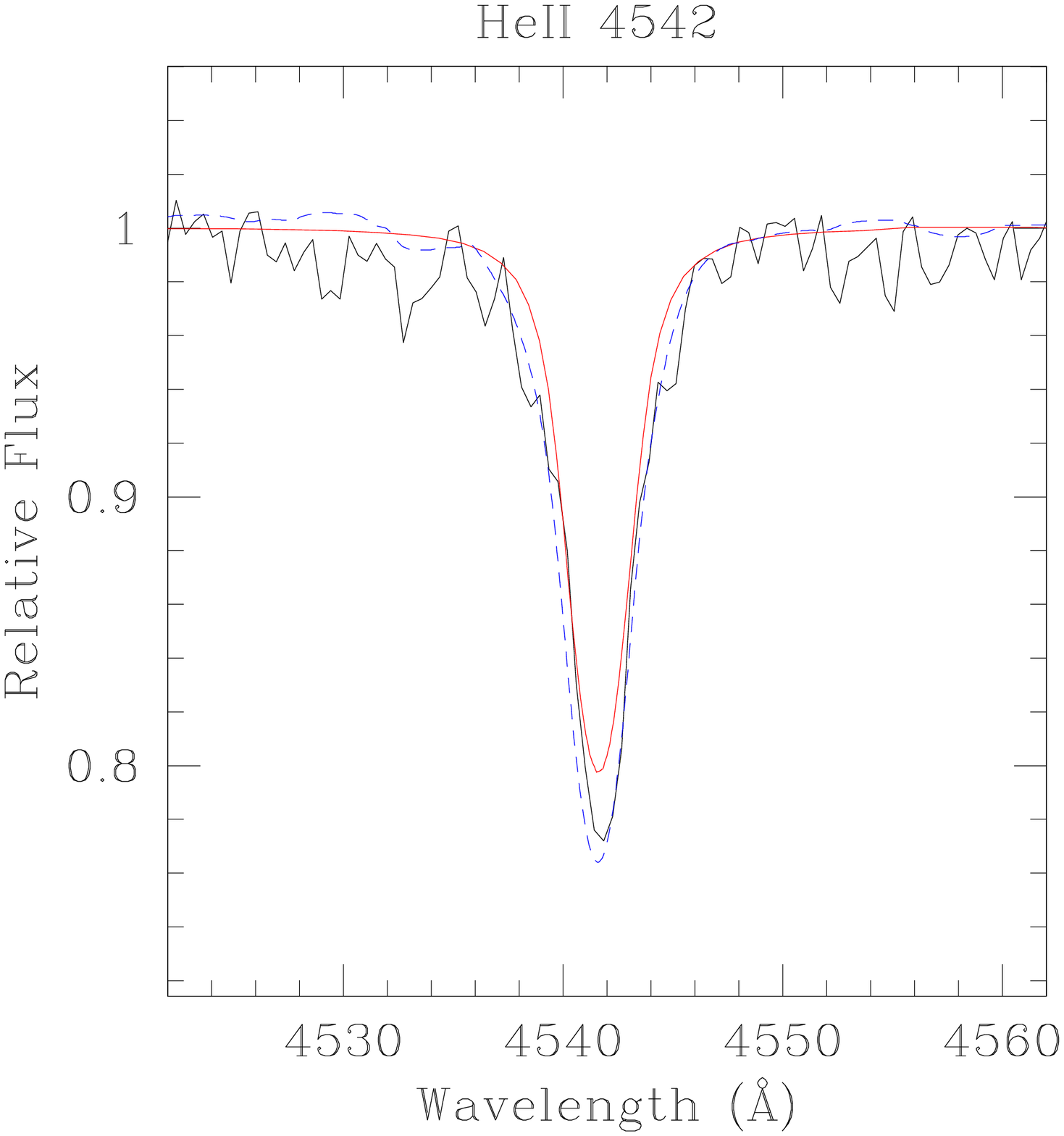}
\plotone{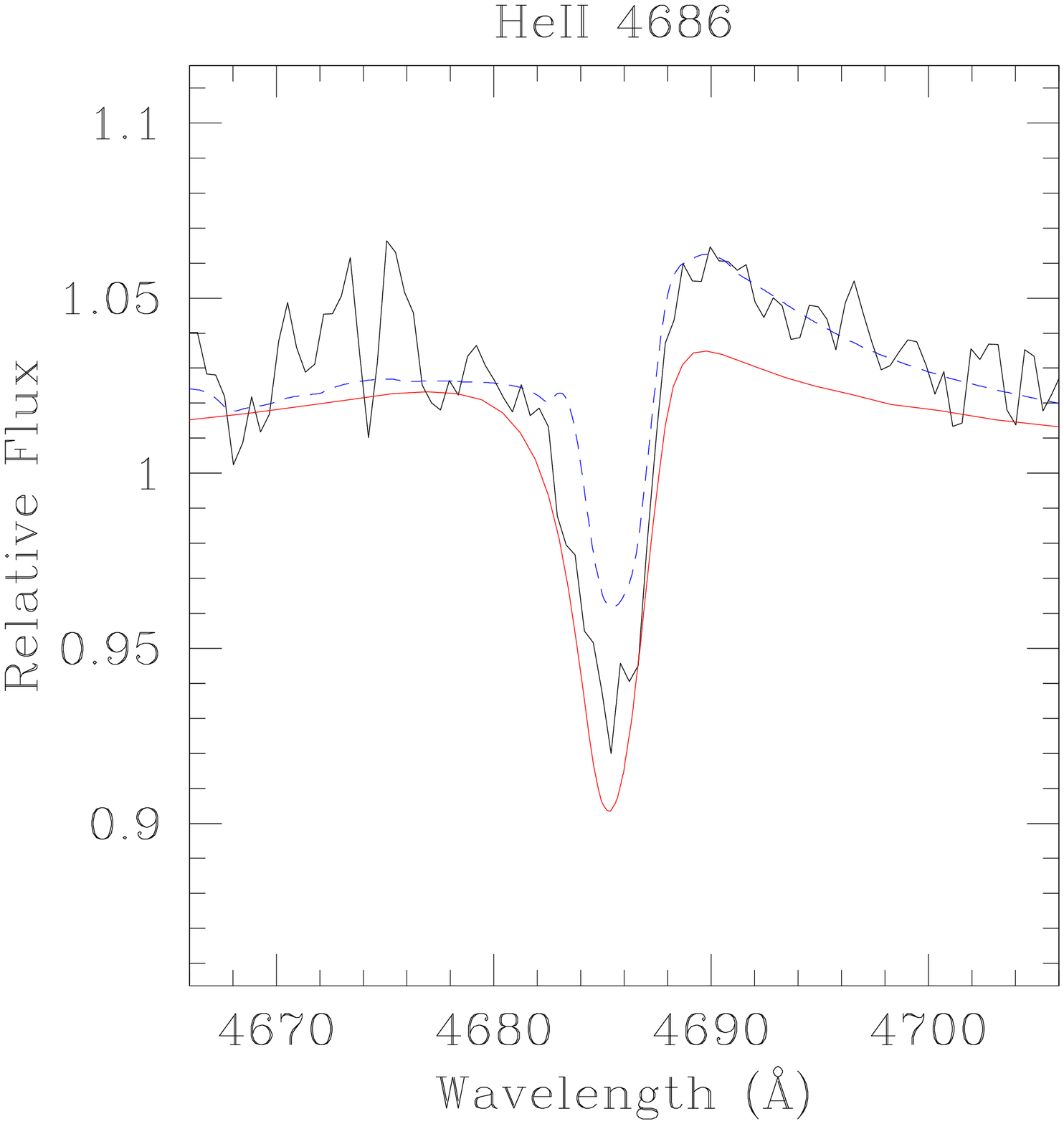}
\caption{\label{fig:LH81:W28-23} Model fits for LH 81:W28-23, an O3.5 V((f+)) star in the LMC.  Black shows the observed spectrum, the red line shows the \fastwind\ fit, and the dashed blue line shows the \cmfgen\ fit. }
\end{figure}

\begin{figure}
\epsscale{0.3}
\plotone{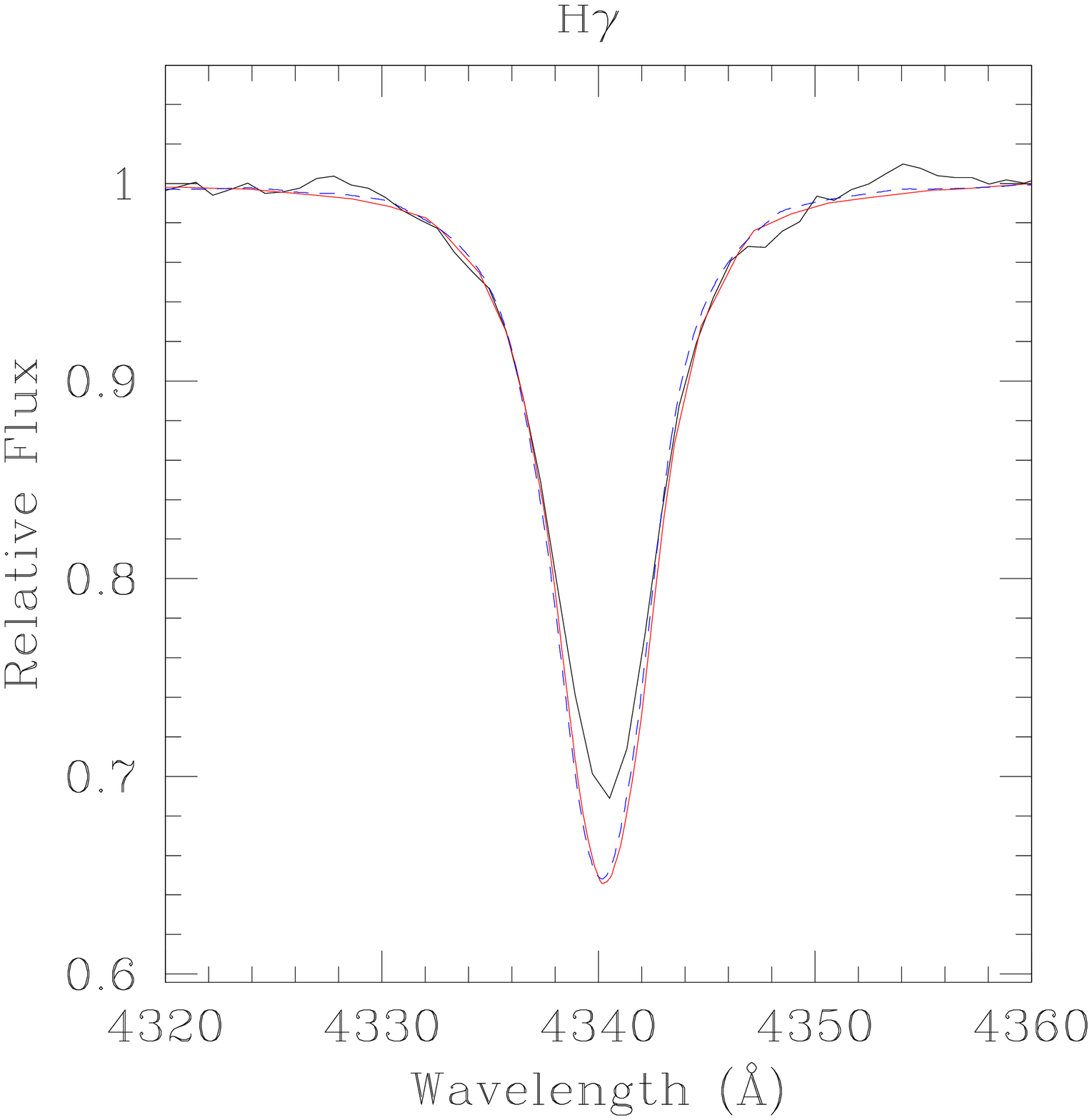}
\plotone{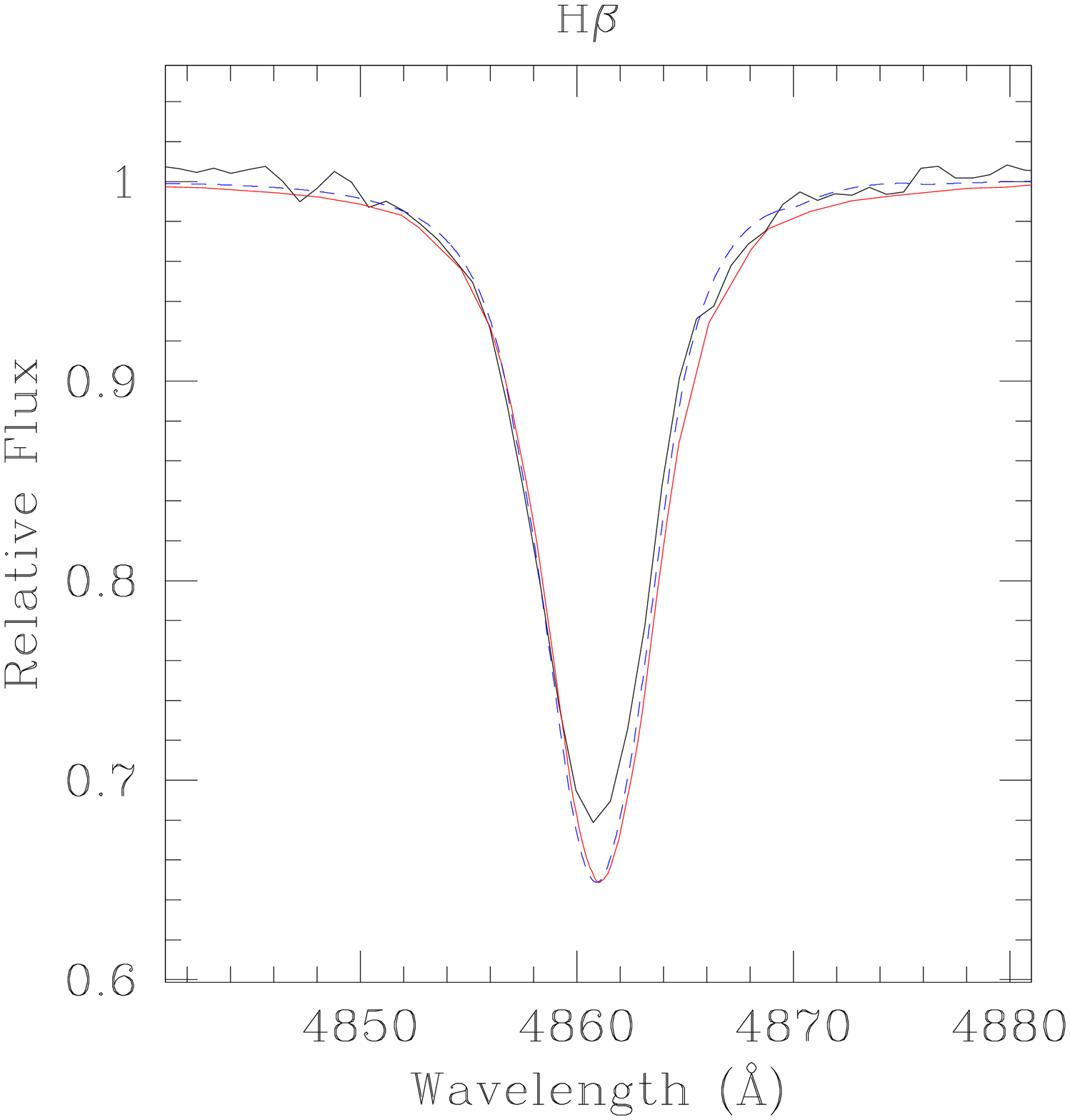}
\plotone{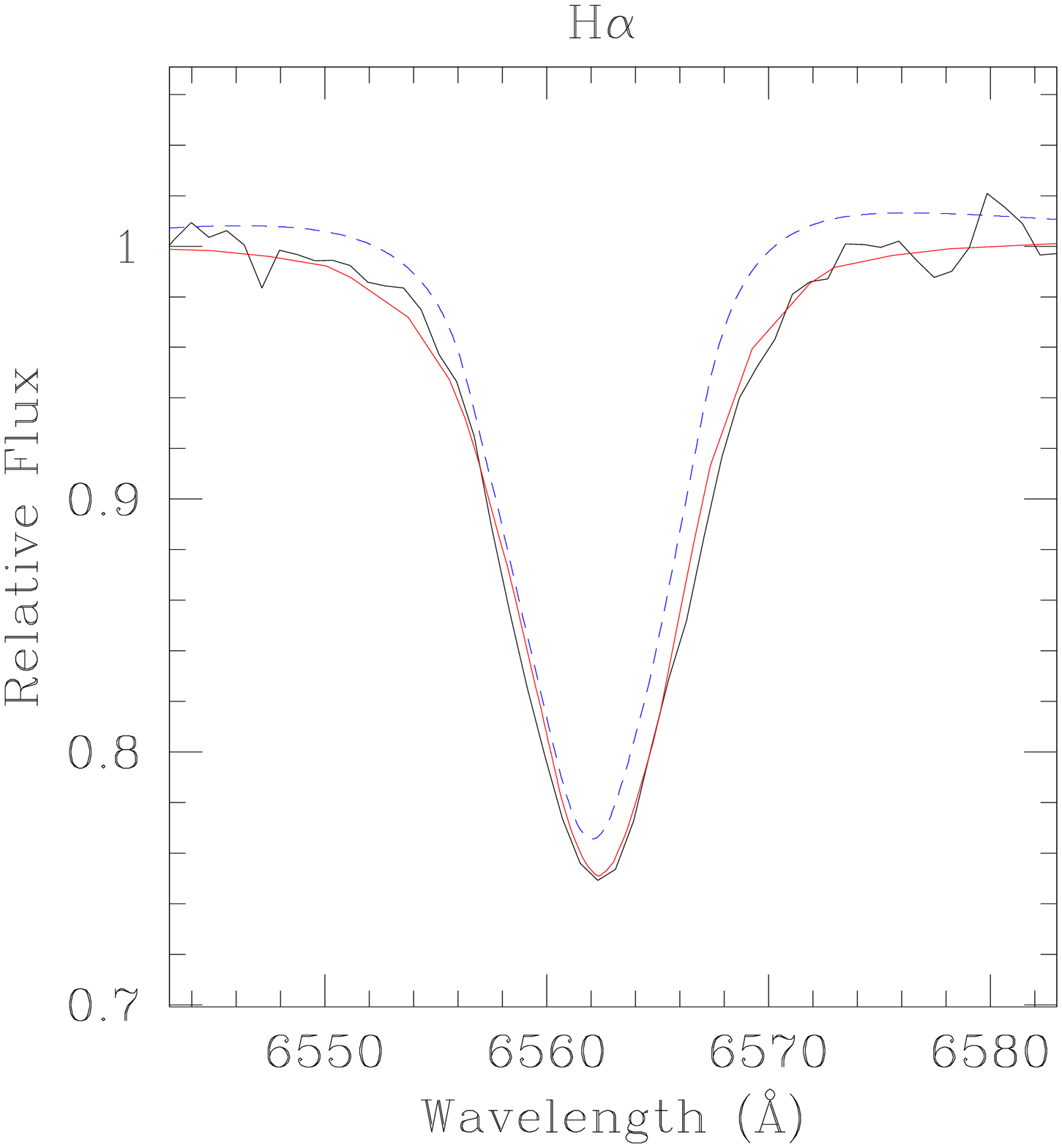}
\plotone{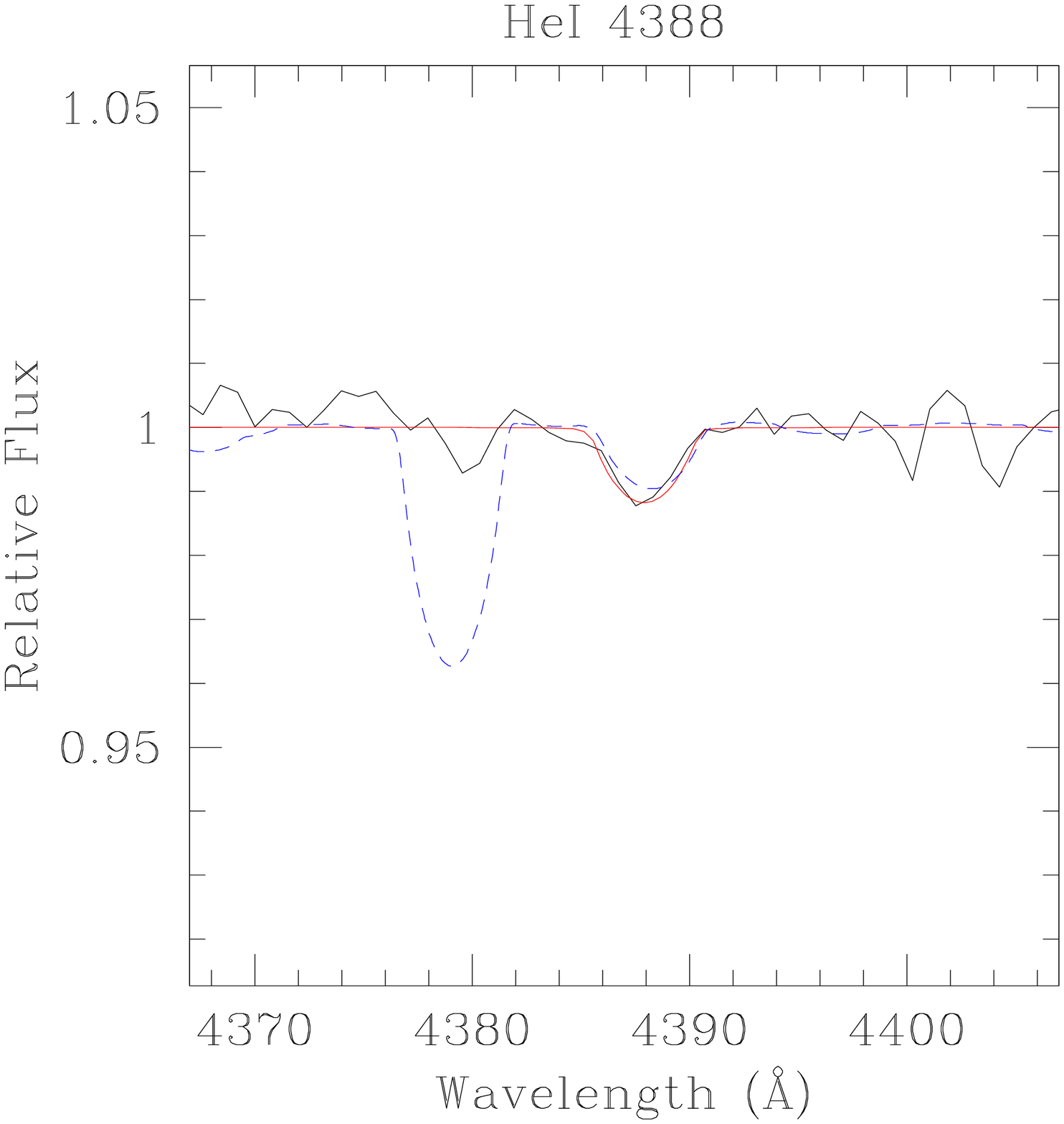}
\plotone{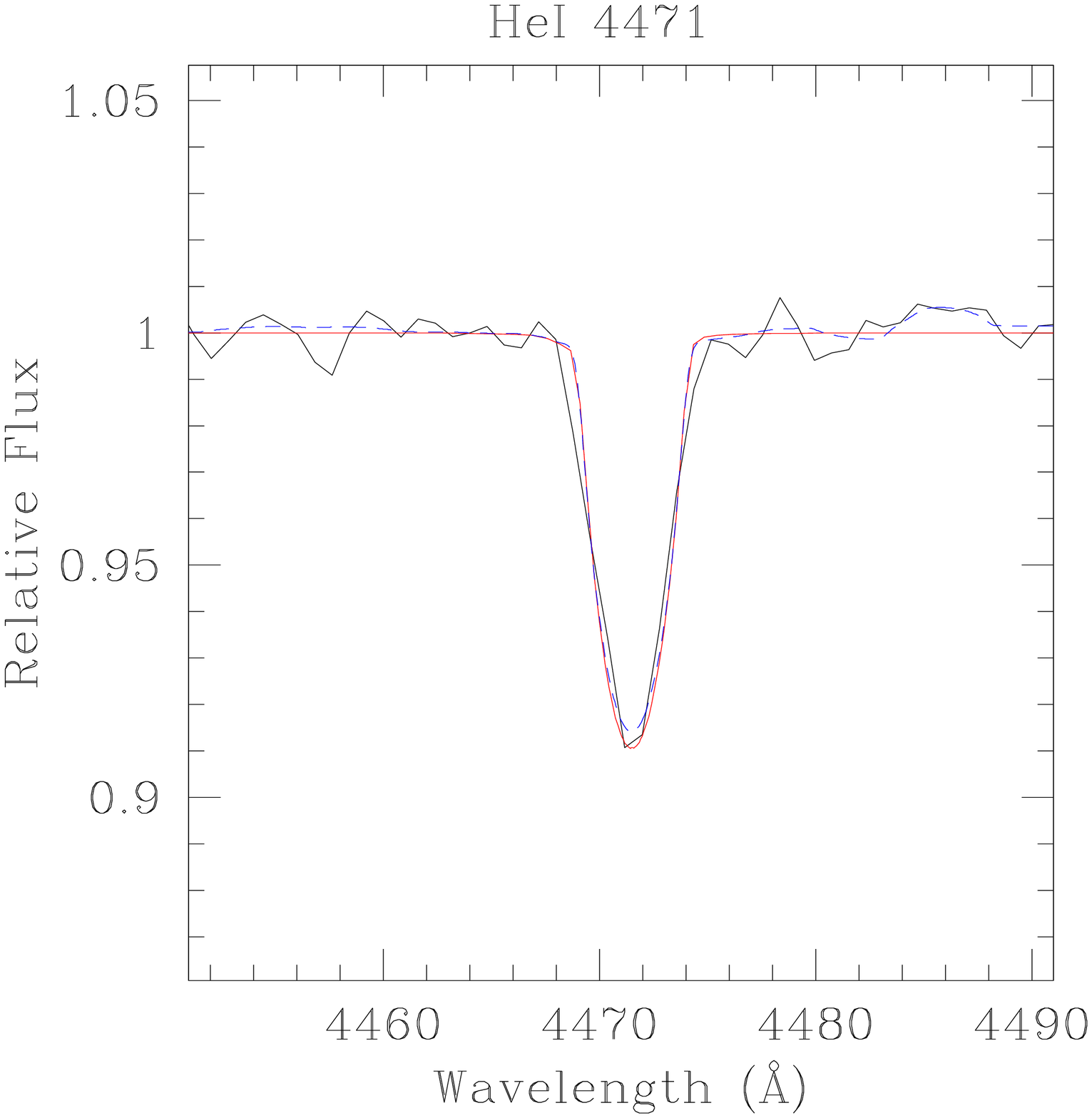}
\plotone{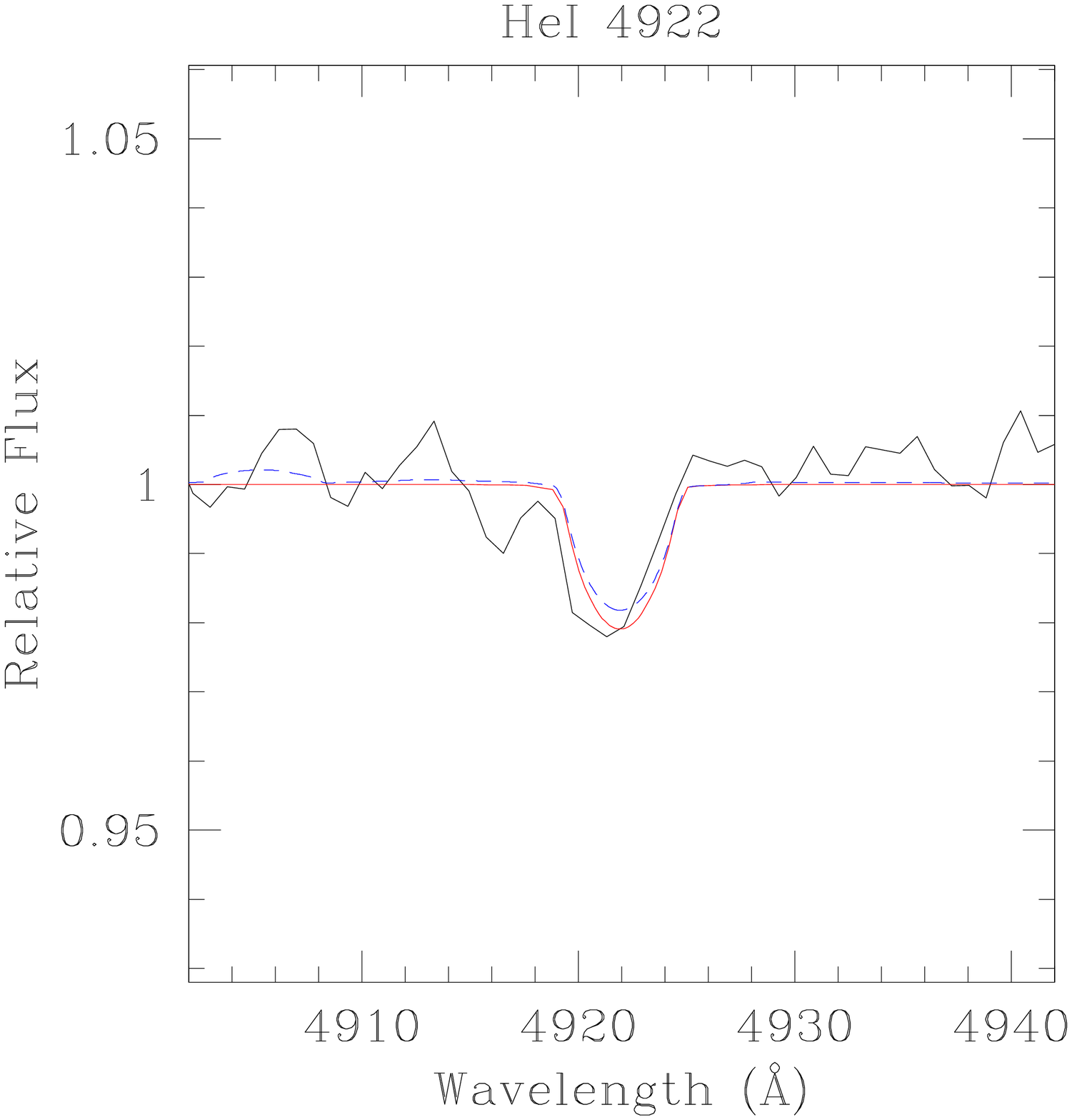}
\plotone{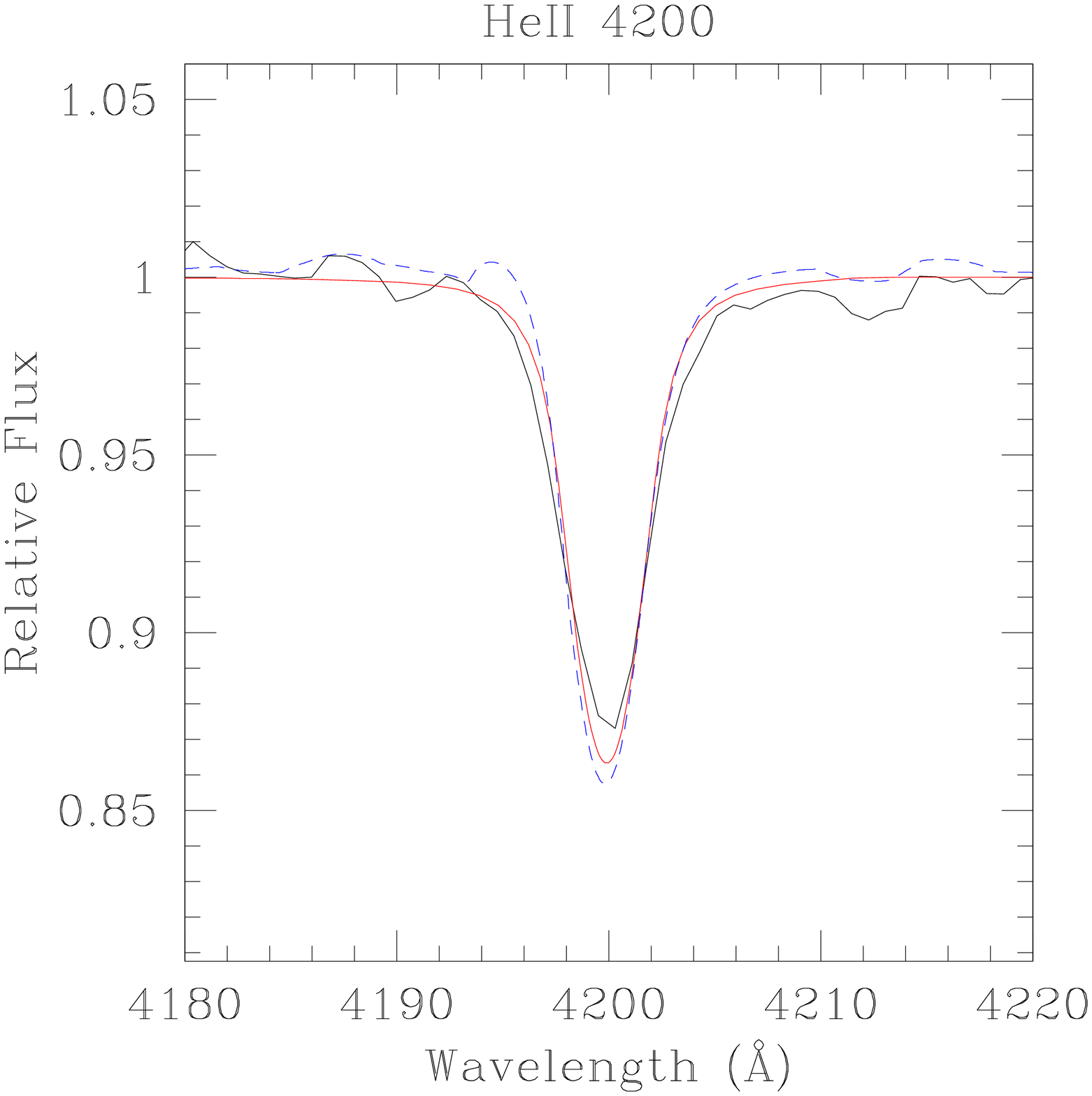}
\plotone{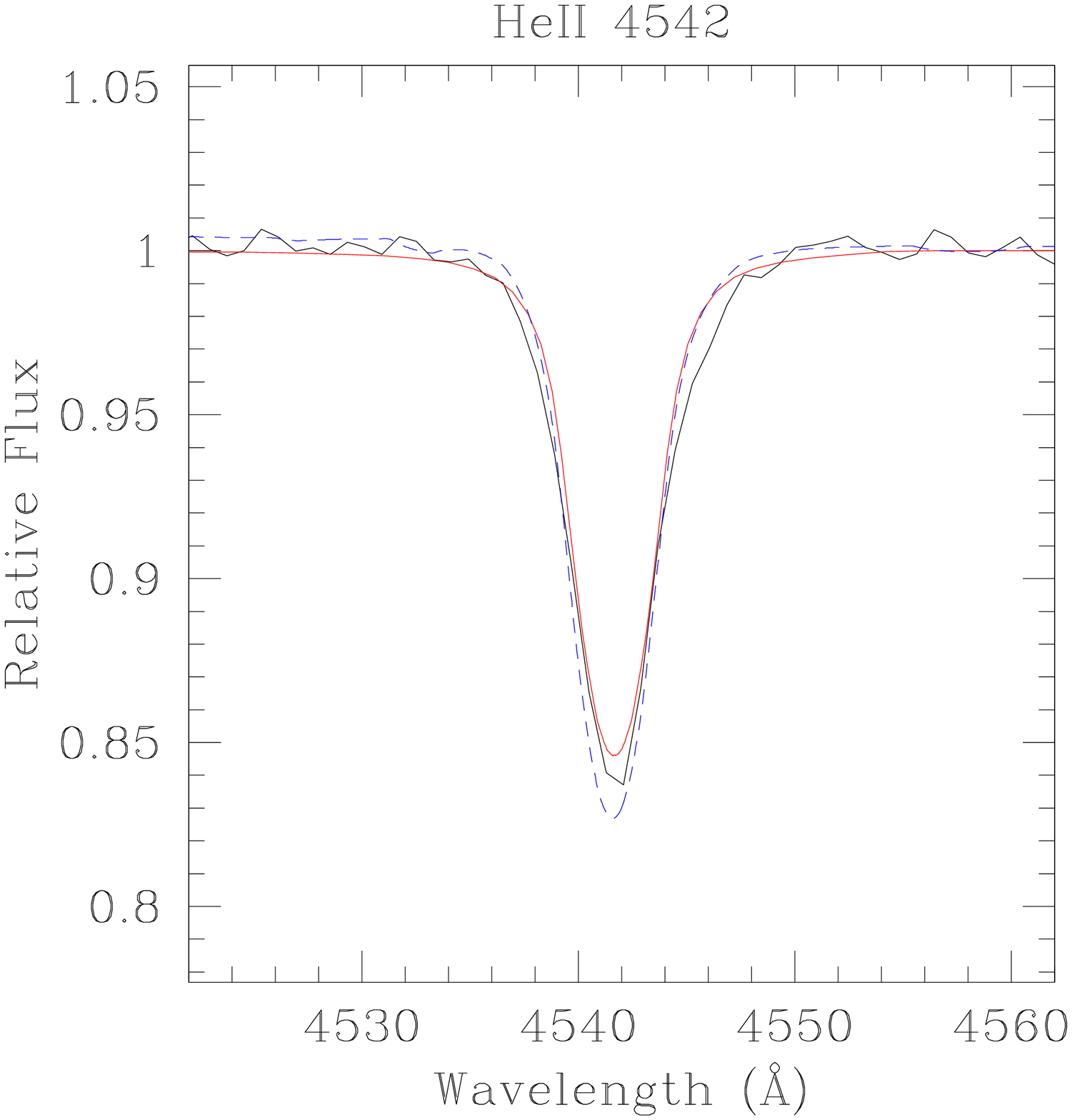}
\plotone{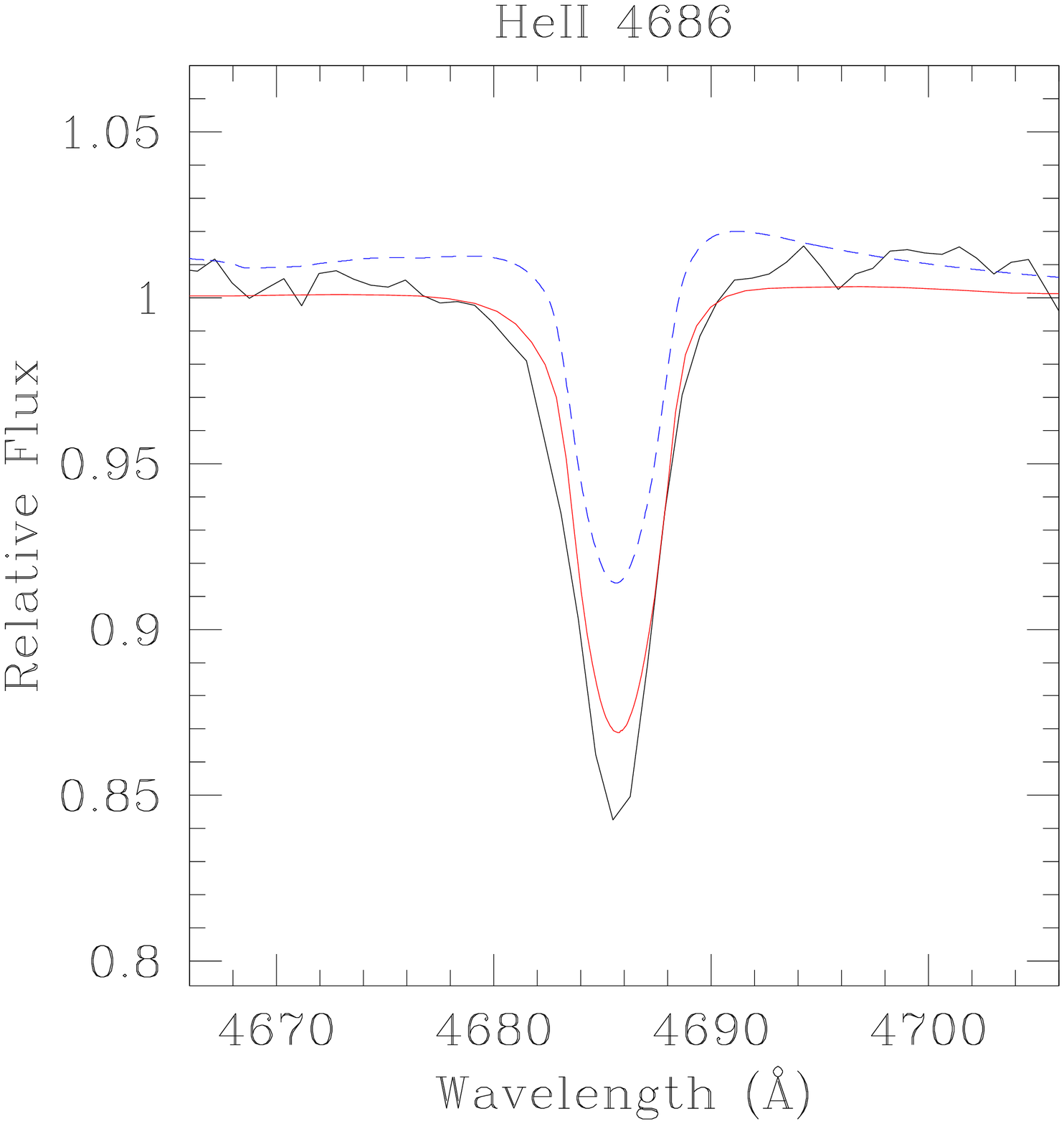}
\caption{\label{fig:Sk-70d69} Model fits for Sk $-70^\circ$69, an O5.5 V((f)) star in the LMC.  Black shows the observed spectrum, the red line shows the \fastwind\ fit, and the dashed blue line shows the \cmfgen\ fit. }
\end{figure}

In Figures~\ref{fig:AzV177}-\ref{fig:Sk-69d124} we show our final model fits for the optical H and He lines (solid red for \fastwind, and dashed blue for \cmfgen), and  in Table~\ref{tab:results} we list the physical parameters derived from our \fastwind\  and \cmfgen\ fits, 
along with the differences.  Rather than discuss the fits for each model, we instead want to call attention to the trends that emerge in the differences between the two codes.  We note in general the fits with both codes are quite good, and despite our excellent
signal-to-noise (several hundred per 1-2~\AA\ spectral resolution element), we may have reached the point where 
much higher signal-to-noise ($\sim$ 1000-2000) would be beneficial in testing the models.

\subsection{Agreements and Disagreements}
\label{Sec:diffs}

The primary driver behind the current undertaking was to see how well the physical properties agreed when the same spectra were fit by the two codes.  Table~\ref{tab:results} provides the answer, at least for this particular sample of ten O-type stars.  

In general, there is no difference in the average effective temperatures determined from the two codes: the mean difference (in the sense of \fastwind\  minus \cmfgen) is -80 K, and the median difference is 0 K.  However, what we think is quite significant is the scatter in the effective temperatures, with a sample standard deviation of 1300 K. This scatter is 2.6$\times$ greater than our fitting precision with each spectrum, which we (perhaps overly optimistically) estimate as 500 K.   The external accuracy of \fastwind\  is usually taken as 1000~K (see, for example, Repolust et al.\ 2004, and Papers I-III), but our confidence fitting the same spectra
is considerably better than that.
In Section~\ref{Sec-HeI} we discuss part of the reason for the occasionally large differences in our temperature determinations.  

In contrast, we find a systematic difference of $-0.12$~dex in the determination of $\log g$ between the two codes, with a scatter (0.07~dex) that is quite similar to our fitting precision (0.05~dex).  We discuss this below in Section~\ref{Sec-logg}, along with the implications for the mass discrepancy. 

Other fitting parameters, i.e., the mass-loss rates and the He/H ratios, agree well between  the two codes, as shown in Table~\ref{tab:results}.   The \cmfgen\ mass-loss values are slightly higher than those of \fastwind\, as we might expect for the difference in how the stratification of clumping is treated, as discussed above.  Still, the differences are small, and the scatters are similar to the uncertainties, both for the mass-loss rates and the He/H ratios.

To better understand how well our results from \cmfgen\ and \fastwind\  agreed, it may be instructive to examine
the differences in the results we find by using two different versions of the same code; i.e., comparing our results with the latest version of \fastwind\  to those we had  previously determined with the same spectra but older versions of the code.  We show the results in Table~\ref{tab:FWresults}.   We see that there is little difference between our old and new values.  It is interesting to note that the scatter in the effective temperature determinations, $\sigma=560$~K, is essentially the same as what we consider our typical uncertainty in determining these values, about 500~K.  In our comparison between 
\fastwind\  and \cmfgen\ in Table~\ref{tab:results} we find a $\sigma$ that is 2.5$\times$ larger.
This underscores that the differences we find between \fastwind\  and \cmfgen, while not systematic, are much greater than our precision in determining these quantities.  As discussed below (Section~\ref{Sec-HeI}) we feel these differences are due to the different behaviors of the code with regard to the He~I singlets and triplets.  Note too, in Table~\ref{tab:FWresults}, the utter consistency in the determination of the surface gravities, which significantly bears on the discussion in Section~\ref{Sec-logg} concerning the mass discrepancy.    We also note that the new values of the mass-loss rates are a bit higher than we would get
by decreasing the old values by a factor of $3.3$, $1/\sqrt{f}$, as previously described.

\subsection{Problems with He~I and the Implications for Determining Effective Temperatures}
\label{Sec-HeI}

\subsubsection{The Problem with Triplets and the Effects of Microturbulence}

First, consider the He~I triplet \fastwind\  problem.   Repolust et al.\ (2004) invoke the  ``generalized dilution effect" (Voels et al.\ 1989)
to explain the weakness of the \fastwind\  He~I $\lambda 4471$ model profile for mid-to-late type giants and supergiants. The explanation offered by Voels et al\ (1989) would only apply to plane-parallel models without winds, and so would not strictly be applicable to \fastwind\ 's problem.  Repolust et al.\ (2004) of course realized this, but could offer no physical explanation and kept the terminology to describe the effect.  Here we confirm that the problem is not present in \cmfgen\ models.   Compare the He~I $\lambda 4471$ profile fits in our three late O-type supergiants, AzV 233 (O9.5~II, Figure~\ref{fig:AzV223}), BI 170 (O9.5~I, Figure~\ref{fig:BI170}) and 
Sk $-69^\circ$124\footnote{Below we conclude that Sk $-69^\circ$124 is likely a binary.} (O9.7~I, Figure~\ref{fig:Sk-69d124}).  The \fastwind\  profile (solid red) for He~I $\lambda 4471$ is significantly weak in all three cases,  but the \cmfgen\ profiles (dashed blue) are spot on.  Thus, the problem cannot be due to something common to both codes.  In most other cases, the \fastwind\  and the \cmfgen\ $\lambda 4471$ profiles agree, and match the spectra well, except for the O5.5 I(f) star AzV~75 (Figure~\ref{fig:AzV75}), where neither give a particularly good fit. Whatever the problem is with the \fastwind\  $\lambda 4471$ profile, the problem seems to begin to occur after O6~I, as \fastwind\  still does a good job for the O6 I(f) star AzV 26 (Figure~\ref{fig:AzV26}). 

Is the problem that \fastwind\  has with He I $\lambda 4471$ restricted to just this one line, or does it affect other triplets?
The He~I $\lambda 4471$ transition
is $1s2p$ $^3$P$^o - 1s4d$ $^3$D.  (For convenience we have reproduced the Moore \& Merrill (1968) version of the Grotrian diagram for He I in Figure~\ref{fig:Grot}.) The He~I~$\lambda 4026$ line is a similar transition ($1s2p$ $^3$P$^o - 1s5d$ $^3$D), but is a blend  with the He~II~$\lambda 4026$ line.  The He~I $\lambda 5876$ line is also similar
($1s2p$ $^3$P$^o - 1s3d$ $^3$D) and is usually quite strong, but \fastwind\  doesn't include this line in its formal solution\footnote{Such lines were not included as spectroscopic studies have
often focused on the blue part of the spectrum ($<$5000~\AA), although
in principle such lines could be readily added to the \fastwind\ 
output.}. In fact, the only other He~I triplet line profile that is normally computed by \fastwind\  is He~I $\lambda 4713$ ($1s2p$ $^3$P$^o - 1s4s$ $^3$S), and that is relatively weak.  (It is for that reason that we did not include it in our assessment of the fits.) We show the fits of \fastwind\  and \cmfgen\ for He~I $\lambda 4713$ in Figures~\ref{fig:tripletsA} and \ref{fig:tripletsB}, along with
the \cmfgen\ fit for He I $\lambda 5876$. 

We see in Figure~\ref{fig:tripletsA} that both \fastwind\  and \cmfgen\ do a relatively good job of fitting the He~I triplets for the early and intermediate O dwarfs and supergiants\footnote{We do not show the fits for the earliest two stars, as the He~I lines are too weak for the purposes of this comparison; also simply to save space, we do not include that of the O5.5 V((f)) Sk $-70^\circ$69, but its fit is every bit as good as that of the O5.5 V((f)) star AzV 388, which is shown.}.  None of that is surprising.  When we turn to the late-type supergiants in Figure~\ref{fig:tripletsB} we are also not surprised to find that when \fastwind\  produces too weak a model profile for He~I $\lambda 4471$ it also produces too weak a profile for He~I $\lambda 4713$.   What we do find surprising is that although \cmfgen\ does a good job of matching the He~I $\lambda 4471$ profile, it also produces too weak a profile for the He~I $\lambda 4713$ and He~I $\lambda 5876$ profiles for the three late-type supergiants, AzV 223, BI 170, and Sk $-69^\circ$124. 

In experimenting with possible cures, we investigated the effect of increasing the microturbulence velocity in computing the emergent profiles.  In this temperature regime, we had used 10 km s$^{-1}$ both for the \fastwind\  and \cmfgen\ formal calculations.  A few experiments showed that increasing the microturbulence velocity strengthened  He I $\lambda 5876$ significantly more than it strengthened He I $\lambda 4471$.    We thus tried using the same \fastwind\  and \cmfgen\ atmosphere models but recomputed the synthetic spectra using values of 15 and 20 km s$^{-1}$.  In Figures~\ref{fig:MicroAzV223} through \ref{fig:MicroSk-69d124} we compare these with the original fits for the He I triplet lines.  Indeed, increasing the microturbulence to 15 km s$^{-1}$ fixes the problem  with the \cmfgen\ fits (shown in dashed blue) of  the He I $\lambda 5876$ lines in AzV 223 (Figure~\ref{fig:MicroAzV223}) and BI 170 (Figure~\ref{fig:MicroBI170}), while still maintaing excellent fits for He I $\lambda 4471$ and $\lambda 4713$. For Sk $-69^\circ$124 we would have to use a slightly higher value (but not as high as 20 km s$^{-1}$), and the solution would still be a bit of a compromise, with He I $\lambda 4471$ slightly too strong and He I $\lambda 5876$ slightly too weak, but still quite acceptable. (We argue below that Sk $-69^\circ$124 is a binary.)

Even more surprising, however, is the effect that the microturbulence has on the He I $\lambda 4471$ profile problem  \fastwind\  has
with the late O-type supergiants.  With a value of 20 km s$^{-1}$ for the microturbulence, the \fastwind\  fits (shown in red) are excellent.  In the next section we will show that this does not come at the expense of any other lines, and in fact it improves the fits in some cases.  It remains an unexplored mystery why the He I $\lambda 4471$ profile is more affected by microturbulence with the \fastwind\  formal solution than with that of \cmfgen.  Note, however, that the improvement of the \cmfgen\ fit for He I $\lambda 5876$ with a $\sim 15$ km s$^{-1}$ value for the microturbulence does suggest that a large value (i.e., 15-20 km s$^{-1}$) may be a more appropriate choice when modeling late-type O supergiants. 

Hubeny et al.\ (1991) have argued that the presence of microturbulence should result in a turbulence pressure term 
in the equation of hydrostatic equilibrium, which would lead to increased surface gravities.  (See also the 3D convective simulations of the sun by Trampedach et al. 2013, which shows that turbulence affects the sun's hydrostatic structure.) Neither \cmfgen\ nor \fastwind\ take this approach.  It is not clear at this time if the need for microturbulence is due to real turbulence, or if it is due to a combination
of pulsations, wind-velocity fields, or even simply errors in our still rather simplified treatment of the complex physics involved.

\subsubsection{Singlets}

\begin{figure}
\epsscale{0.3}
\plotone{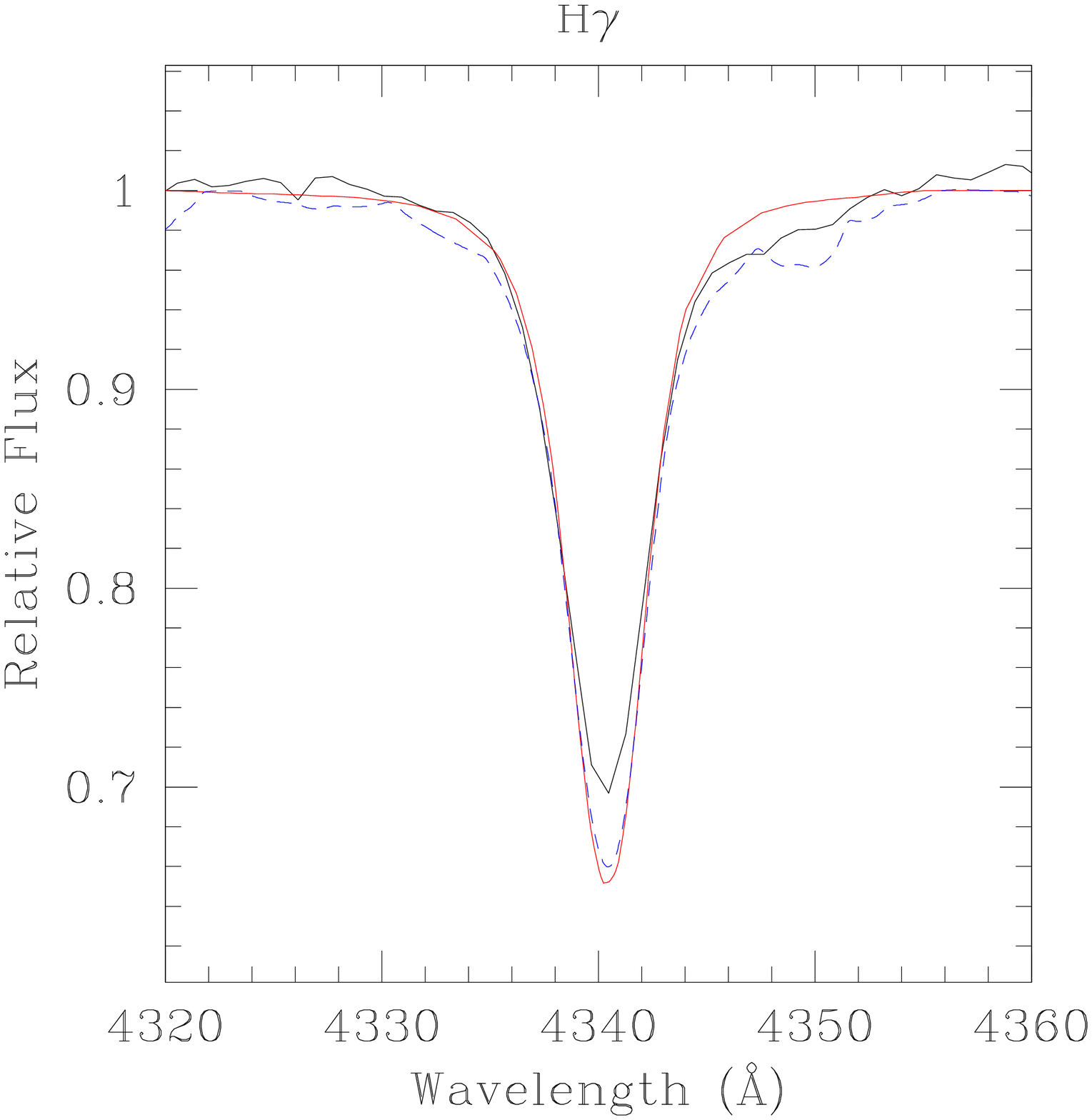}
\plotone{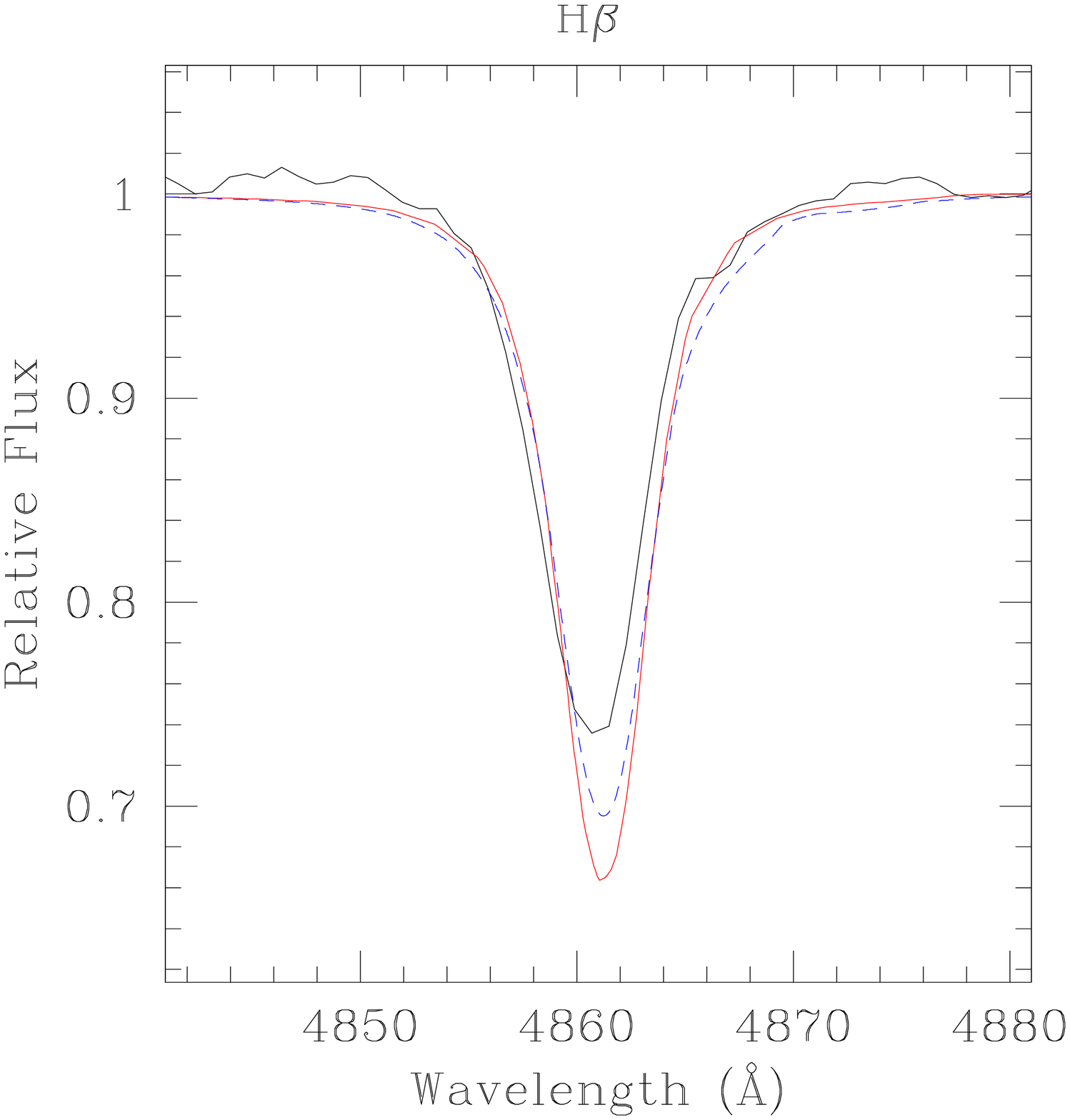}
\plotone{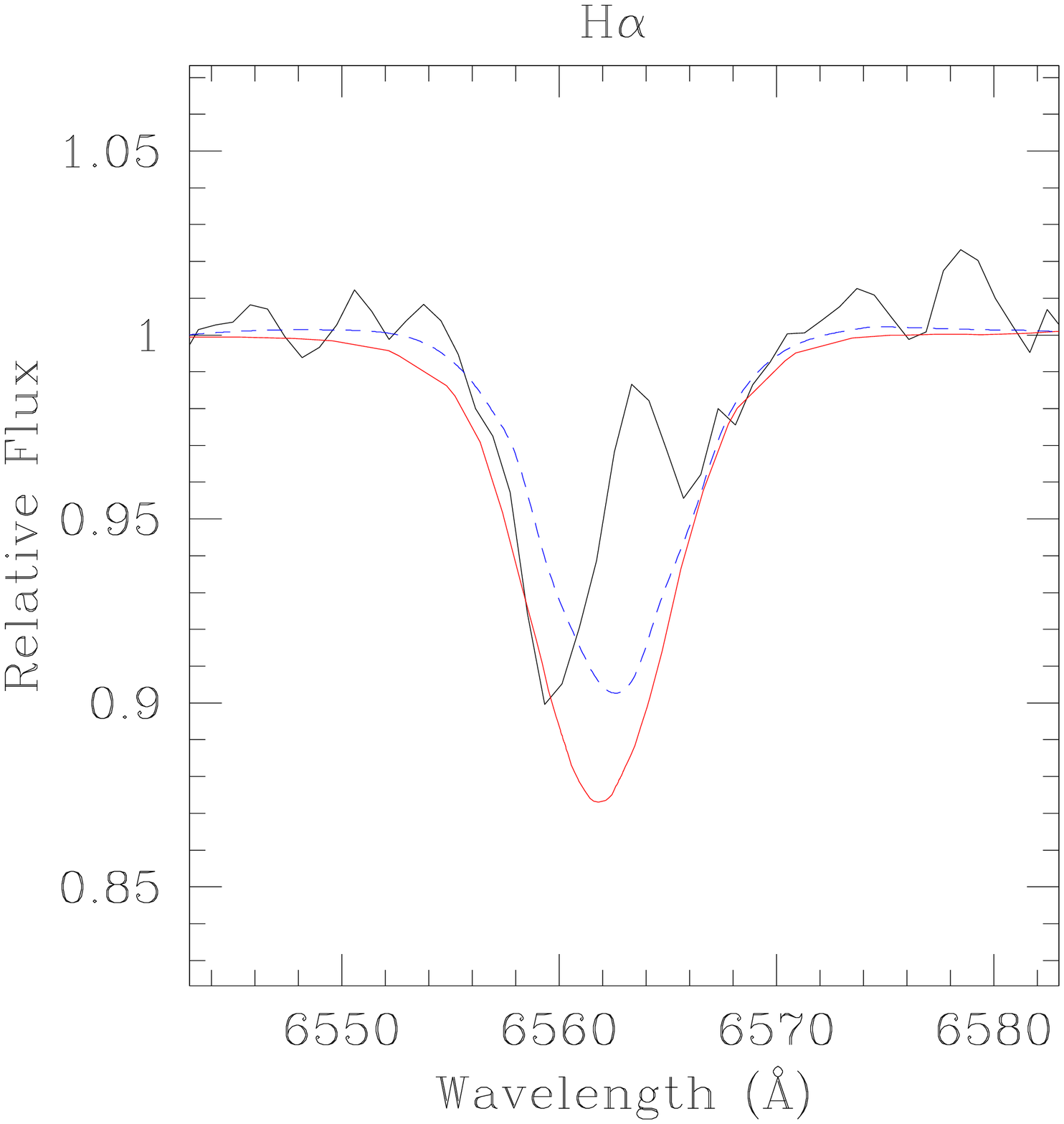}
\plotone{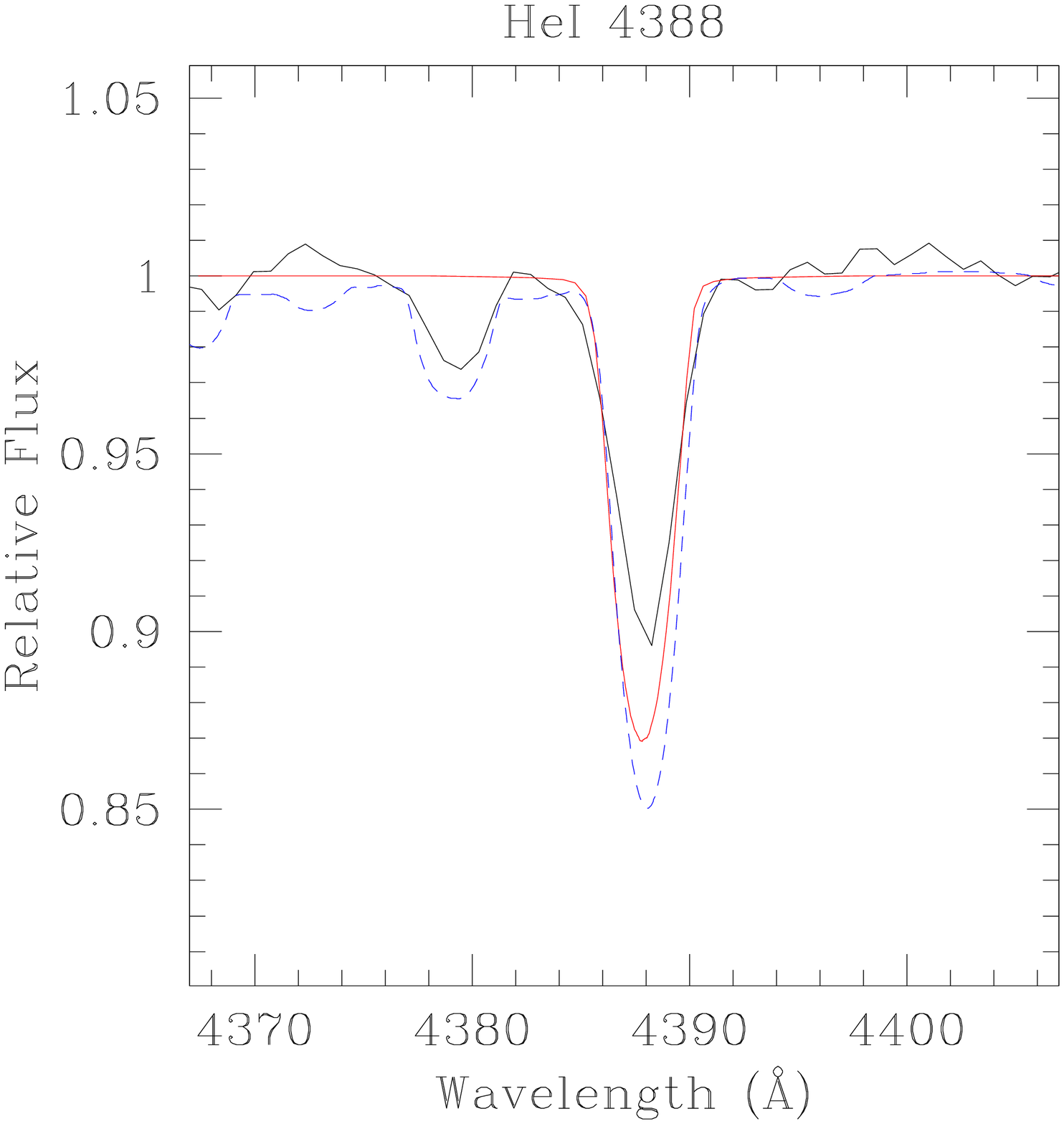}
\plotone{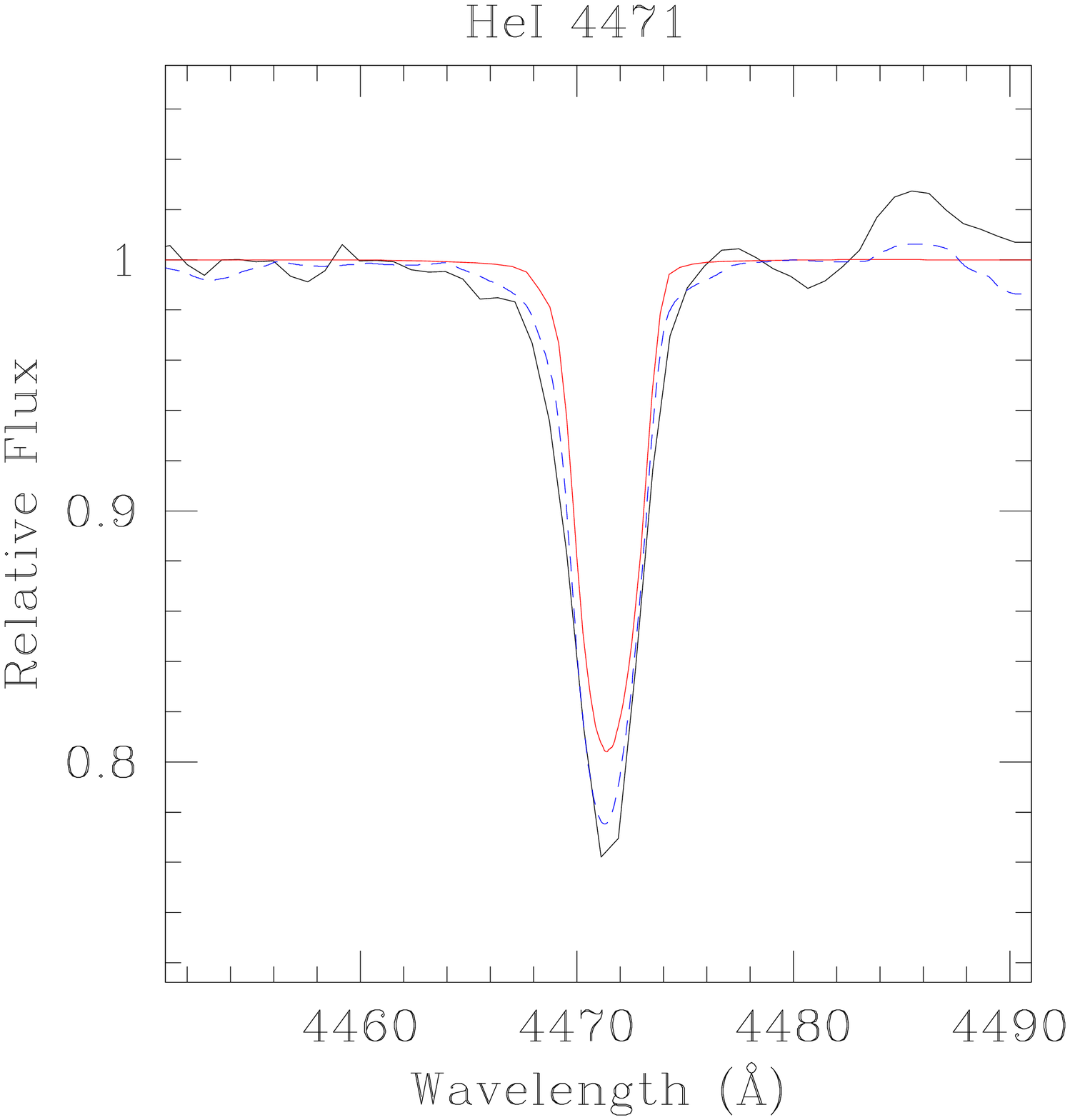}
\plotone{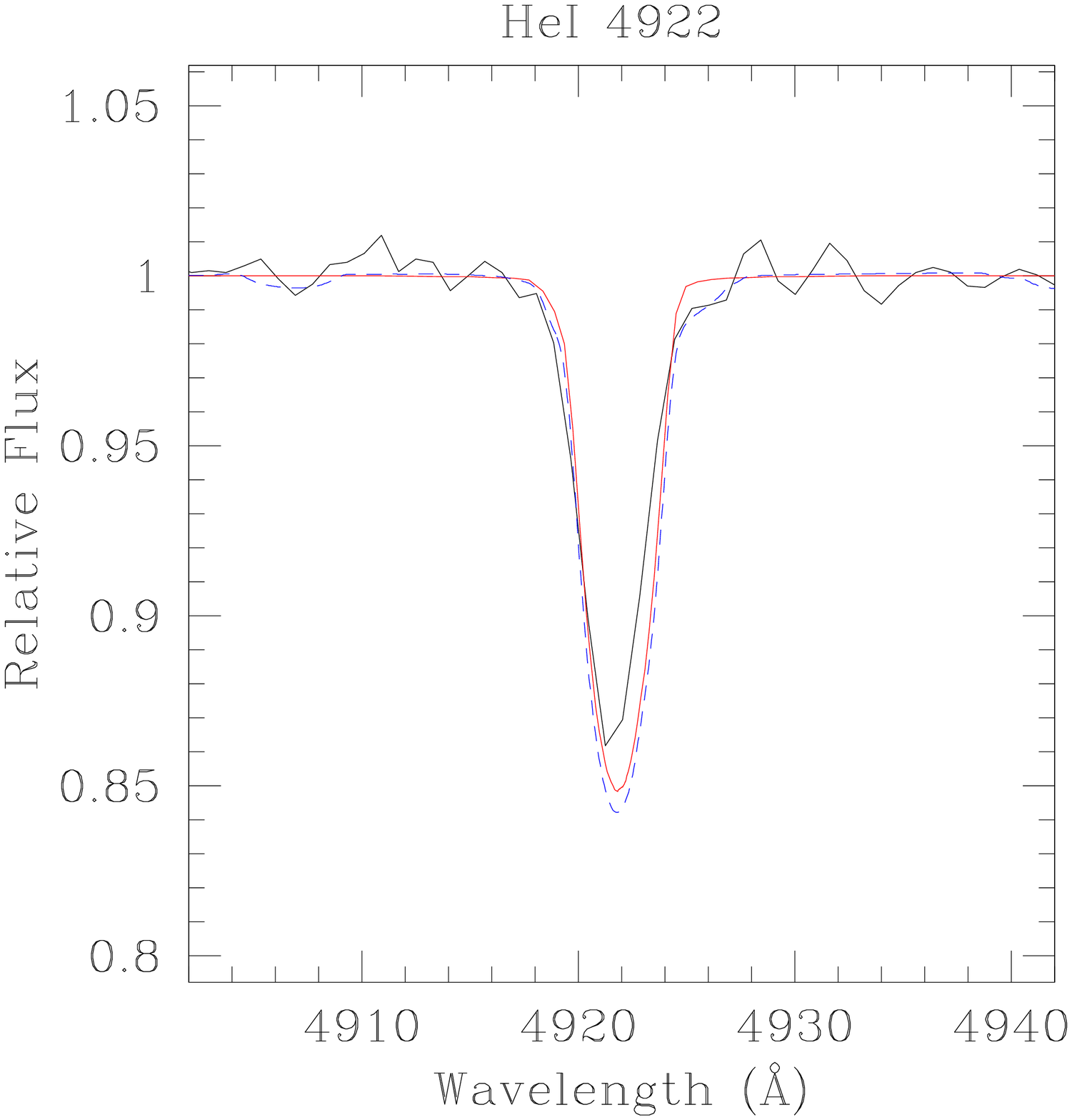}
\plotone{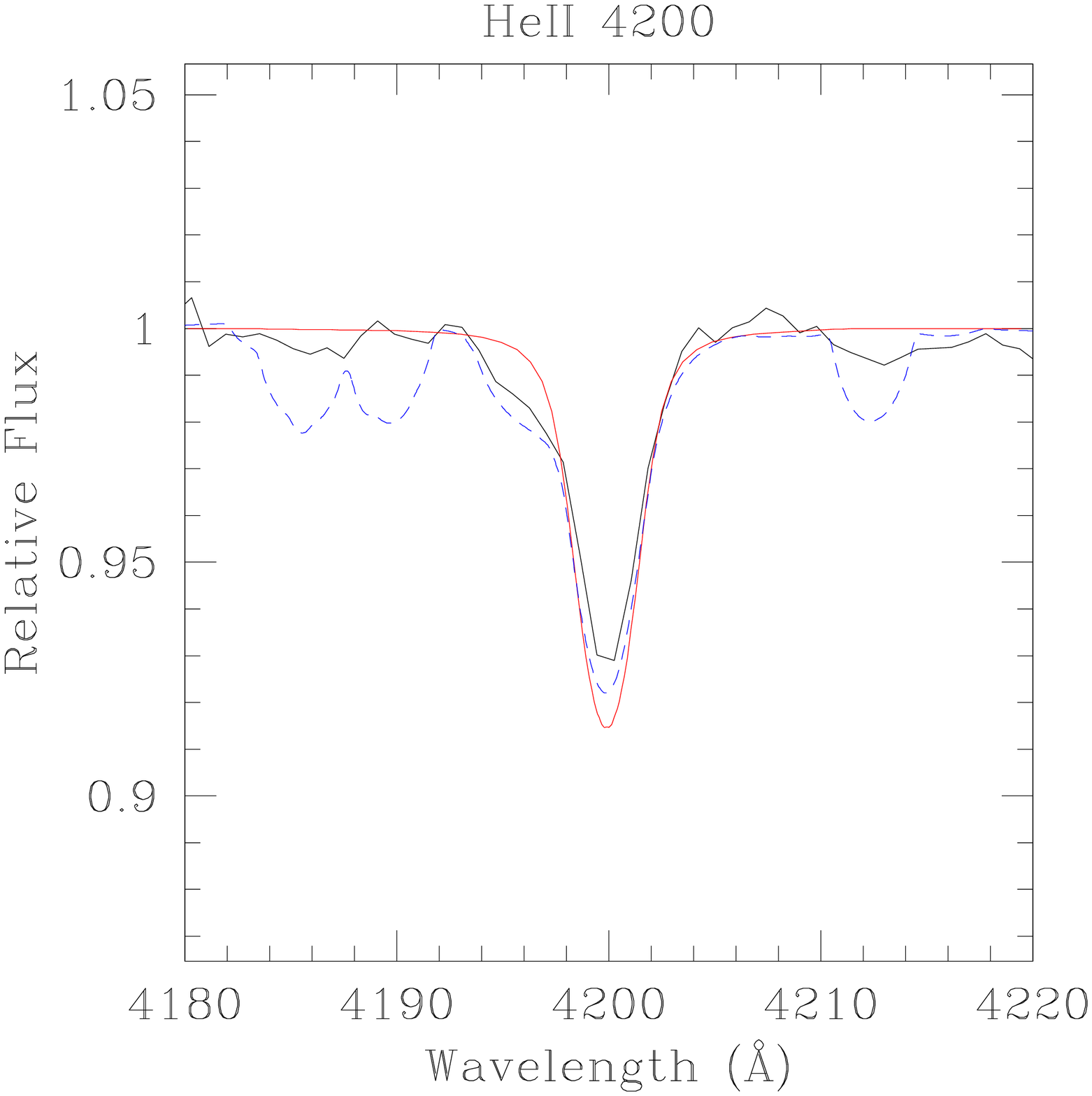}
\plotone{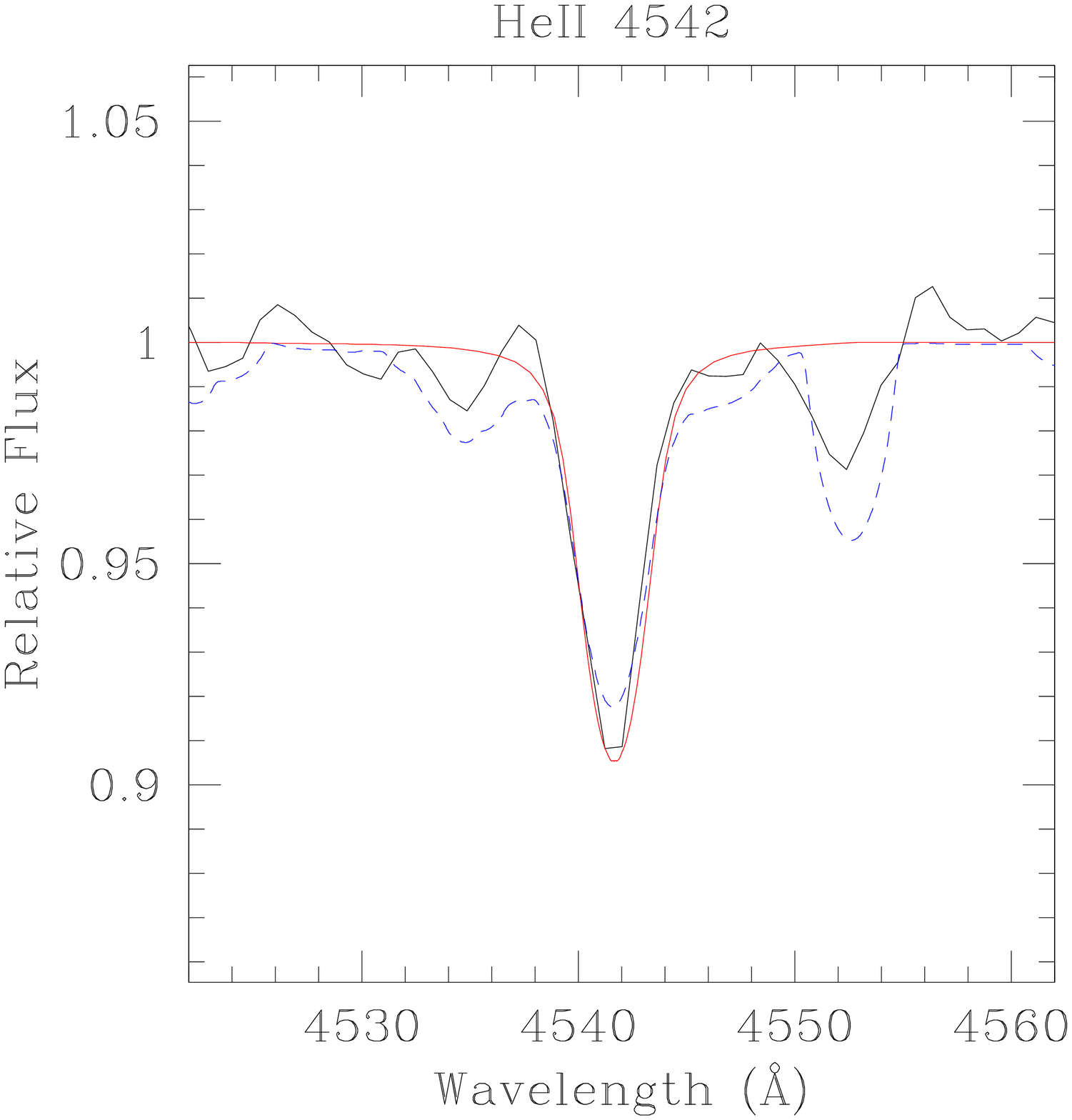}
\plotone{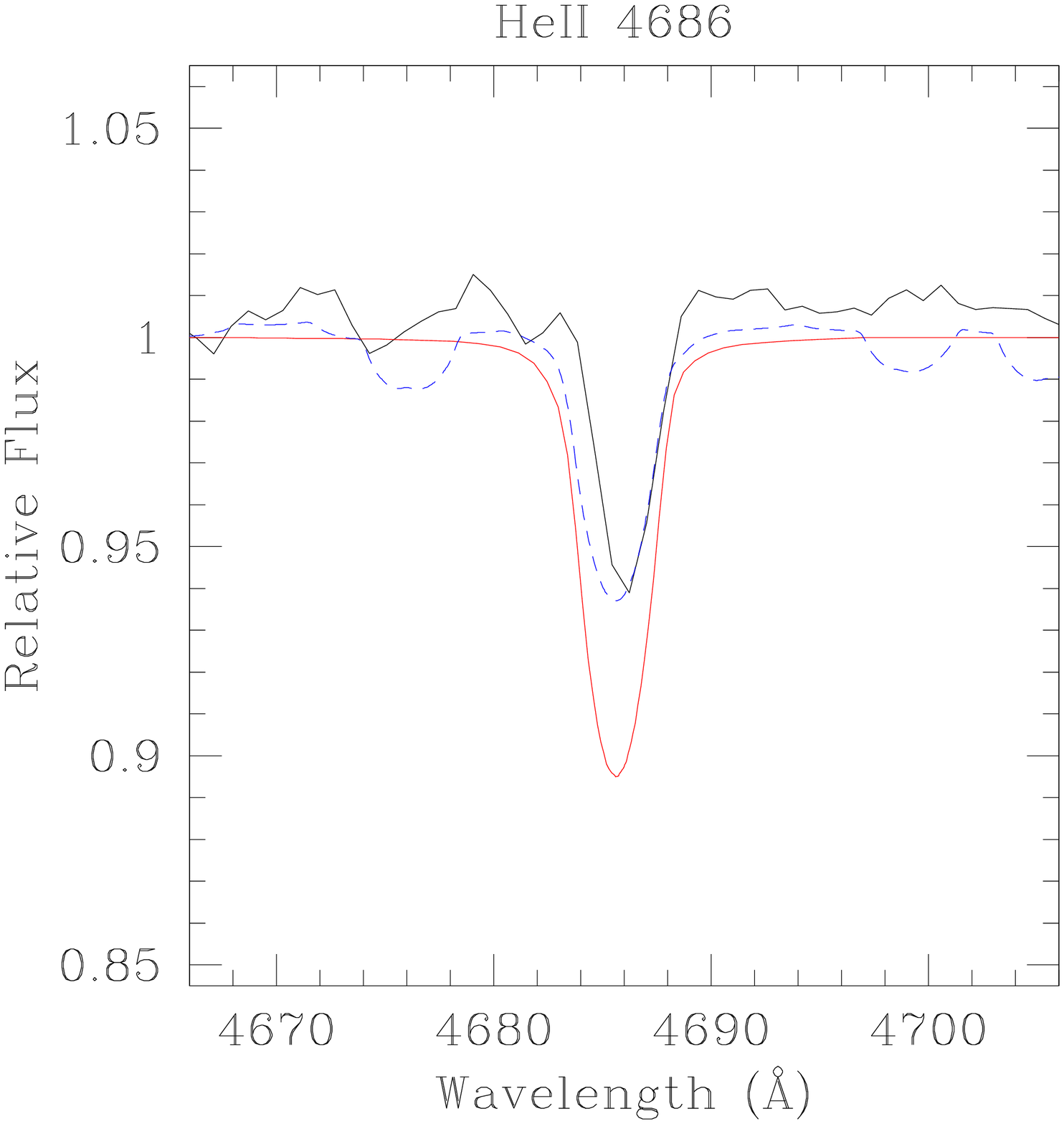}
\caption{\label{fig:BI170} Model fits for BI 170, an O9.5 I star in the LMC.  Black shows the observed spectrum, the red line shows the \fastwind\ fit, and the dashed blue line shows the \cmfgen\ fit. }
\end{figure}

\begin{figure}
\epsscale{0.3}
\plotone{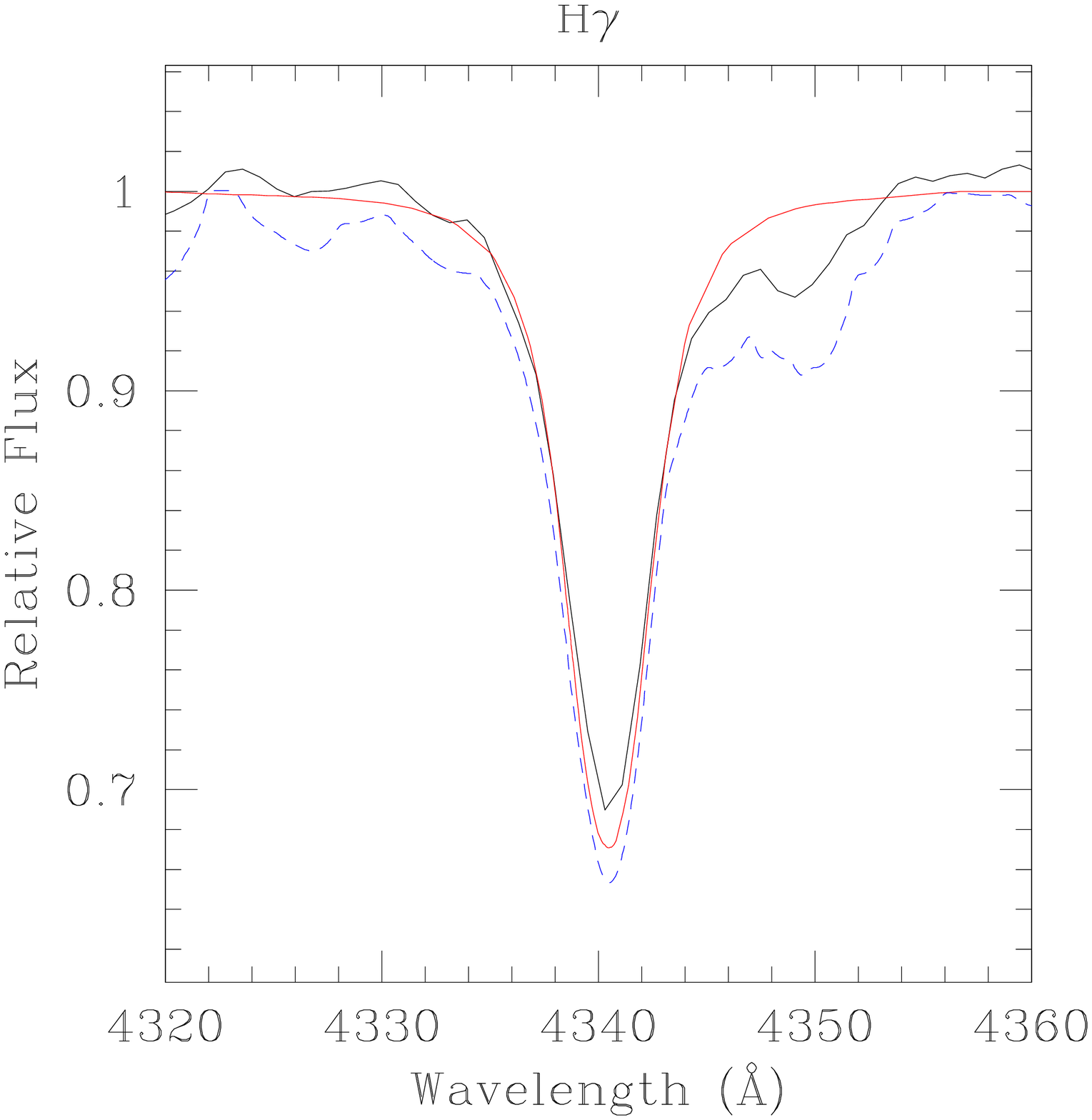}
\plotone{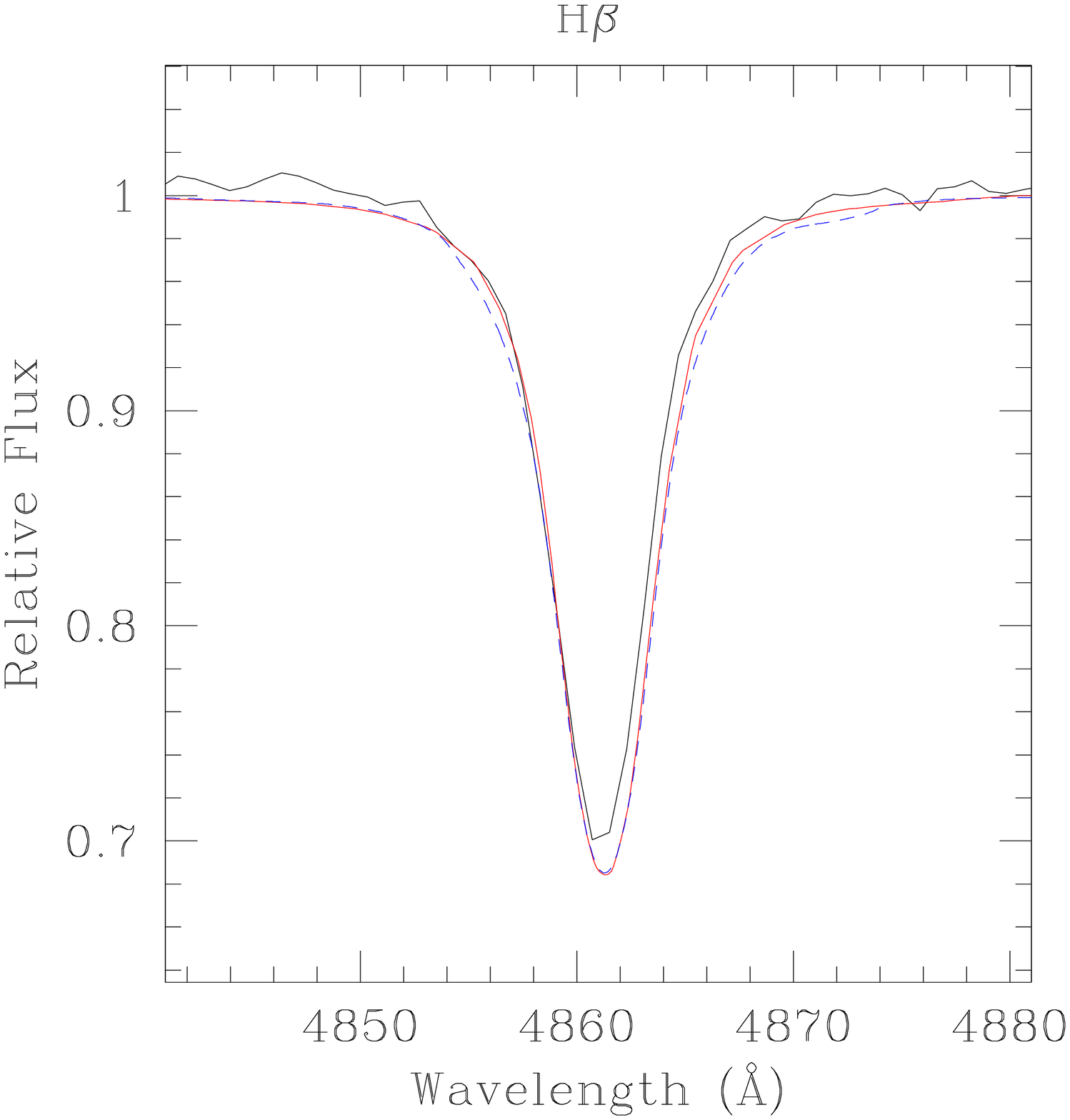}
\plotone{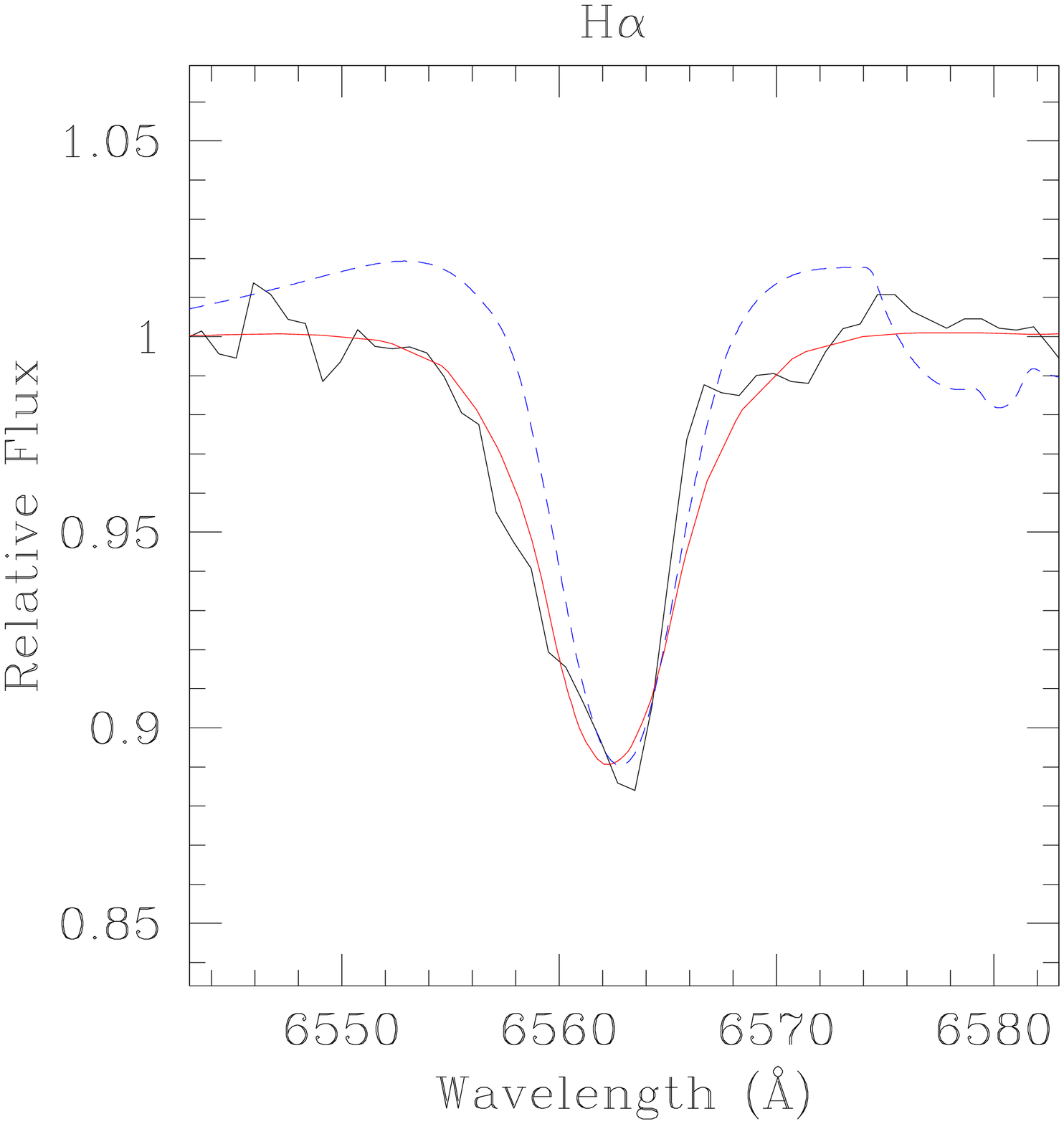}
\plotone{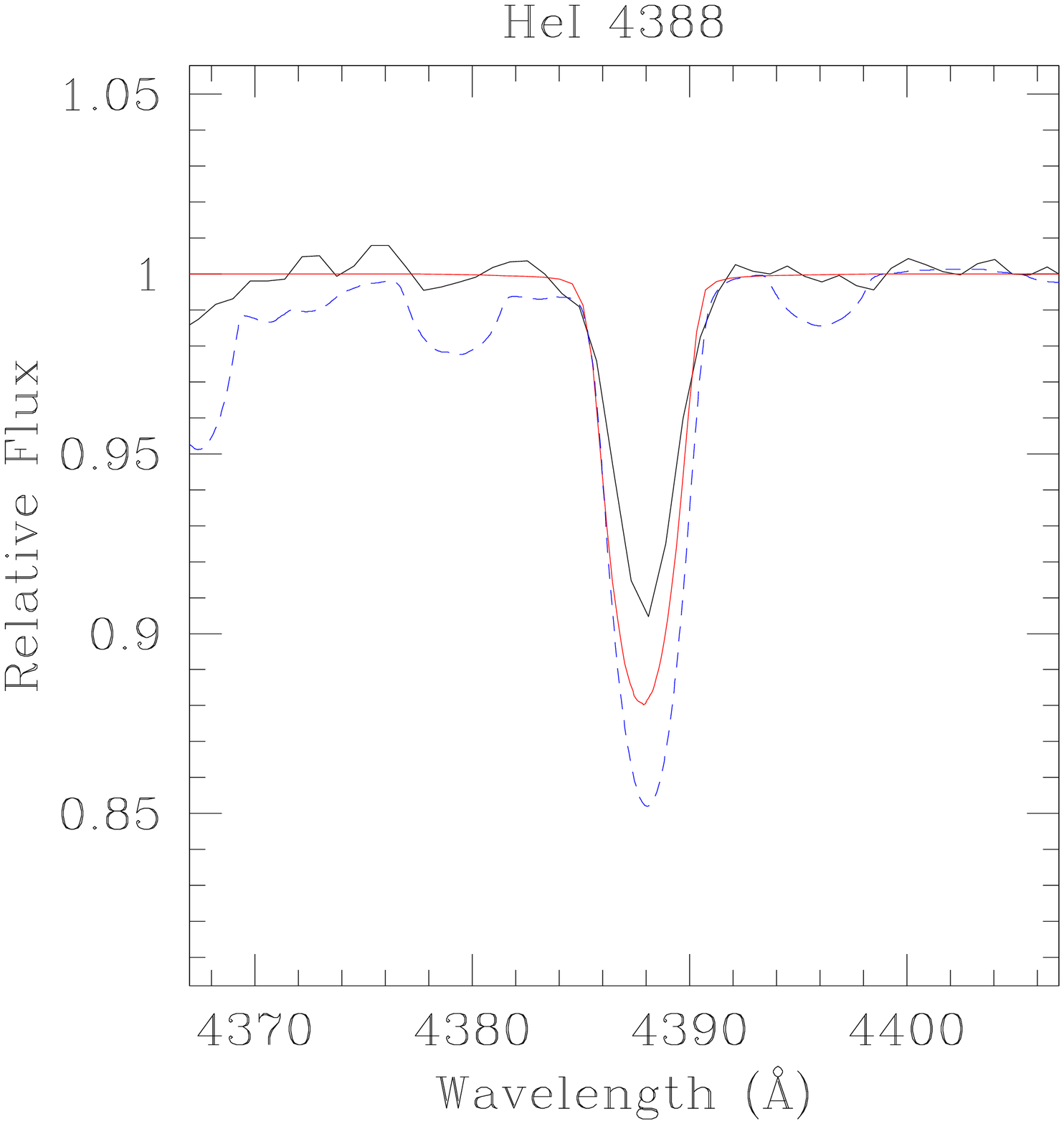}
\plotone{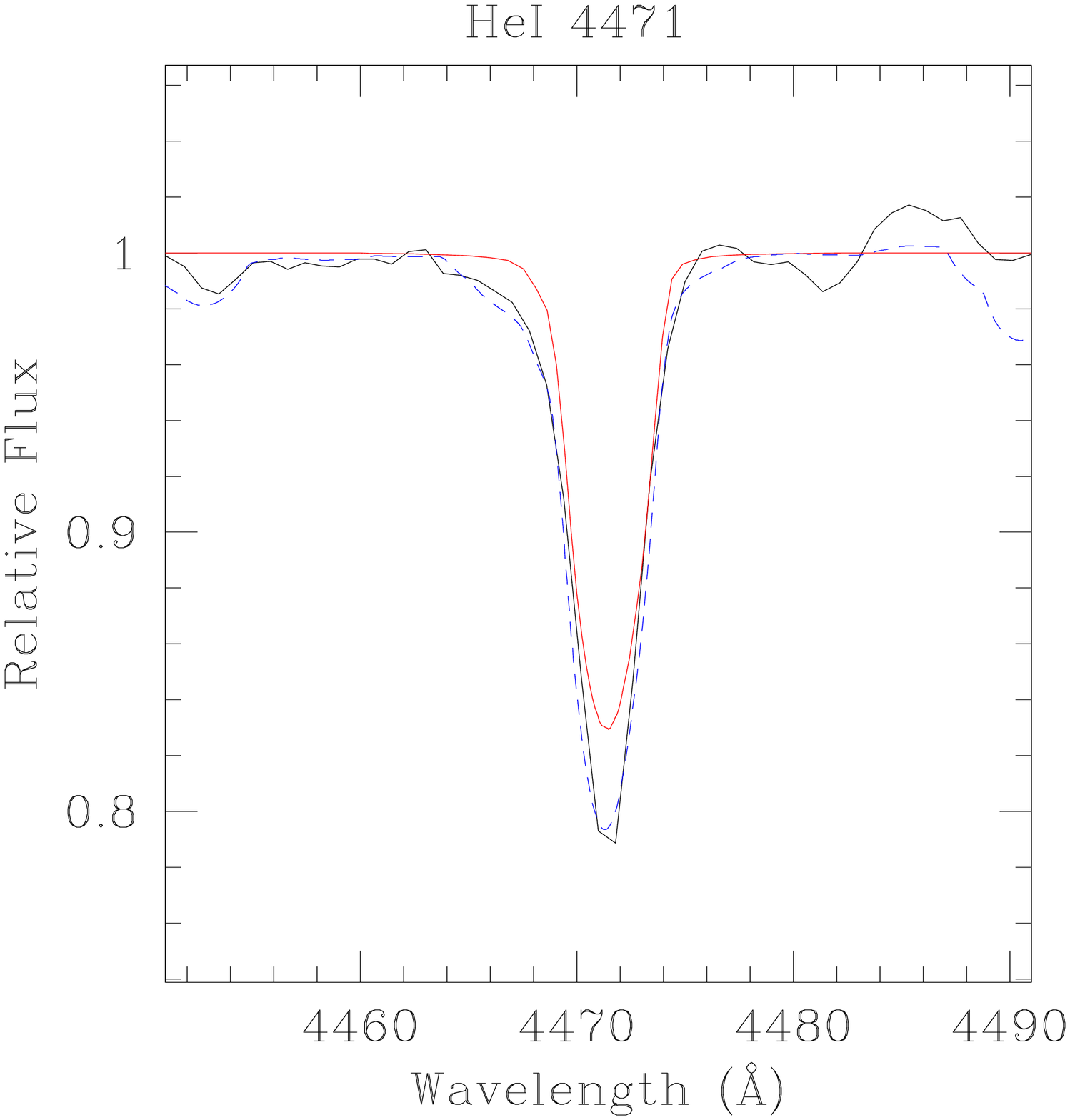}
\plotone{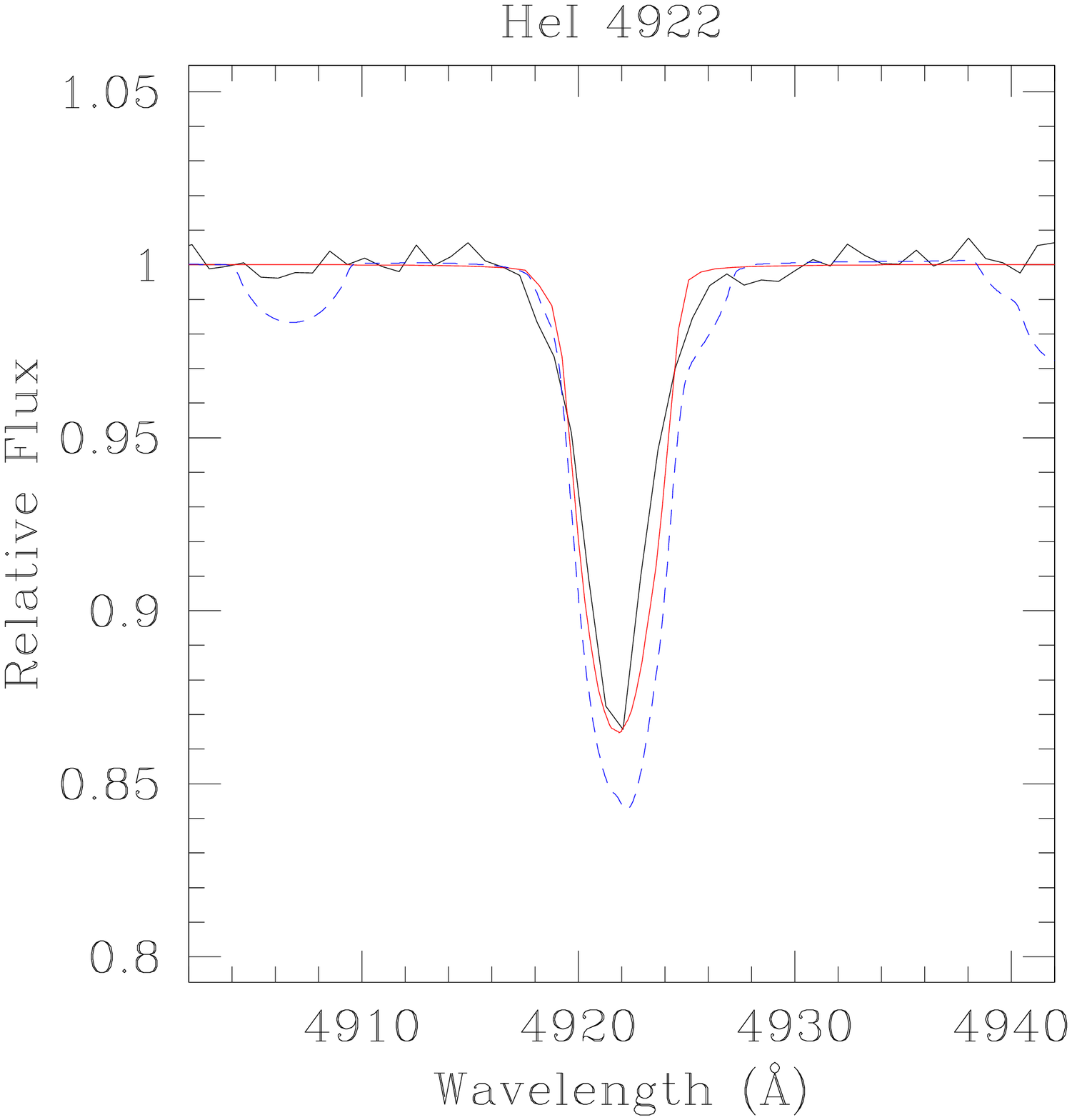}
\plotone{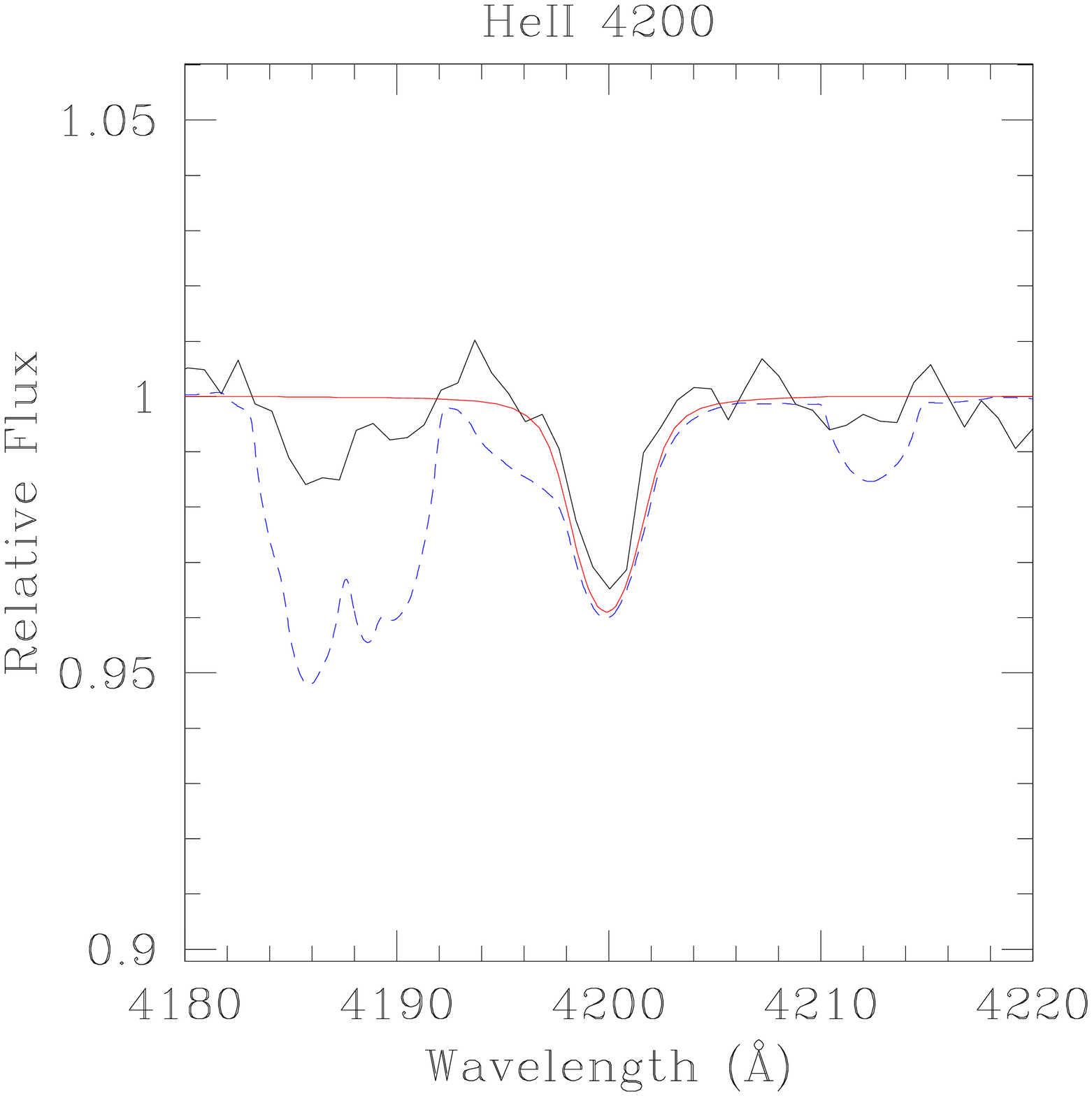}
\plotone{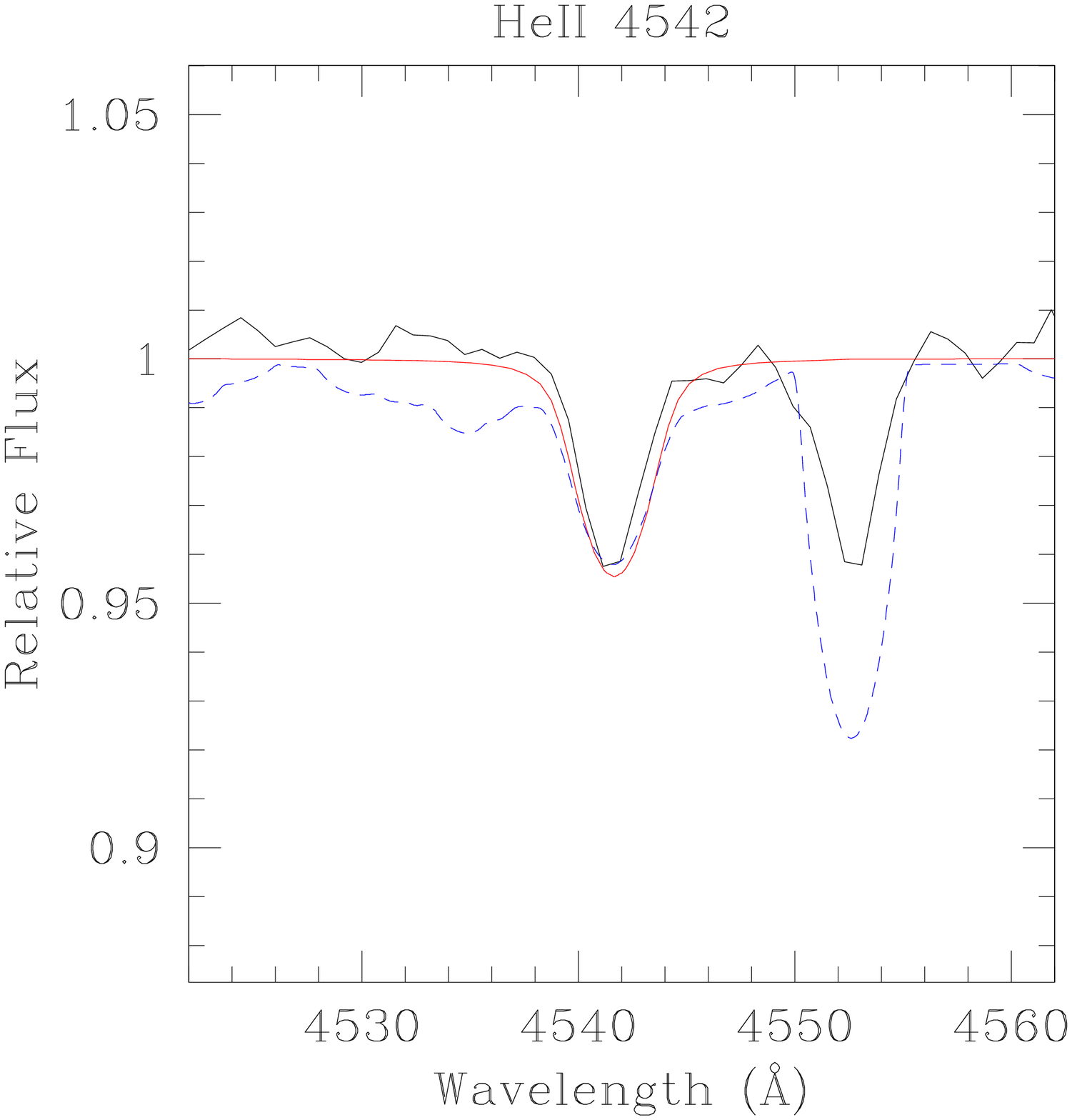}
\plotone{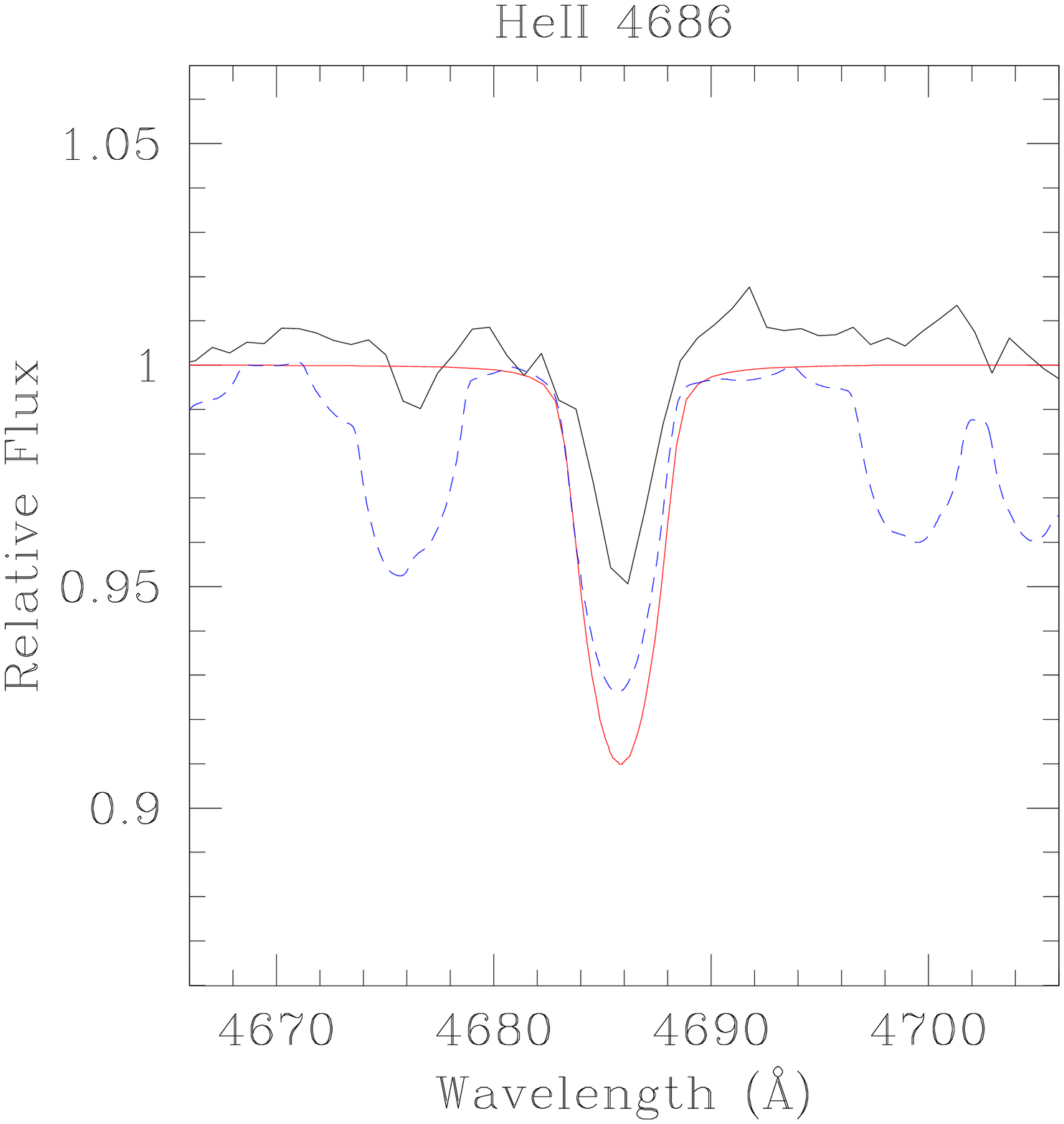}
\caption{\label{fig:Sk-69d124} Model fits for Sk $-69^\circ$124, an O9.7 I star in the LMC.  Black shows the observed spectrum, the red line shows the \fastwind\ fit, and the dashed blue line shows the \cmfgen\ fit. }
\end{figure}

Next, we turn to the He~I singlet fits.  Consider the He~I $\lambda 4388$ model profiles in the same late O-type supergiants stars (i.e., AzV 233 in Figure~\ref{fig:AzV223}, BI 170 in Figure~\ref{fig:BI170}, and Sk $-69^\circ$124 in Figure~\ref{fig:Sk-69d124}). 
\cmfgen\ produces too strong a profile, and while the match by \fastwind\  is better, it is certainly not spectacular, at least for the second two stars.
Najarro et al.\ (2006) explained the poor fits to He~I $\lambda 4388$ by \cmfgen\ as being due to uncertainties in the atomic data of the Fe IV lines overlapping with the He~I $1s^2$ $^1$S$- 1s2p$ $^1$P$^o$ resonance line at 584~\AA, whose upper level is the lower level of 
He I $\lambda 4388$ ($1s2p$ $^1$P$^o-1s5d$ $^1$D), as shown in Figure~\ref{fig:Grot}.
 According to Najarro et al.\ (2006), a similar effect is also seen for the singlet He~I $\lambda 4922$ ($1s2p$ $^1$P$^o-1s4d$ $^1$D) by Najarro et al.\ (2006), which is not surprising given that both lines have the same lower level.  
 We can test this explanation further, as we might expect the He I $\lambda 5016$ ($1s2s$ $^1$S$-1s3p$ $^1$P$^o$) fit to
 be better, since it has a different lower level, as argued by Najarro et al.\ (2006).  In Figures~\ref{fig:singletsA} and \ref{fig:singletsB} we compare the profile fits by \cmfgen\ and \fastwind\  to these three He I singlet lines.  We find
 that even when the \cmfgen\ fits to $\lambda 4388$ and $\lambda 4922$ are very poor (e.g., for the late O supergiants in
Figure~\ref{fig:singletsB}), the fit to He I $\lambda 5016$ is quite good.  Unfortunately, the He I $\lambda 5016$ line is not computed in the formal solution by \fastwind.

In Paper III we argued that because of its approximate treatment of line blanketing and blocking, \fastwind\  should be less sensitive to the problem with the uncertain atomic data for the Fe IV lines that overlap with the 584~\AA\ He I resonance line, and hence should give
better fits to He I $\lambda 4388$ than with the more exact treatment by \cmfgen. Indeed this seems to be the case. Nevertheless, \fastwind\  isn't immune to the problem.  For BI 170 (Figure~\ref{fig:BI170}) and Sk $-69^\circ$124 (Figure~\ref{fig:Sk-69d124}) we get adequate fits to He~I $\lambda 4922$ but  the He~I $\lambda 4388$ \fastwind\  profile lines are too strong, albeit not as badly as with \cmfgen.  We note that in both cases we adopted a significantly hotter temperature based on the \fastwind\  fits, as we knew to expect
the triplets (such as He~I $\lambda 4471$) to give inconsistent answers and we tried to get the singlet lines weak enough to fit the profile.   However, we can see that for {\it some} stars \cmfgen\ and \fastwind\  both give satisfactory fits to the singlet lines, i.e.,  the O5.5 I(f) star AzV 75 (Figure~\ref{fig:AzV75}), the O6 I(f) star AzV 26 (Figure~\ref{fig:AzV26}), the O8~V star NGC346-682 (Figure~\ref{fig:NGC346-682}), and the O5.5 V((f)) star Sk $-70^\circ$69 (Figure~\ref{fig:Sk-70d69}).   The fact that the He I singlet fits only fail in \cmfgen\ for the late-type supergiants in our sample is not surprising, as the influence of the iron lines depends upon the effective temperature and surface gravities.  

Finally, we investigate the effect of microturbulence on the He I singlet profiles.  We argued above that a higher value than 10
km s$^{-1}$ was needed in order to obtain satisfactory fits for all of the He I triplet lines (particularly He I $\lambda 5016$) for the three late-O supergiants.  In Figures~\ref{fig:MicroSingAzV223} through \ref{fig:MicroSingSk-69d124}  we show the effect of increasing the microturbulence velocity on the He I singlets for these
three stars.  For AzV 223 (Figure~\ref{fig:MicroSingAzV223}) there is actually a slight improvement in the reasonably good fits obtained with \cmfgen\ and \fastwind.  For BI 170 and Sk $-69^\circ$124, where the singlet fits for He I $\lambda 4388$ and $\lambda 4922$ were already poor, increasing the microturbulence certainly doesn't improve the situation, as the
model lines were already too deep.  Note, however, that the \cmfgen\ fit for He I $\lambda 5016$ remains quite good even at the higher microturbulence values.   

For convenience, we show the fits for the H, He I, and He II lines used when originally modeling these three stars but now computed using a 15 km s$^{-1}$ microturbulence.  The improvement in the \fastwind\  He I $\lambda 4471$ fits are dramatic.  Note that we did not have to change the effective temperatures or any other physical parameters to achieve this improvement.  This improvement is not completely surprising, as it well known that macroturbulence is larger
in supergiants (see, e.g., Howarth et al. 1997), and hence it might be expected that microturbulence be larger
as well.

\subsubsection{Summary}

The prevailing wisdom has been that \fastwind\  does a poor job of fitting the He I triplets (particularly He I $\lambda 4471$) for late O-type supergiants for poorly understood reasons (Repolust et al.\ 2004, and Papers I-III).  At the same time, Najarro et al.\ (2006) recommend eschewing the He I singlets because of the overlap of the Fe IV lines with the He~I $1s^2$ $^1$S$- 1s2p$ $^1$P$^o$ resonance line at 584~\AA, whose upper level is the lower level of He I $\lambda 4388$ and $\lambda 4922$ lines.  What we found here modifies this significantly, at least for our sample of SMC and LMC stars\footnote{The effect on the He I lines may well be metallicity dependent.}:

\begin{enumerate}
\item The \cmfgen\ problem with the He I singlets occurs only for the late O-types; we obtain quite satisfactory fits for earlier types.  This is not really surprising as the strengths of the Fe IV transitions will also be temperature dependent.  However, even when the fits for He I $\lambda 4388$ and $\lambda 4922$ fail, the He I singlet $\lambda 5016$ is very well fit by \cmfgen, in accord with Najarro et al.\ (2006), as the lower level of this line will be less affected.  Although \fastwind\  does a better job of matching the He I $\lambda 4388$ and $\lambda 4922$ line (since its blanketing calculation is not exact), it is nevertheless affected.  Including He~I~$\lambda 5016$ in the formal solution for \fastwind\  would be a useful addition.

\item The \fastwind\  problem with the He I triplets (such as He I $\lambda 4471$) for the late O-type supergiants largely goes away if we use a microturbulence of 15-20 km s$^{-1}$ in computing the line profiles.  At the same time, a value of 15 km s$^{-1}$ for the late O-type supergiants causes \cmfgen\ to give more consistent results for the He I $\lambda 5876$ triplet, which is otherwise too weak
compared to He I $\lambda 4471$, suggesting that a large value (15-20 km s$^{-1}$) may be more appropriate for late O-type supergiants.  Inclusion of He~I~$\lambda 5876$ in the formal solution for \fastwind\  would also be useful.

\end{enumerate}

We emphasize that we cannot achieve the same good effect by simply lowering the effective temperature for the \fastwind\  models to strengthen the He I $\lambda 4471$ line.  To do so renders the He II lines ridiculously weak.  We demonstrate this in Figure~\ref{fig:compare}.  For AzV 223, if we keep the microturbulence velocity at 10 km s$^{-1}$ and lower the effective temperature of the model from 31,600 K (shown in dashed red) to 29,000 K (shown in dashed green) the He I $\lambda 4471$ line gets a bit stronger, but
still much weaker than the actual observed spectrum.  The lower two panels though show the drastic change in the He II lines.
Thus, our temperatures for the late O-type supergiants (where the He I singlet/triplet behaviors are undependable) are nevertheless well constrained by the He II strengths.

Given these difficulties, we would conservatively estimate the {\it actual} uncertainty in the determination of the effective temperature of a star with either code
to be 1000~K, rather than our 500~K fitting precision.  This is consistent with the errors usually quoted, e.g., Papers I-III. Still, this 2-3\% uncertainty is remarkable,
given the physical complexities involved.

\subsection{Systematic Differences with the Surface Gravities: the Mass Discrepancy Resolved Again?}
\label{Sec-logg}

We turn next to our results for the surface gravity.  Our experience in the fitting is that the wings of the H$\gamma$ profiles allow fitting $\log g$ to a precision of about 0.05~dex.  Yet, it is clear from Table~\ref{tab:results} that there is a systematic difference between the \cmfgen\ and \fastwind\  results, with \fastwind\  consistently yielding a surface gravity that is about 0.10~dex lower.  
This problem was first reported by Neugent et al.\ (2010) based upon preliminary results of the current project, and a similar systematic effect can be inferred from Najarro et al.\ (2011) using a smaller sample of Galactic stars.  

In general, \cmfgen\ does a better job of fitting the H$\gamma$ wings, because it includes metal lines that might affect the wings, such as OII $\lambda 4350$.  But, that can't be the entire explanation, as one would expect to find O II lines primarily among the later O-type supergiants, and the systematic difference in $\log g$ seems to be independent of effective temperature or luminosity class.  Consider the excellent fits shown from H$\gamma$ for the O5.5 V((f)) star Sk $-70^\circ$69 in Figure~\ref{fig:Sk-70d69}.  The \cmfgen\ and \fastwind\  H$\gamma$ profiles are nearly indistinguishable, and yet the former corresponds to $\log g=3.80$ and the latter to $\log g=3.70$, albeit with rather different effective temperatures.  Perhaps a more convincing case is that of the O6 I(f) star AzV 26, where the adopted effective temperatures are the same and the H$\gamma$ profile fits are hard to distinguish in the wings (Figure~\ref{fig:AzV26}), but the \cmfgen\ fit is based on $\log g=3.60$ and the \fastwind\  fit on $\log g=3.50$.  

How sensitive are the fits to $\log g$?  In Figure~\ref{fig:compHgam} we show two \fastwind\  models, one computed with the value of $\log g=3.50$ we obtained using \fastwind\, and one computed with the $\log g=3.60$ value we derived using \cmfgen.  There is a clear difference, and based on the (uncontaminated) blue side of the line, there is no question of mistaking which fit is better with \fastwind.  Thus, the difference between the two codes in the resulting value for the surface gravity is quite real.  At the same time, the exact value we obtain with either code is clearly dependent upon the normalization, and it would take a very brave person to say that we can do better (in an absolute sense) than 0.1~dex in $\log g$.  Thus, for the purposes of computing errors (below) we will adopt 0.1~dex as $\sigma_{\log g}$, about twice the precision we achieved with a particular spectrum.  (Note that the 0.1~dex value is consistent with what is usually quoted anyway; see, for example, Repolust et al.\ 2004 and Papers I-III.)

We were naturally curious as to the source of the discrepancy in $\log
g$ between the fits made with the two codes. D. J. H. computed a set
of \cmfgen\ models (pure H and He) spanning the relevant range in
surface gravities and effective temperatures with various Stark
broadening profiles, and the theoretical H lines were then compared to
those of \fastwind, computed by J.P. with the same physical parameters.
The emergent profiles were computed both with no rotation and with a
projected rotational broadening of $v \sin{i}=80$ km s$^{-1}$. It was immediately
evident that there was a systematic difference in the wings of
H$\gamma$ (and other H lines) in the same sense shown by our fitting
methods: the wings of H$\gamma$ were always, and mostly 
significantly, lower with the \fastwind\ models, requiring lower surface
gravities (by typically 0.1 dex, consistent with our findings from the
profile fitting) to achieve the same effect\footnote{Note that this
systematic difference has never been noticed before, though many
comparisons between \fastwind\ and other codes, notably \cmfgen, had been
performed previously; see for example, Santolaya-Rey et al. (1997),
Puls et al. (2005) and Rivero-Gonzalez et al. (2012b).  To some extent, this is due to the complexities of the codes,
and the fact that ours is the first effort to model the same spectra (involving the same normalization) with the two codes.}. Our Magellanic
Cloud star fitting with \cmfgen\ used the Lemke (1997) Stark profiles,
while \fastwind\ adopts the Sch\"{o}ning \& Butler (1989a, 1989b)
formulation, but the differences in our pure H/He tests were still
prominent -- though slightly smaller -- even when the Sch\"{o}ning \&
Butler Stark profiles were used with \cmfgen\footnote{The two types of
Stark profiles are of similar quality, and the differences between
them probably reflects our current state of knowledge.}. Investigation
showed that the difference between the two codes probably traces to
the treatment of electron scattering, at least for objects with $\log
g$ below 4.0. As described in the Appendix, \cmfgen\ uses coherent scattering to compute the level 
populations, but incoherent scattering when computing the spectrum.  
In contrast, \fastwind\, 
doesn't explicitly use any scattering model, but instead uses constant
Thomson scattering emission over the line (with the mean intensity
from the neighboring continuum) to approximate the incoherent
scattering approach (see Mihalas et al.\ 1976). However, these tests
showed that the constant emission method used by \fastwind\ actually is
closer to that of coherent scattering, and does not have the intended
effect on the emergent line profiles; i.e., \fastwind's surface
gravities are probably too low.  Further work is planned on
investigating this problem, particularly with respect to fully
blanketed model atmospheres as used in our current analyses.

Does this have any implication for the ``mass discrepancy"? Groenewegen et al.\ (1989) and Herrero et al.\ (1992) called attention to the fact that the masses for O-type stars derived from spectroscopic analysis via stellar atmosphere modeling were often smaller than those derived by evolutionary tracks using the same effective temperatures and bolometric luminosities, sometimes by as much as a factor of 2. The ``spectroscopic mass" depends upon the derived surface gravity and the stellar radius ($m_{\rm spect}\sim gR^2$), with the value of $R$ derived from the bolometric luminosity and effective temperatures, the same quantities used with the evolutionary models to find the masses.   Thus, the determination of the surface gravity has long been considered a possible culprit if the problem lay with the atmosphere models.  Improvements both in the stellar evolutionary models and stellar atmosphere models have resulted in reducing the mass discrepancy.  For instance, Lanz et al.\ (1996) found that using fully blanketed atmosphere models greatly alleviated the problem.  However, most modern studies find that the problem has not been completely eliminated; see
discussion in Paper II and Repolust et al.\ (2004) for instance.  The analysis of the current sample of stars in Papers I-III found stars with significant mass discrepancies and stars without, with the stars with the largest discrepancies tending to be those of earliest type.    

Consider the O3.5 V((f+)) star LH 81:W28-23.  The spectroscopic mass derived in Paper II was 24$M_\odot$, while its evolutionary mass was 55$M_\odot$.  Increasing the surface gravity by 0.15~dex (as indicated by the \cmfgen\ fit) by itself would raise the spectroscopic mass to 34$M_\odot$, significantly reducing the mass discrepancy.   The fact that for this star we also find a significantly cooler effective temperature from the \cmfgen\ fit also helps reduce the mass discrepancy, as it both lowers the derived evolutionary mass (as the bolometric correction becomes less negative, lowering the luminosity), and it also increases
the stellar radius, further increasing the spectroscopic mass. 
 
In Table~\ref{tab:MD} we reconsider the mass discrepancy for all of these stars.  We begin by computing the bolometric luminosities ($\log L$, in solar units) using the approximation that the bolometric correction ${\rm BC}=-6.90 \log T_{\rm eff} +27.99$ from Paper II: $M_{\rm bol}=M_V+{\rm BC}$, where the values for $M_V$ come from Table~\ref{tab:fixed}.  Then
$\log L=-(M_{\rm bol}-4.75)/2.5$, where we have taken the bolometric magnitude of the sun to be 4.75.  The effective radius $R$ (in solar units) then follows from the Stefan-Boltzmann equation: 
$\log L = 4 \log T_{\rm eff} -15.05 + 2 \log R$, where we have adopted the effective temperature 
of the sun as 5778 K.  In computing the spectroscopic mass $m_{\rm spec}$ (also in solar units) we have taken the surface gravity we've measured (Table~\ref{tab:results}) and corrected this for the effects of centrifugal acceleration, as the measured (effective) surface gravities have been lessened by the rotation of the star.  For this, we used $$g_{\rm true}=g_{\rm eff} + \frac{(v \sin i)^2}{6.96R},$$ following Repolust et al.\ (2004), where $g$ is in cgs units and $v \sin i$ is in km s$^{-1}$.   Two caveats are worth noting: first, this correction is valid only in a statistical sense, as it assumes randomly distributed rotational axis orientations.  Secondly, the rotational velocities $v \sin i$ we use from Table~\ref{tab:fixed} are really too large on average, as the broadening of our profiles include (a) the actual rotation, (b) the instrumental resolutions (of order 100-150 km s$^{-1}$), and (c) the macroturbulence velocity\footnote{Macroturbulence is a large-scale velocity field invoked to explain the lack of slow rotators among early-type stars; Conti \& Ebbets (1977) and Ebbets (1979) demonstrated that they could separate the signature of rotation from other large-scale atmospheric motions using Fourier techniques (see also Sim\'{o}n-D\'{i}az et al.\ 2010 and references therein).  Aerts et al.\ (2009) and Sim\'{o}n-D\'{i}az et al.\ (2010)  argue that the physical explanation for macroturbulence, at least in OB-type supergiants, is non-radial pulsations.}  (of order 30-100 km s$^{-1}$).   Still, the corrections are small (typically 0.01-0.02~dex, and 0.05~dex at the most extreme), and over-estimating $v\sin i$ would increase the value of $m_{\rm spect}$, reducing the mass discrepancy.  Finally, $$m_{\rm spec}= (g_{\rm true}/g_\odot) R^2,$$ where we take $\log g_\odot=4.438$.  

To determine the evolutionary masses for the LMC stars, we used the newest Geneva models
of V. Chomienne et al.\ (2013, in preparation) based upon the latest improvements described by Ekstr\"{o}m et al.\ (2012) but computed for a metallicity of $z=0.006$.   For the SMC stars, we used the older Geneva models described by Maeder \& Meynet (2001).   Both models incorporate the effects of stellar rotation, albeit with somewhat different starting assumptions.  To obtain the evolutionary masses, we performed a linear interpolation for $\log m_{\rm evol}$ using the
observed $\log L$ and whichever two tracks bounded the star's luminosity.  This process was mainly straight forward\footnote{We performed the interpolations along isochrones in general, although for Sk $-69^\circ$124 the lifetime of the hydrogen burning phase of the upper track was too short and we instead did the interpolation along a line of constant effective temperature.  For AzV~26 and AzV~75 we did the interpolation both ways, finding negligible differences.}, but for BI~170 there was an ambiguity in the
mass predicted by the evolutionary tracks, as the tracks zig-zag at the end of core H-burning, which  left us unsure as to its evolutionary mass.

Of course, in evaluating the significance of these comparisons we must understand the size of the errors.  
Our fitting uncertainties in the  effective temperatures are 500 K; we assume that the actual error is 1000 K, in keeping with the above discussion.  This leads to an error in $\log T_{\rm eff}$ of $\pm0.015$~dex for our coolest star and $\pm0.009$~dex for our hottest star.  The uncertainty in $\log L$ will include a component that is due to the propagation of the error in $\log T_{\rm eff}$ due to the bolometric correction ($\Delta{\rm BC}=-6.9\Delta\log T_{\rm eff}$; i.e., $\sim 0.10$~mag) but will be dominated by the uncertainty 
in the absolute visual magnitude, which we estimate to be 0.18~mag, based upon a 0.15~mag uncertainty in $A_V$ and the uncertainty in the distances to the LMC and SMC, which we take to be 0.1~mag based upon discussion in van den Bergh (2000).  
Thus the typical uncertainty in $\log L$ is 0.08~dex.
This error in $\log L$ dominated the uncertainty in our determination of 
$m_{\rm evol}$, and results in a typical error of 0.04~dex in 
$\log m_{\rm evol}$.   The uncertainty in $\log m_{\rm spec}$ is then
$$\sigma_{\log m_{\rm spec}}=\sqrt{4\sigma^2_{\log R}+\sigma^2_{\log g}}.$$   
The uncertainty in $\log R$ then follows from propagation of errors through the Stefan-Boltzman equation, i.e., 
$$\sigma_{\log R}=
\sqrt{0.25 \sigma^2_{\log L} 
+ 4\sigma^2_{\sigma_{\log T_{\rm eff} }} },$$ 
or $\sigma_{\log R} \sim $0.04~dex for all of our stars.   Our fitting uncertainty in $\log g$ is 0.05~dex, but as argued above,
we will assume an ``accuracy" of 0.10~dex for the surface gravities determined by each code.  This leads to a nearly constant
uncertainty in $\log m_{\rm spec}$ of 0.13~dex. 

However, in measuring the size of the mass discrepancy we do better to consider the quantity $\Delta \log m= \log m_{\rm spec}/m_{\rm evol}$, and its associated error.  This is because both $m_{\rm spec}$ and $m_{\rm evol}$ depend upon $\log L_\odot$, and
any error in the luminosity will be the same for a given star, be it due to the uncertainty in $A_V$ or in the BC.   In evaluating the
error we assumed that the error $\Delta m_{\rm evol}$ was roughly 0.5 $\Delta L$.  We can calculate this error straightforwardly, since
$$\log m_{\rm spec} \sim \log g + \log R^2 \sim \log g + \log L - 4 \log T_{\rm eff}$$
and
$$\log m_{\rm evol} \sim 0.5 \log L$$
then
$$\log \frac{m_{\rm spec}}{m_{\rm evol}} \sim \log g - 0.5 \log L - 4 \log T_{\rm eff}$$
and hence
$$\sigma_{\log \frac{m_{\rm spec}}{m_{\rm evol}}} = \sqrt{\sigma^2_{\log g} + 0.25\sigma^2_{\log L} + 16 \sigma^2_{\log T_{\rm eff}}},$$
i.e., greatly reducing the dependence on the error in $\log L$.  The errors on $\log \frac{m_{\rm spec}}{m_{\rm evol}}$ are thus about 0.12~dex.    In other
words, mass discrepancies smaller than $\sim$30\% for a single object are not significant. If we had instead adopted our fitting precision (rather than twice the precision) on the uncertainties, then the
error on this quantity would have been 0.07~dex rather than 0.12~dex. 

We should also comment further on the actual ``error" in the evolutionary models.  What if we had chosen different models?  The LMC models assume an initial rotation velocity that is 40\% of the breakup speed (Ekstr\"{o}m et al.\ 2012) as this agrees well with the peak
rotational velocities observed in young B stars (Huang et al.\ 2010).  (Of course, this is a statistical average, and any individual star may have an initial rotation that is higher or lower than this value.) However, some evidence suggests that this rotation speed may be a bit high; see discussion in Neugent et al.\ (2012).  For the mass range and ages considered here, there is generally
little difference in the evolutionary masses we would compute using the tracks with and without rotation: for instance, rather than
$m_{\rm evol}=35.2 M_\odot$ we obtain with the \cmfgen\ results and the rotating models, we would instead obtain
$m_{\rm evol}=35.1 M_\odot$, a completely negligible difference.   If instead we had used the older rotation models at LMC metallicity
described by Meynet \& Maeder (2005), we would have obtained $m_{\rm evol}=37.9^{+3.0}_{-2.9} M_\odot$, slightly higher that what we obtain from the newer models, but still within the errors.  We'll note that Massey et al.\ (2012) recently compared the evolutionary masses to the masses determined in two binary systems, finding very good agreement, with the evolutionary masses higher by about 11\% (i.e., 0.05~dex) compared to the dynamical masses.  

The comparisons are shown in Figure~\ref{fig:md}, where we have plotted $\log \frac {m_{\rm spec}}{m_{\rm evol}}$ against various properties.  Consider the left
panels in the figure.  We see that with our \fastwind\ results there is a systematic tendency for the masses to be too low.  The average difference in $\log m$ between the spectroscopic mass and the evolutionary mass is $-0.09\pm0.04$~dex, where the uncertainty quoted is the error-weighted sigma of the mean.    The median difference (which helps avoid outliers) is also $-0.09$~dex.  This is in the classic sense of the mass discrepancy, with the model atmosphere results
deriving too small a mass compared to that predicted by the evolutionary tracks.  By contrast, the values in the right panel, which show the \cmfgen\ results, are more evenly distributed about the 0 line, and the average difference in $\log m$ is $+0.01\pm0.04$, with a median of +0.09~dex.  (Recall that we expect a small positive bias due to overestimating $v \sin i$.)     Another way to consider these data is from the point of view of outliers: in the case of \fastwind, there are two stars whose masses are more than 2.5$\sigma$ away from the masses of the evolutionary tracks (both in a negative direction, i.e., in the classic sense of the mass discrepancy), while there
are none more than 1.9$\sigma$ away with the \cmfgen\ results.  Thus, the higher surface gravities found with \cmfgen\ are more consistent with the evolutionary
masses.  Still, it should be noted that in even in the case of the two \fastwind\ ``outliers" the spectroscopic masses found by \cmfgen\ also show relatively large differences with the evolutionary masses.  
Note from the bottom panel that these two stars are the ones with the highest temperatures, consistent with the findings of Paper III, that the
\fastwind\ mass discrepancy was most significant for the hotter stars in the sample.  Further investigation of these stars may be instructive.

\subsection{Fitting the Metal Lines}
\label{Sec-metals}

\cmfgen\ automatically computes the entire spectrum, including lines of C, N, and O.  These lines are not as useful for determining the fundamental physical properties of the stars (such as effective temperatures), as the behavior even of the relative strength of two ions of the same element (i.e.,  N III vs N IV) depends strongly not only on the effective temperature but also surface gravity, mass-loss rates, and especially the assumed N abundance (Rivero Gonz\'{a}lez et al.\ 2012b).  However, it is reassuring to see how well these metal lines are fit.  After we had obtained a good model fit with \cmfgen\ paying attention to only the He and H lines (as we did with \fastwind) we then varied the CNO abundances (if needed) to improve the fit to selected spectral lines of these elements. 
We list the final adopted CNO mass fractions in Table~\ref{tab:abund}, and illustrate the fits in Figures~\ref{fig:azv177metal} through \ref{fig:sk69124metal}.  The final \cmfgen\ fits are shown by the dashed blue lines; if these reflected a change in abundances, we show
the original fits in red.  

In all cases the final fits are better than the original, but the final adopted CNO abundances in Table~\ref{tab:abund} should be taken as uncertain by factors of several.  In principle, changing the CNO abundances is not an unreasonable thing to do, particularly for the supergiants, as we expect evolution of the surface composition of these elements as a star ages.  Certainly in cases where we needed a higher He/H to fit the He lines, we also see strong evidence of changes in the surface composition.    Still, there are complexities to using many of these lines for abundance determinations; Martins \& Hillier (2012) demonstrate complexities of using the C III 4647-51-50 complex, and argue that abundances determined using this line are suspect.   We suspect these cautions may extend to other lines.

Rivero Gonz\'{a}lez et al.\ (2012b) derive nitrogen abundances\footnote{Rivero Gonz\'{a}lez et al.\ (2012b) report the nitrogen abundances in terms of a logarithmic number ratio, with  [N]=log N/H + 12.  If He/H is not enhanced (i.e., the He/H number ratio is 0.10) then our assumed SMC and LMC abundance ratios would correspond to [N]=7.34 and 7.74, respectively.  If He/H is enhanced to 0.15, then the unenhanced [N] abundances would be 7.40 and 7.80, respectively.  The advantage of using mass fractions rather than number ratios is that it is easier to then separate metal abundance enhancements from the expected increase in He/H as a star evolves.} using a modified version of \fastwind\ for two of our stars (using the identical spectra), namely AzV~177 and LH 81:W28-23.  For AzV~177 they find an enhancement of about a factor of 3, while we find no enhancement necessary with \cmfgen\ for a good fit.  For LH81:W28 they find an enhancement of a factor of 4, while we find a similar enhancement with \cmfgen\ of a factor of 5.  Clearly more tests are needed to know if \fastwind\ N abundances are consistent with those of \cmfgen\ or not.

Sk $-69^\circ$124 (Figure~\ref{fig:sk69124metal}) represents a special case.  In our initial modeling (red) we did not change the abundances but simply noted that all the CNO lines were poorly fit, and are inconsistent.   D. J. H. attempted to refit the star and found improved fits by lowering the CNO abundances  to $\sim$1/3rd that of the LMC, almost SMC-like\footnote{It was also necessary to lower the rotation velocity to 100 km s$^{-1}$, Gaussian smooth the result, and use a 15 km s$^{-1}$ microturbulence.}.  This provides some improvement to the optical CNO lines, but it cannot be understood from the point of stellar evolutionary theory, as may be inferred from below. An intriguing possibility would be that this star formed out of gas that was accreted from the SMC  (see, e.g., Olsen et al.\ 2011).  The radial velocity of Sk $-69^\circ$124 is also abnormally small (190 km s$^{-1}$) for LMC membership, about 90 km s$^{-1}$ more negative than what is expected given the star's position in the LMC.   This might support the possibility that it originated from gas drawn in from the SMC, but even so, the velocity is quite abnormally small\footnote{We are grateful to Knut Olsen for correspondence about the expected radial velocities of this star.}.  We think a more likely explanation is that the star is a binary and that the difficulties in fitting the CNO abundances and the strain to fit even the He lines (Section \ref{Sec-HeI}) are due to that.  We therefore have not included Sk $-69^\circ$124 in computing the differences in
the tables.

The Geneva evolutionary models also predict the evolution of the surface abundances; in Figure~\ref{fig:evolabund}  we show the
expected change with age over the course of main-sequence evolution. The exact details depend
upon how much initial rotation is assumed, but generally we expect to see a strong N enhancement (along with a modest He enhancement)  at the expense of C and N.   For the SMC results, things are pretty much as expected for the dwarfs,
which all have normal SMC abundances.  For the two supergiants, the enhancement for AzV 75 makes sense; we are surprised
though that AzV 233 doesn't require a higher N abundances, but inspection of Figure~\ref{fig:CNOAzV233} shows that at this
cooler temperature NIV $\lambda 4058$ is nonexistence, and the only constraint comes from NIII $\lambda \lambda 4634,42$.
For the LMC stars,  there are three surprising cases. LH81:W28-23 shows a very depleted C abundance relative to O, and that is hard to understand from an evolutionary point of view.  With a high He enhancement, one would
expect both C {\it and} O to be similarly depleted, and in any case, such enhancements are not expected early in a star's life, which is what the effective temperature and luminosity of LH81:W28-23 suggests.  For BI 170 it is unexpected to see C and O depletion without increased N.   Better constraints on the abundances may be obtained by including the UV lines, and such studies are planned.

\section{Conclusions}

\begin{figure}
\epsscale{1.0}
\plotone{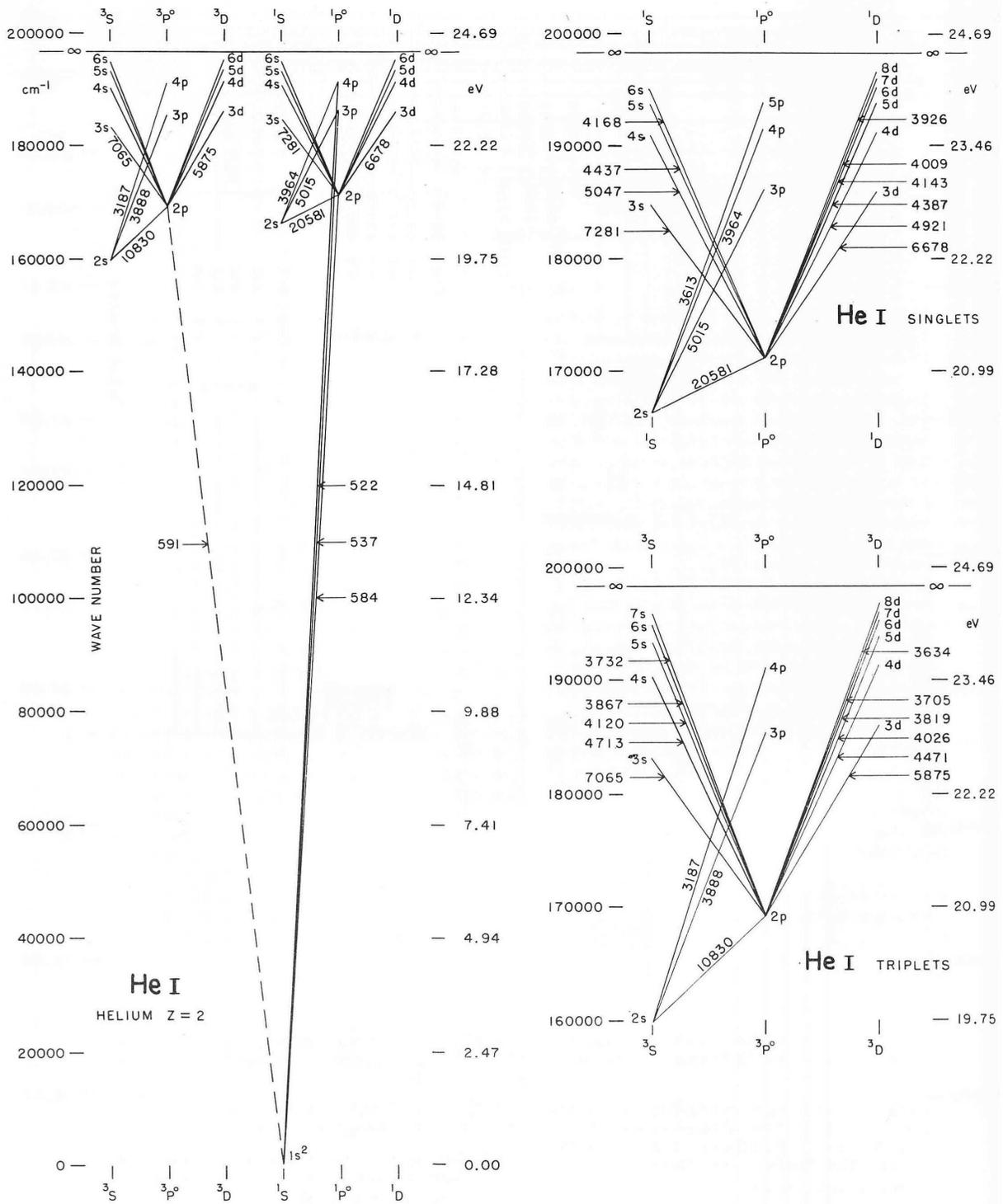}
\caption{\label{fig:Grot} Grotrian diagram for He I, reproduced from Moore \& Merrill (1968).}
\end{figure}

\begin{figure}
\epsscale{0.25}
\plotone{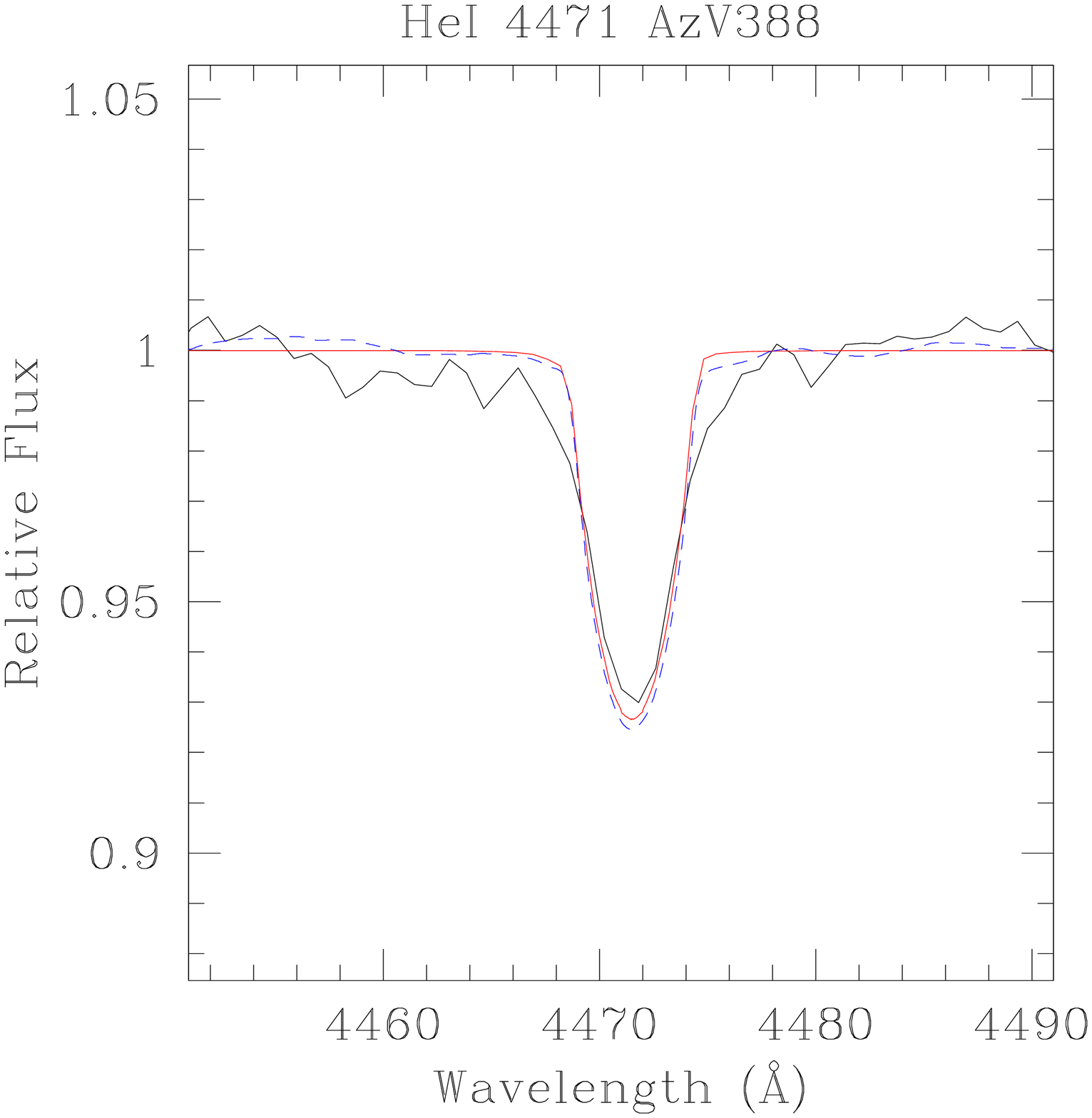}
\plotone{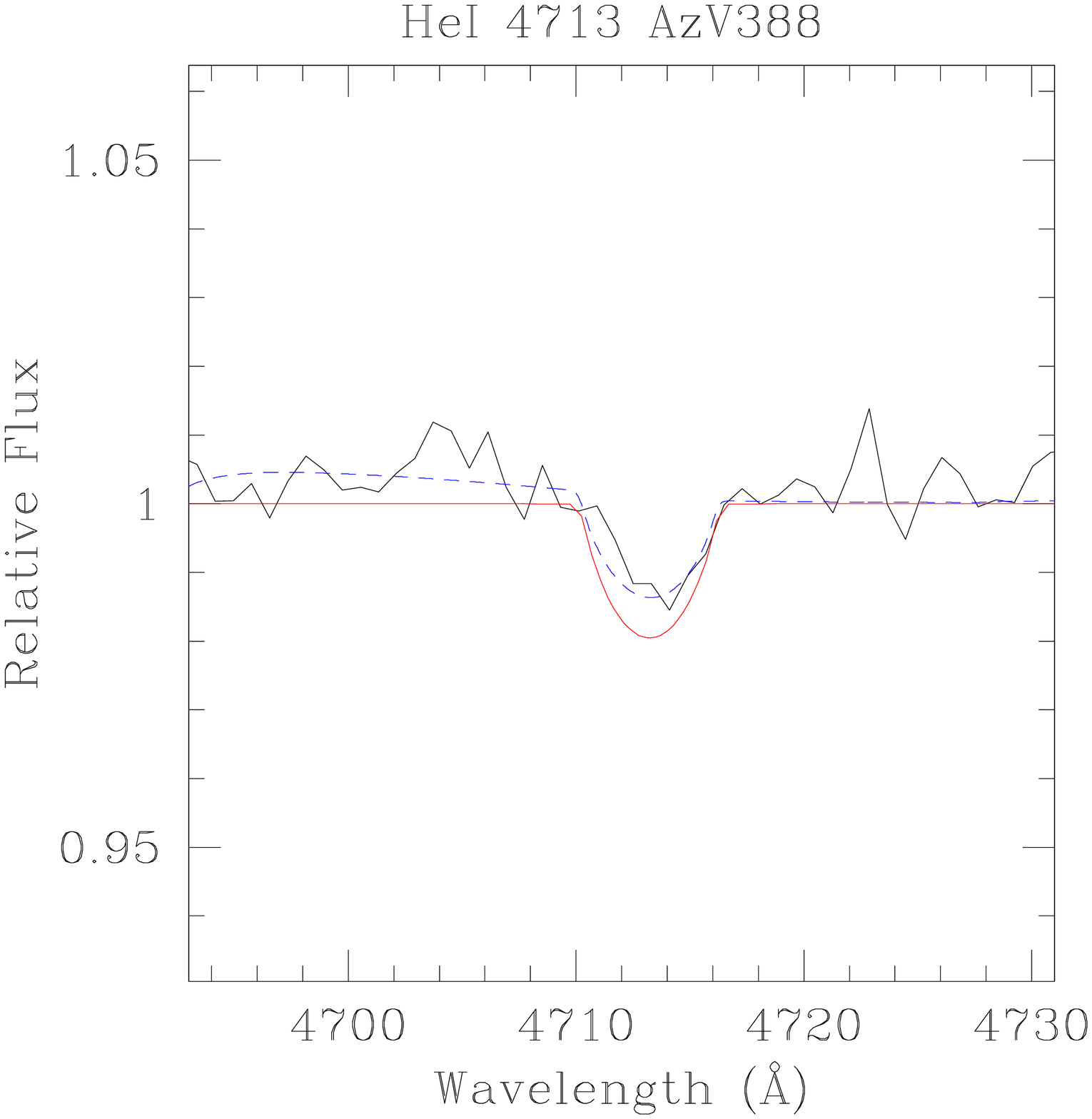}
\plotone{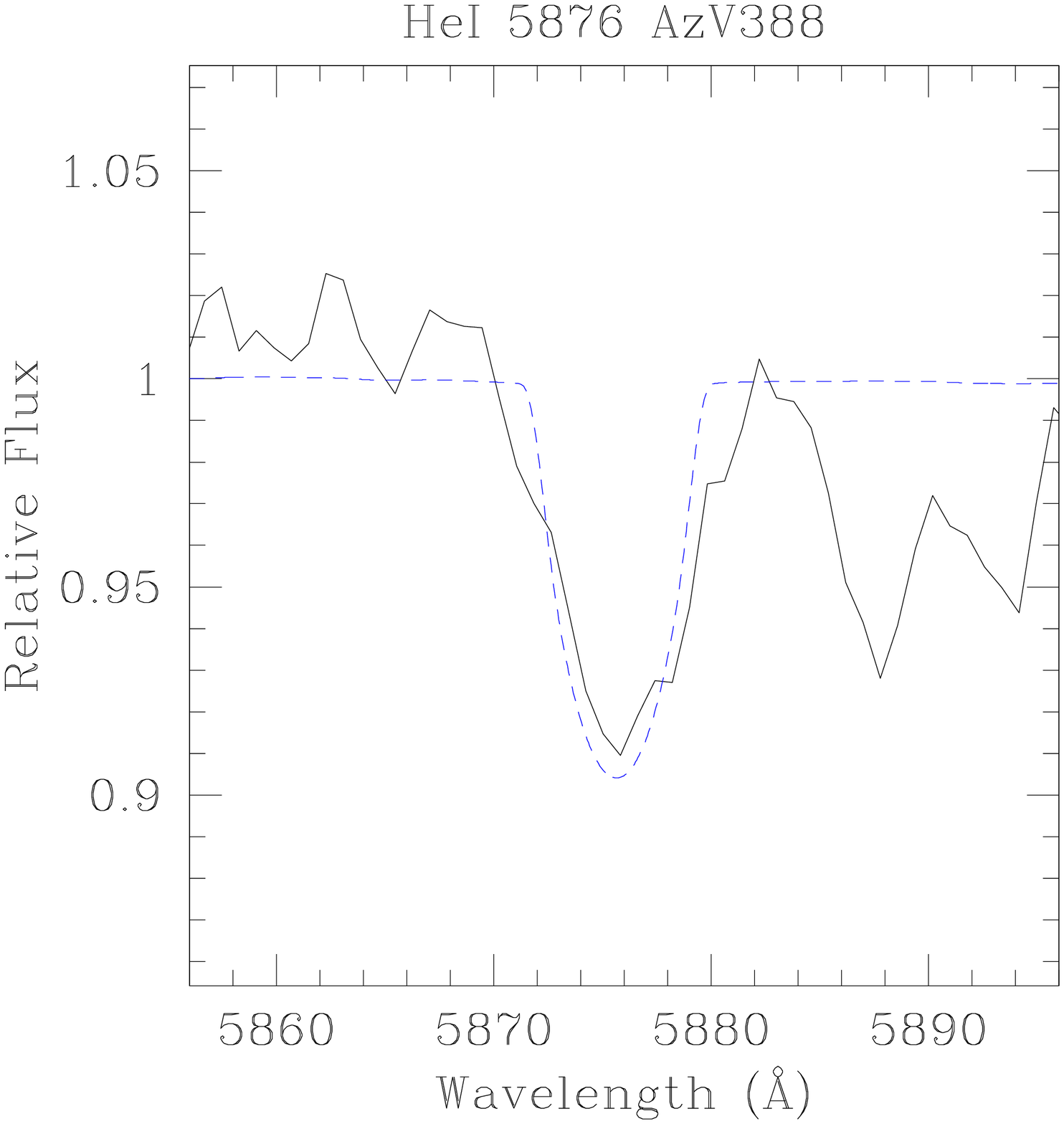}
\plotone{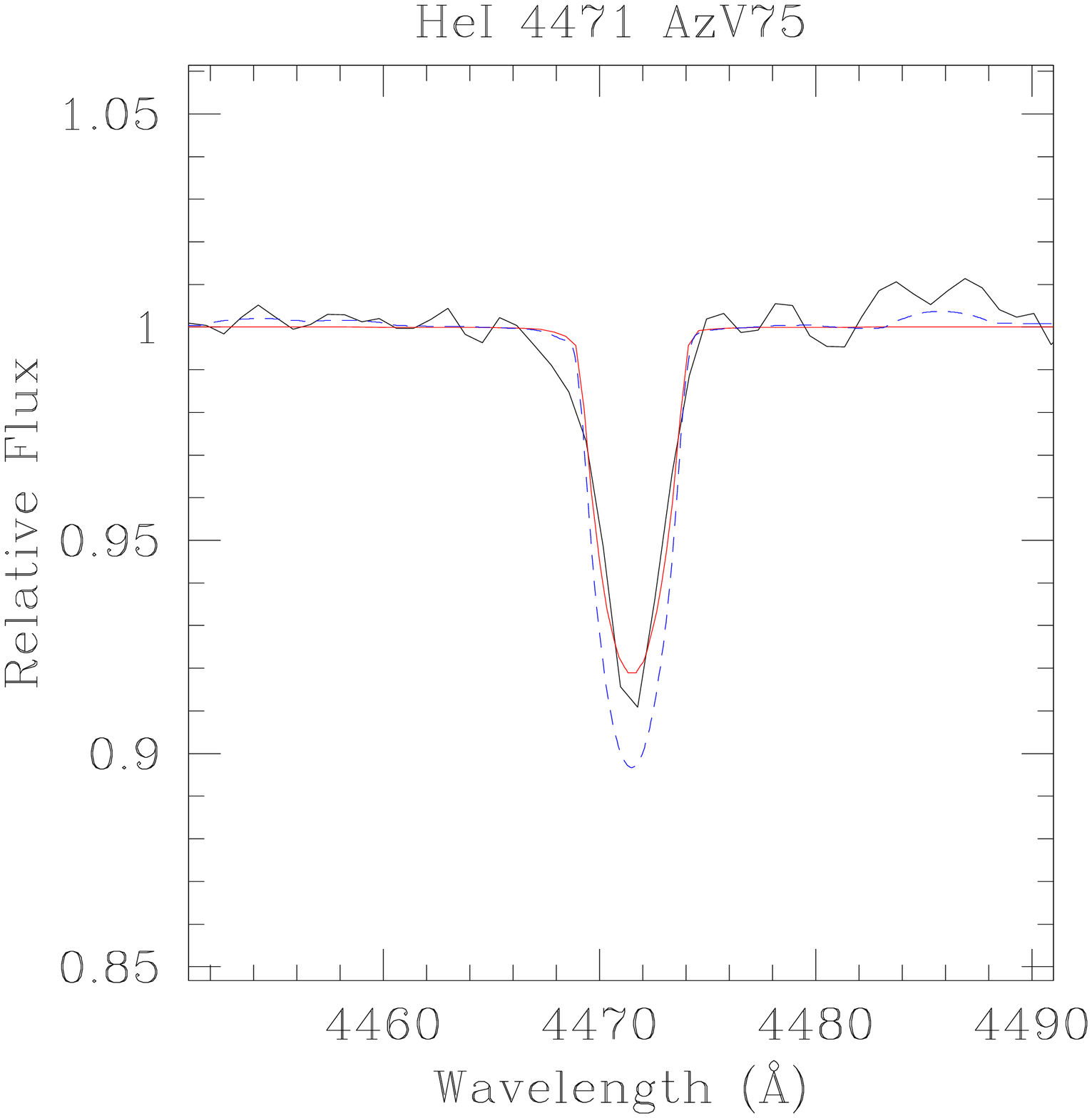}
\plotone{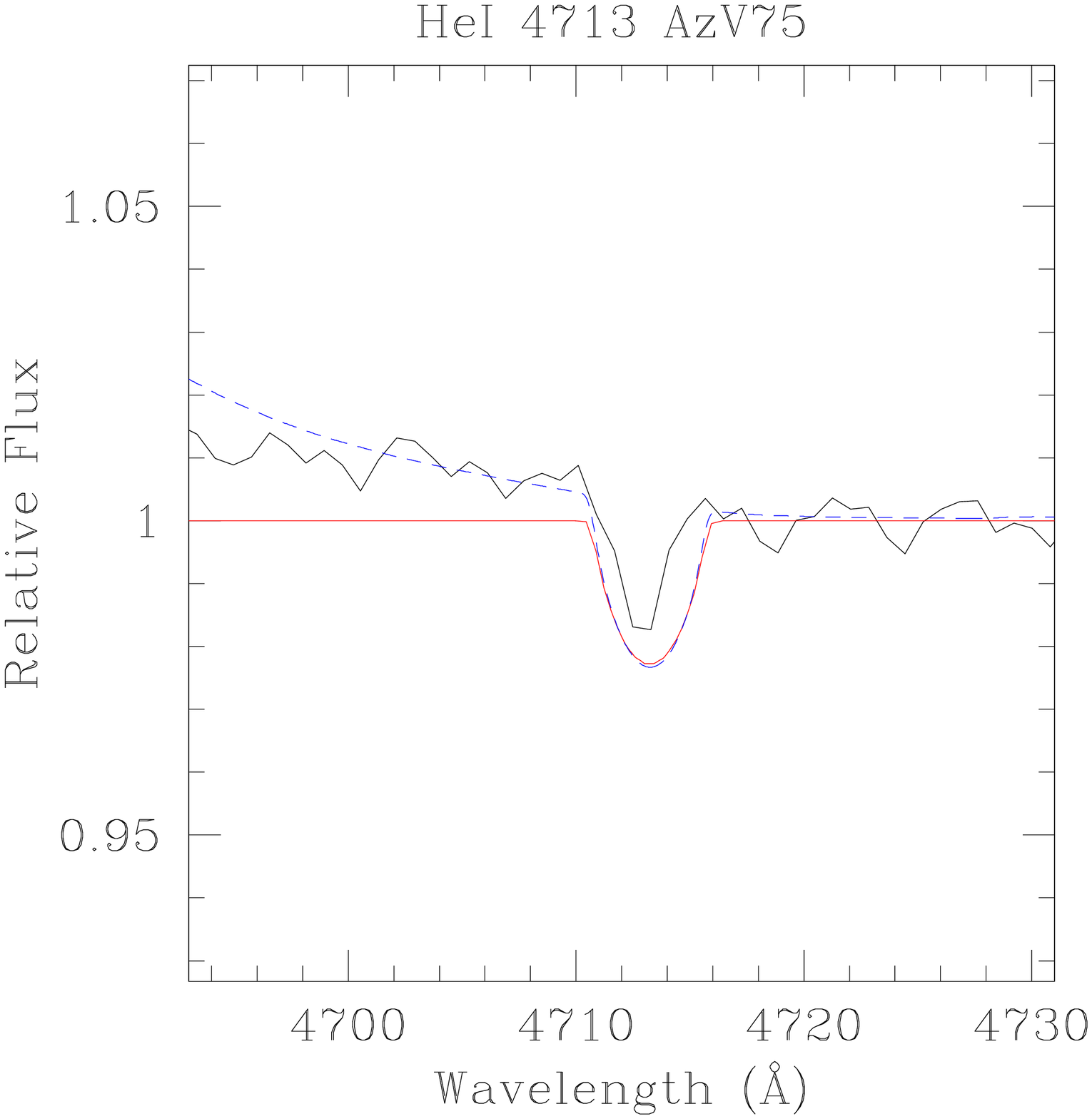}
\plotone{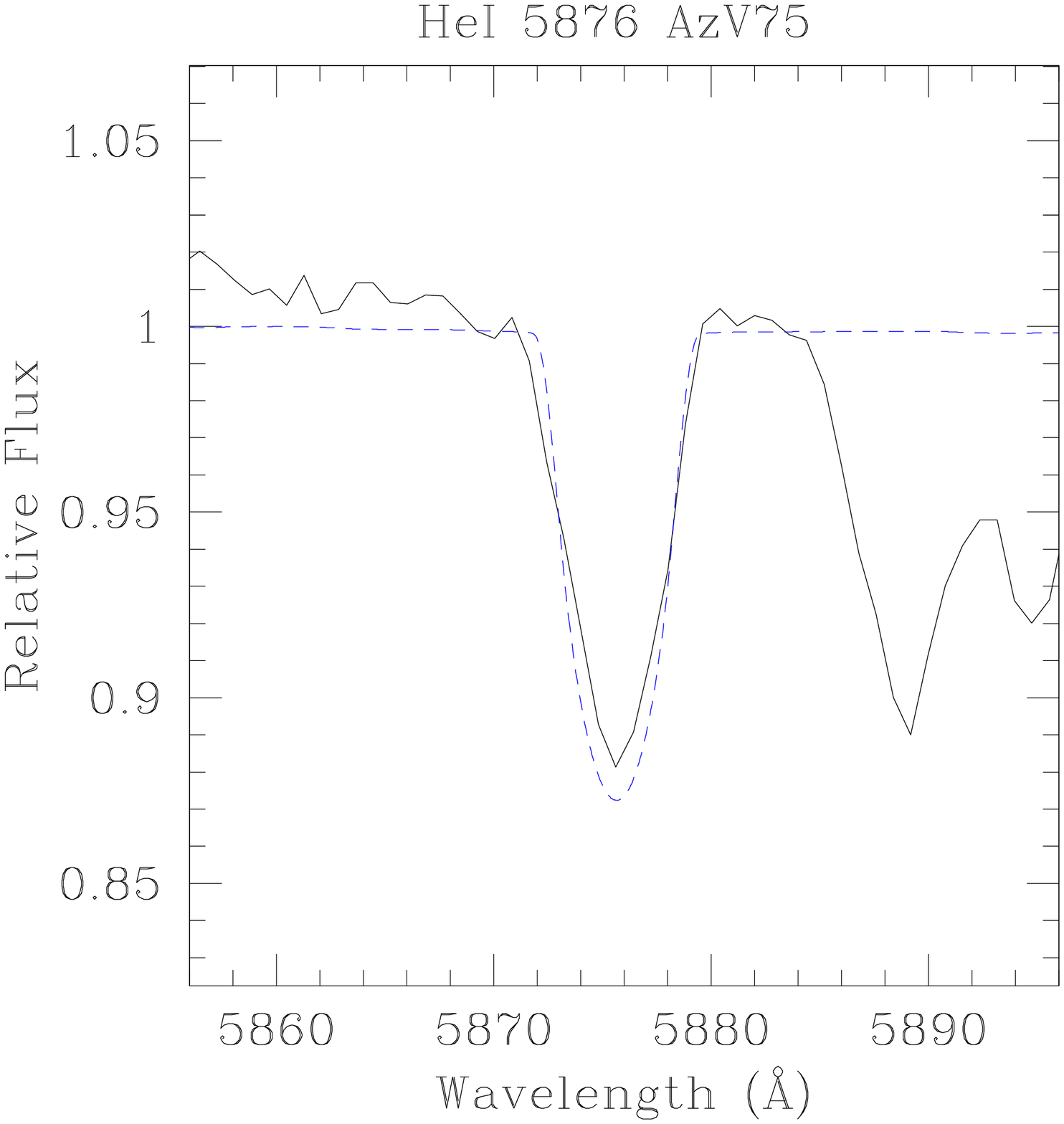}
\plotone{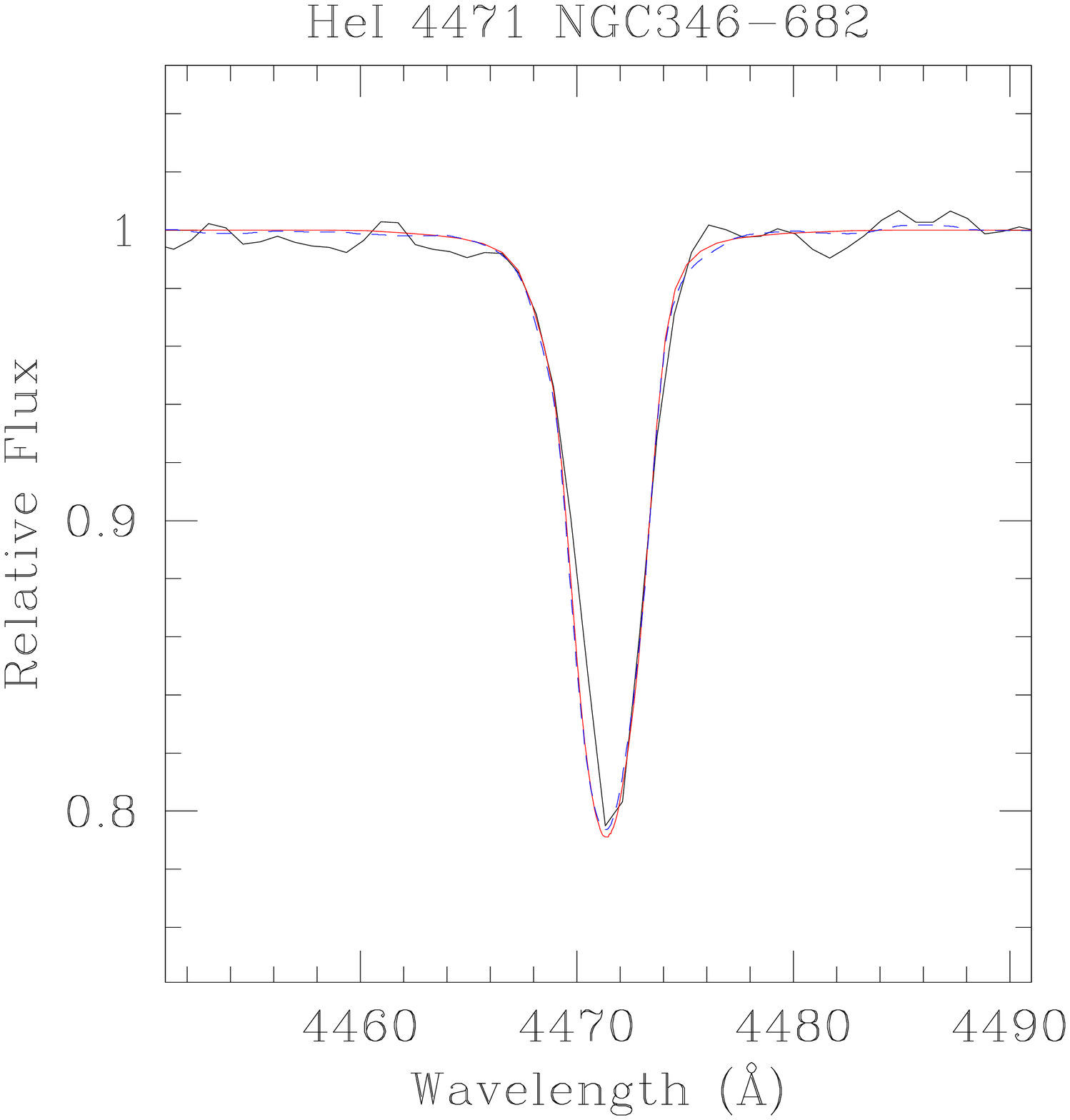}
\plotone{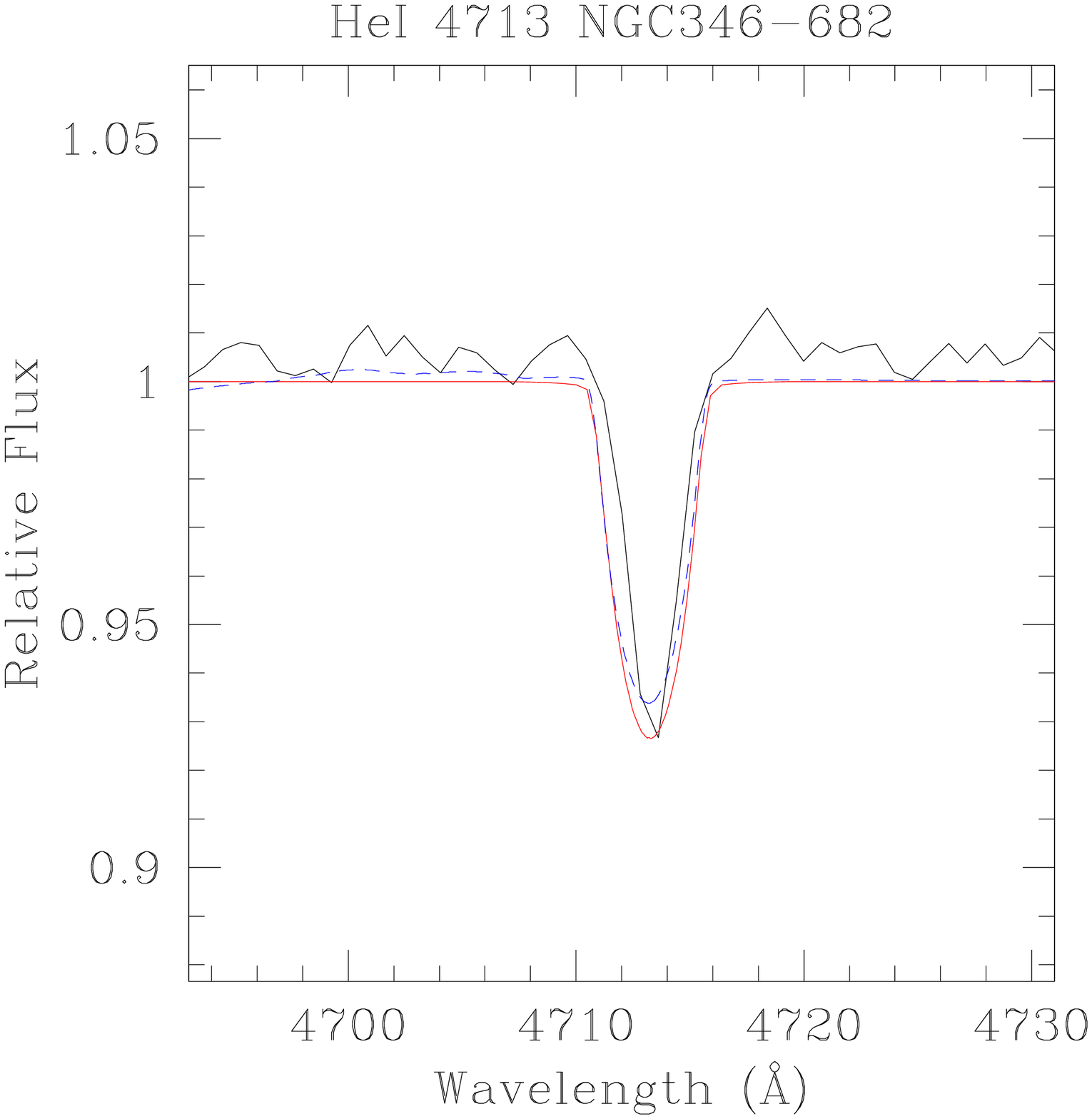}
\plotone{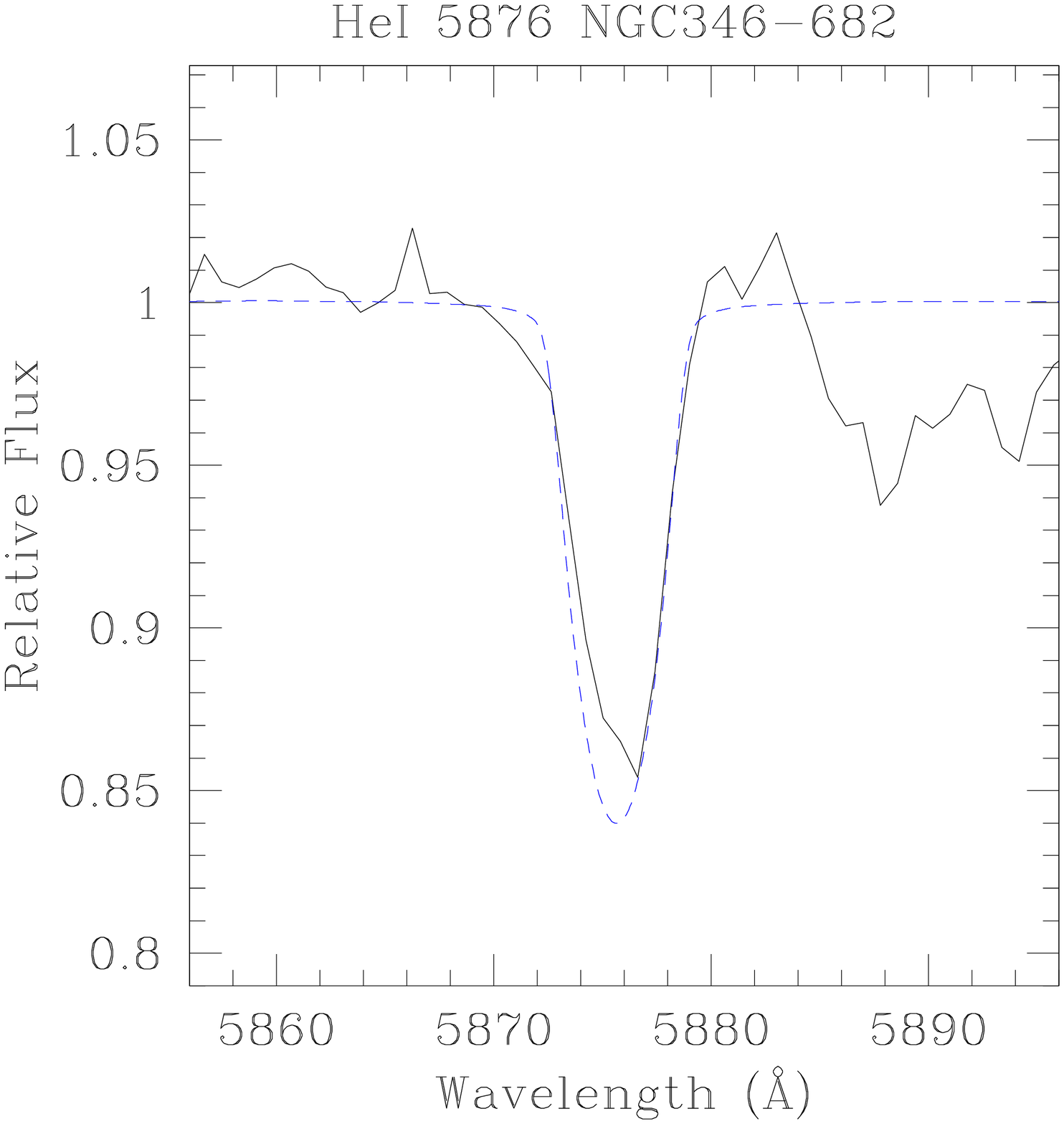}
\caption{\label{fig:tripletsA}  The fits for the He~I triplet lines for a sample of early to intermediate O-type dwarfs and supergiants.
Black shows the observed spectrum, the red line shows the \fastwind\ fit, and the dashed blue line shows the \cmfgen\ fit.
Note that the He~I $\lambda 4471$ and He~I $\lambda 5876$ line (the latter not fit by \fastwind) are $^3$P$^o -^3$D
transitions, while the He~I $\lambda 4713$ is a $^3$P$^o -^3$S transition. (See Figure~\ref{fig:Grot}.)  The stars shown here are AzV 388, an
O5.5 V((f)) star, AzV 75, an O5.5 I(f) star, and NGC 346-682, an O8 V star, all in the SMC. }
\end{figure}

\begin{figure}
\epsscale{0.25}
\plotone{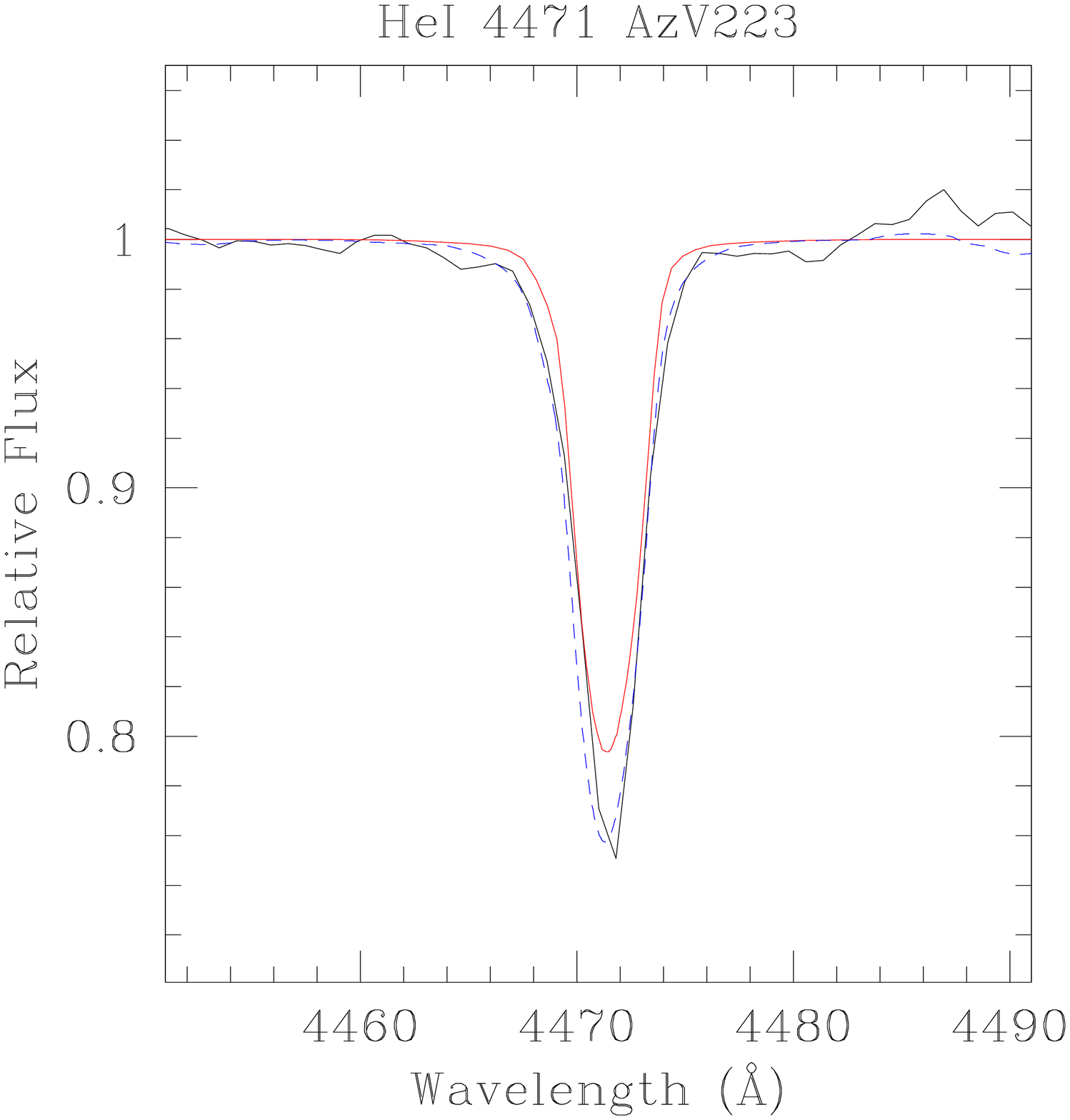}
\plotone{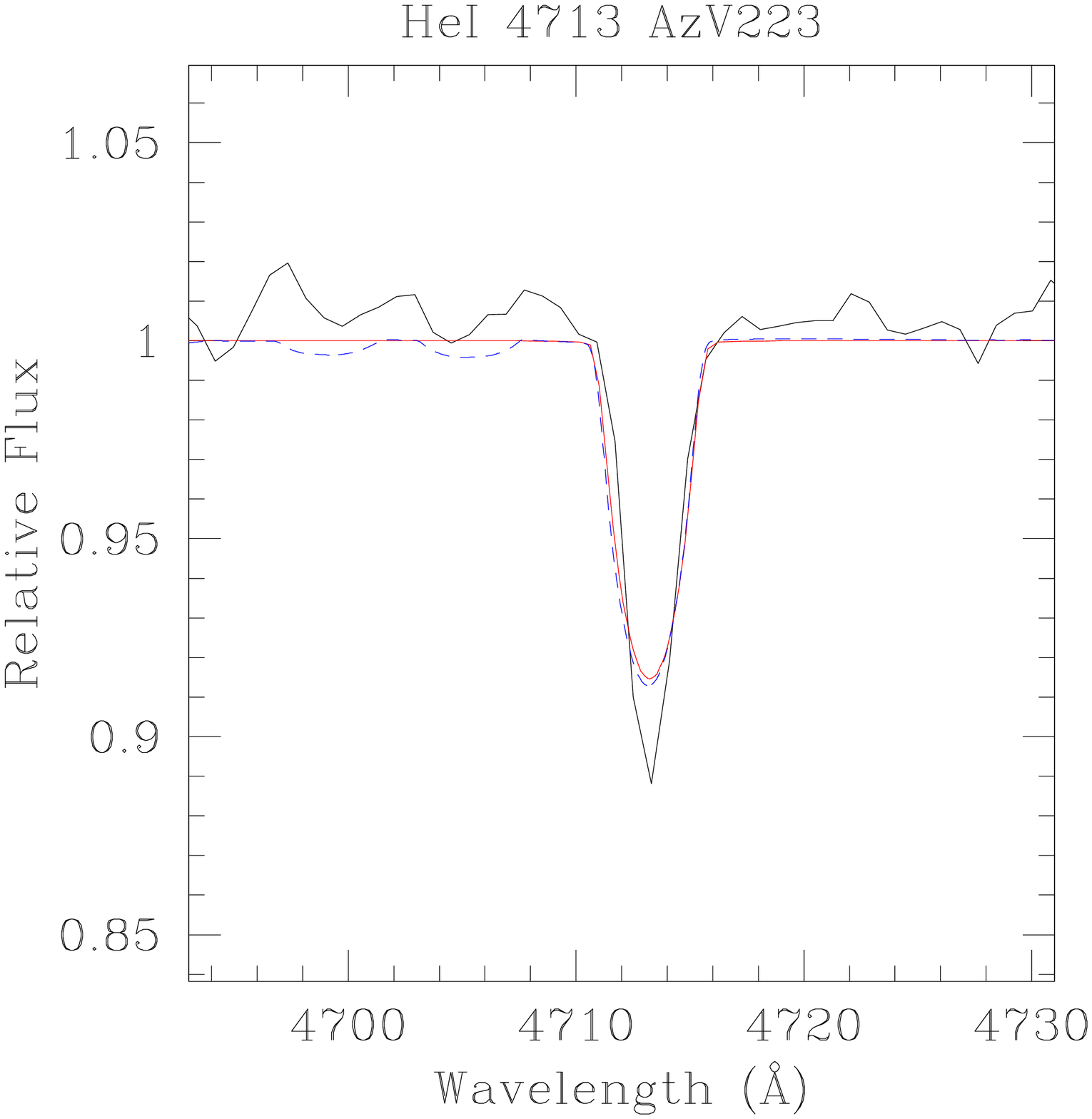}
\plotone{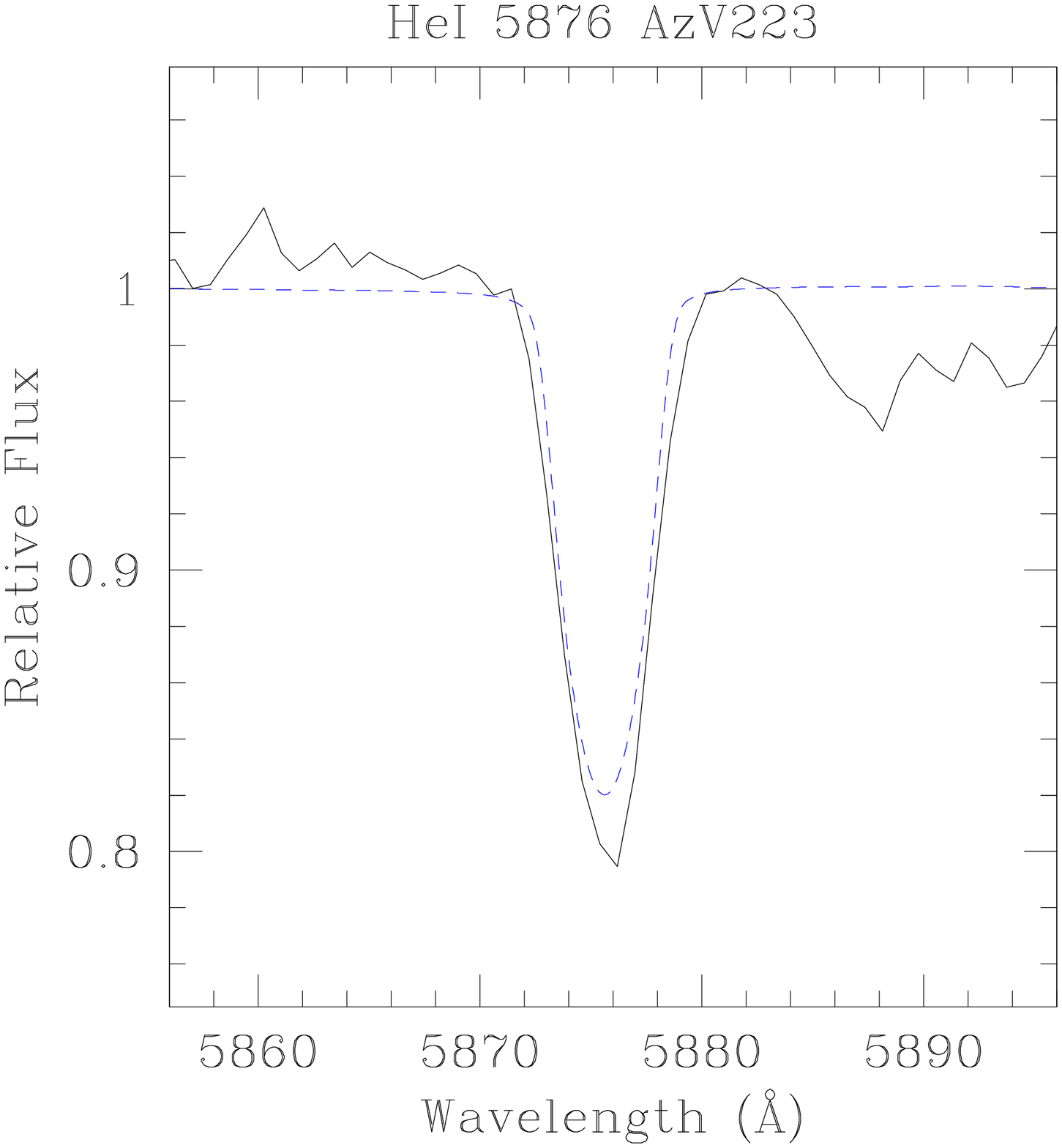}
\plotone{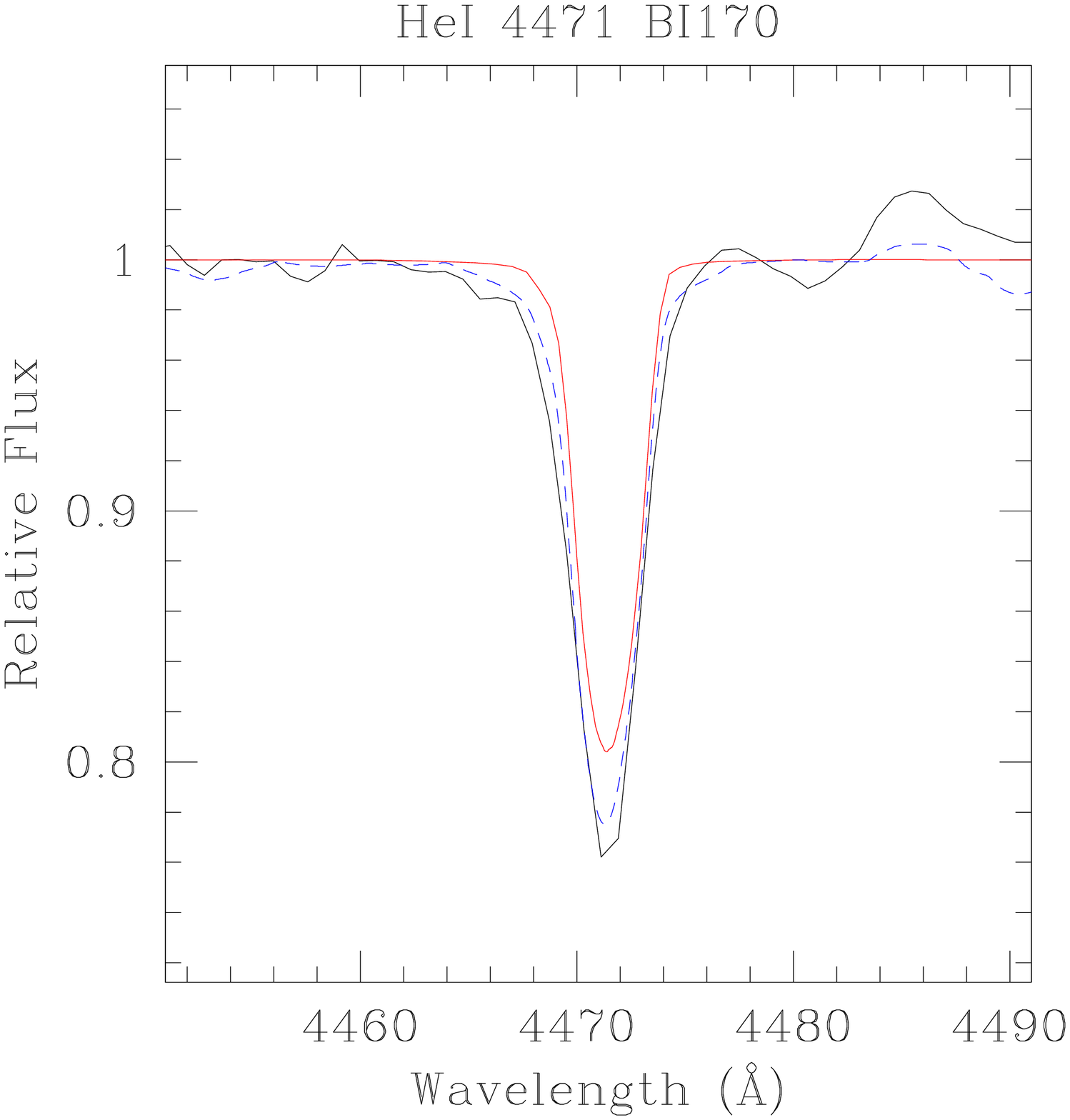}
\plotone{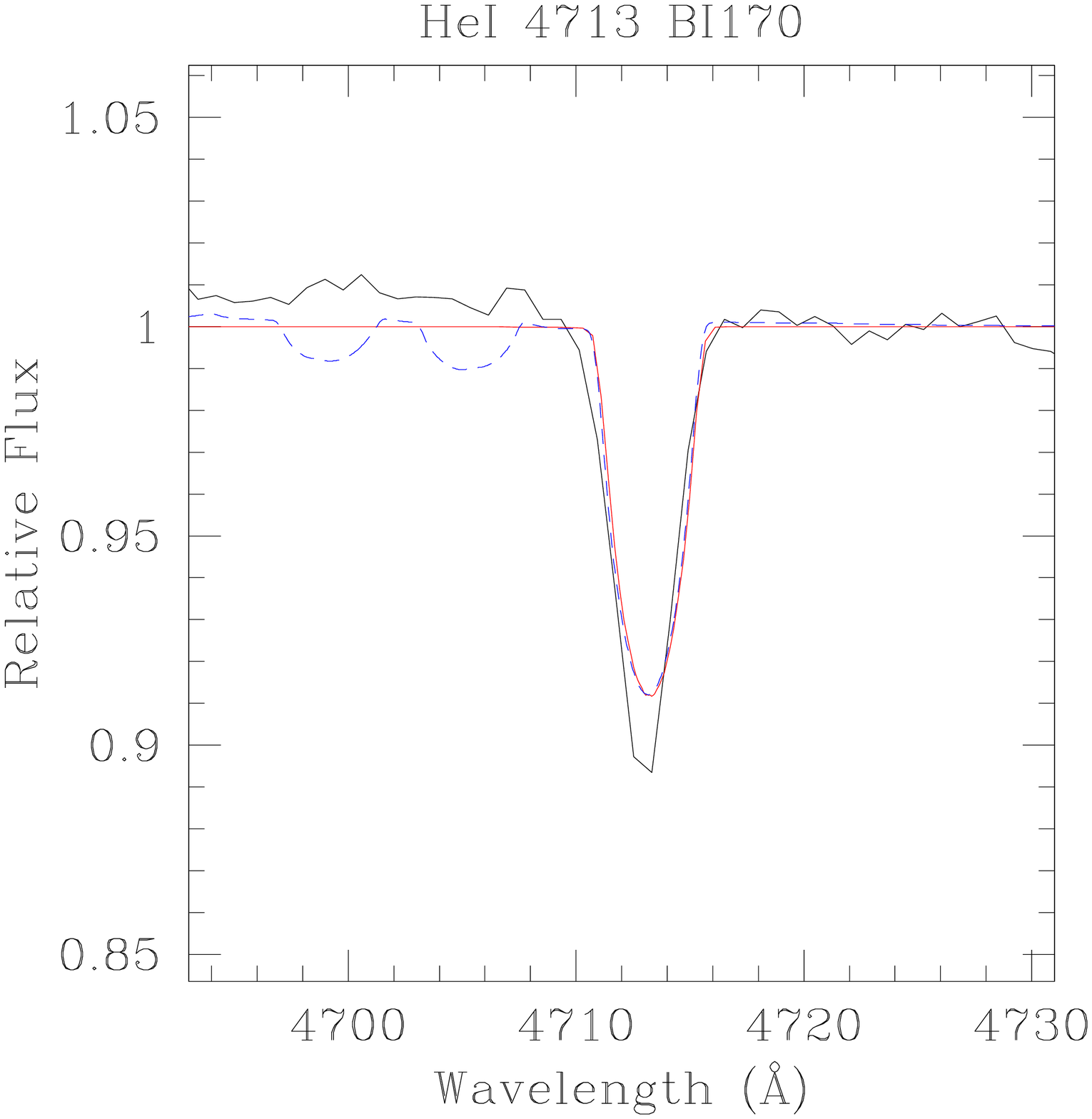}
\plotone{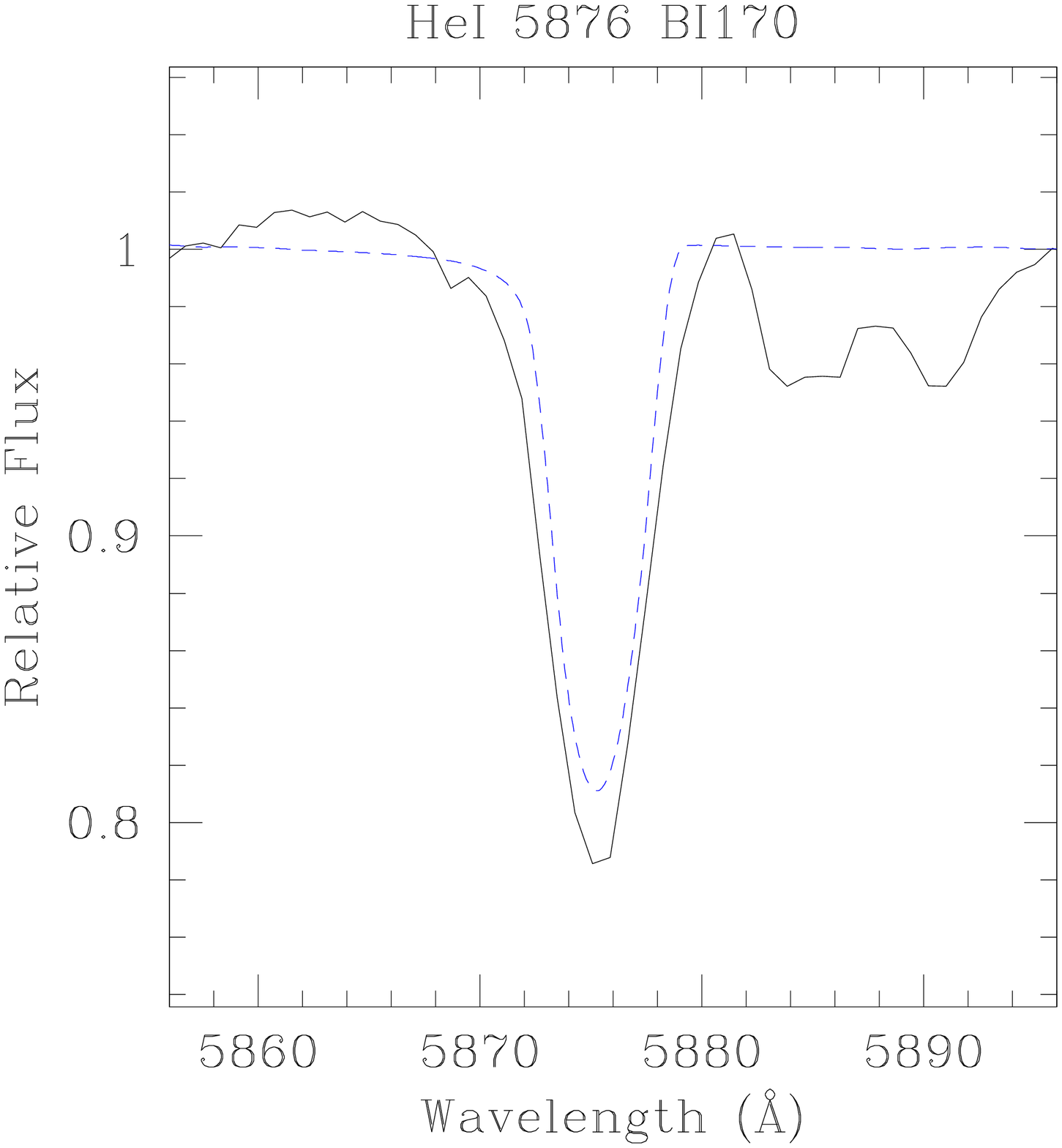}
\plotone{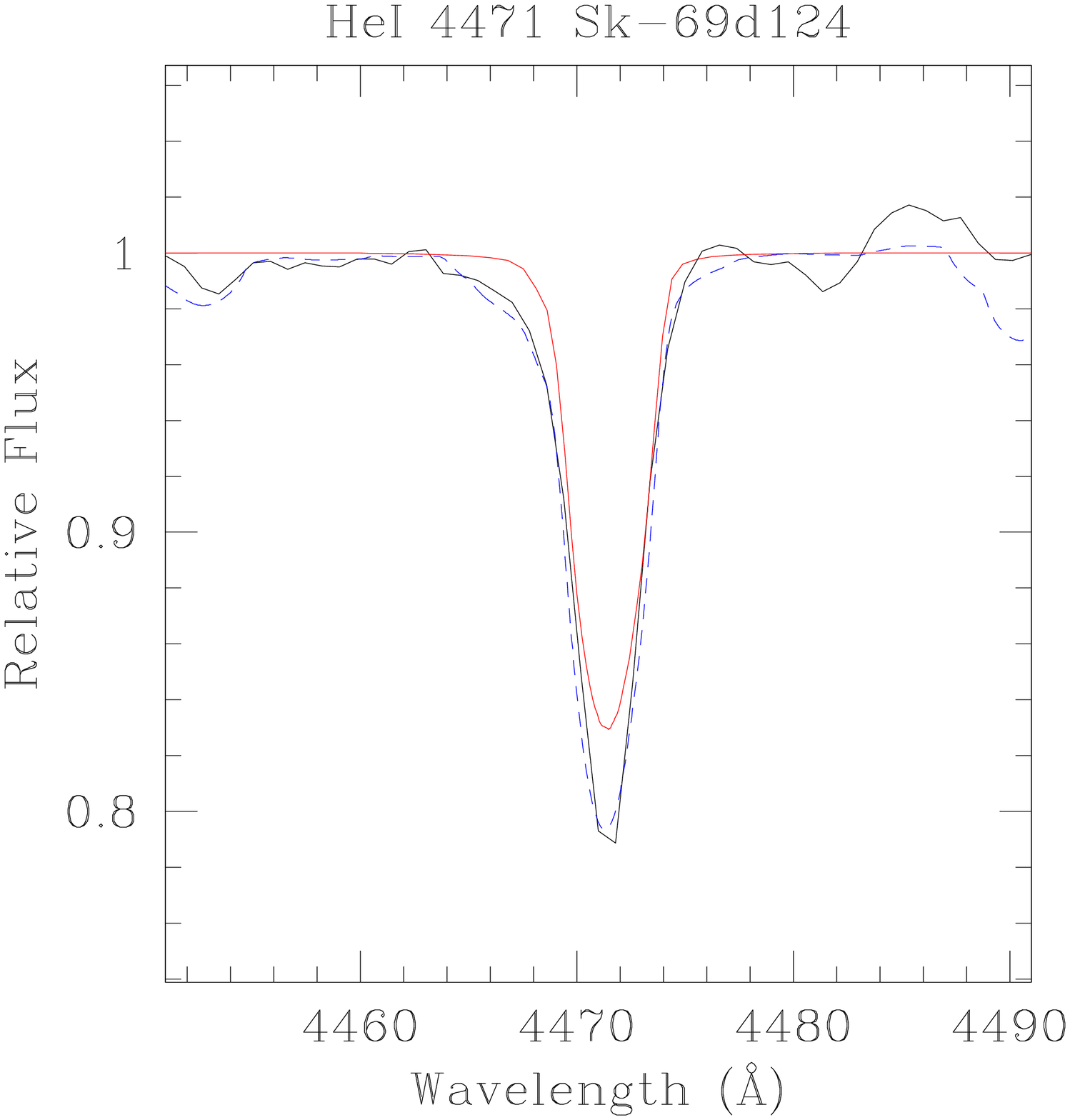}
\plotone{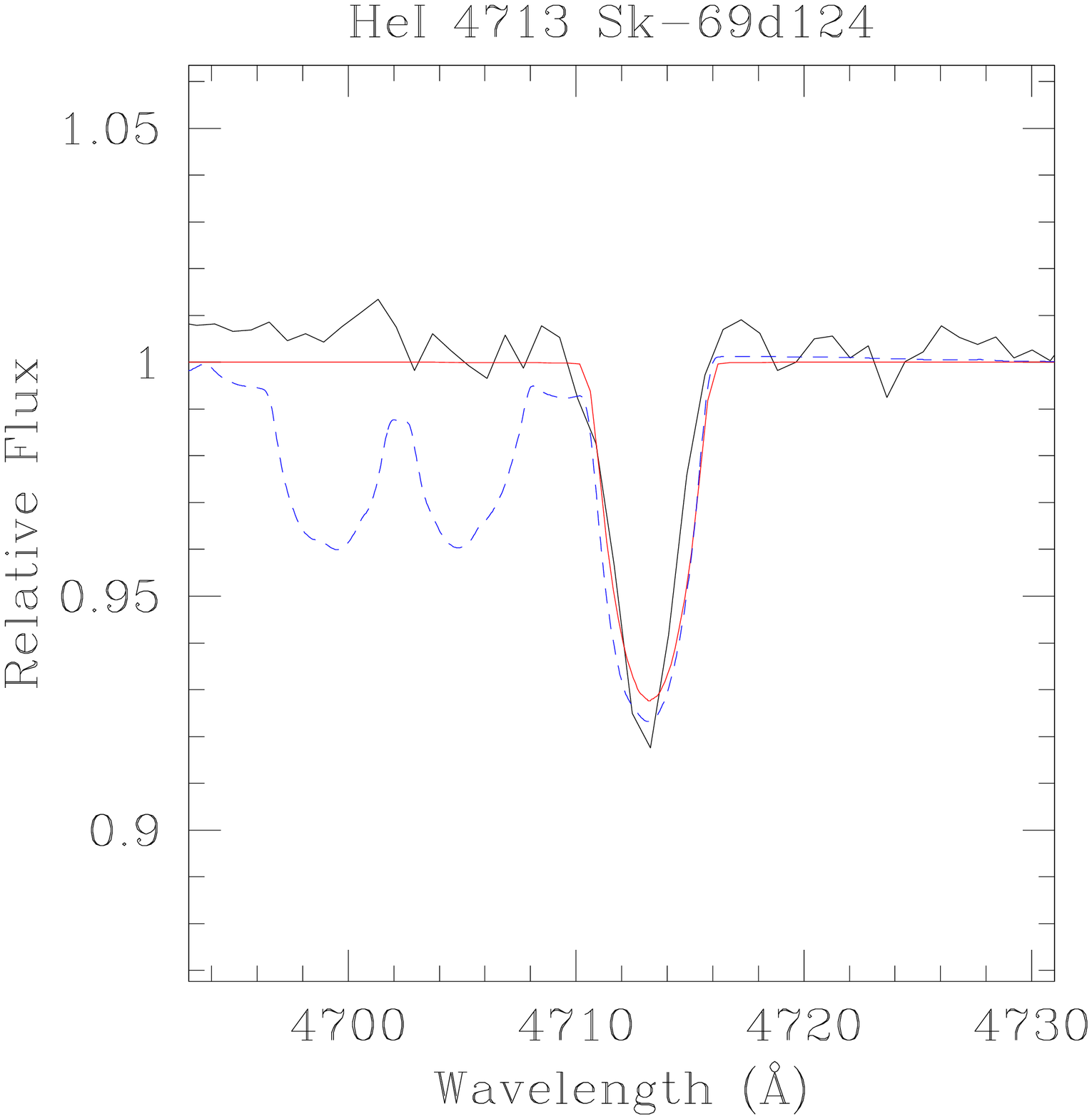}
\plotone{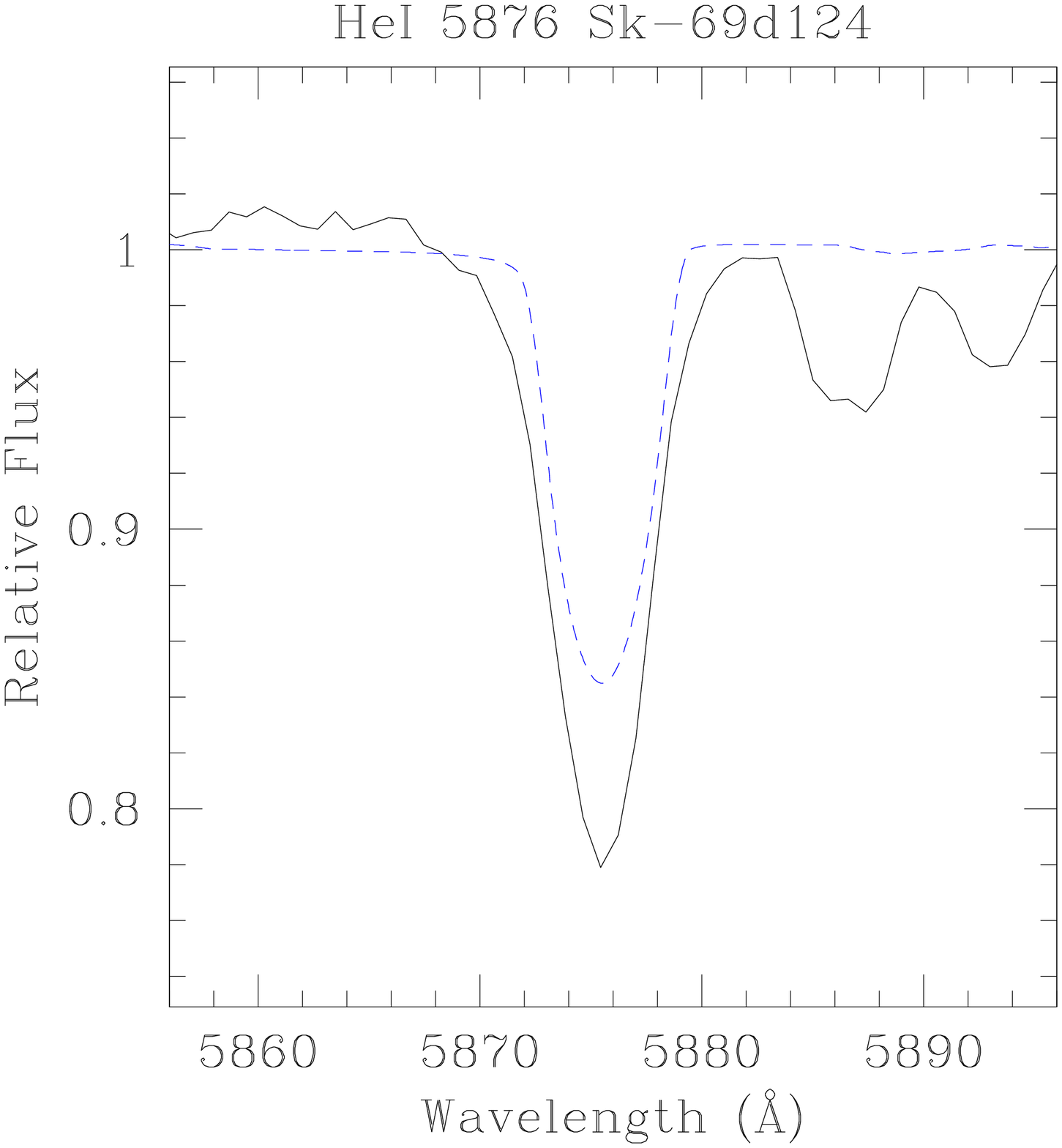}
\caption{\label{fig:tripletsB}  The fits for the He~I triplet lines for the late-type O supergiants. Black shows the observed spectrum,
the red line shows the \fastwind\ fit, and the dashed blue line shows the \cmfgen\ fit. Note that the He~I $\lambda 4471$ and
He~I $\lambda 5876$ line (the latter not fit by \fastwind) are $^3$P$^o - ^3$D  transitions, while the He~I $\lambda 4713$ is
a $^3$P$^o - ^3$S transition. (See Figure~\ref{fig:Grot}.)  The stars shown here are AzV 233, an O9.5 II star in the SMC, BI 170, an O9.5 I star in the LMC,
and Sk $-69^\circ$124, an O9.7 I star in the LMC.}
\end{figure}

\begin{figure}
\epsscale{0.25}
\plotone{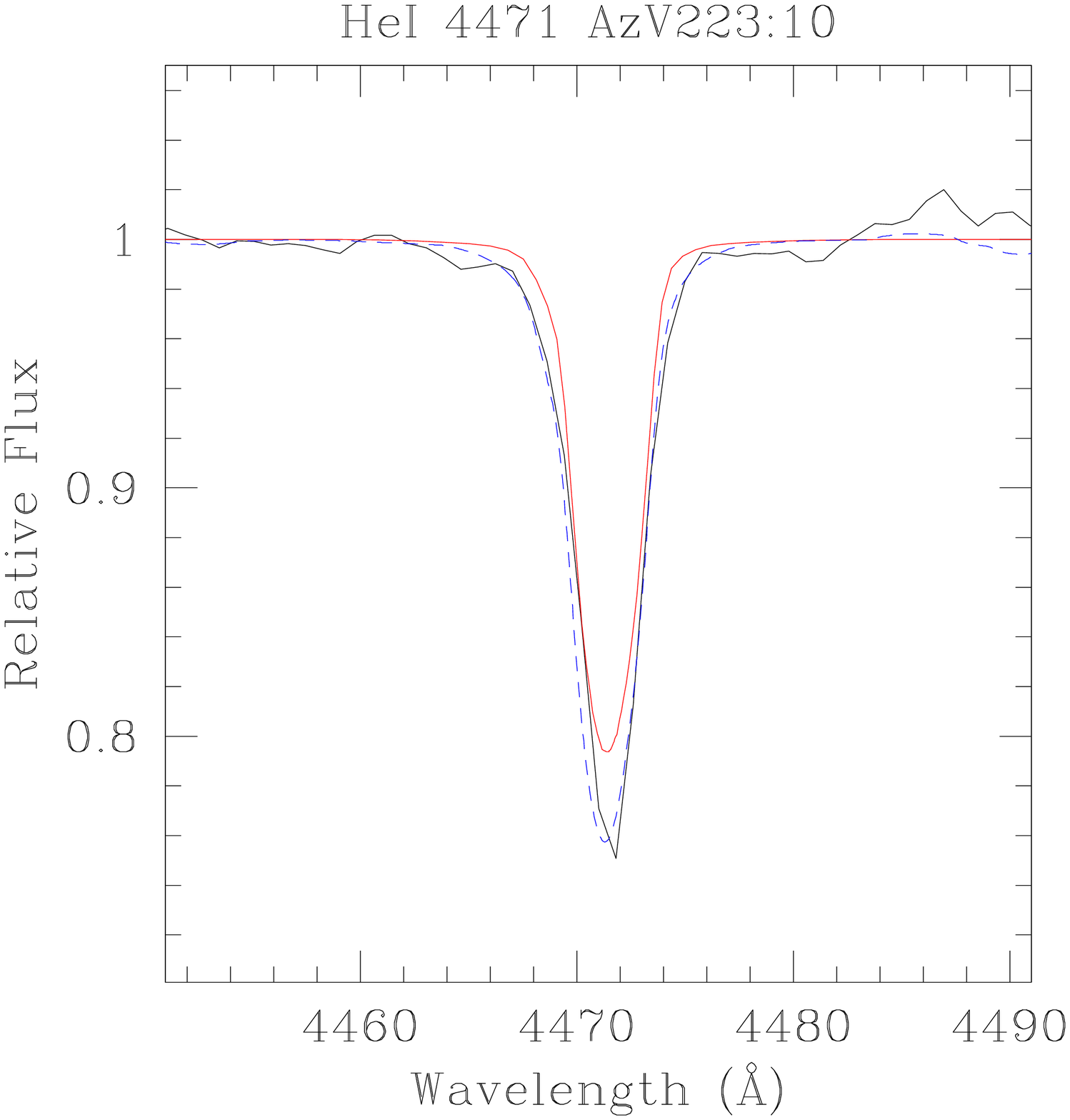}
\plotone{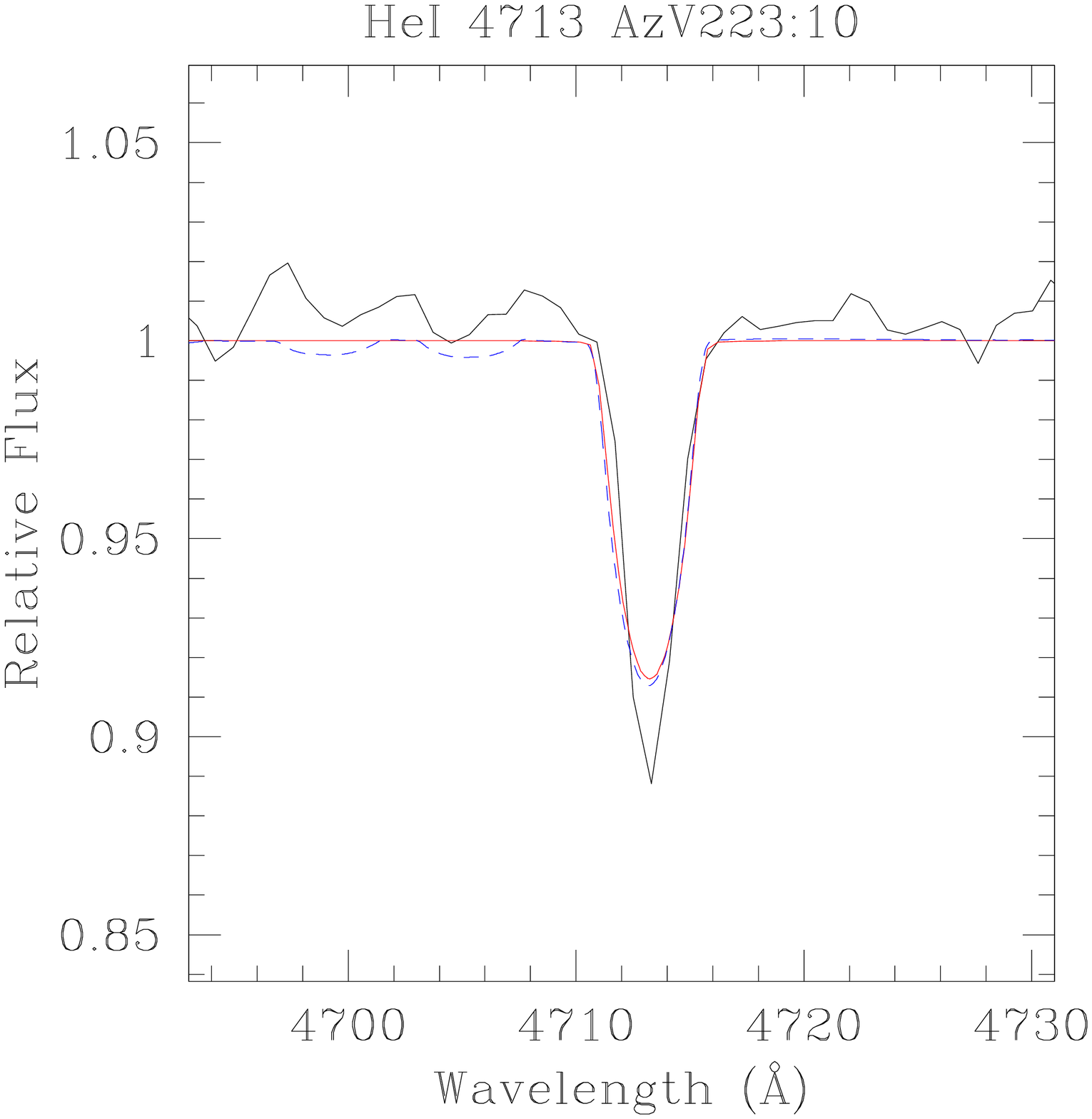}
\plotone{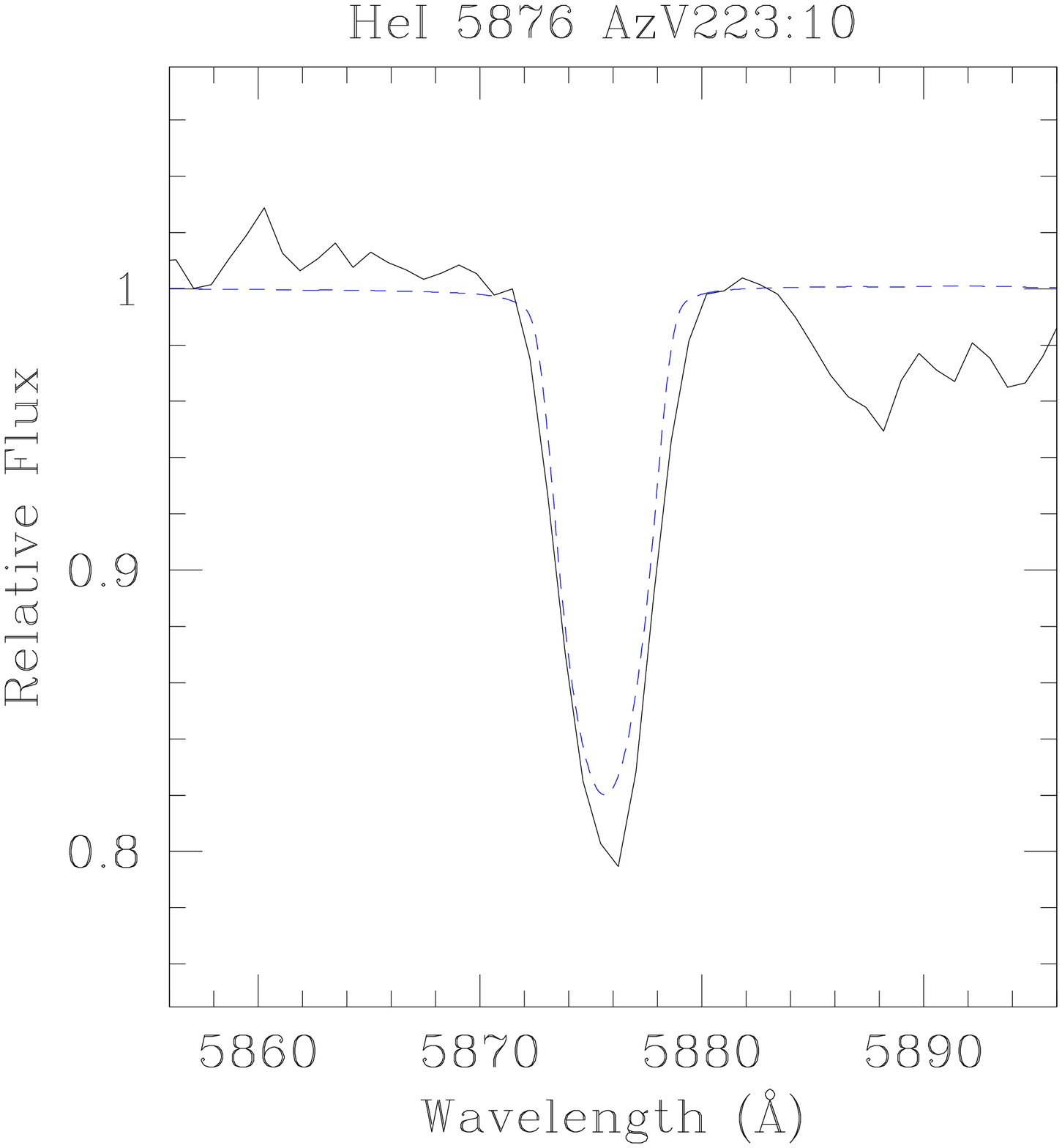}
\plotone{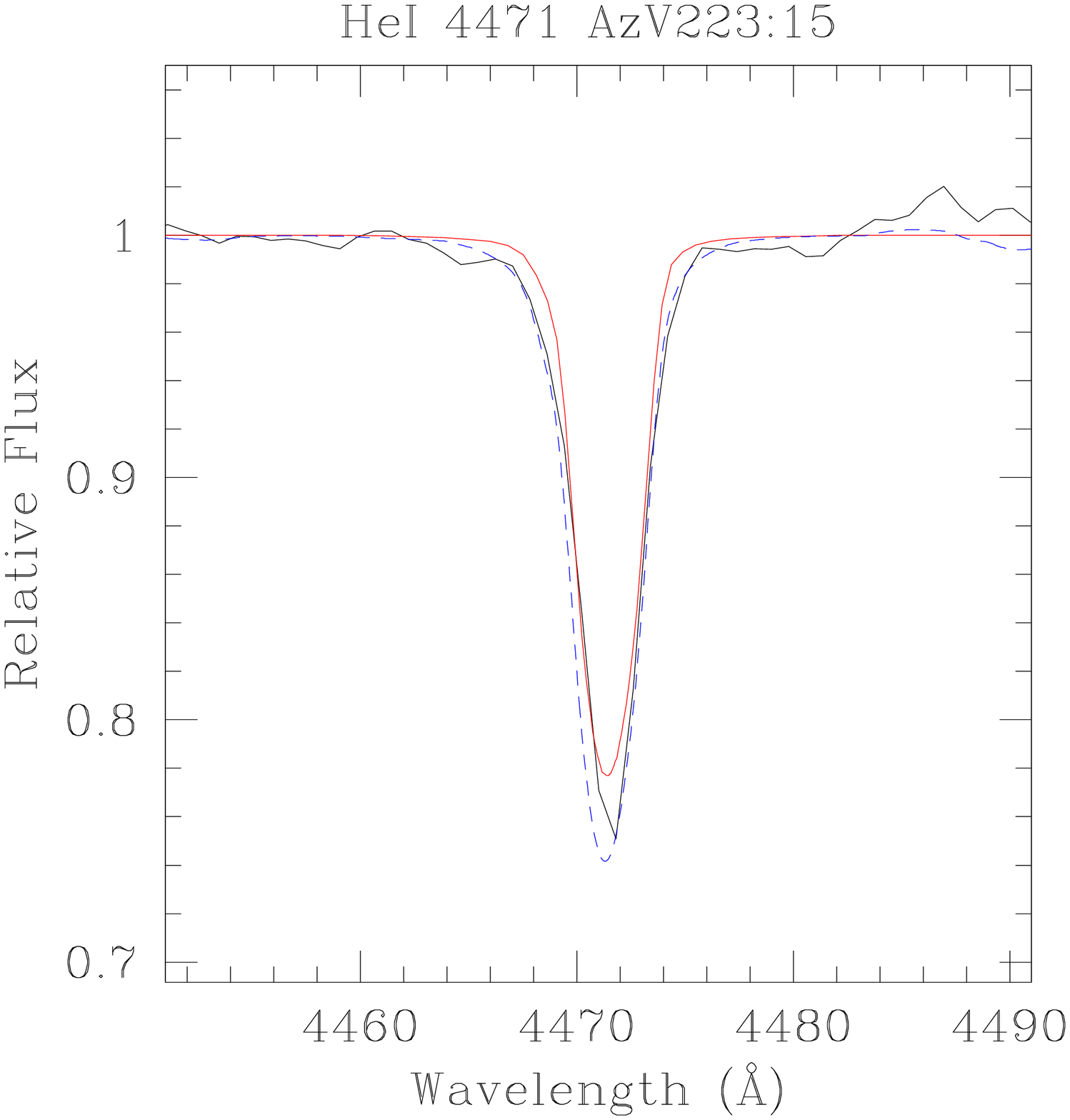}
\plotone{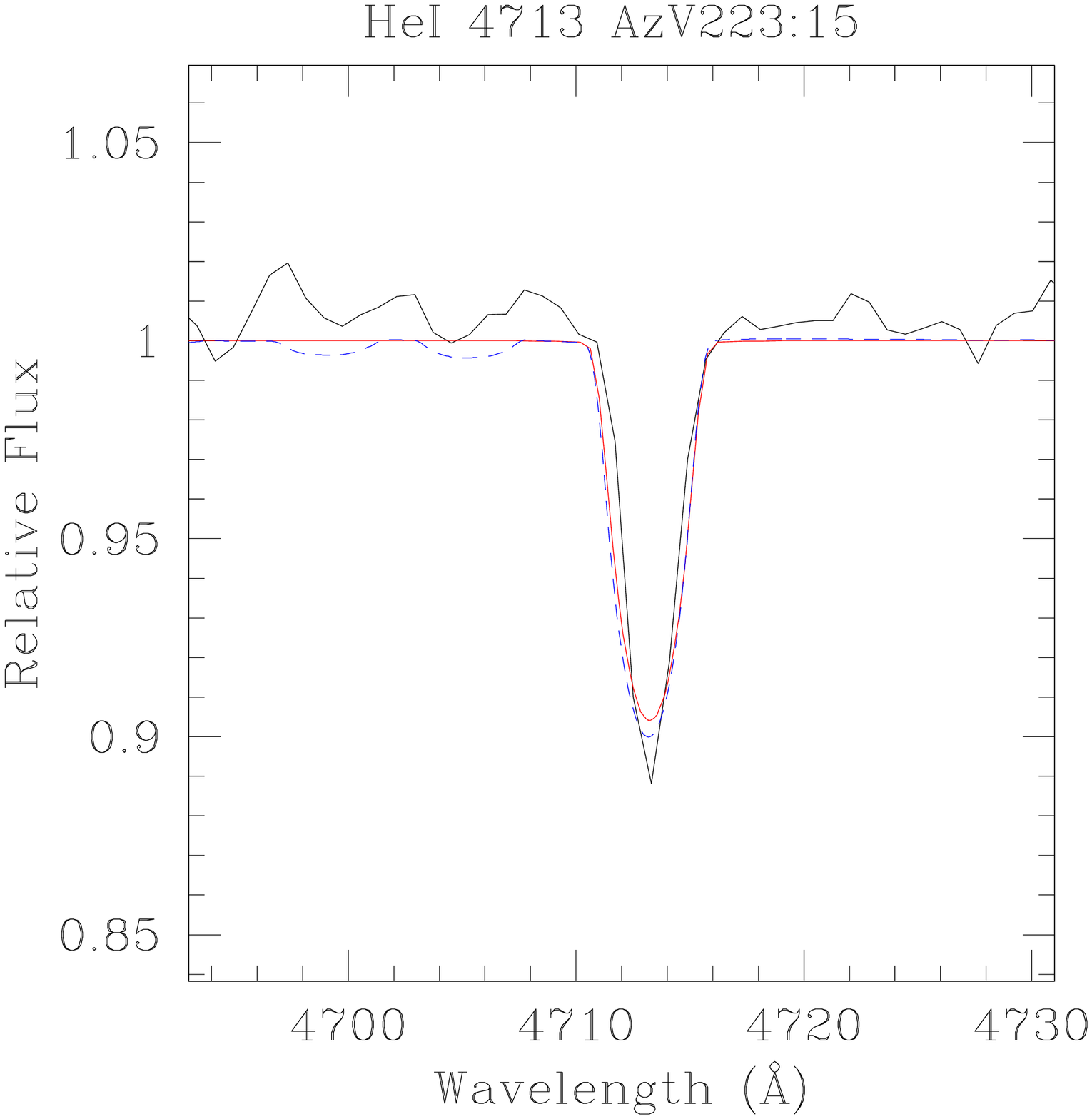}
\plotone{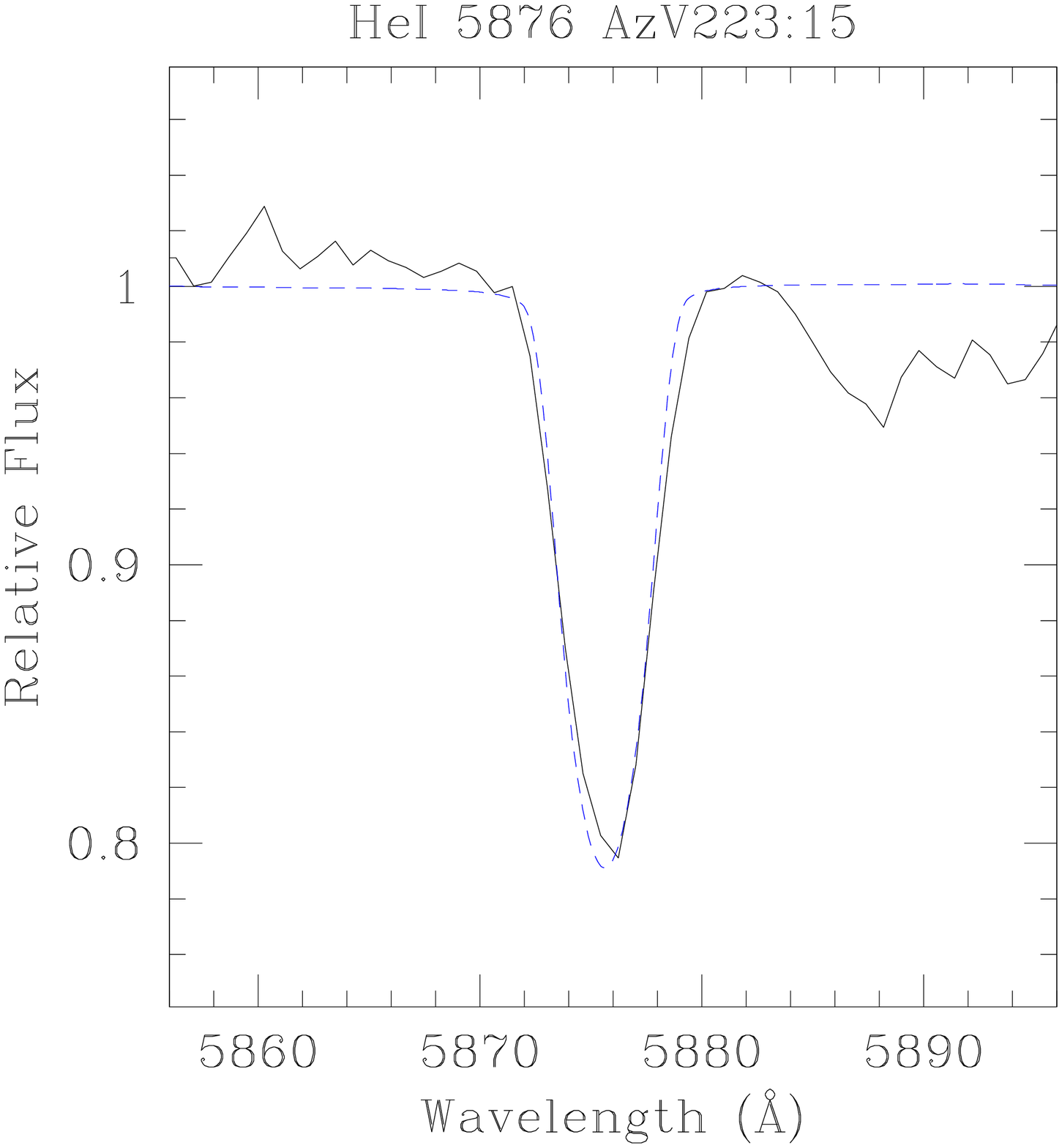}
\plotone{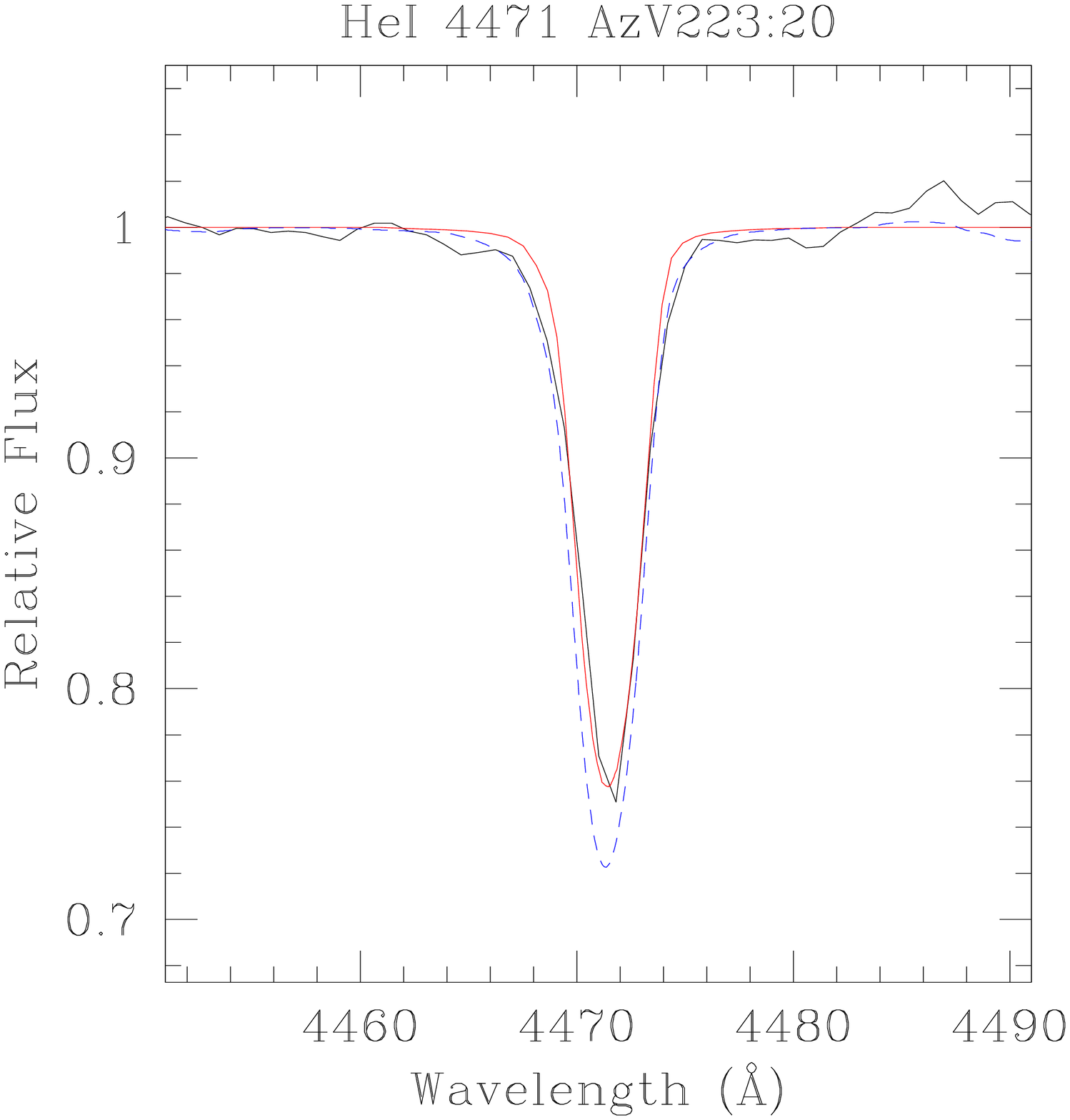}
\plotone{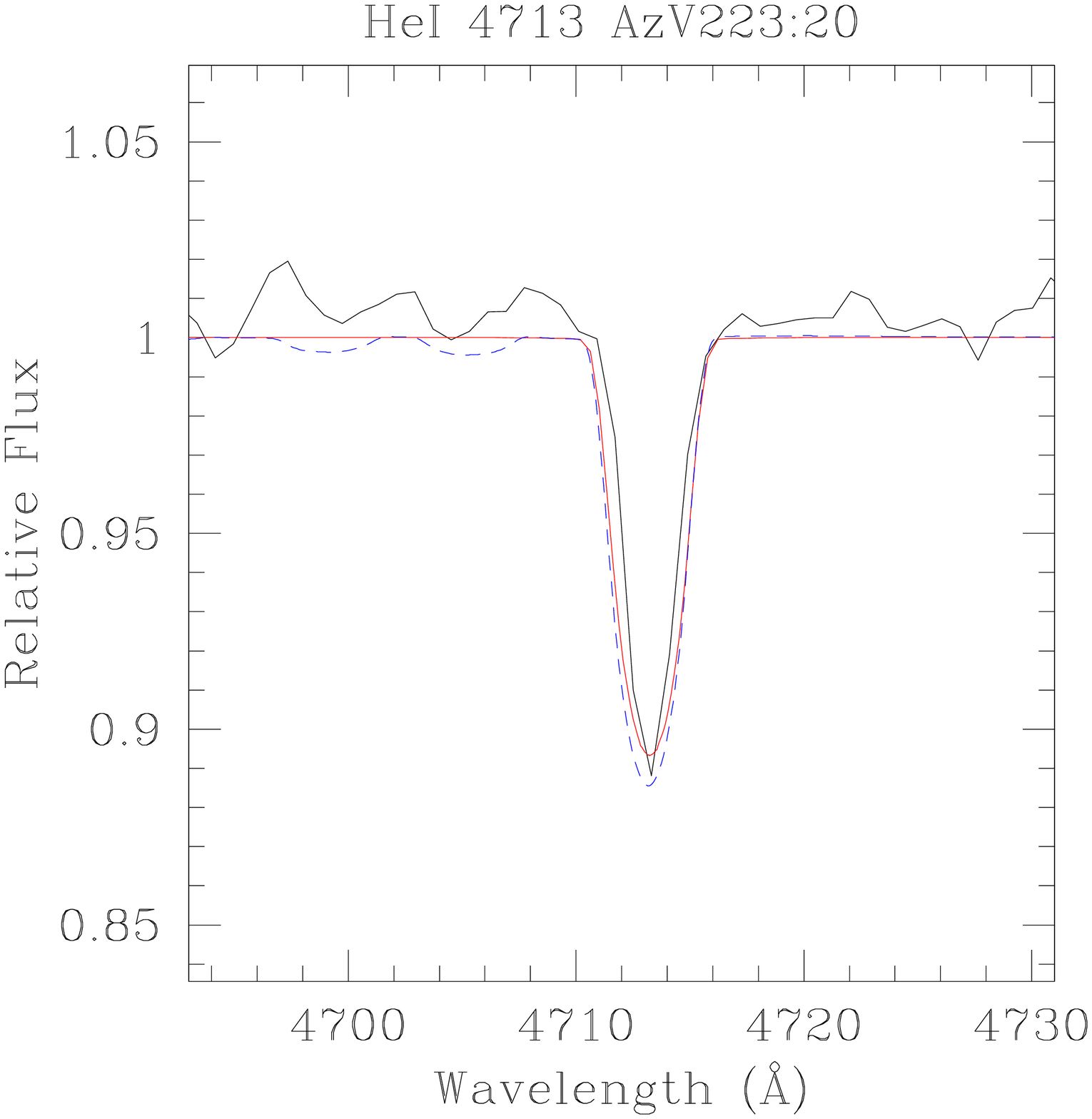}
\plotone{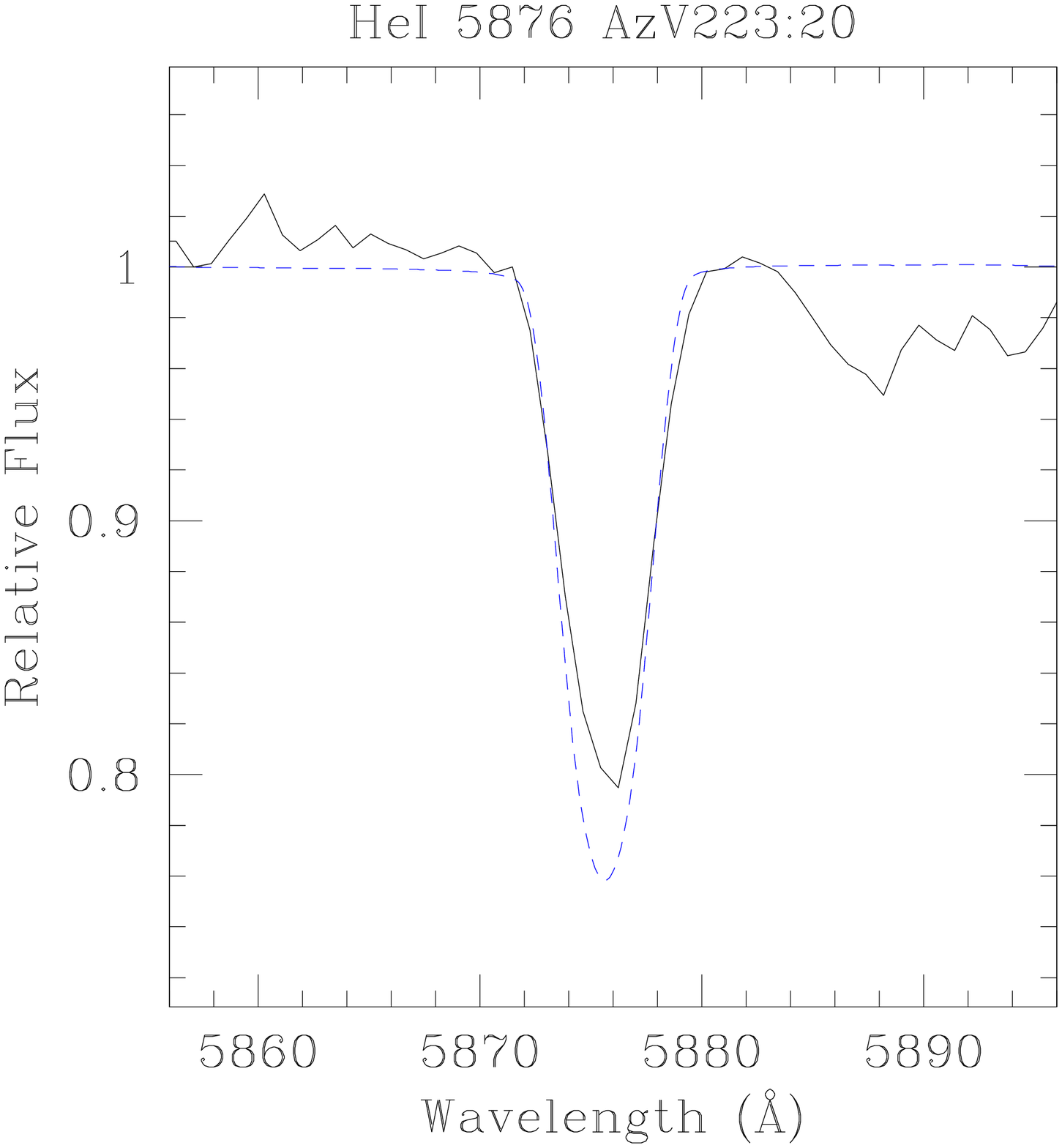}
\caption{\label{fig:MicroAzV223}  The effect of  microturbulence on the He I triplets in AzV223, an O9.5 II star in the SMC.
Black shows the observed spectrum, the red line shows the \fastwind\ fit, and the dashed blue line shows the \cmfgen\ fit.
The upper three panels show the model profiles computed using the  ``standard" 10 km s$^{-1}$ microturbulent velocities, the middle three panels show that obtained using 15 km s$^{-1}$, and the bottom three panels show the model profiles obtained using 20 km$^{-1}$.}
\end{figure}

\begin{figure}
\epsscale{0.25}
\plotone{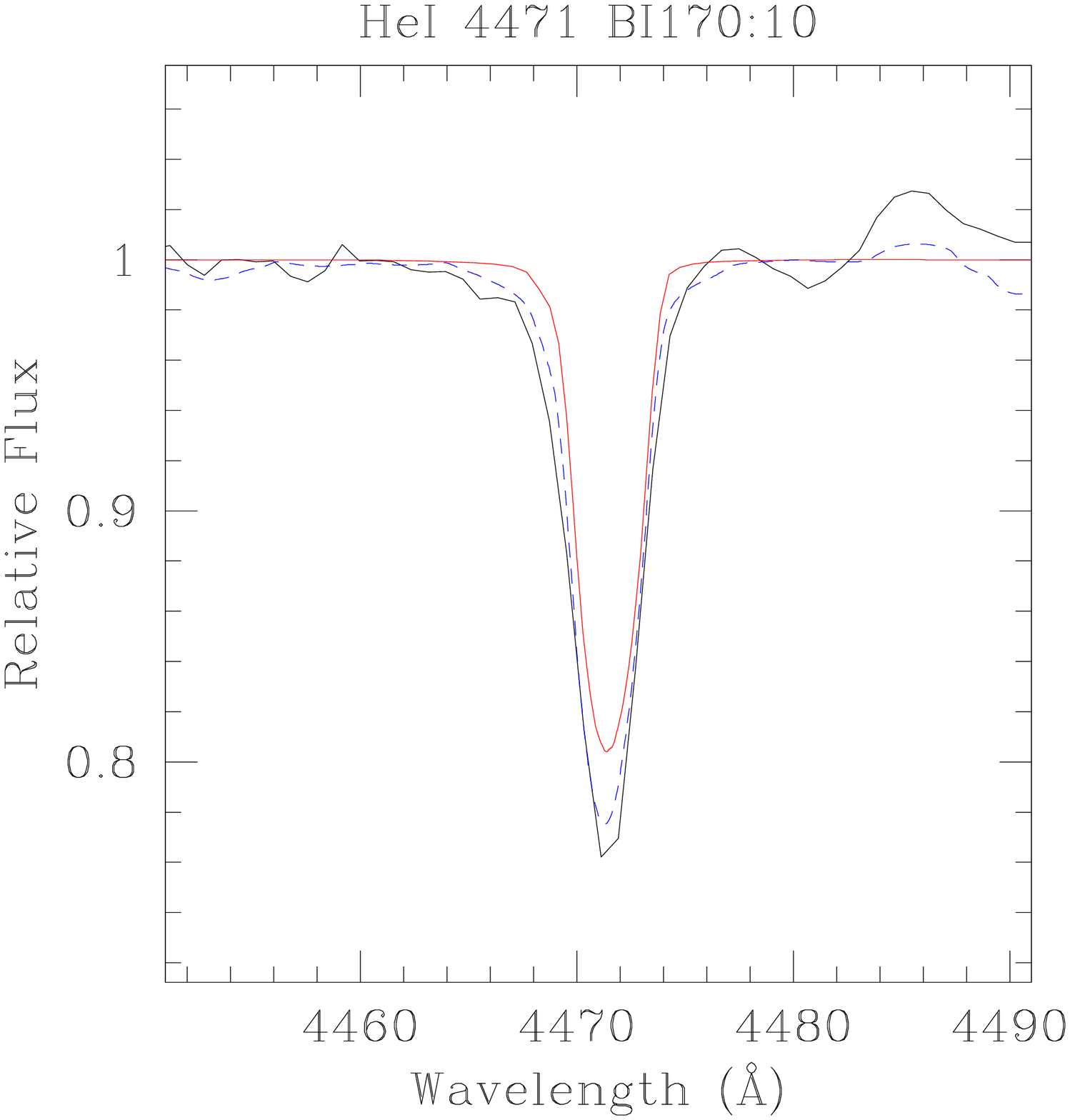}
\plotone{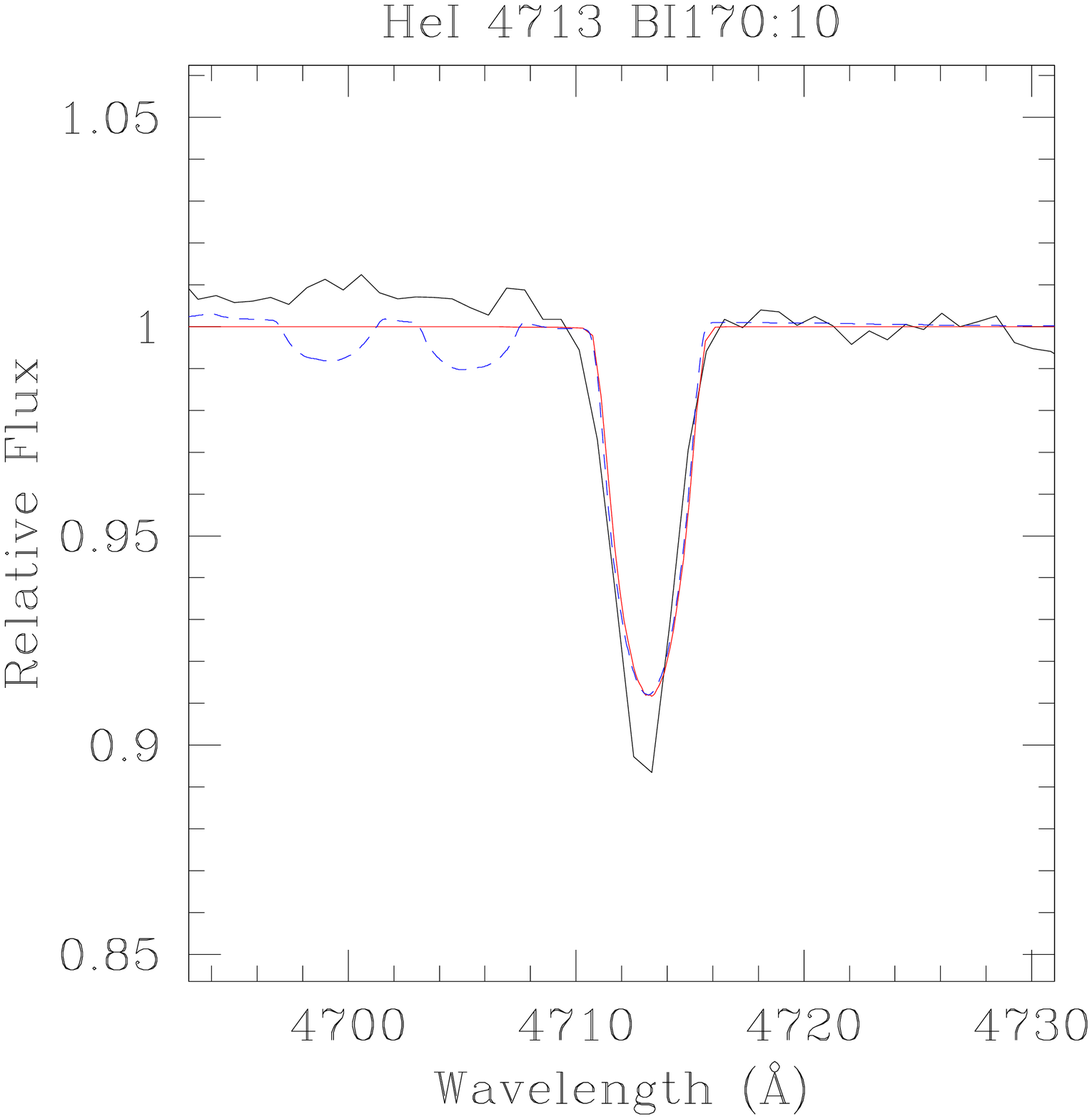}
\plotone{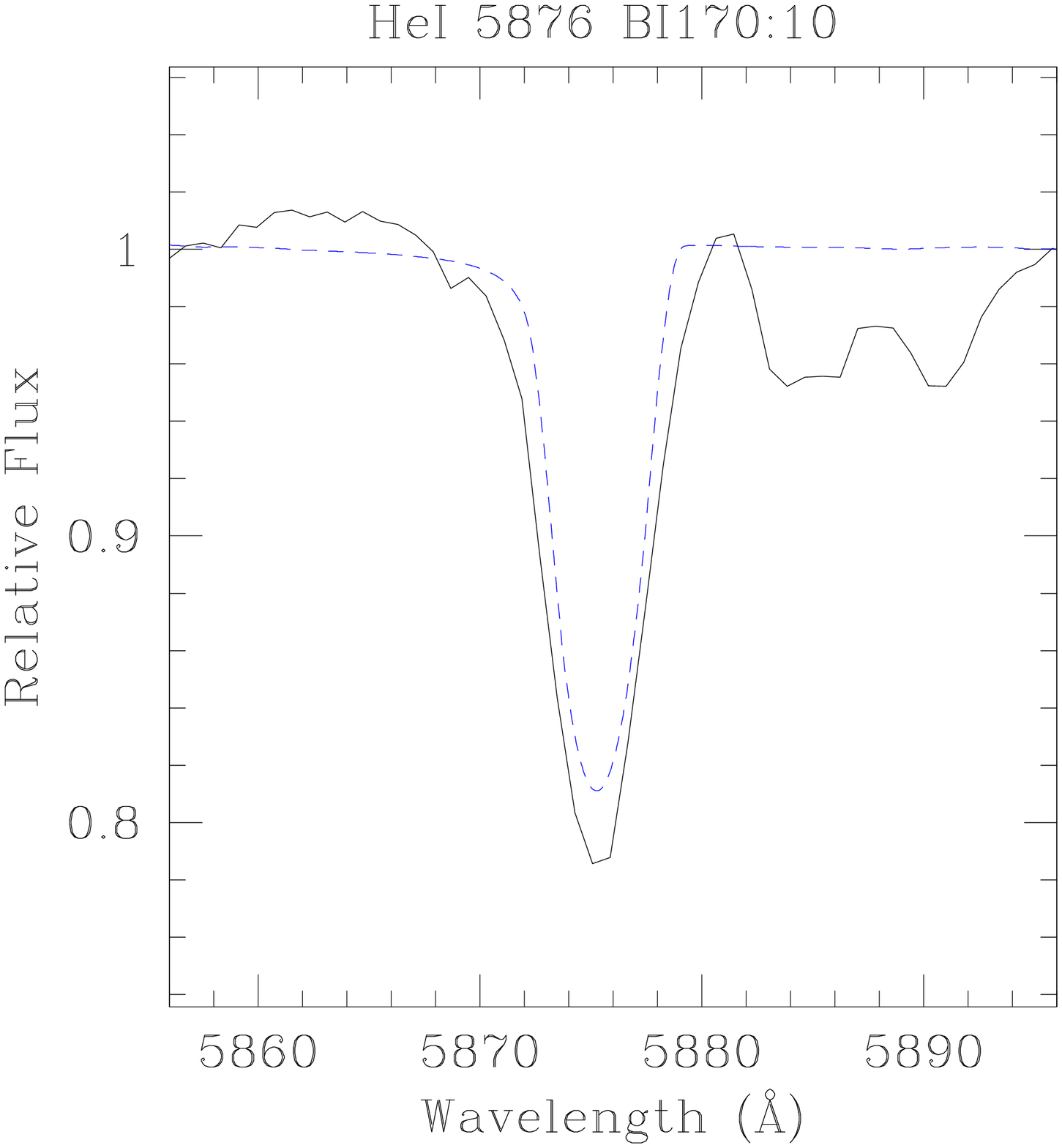}
\plotone{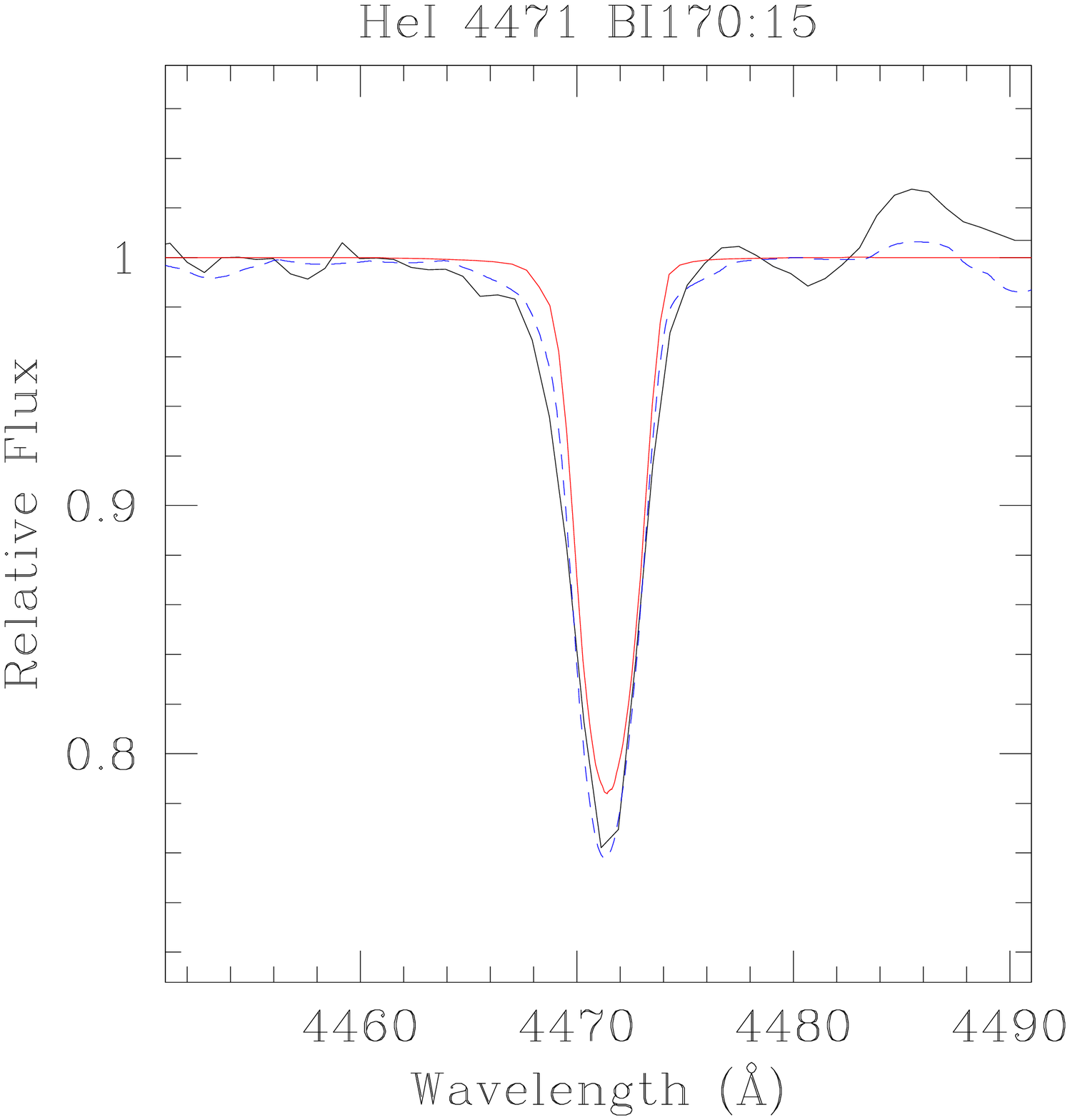}
\plotone{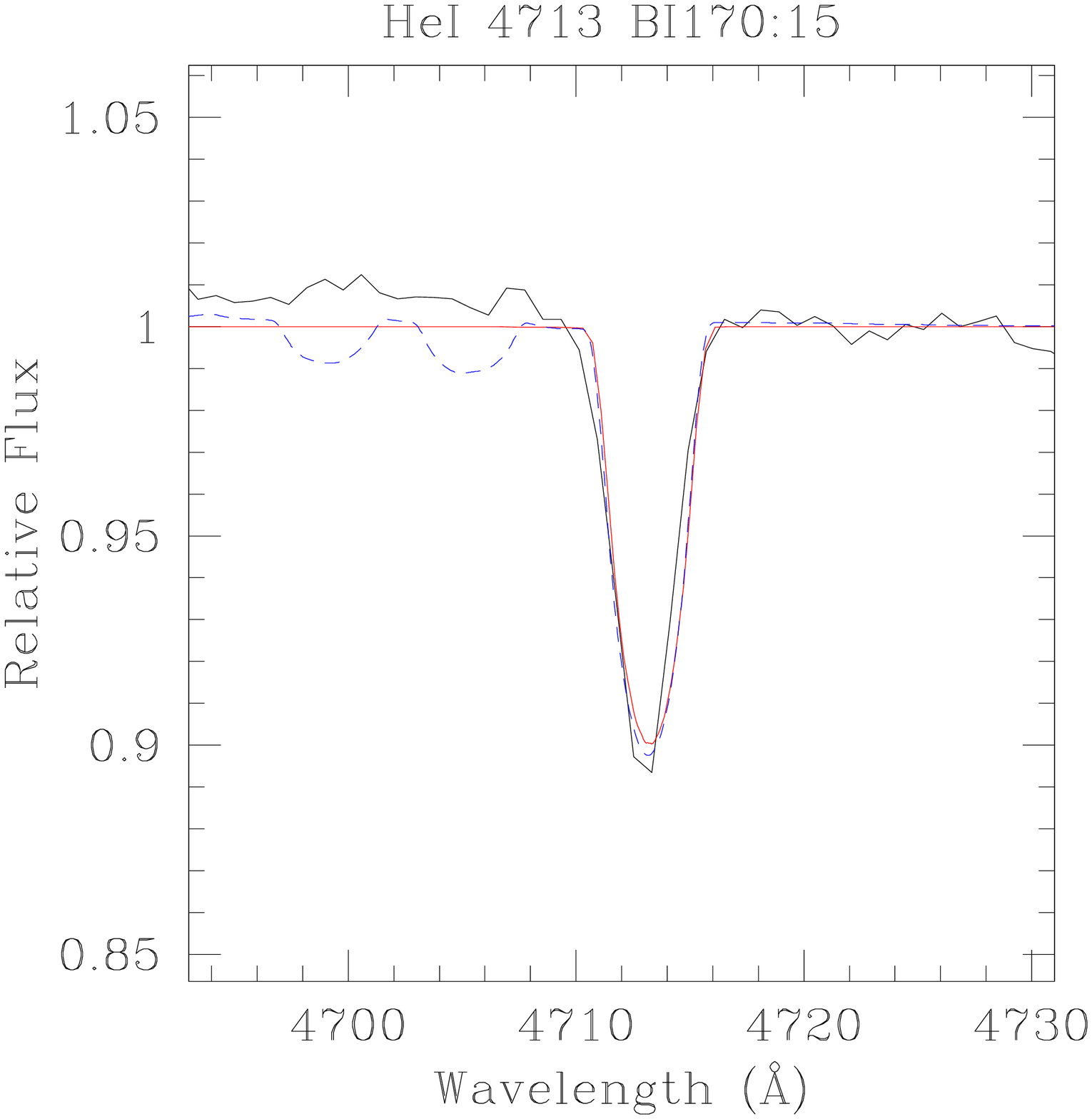}
\plotone{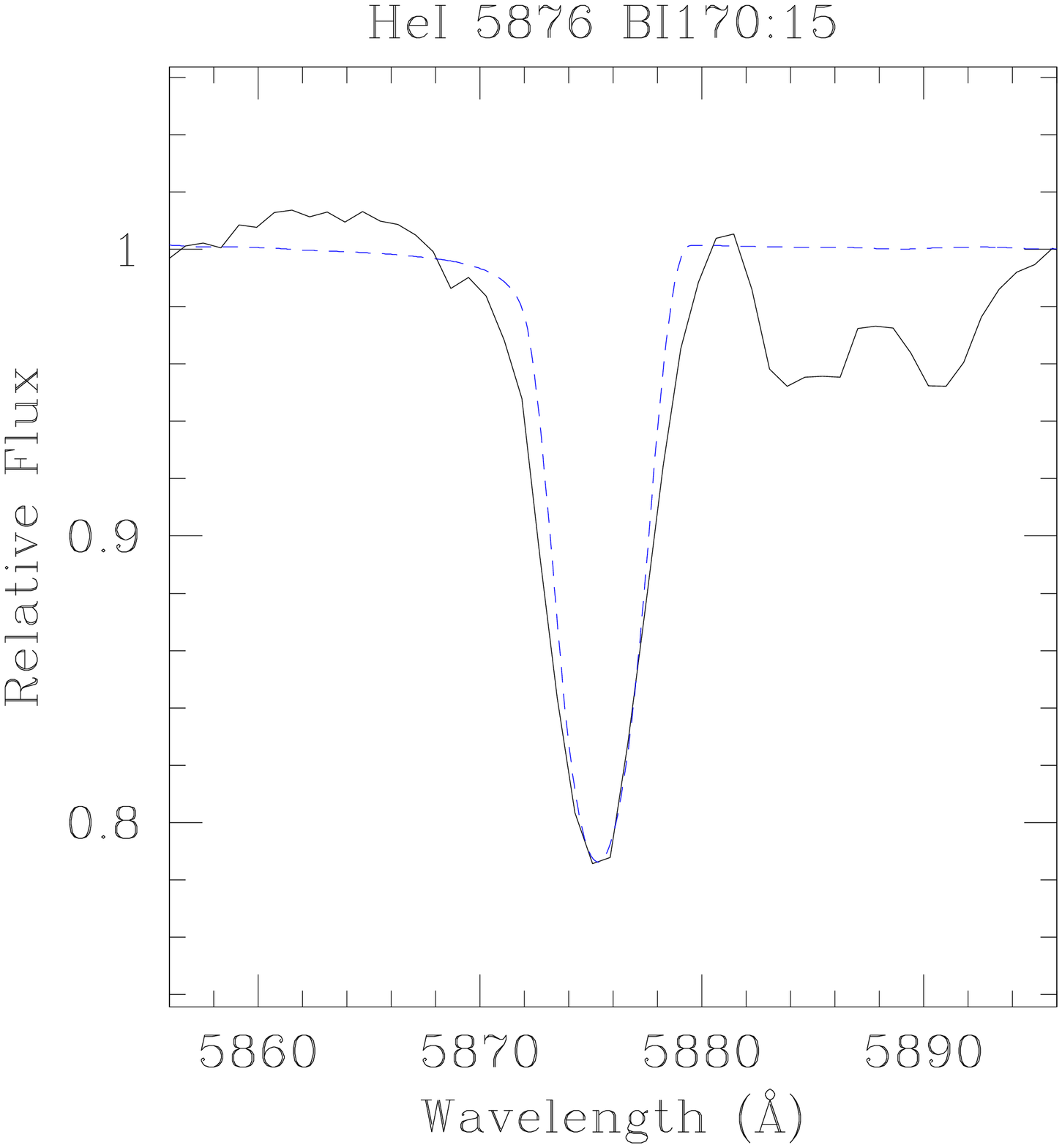}
\plotone{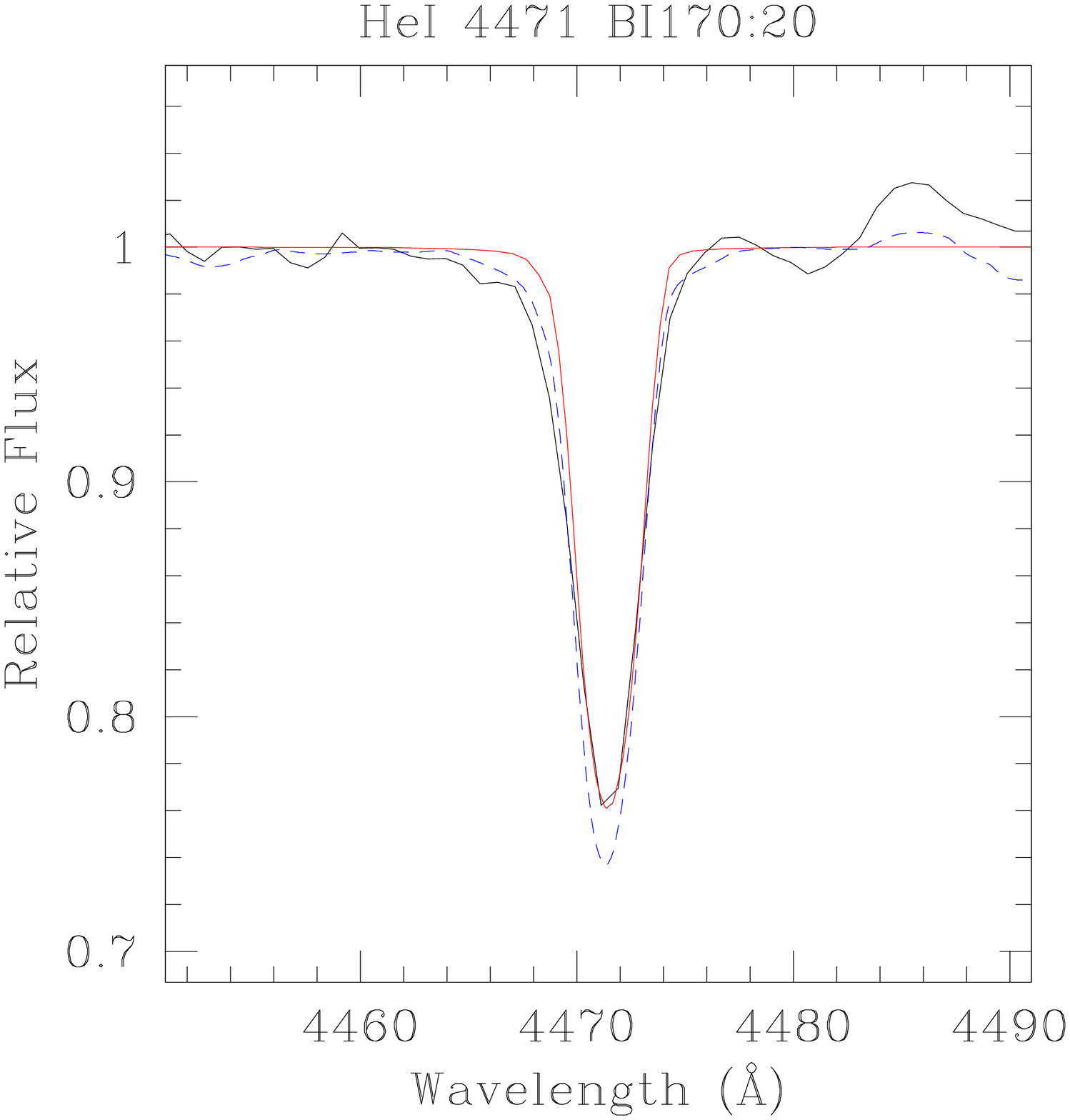}
\plotone{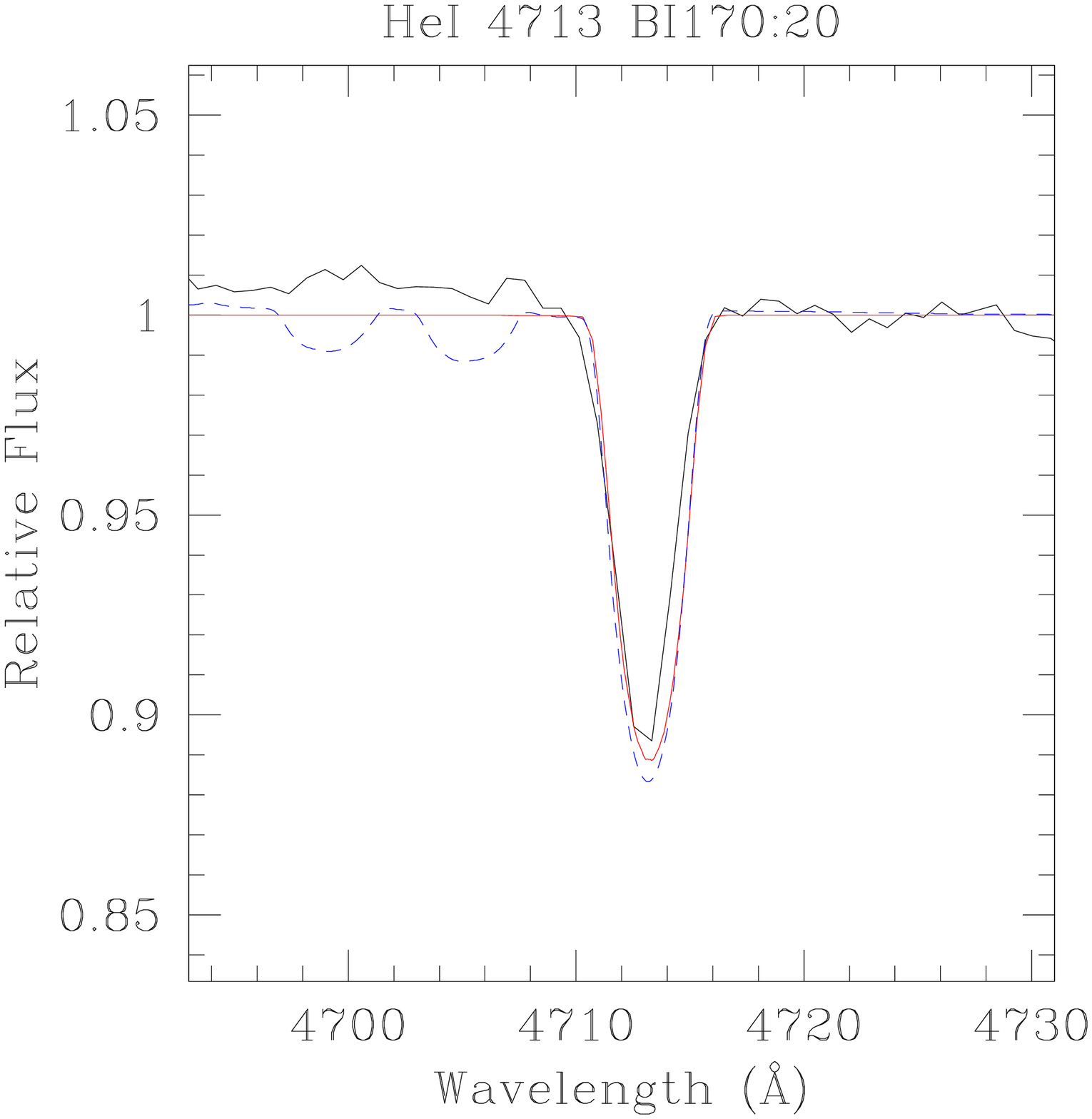}
\plotone{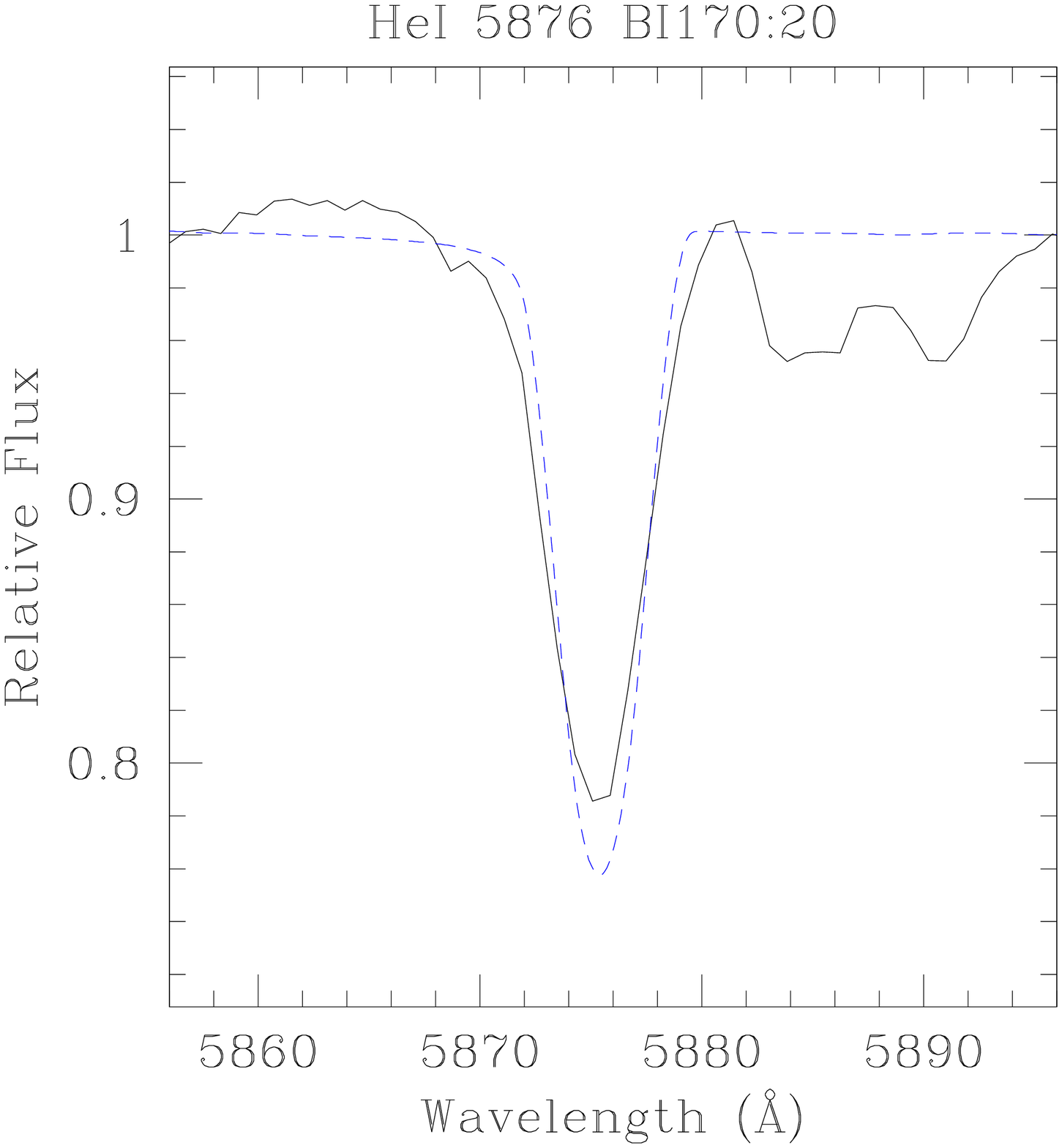}
\caption{\label{fig:MicroBI170}  The effect of  microturbulence on the He I triplets in BI 170, an O9.5 I star in the LMC.
Black shows the observed spectrum, the red line shows the \fastwind\ fit, and the dashed blue line shows the \cmfgen\ fit.
The upper three panels show the model profiles computed using the  ``standard" 10 km s$^{-1}$ microturbulent velocities, the middle three panels show that obtained using 15 km s$^{-1}$, and the bottom three panels show the model profiles obtained using 20 km$^{-1}$.}
\end{figure}

\begin{figure}
\epsscale{0.25}
\plotone{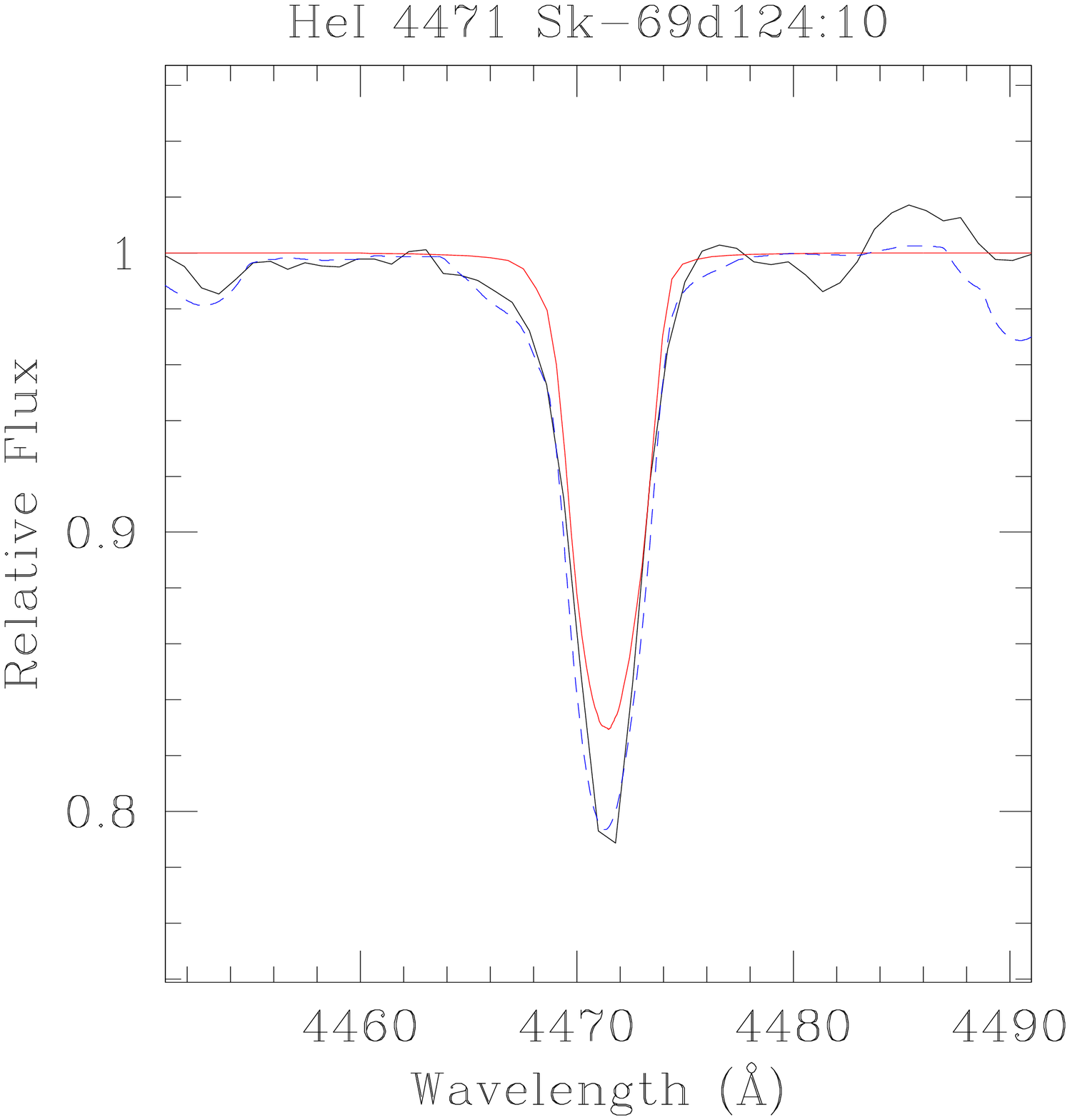}
\plotone{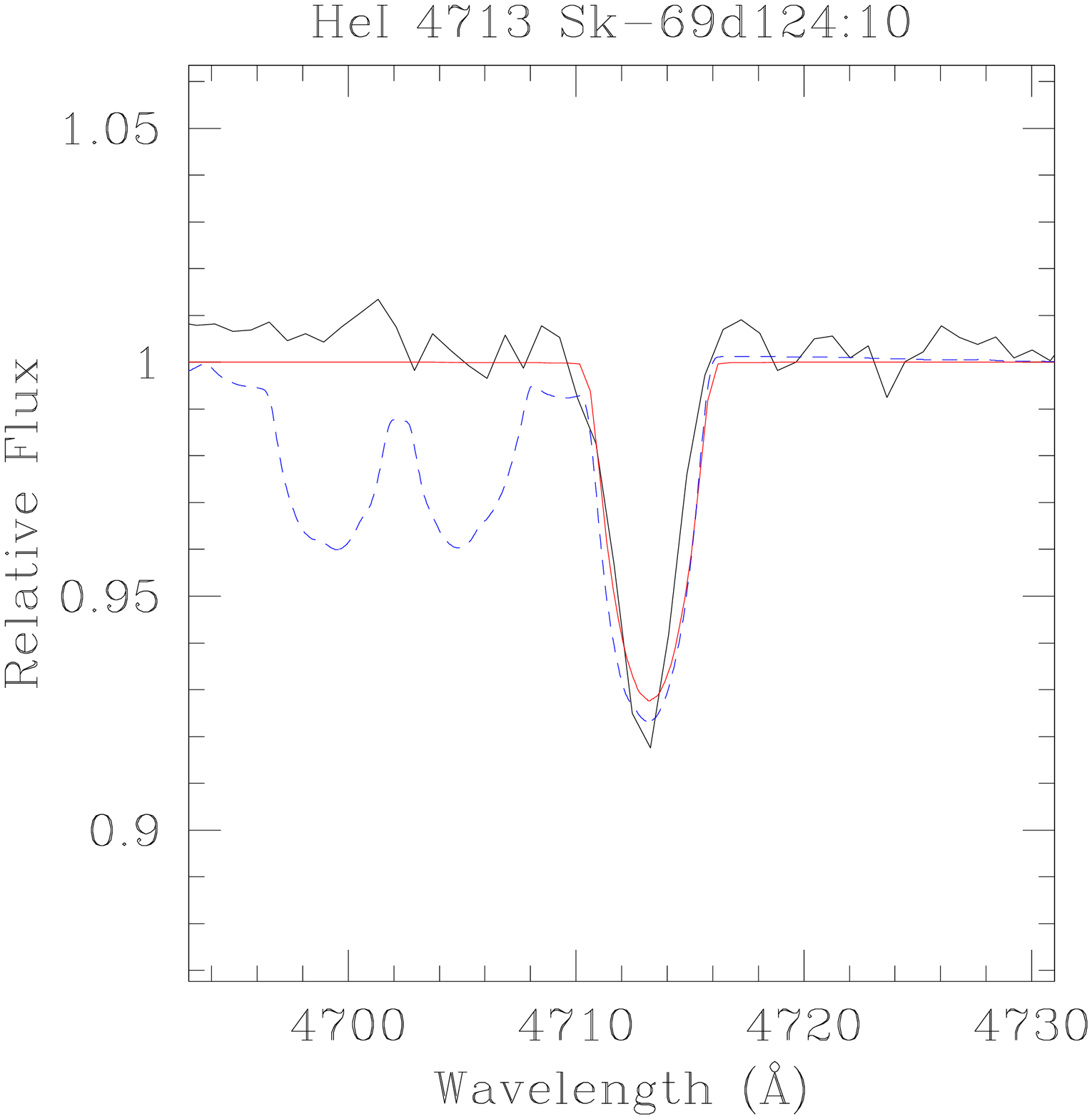}
\plotone{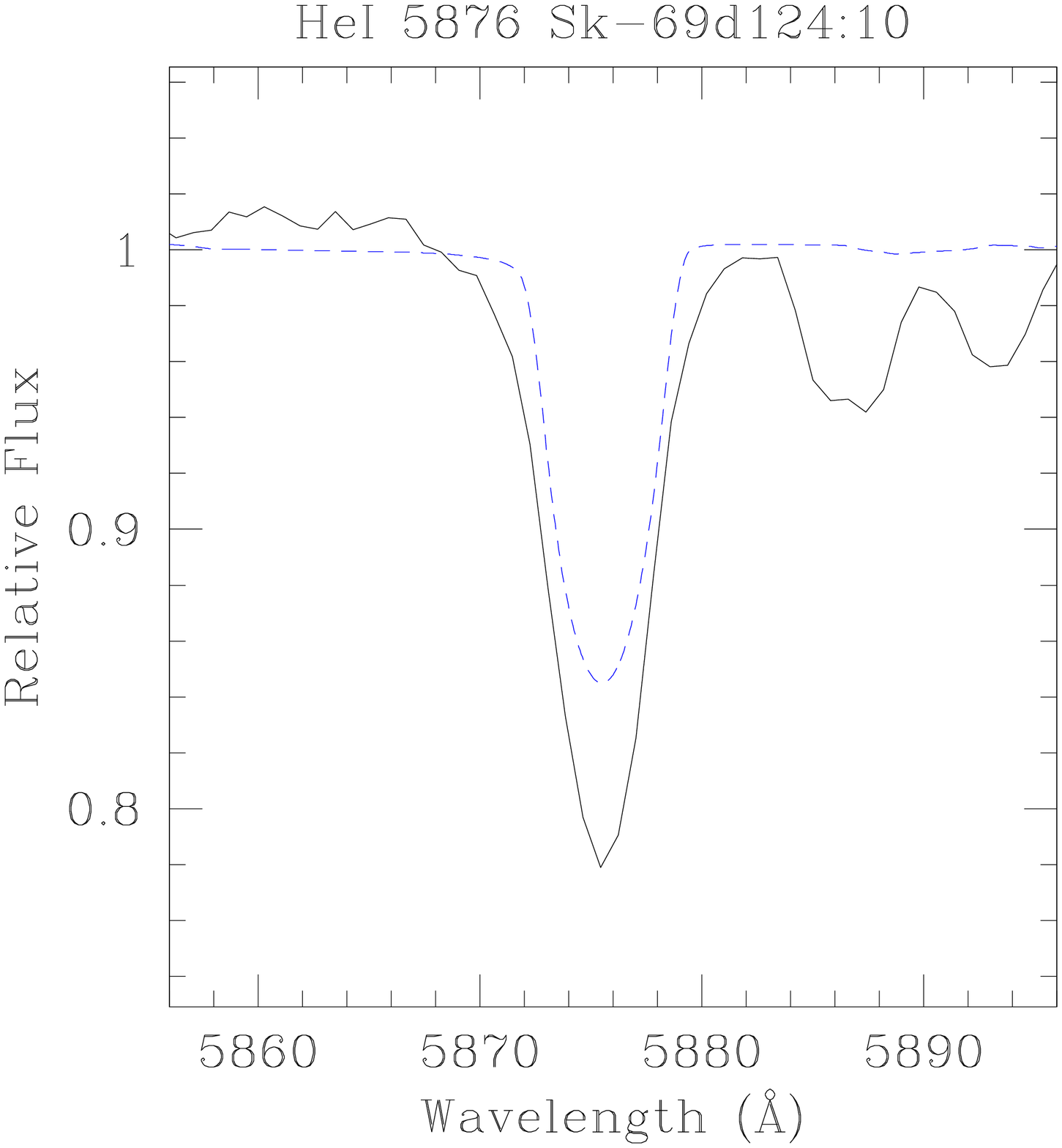}
\plotone{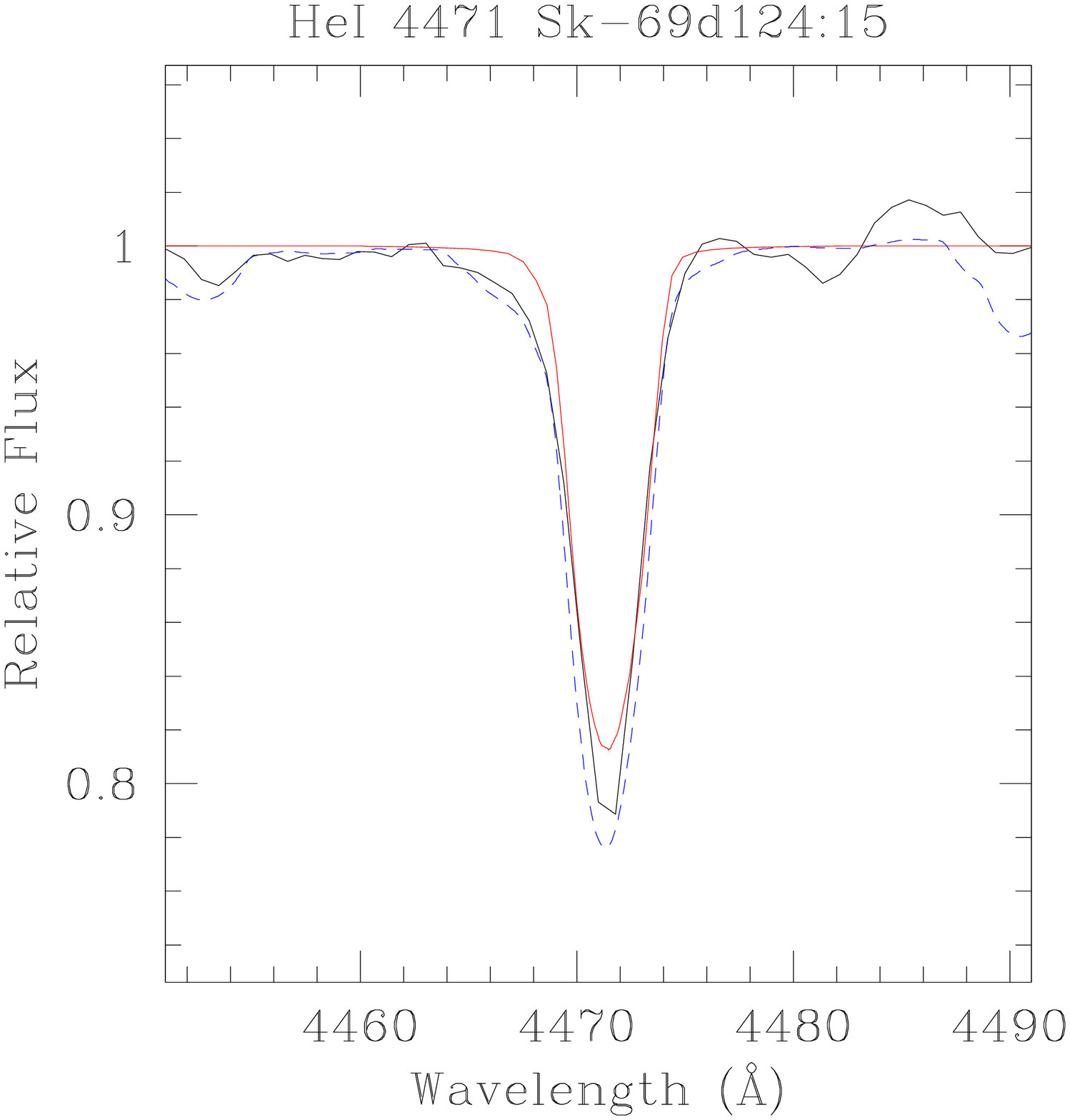}
\plotone{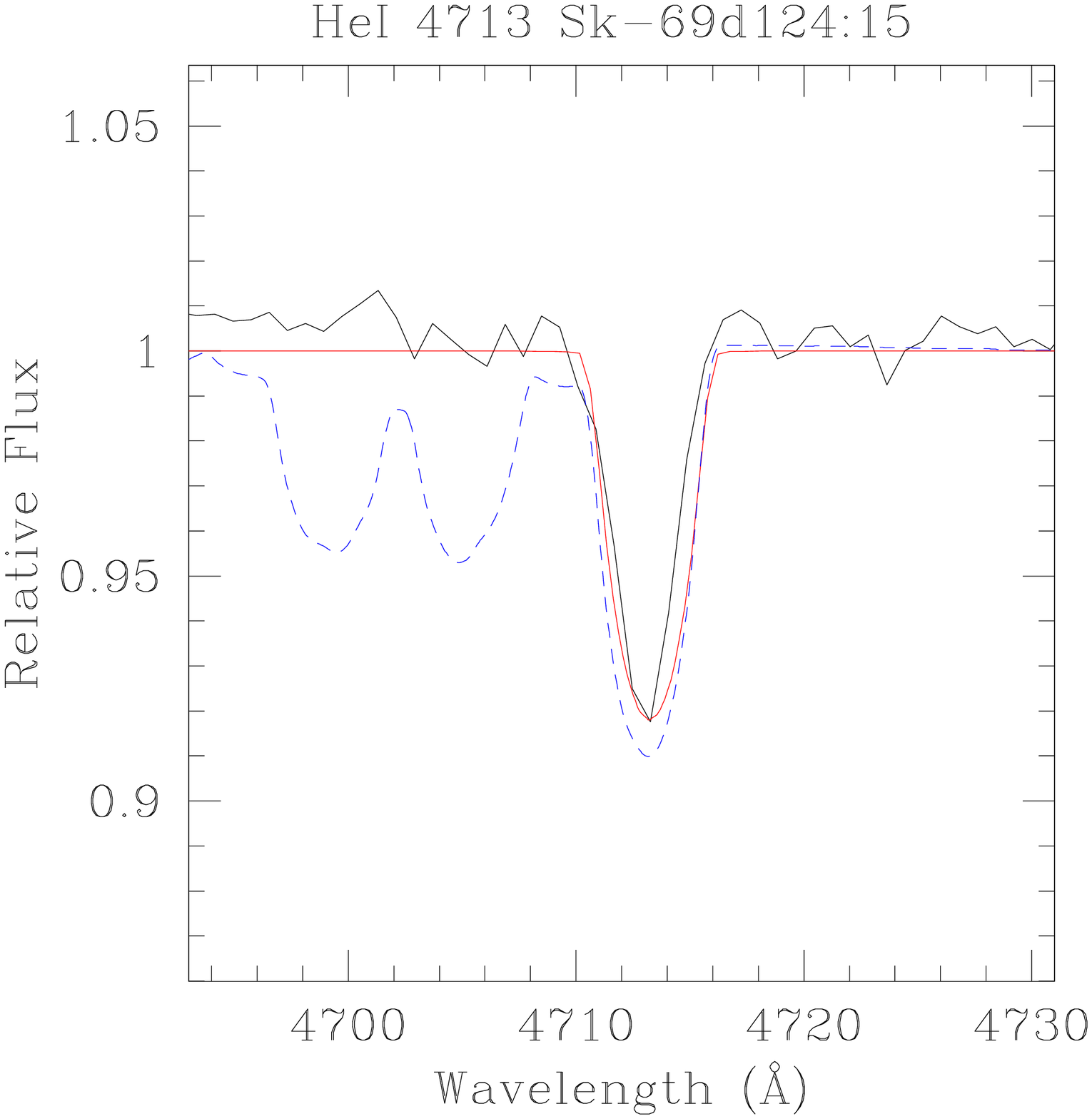}
\plotone{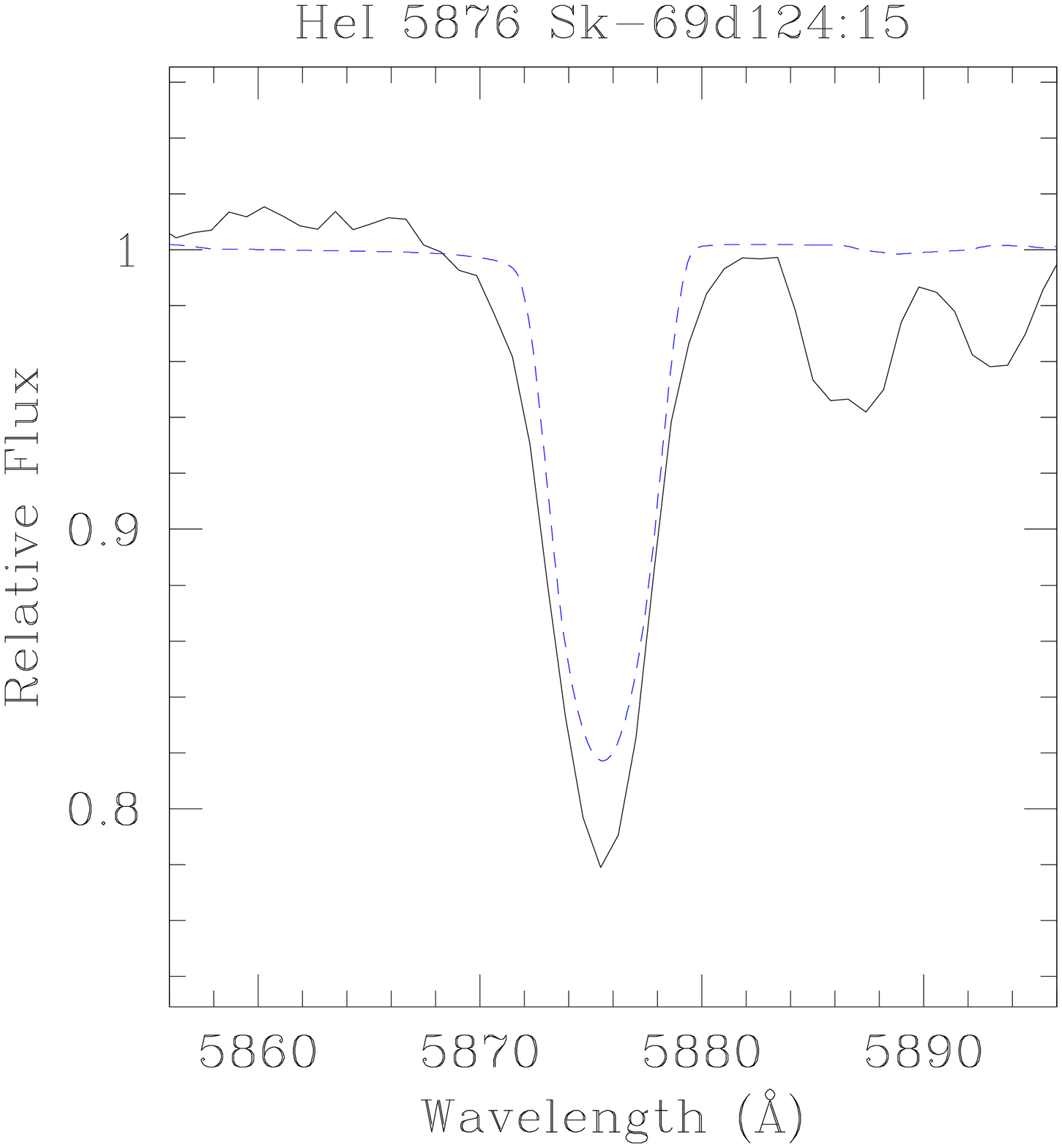}
\plotone{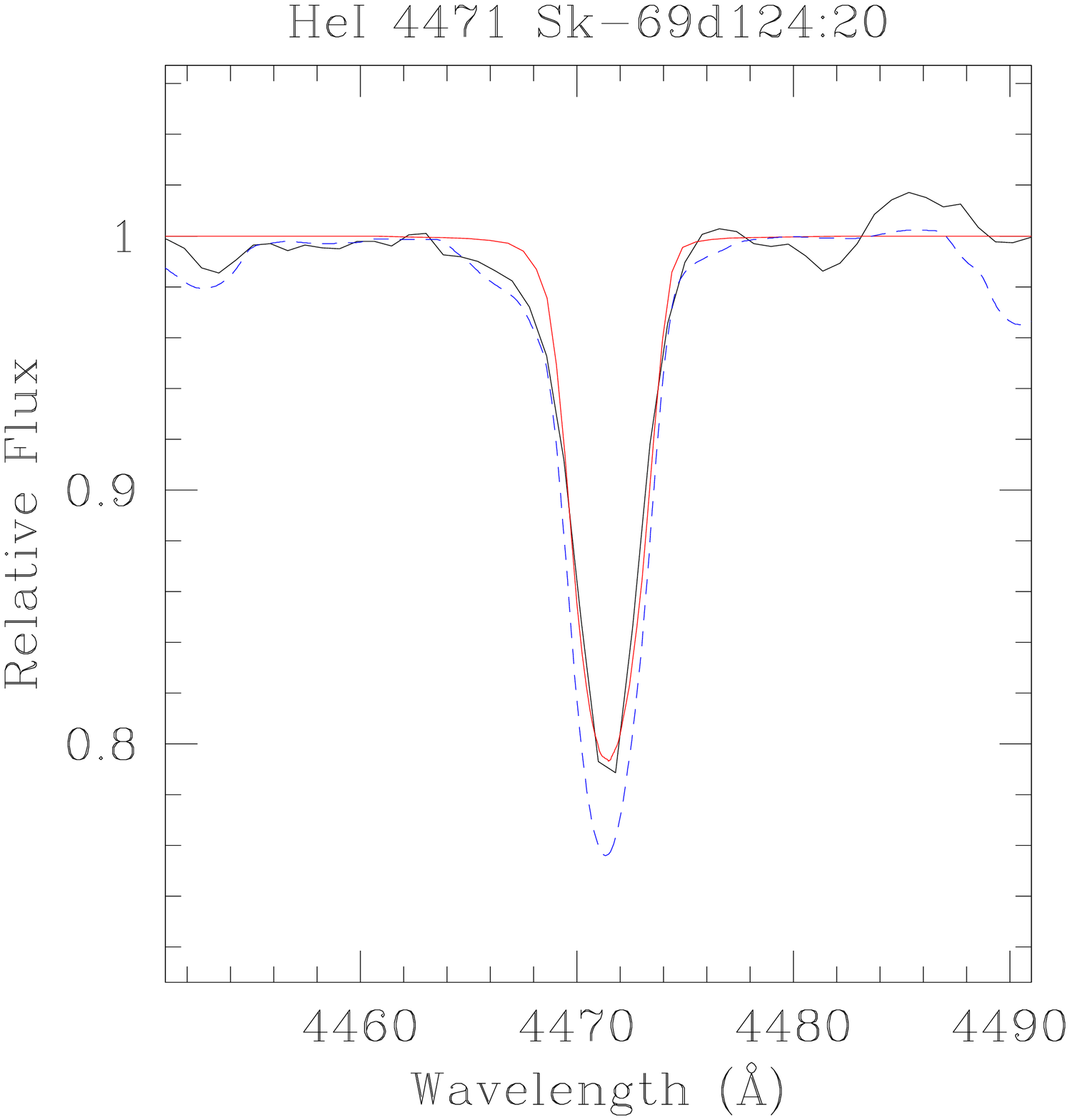}
\plotone{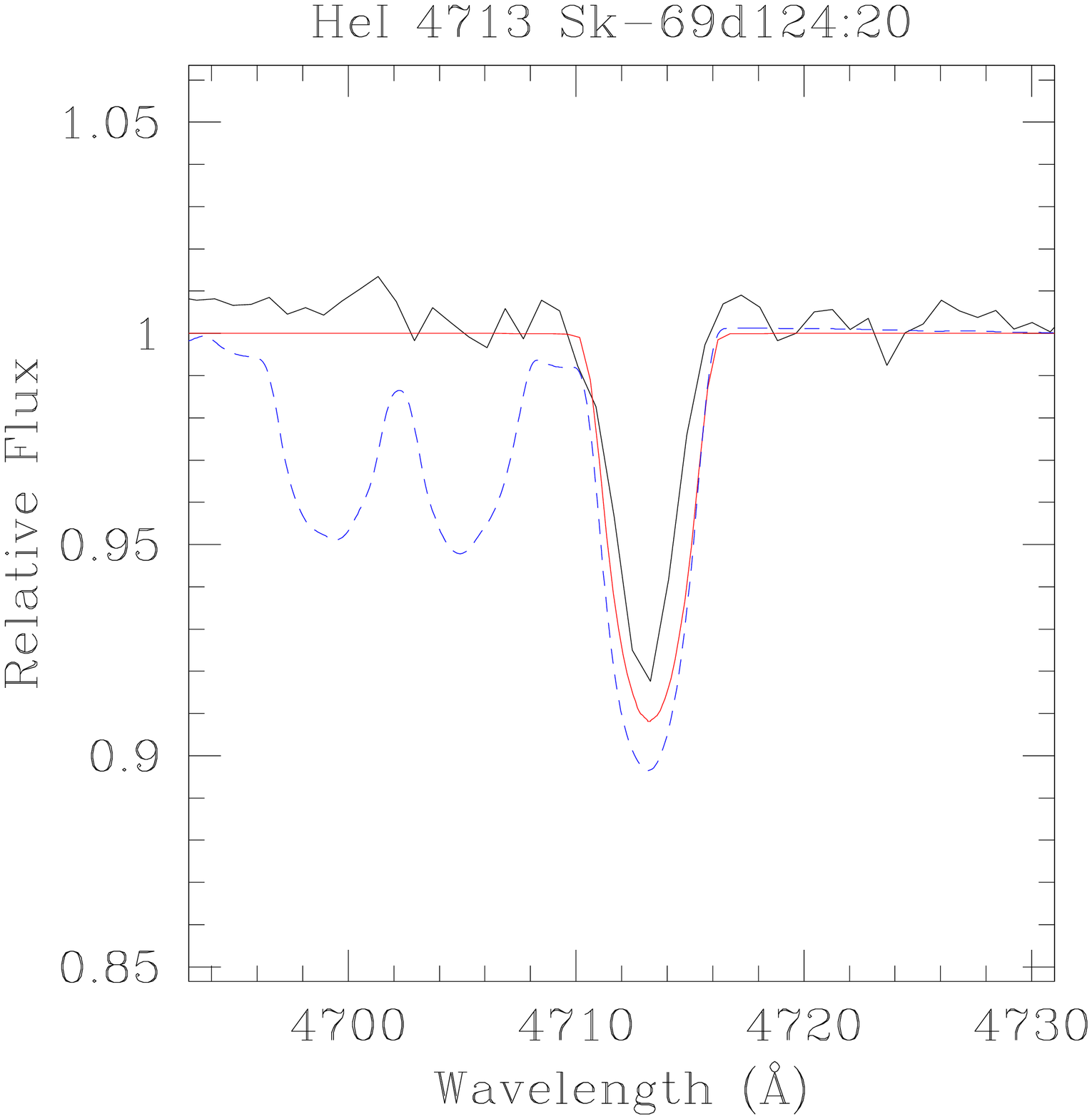}
\plotone{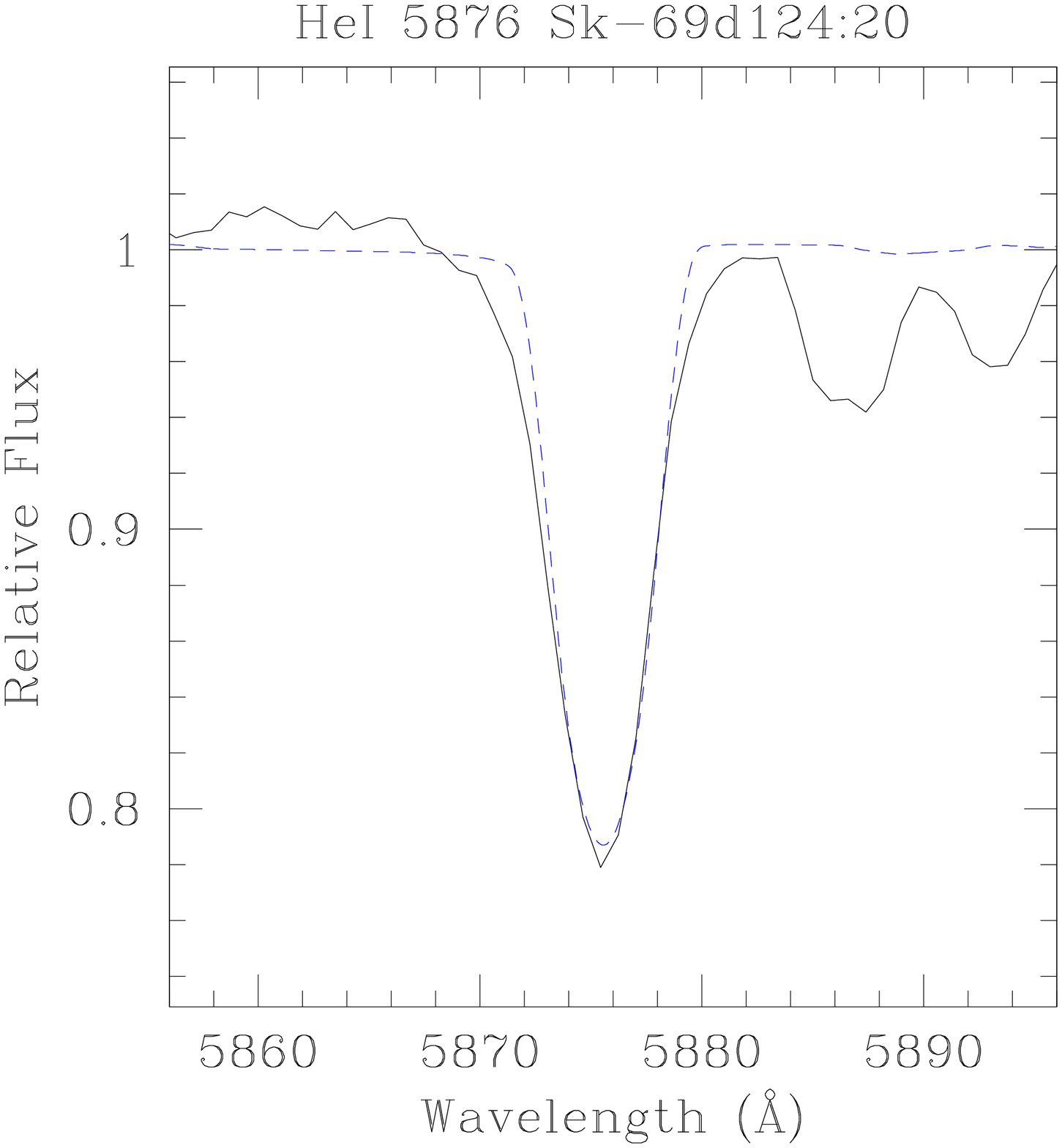}
\caption{\label{fig:MicroSk-69d124}  The effect of  microturbulence on the He I triplets in Sk $-69^\circ$124, an O9.7 I star in the LMC.
Black shows the observed spectrum, the red line shows the \fastwind\ fit, and the dashed blue line shows the \cmfgen\ fit.  The upper three panels show the model profiles computed using the  ``standard" 10 km s$^{-1}$ microturbulent velocities, the middle three panels show that obtained using 15 km s$^{-1}$, and the bottom three panels show the model profiles obtained using 20 km$^{-1}$.}
\end{figure}

\begin{figure}
\epsscale{0.25}
\plotone{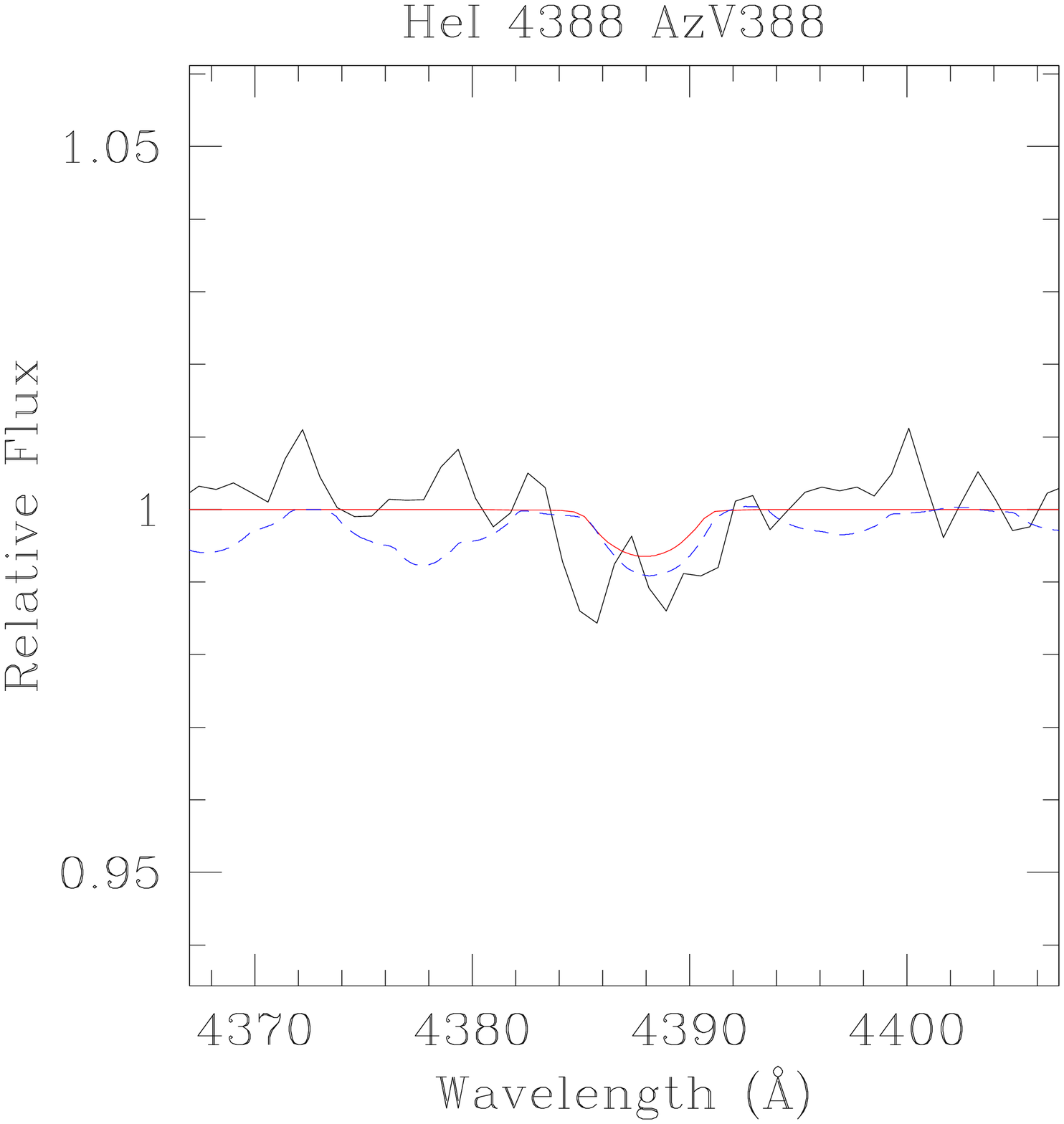}
\plotone{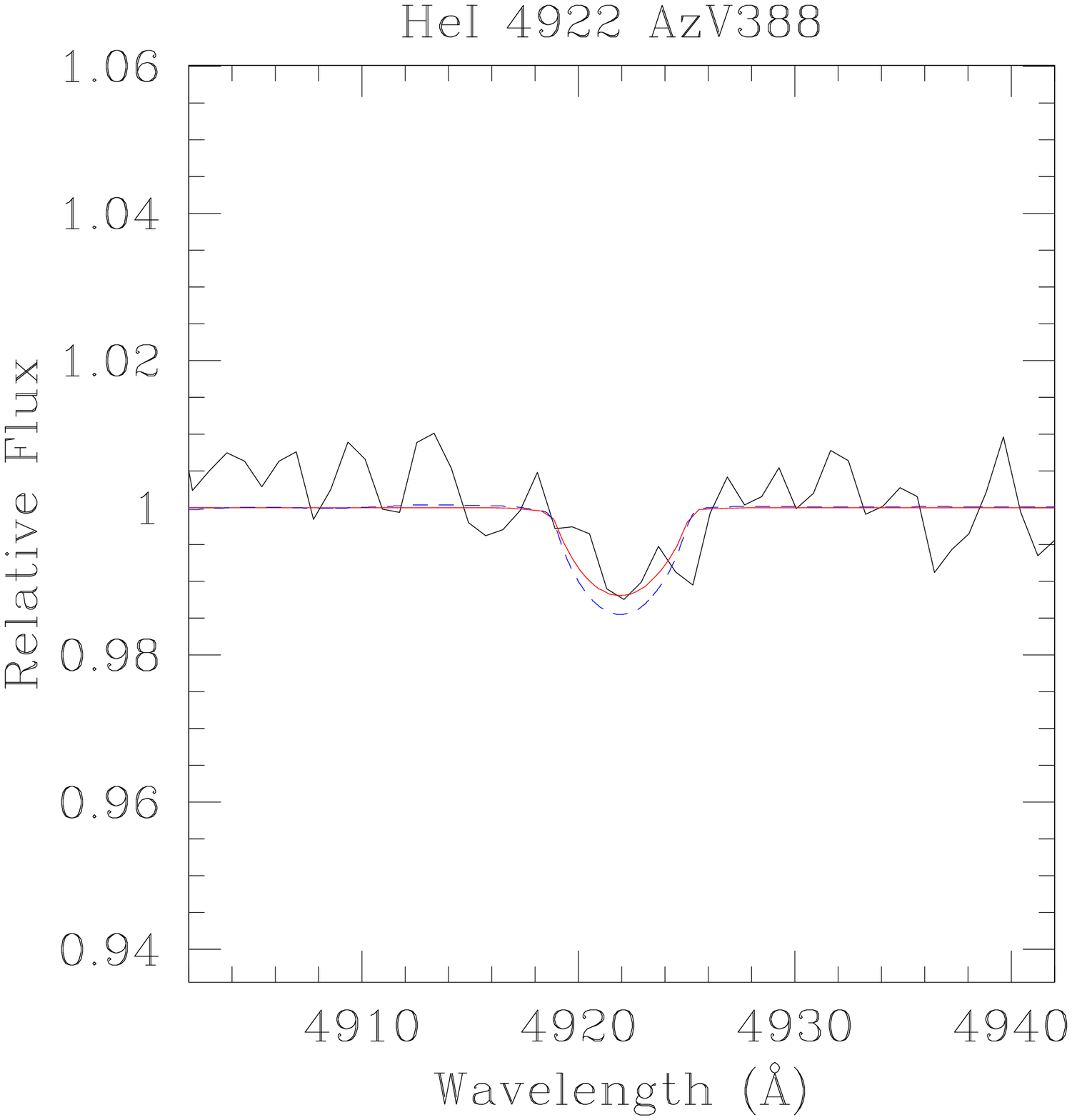}
\plotone{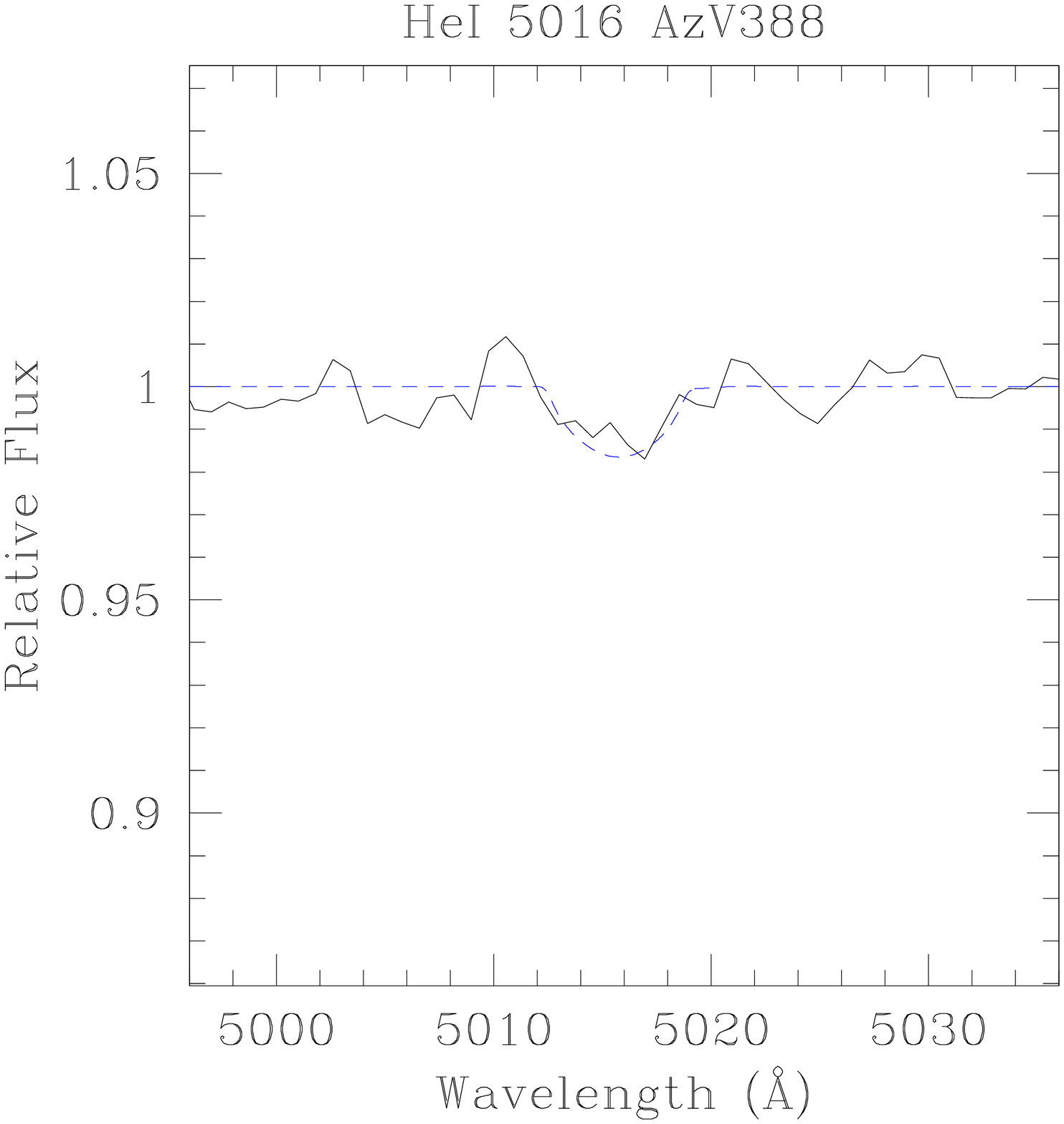}
\plotone{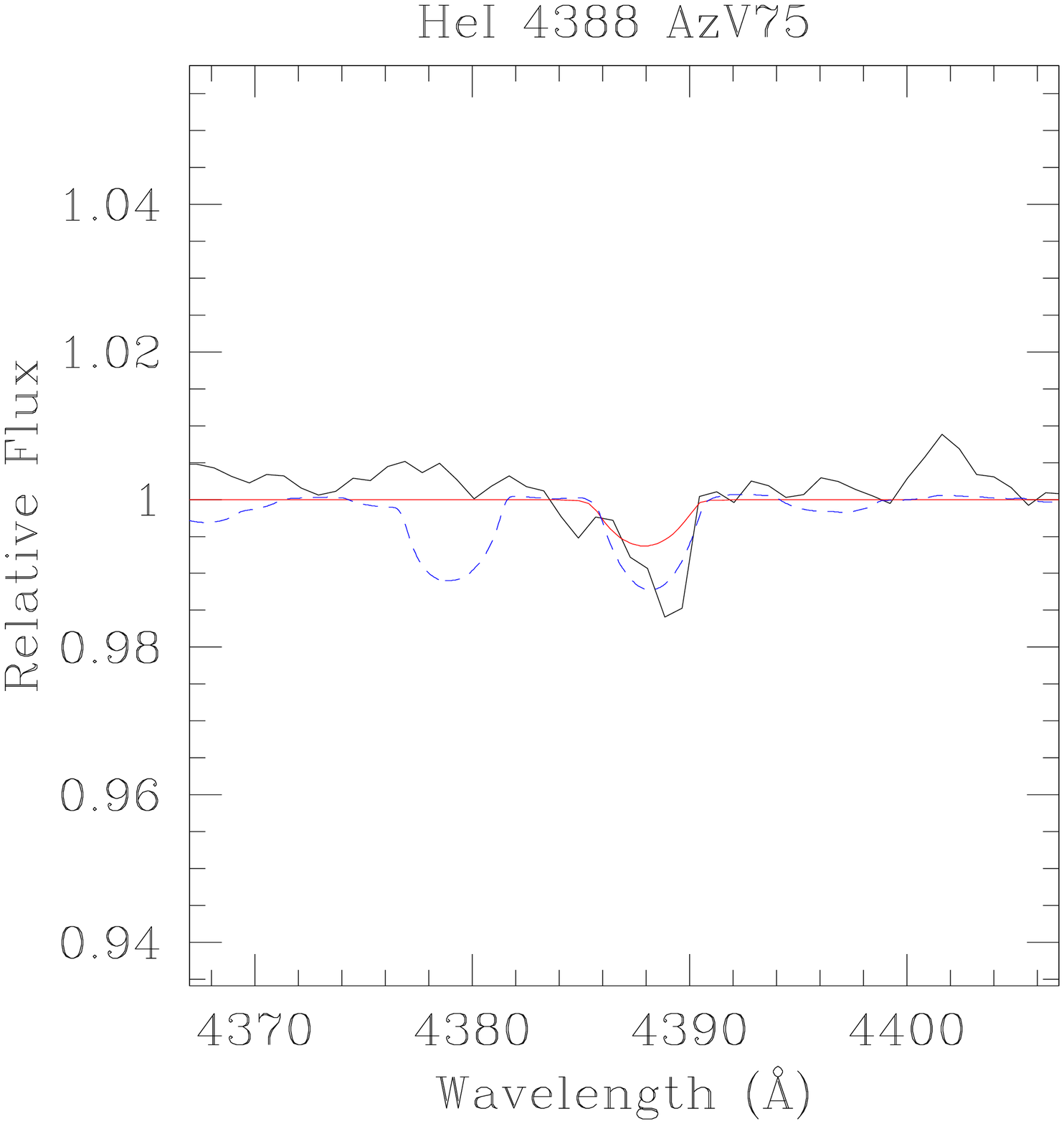}
\plotone{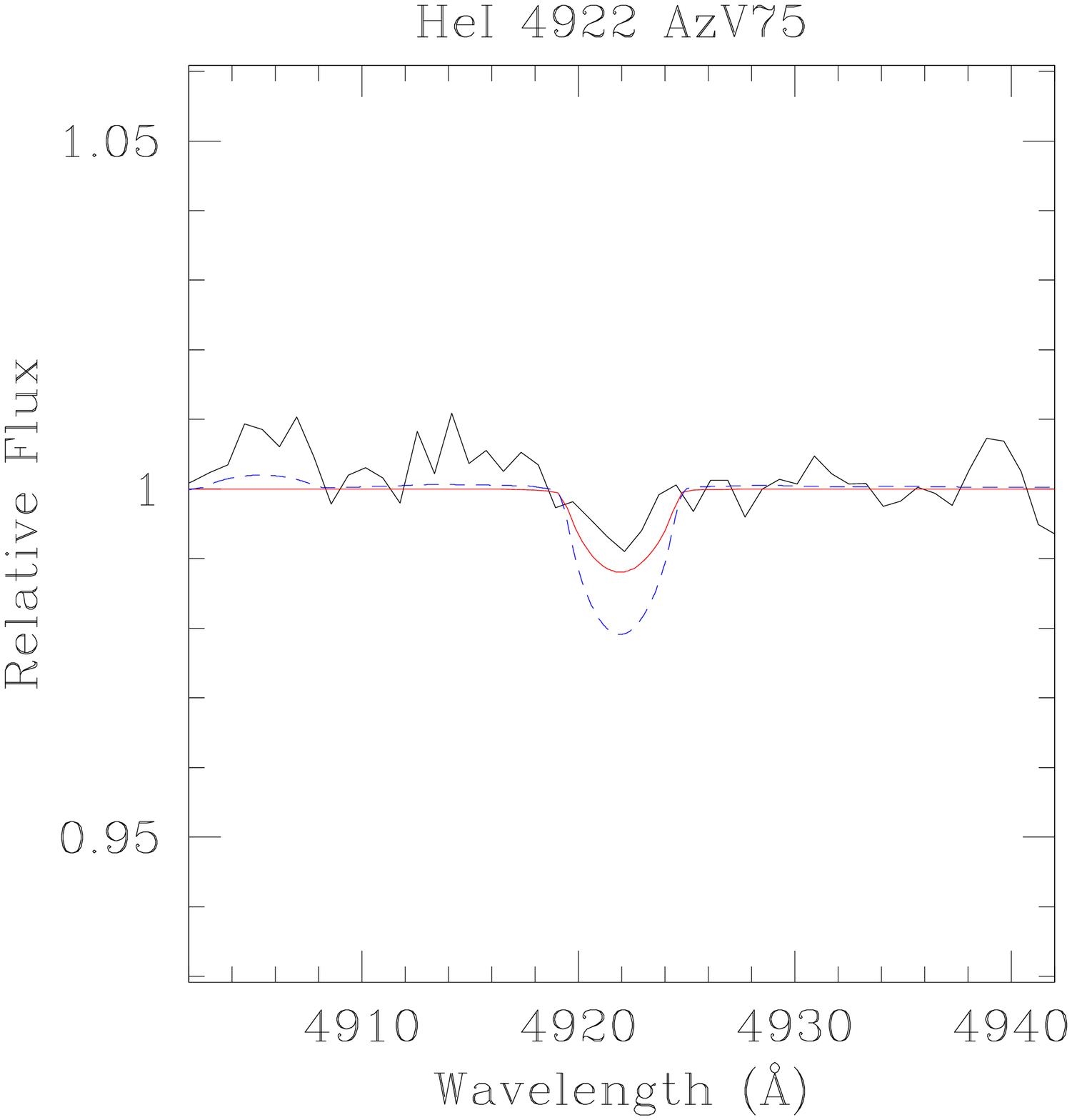}
\plotone{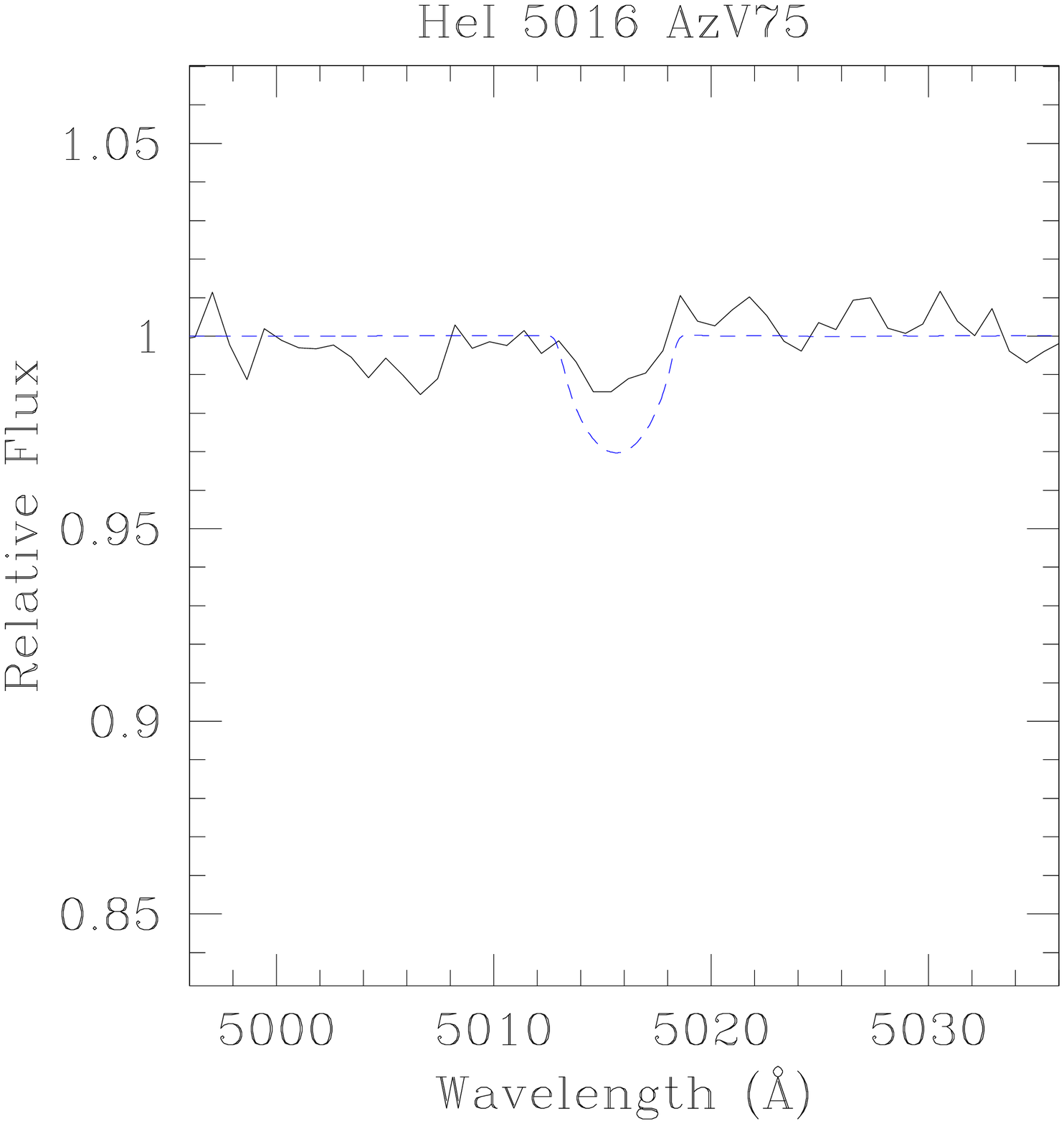}
\plotone{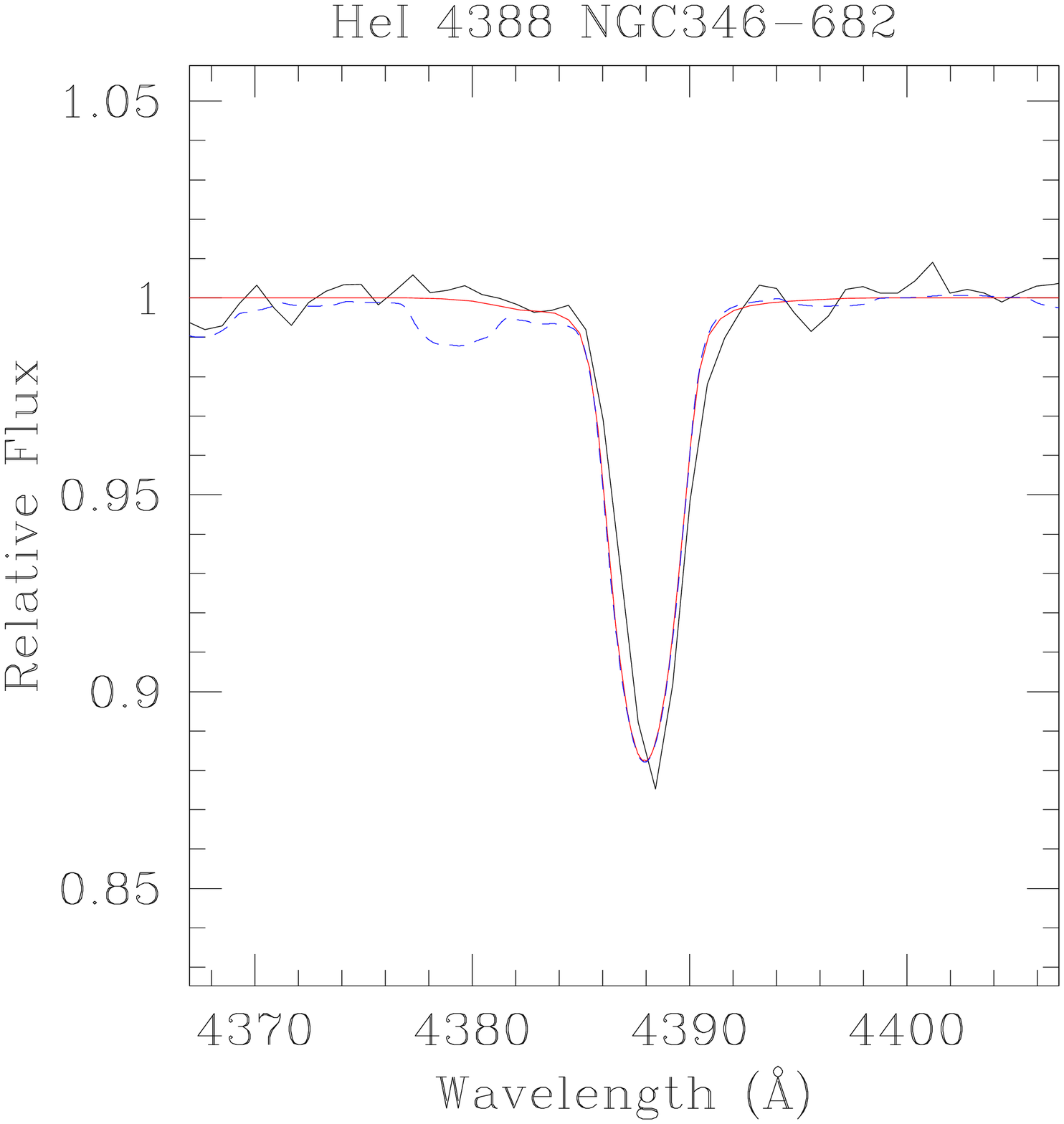}
\plotone{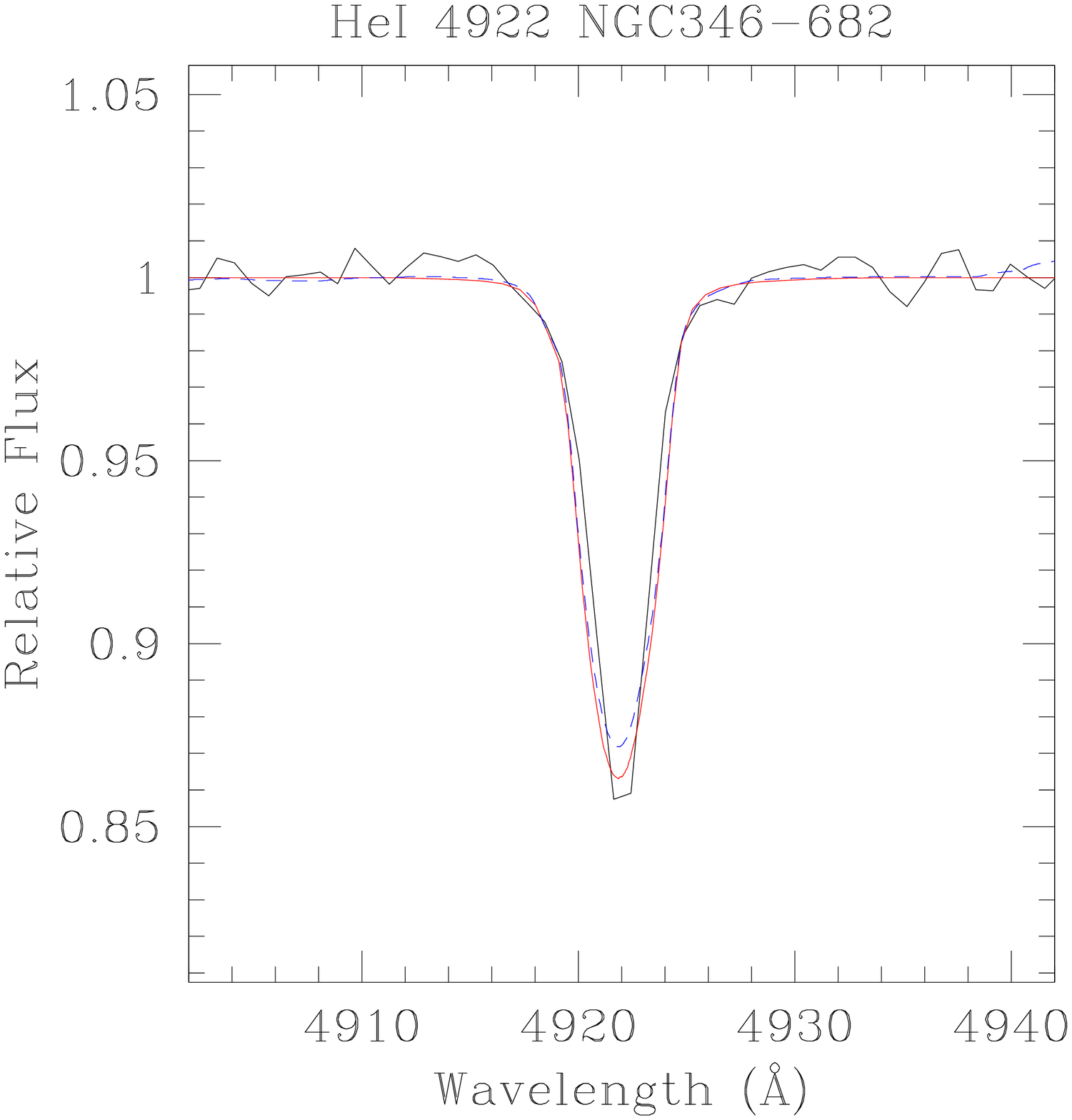}
\plotone{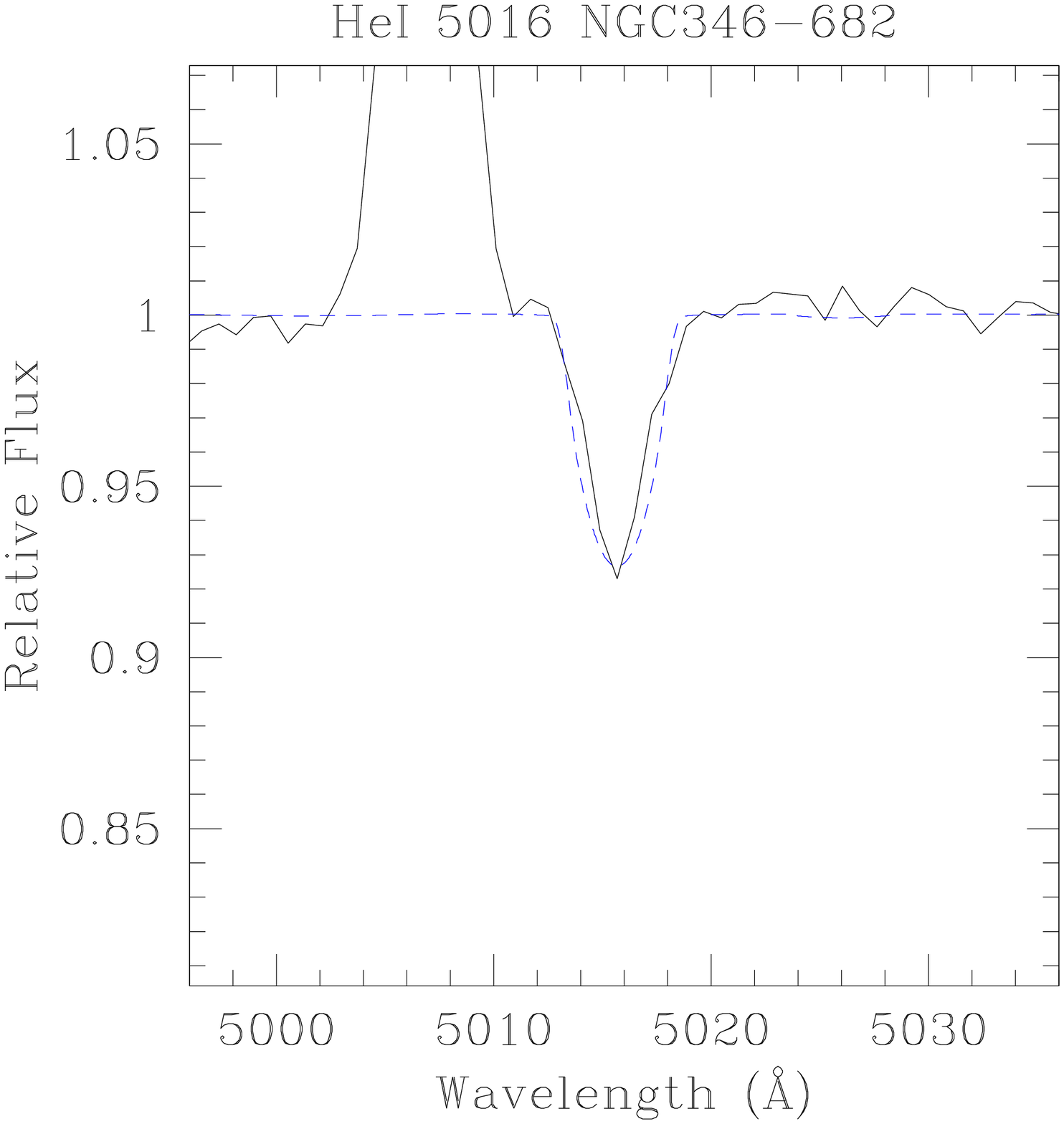}
\caption{\label{fig:singletsA}  The fits for the He~I singlet lines for a sample of early to intermediate O-type dwarfs and supergiants.  Black shows the observed spectrum, the red line shows the \fastwind\ fit, and the dashed blue line shows the \cmfgen\ fit.
 Note that the He~I $\lambda 4388$ and He~I $\lambda 4922$ lines are both $^3$P$^o-^3$D  transitions, while the He~I $\lambda 5016$ line (not fit by \fastwind) is a $^1$S$-^1$P$^o$ transition, and thus should be less affected by the problem with Fe IV transition overlapping the $^1$S$-^1$P$^o$ resonance line. (See Figure~\ref{fig:Grot}.)
The stars shown here are AzV 388, an O5.5 V((f)) star, AzV 75, an O5.5 I(f) star, and NGC 346-682, an O8 V star, all in the SMC.  }
\end{figure}

\begin{figure}
\epsscale{0.25}
\plotone{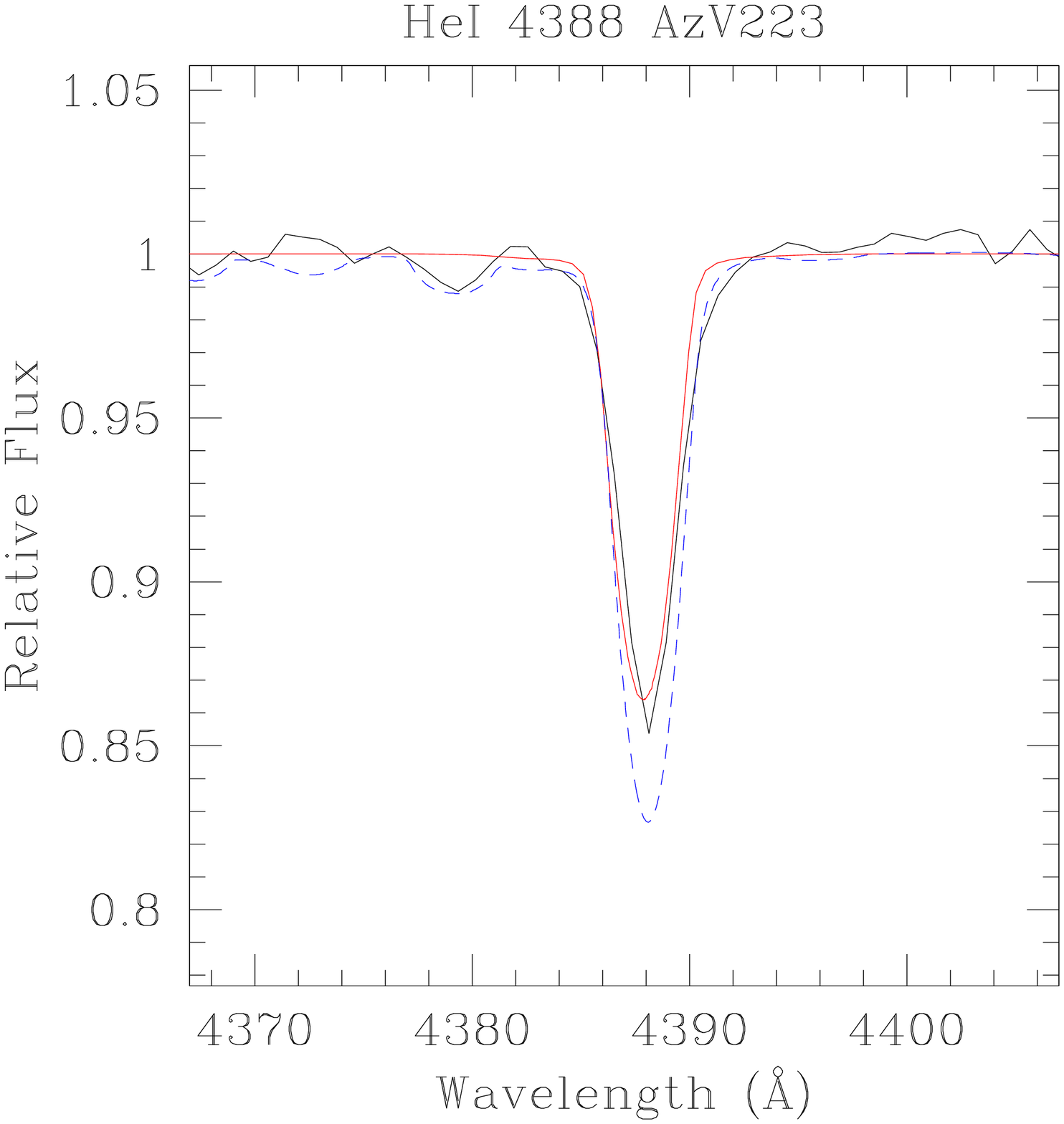}
\plotone{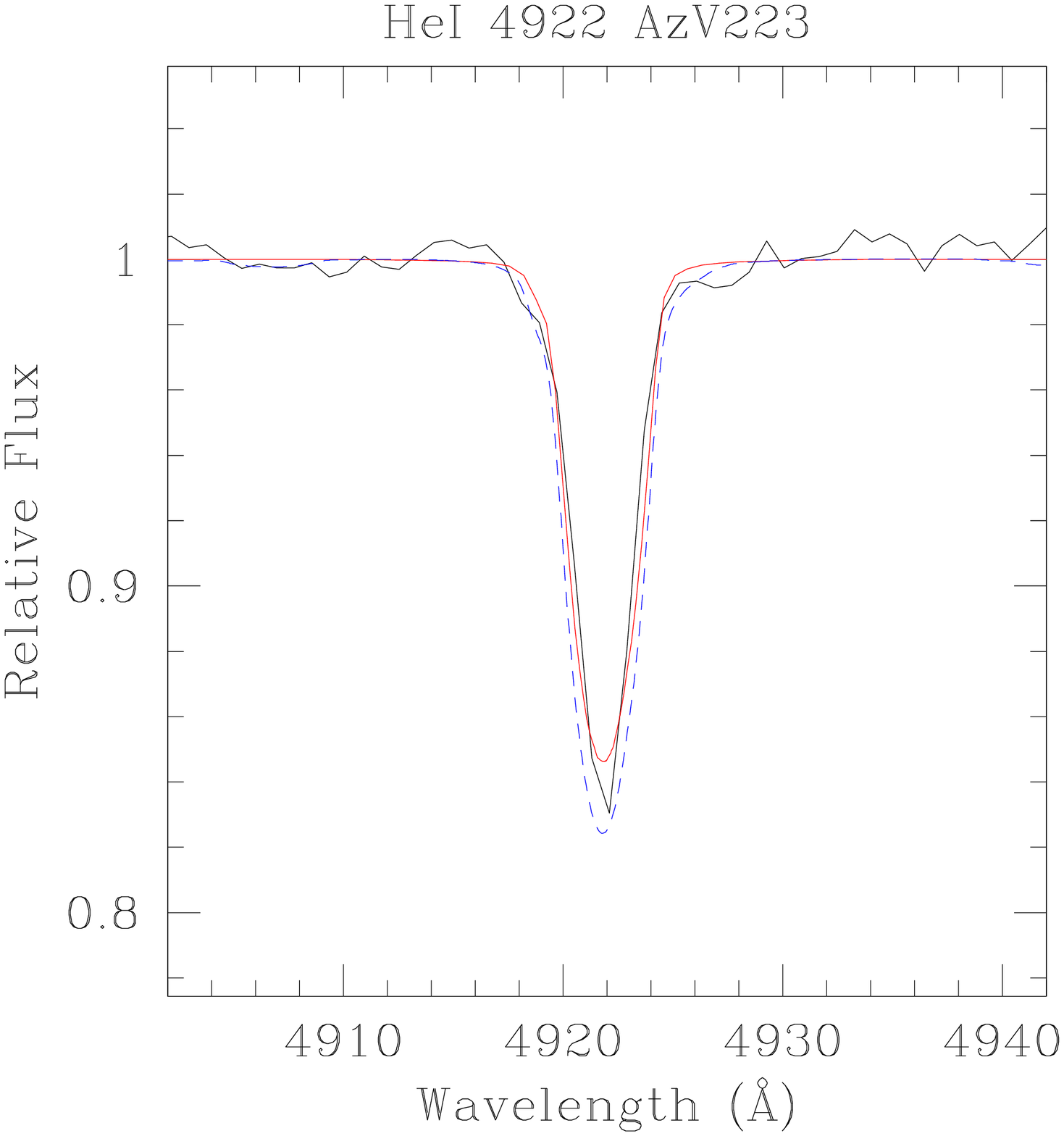}
\plotone{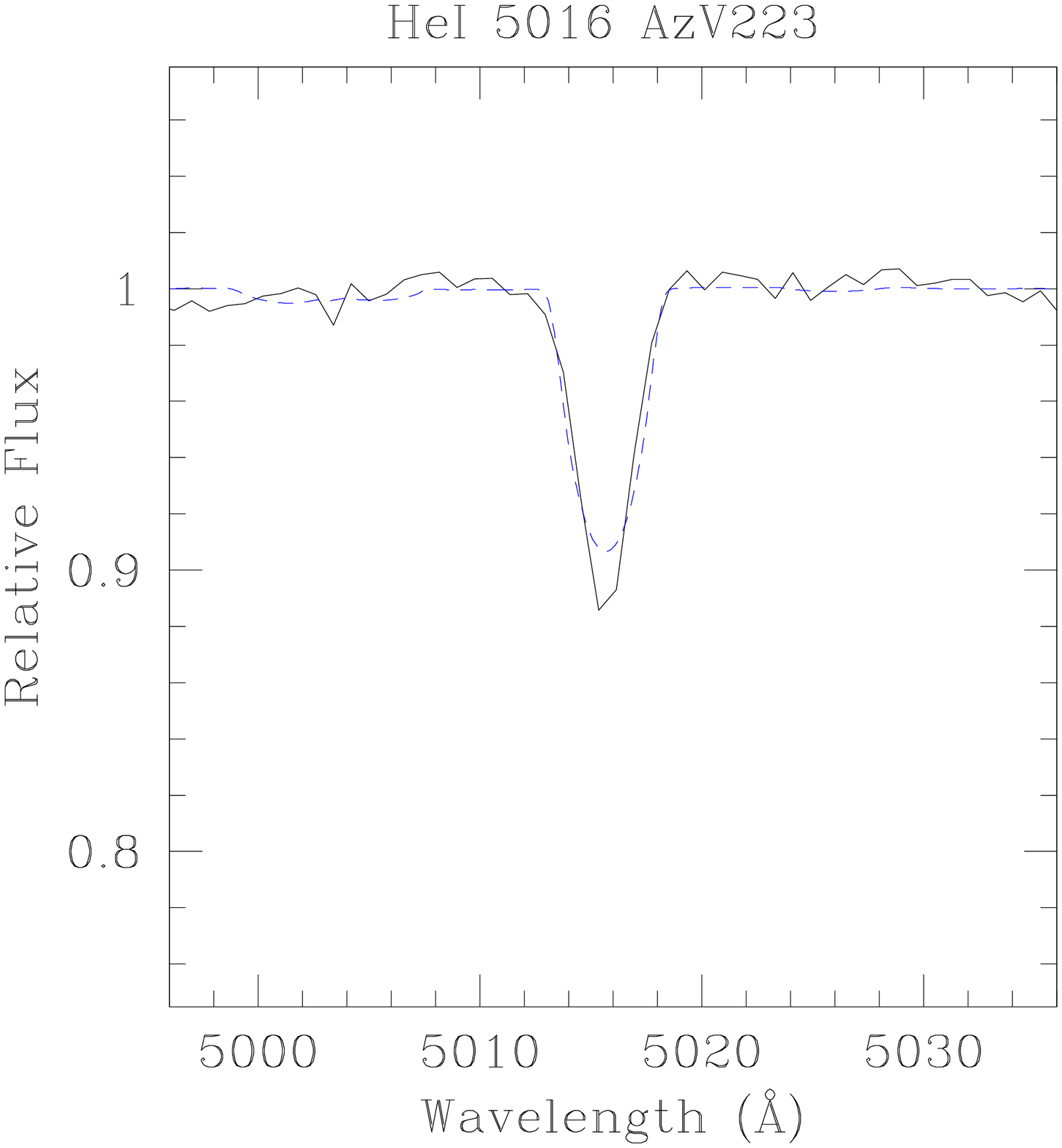}
\plotone{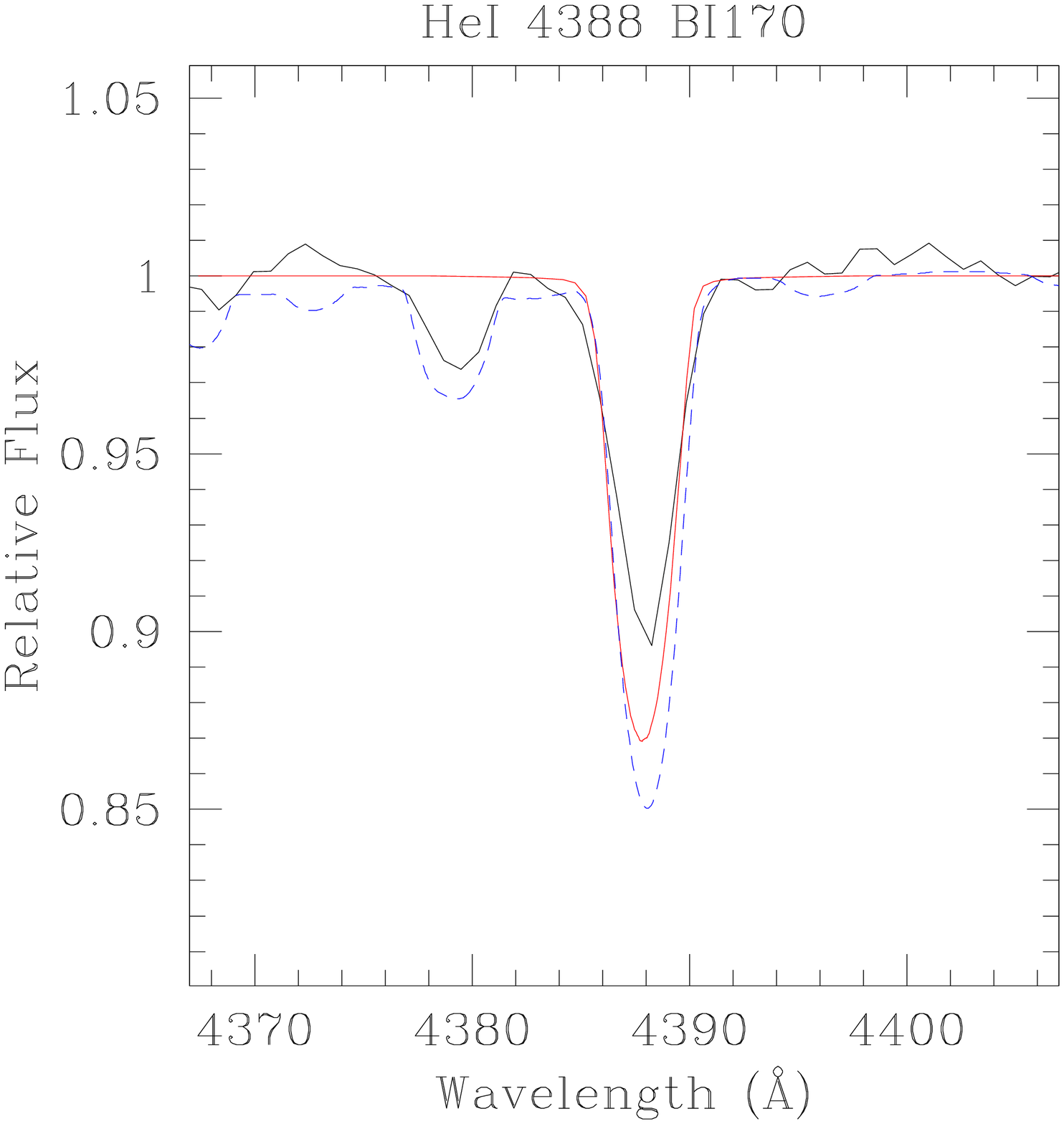}
\plotone{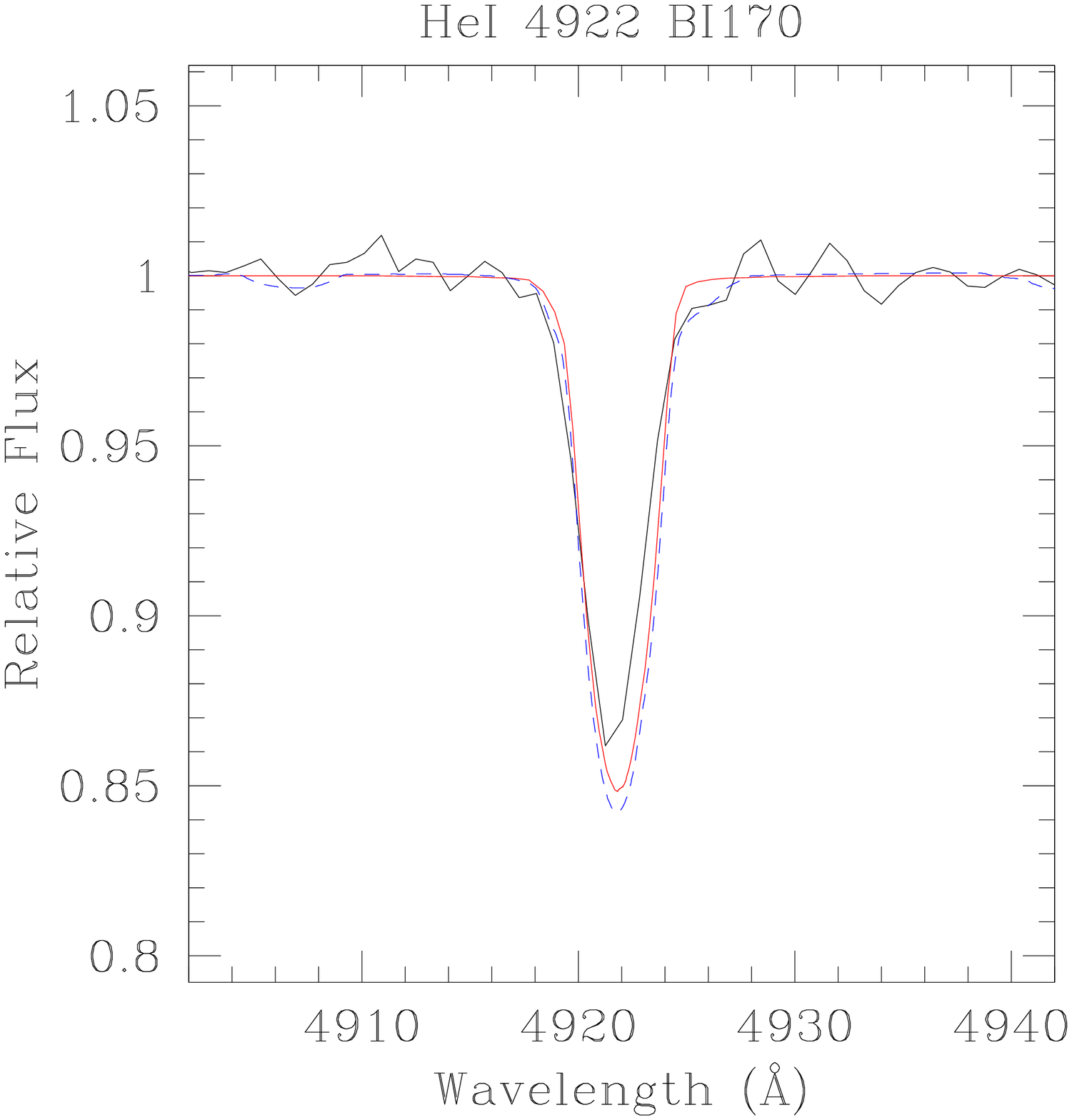}
\plotone{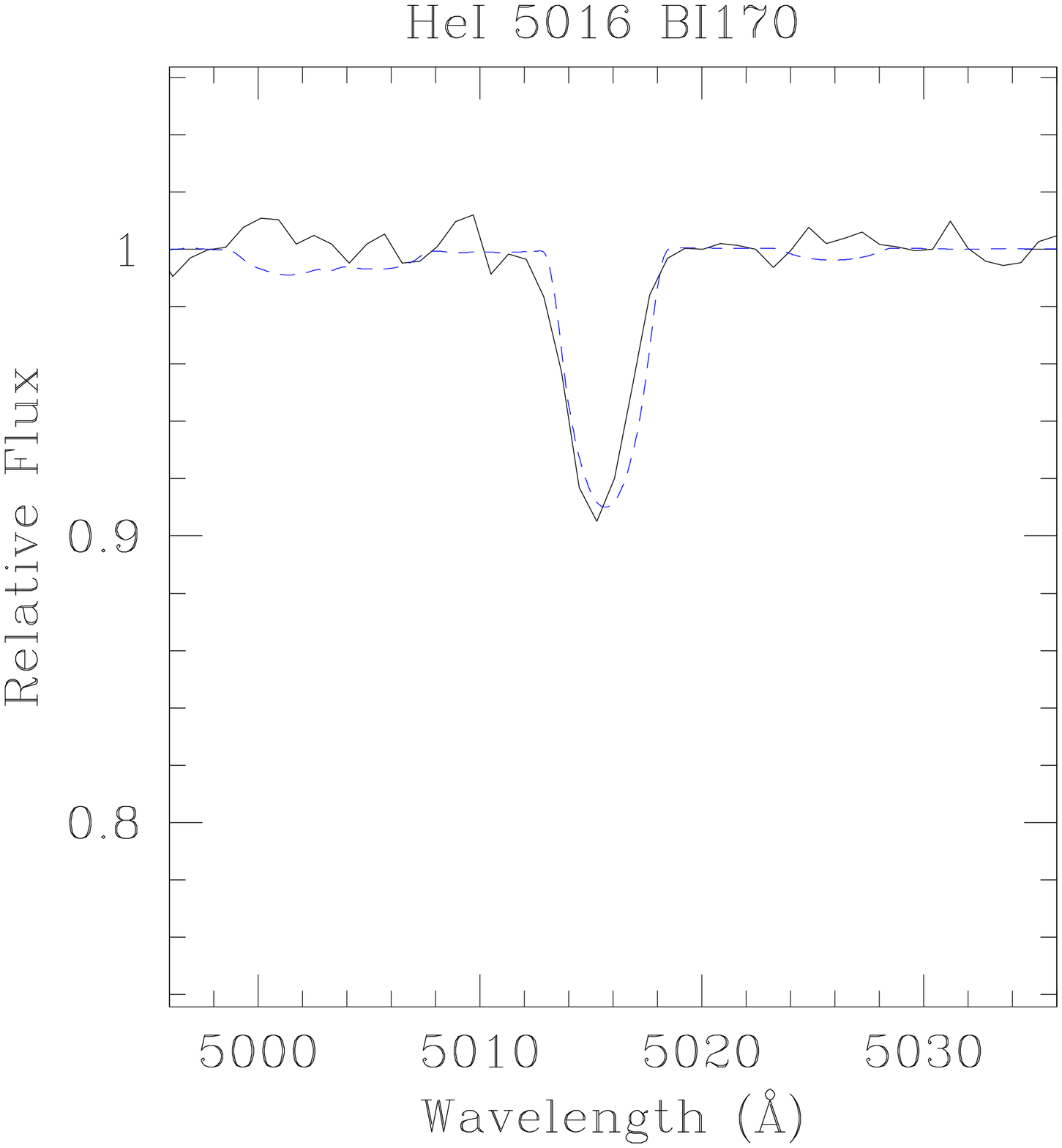}
\plotone{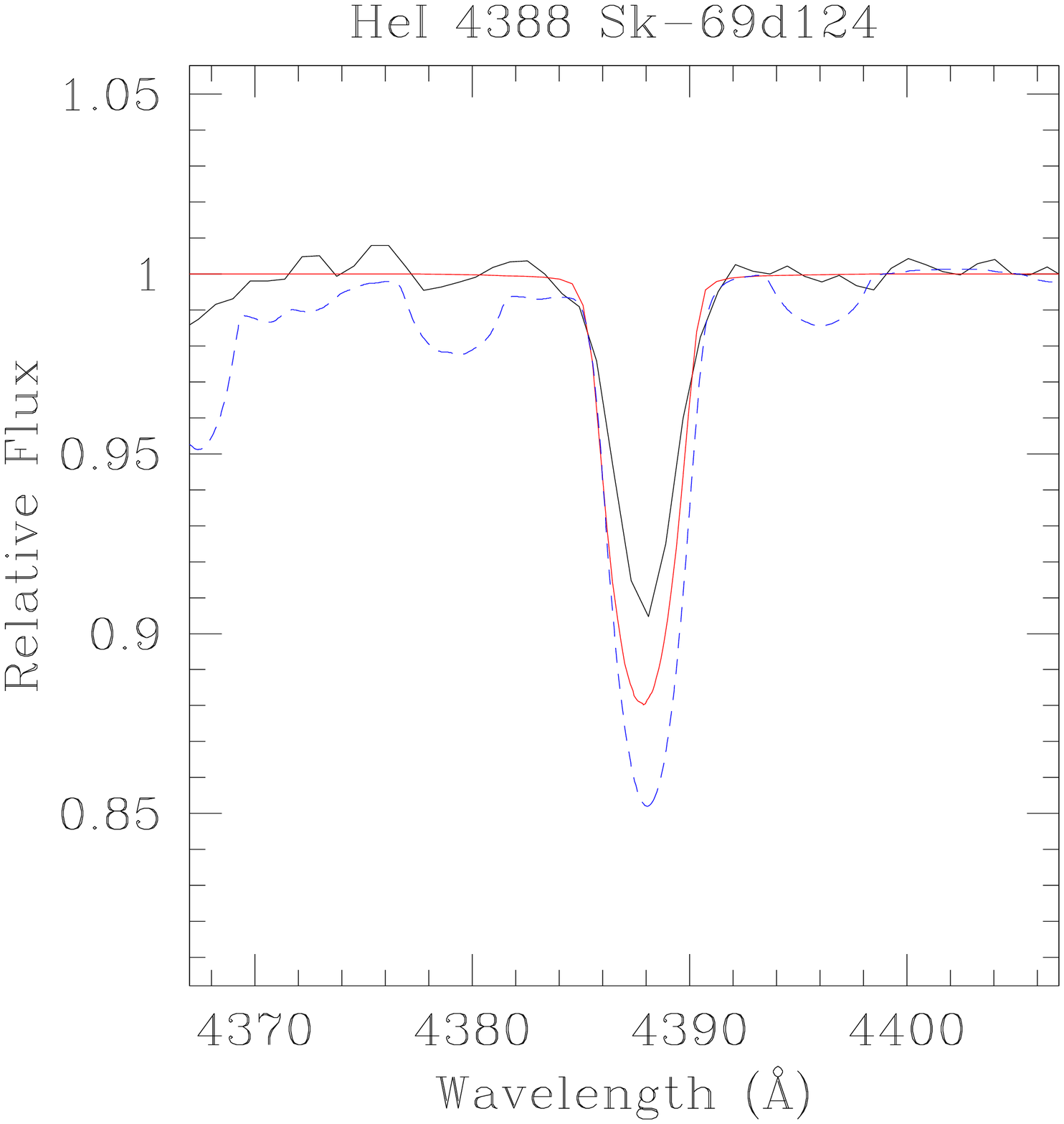}
\plotone{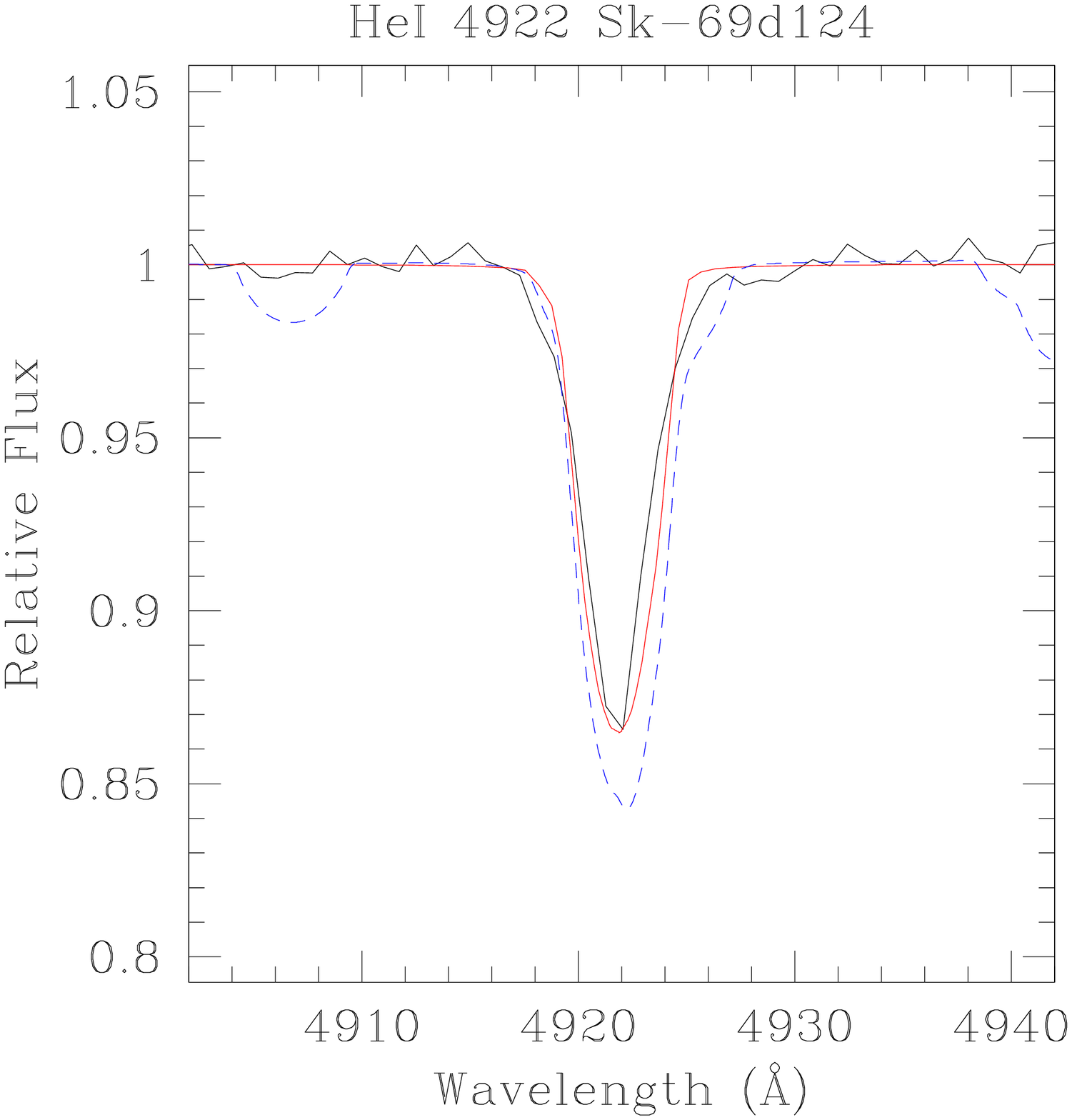}
\plotone{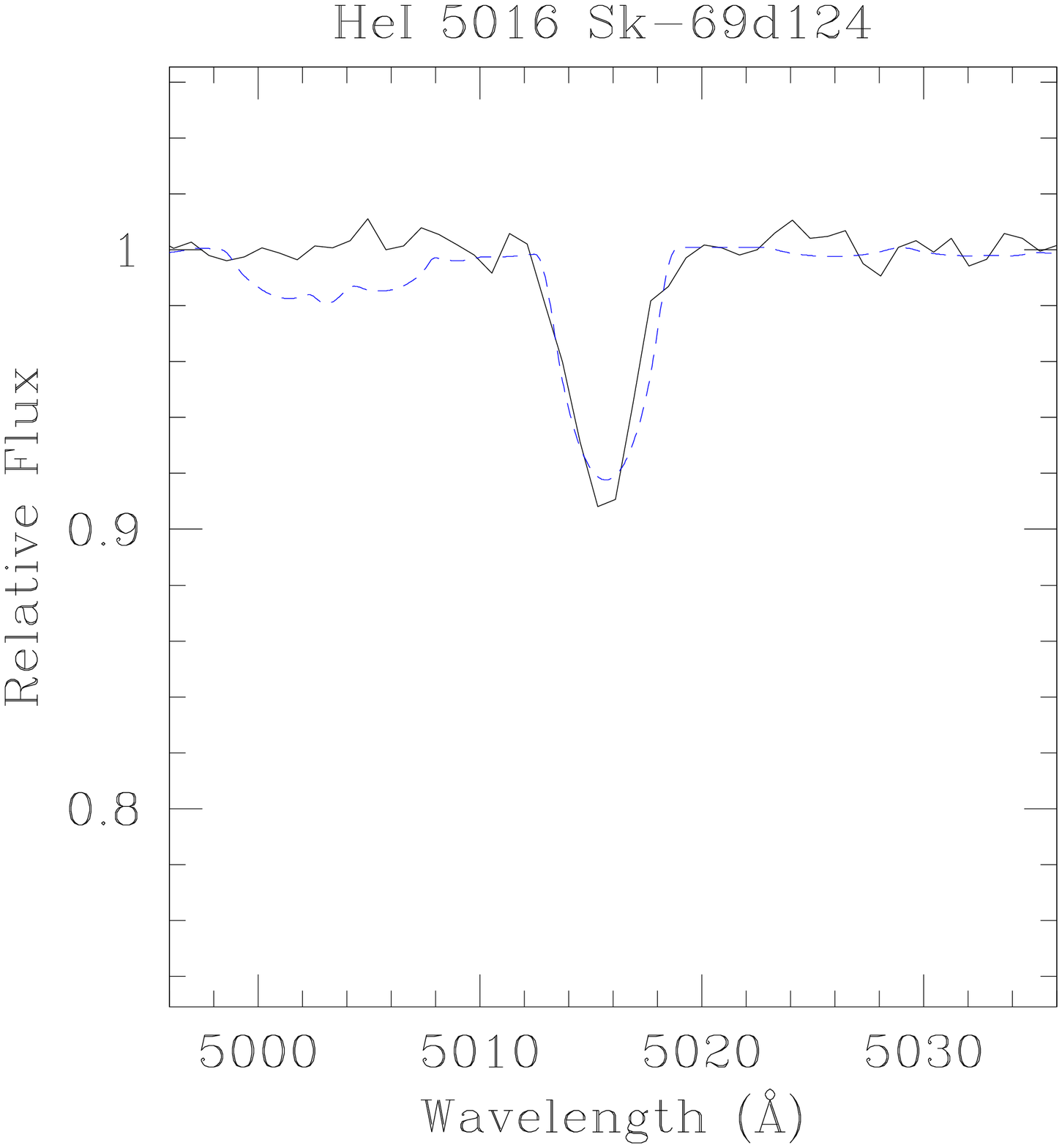}
\caption{\label{fig:singletsB}  The fits for the He~I singlet lines for the late-type O supergiants.  Black shows the observed spectrum, the red line shows the \fastwind\ fit, and the dashed blue line shows the \cmfgen\ fit. Note that the He~I $\lambda 4388$ and He~I $\lambda 4922$ lines are both $^3$P$^o-^3$D  transitions, while the He~I $\lambda 5016$ line (not fit by \fastwind) is a $^1$S$-^1$P$^o$ transition, and thus should be less affected by the problem with Fe IV transition overlapping the $^1$S$-^1$P$^o$ resonance line. (See Figure~\ref{fig:Grot}.)  The stars shown here are AzV 233, an O9.5 II star in the SMC, BI 170, an O9.5 I star in the LMC, and Sk $-69^\circ$124, an O9.7 I star in the LMC.}
\end{figure}

\begin{figure}
\epsscale{0.25}
\plotone{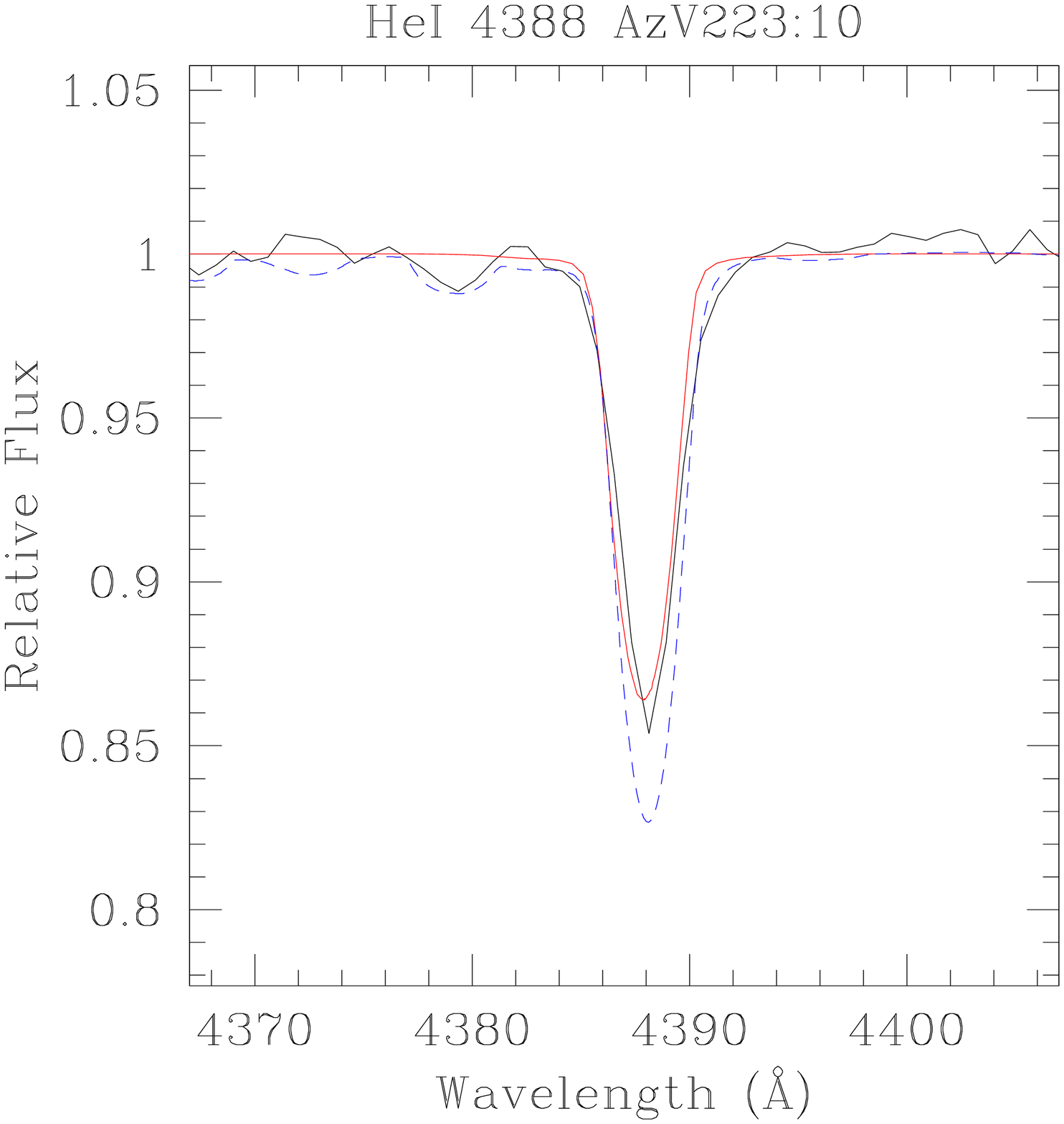}
\plotone{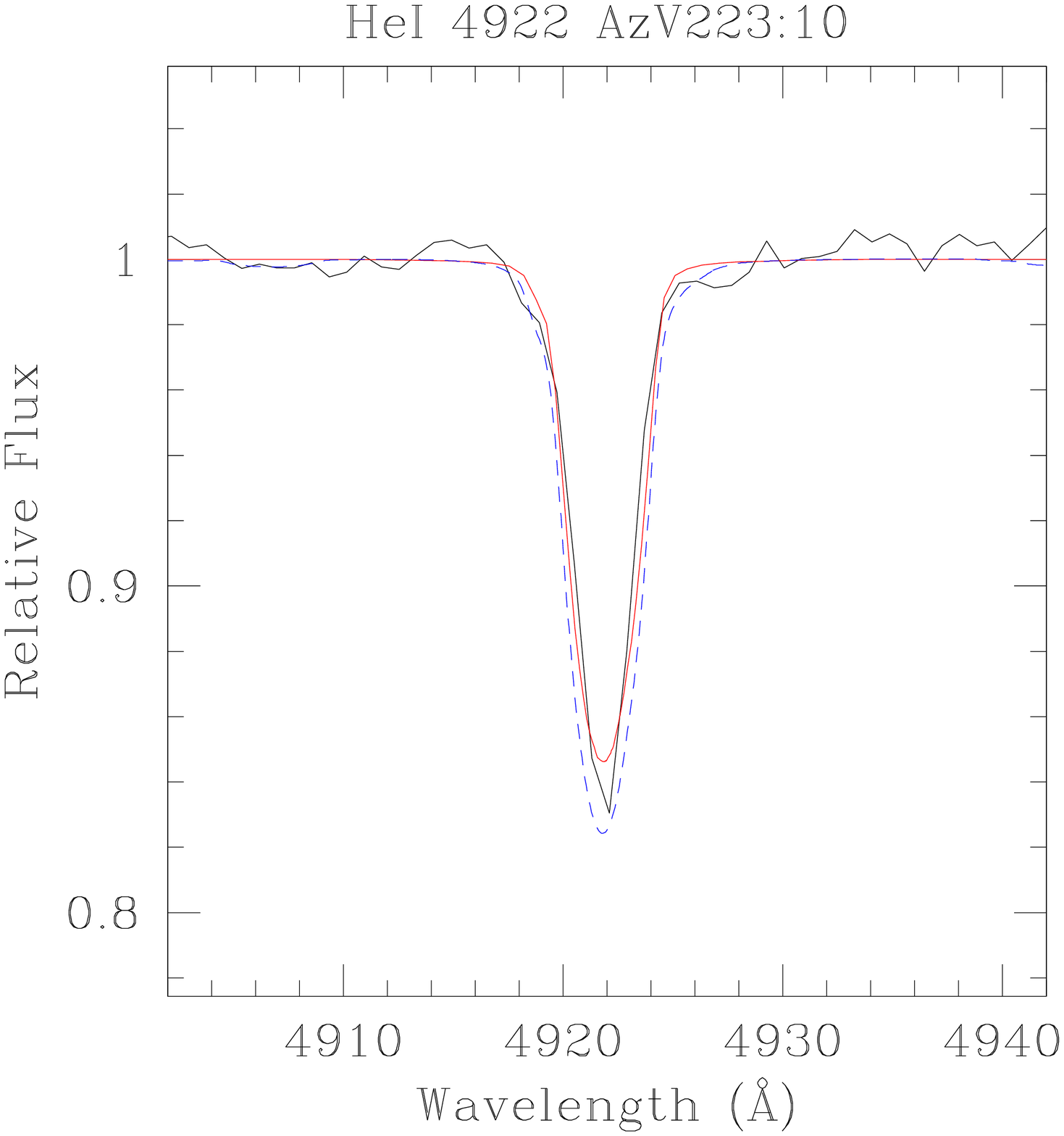}
\plotone{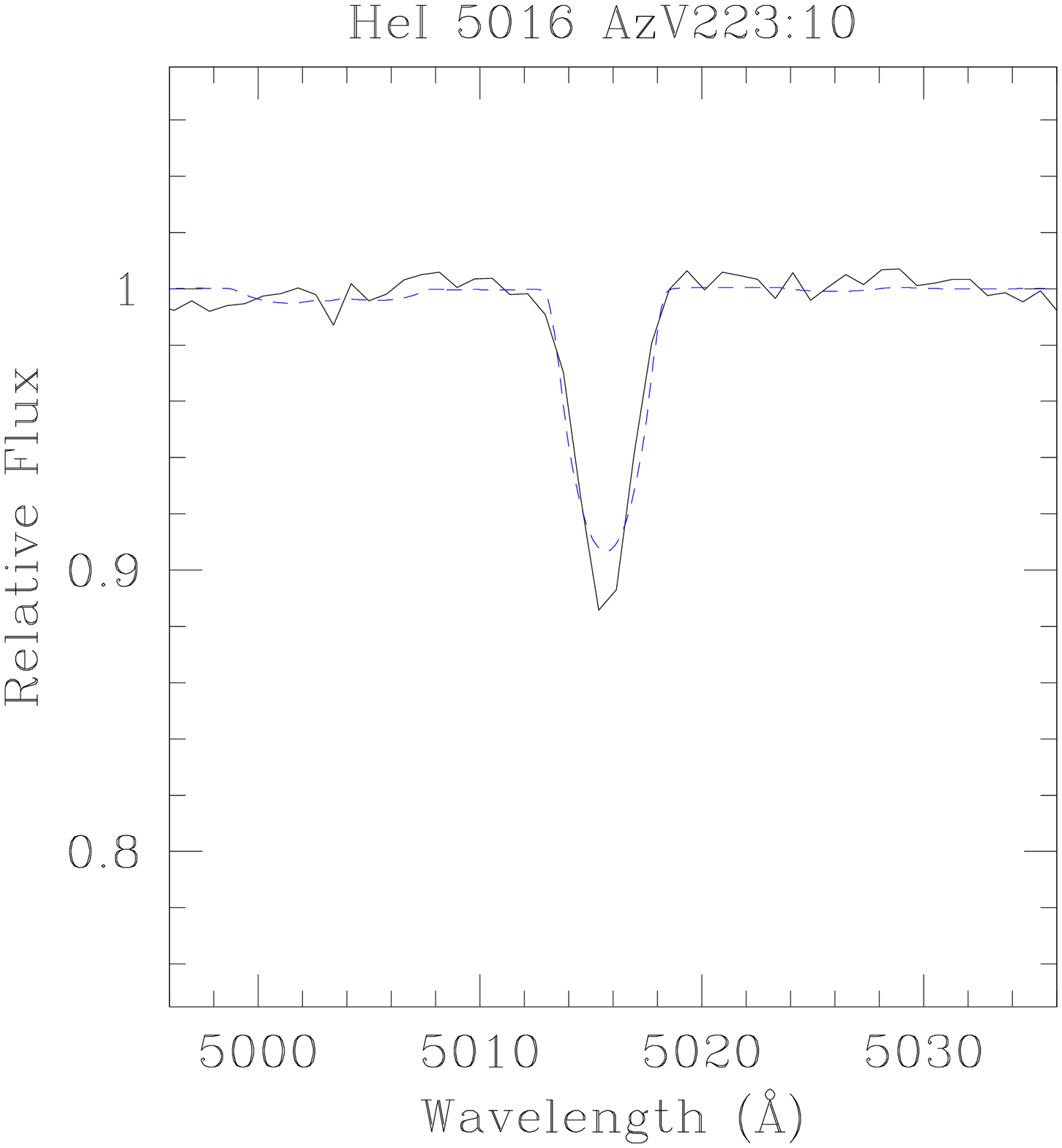}
\plotone{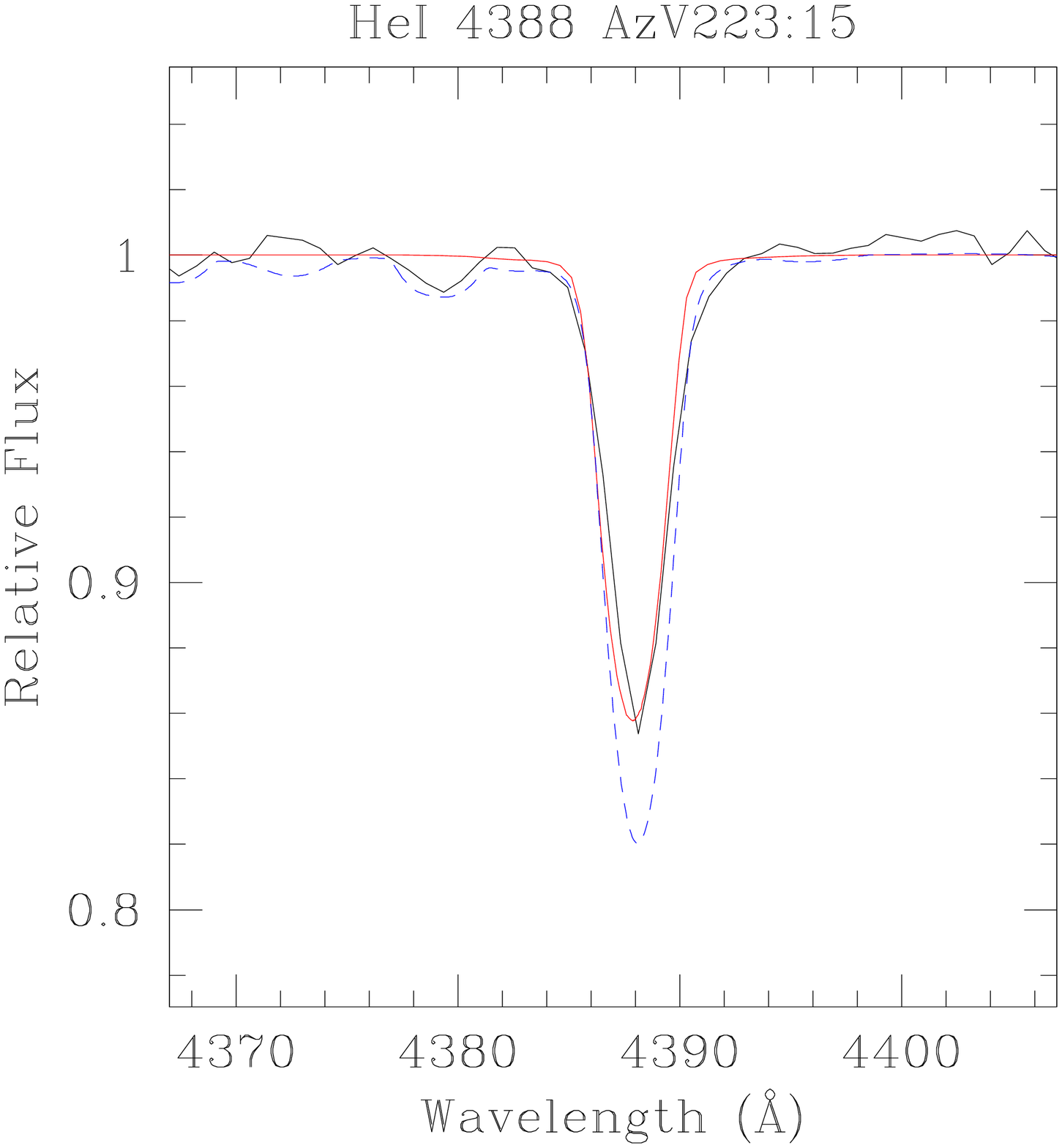}
\plotone{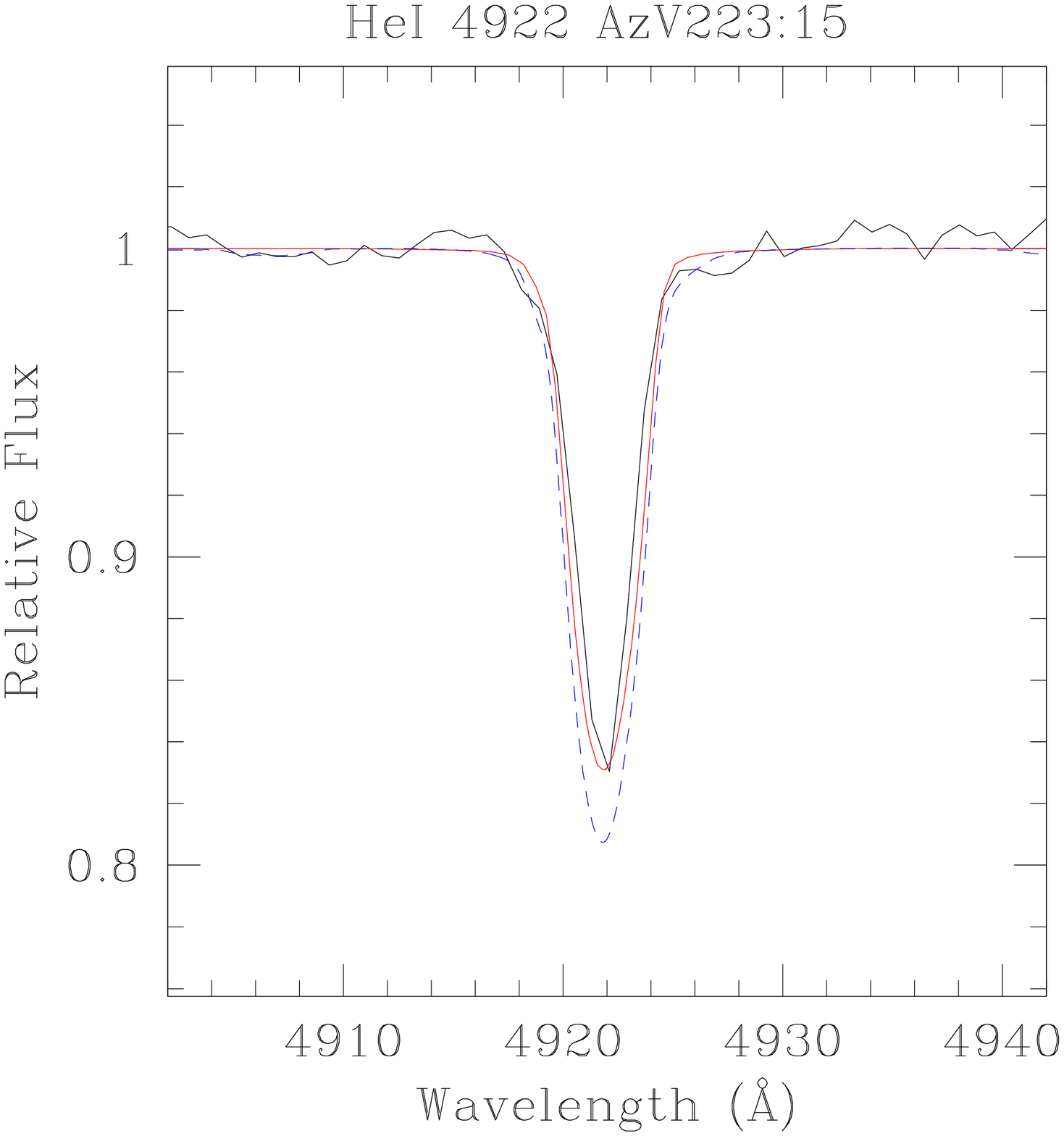}
\plotone{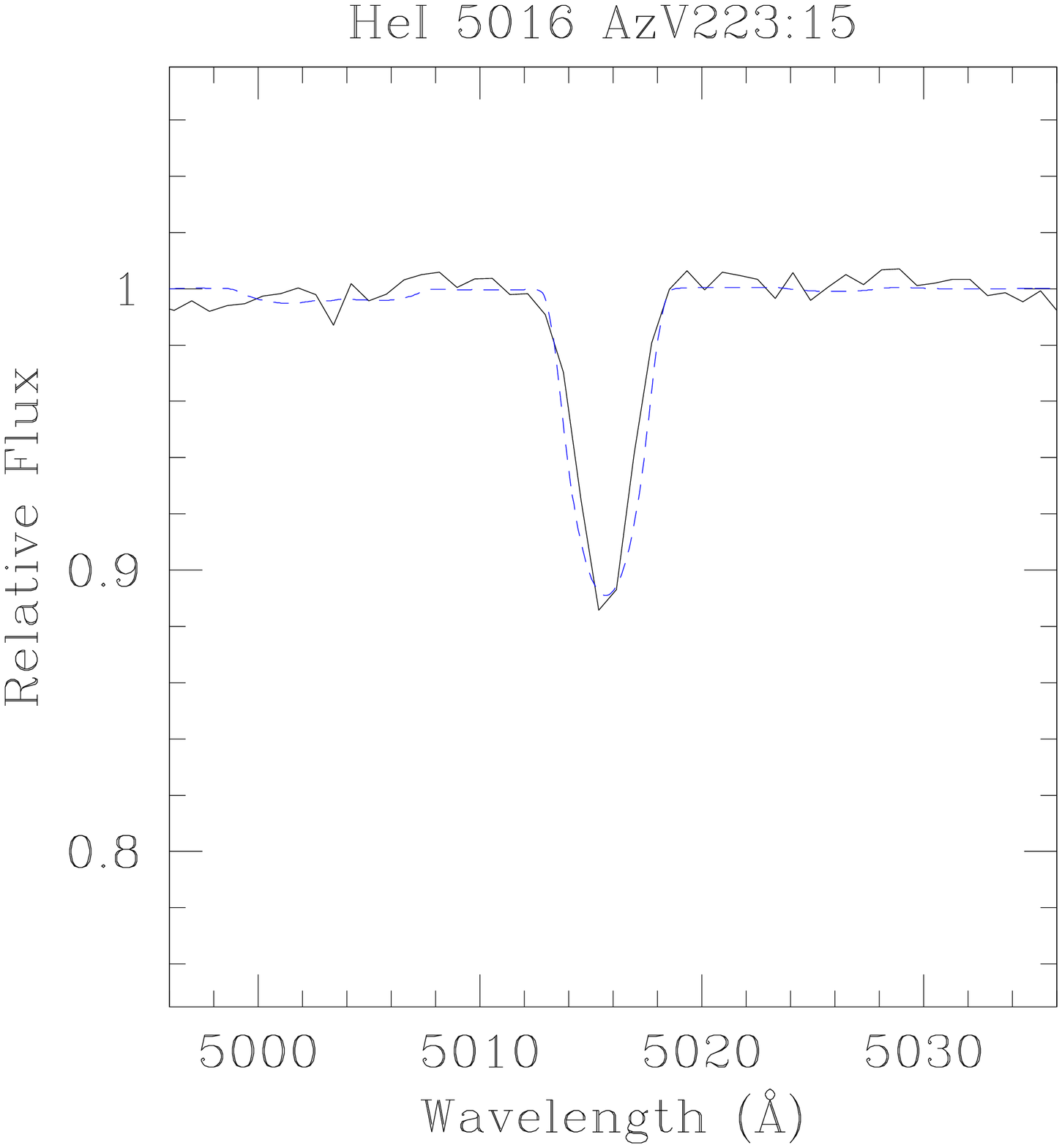}
\plotone{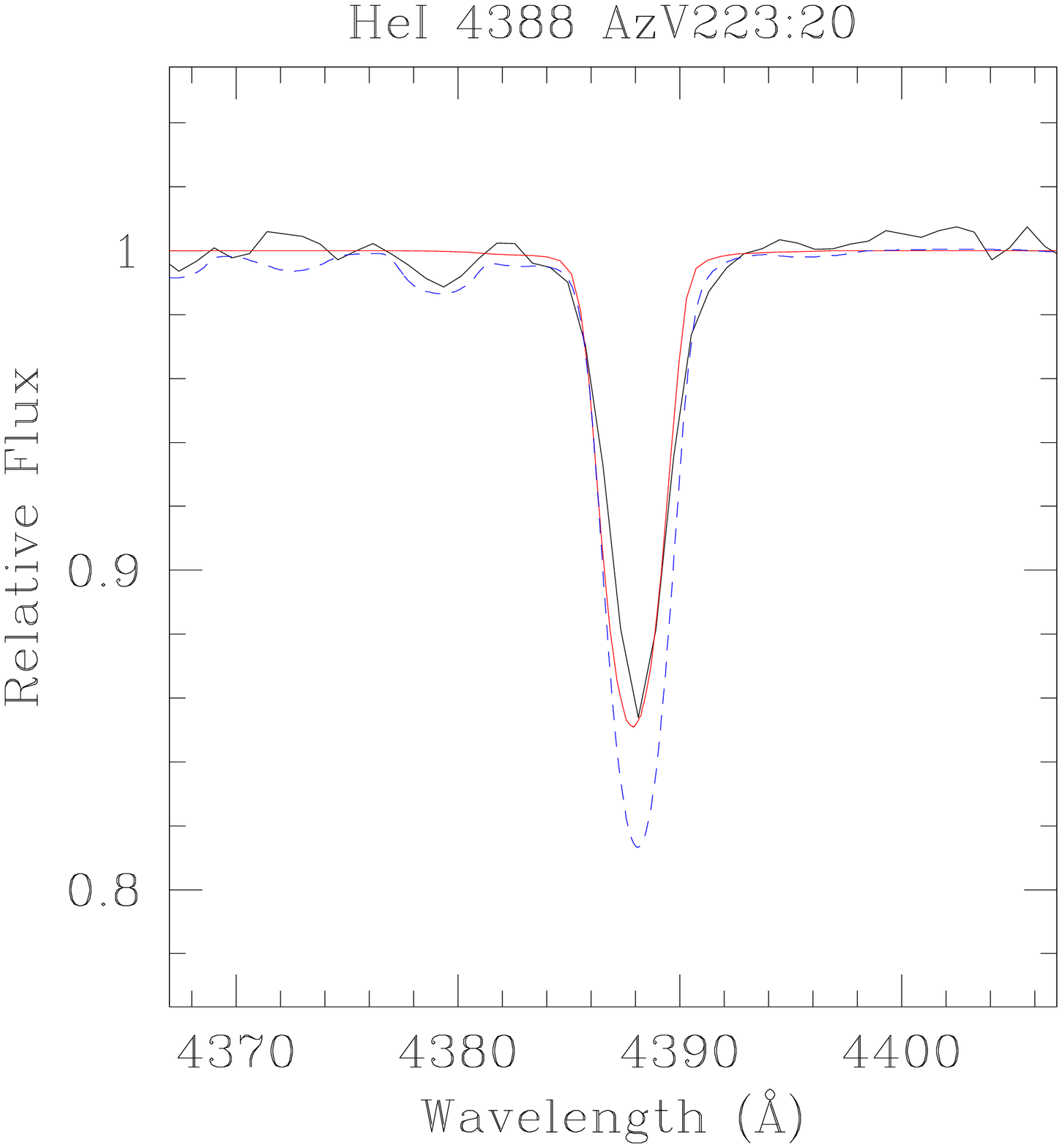}
\plotone{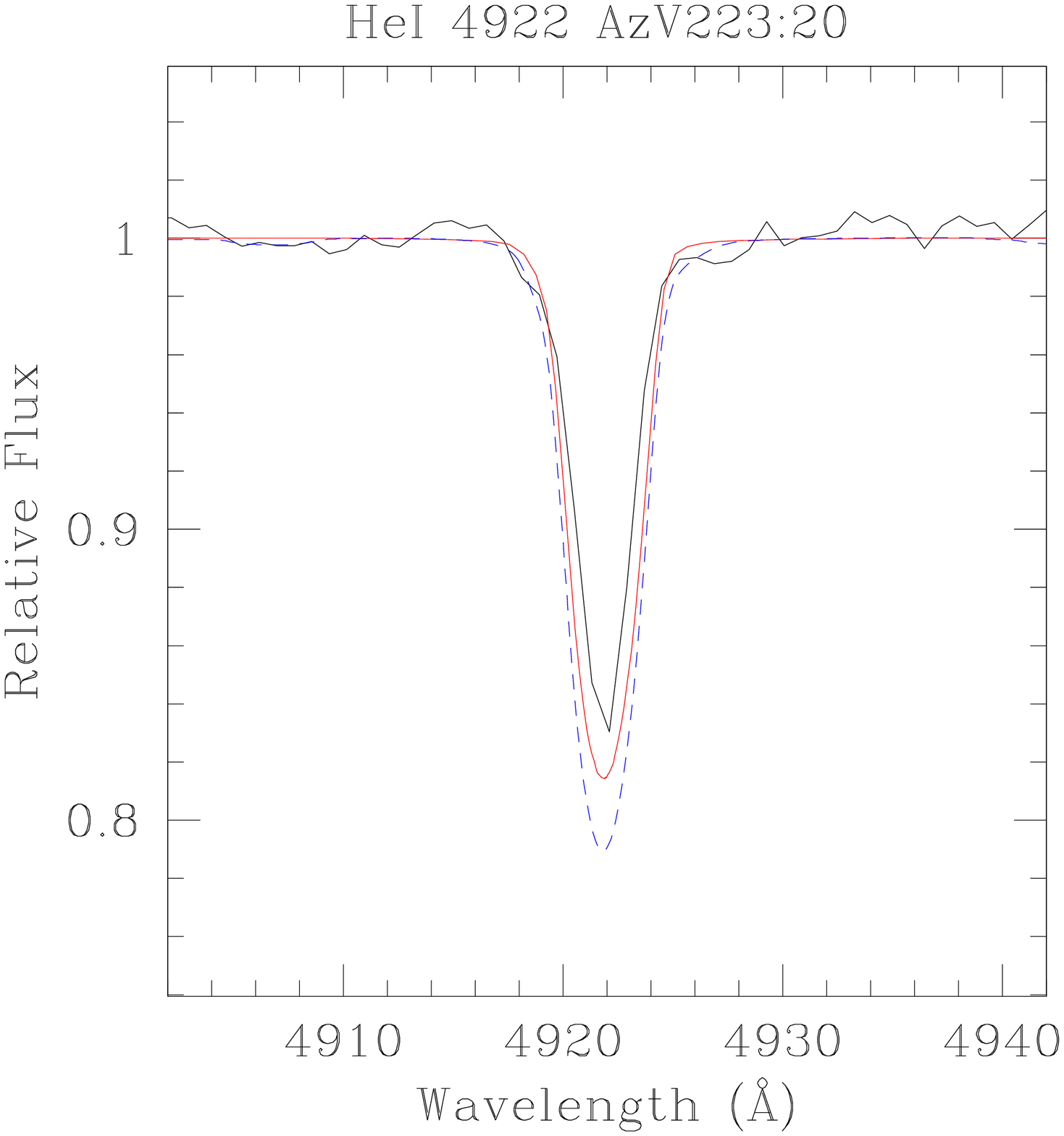}
\plotone{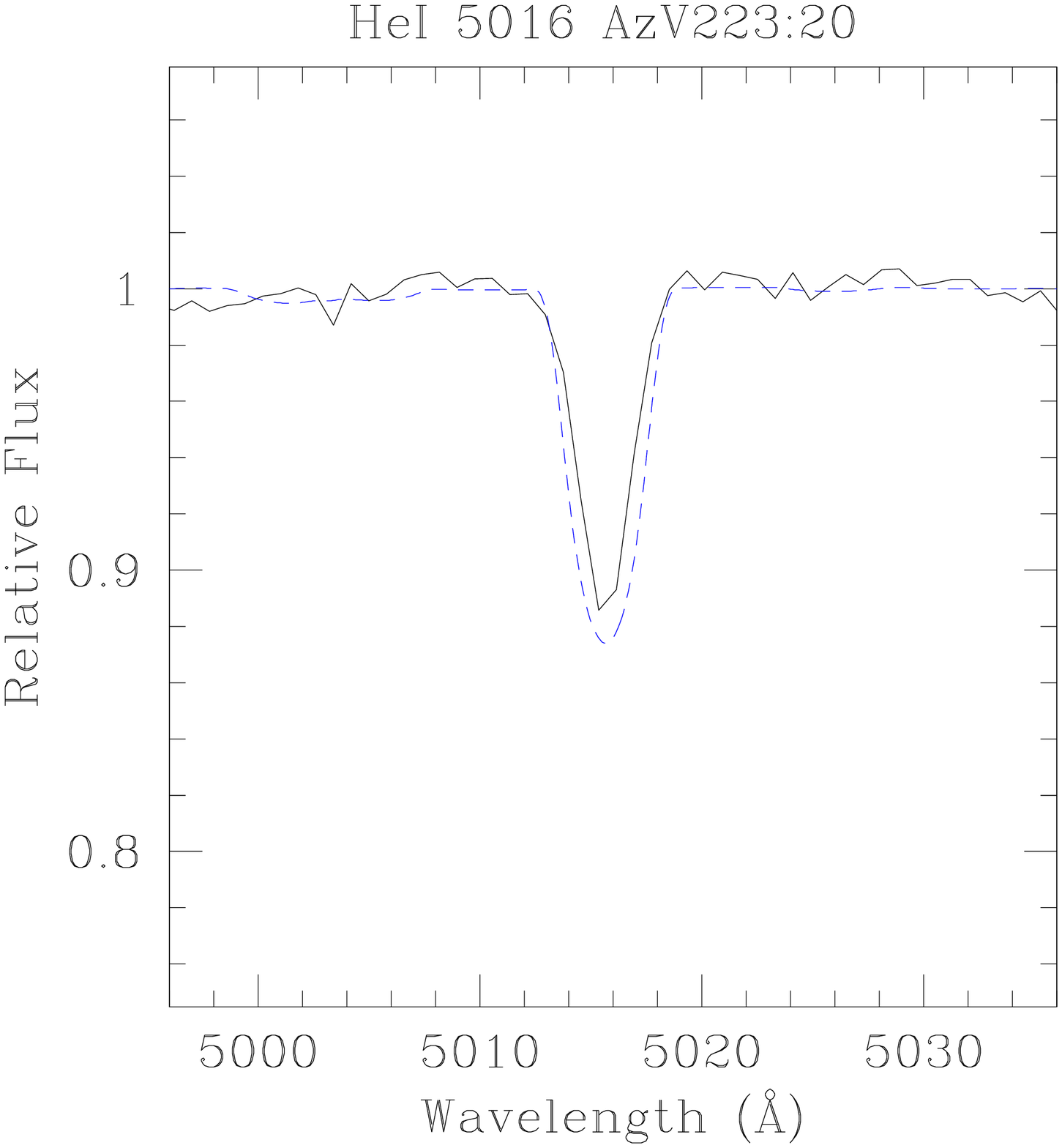}
\caption{\label{fig:MicroSingAzV223} The effect of  microturbulence on the He I singlets in AzV 223, an O9.7 II star in the SMC.
Black shows the observed spectrum, the red line shows the \fastwind\ fit, and the dashed blue line shows the \cmfgen\ fit.
The upper three panels show the model profiles computed using the  ``standard" 10 km s$^{-1}$ microturbulent velocities, the middle three panels show that obtained using 15 km s$^{-1}$, and the bottom three panels show the model profiles obtained using 20 km$^{-1}$.}
\end{figure}

\begin{figure}
\epsscale{0.25}
\plotone{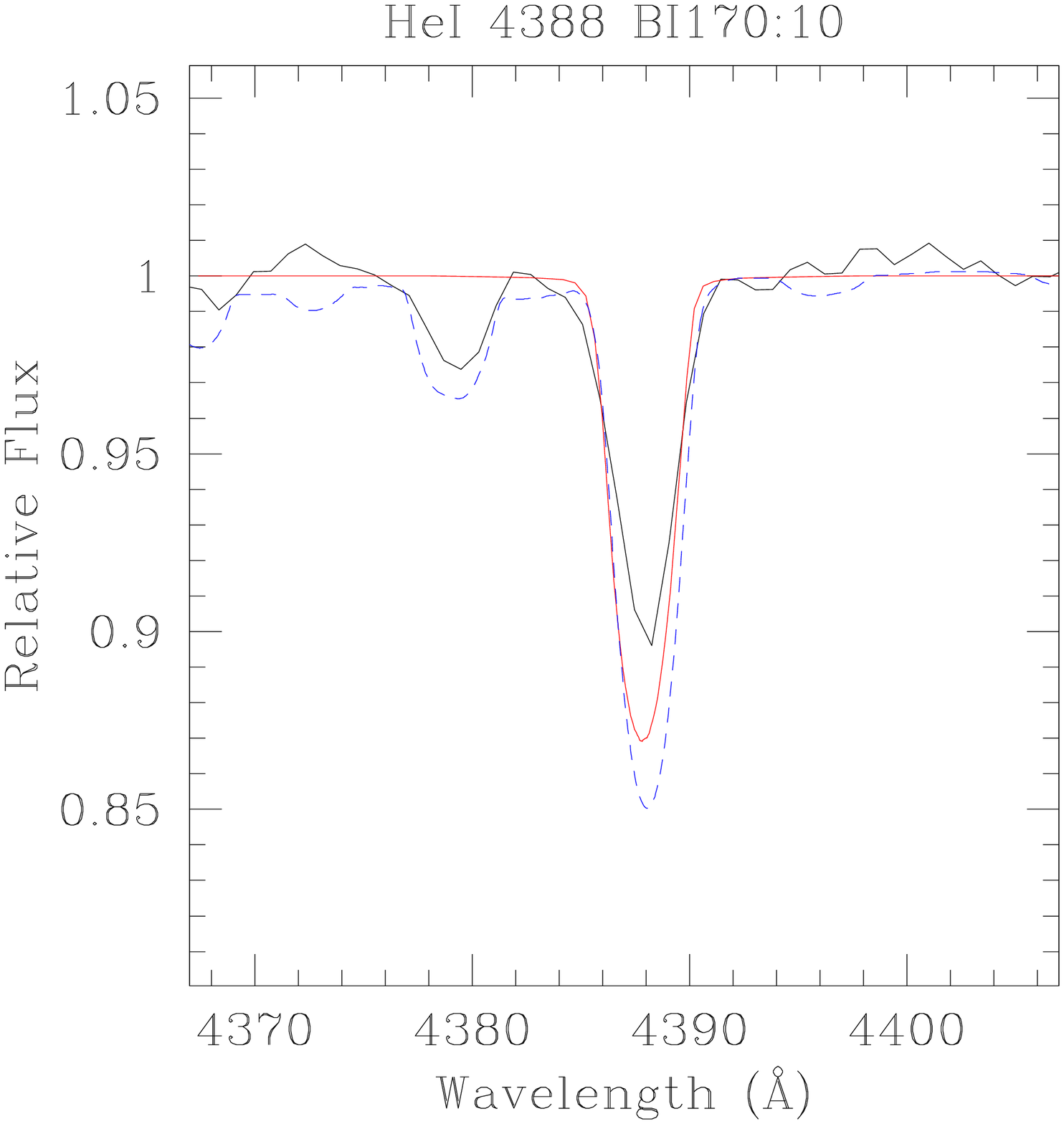}
\plotone{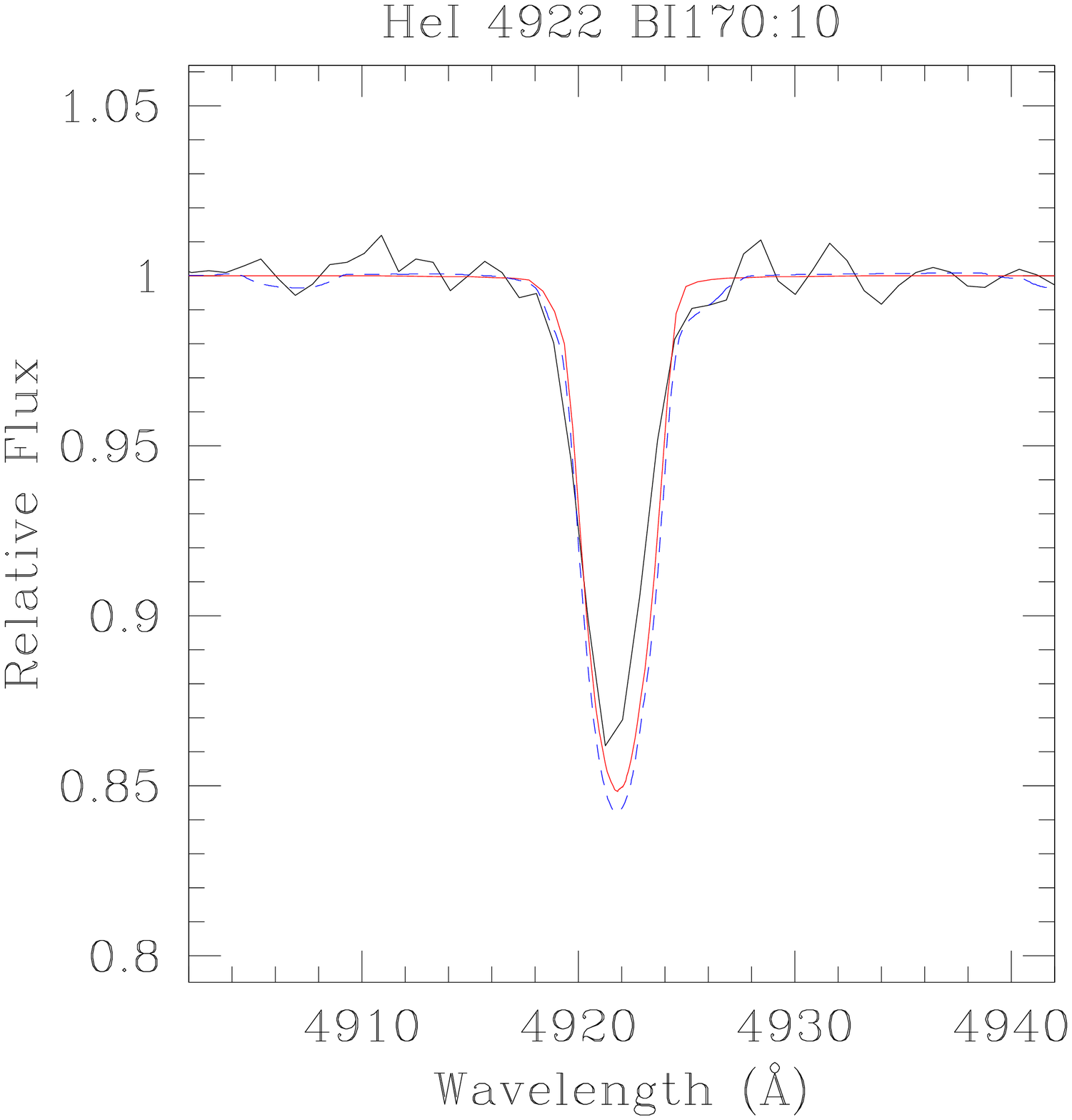}
\plotone{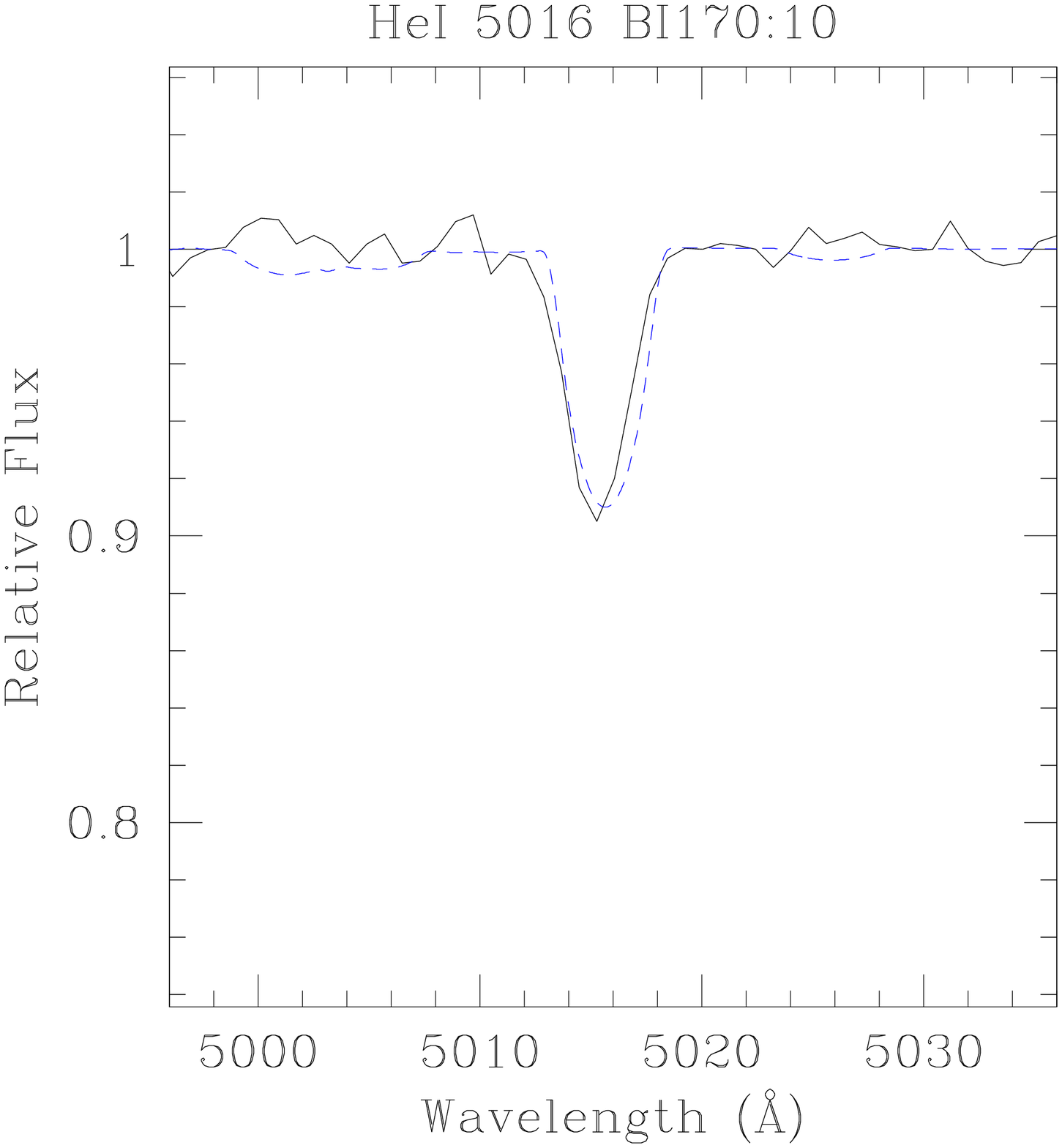}
\plotone{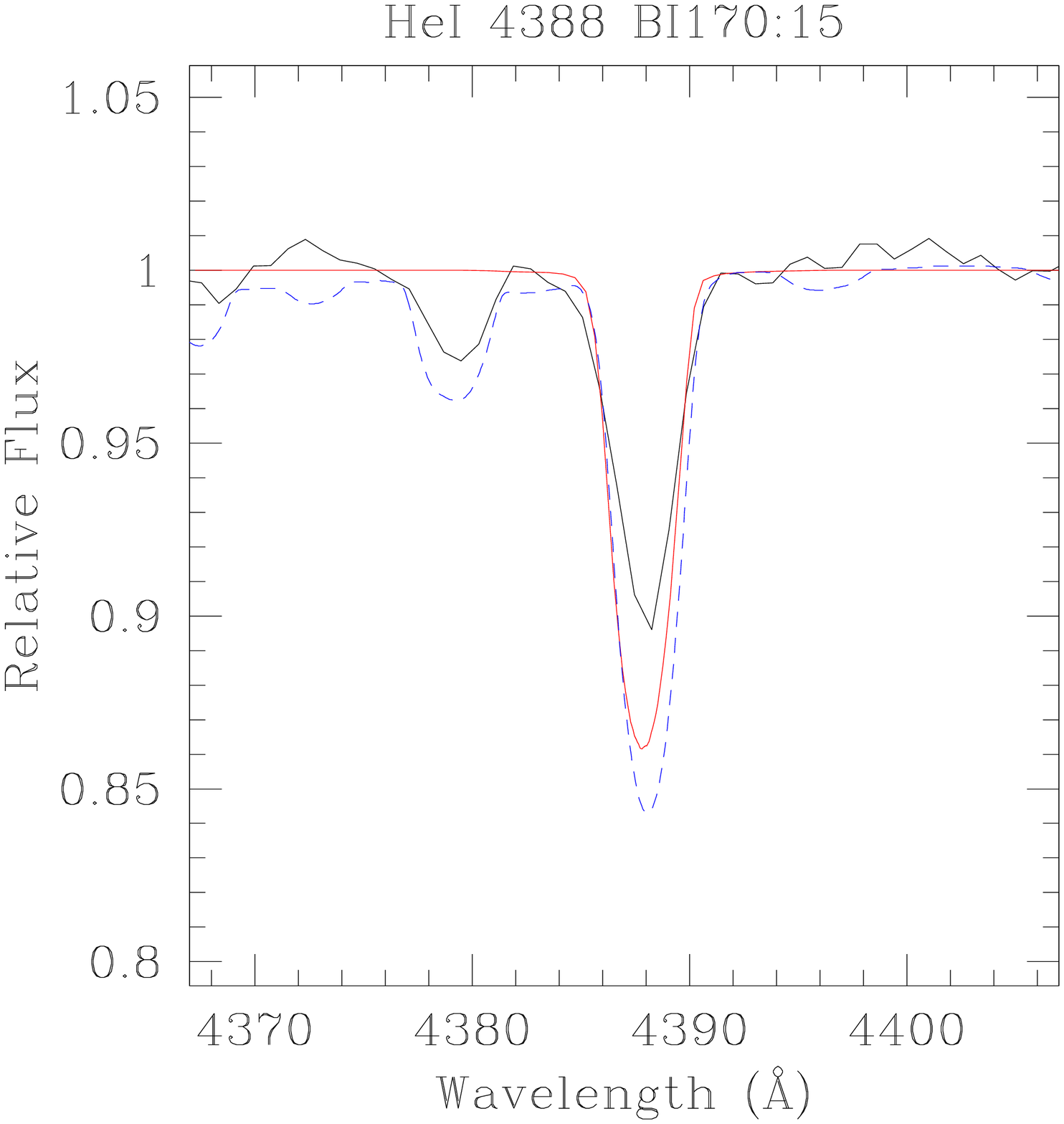}
\plotone{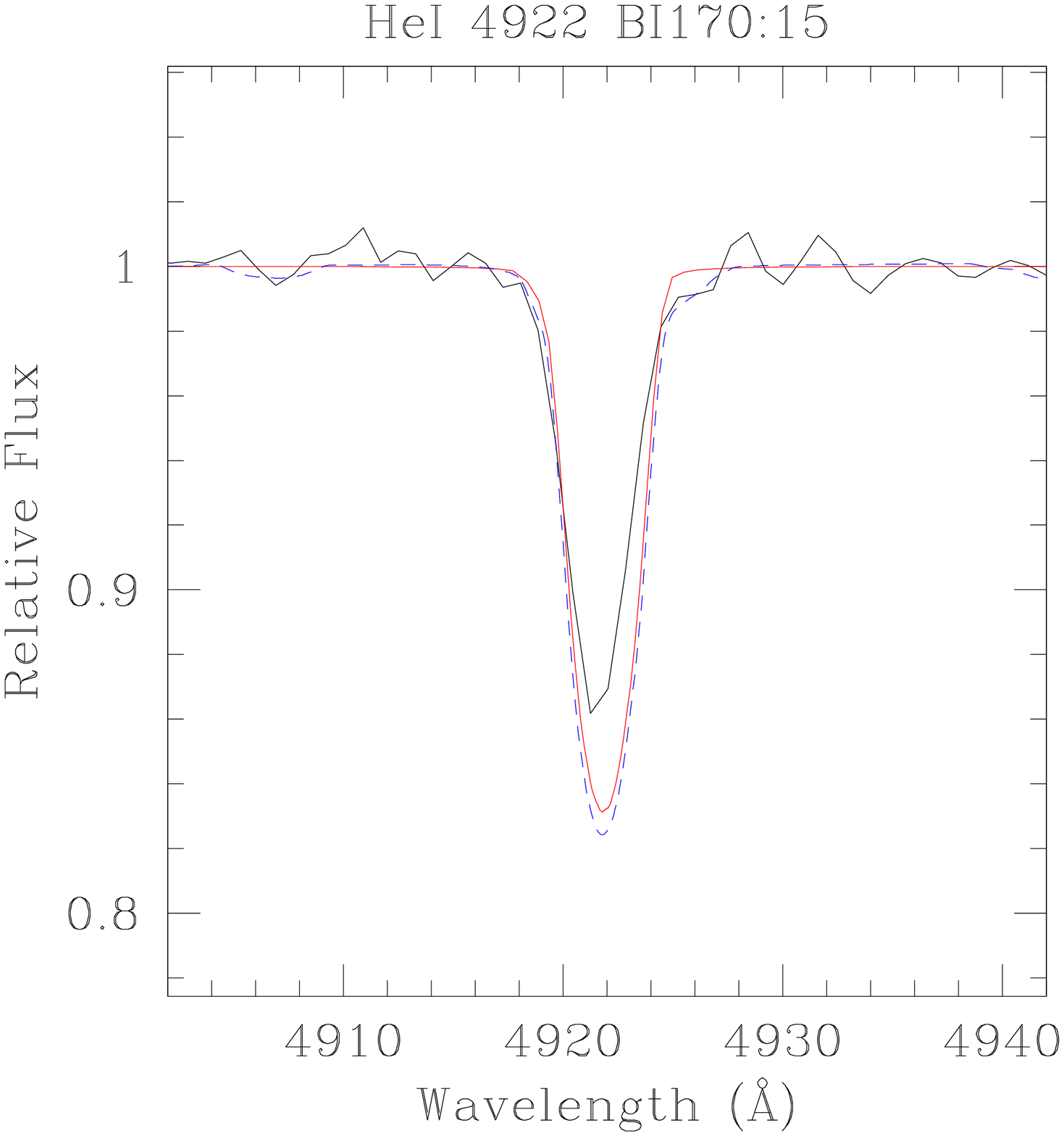}
\plotone{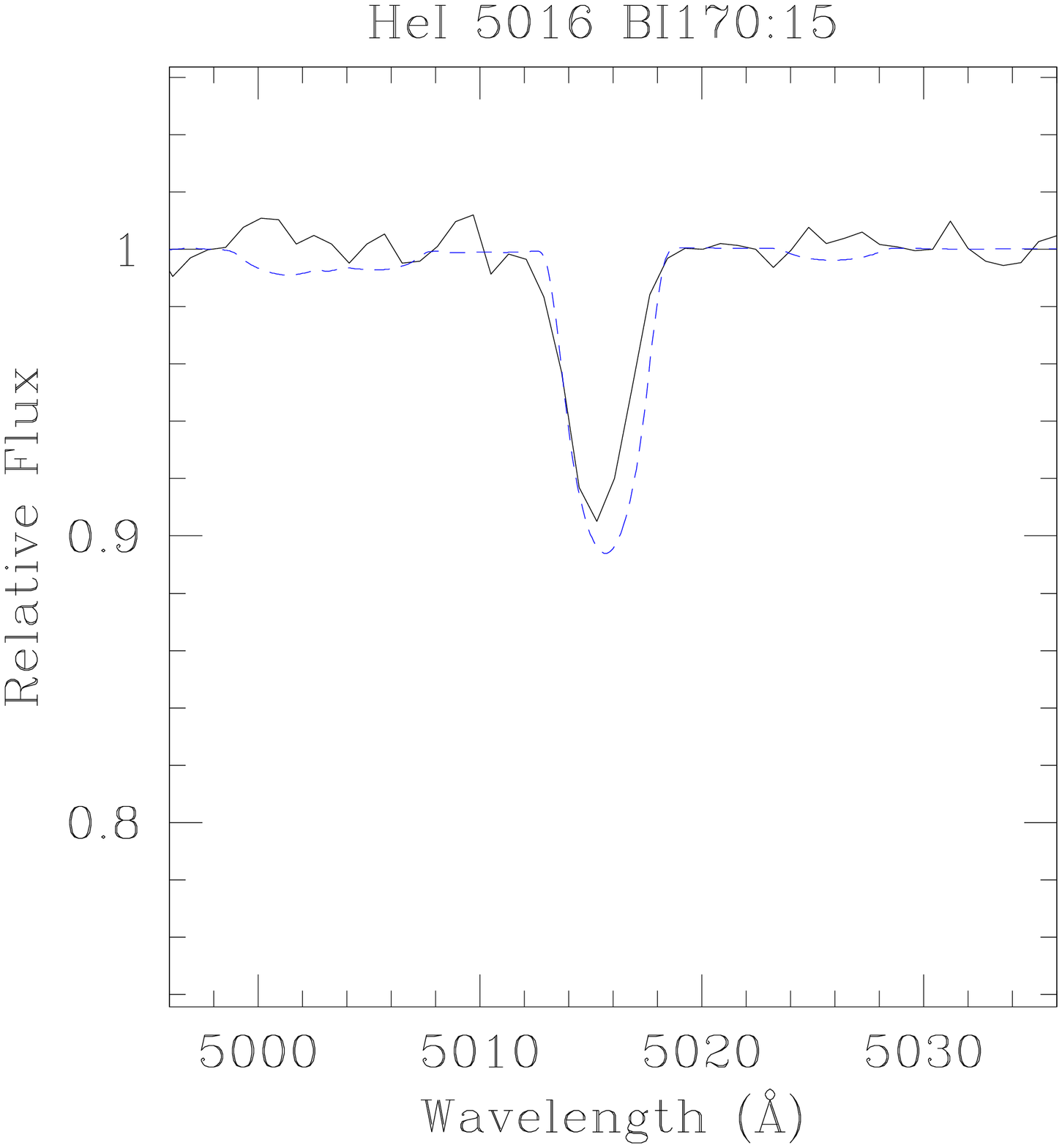}
\plotone{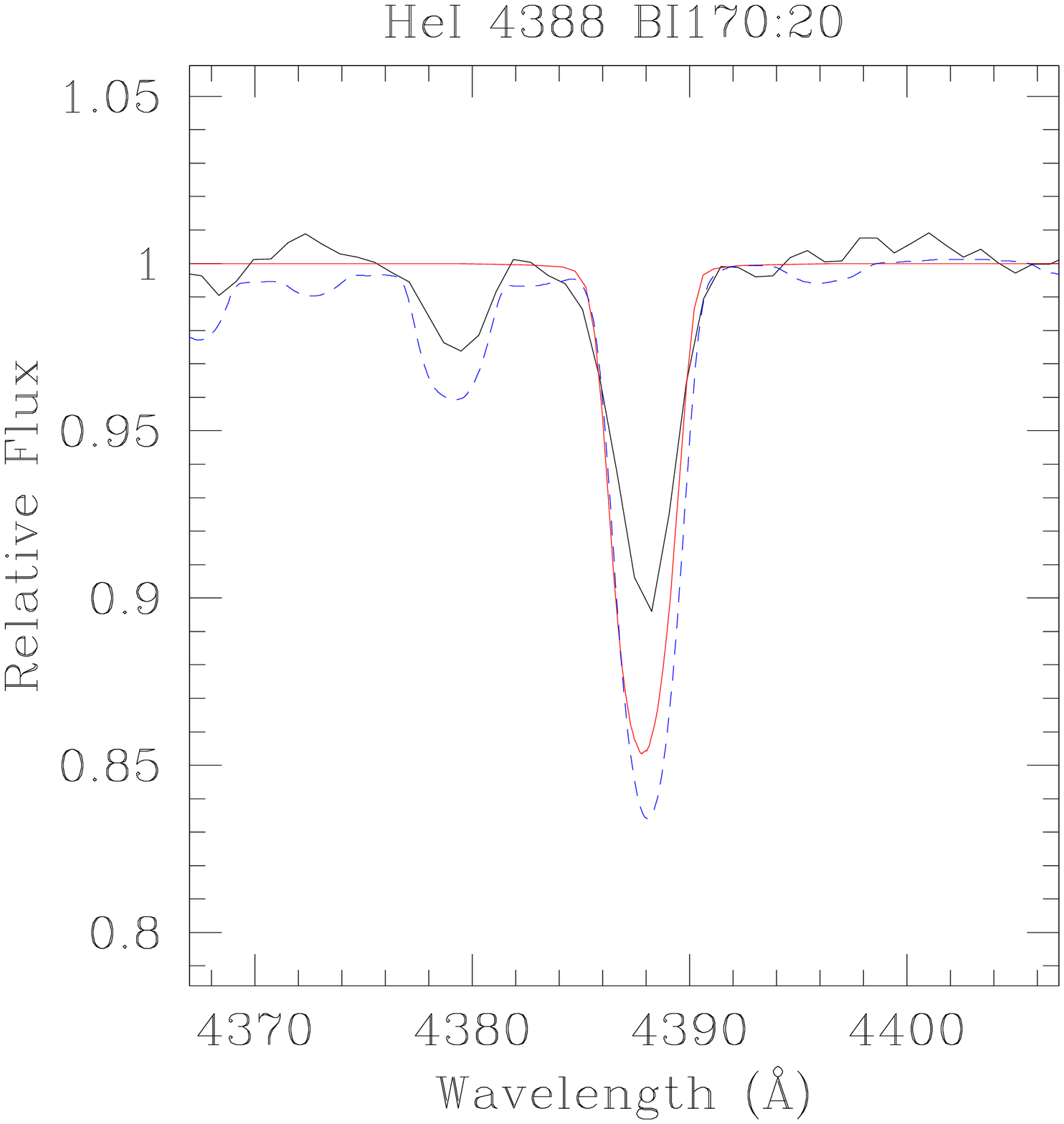}
\plotone{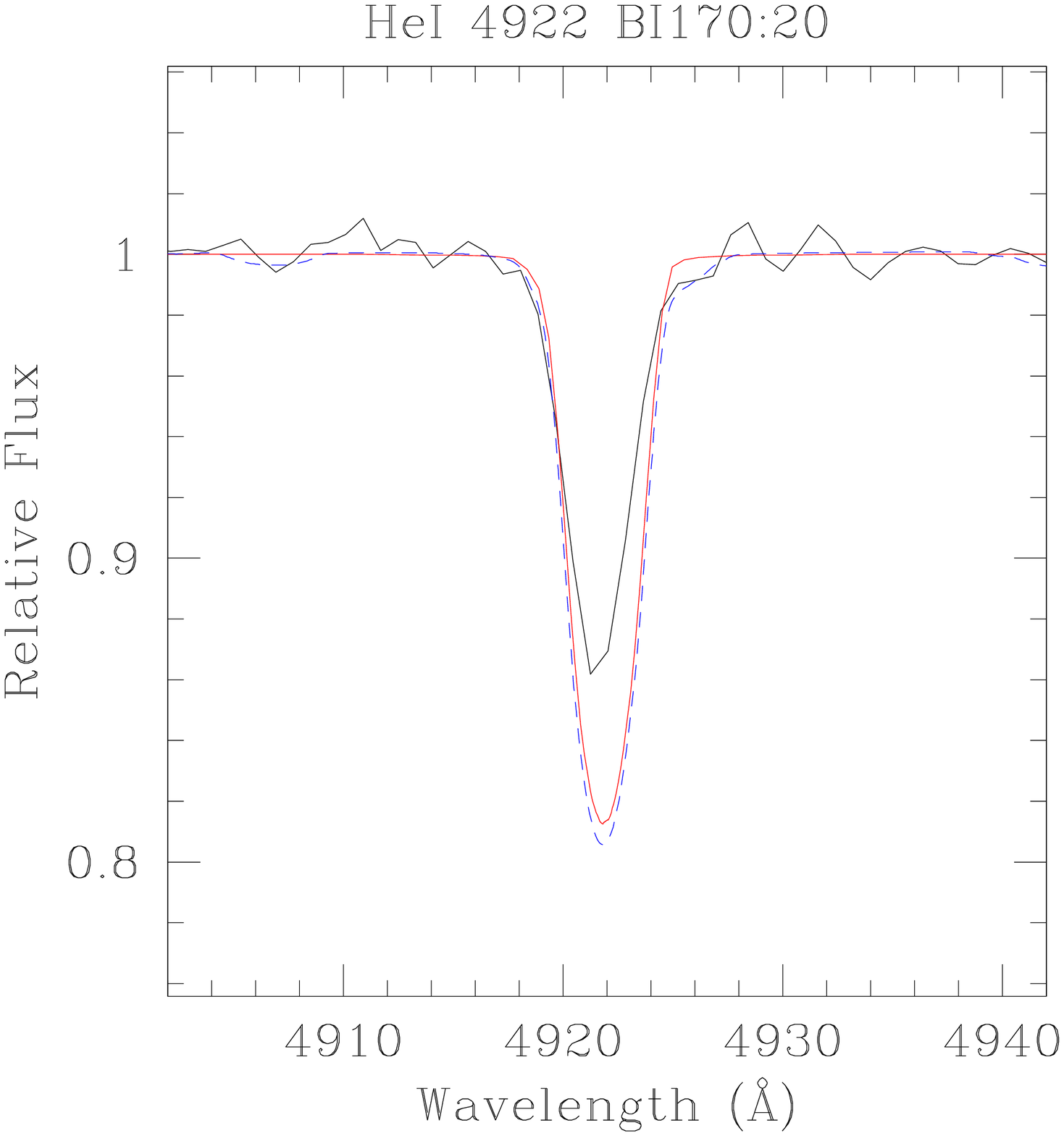}
\plotone{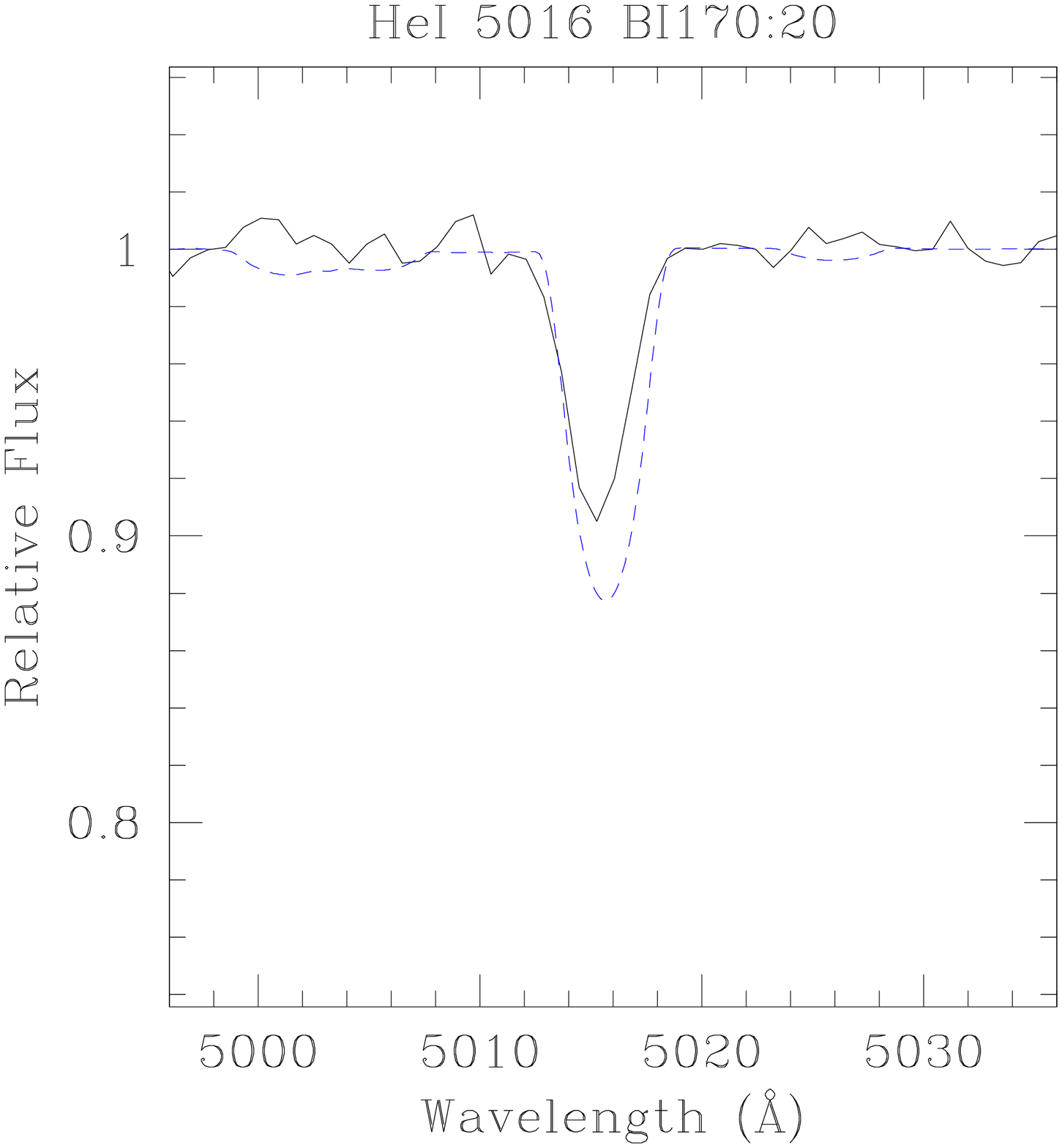}
\caption{\label{fig:MicroSingBI180}The effect of  microturbulence on the He I singlets in BI 170, an O9.5 I star in the LMC.
Black shows the observed spectrum, the red line shows the \fastwind\ fit, and the dashed blue line shows the \cmfgen\ fit.
The upper three panels show the model profiles computed using the  ``standard" 10 km s$^{-1}$ microturbulent velocities, the middle three panels show that obtained using 15 km s$^{-1}$, and the bottom three panels show the model profiles obtained using 20 km$^{-1}$.}
\end{figure}

\begin{figure}
\epsscale{0.25}
\plotone{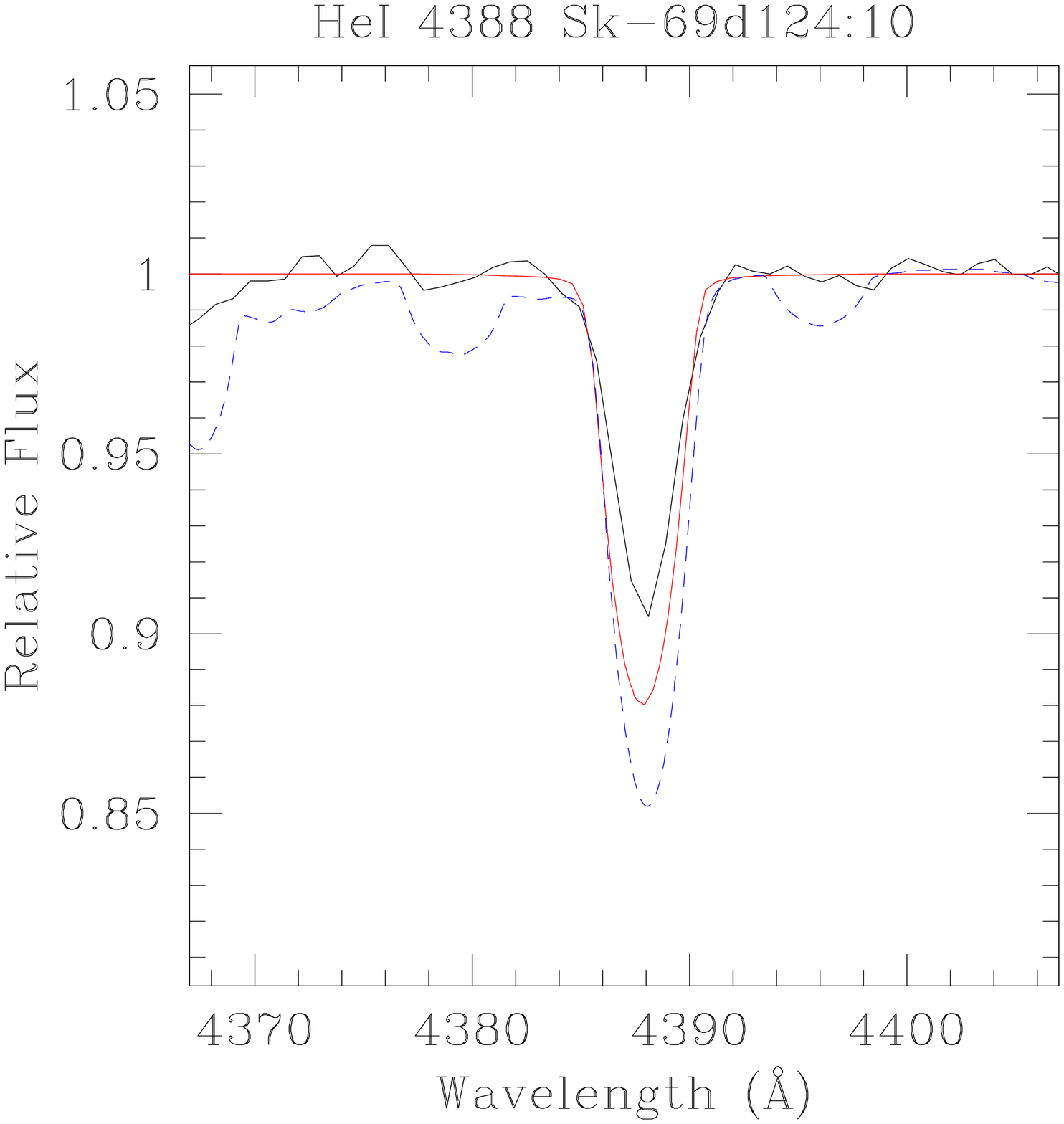}
\plotone{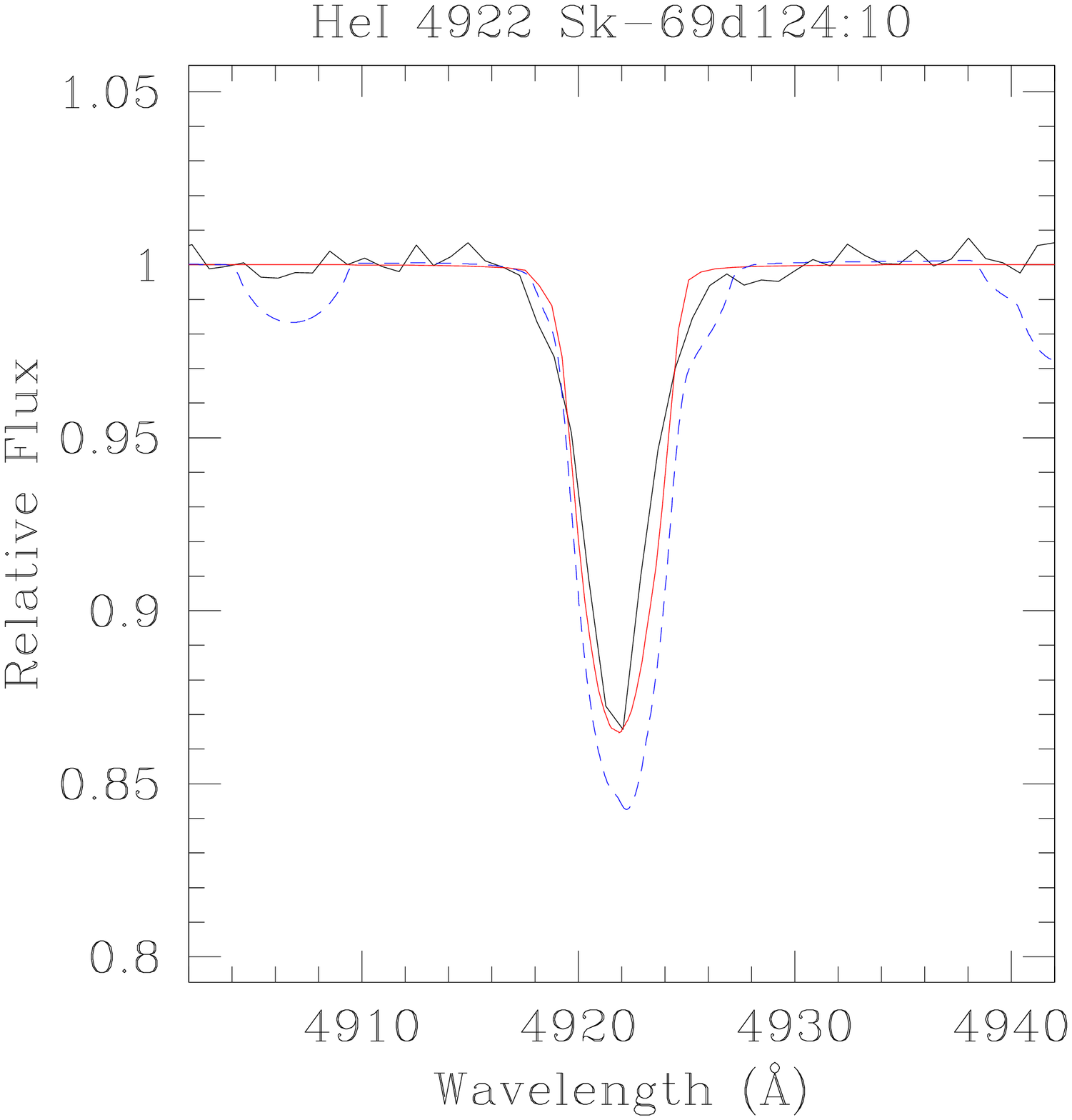}
\plotone{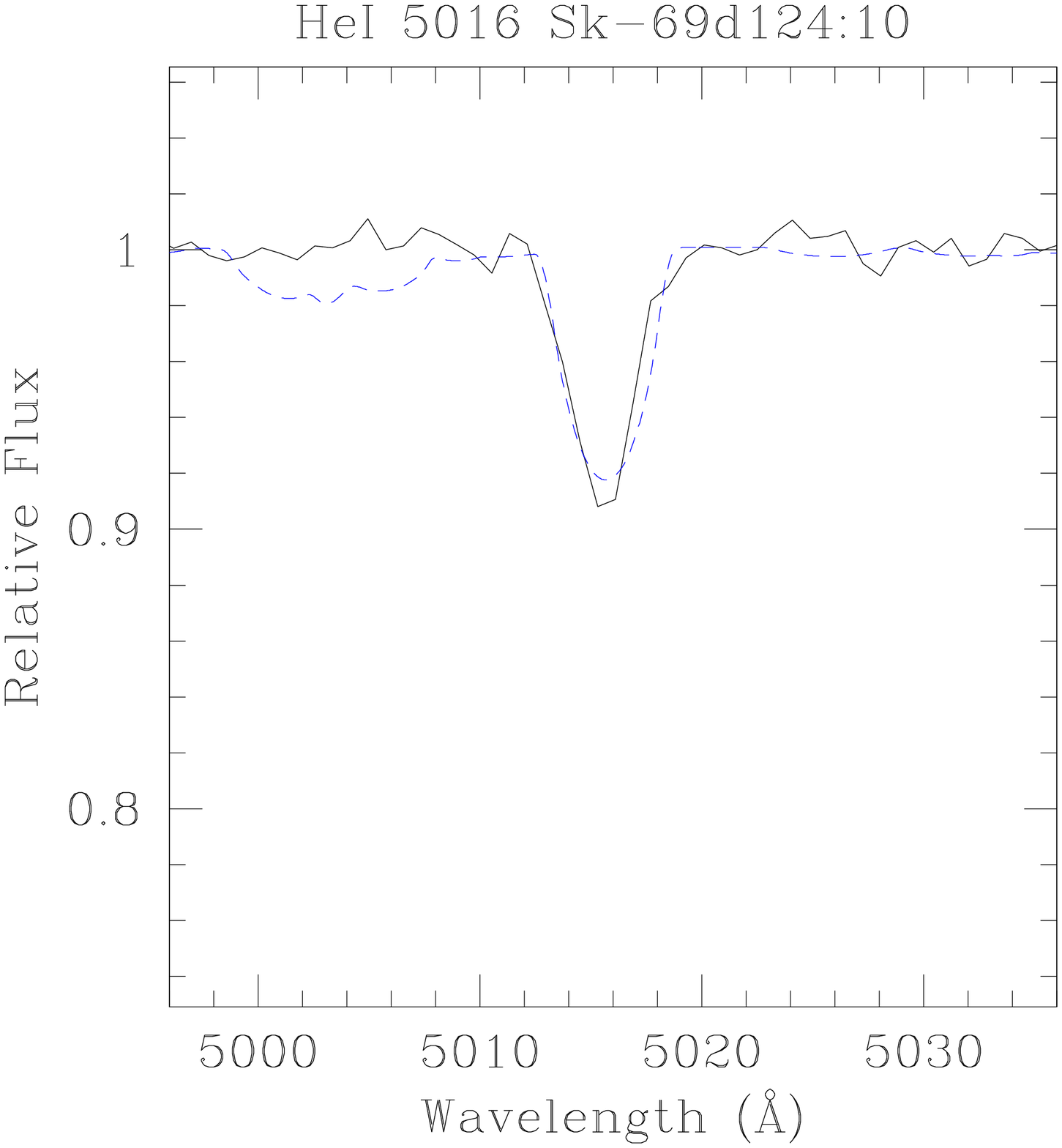}
\plotone{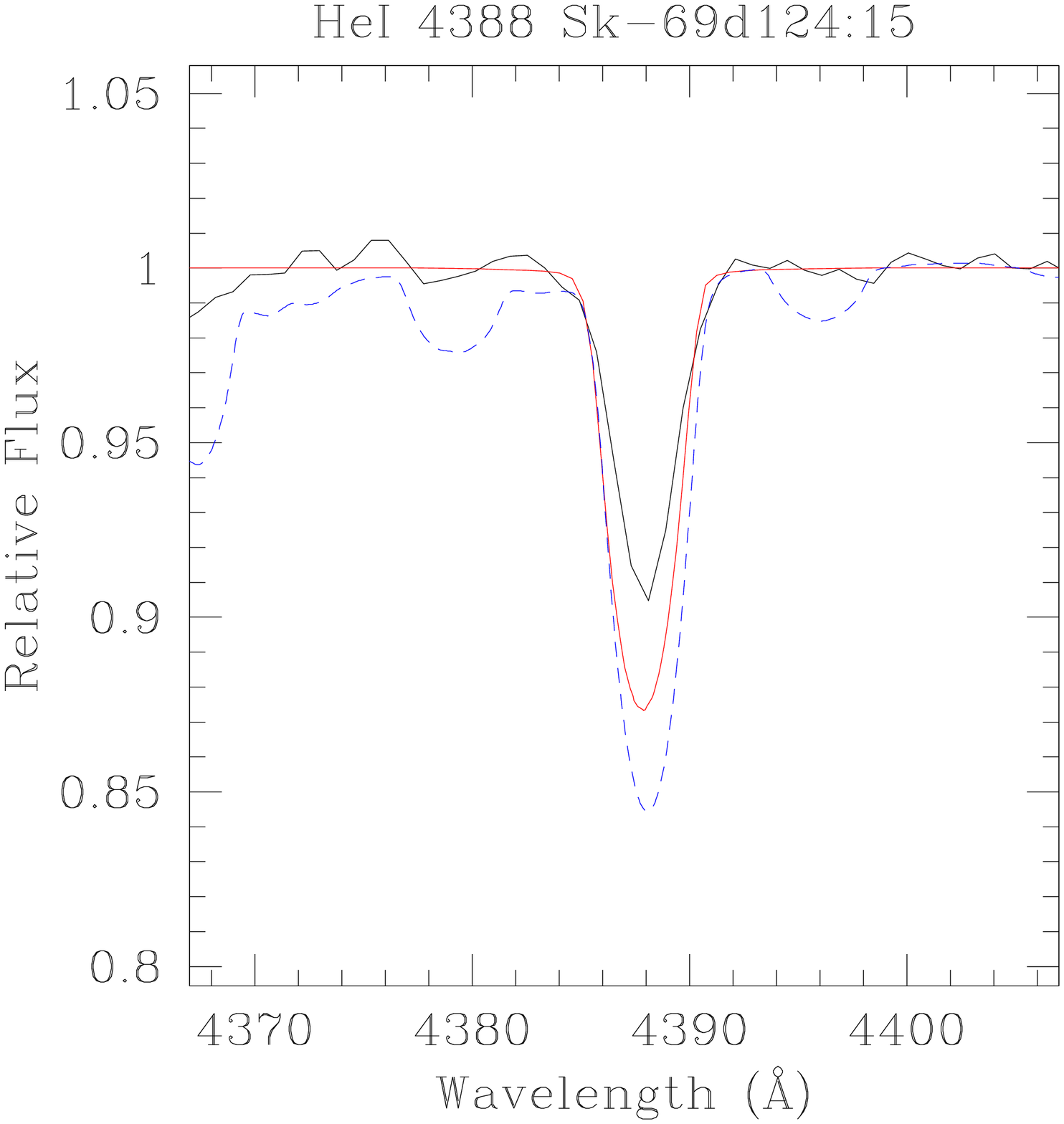}
\plotone{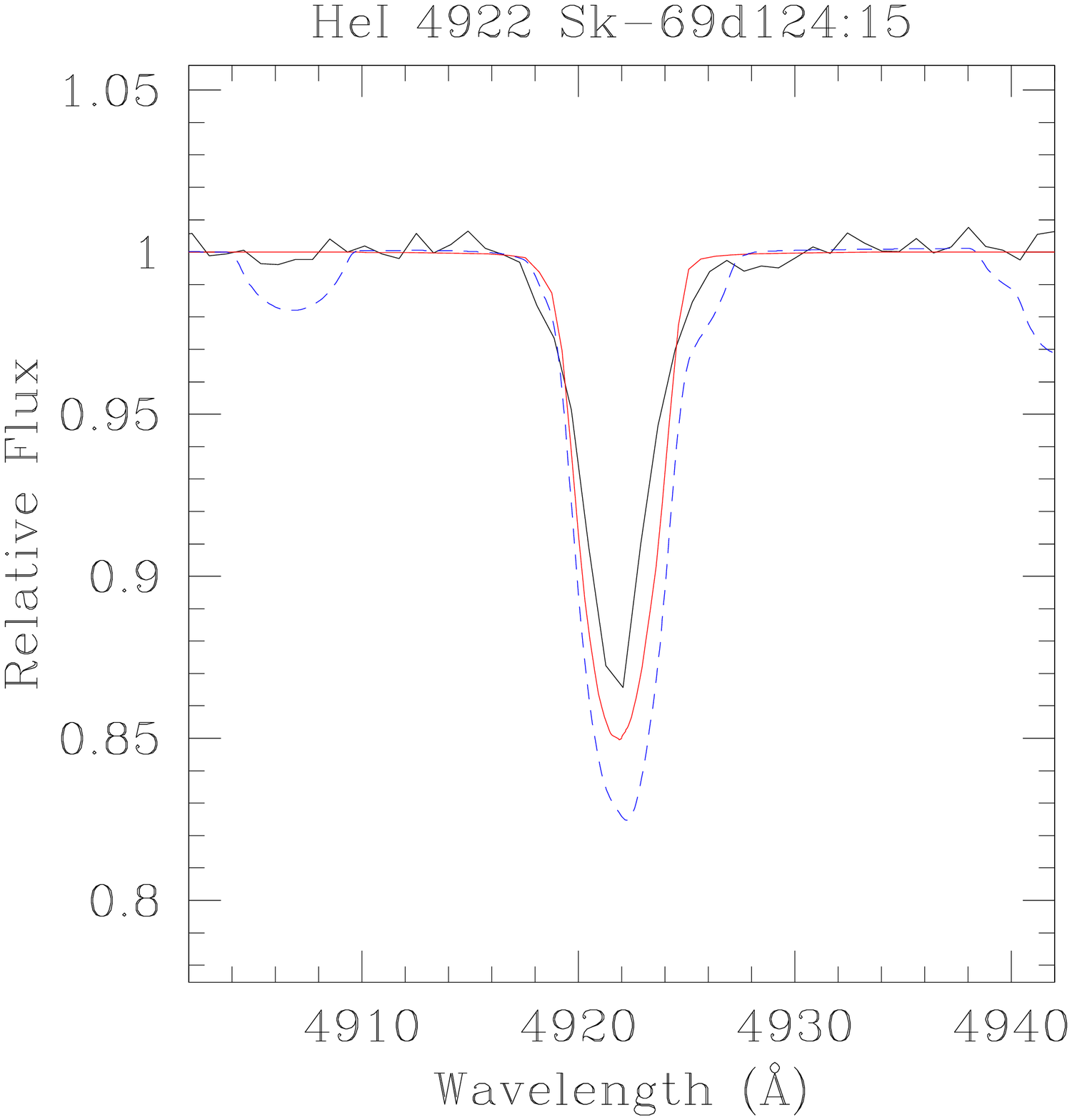}
\plotone{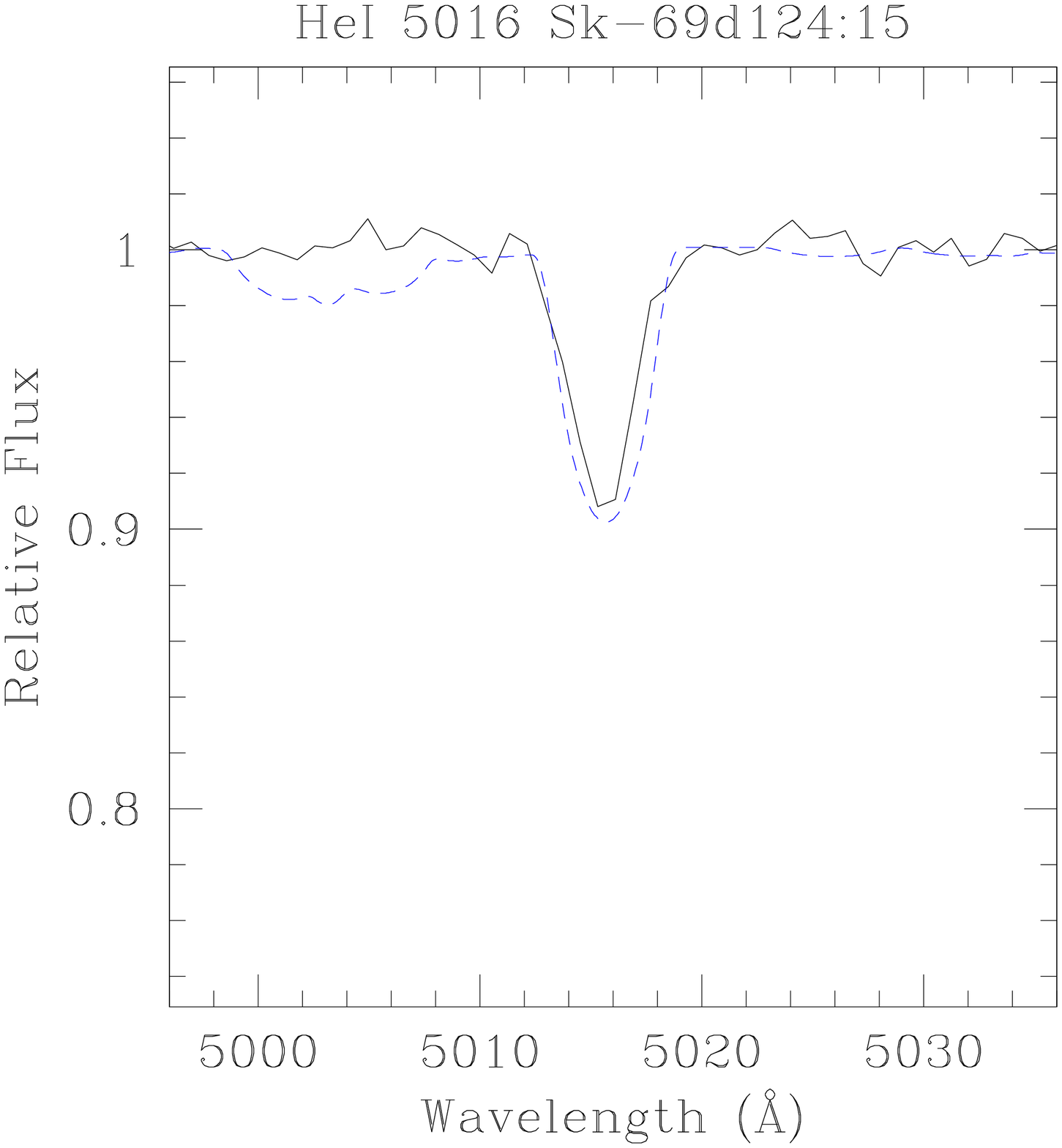}
\plotone{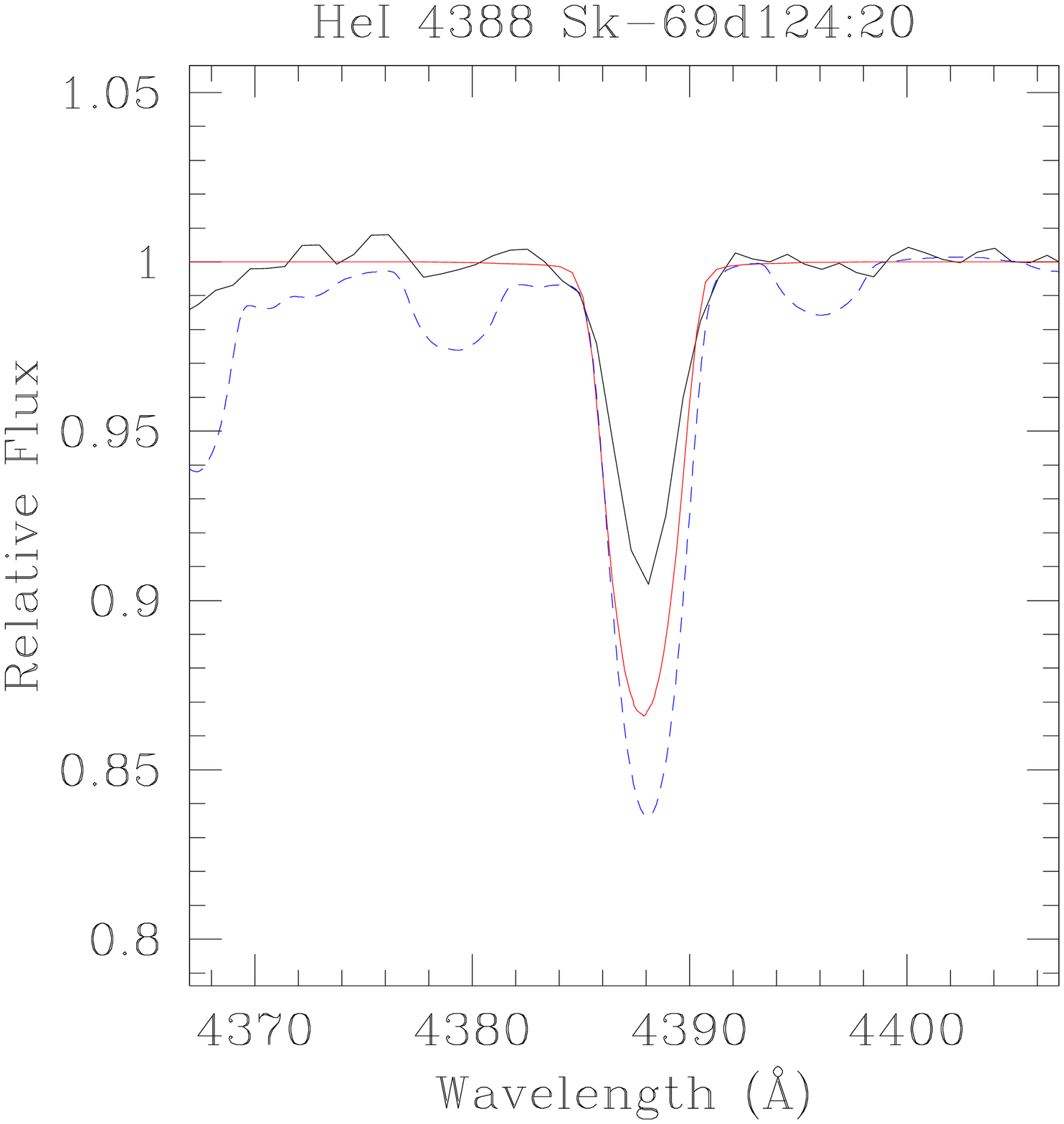}
\plotone{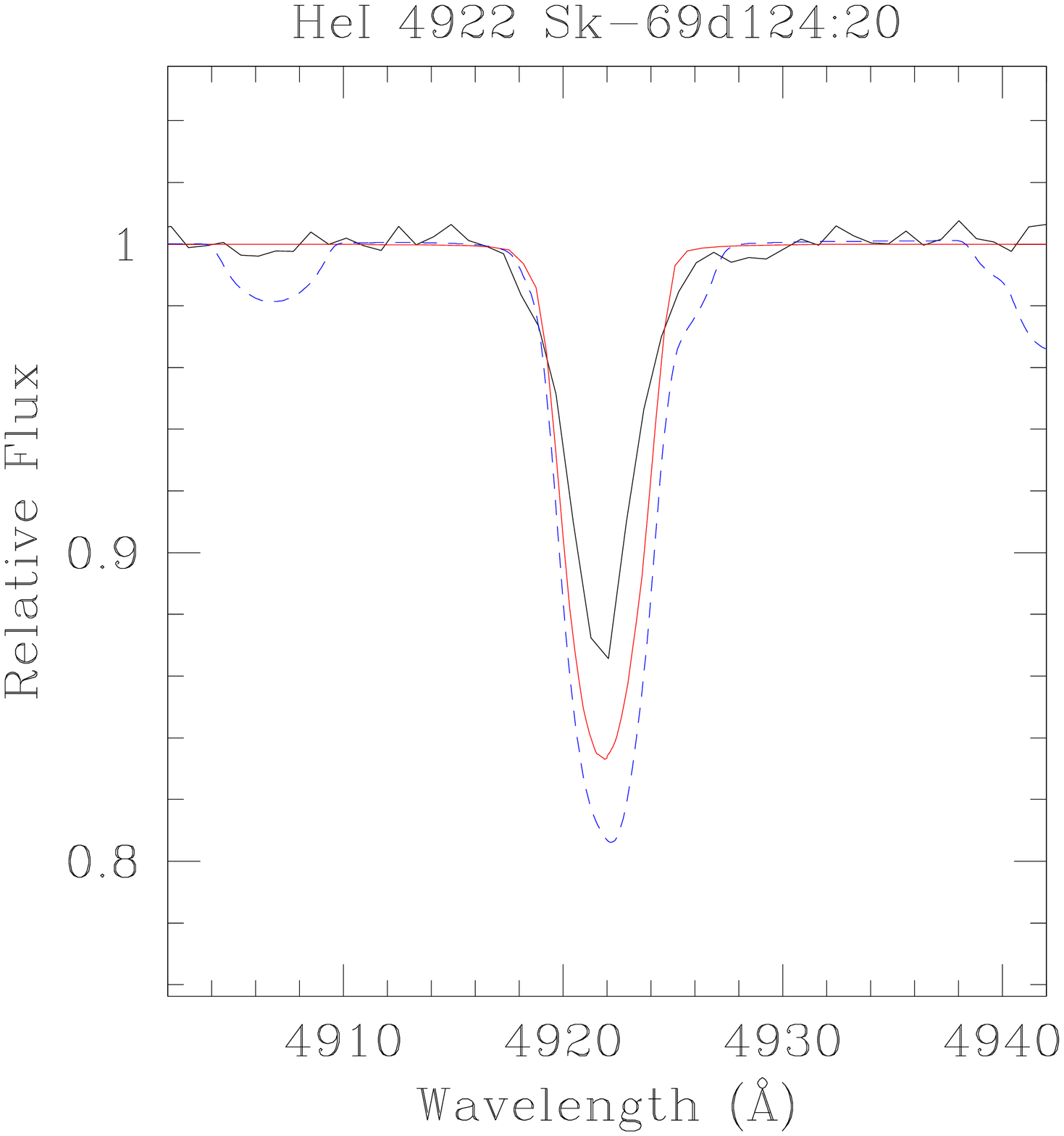}
\plotone{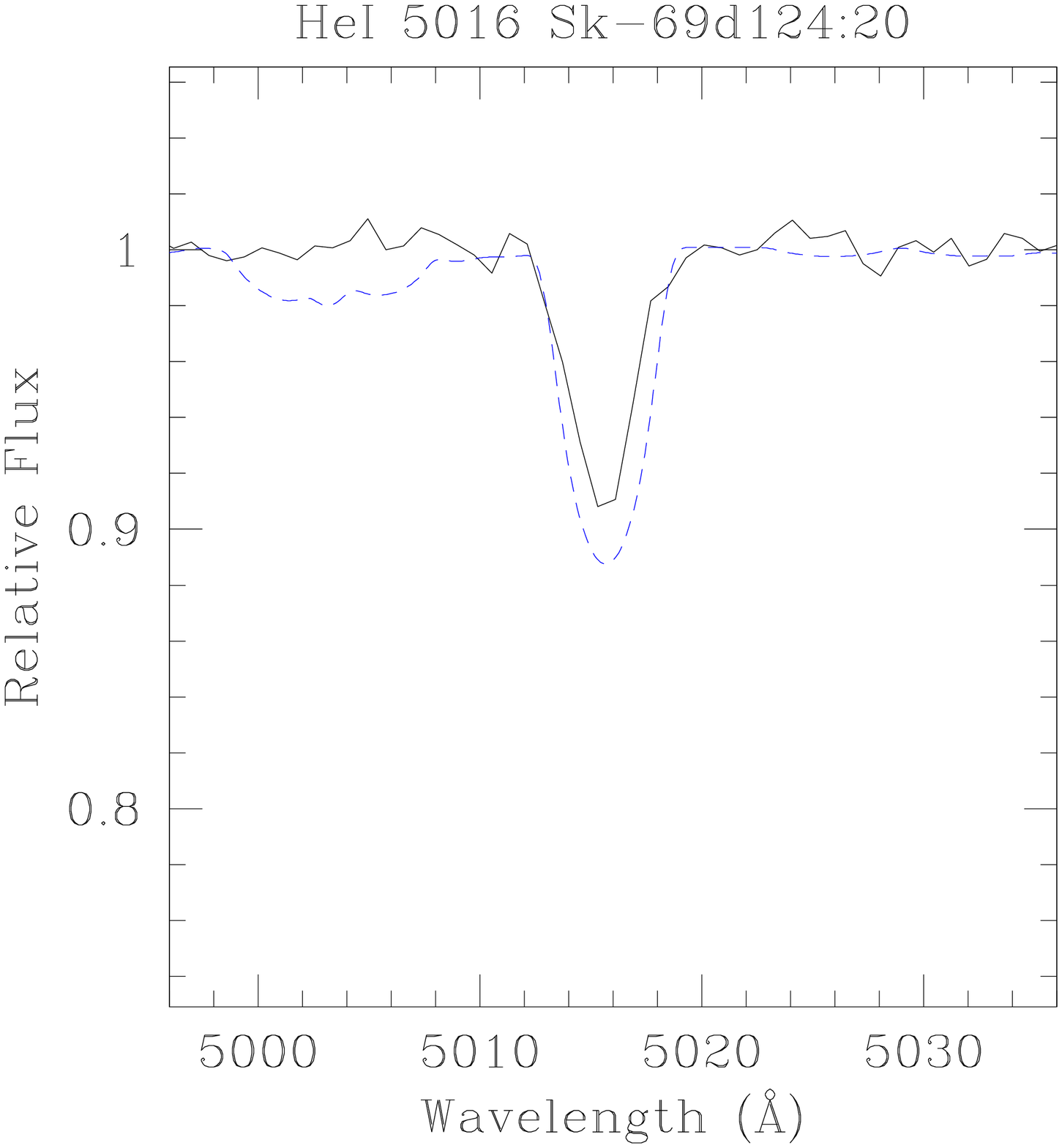}
\caption{\label{fig:MicroSingSk-69d124}The effect of  microturbulence on the He I singlets in Sk $-69^\circ$124, an O9.7 I star in the LMC.
Black shows the observed spectrum, the red line shows the \fastwind\ fit, and the dashed blue line shows the \cmfgen\ fit.  The upper three panels show the model profiles computed using the  ``standard" 10 km s$^{-1}$ microturbulent velocities, the middle three panels show that obtained using 15 km s$^{-1}$, and the bottom three panels show the model profiles obtained using 20 km$^{-1}$.}
\end{figure}

We find the He I singlet lines linked to the $2p$ (Figure~\ref{fig:Grot}) lower level (for instance,  He I $\lambda 4388$) are poorly fit
by \cmfgen\ for the late O-type stars, but somewhat better (but not spectacularly so) by \fastwind.  The He I singlet line
$\lambda 5016$, whose lower level is $2s$ (Figure~\ref{fig:Grot}), is however well modeled by \cmfgen.  
(The line is not included in the formal output
of \fastwind, a lack which should be addressed.)   
The He I triplets $\lambda 4471$ and $\lambda 4713$ are well modeled by \cmfgen\ with a microturbulance of 10
km s$^{-1}$, but with \fastwind\ a value of 15-20 km s$^{-1}$ is needed to produce a strong enough He I $\lambda 4471$ line for  the
late-type O supergiants.   
Increasing the microturbulence velocity to 15 km s$^{-1}$ for the late-type O supergiants in \cmfgen\ results in excellent fits not only
for He I $\lambda 4471$ and $\lambda 4713$ but also for $\lambda 5876$.   Including the He I $\lambda 5876$ line in 
\fastwind would help check for consistency in choosing a microturbulence velocity.    To summarize:
\begin{enumerate}
\item Both codes do well at fitting the He I singlets and triplets using 10 km s$^{-1}$ microturbulence (\cmfgen) and 0 km s$^{-1}$ (\fastwind)for the hotter
O supergiants and dwarfs.
\item For the late-type O supergiants, a value of 15 km s$^{-1}$ produces good fits for the He I triplet lines both by \cmfgen\ and \fastwind. 
\item For the late-type O supergiants, \cmfgen\ does a poor job of fitting the He I singlets coupled to the $2p$ lower level (i.e., He I $\lambda 4388$ and $\lambda 4922$) but does very well with He I $\lambda 5016$, whose lower level is $2s$. \fastwind\ does a somewhat better job for the $2p$ He I singlets, but is not immune from
the problem. 
\end{enumerate}

There is, in general, very good agreement between the physical properties of massive O stars derived by modeling the same data both by \fastwind\ and \cmfgen.  There is no significant difference in the mean or median effective temperature found, although there is 
quite a bit of scatter. More interesting, perhaps, is the systematic 0.1~dex difference in the $\log g$'s obtained by our fits using the two programs, with \fastwind\ producing a lower surface gravity.  We demonstrate that this 0.1~dex difference is quite significant to the
mass discrepancy problem.  The \cmfgen\ spectroscopic masses are in better agreement with the evolutionary masses, while \fastwind's spectroscopic masses tend to be too low, consistent with the long-standing mass discrepancy.

\acknowledgements
Support for program number HST-AR-11270 was provided by NASA through a grant from the Space Telescope Science Institute, which is operated by the Association of Universities for Research in Astronomy, Incorporated, under NASA contract NAS5-26555.  Partial support was also provided under the National Science Foundation from grant AST-1008020. We're grateful to Vincent Chomienne and Georges Meynet for making the $z=0.006$ Geneva evolutionary models available to us prior to publication, and
to Knut Olsen for providing the calculation on the expected radial velocity of Sk $-69^\circ$124.  An anonymous referee made many useful suggestions for improving the manuscript.  Our esteemed colleagues Deidre Hunter and Sumner Starrfield were kind enough to also provide critical readings and comments.

\appendix
\section{A Brief Discussion of \cmfgen\ and \fastwind}

Although both \cmfgen\ and \fastwind\ have been extensively described in the literature, we repeat some of the salient theoretical considerations here.

\subsection{\cmfgen}
\cmfgen is a general purpose non-LTE radiative transfer code whose properties have been outlined by Hillier (1990), Hillier \& Miller (1998), and Hillier \& Dessart (2012).  In the context of OB stars, it solves the time-independent spherical radiative transfer code in the co-moving frame subject to the constraints on the level populations imposed by the equations of statistical equilibrium. To simplify the atomic models, super-levels are utilized. Levels within a super-level are assumed to have the same departure coefficient (for a few cases a slightly more sophisticated assumption is utilized). The number of super-levels is model and ion dependent and is easily changed. To compute atomic level populations we adopt a Doppler profile of fixed width (usually defined by the microturbulent velocity).  The temperature is computed using the assumption of radiative equilibrium. The accuracy of the temperature structure is checked using flux conservation, and the electron-energy balance. The later two are not exactly satisfied because of numerical discretization errors, and the use of super-levels (Hillier 2003). To treat electron scattering we assume coherent scattering in the co-moving frame. Test models in which this assumption is relaxed (in \cmfgen\ only) show very similar spectra to models computed assuming coherent scattering.

\cmfgen\ computes the hydrostatic structure using an iterative procedure which is undertaken during the general \cmfgen\ procedure. For O stars, only a few hydrostatic iterations are needed to obtain a well converged hydrostatic structure. The initial hydrostatic structure is generally taken from an earlier model (but can be generated), although in earlier versions of \cmfgen\ we often utilized the density structure from a \tlusty\ (Lanz \& Hubeny 2003)  model. We attach a velocity law to the hydrostatic structure at 0.5 times the isothermal sound speed, although this was recently adjusted to 0.75 times the isothermal sound speed. Below the sonic point,
the $v dv/dr$ term (see, for example, Lamers \& Cassinelli 1999)  is taken into account. While possible, a contribution to the hydrostatic structure from turbulent pressure was not taken into account in the models presented in this paper.

With only a few exceptions, atomic data and super-level assignments are made available to \cmfgen\ via ASCII  files. For many species, more than one data set is available. Photoionization cross-sections are smoothed by a Gaussian (FWHM of 3000\,\kms) although for most species this can be varied. For dipole-forbidden transitions, without published data, we adopt a fixed collision strength of 0.1.  In general, D. J. H.  has not found major changes in the He I triplet or H lines when different atomic data for Fe has been adopted, although this may be regime dependent. Level dissolution is taken into account for levels higher than $n_{\hbox{\small eff}} > 2 z$ where $z$ is the effective charge on the core (e.g., 3 for C III). Collision rates for He I are from Berrington \& Kingston (1987). For H we have three different data sets, but for the present calculations we utilized collision strengths adapted from
Mihalas et al.\ (1975), which are similar to those utilized in \tlusty. 

Atomic data is taken from a wide variety of sources with the major sources being NIST (Ralchenko et al.\ 2010, Kframida et al.\ 2012), the Opacity Project (Seaton 1987, Cunto et al.\ 1993), Bob Kurucz\footnote{http://kurucz.harvard.edu/atoms.html},  and Sultana Nahar's NORAD web site\footnote{http://www.astronomy.ohio-state.edu/$\sim$nahar/nahar\_radiativeatomicdata/index.html}.

Once the atmospheric structure and level populations are known, we undertake a spectrum calculation using {\sc cmf\_flux} (Busche \& Hillier 2005) --- this is typically done only for the observable wavelength range of 1000\AA\ to 4$\micron$. Stark broadening is considered for H and He, Voigt profiles are used for UV resonance lines, while pure Doppler profiles are adopted for all other lines. For this calculation, we compute the opacities and emissivities in the co-moving frame. We also compute the radiation field in the co-moving frame which allows us to compute the frequency-dependent emissivity due to electron scattering. To take into account thermal broadening due to the electron motions, we need to iterate -- only two iterations (if that) are needed for O stars. For the computation of the observer's frame spectrum, we adopt the usual $(p,z)$ coordinate system. Along each ray, we have enough points (typically every 0.25 Doppler widths) to fully sample the rapidly varying line opacity and emissivity, and to adequately sample optical depth space. We use the same spatial grid for all frequencies. The opacity and emissivity are taken from the co-moving frame calculation, and are interpolated into the observer's frame.

The H and He atomic models used in \cmfgen\ can be briefly described as follows. 
H I has 30 levels with 20 super levels; its super levels begin at level 16.
Similarly, He II has 30 levels with 22 super levels; its super levels begin at level 21.
The He I model atom has 69 levels (up to $n=20$). Levels with $n < 8$ have their $l$ states split,
except  that $l=3$ and higher $l$ states are treated as a single state. Singlet and triplet states are
distinct in the model atom, and 45 super levels were used with the present calculations. 

In general, \cmfgen\ provides excellent matches to observed optical and UV spectra as can be seen, for example, in fits to O supergiants (Bouret et al.\ 2012), O dwarfs in the Galaxy and SMC (Martins et al.\ 2012, Bouret et al.\ 2013), and to the LBV AG Carinae (Groh et al.\ 2009).

\subsection{\fastwind}
\fastwind\ is specifically designed for modeling the optical and NIR spectra of O-type stars;
details are given in Santolaya-Rey et al.\ (1997), Puls et al. 2005, and 
Rivero Gonzalez et al.\ (2011b).  The latter reference is especially pertinent for the most
recent version of \fastwind, which has been used in the present study.
\fastwind\ consists of two program packages. The first one
calculates the atmospheric structure and the NLTE
occupation numbers. To set up the atmospheric stratification, \fastwind\ adopts a smooth transition between an analytical wind
structure, described by a {$\rm \beta$} velocity law (Castor \& Lamers 1979; for the actual formulation, see
Santolaya-Rey et al.\ 1997), and a
quasi-hydrostatic photosphere with a velocity law following from the
equation of continuity. The transition between wind and photosphere
can be specified by the user, with a default value (used here) of 0.1
of the local sound speed,
which results in a hydro-structure that compares
well with hydrodynamic models (e.g., Puls 2009). Also the photosphere
is calculated in spherical symmetry, i.e., allows for a potential
extension, and accounts for photospheric line and continuum
acceleration and a consistent temperature structure. In the first
iterations, the radiative acceleration is calculated from the
Rosseland mean, later updated by using the flux-mean from the current
NLTE opacities. The photospheric stratification is calculated without 
accounting for any turbulent pressure.

To solve the NLTE rate equations, a distinction is made between two
groups of elements, called explicit and background elements.

The explicit ones (mainly H and He but also, depending on the purpose
of the analysis, other elements such as N) are those to be used as
diagnostic tools. They are treated with high precision, i.e., by a
complete NLTE approach and a co-moving frame radiative transfer
(only Doppler-broadening), by means of an Accelerated
Lambda Iteration with an Accelerated Lambda Operator designed for
co-moving frame calculations (Puls 1991). The present version of \fastwind\ does not account for level dissolution. 

For these explicit elements, a detailed description of all required
atomic data (levels, transition types, cross-sections in various
parameterizations) needs to be provided in an external file, following
the so-called DETAIL input format (Butler \& Giddings 1985).

The background elements are used to obtain the background radiation
field, accounting for line-blocking and blanketing effects. The occupation
numbers for these elements are also calculated in NLTE, but using a
Sobolev line transfer for the bulk of weaker lines, and a co-moving
frame transfer for only those lines which are still strong in the wind. The
opacities and emissivities from these background elements, together with
those from the explicit ones, are used to calculate the radiation
field and the temperature structure.  The atomic data for the
background elements are provided in a fixed way, from the compilation
by Pauldrach et al.\ (1994, 2001)\footnote {In brief, the atomic
structure code {\sc superstructure} (Eissner \& Nussbaumer 1969; Eissner
1991) has been used to calculate all bound state energies in LS- and
intermediate-coupling, as well as related atomic data, particularly
oscillator strengths including those for stabilizing transitions.}.

The decisive, time-saving approximation within \fastwind\ regards the
calculation of the line opacities and emissivities required for the
calculation of the background radiation field. In brief, suitable
means for the line opacities/emissivities are used, averaged over a
frequency interval on the order of the terminal wind speed ($\sim$1000-500 km s$^{-1}$); see
Puls et al.\ (2005) for further details.

The temperature structure is calculated from the thermal
electron balance (e.g., Kub{\'a}t et al.\ 1999 and references therein),
accounting for all processes from explicit and background elements.
This is done in parallel withe iteration cycle rate-equations and radiative transfer.

The second program package allows to calculate the emergent
line profiles. The radiative line transfer is solved using the occupation
numbers from the explicit elements and the background
opacities/emissivities from the converged NLTE solution. The
integrations are performed in the observer's frame, and on a radial
micro-grid. Basic features are a separation between line and continuum
transport (see Santolaya-Rey et al.\ 1997 and references therein), and
various line broadening options (most important: Stark and Voigt
broadening). An external input file with all information needs to be
provided, notably the broadening functions or parameters.

Finally, let us briefly describe the H and He atomic models used in the
present work.  Our H I and He II models consist of 20 levels each
until principal quantum number $n$ = 20, and He I includes 49
levels until $n$ = 10, where levels with $n$ = 8{\ldots}10 have been
packed. 

H and He atomic data are similar to those used in DETAIL (see above),
and have been described, though for ``smaller" model atoms, in
Santolaya-Rey et al.\ (1997). The only exception are the hydrogen
collisional strengths, where the dataset provided by Giovanardi et al.\
(1987) is now used as a default, though other data are possible as
well. The effects of using different collision strengths have been
tested by Przybilla \& Butler (2004) and Najarro et al.\ (2011), and
both publications conclude that the differences are negligible for UV
and optical lines, but may become important at infrared wavelengths
(from the Paschen series on). 

The line-blanketed version of \fastwind\ has been used in with good success to model the physical parameters
of Galactic O stars in both the optical  (Repolust et al.\ 2004) and NIR (Repolust et al.\ 2005),
Magellanic Cloud O-type stars (e.g., Papers I-III),  B-type supergiants in M33 (Urbaneja et al.\ 2005), 
and a very low-metallicity Of type star in IC 1613 (Herrero et al.\ 2012), to name just a few applications.
Usage with an automatic fitting algorithm resulted in the analysis of many hundreds of stars as part
of the VLT-FLAMES project (e.g., Mokiem et al.\ 2006, 2007).

\clearpage
\clearpage
\clearpage
\clearpage
\clearpage
\clearpage
\clearpage
\clearpage
\clearpage
\clearpage

\clearpage

\clearpage
\clearpage
\clearpage
\clearpage
\clearpage
\clearpage
\clearpage
\clearpage
\clearpage
\clearpage
\begin{figure}
\epsscale{0.3}
\plotone{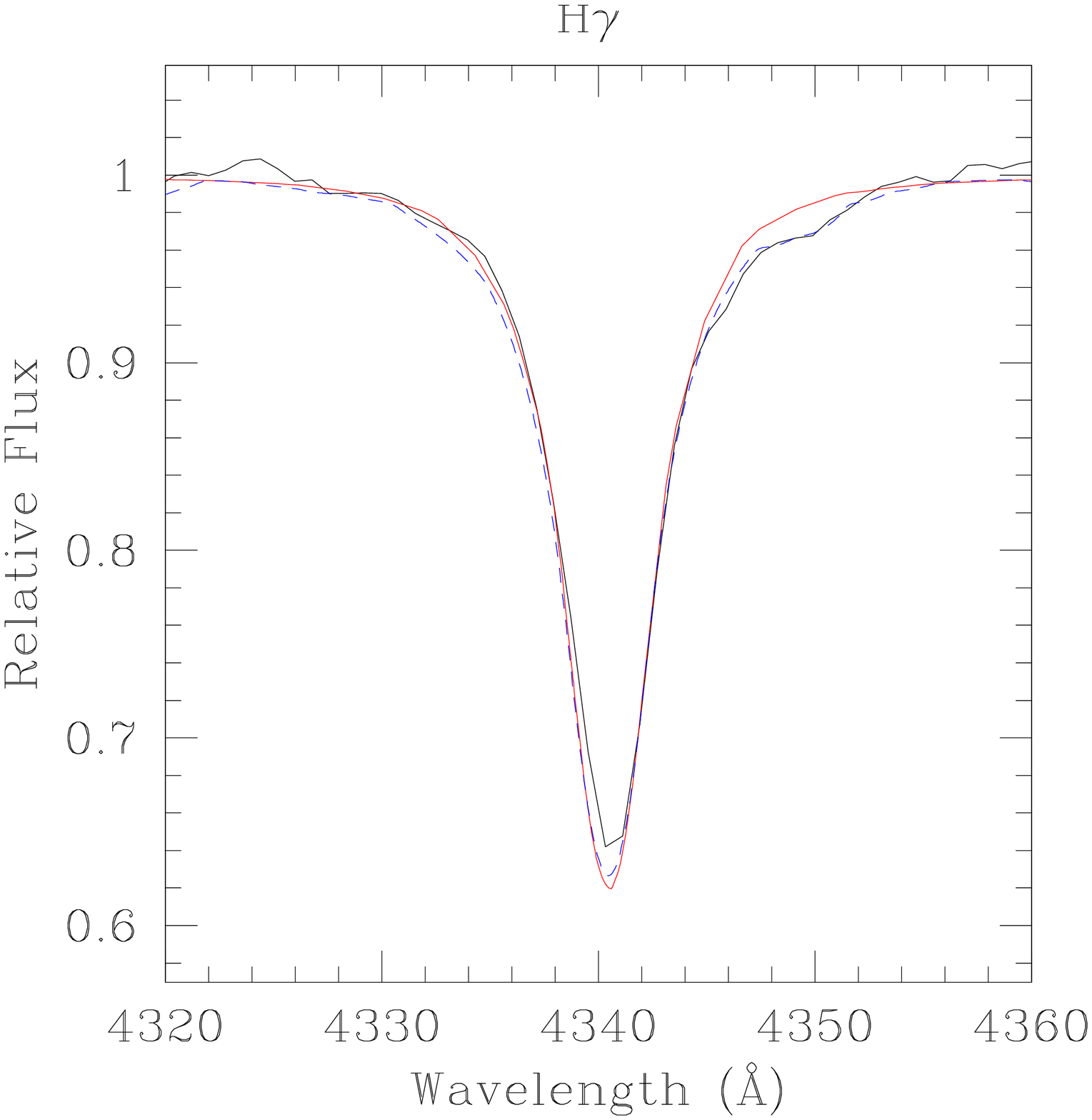}
\plotone{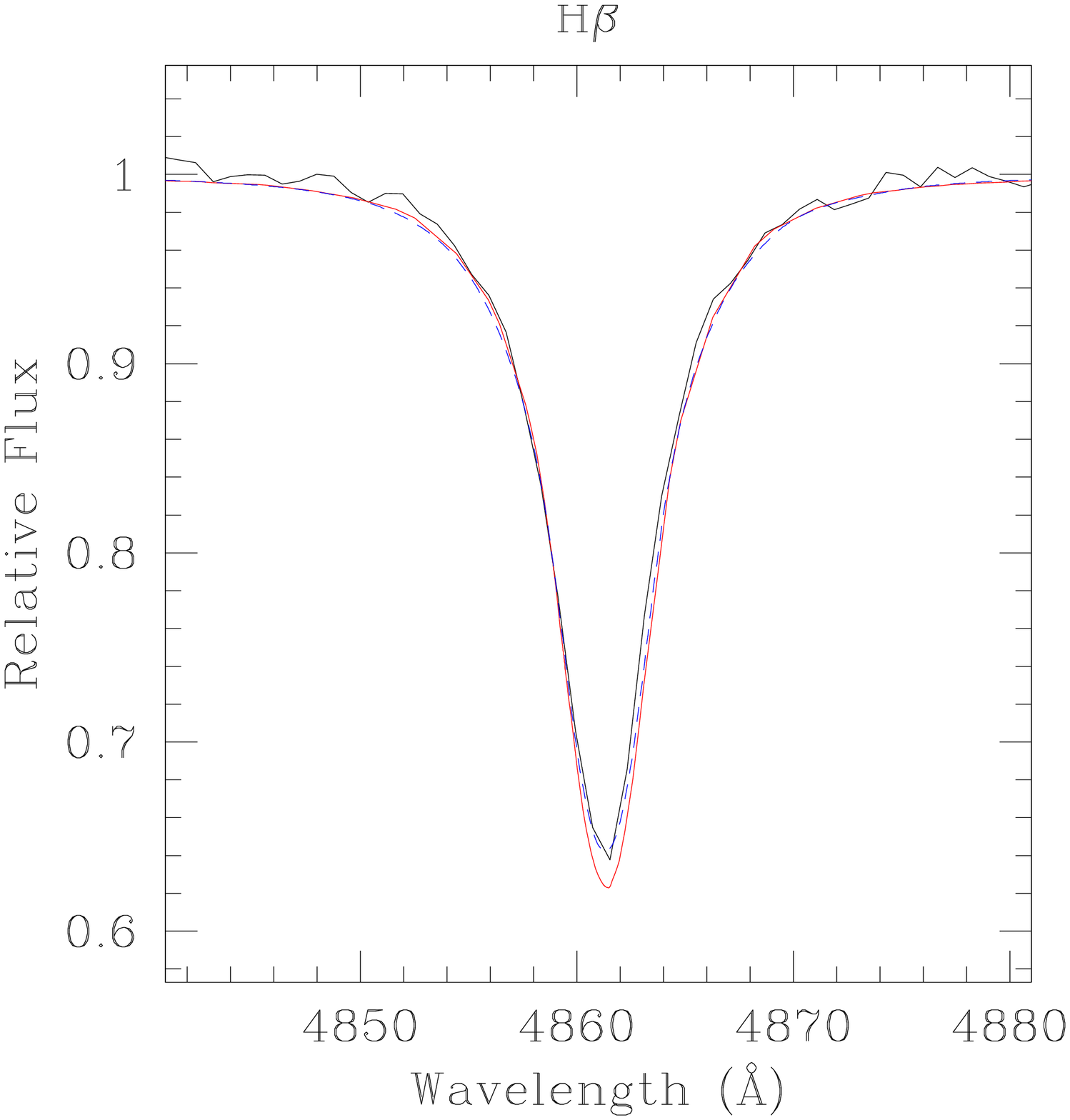}
\plotone{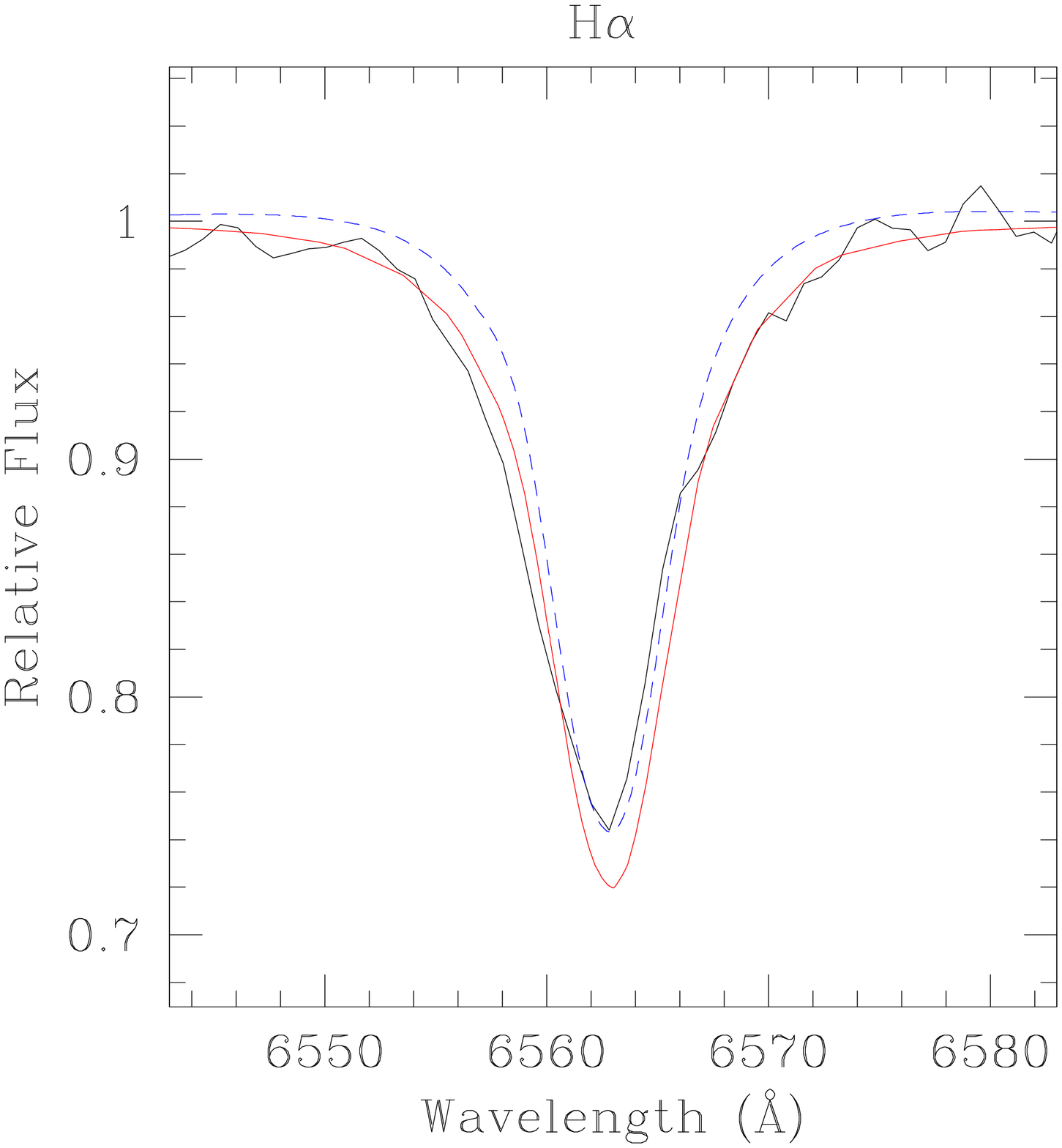}
\plotone{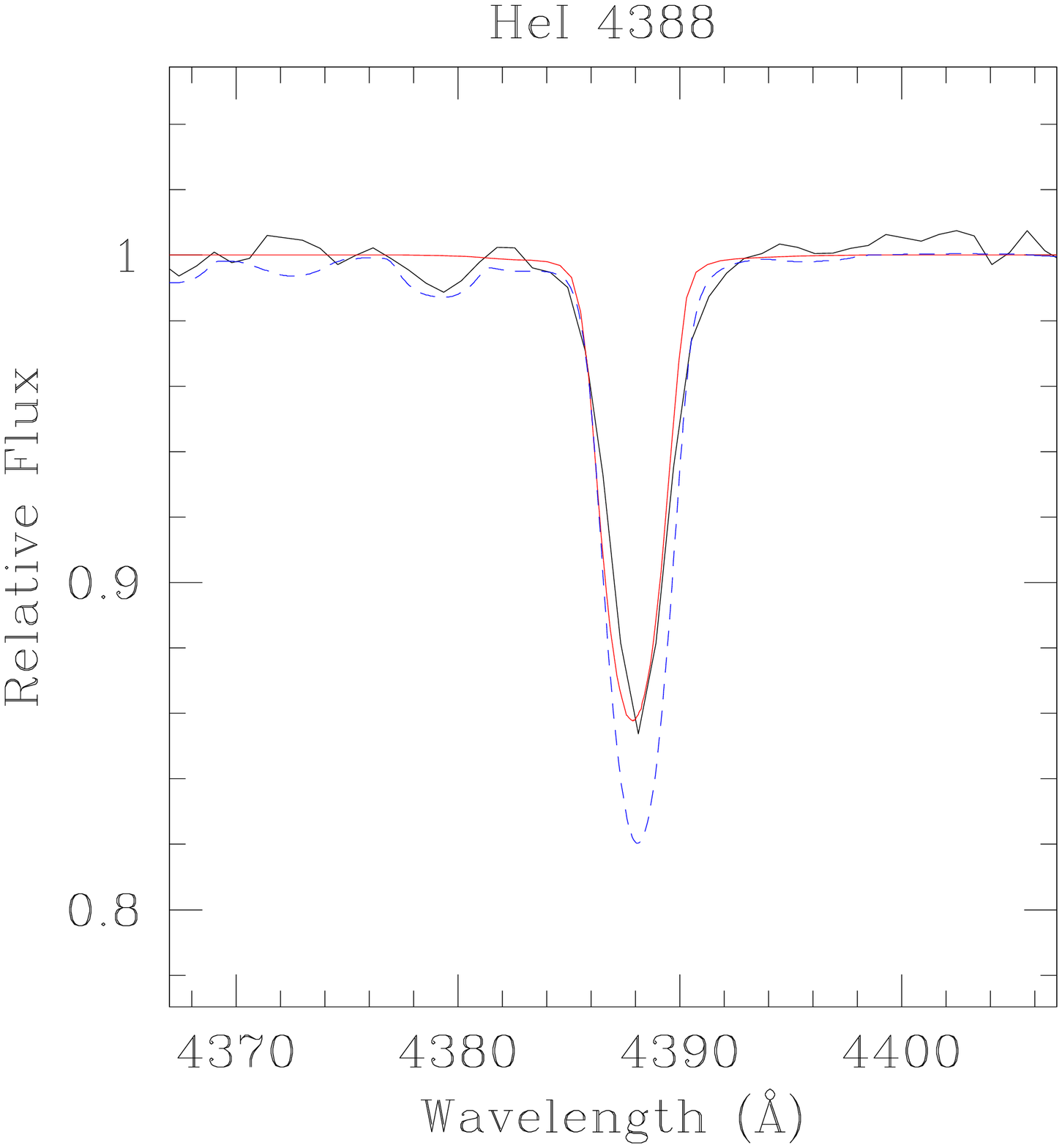}
\plotone{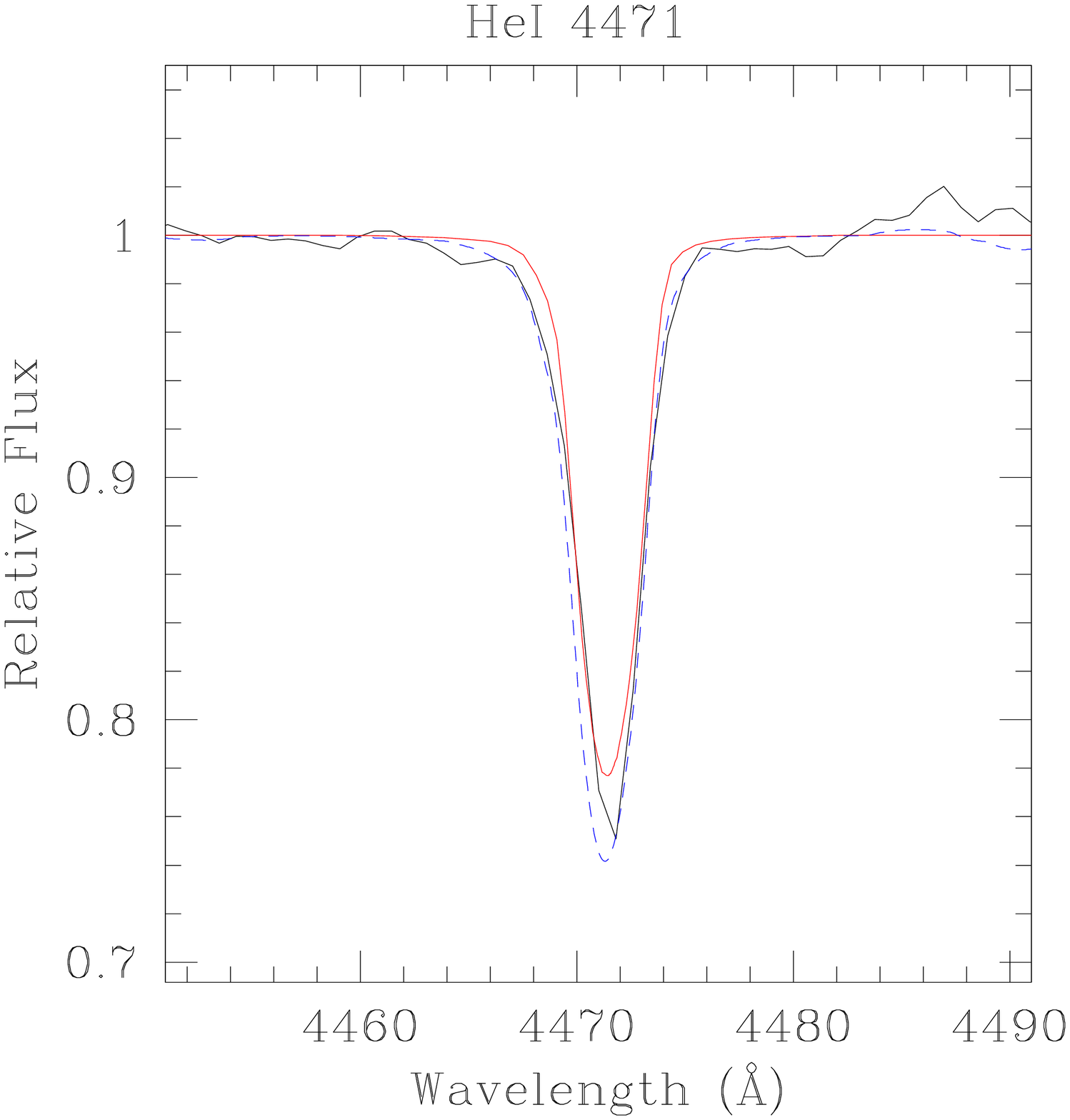}
\plotone{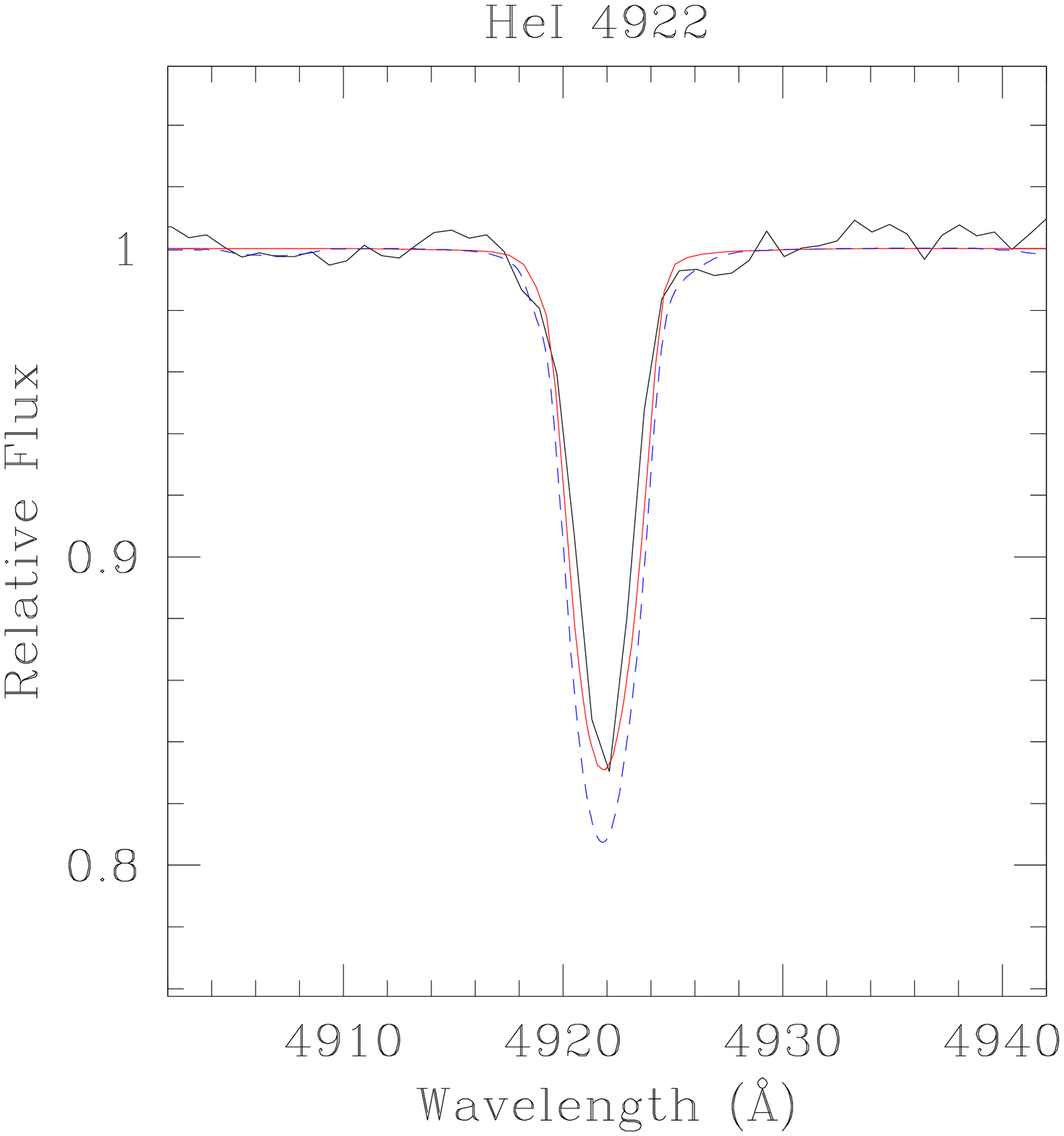}
\plotone{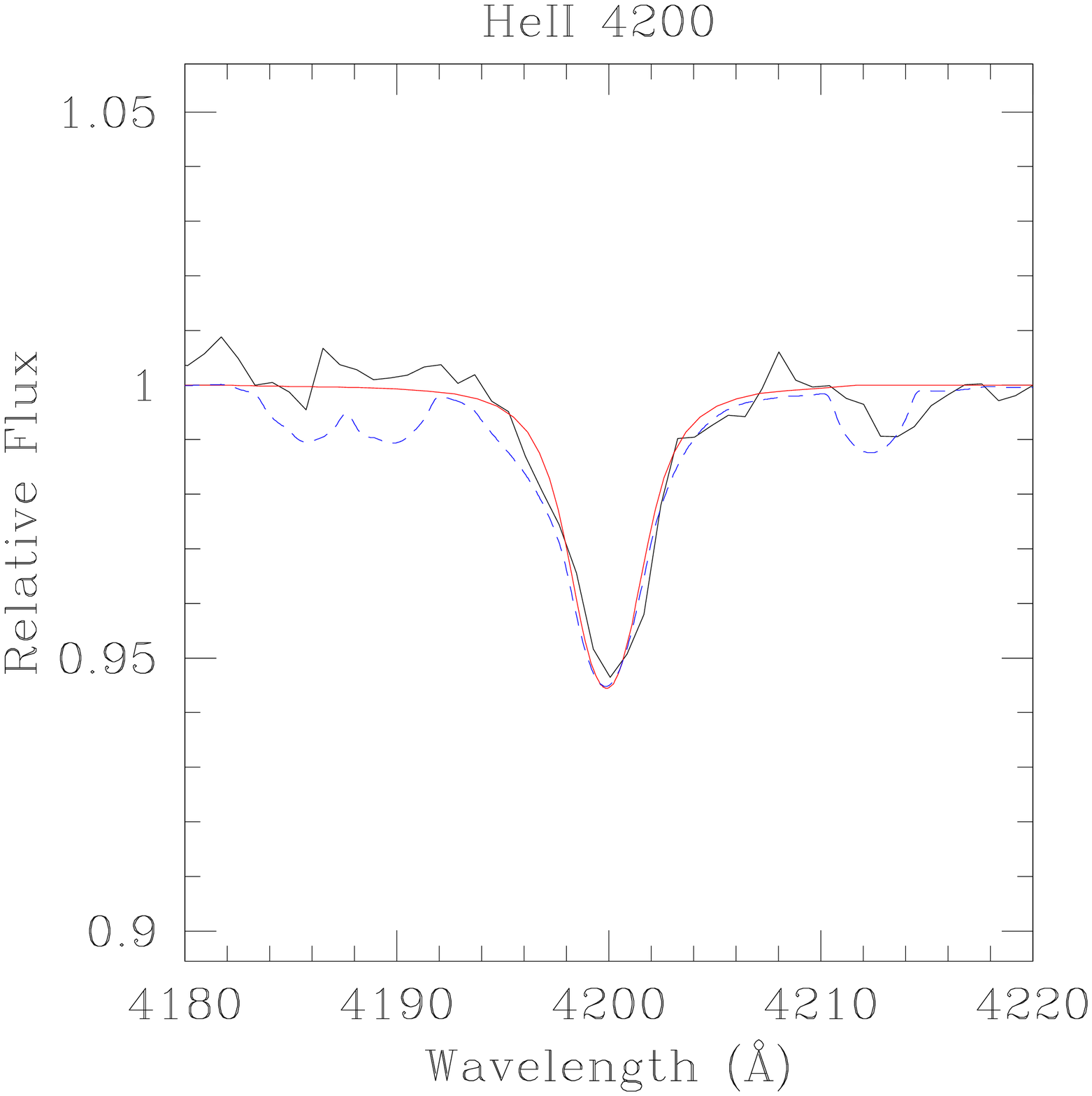}
\plotone{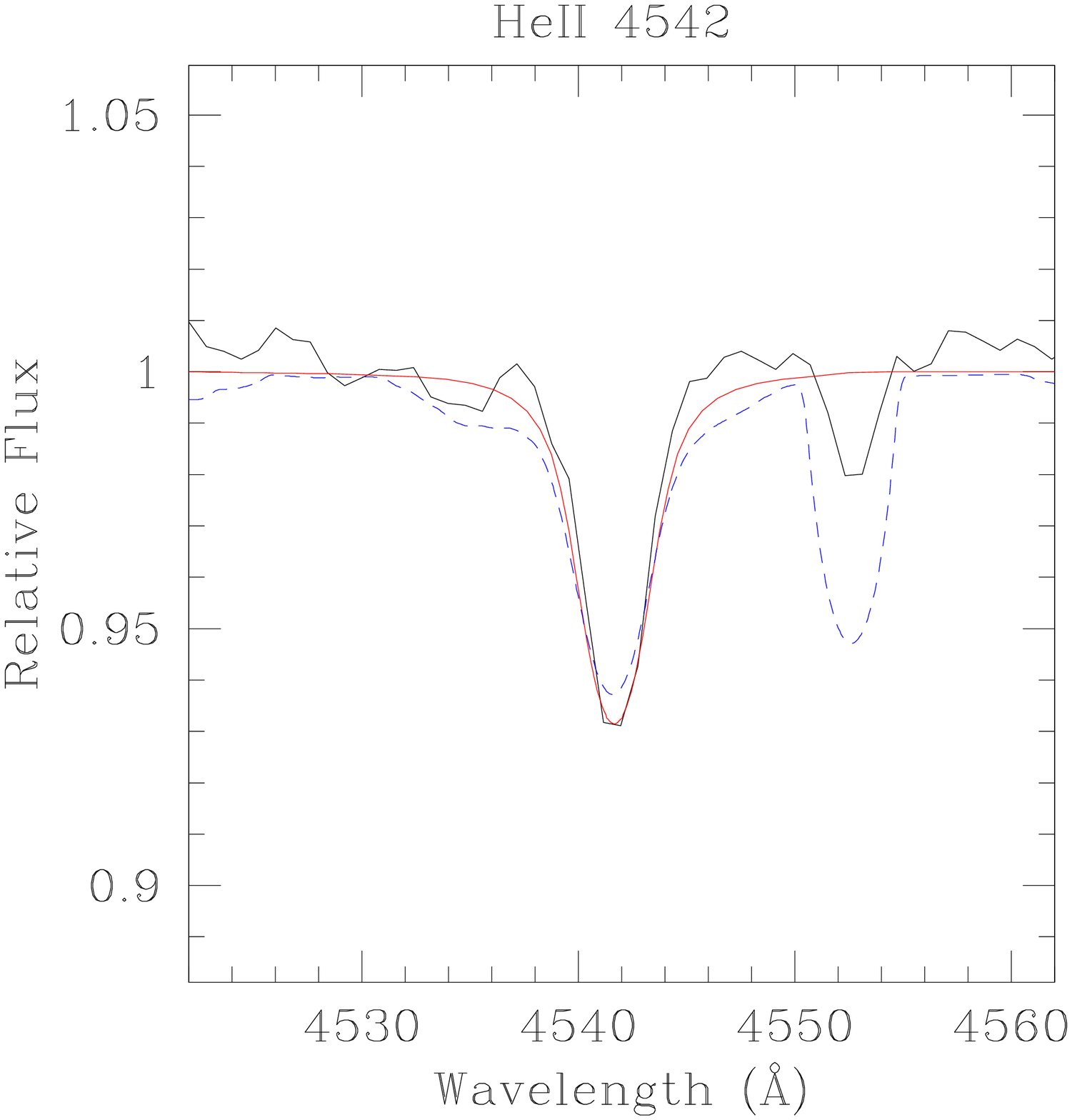}
\plotone{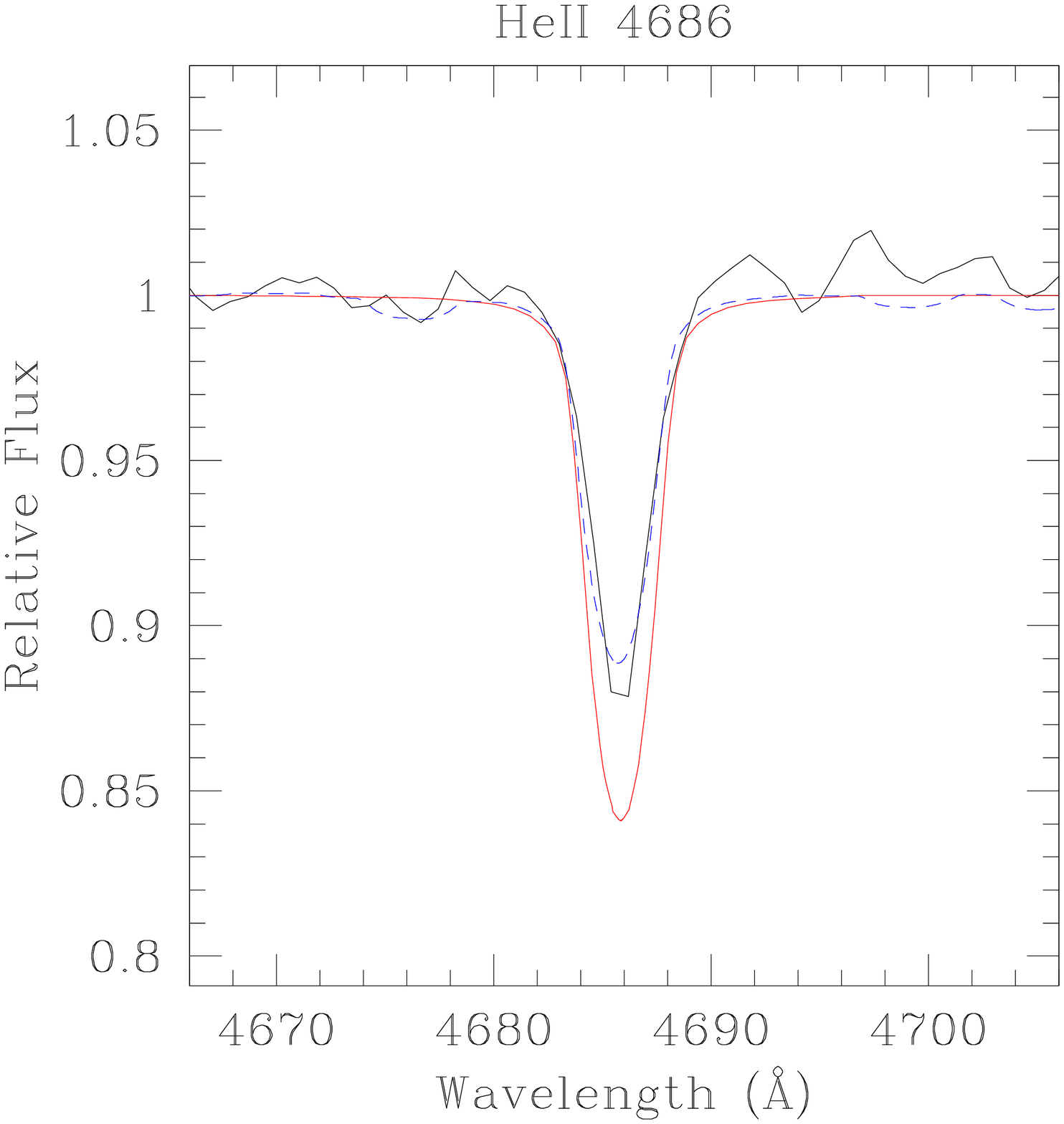}
\caption{\label{fig:AzV22315} Model fits for AzV 223, an O9.5 II star in the SMC computed with 15 km s$^{-1}$ microturbulence.  Black shows the observed spectrum, the red line shows the \fastwind\ fit, and the dashed blue line shows the \cmfgen\ fit.  Compare to Figure~\ref{fig:AzV223}.}
\end{figure}
\clearpage
\begin{figure}
\epsscale{0.3}
\plotone{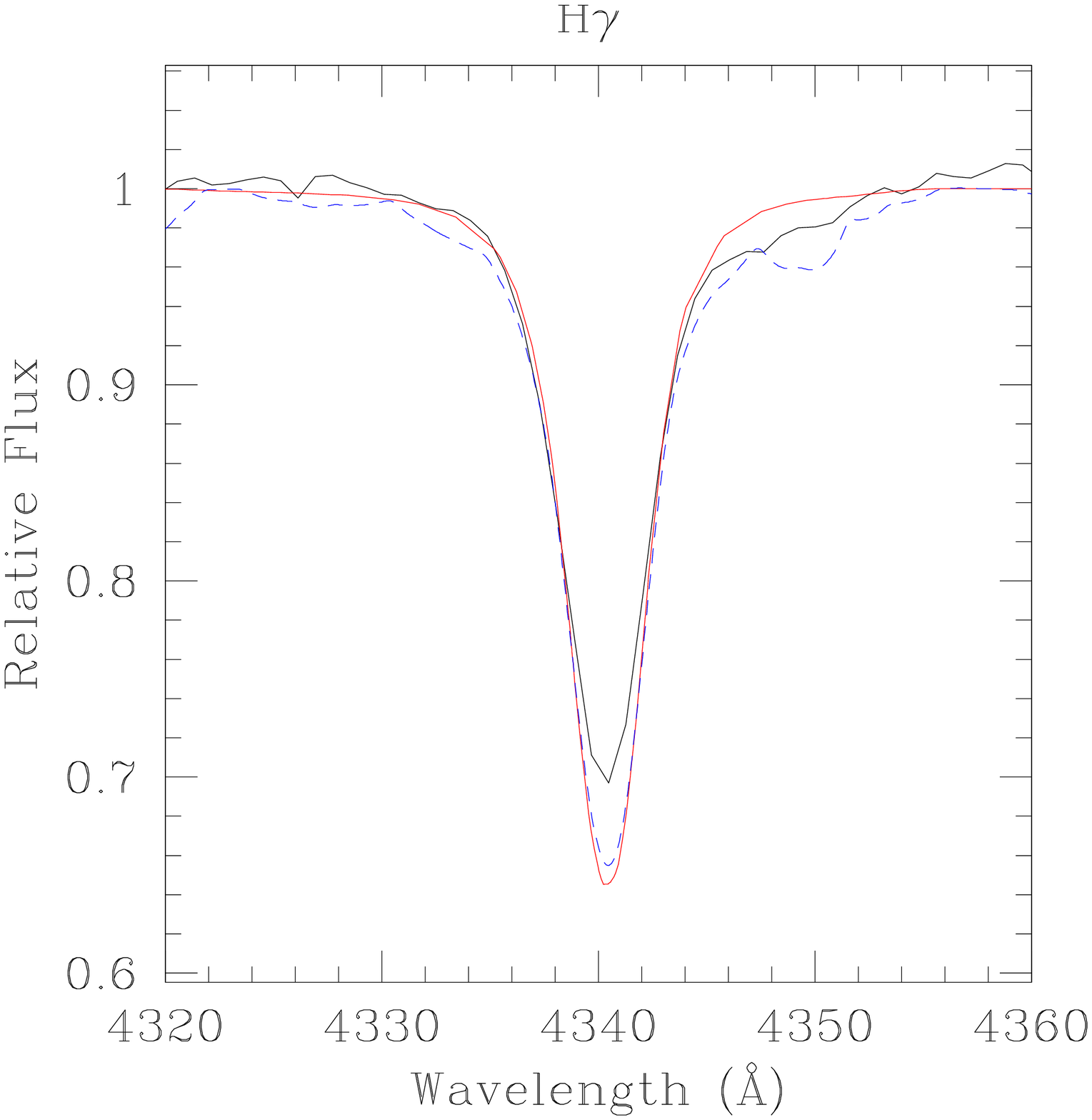}
\plotone{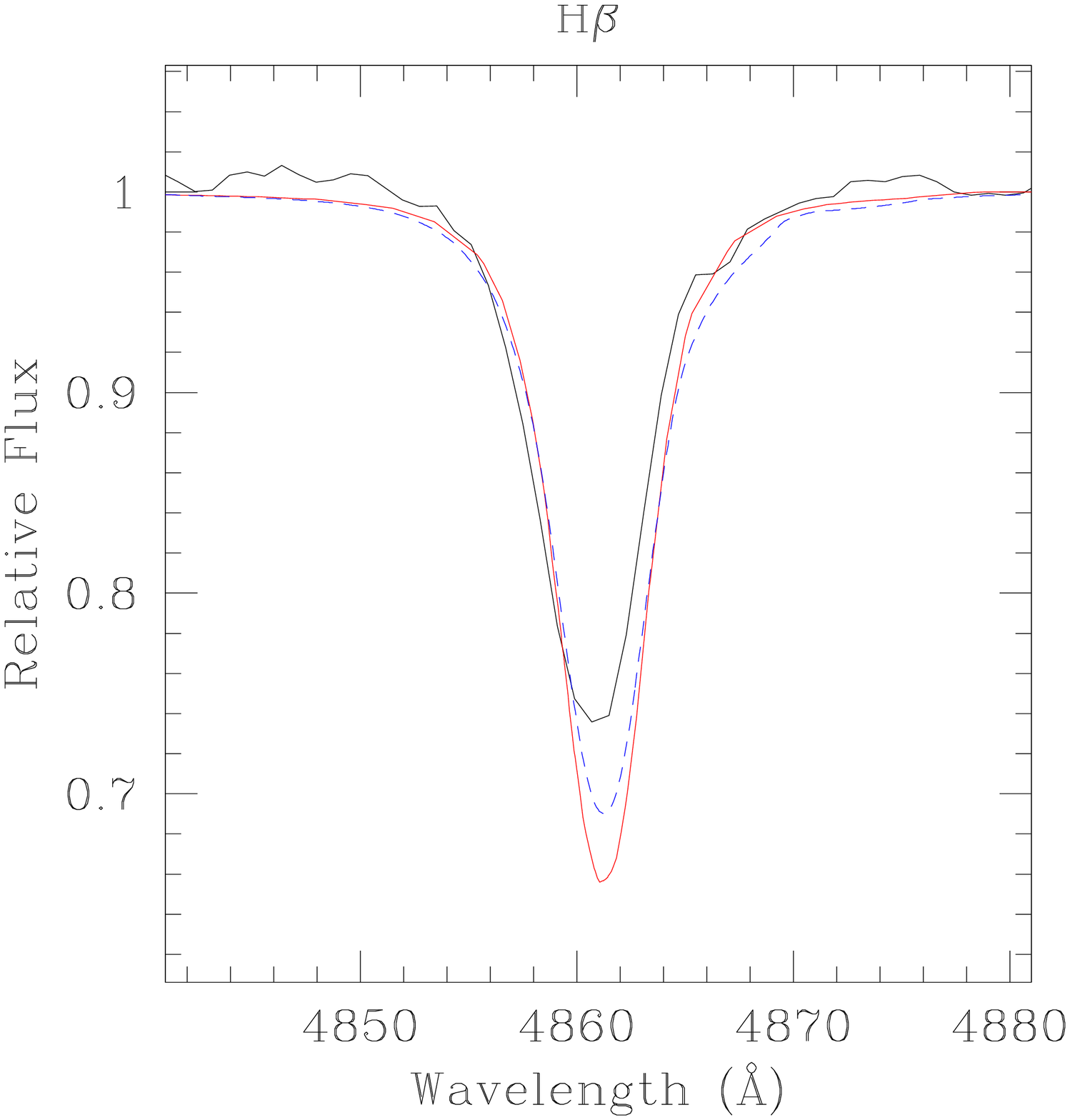}
\plotone{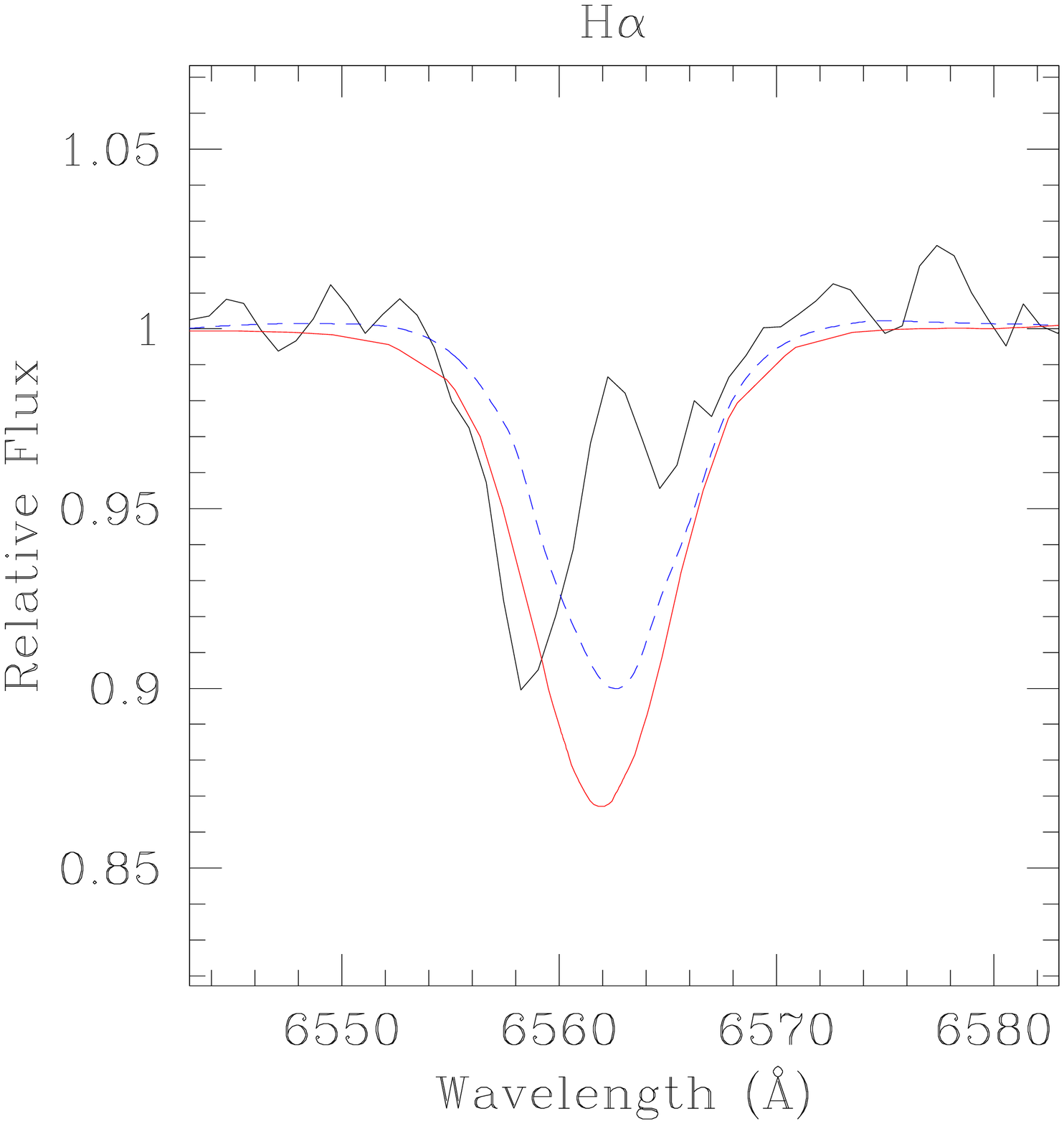}
\plotone{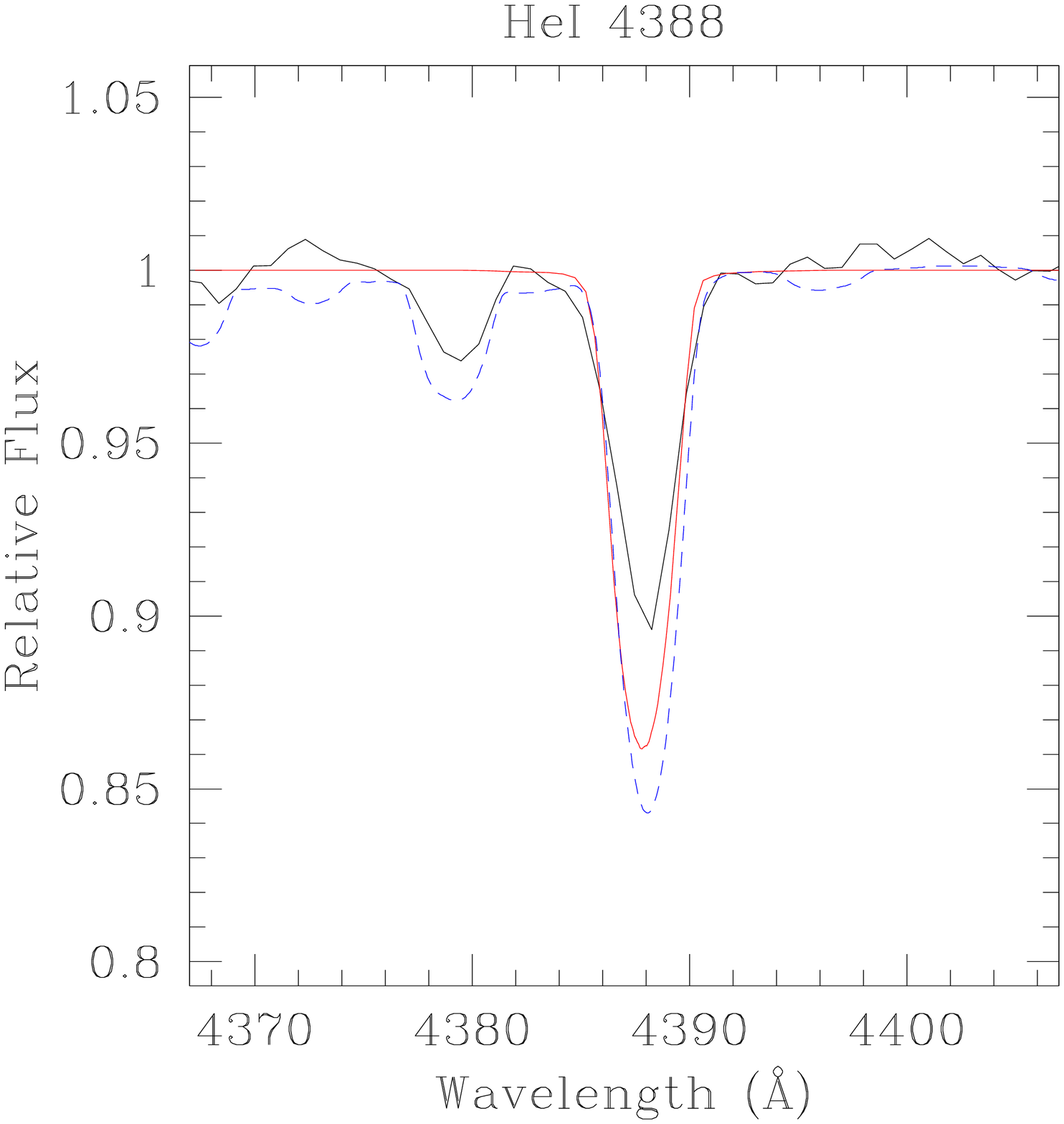}
\plotone{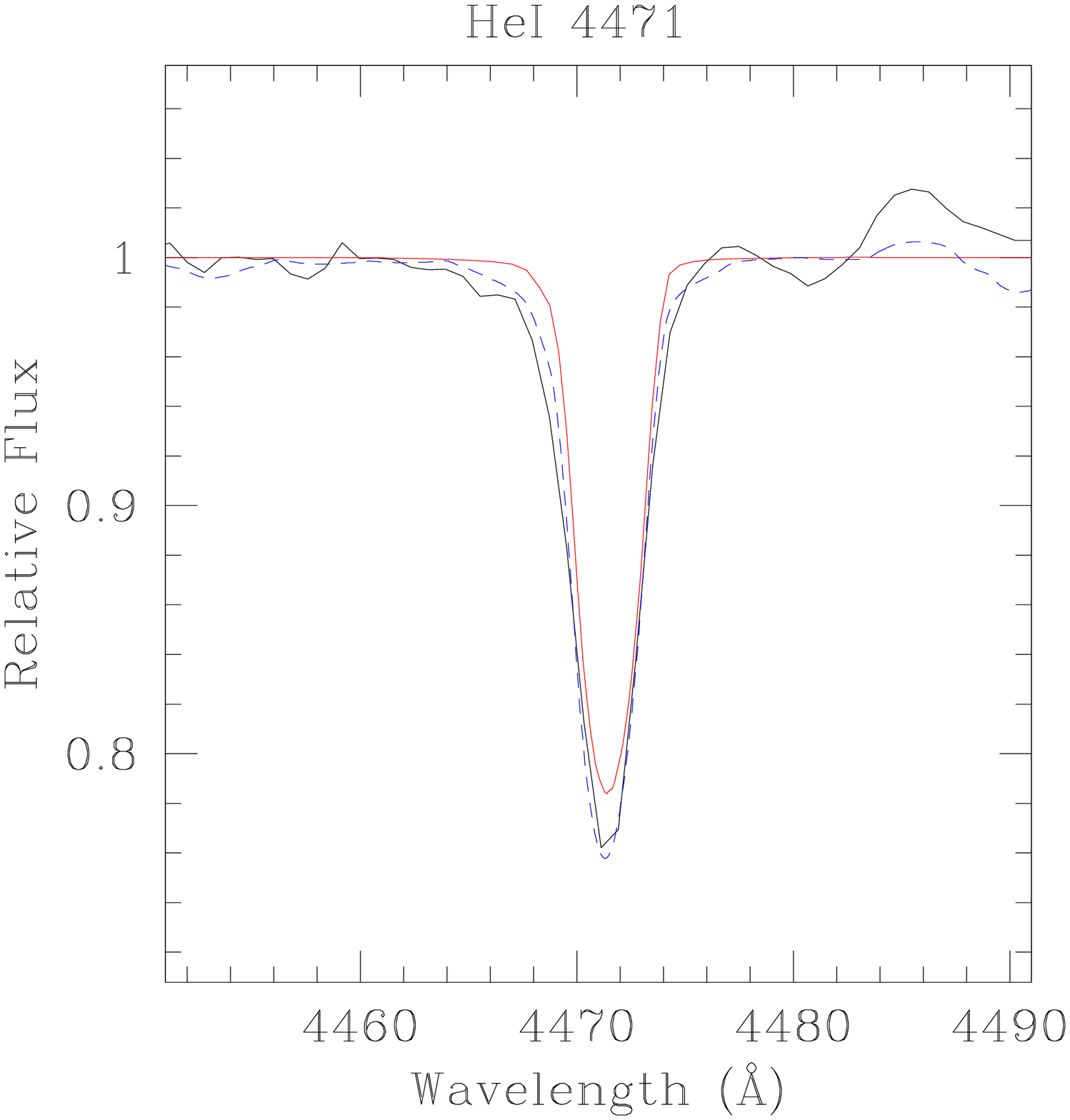}
\plotone{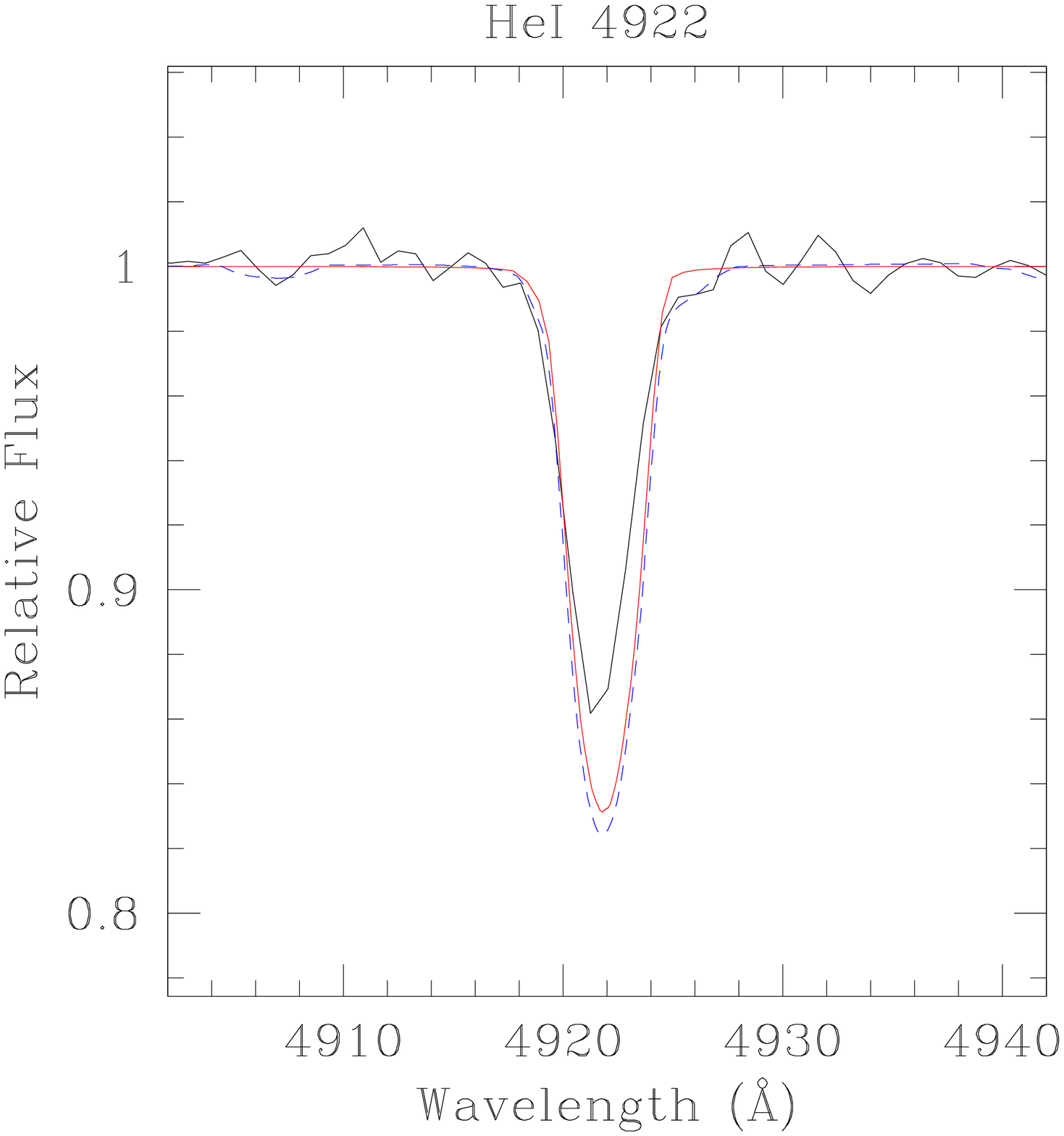}
\plotone{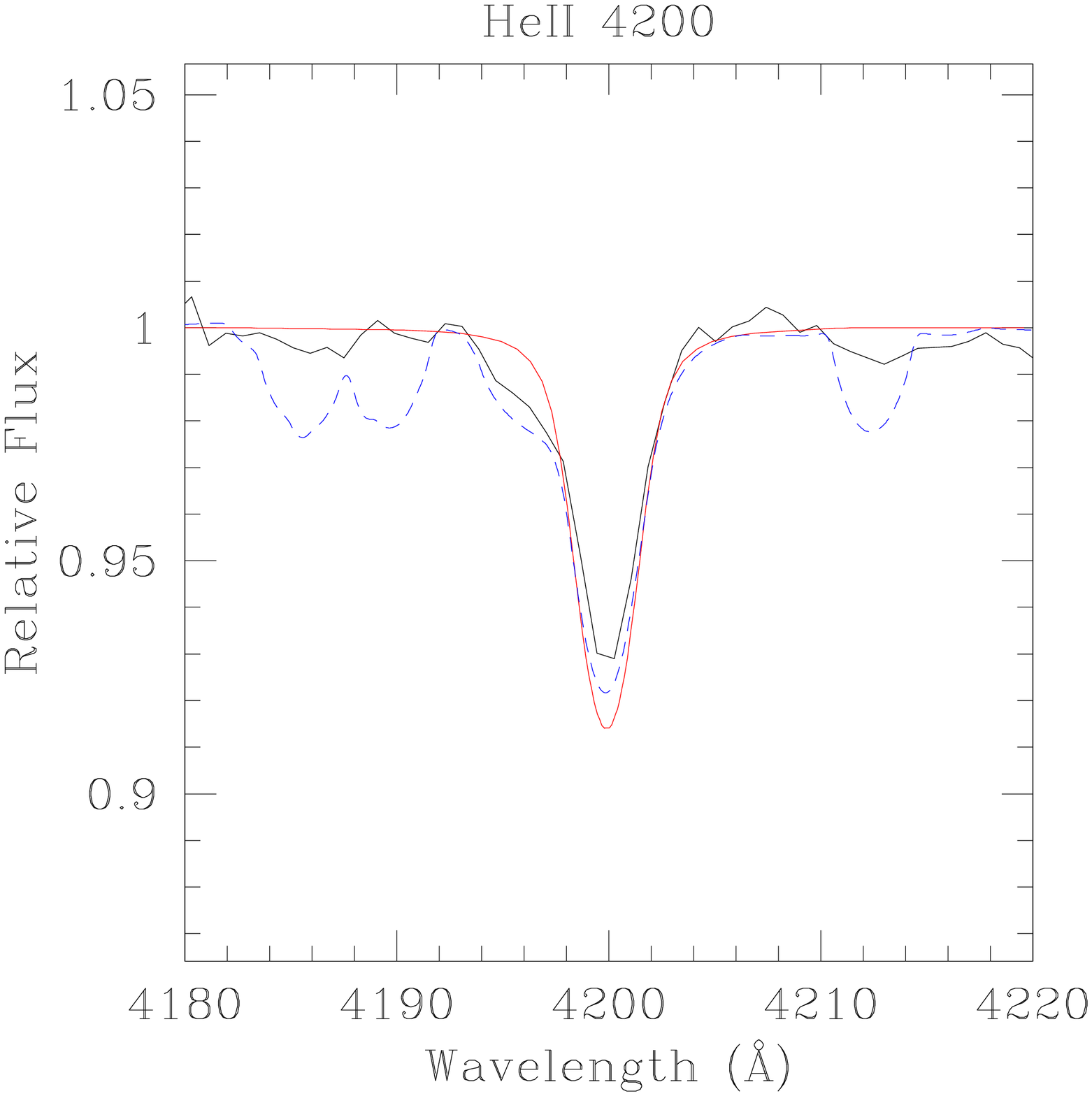}
\plotone{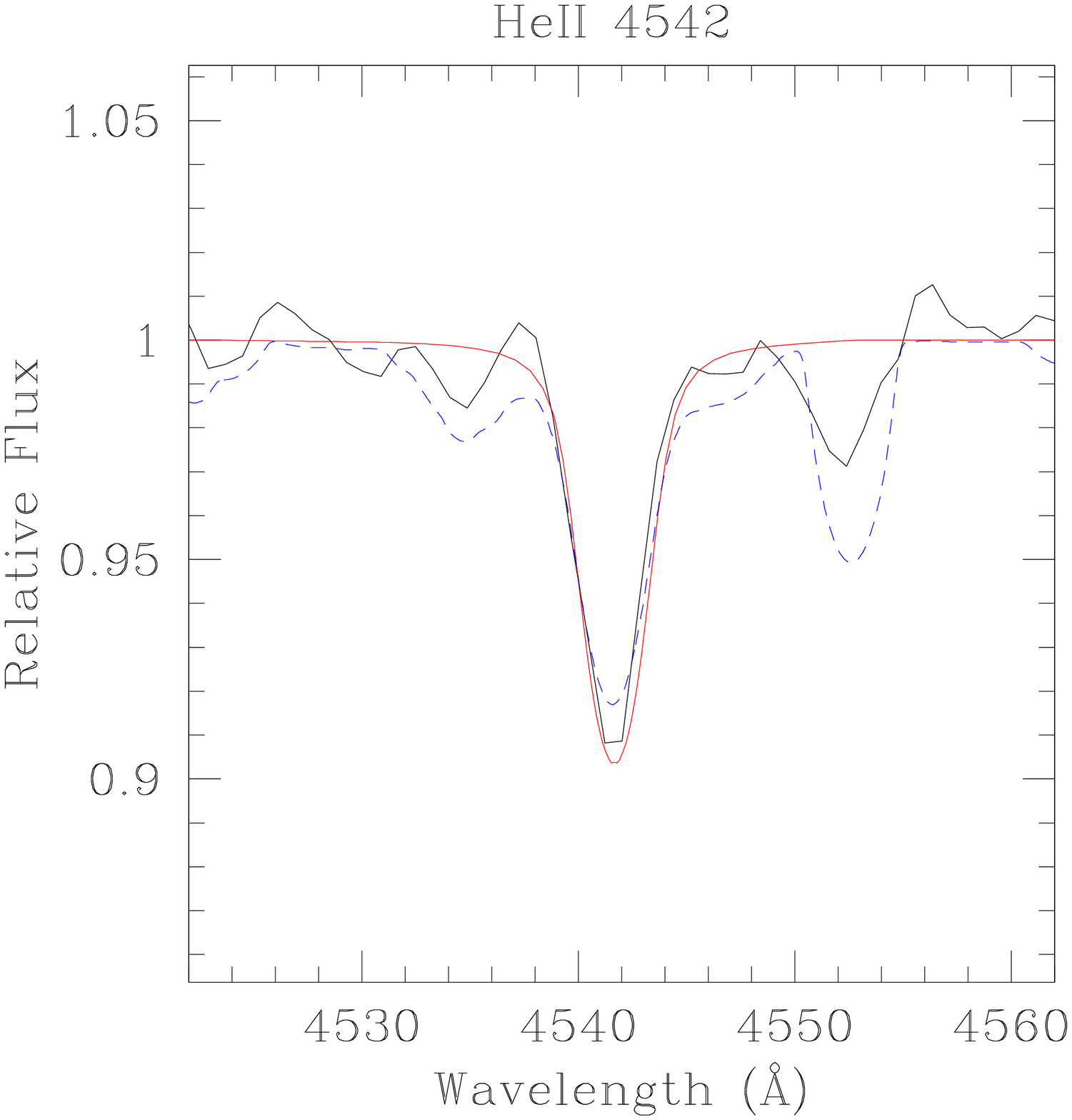}
\plotone{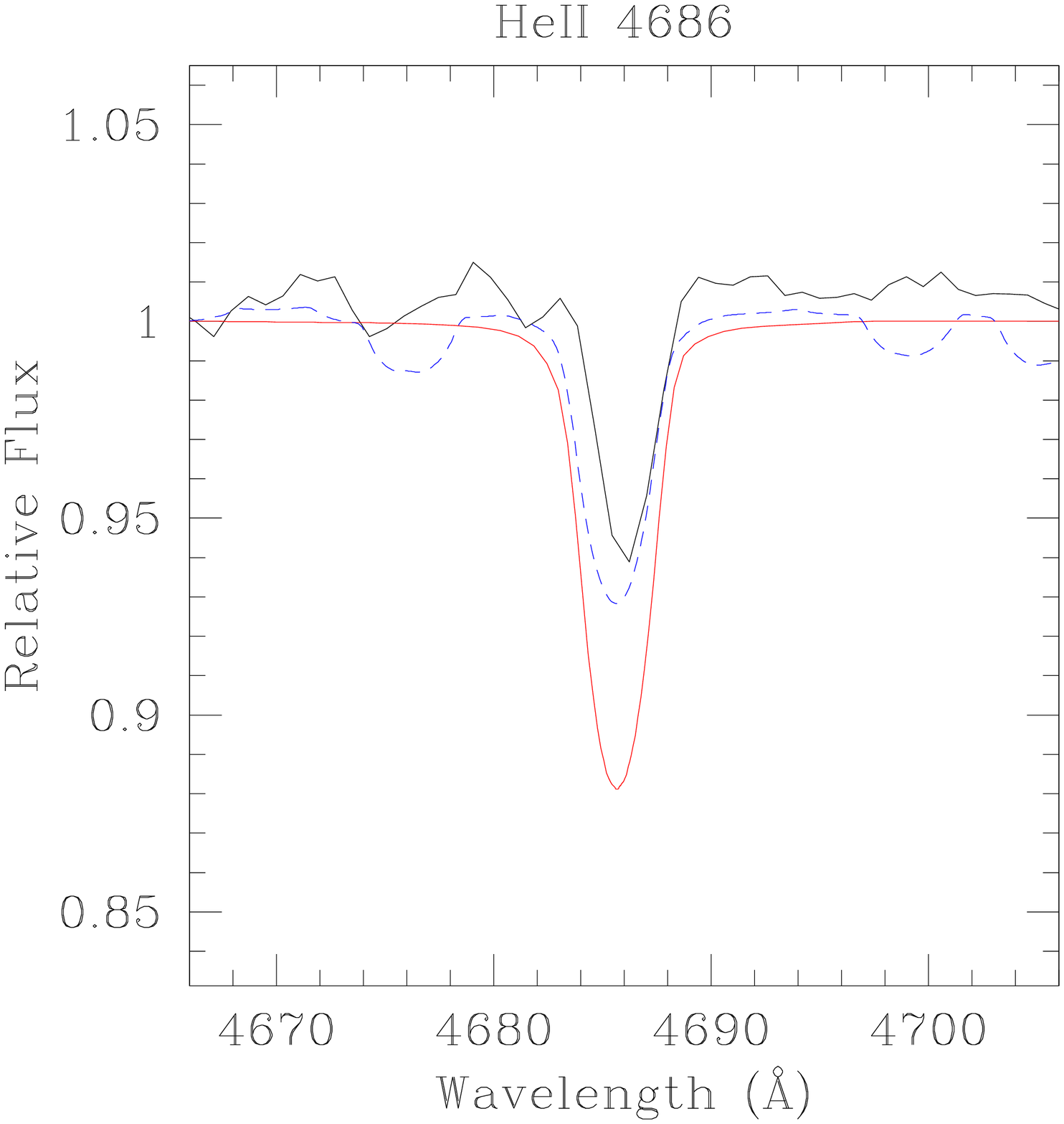}
\caption{\label{fig:BI17015} Model fits for BI 170, an O9.5 I star in the LMC computed with 15 km s$^{-1}$ microturbulence.   Black shows the observed spectrum, the red line shows the \fastwind\ fit, and the dashed blue line shows the \cmfgen\ fit. Compare to Figure~\ref{fig:BI170}.}
\end{figure}
\clearpage
\begin{figure}
\epsscale{0.3}
\plotone{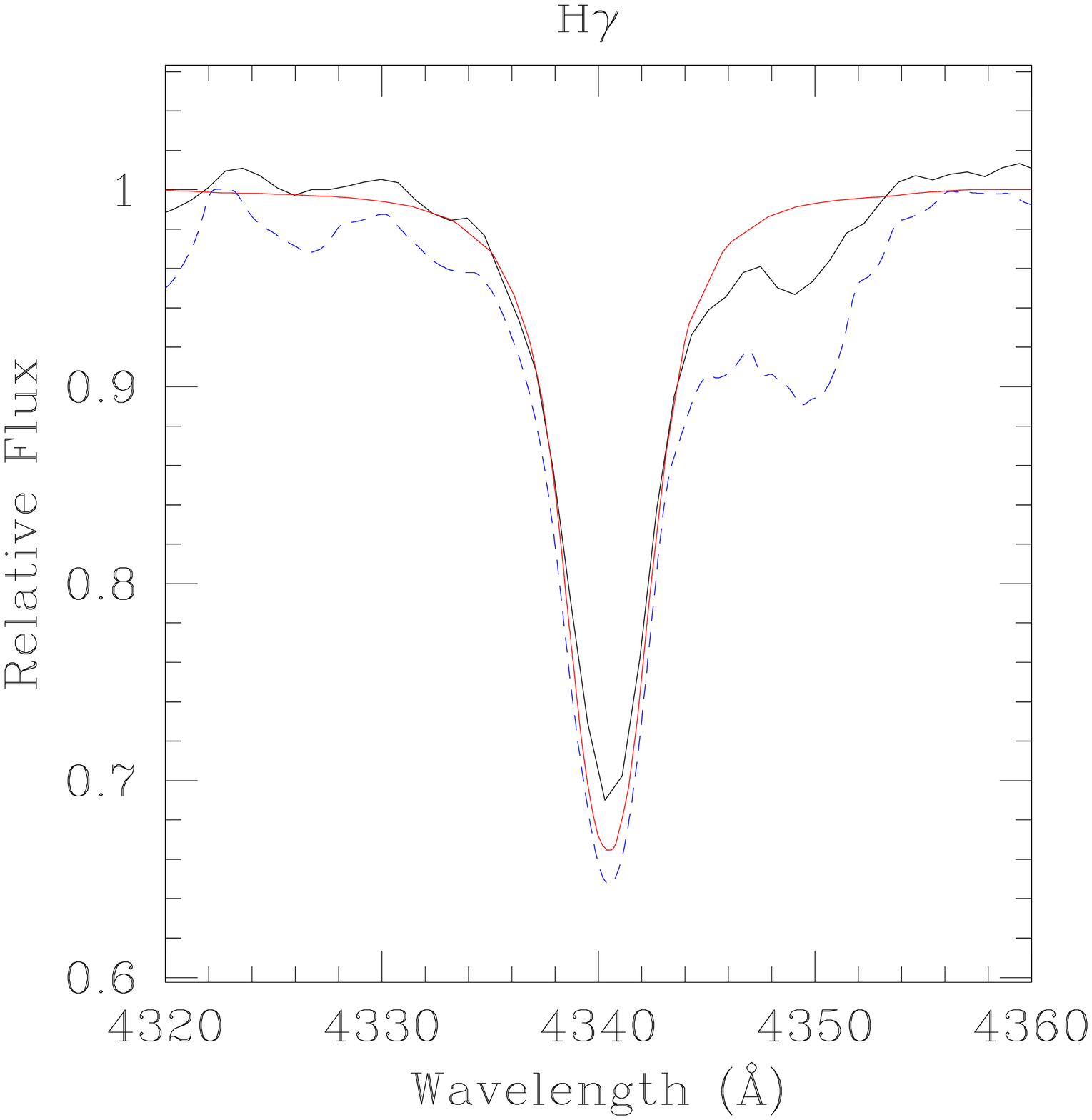}
\plotone{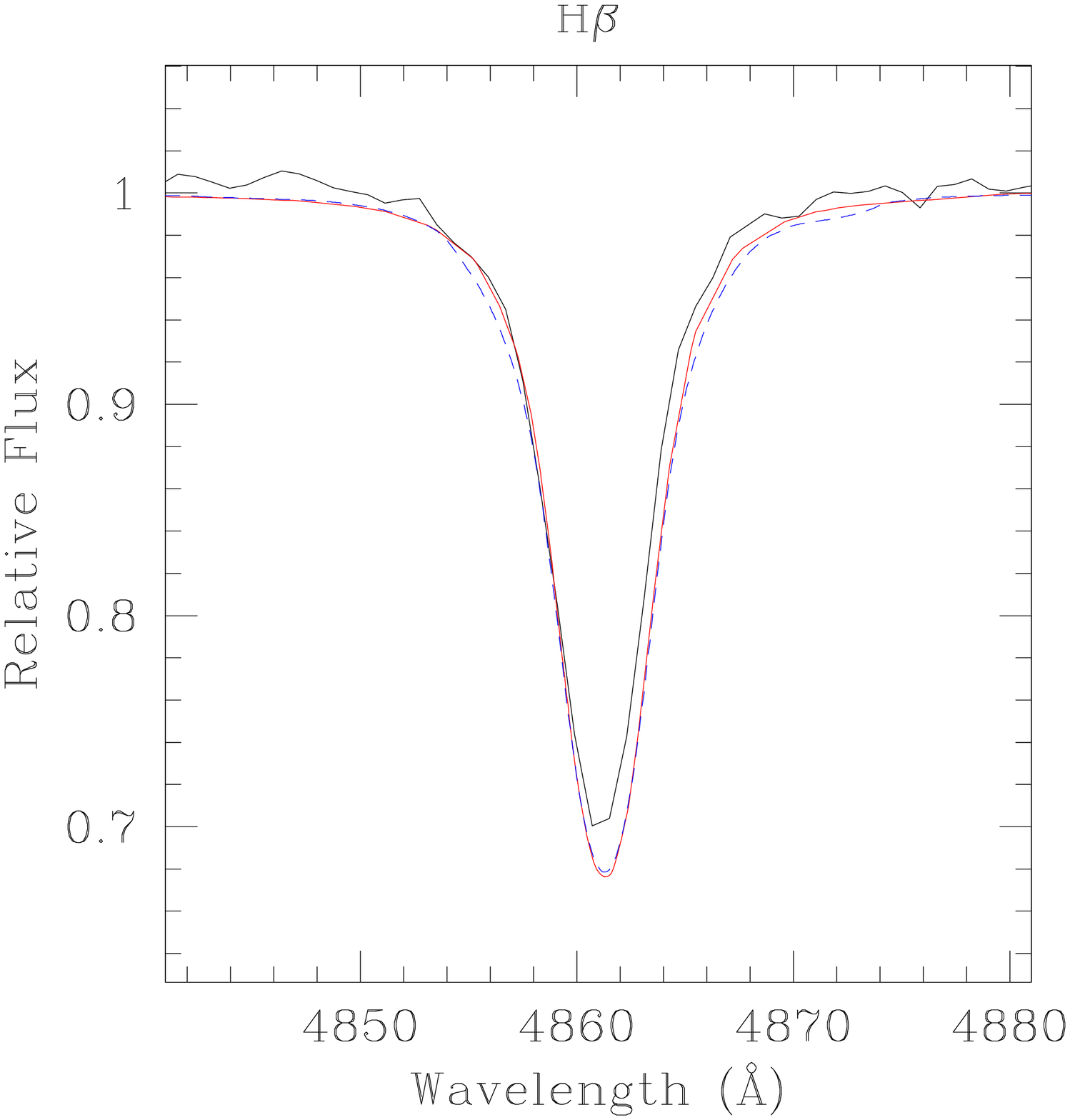}
\plotone{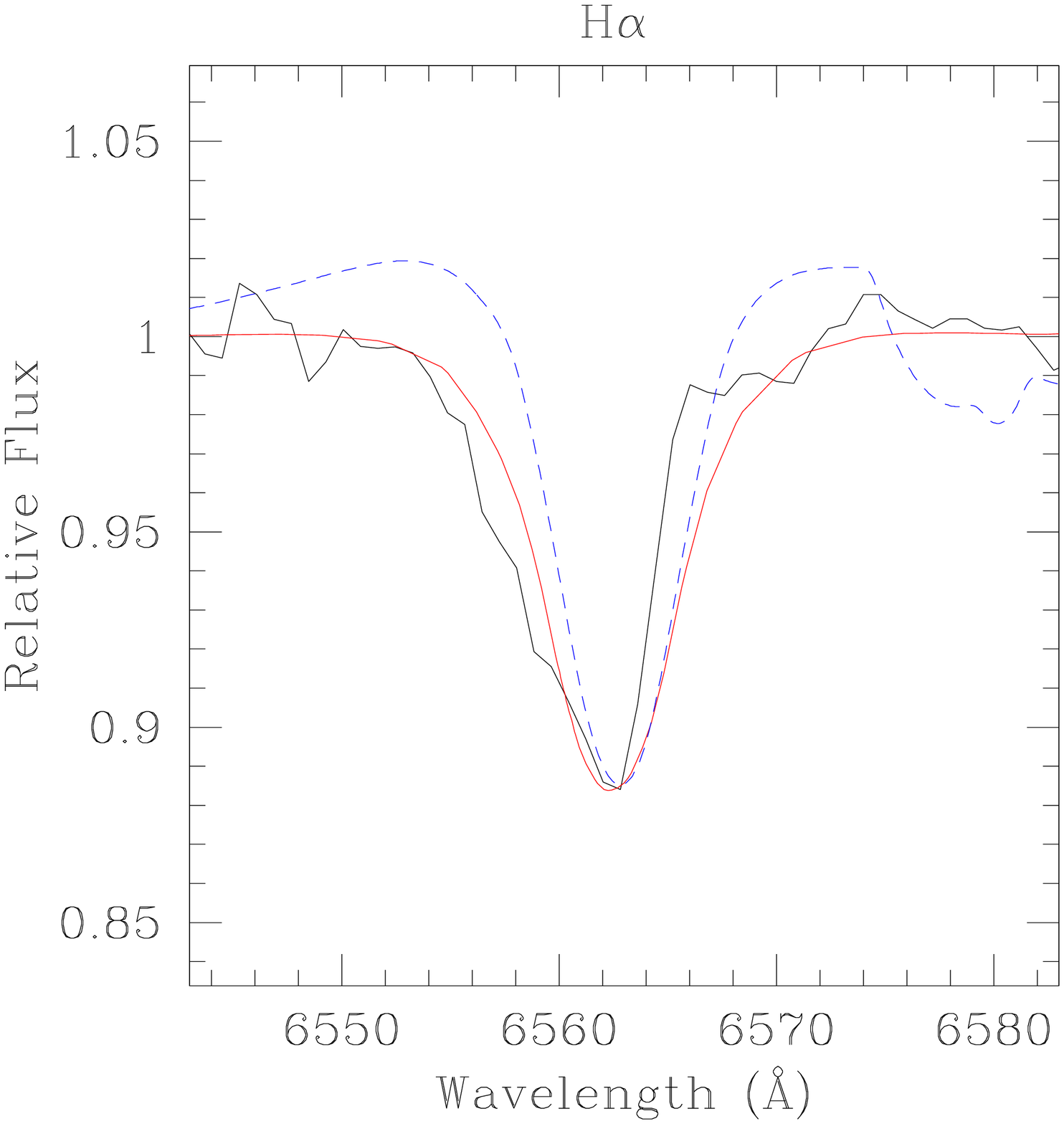}
\plotone{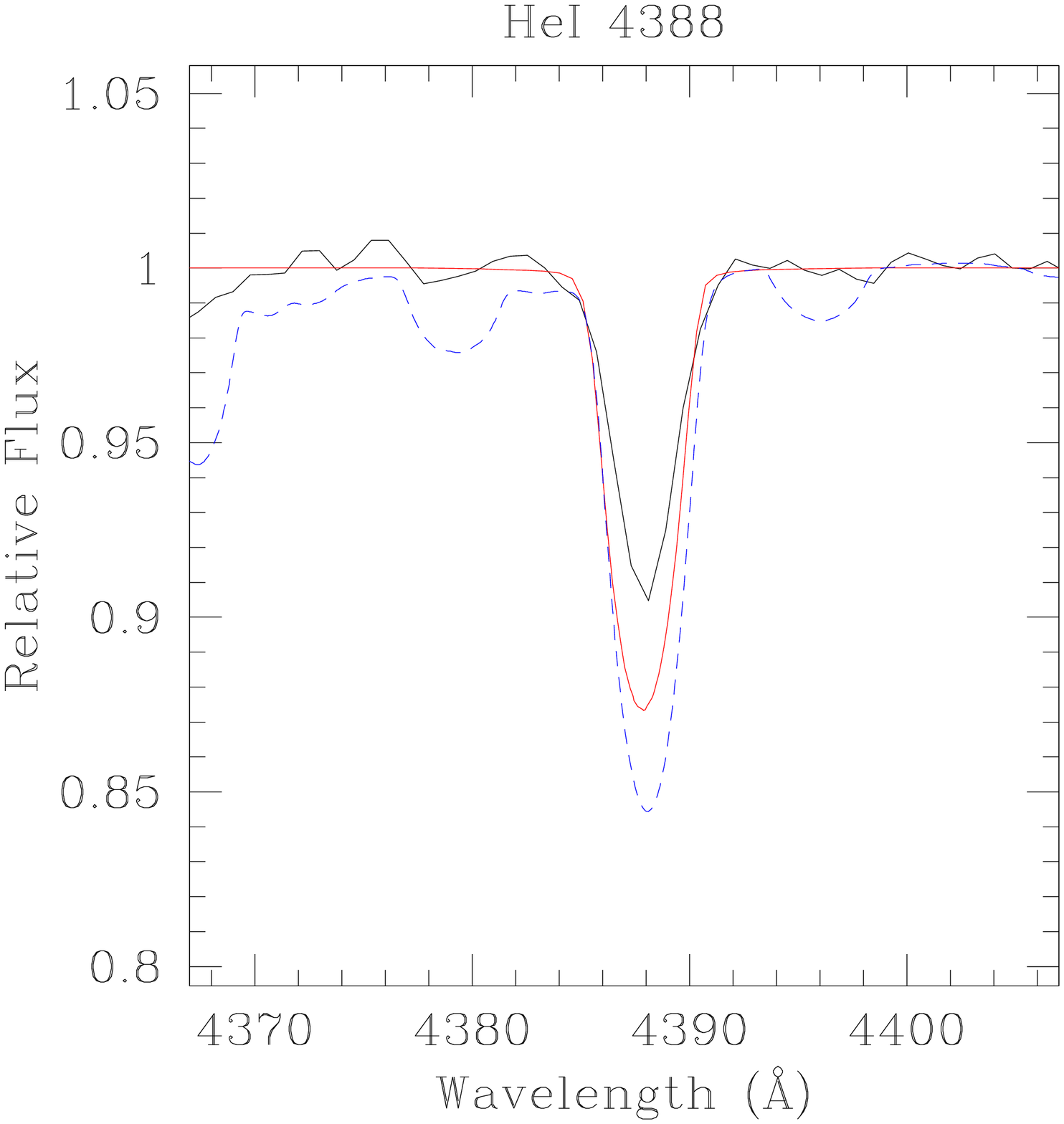}
\plotone{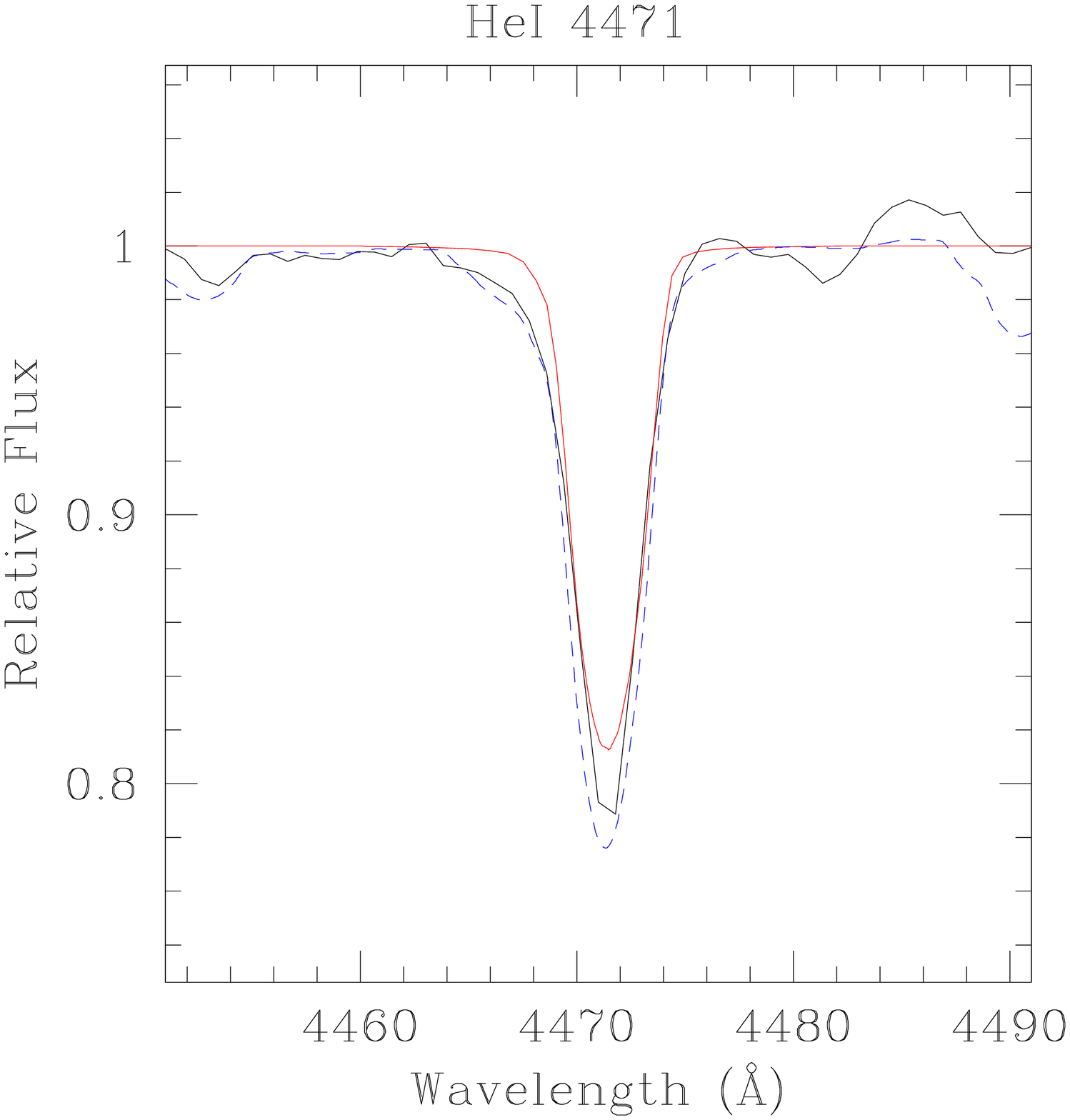}
\plotone{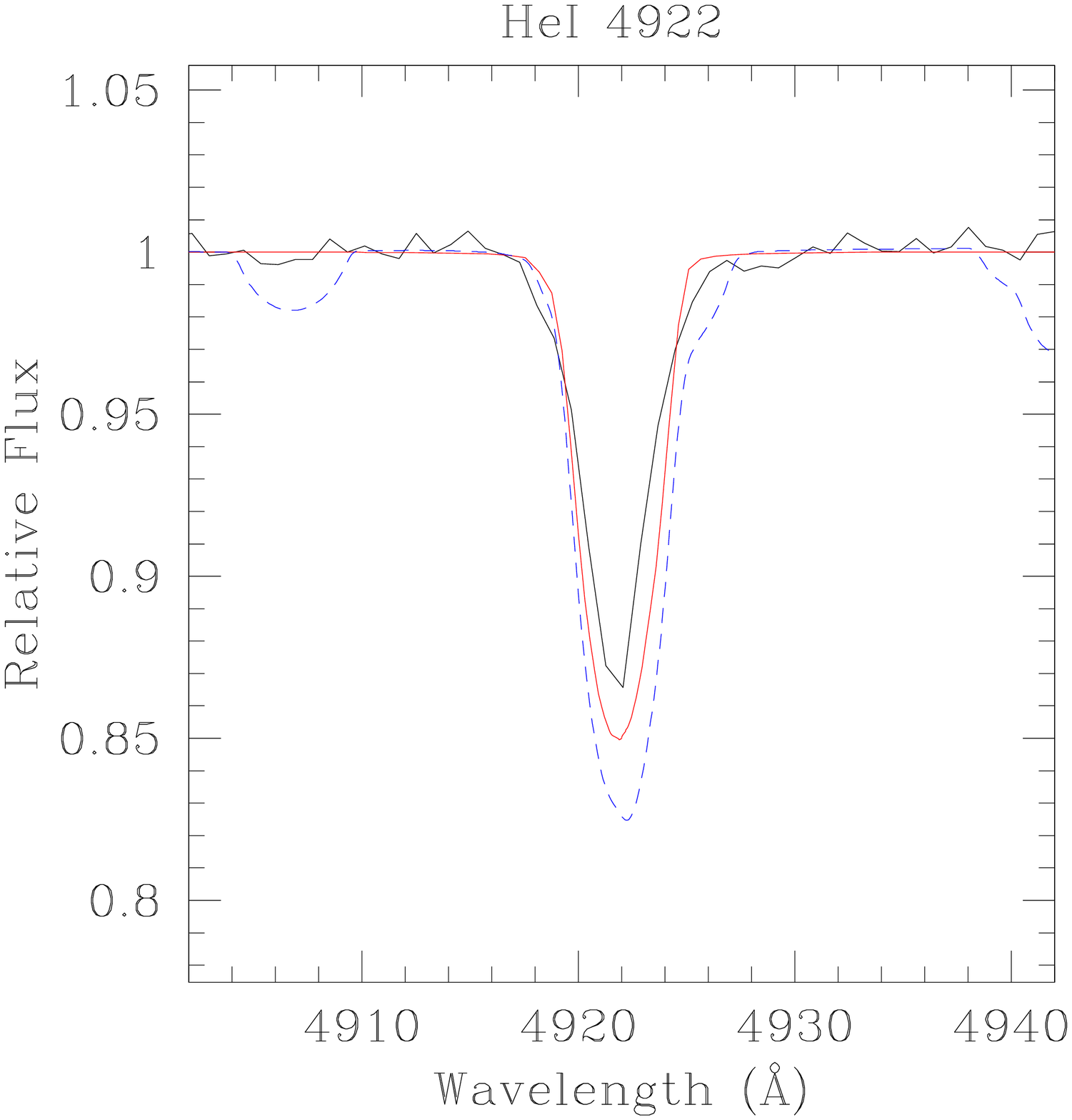}
\plotone{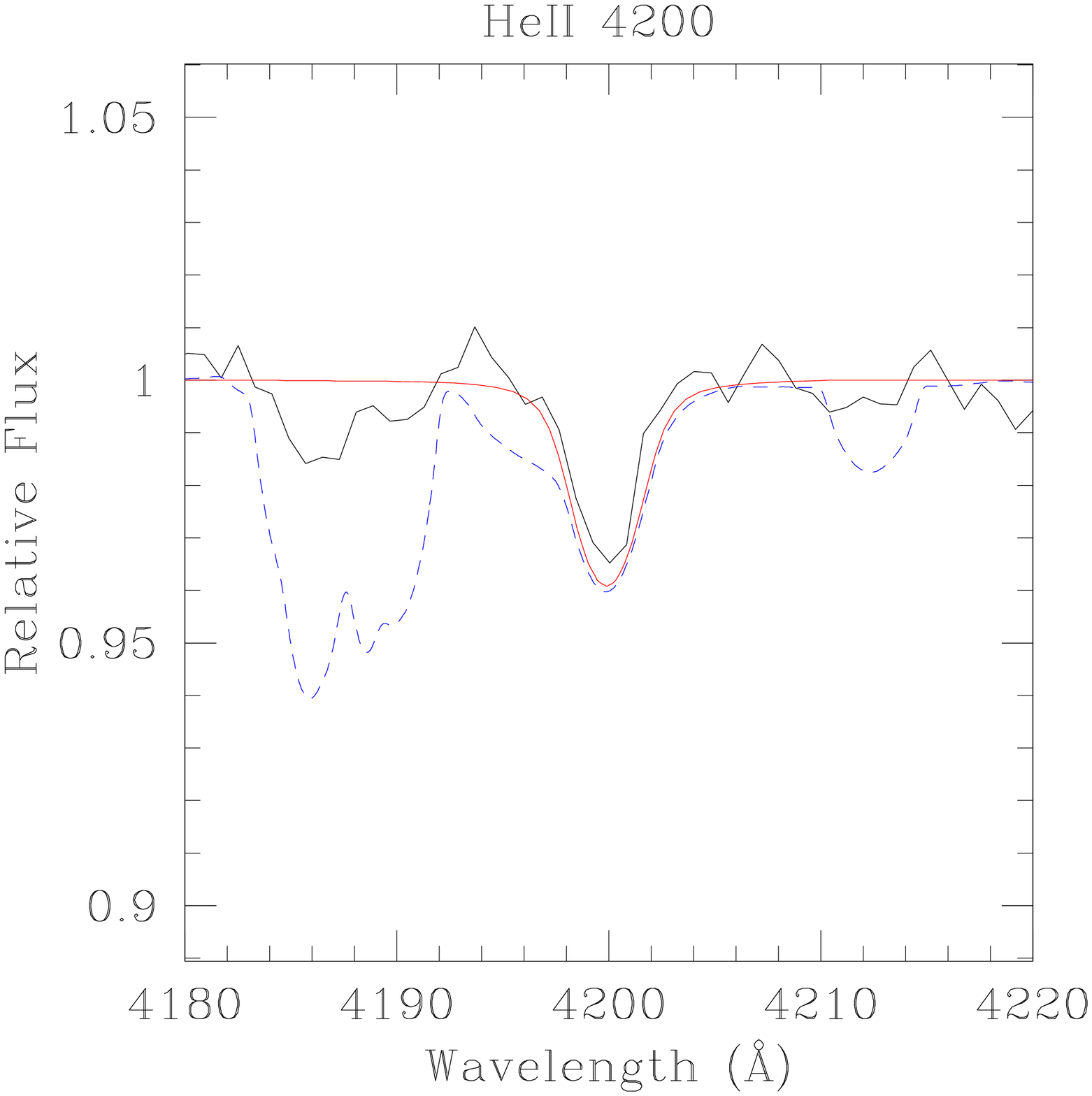}
\plotone{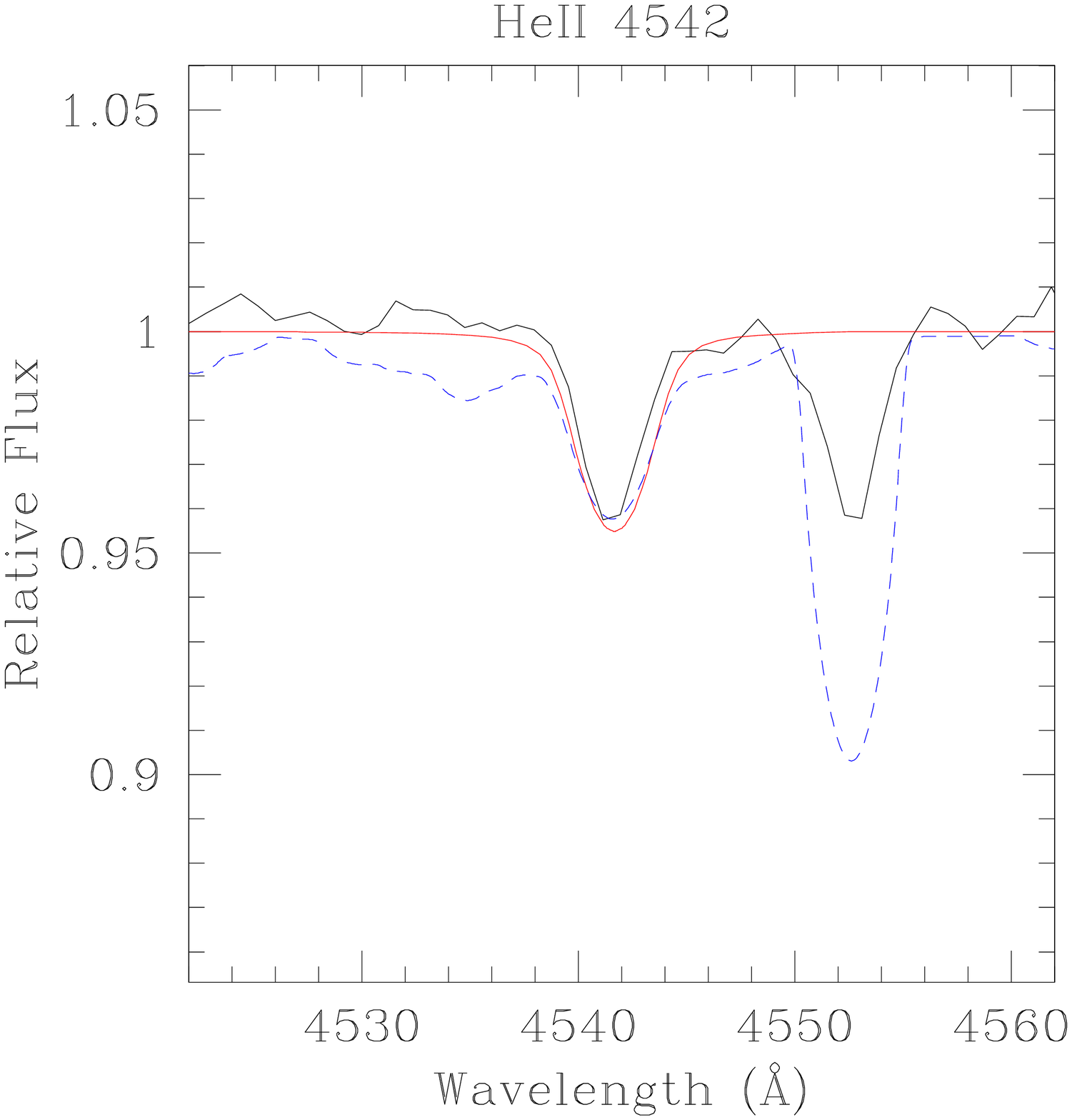}
\plotone{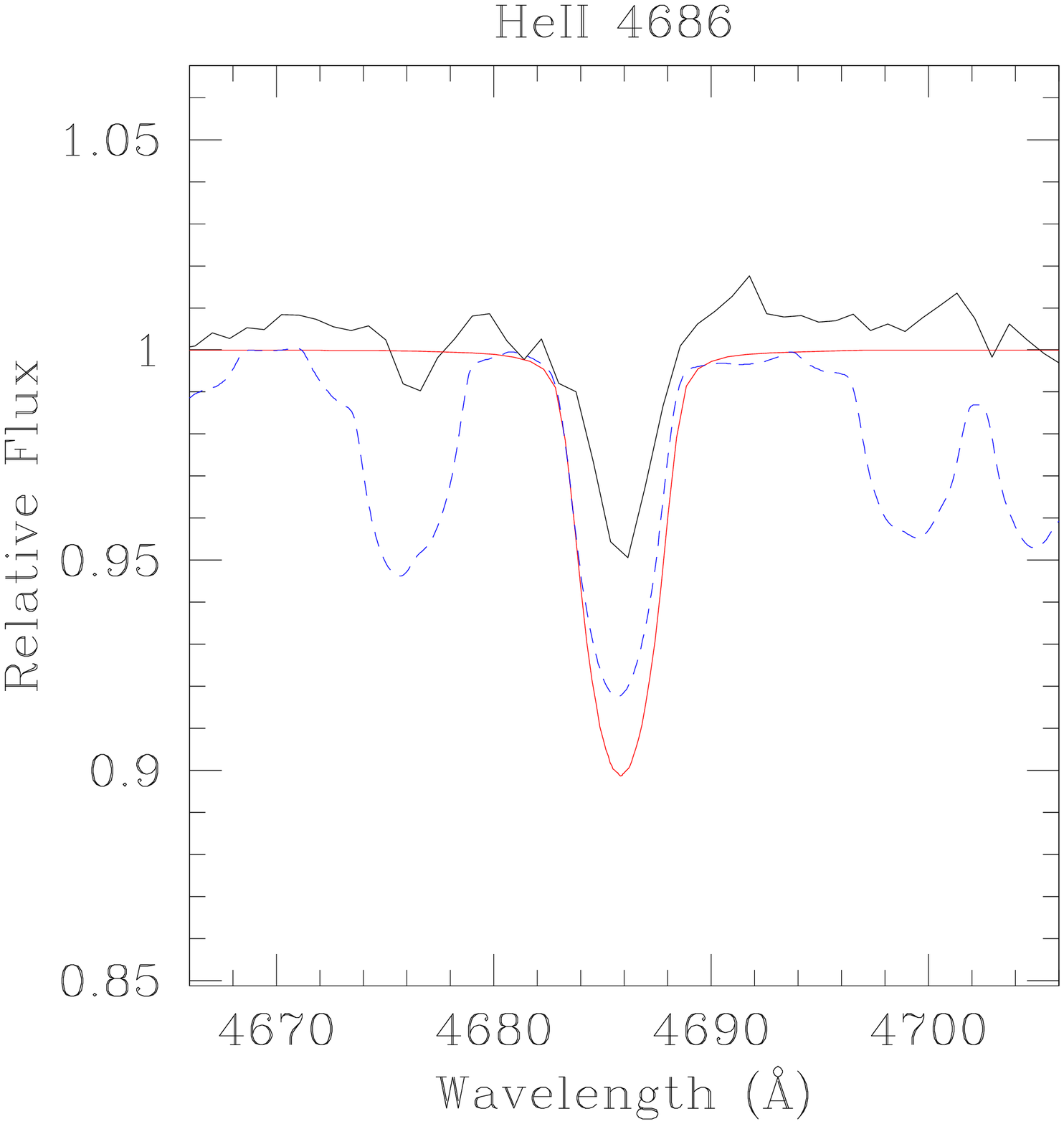}
\caption{\label{fig:Sk-69d12415} Model fits for Sk $-69^\circ$124, an O9.7 I star in the LMC computed with 15 km s$^{-1}$ microturbulence.  Black shows the observed spectrum, the red line shows the \fastwind\ fit, and the dashed blue line shows the \cmfgen\ fit. Compare to Figure~\ref{fig:Sk-69d124}.}
\end{figure}
\clearpage
\begin{figure}
\epsscale{0.45}
\plotone{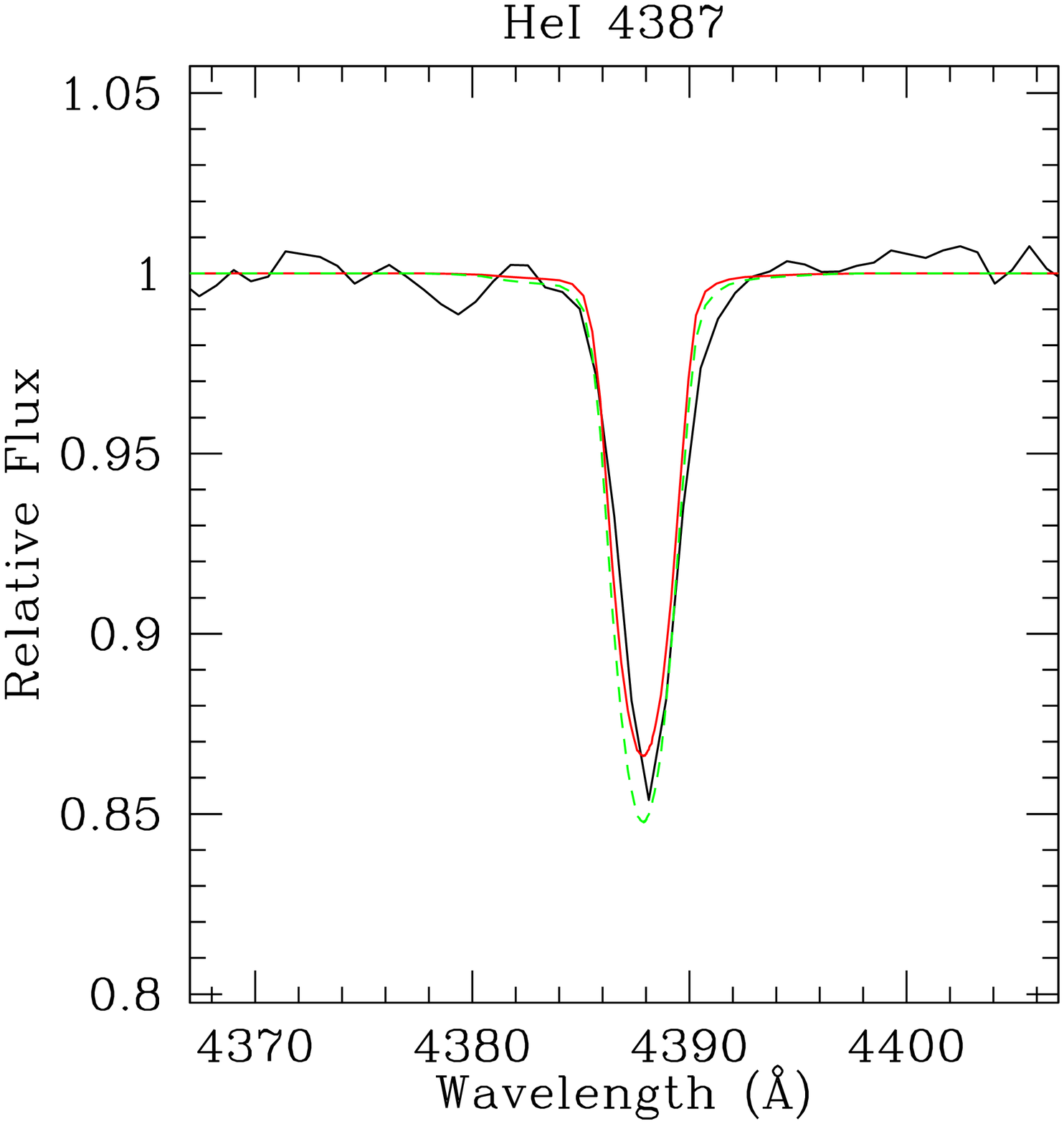}
\plotone{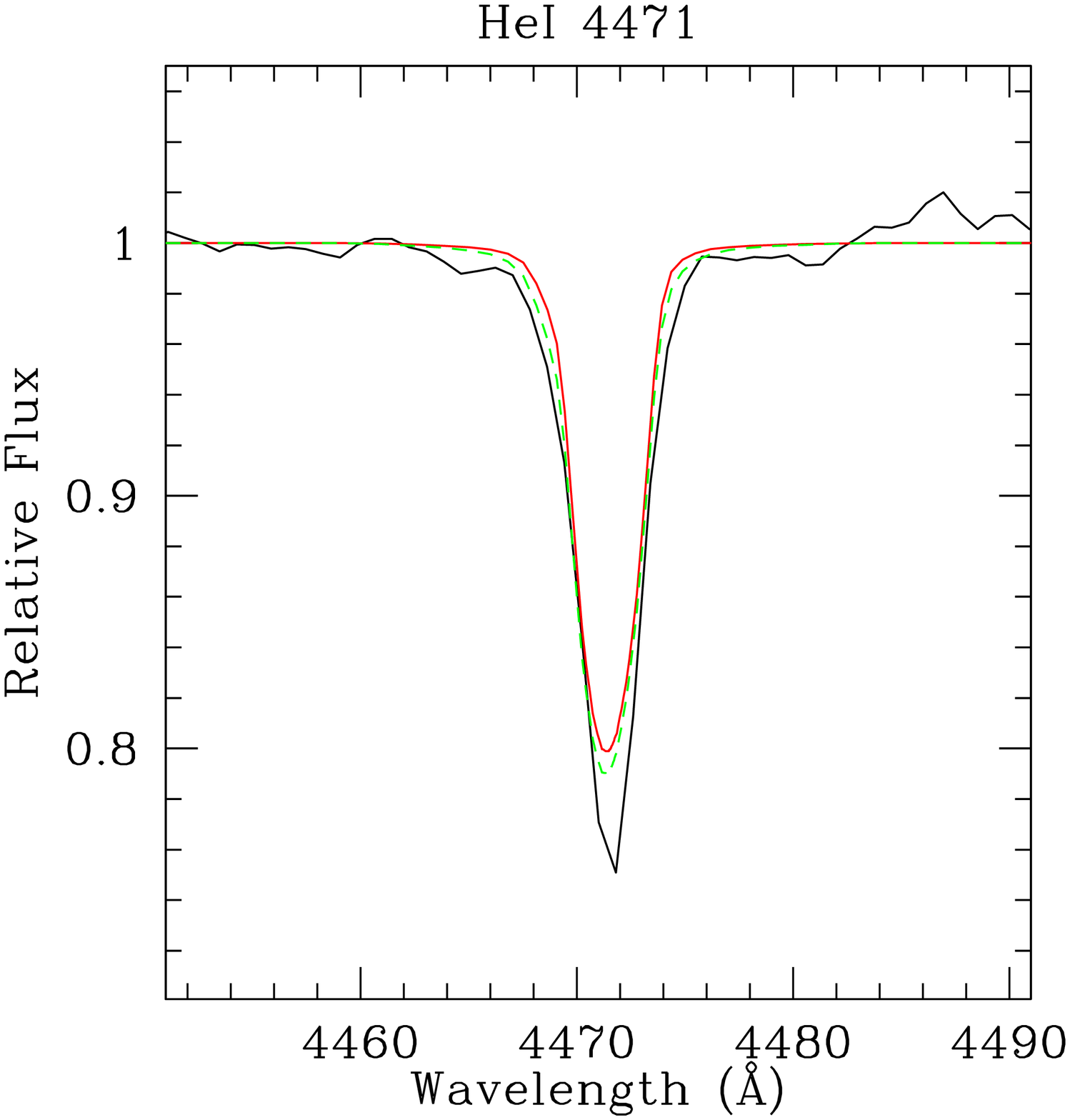}
\plotone{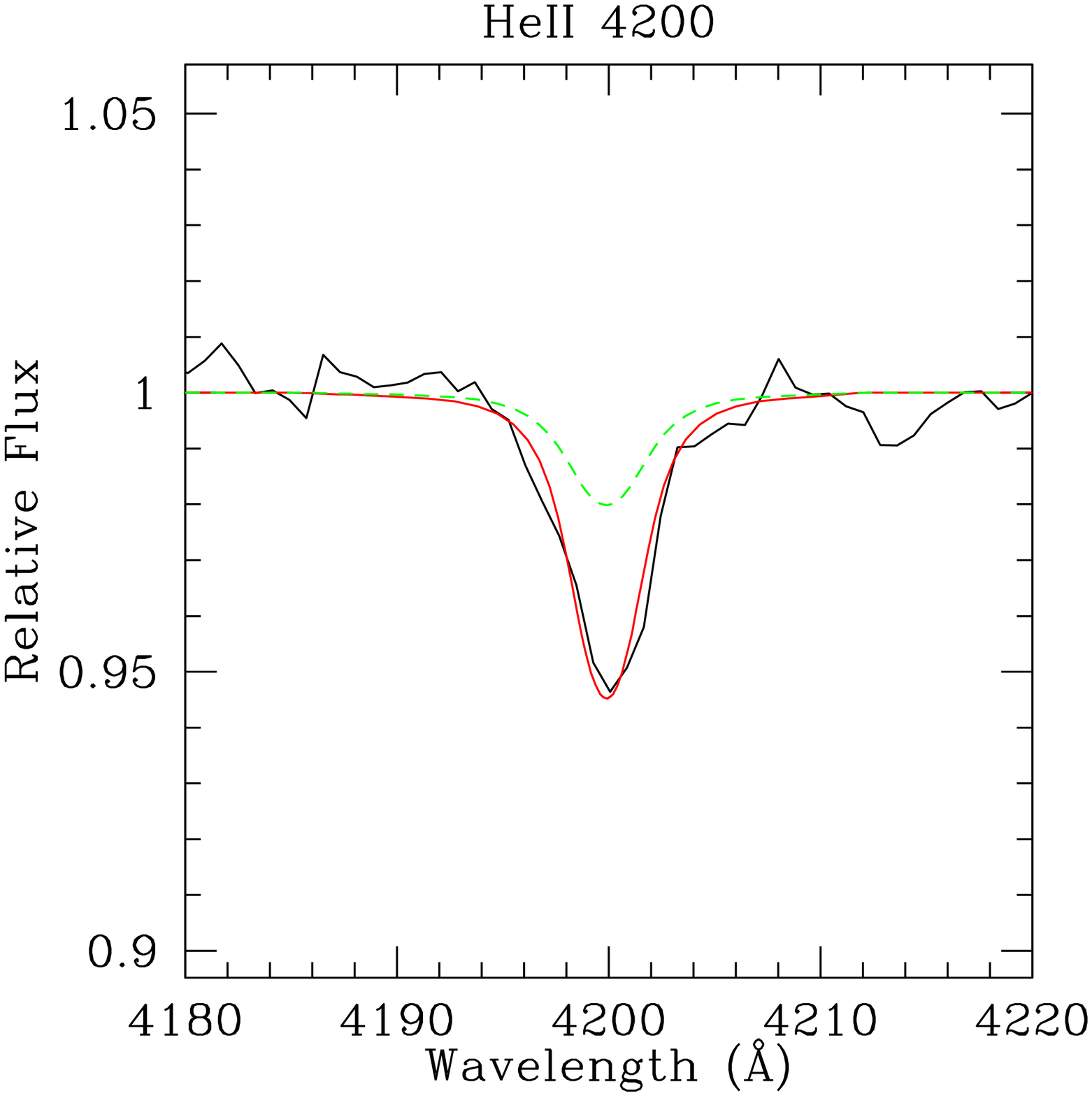}
\plotone{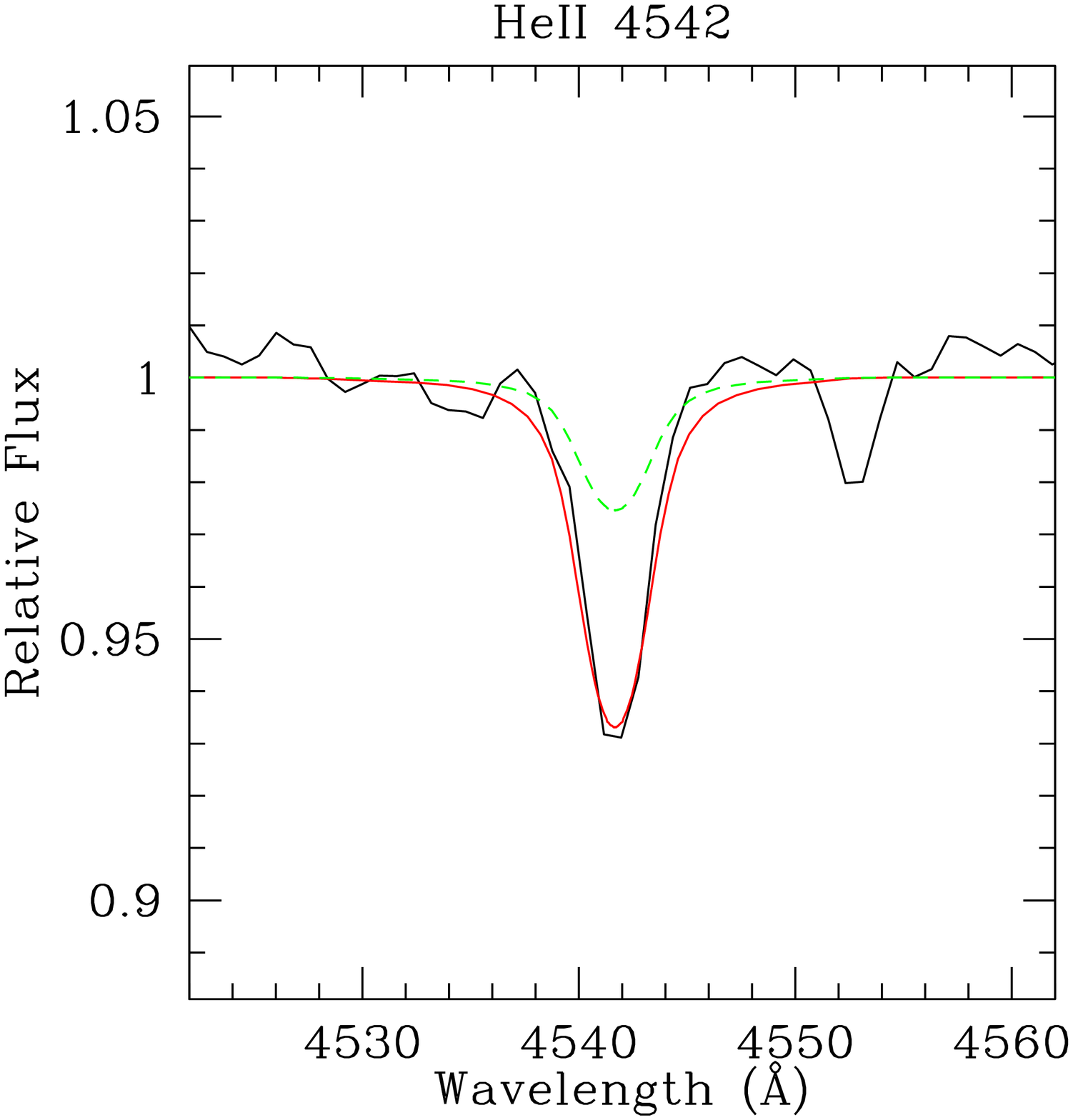}
\caption{\label{fig:compare}  Effect of lowering the effective temperature in \fastwind\ fit of AzV 223, an O9.5 II star in the SMC. Black shows the observed spectrum, and the red line shows the adopted \fastwind\ fit (as in Figure~\ref{fig:AzV223}) with an effective temperature of 31,600~K. The dashed green line shows the effect of lowering the \fastwind\ model temperature to 29,000~K.  The He~I $\lambda 4471$ fit improves only marginally, while the He~II profiles show that such a low temperature is unacceptable.}
\end{figure}
\clearpage
\begin{figure}
\epsscale{0.45}
\plotone{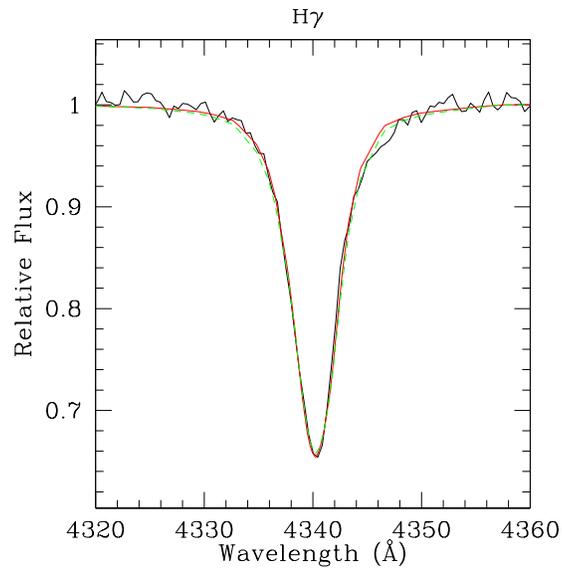}
\caption{\label{fig:compHgam}  Effect of increasing the surface gravity in the \fastwind\ fit for H$\gamma$ in the spectrum of AzV 26, an O6 I(f) star in the SMC. Black shows the observed spectrum, and the red lines shows the adopted \fastwind\ fit (as in Figure~\ref{fig:AzV26}) with a surface gravity of $\log g=3.50$.  The dashed green line shows the effect of increasing the \fastwind\ surface gravity to $\log g=3.60$, to match what we got from our \cmfgen\ modeling.  The fit becomes poorer on the uncontaminated blue side of the line. }
\end{figure}
\clearpage
\begin{figure}
\epsscale{0.35}
\plotone{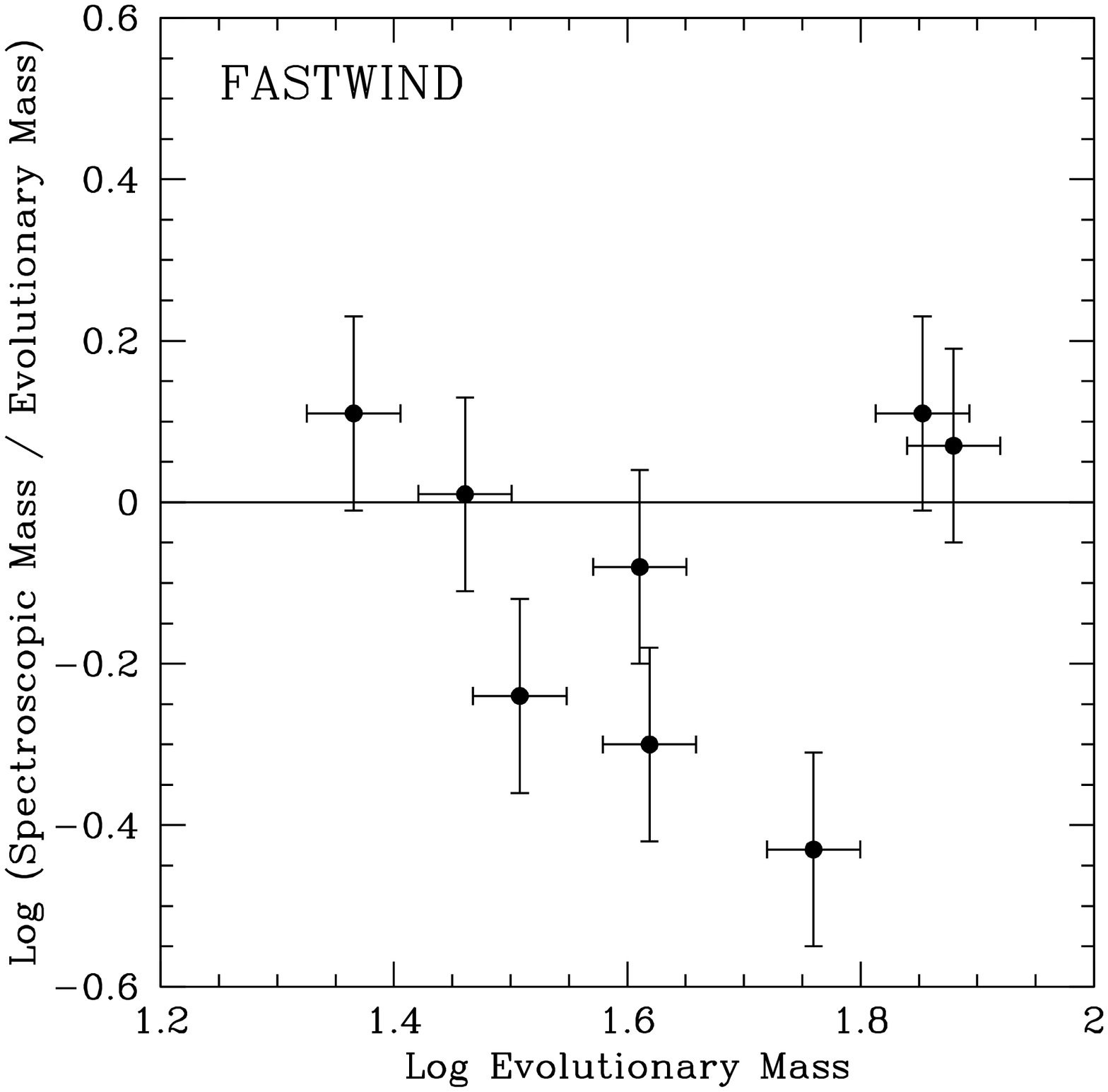}
\plotone{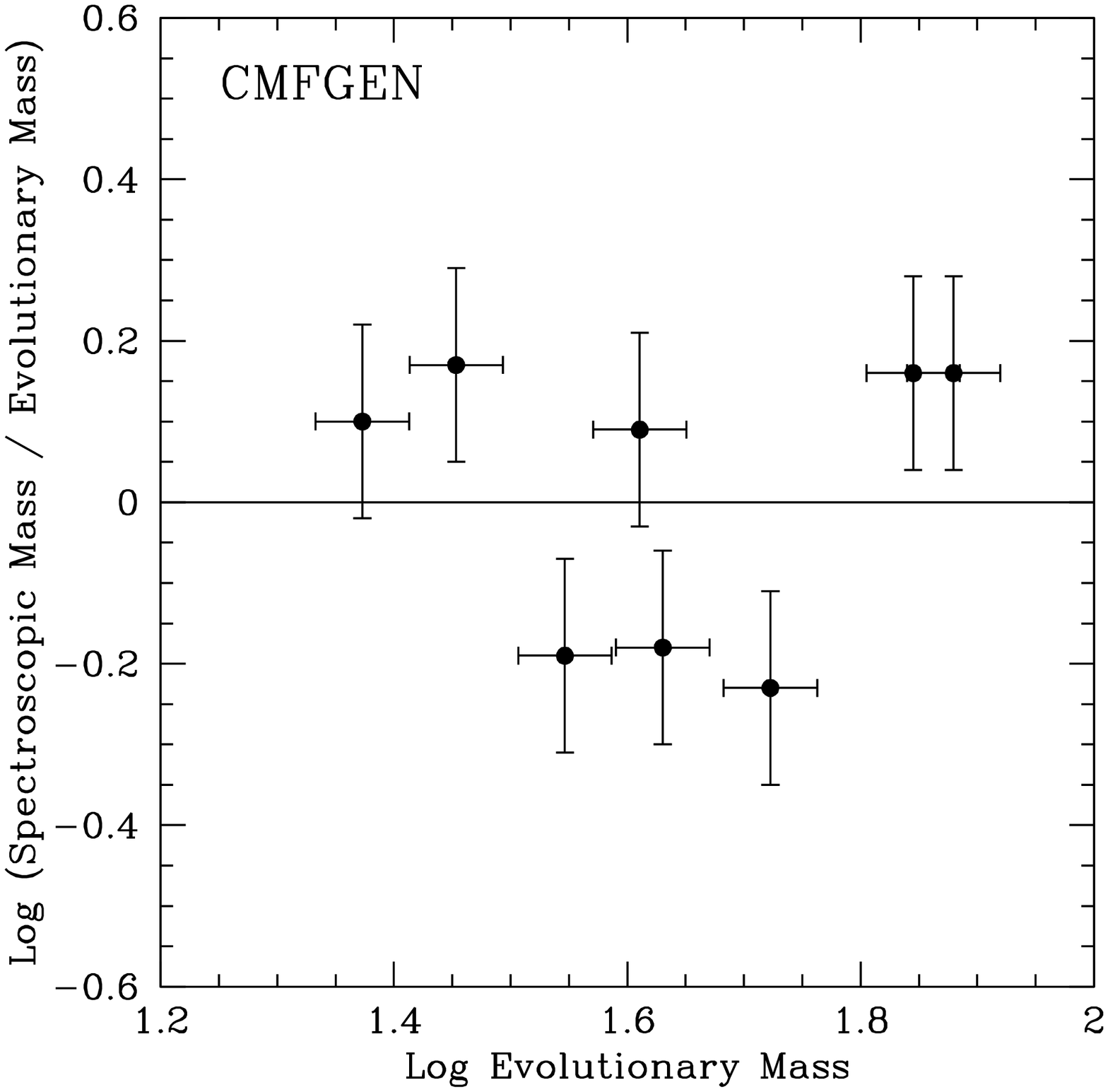}
\plotone{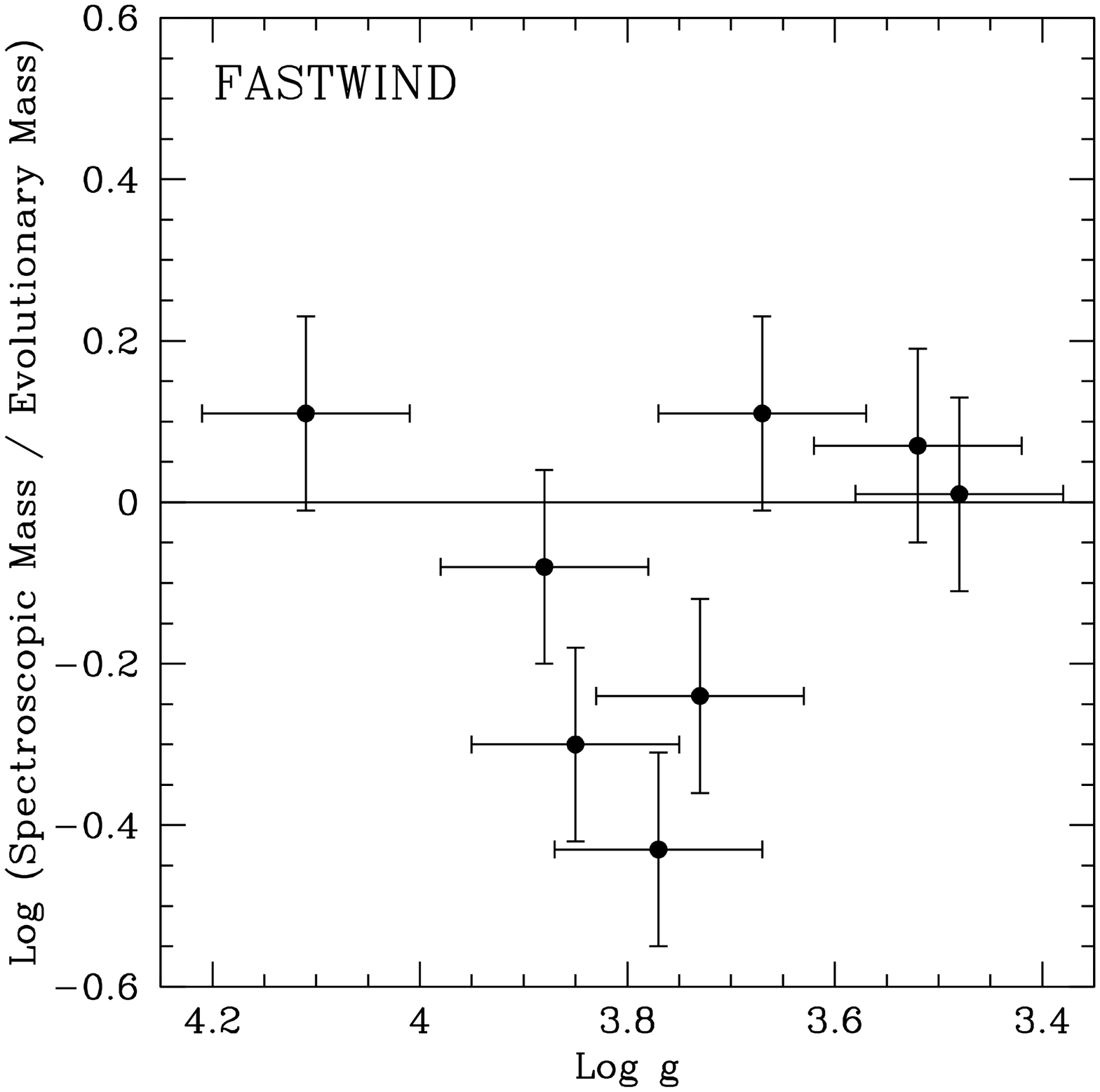}
\plotone{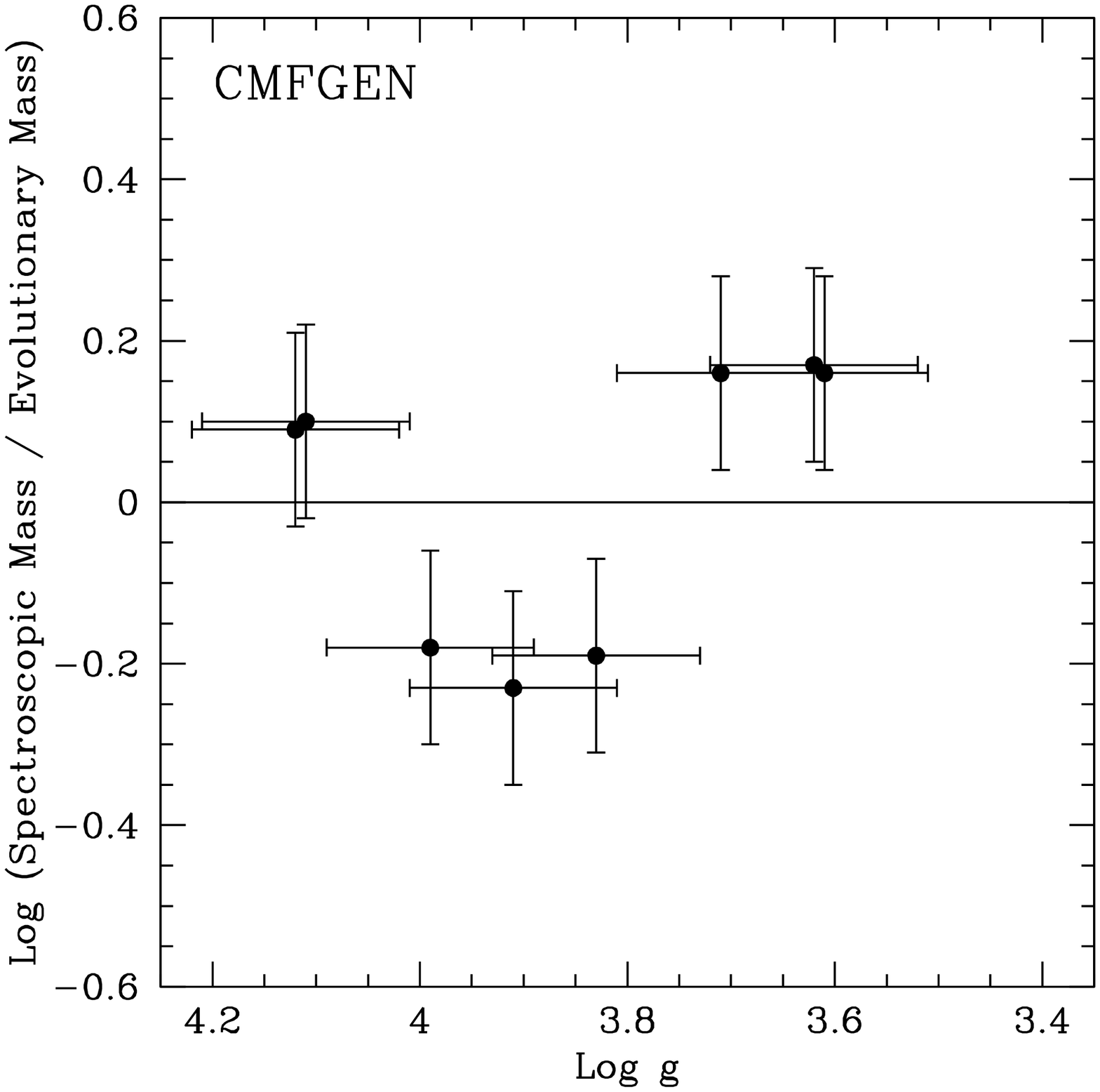}
\plotone{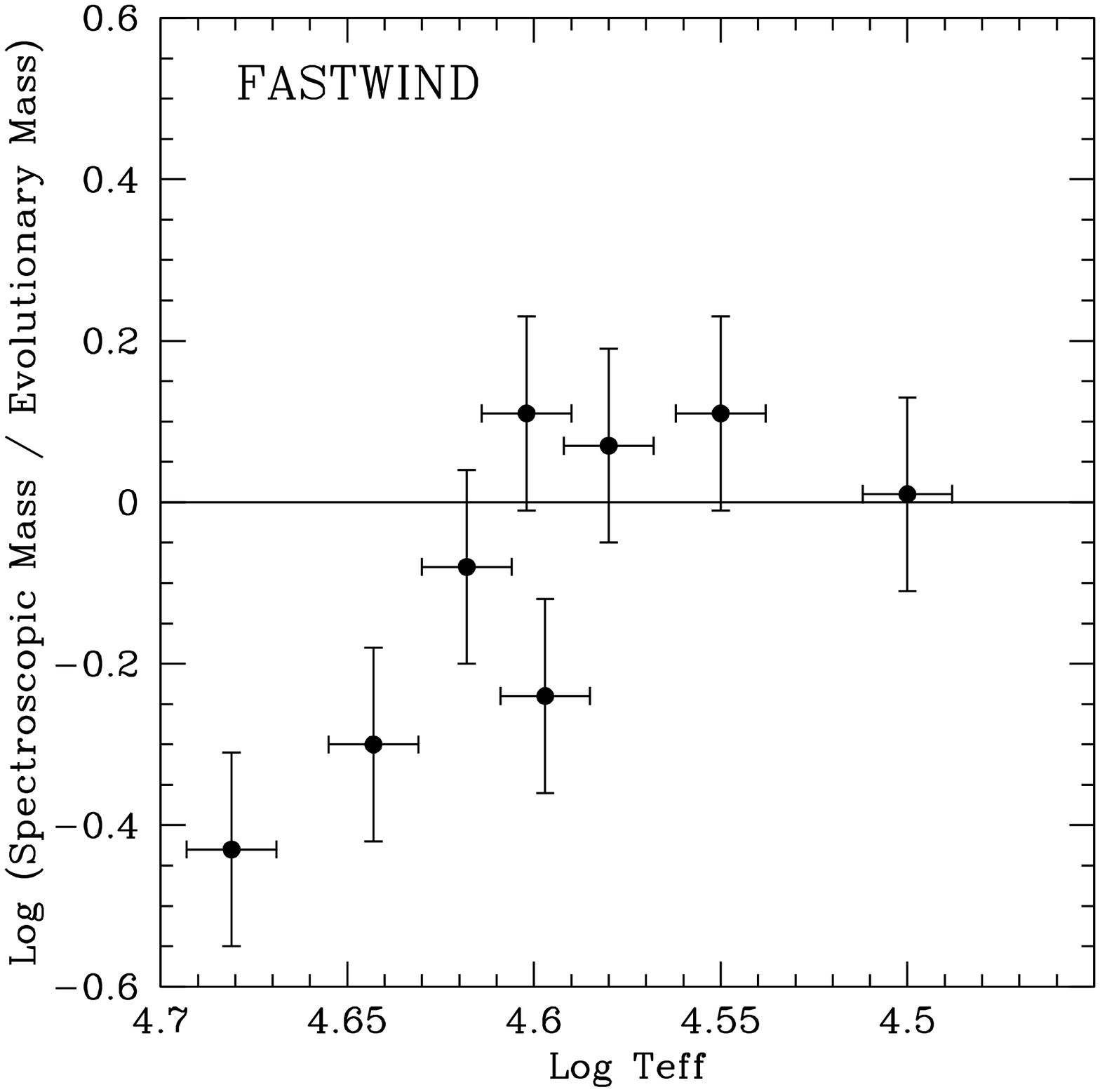}
\plotone{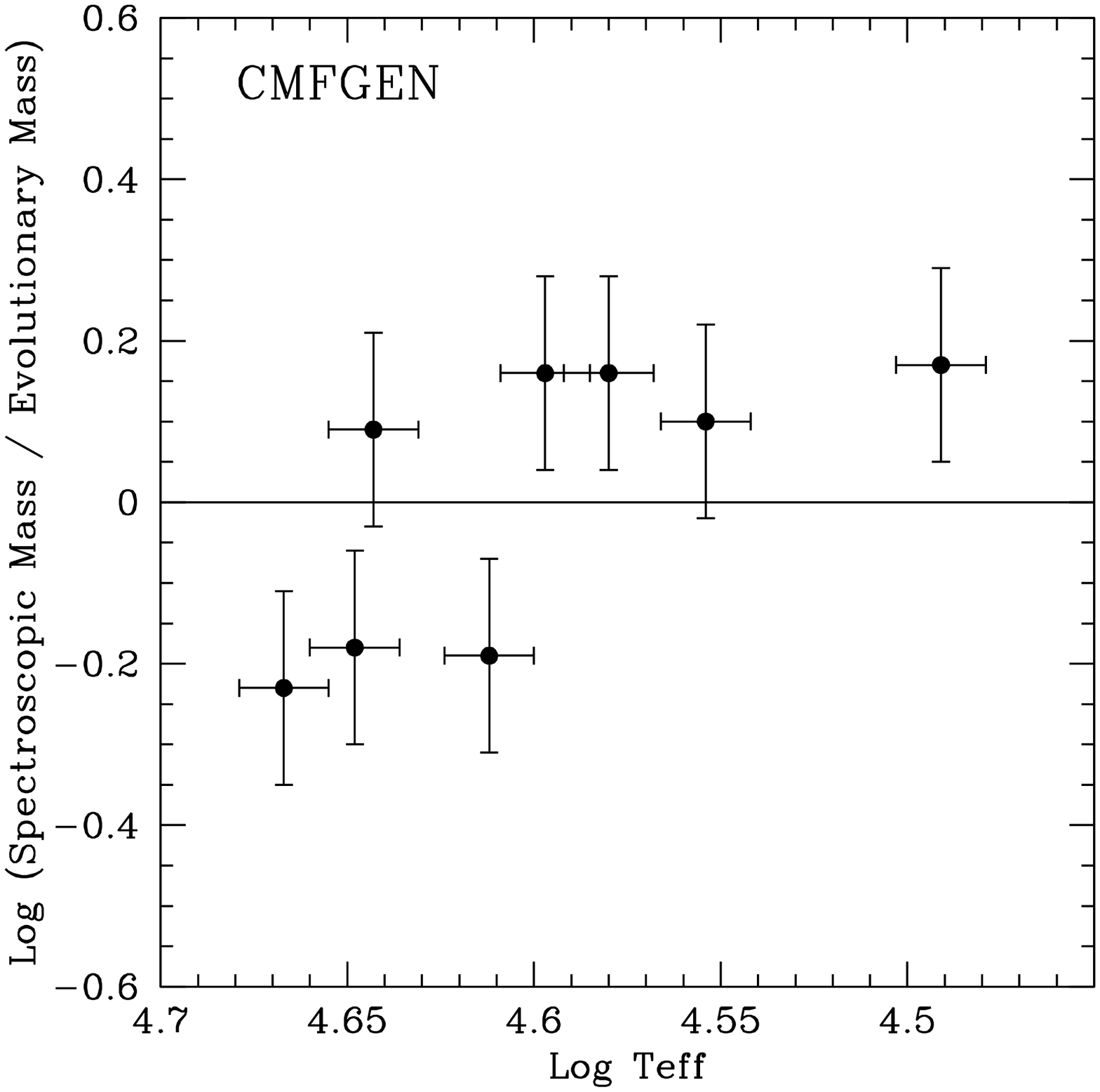}
\caption{\label{fig:md} The mass-discrepancy.  The log of the ratio of the spectroscopic mass to the evolutionary mass $\log \frac{m_{\rm spec}}{m_{\rm evol}}$) is plotted against ({\it top}) the log of the evolutionary mass ($\log m_{\rm evol}$), ({\it middle}) the log of the surface gravity, and ({\it bottom}) the log of the effective temperature,  for both our \fastwind\ results ({\it left}) and \cmfgen\ results ({\it right}).}
\end{figure}
\clearpage
\begin{figure}
\epsscale{0.3}
\plotone{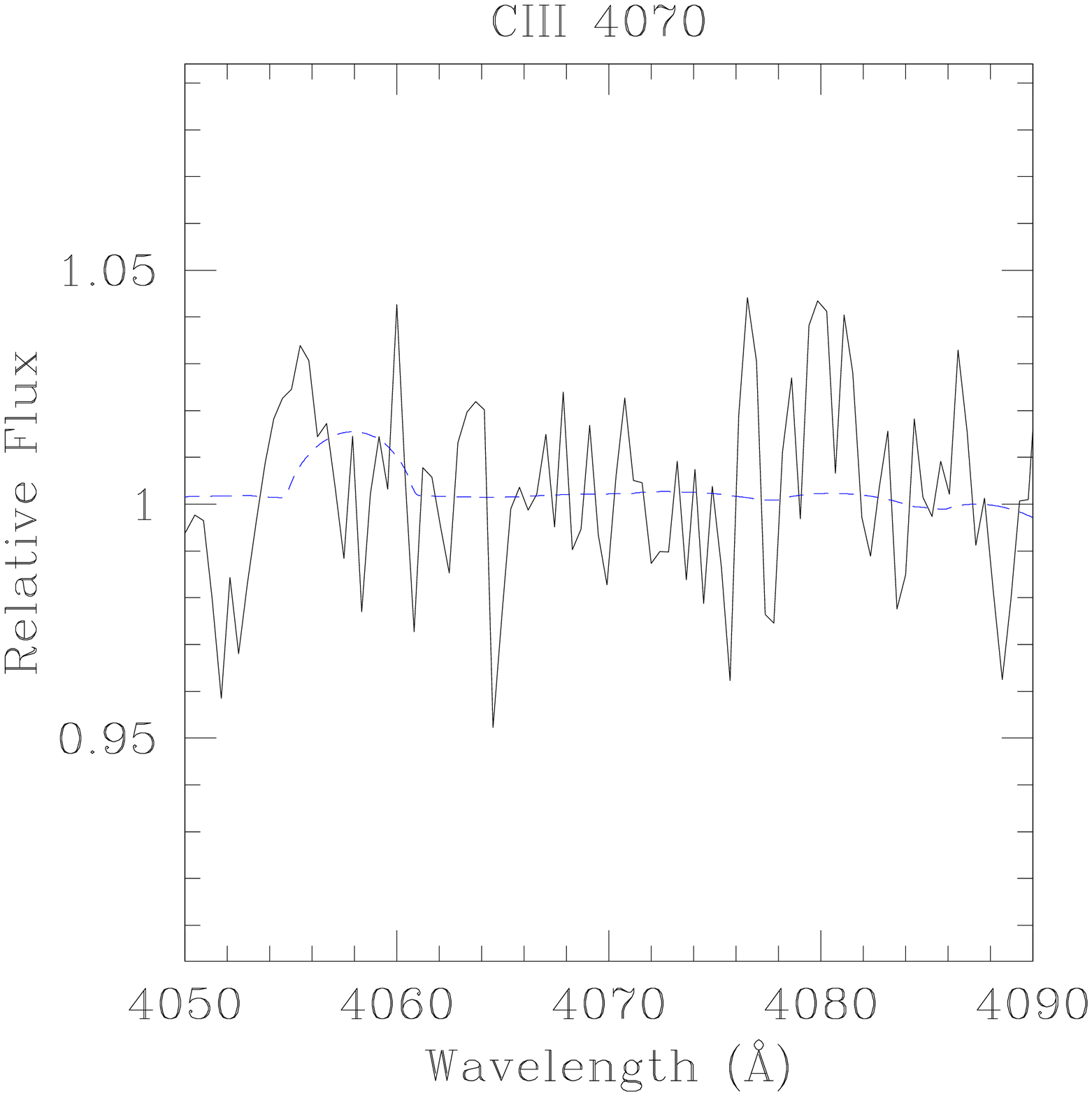}
\plotone{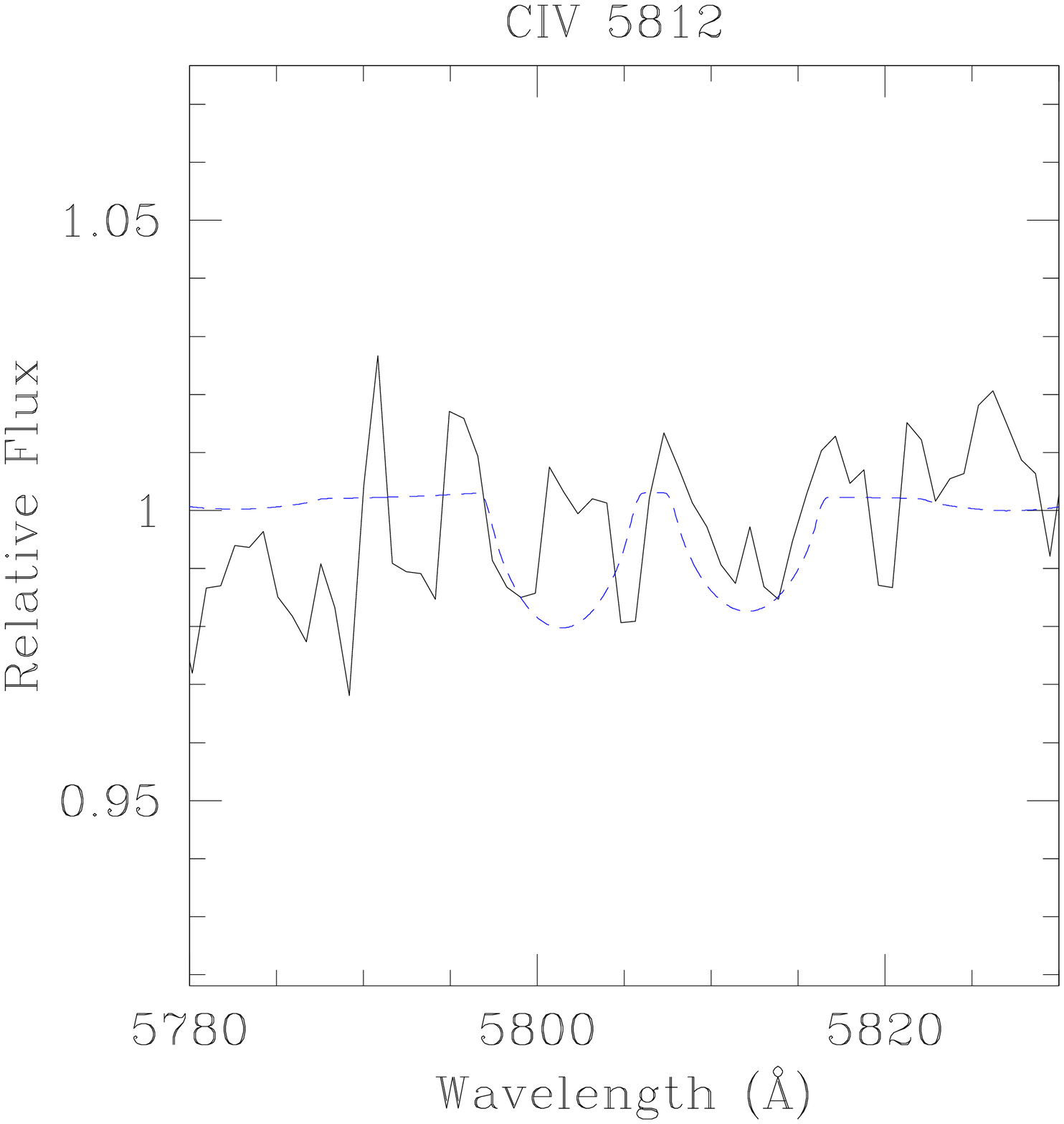}
\plotone{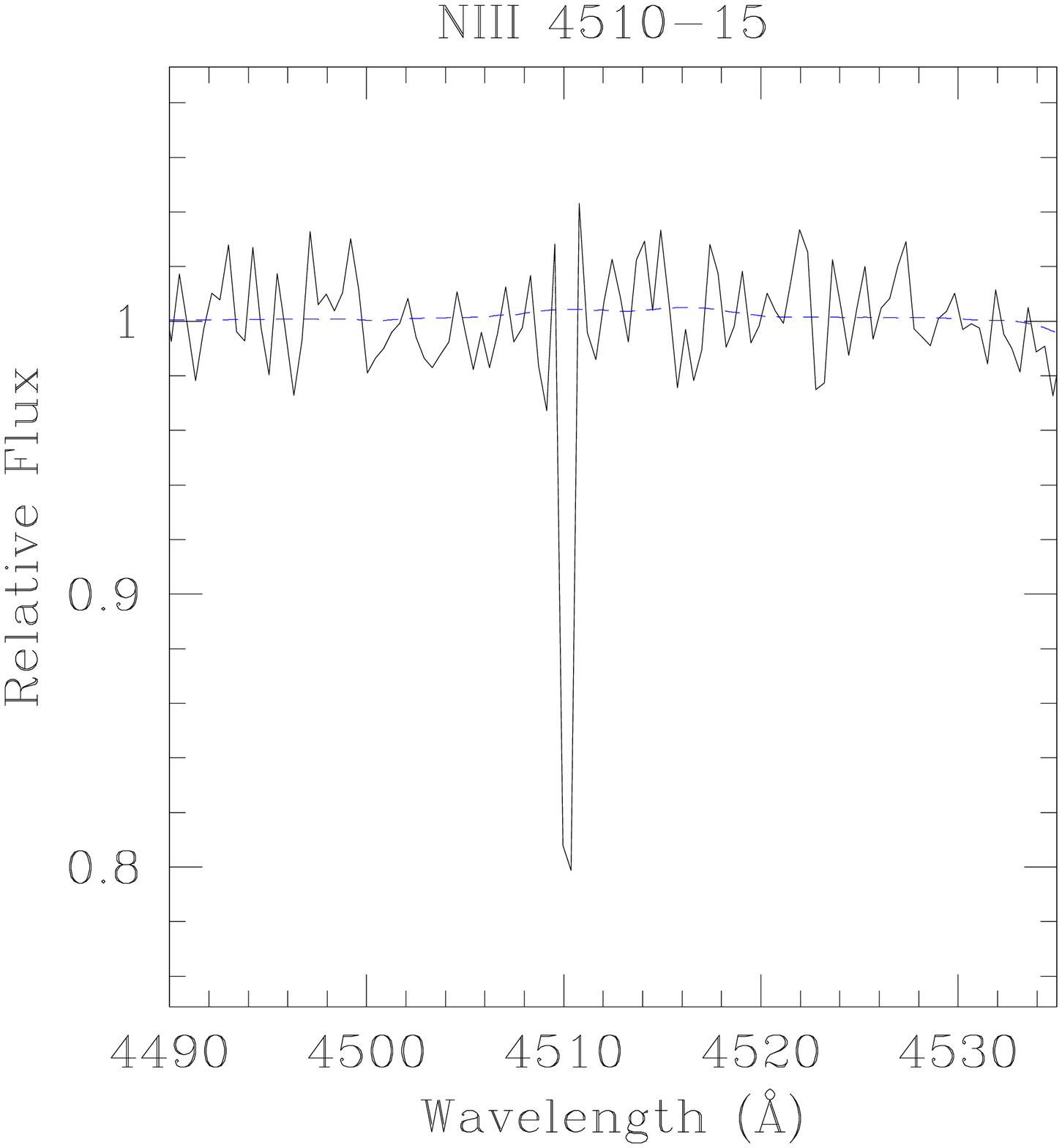}
\plotone{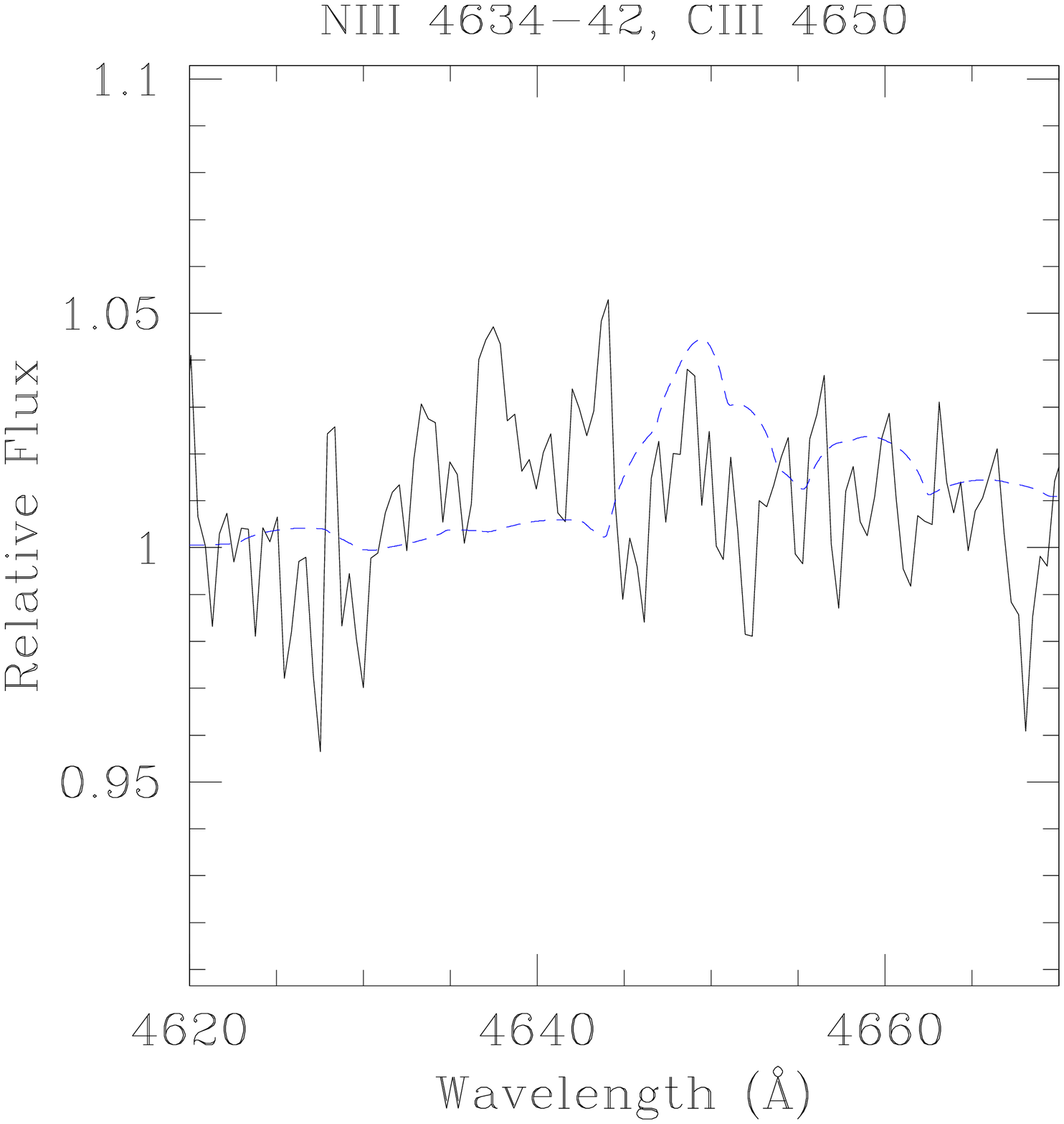}
\plotone{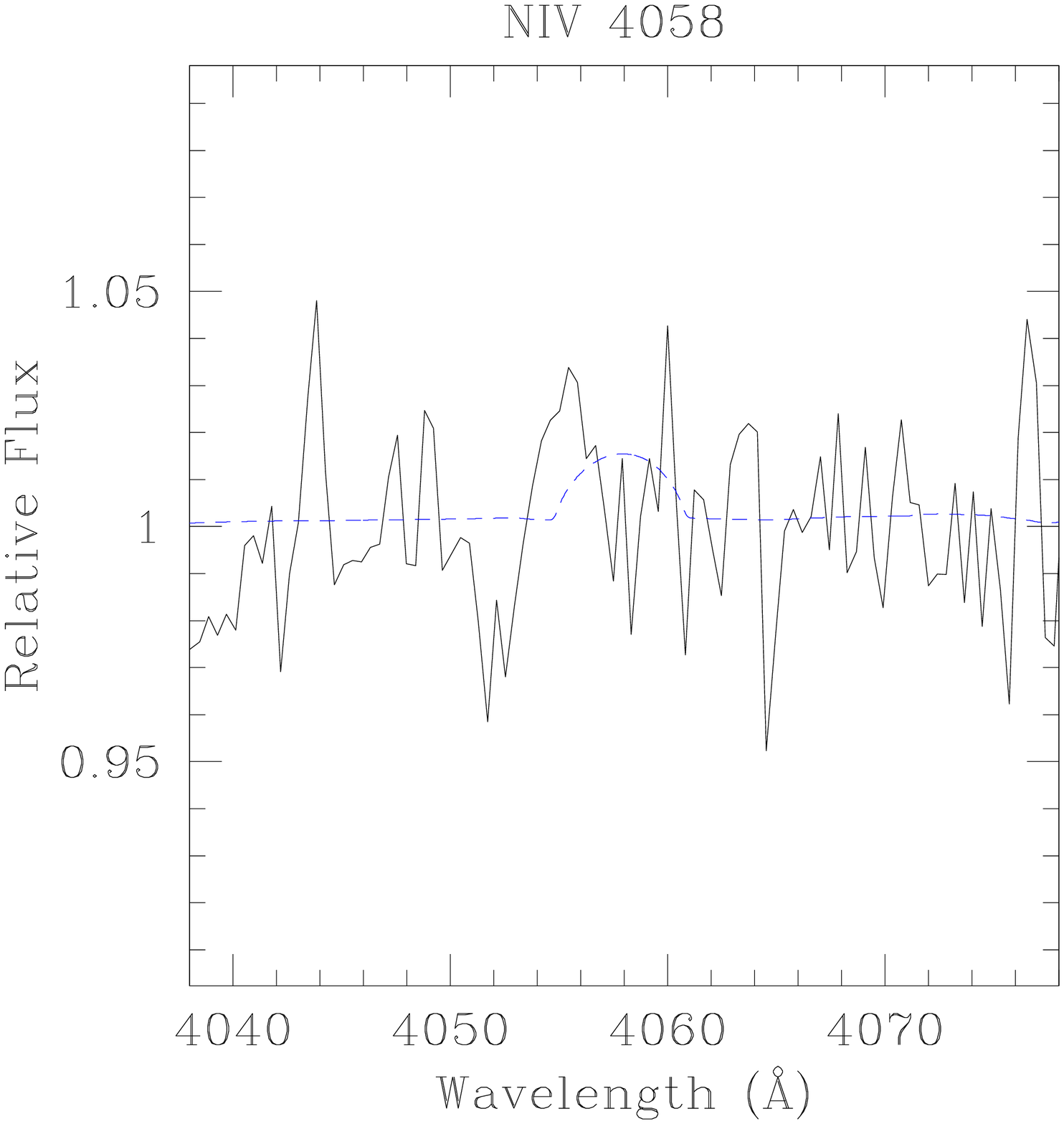}
\plotone{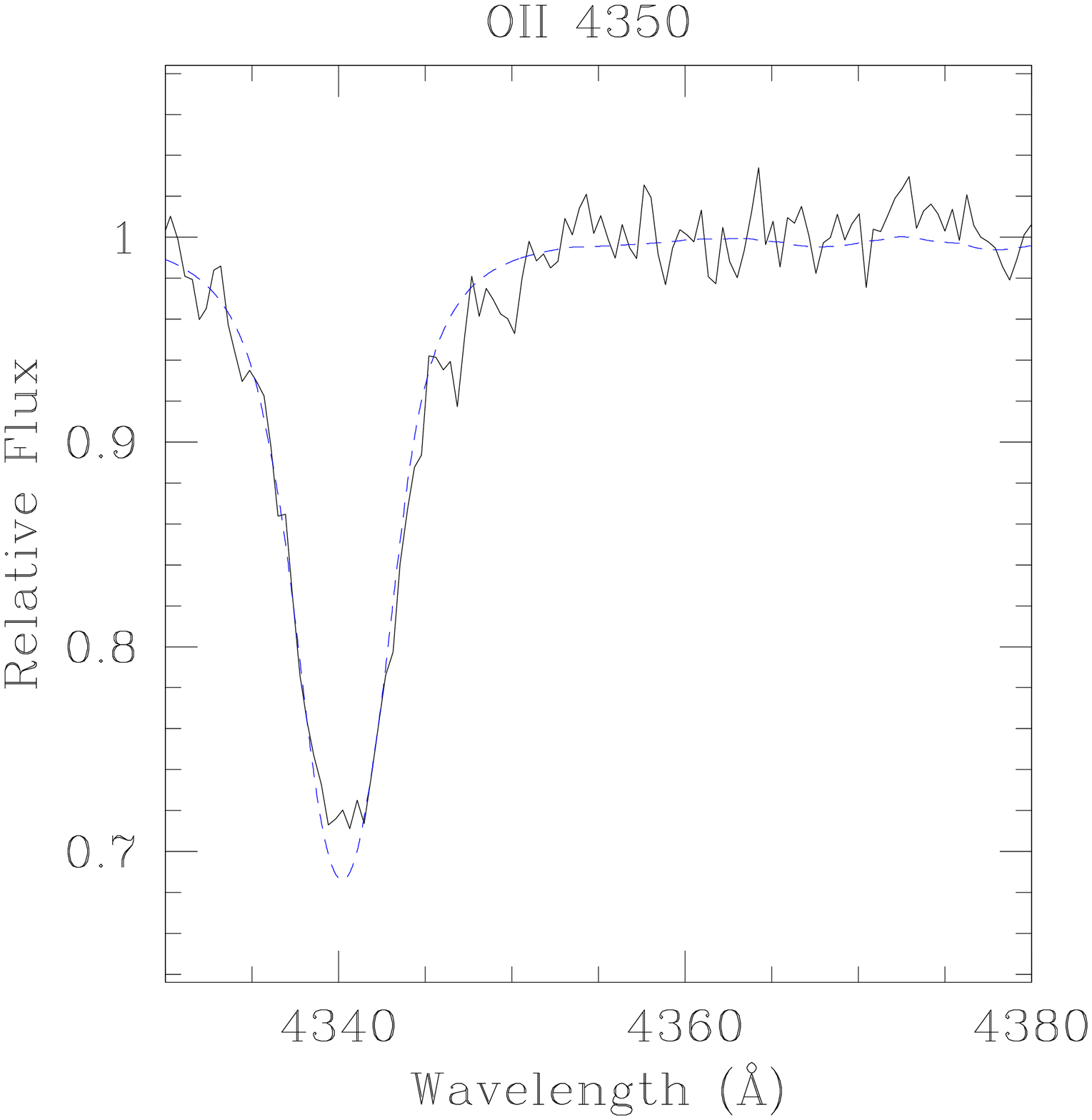}
\plotone{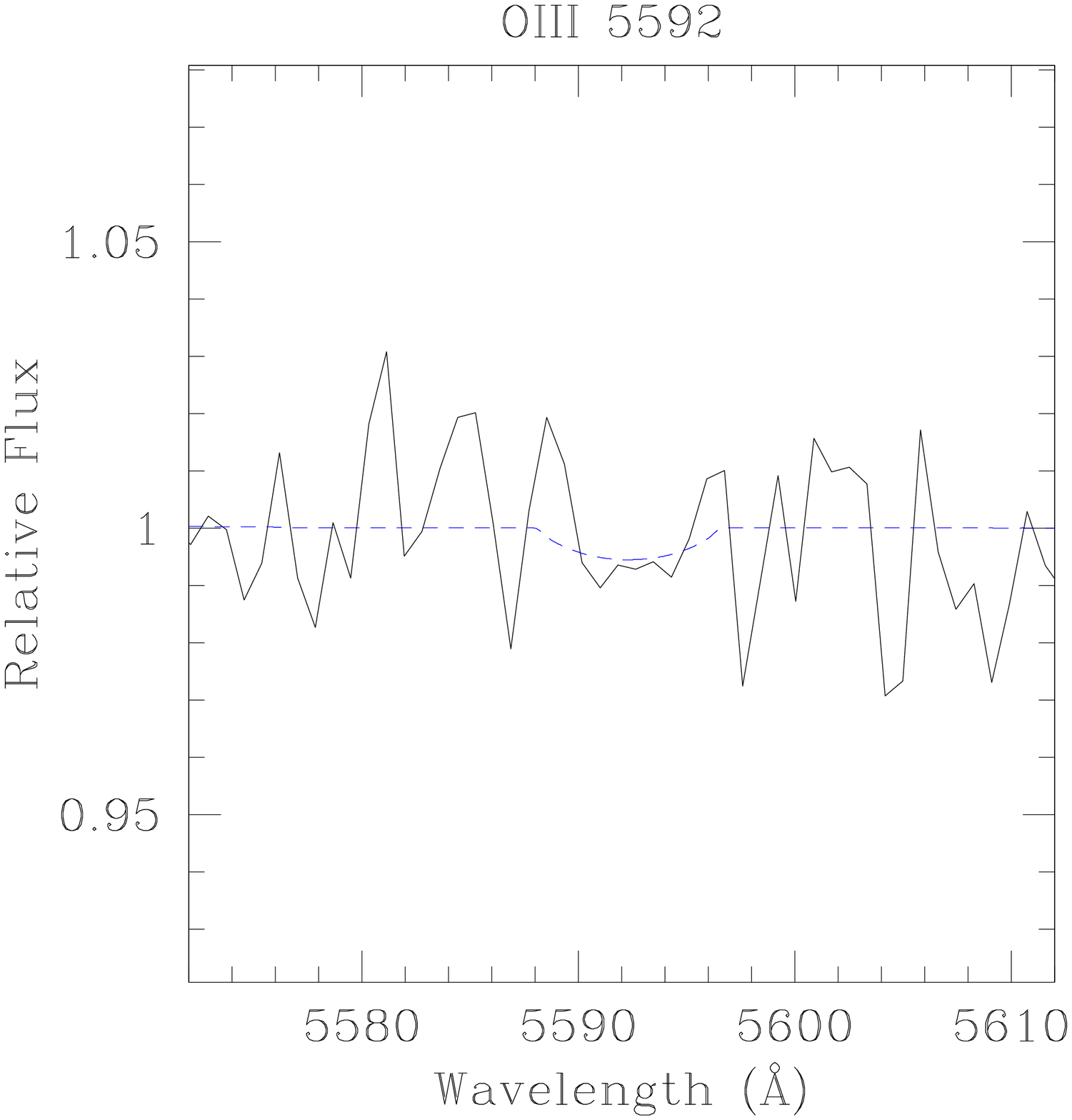}
\caption{\label{fig:azv177metal} \cmfgen\ fits of CNO lines for AzV 177, O4 V((f)). Black denotes the observed spectrum, and dashed blue shows the \cmfgen\ model with SMC abundances ($Z/Z_\odot=0.2$).}
\end{figure}
\clearpage
\begin{figure}
\epsscale{0.3}
\plotone{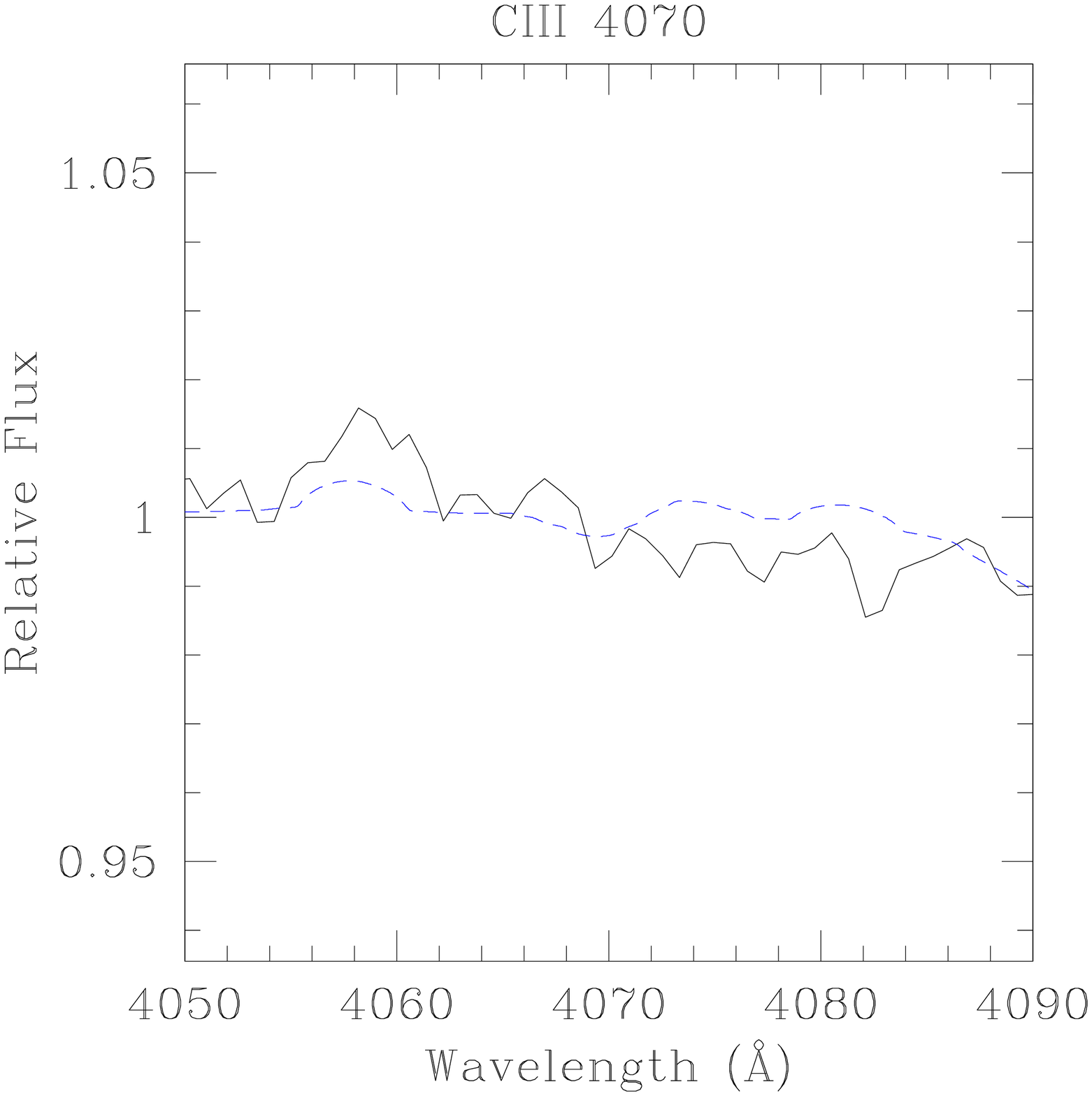}
\plotone{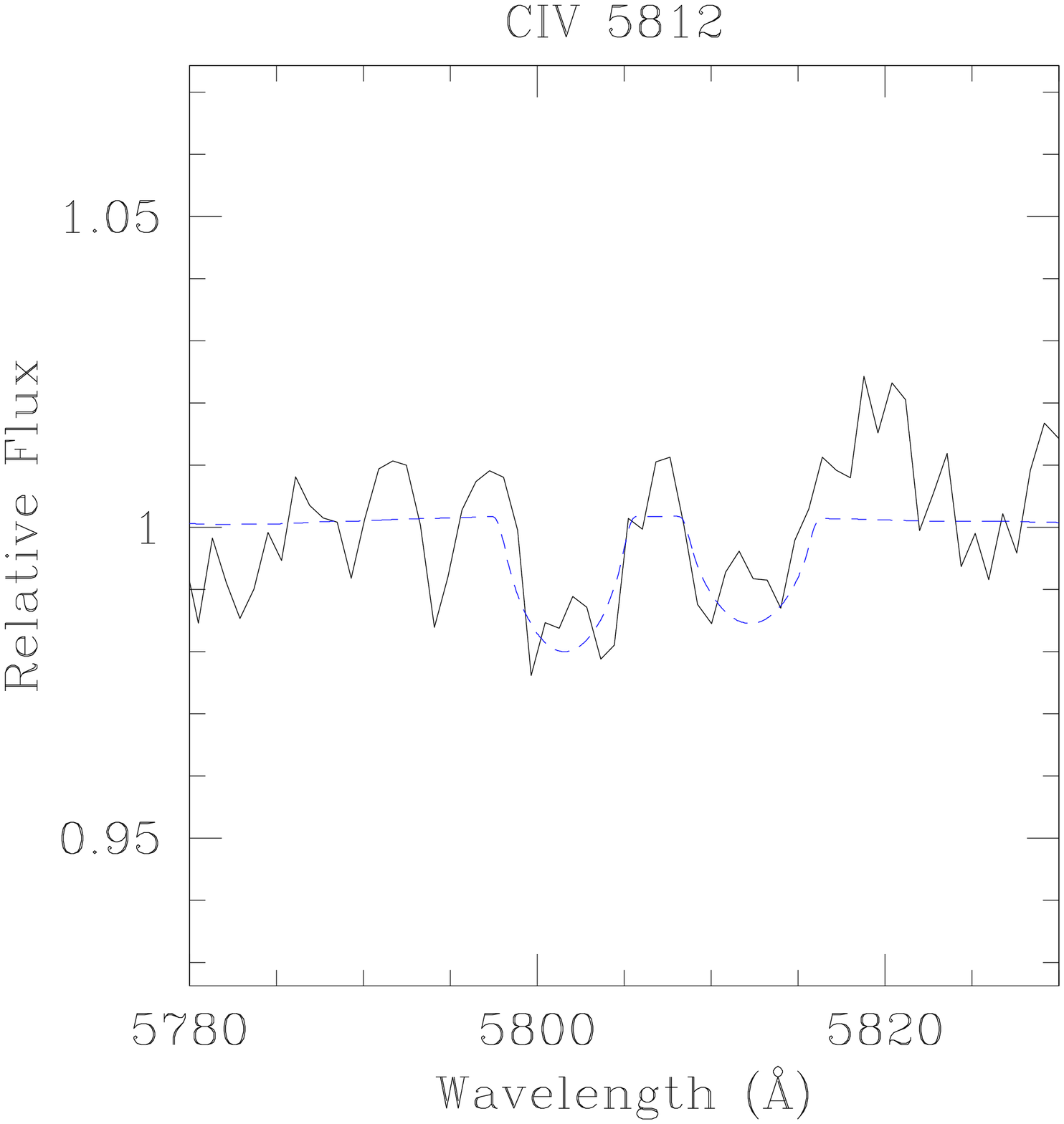}
\plotone{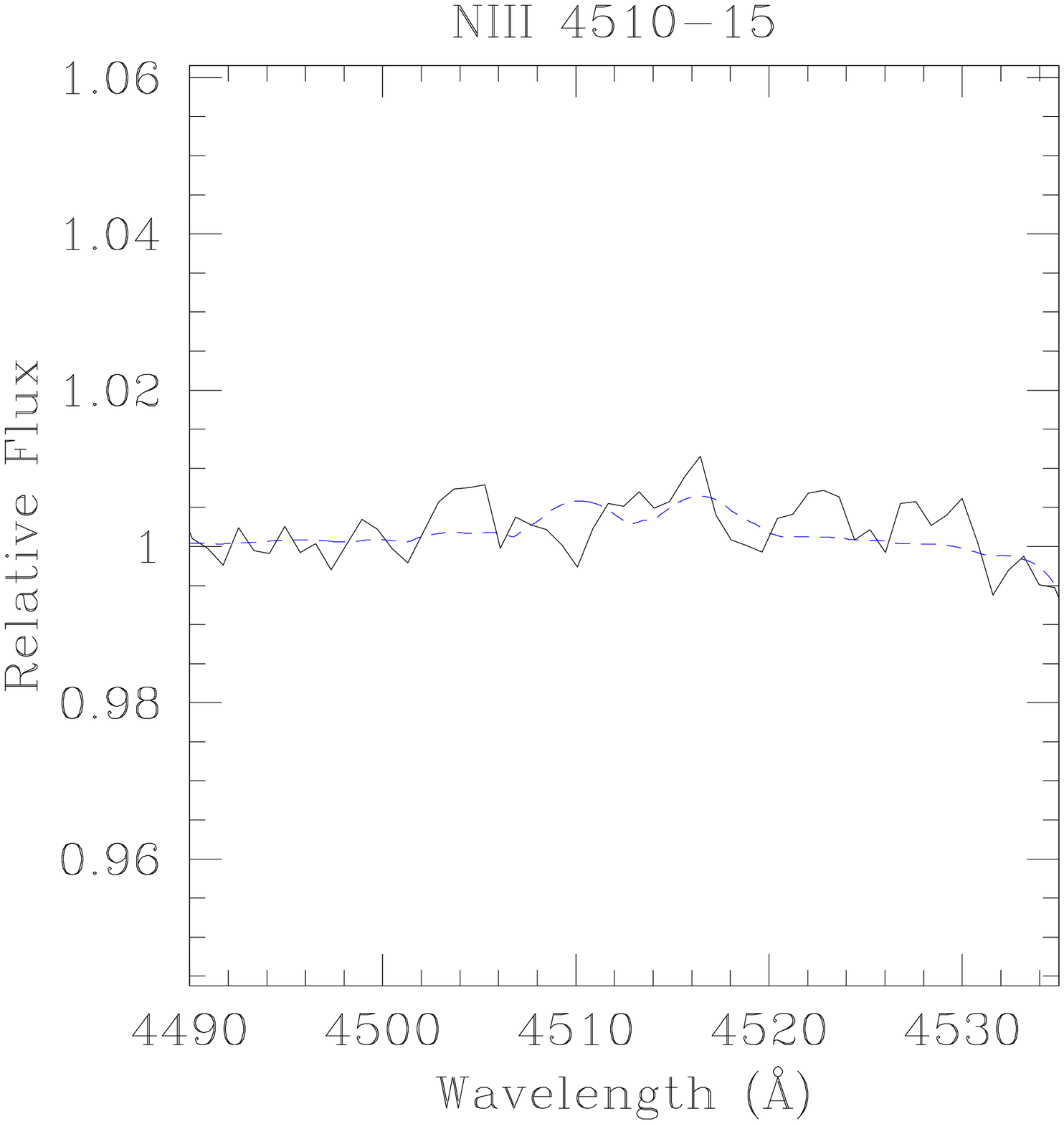}
\plotone{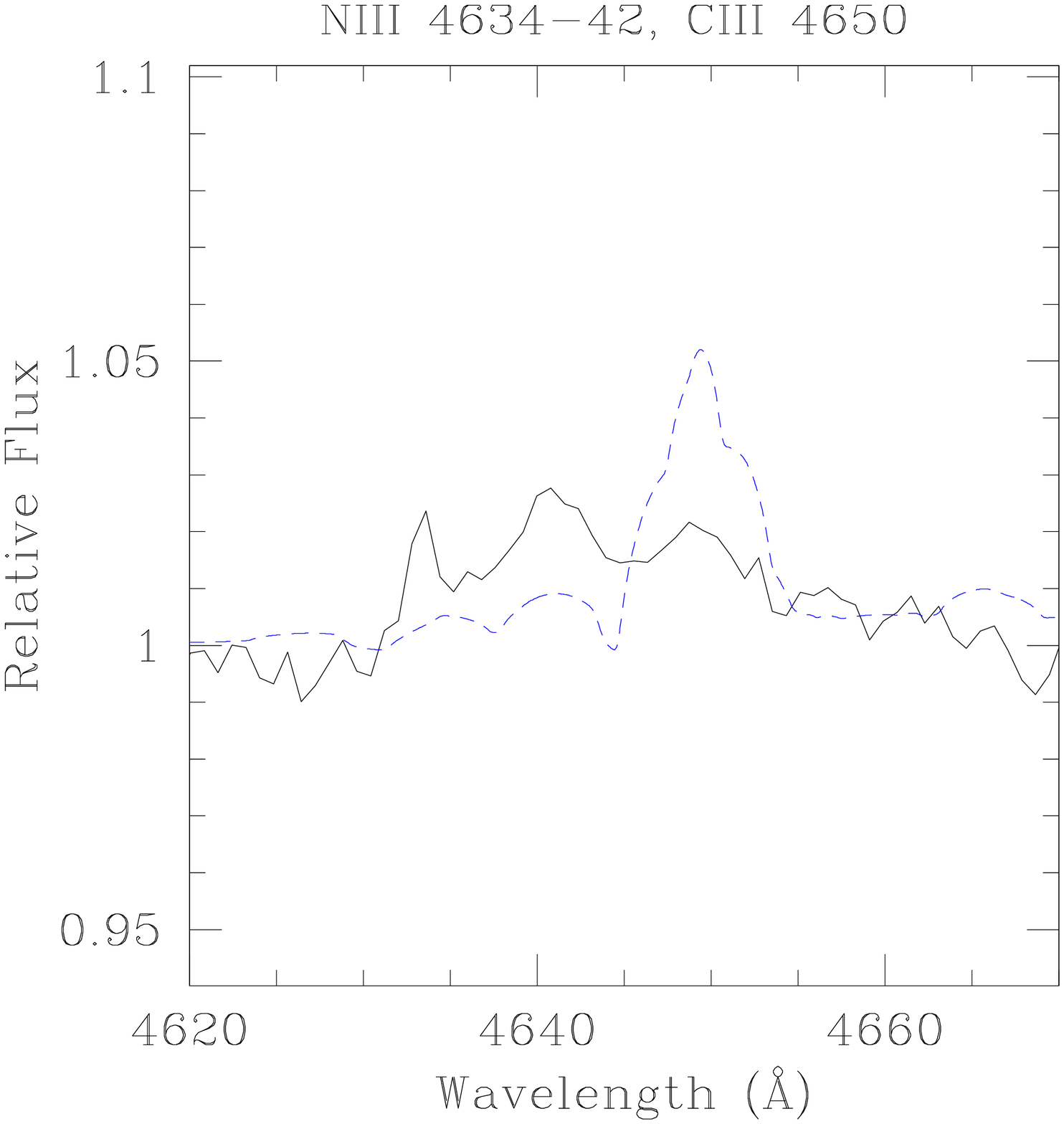}
\plotone{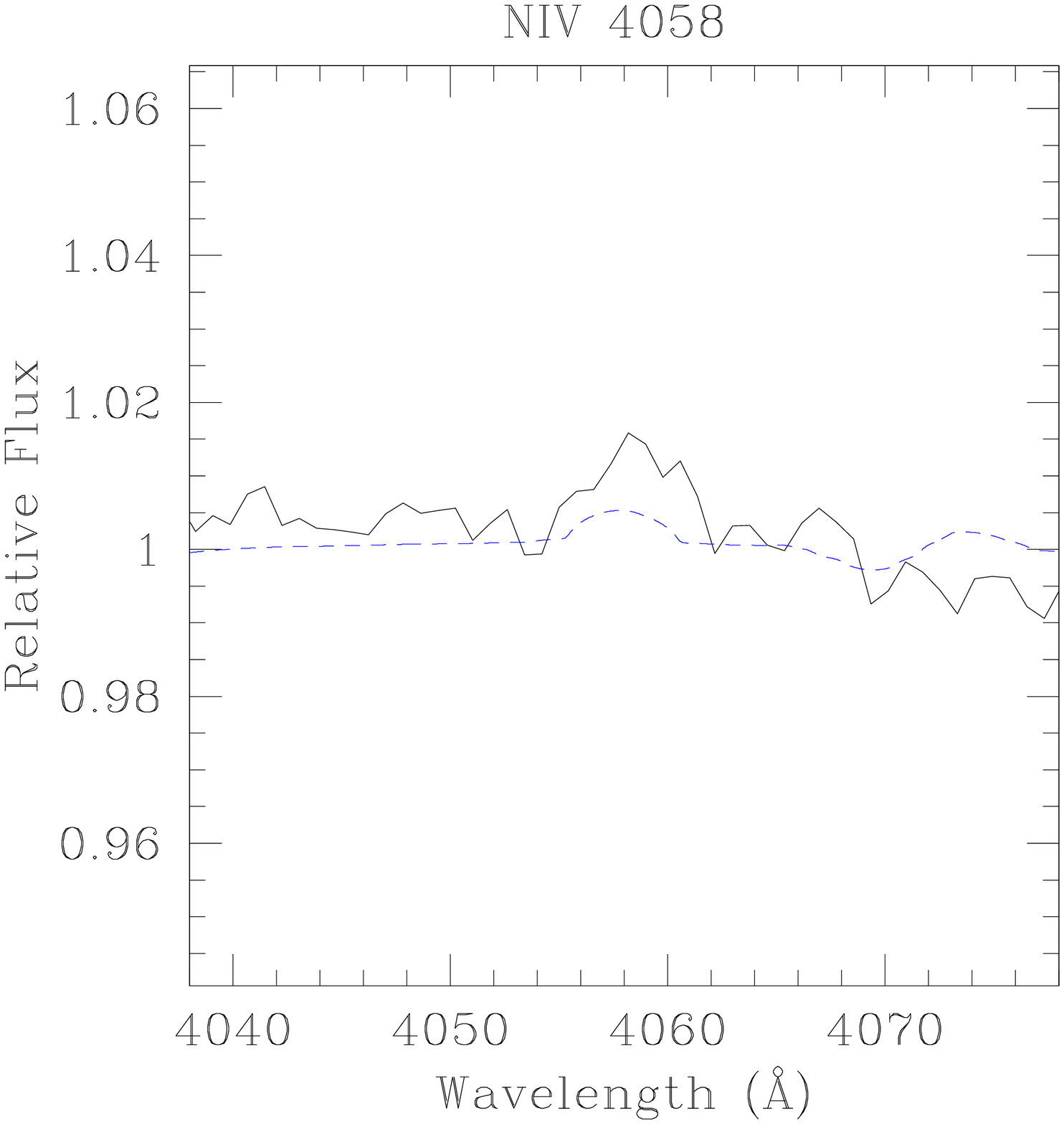}
\plotone{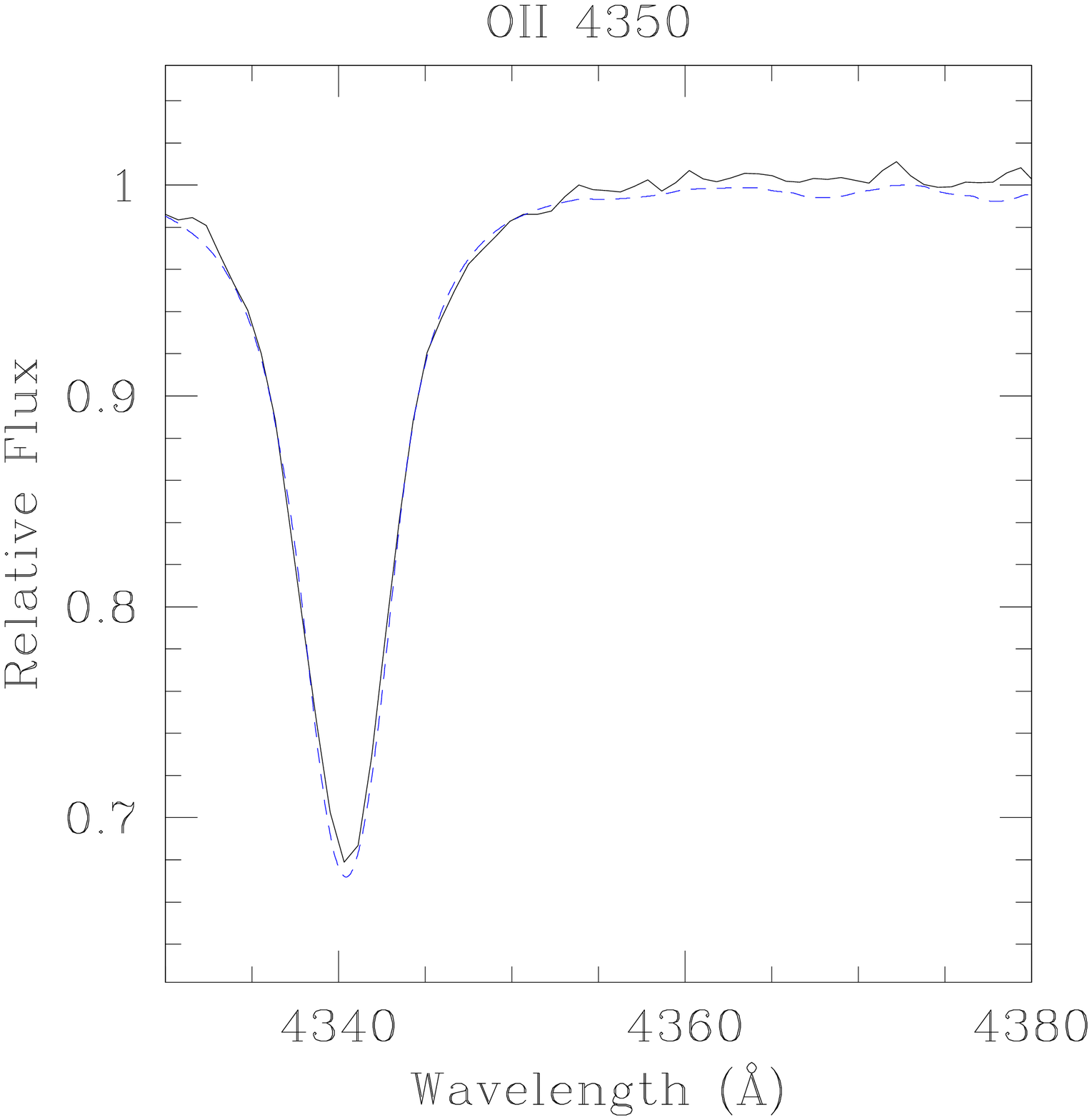}
\plotone{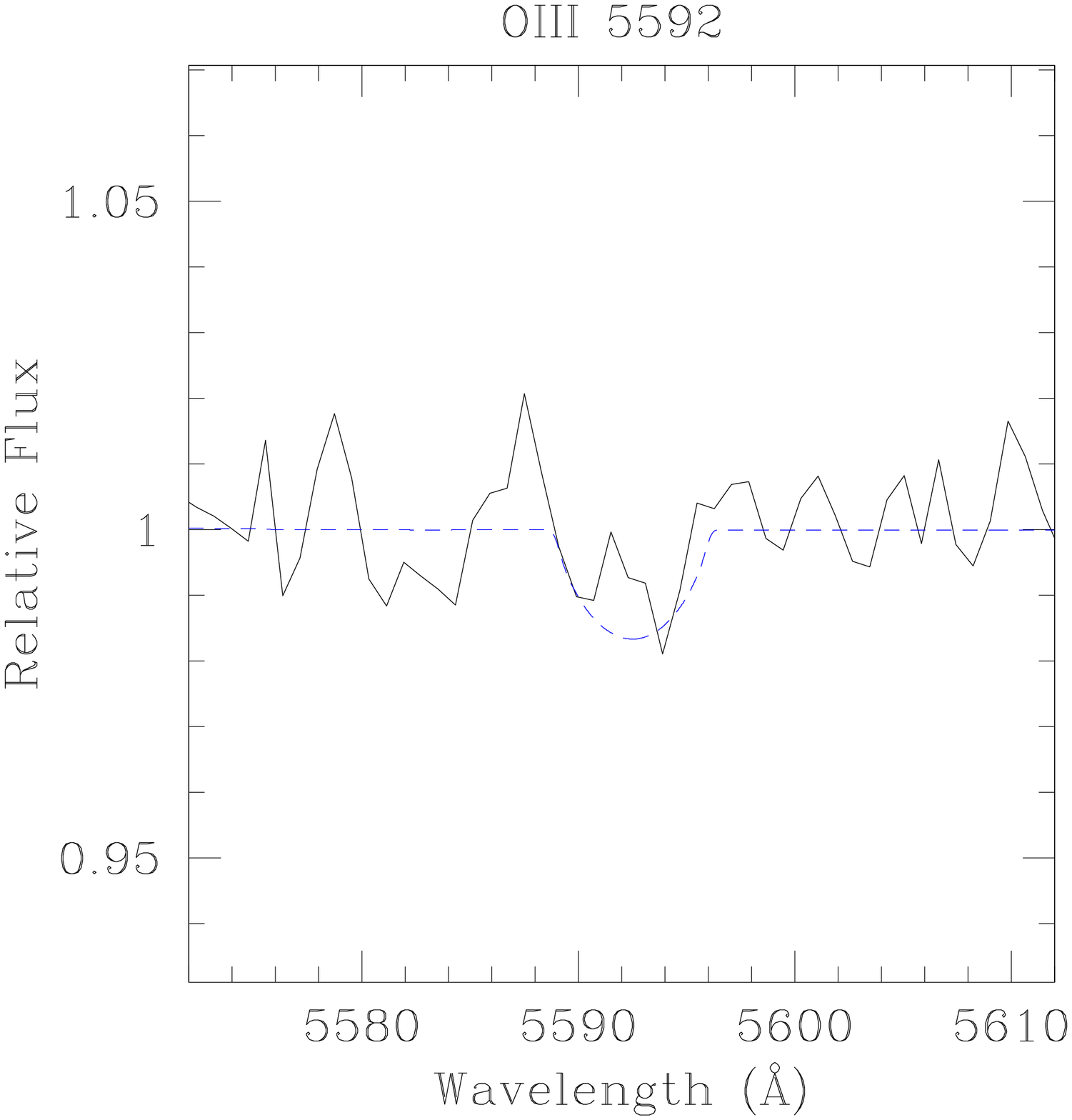}
\caption{\cmfgen\ fits for CNO lines for AzV 388, O5.5 V((f)). Black denotes the observed spectrum, and dashed blue shows the \cmfgen\ model with SMC abundances ($Z/Z_\odot=0.2$).}
\end{figure}
\clearpage
\begin{figure}
\epsscale{0.3}
\plotone{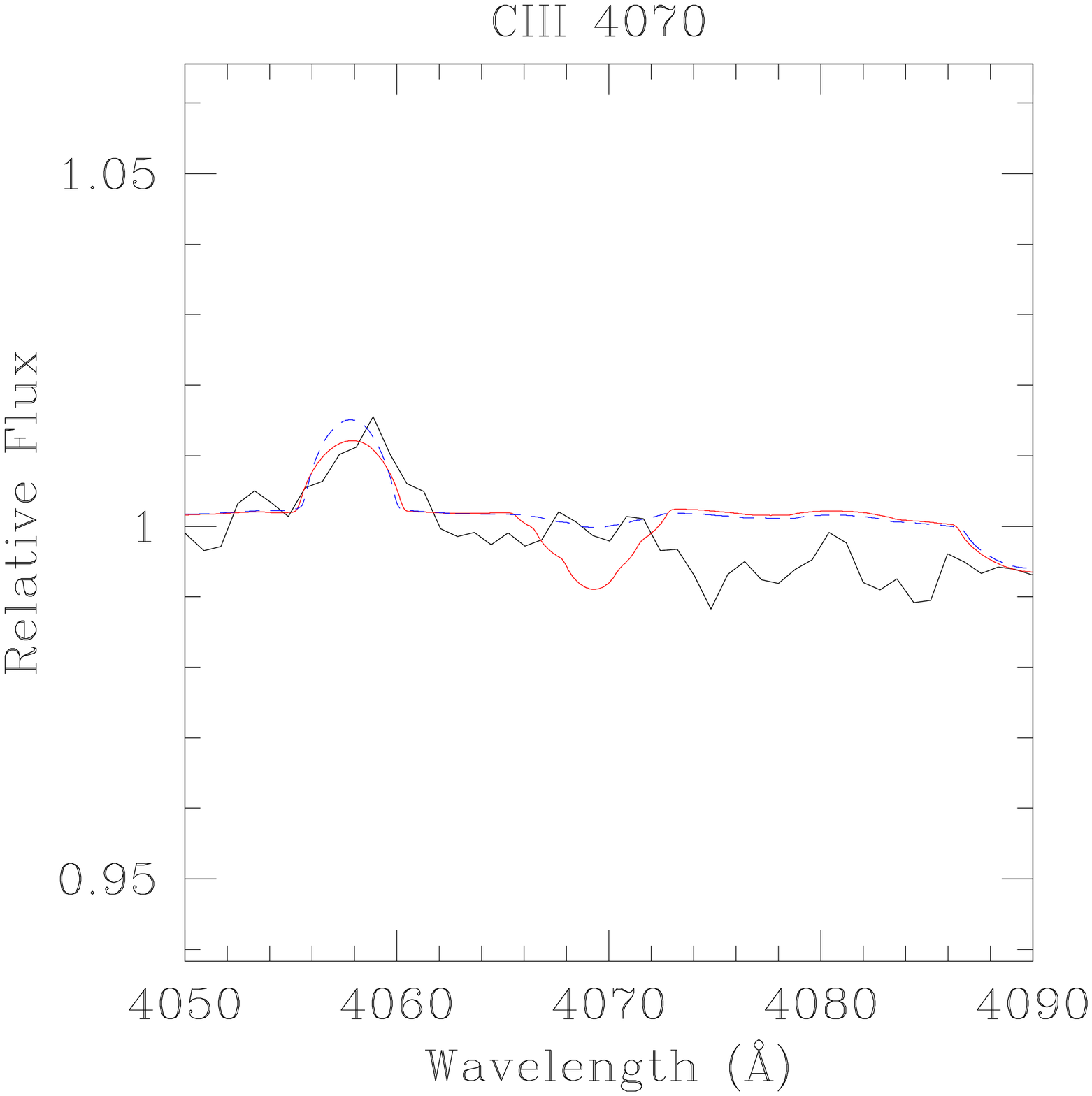}
\plotone{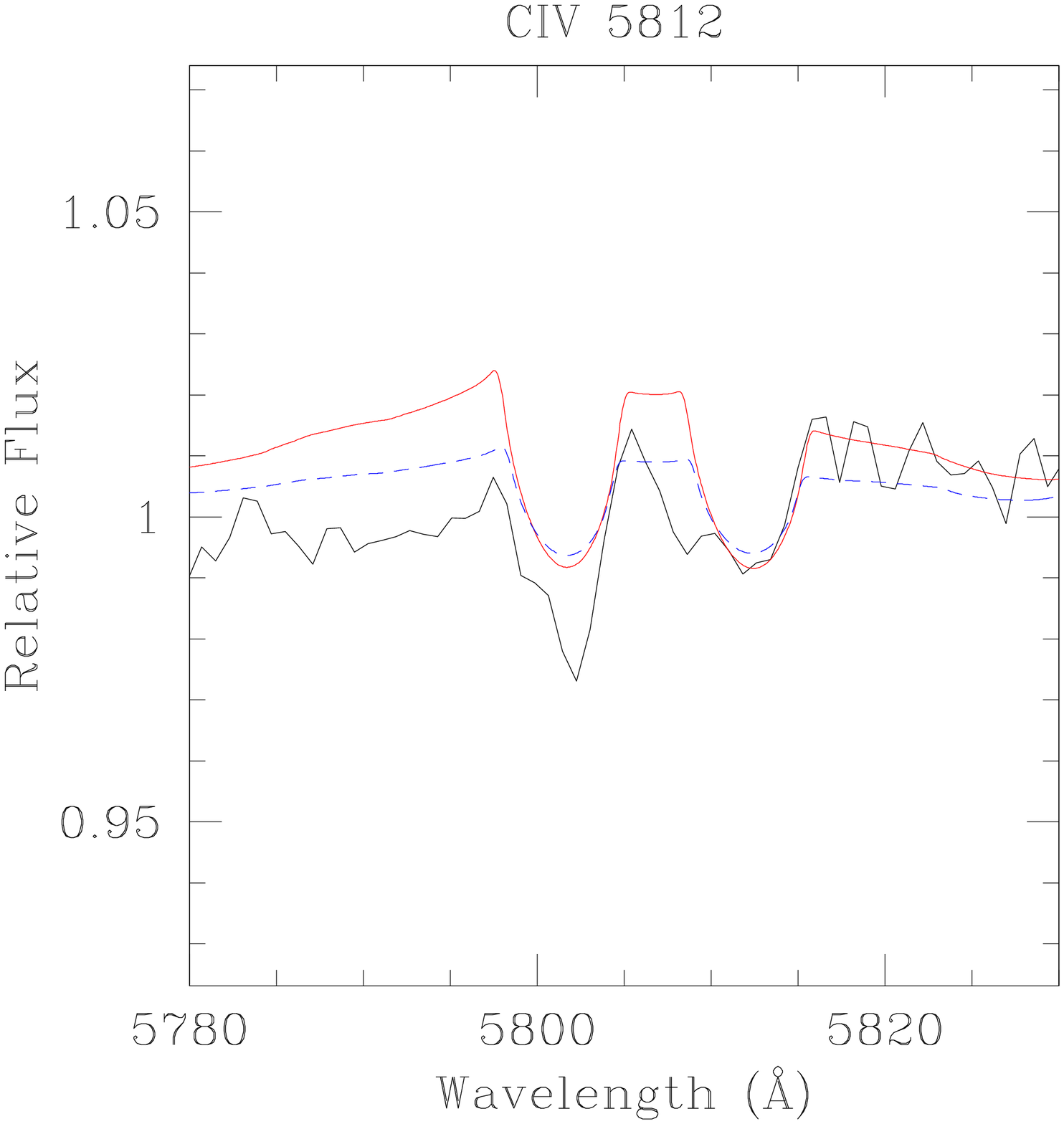}
\plotone{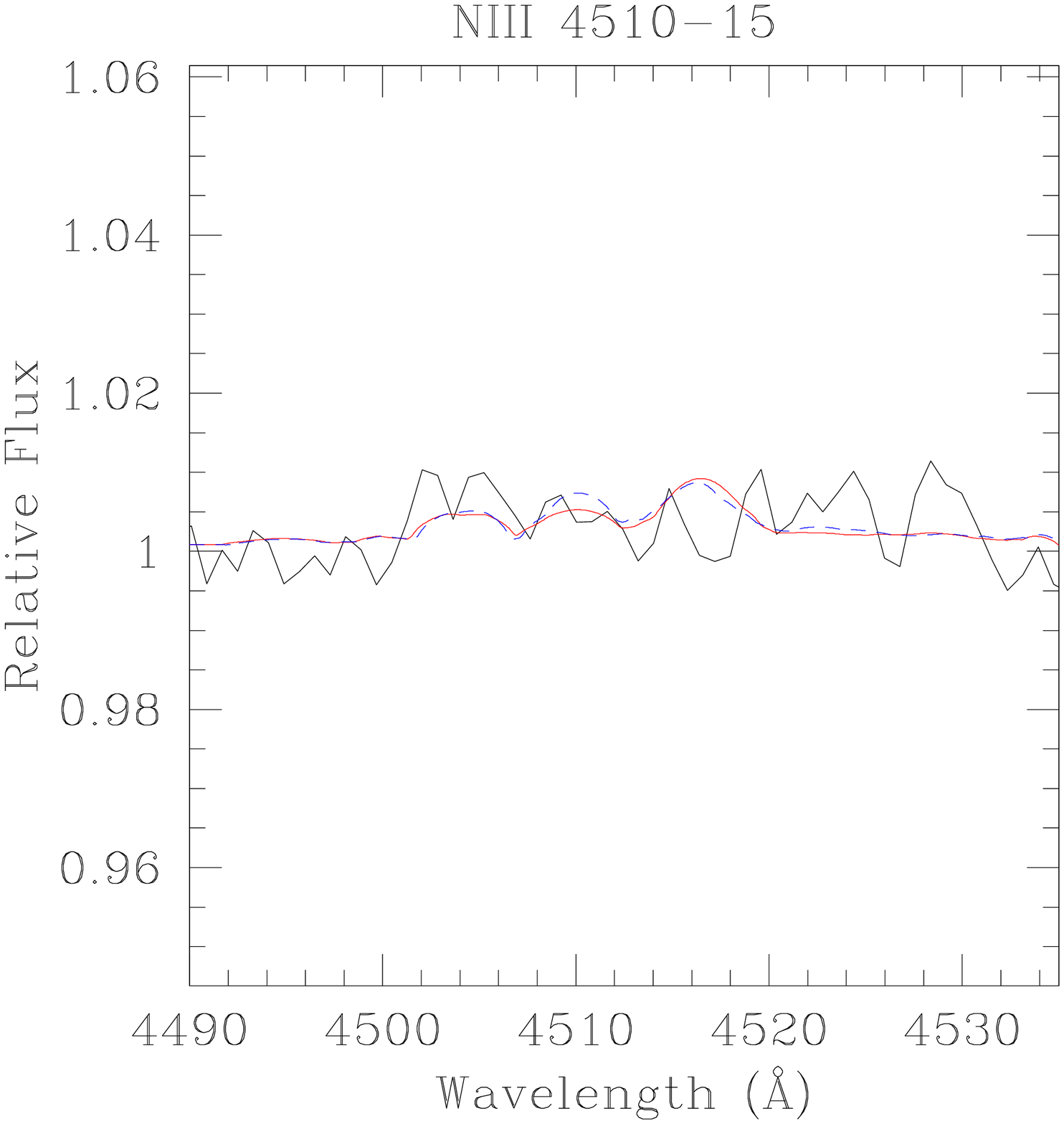}
\plotone{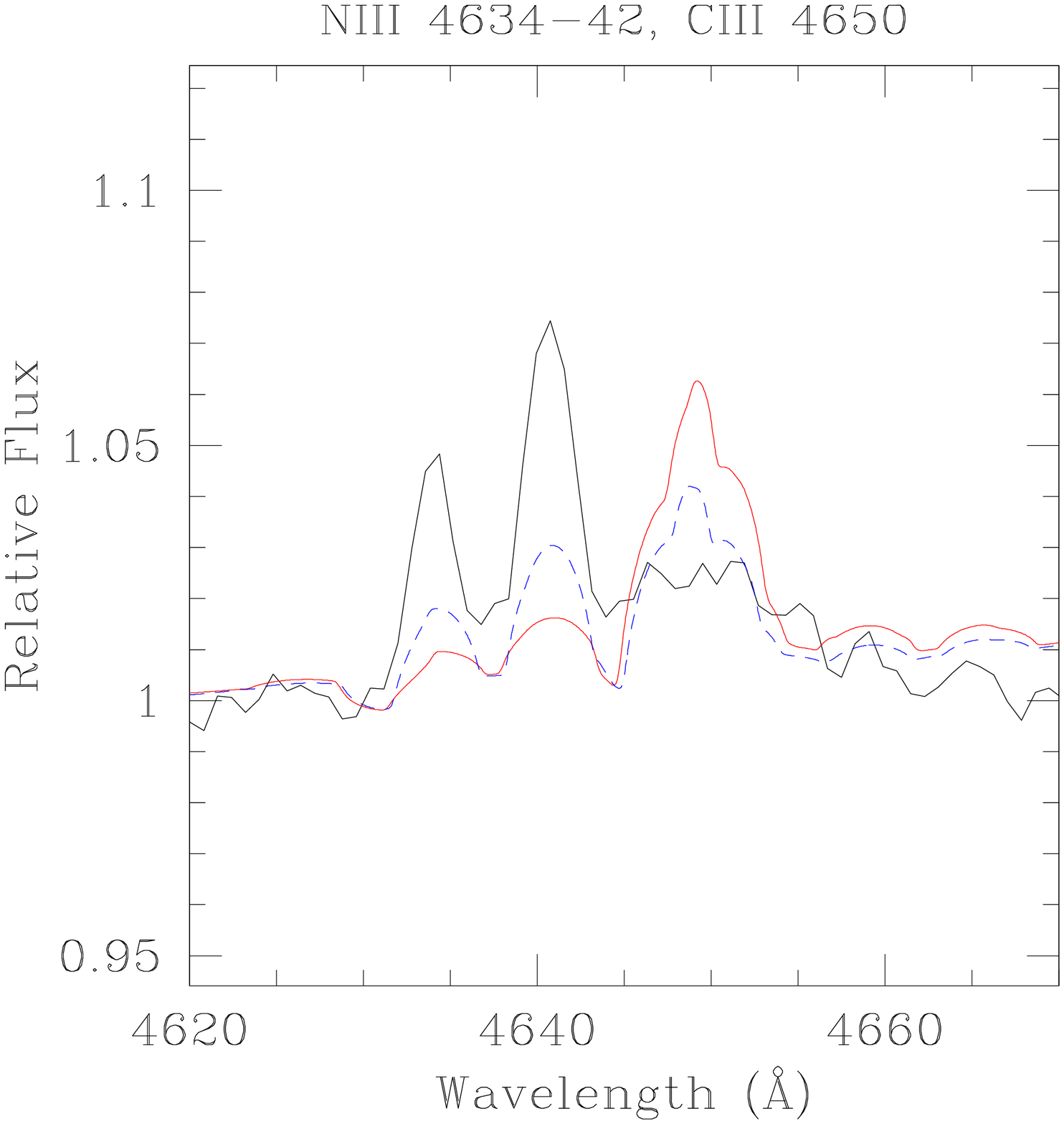}
\plotone{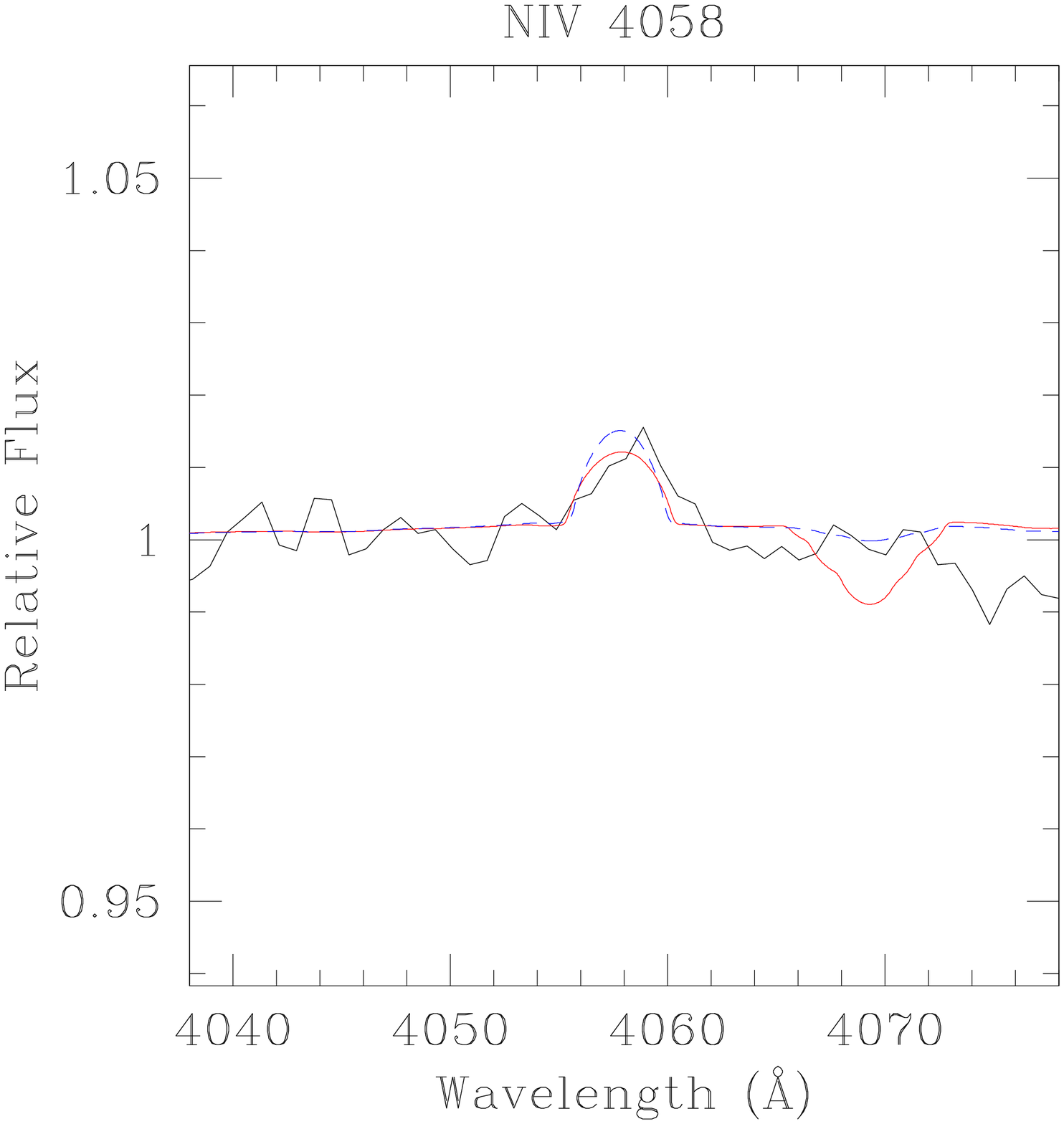}
\plotone{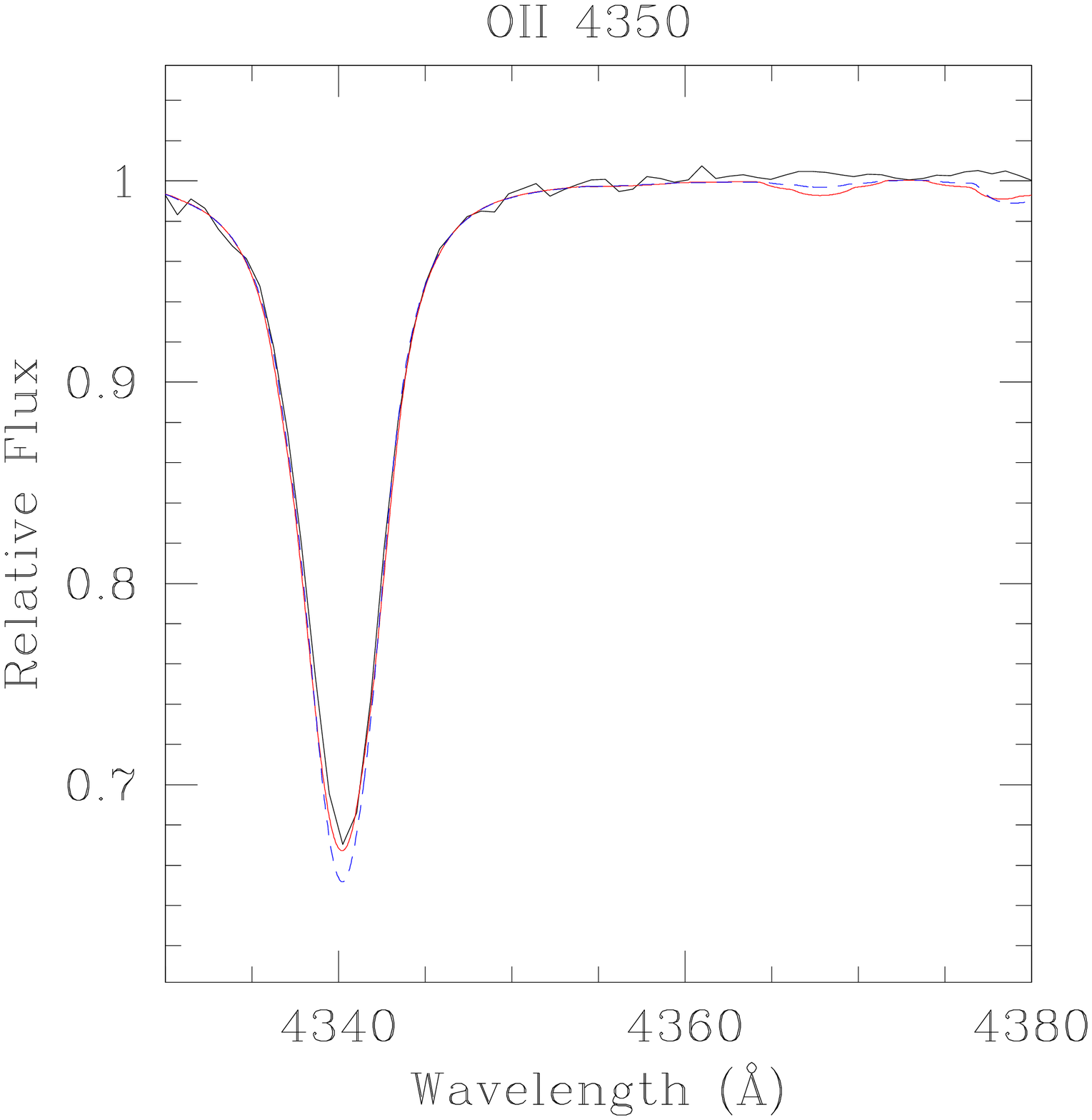}
\plotone{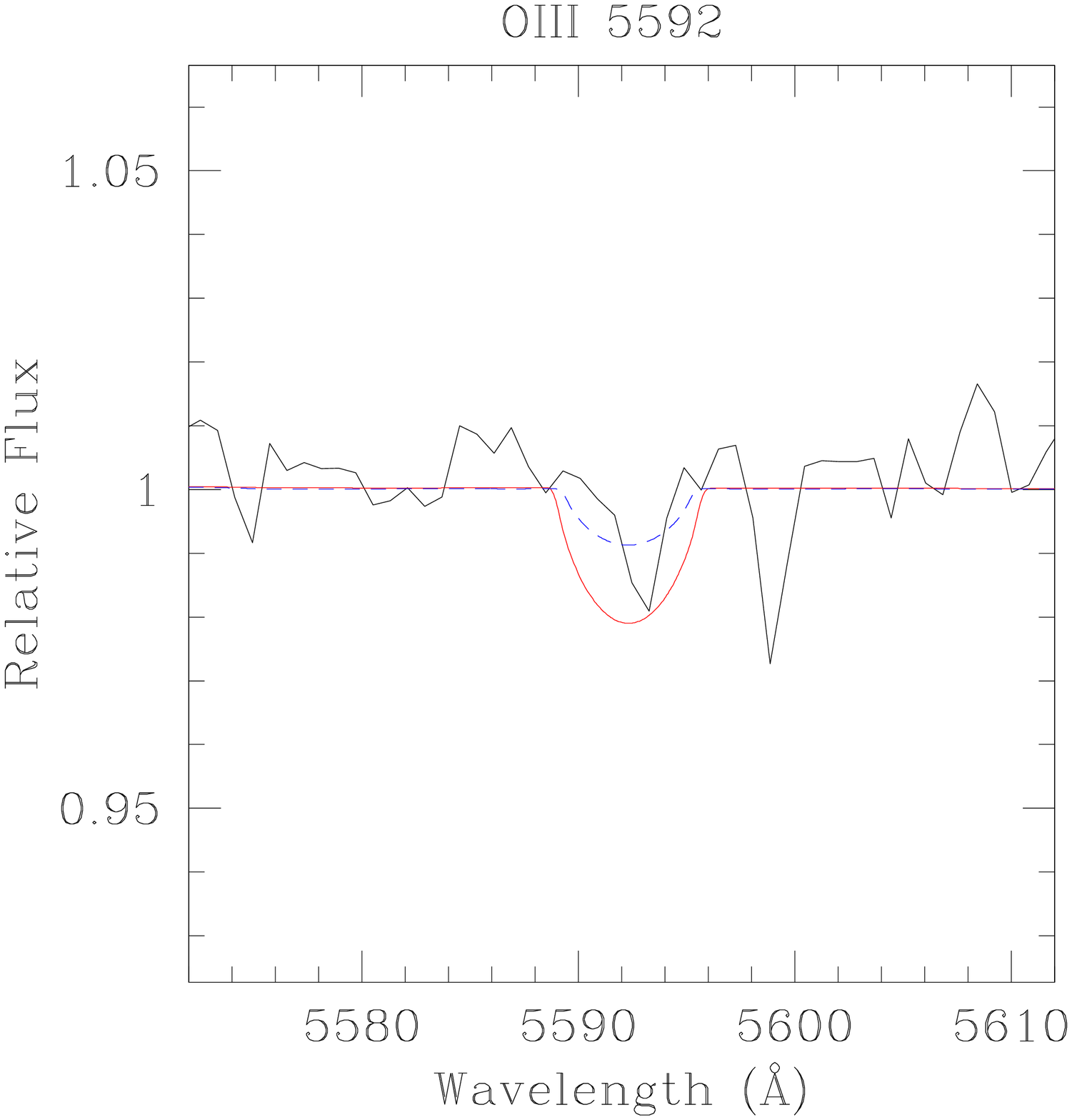}
\caption{\cmfgen\ fits for CNO lines for AzV 75, O5.5 I(f). Black denotes the observed spectrum, solid red indicates the \cmfgen\ model with SMC abundances ($Z/Z_\odot=0.2$), and the dashed blue shows the \cmfgen\ model with C and O decreased by a factor of 2.5 and N increased by a factor of 2.4.}
\end{figure}
\clearpage
\begin{figure}
\epsscale{0.3}
\plotone{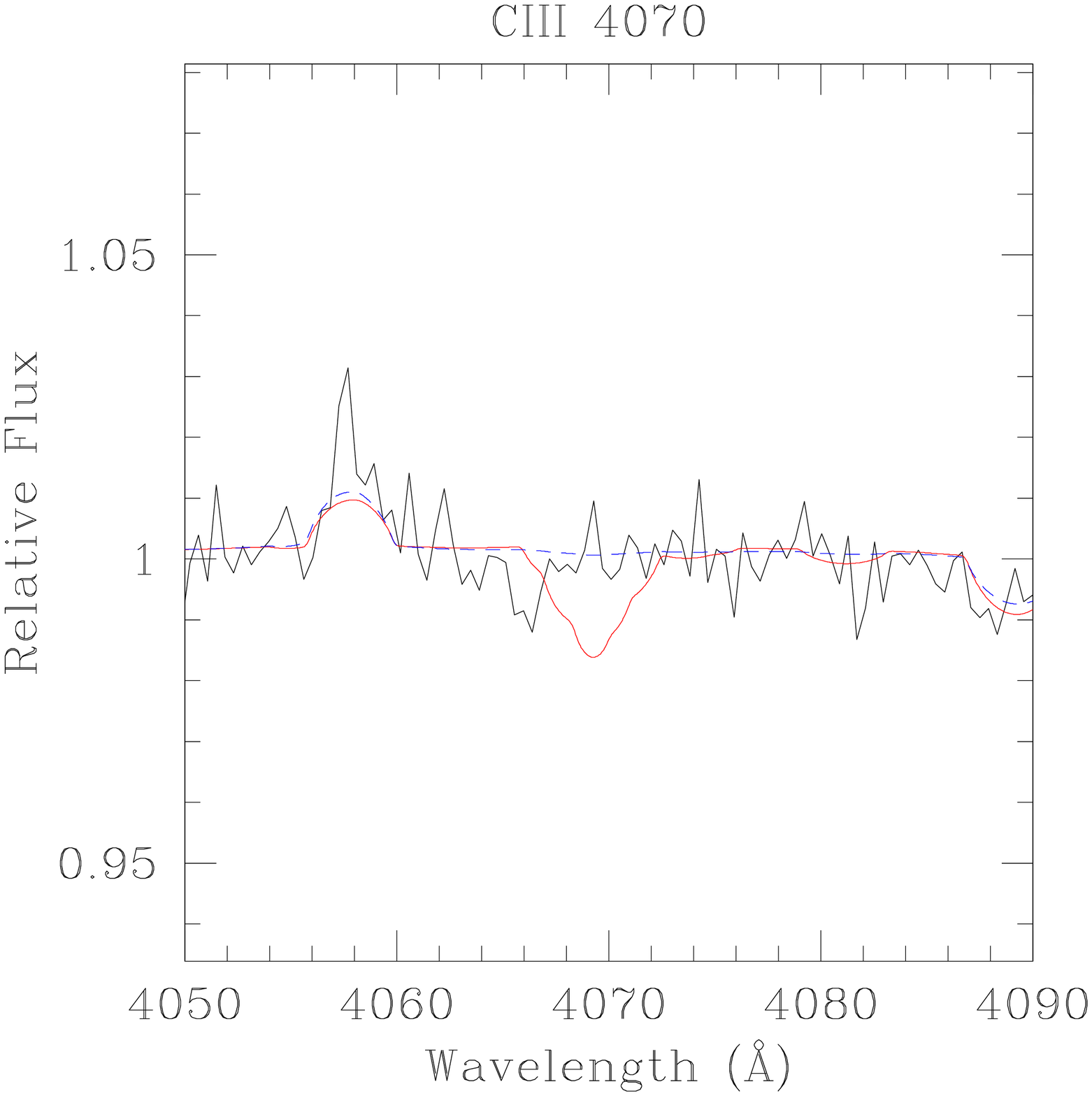}
\plotone{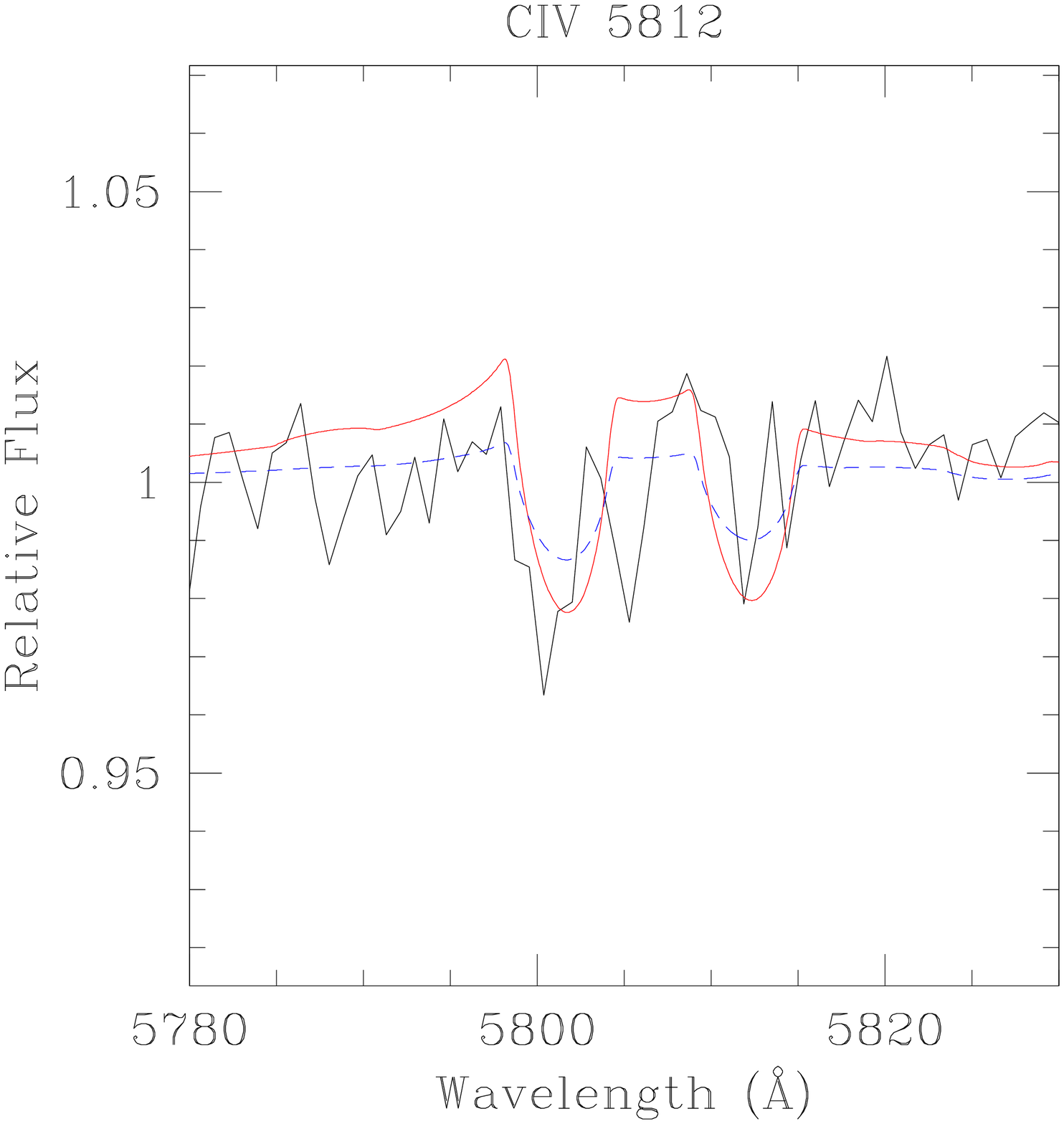}
\plotone{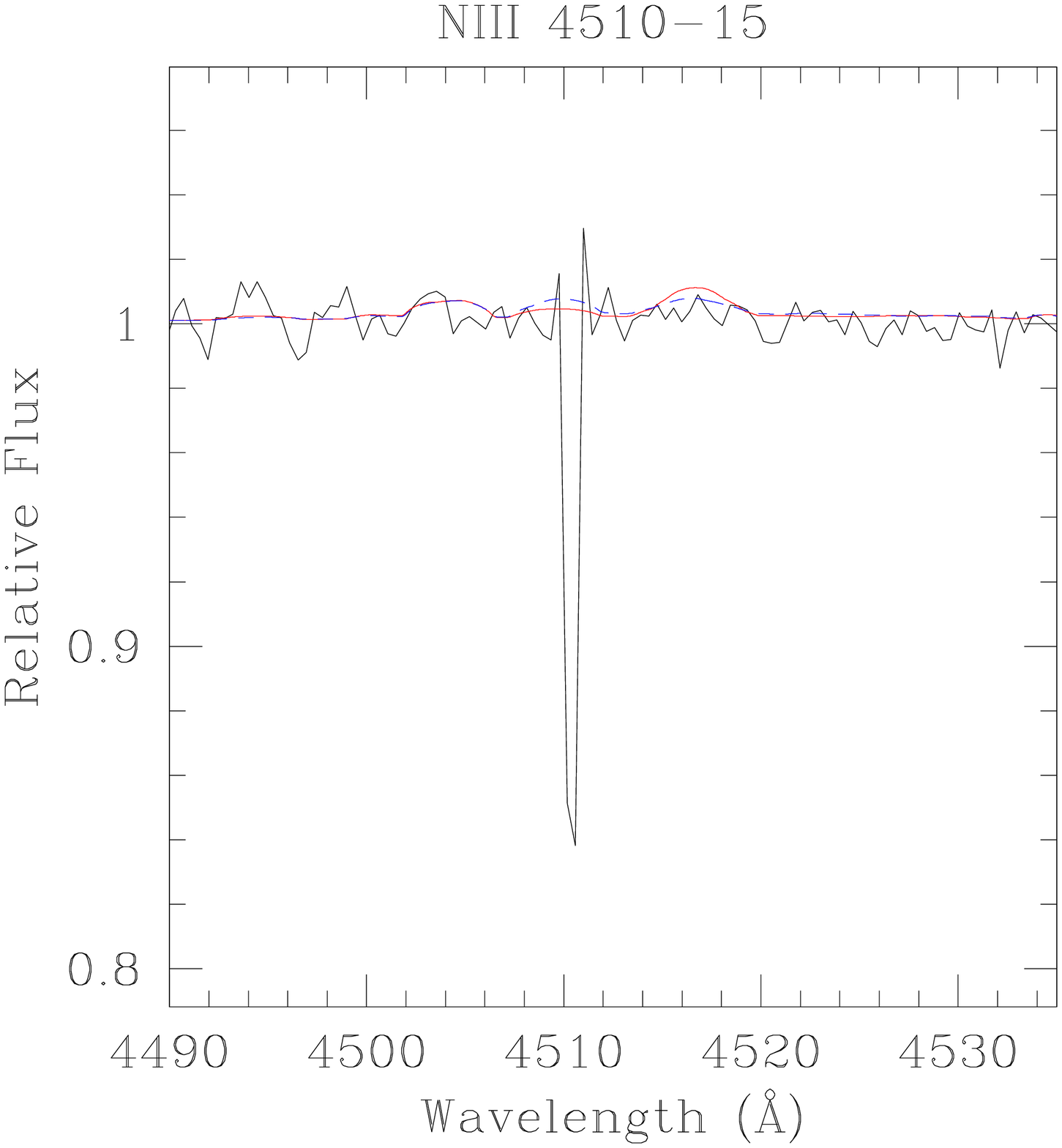}
\plotone{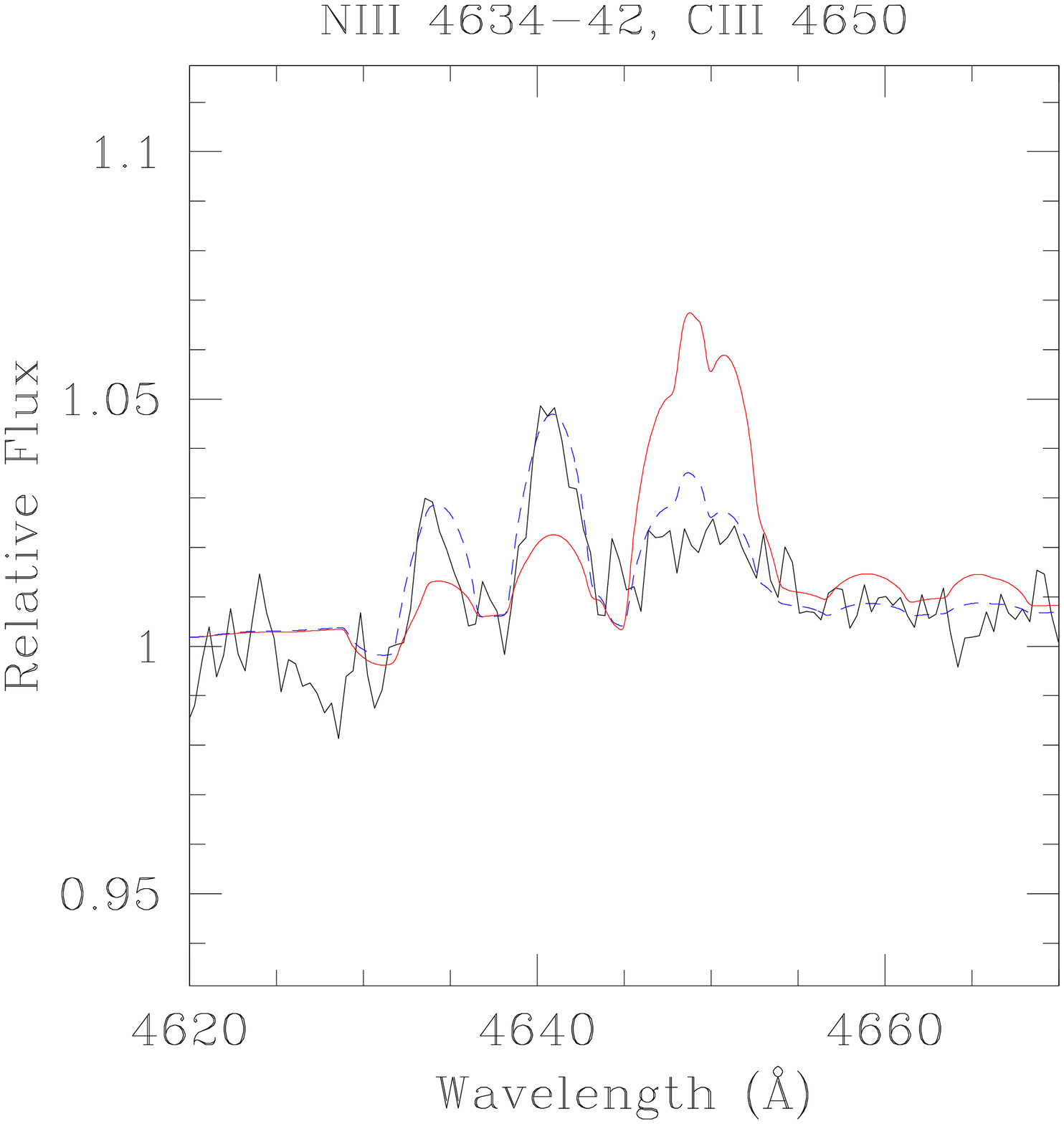}
\plotone{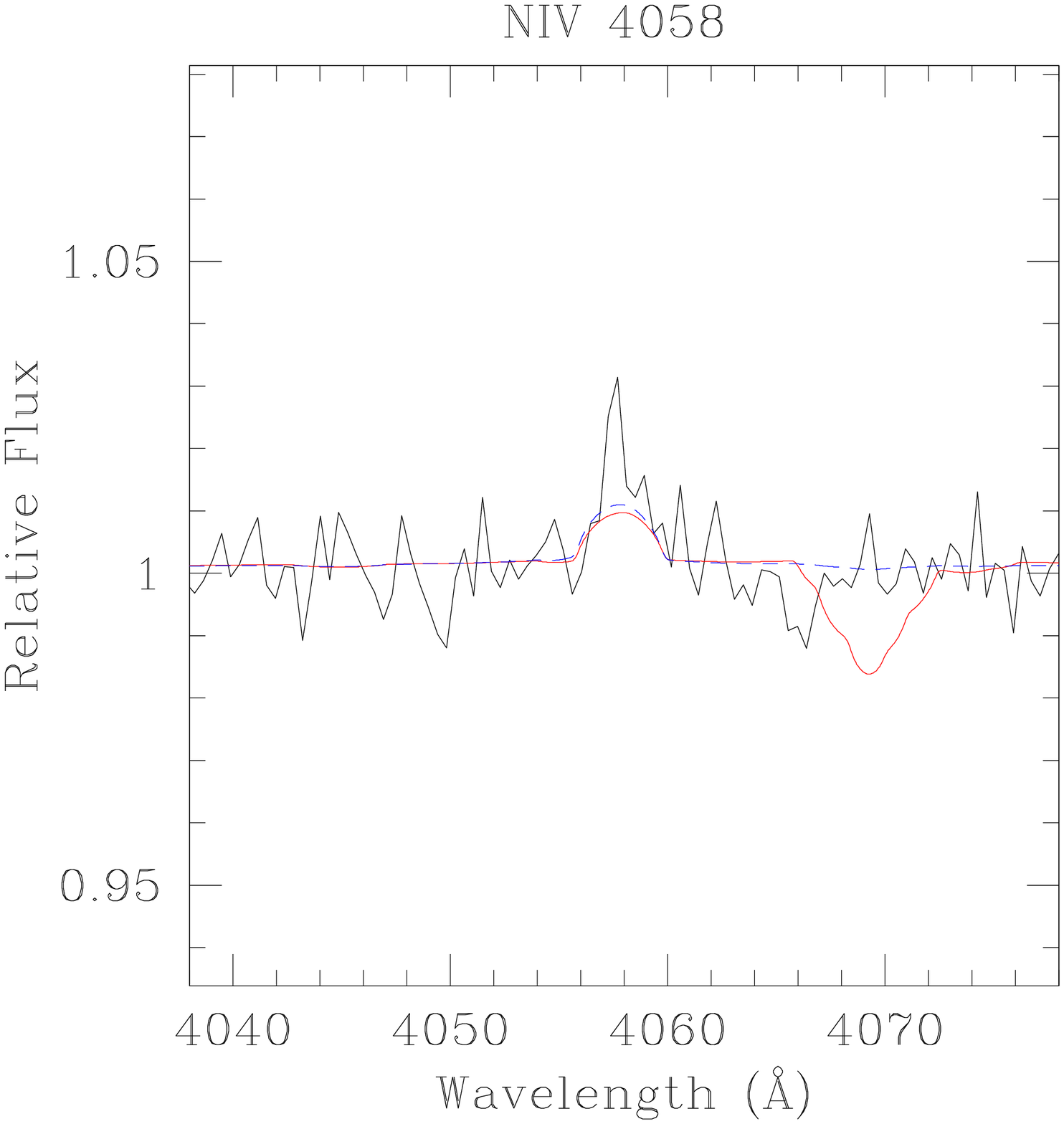}
\plotone{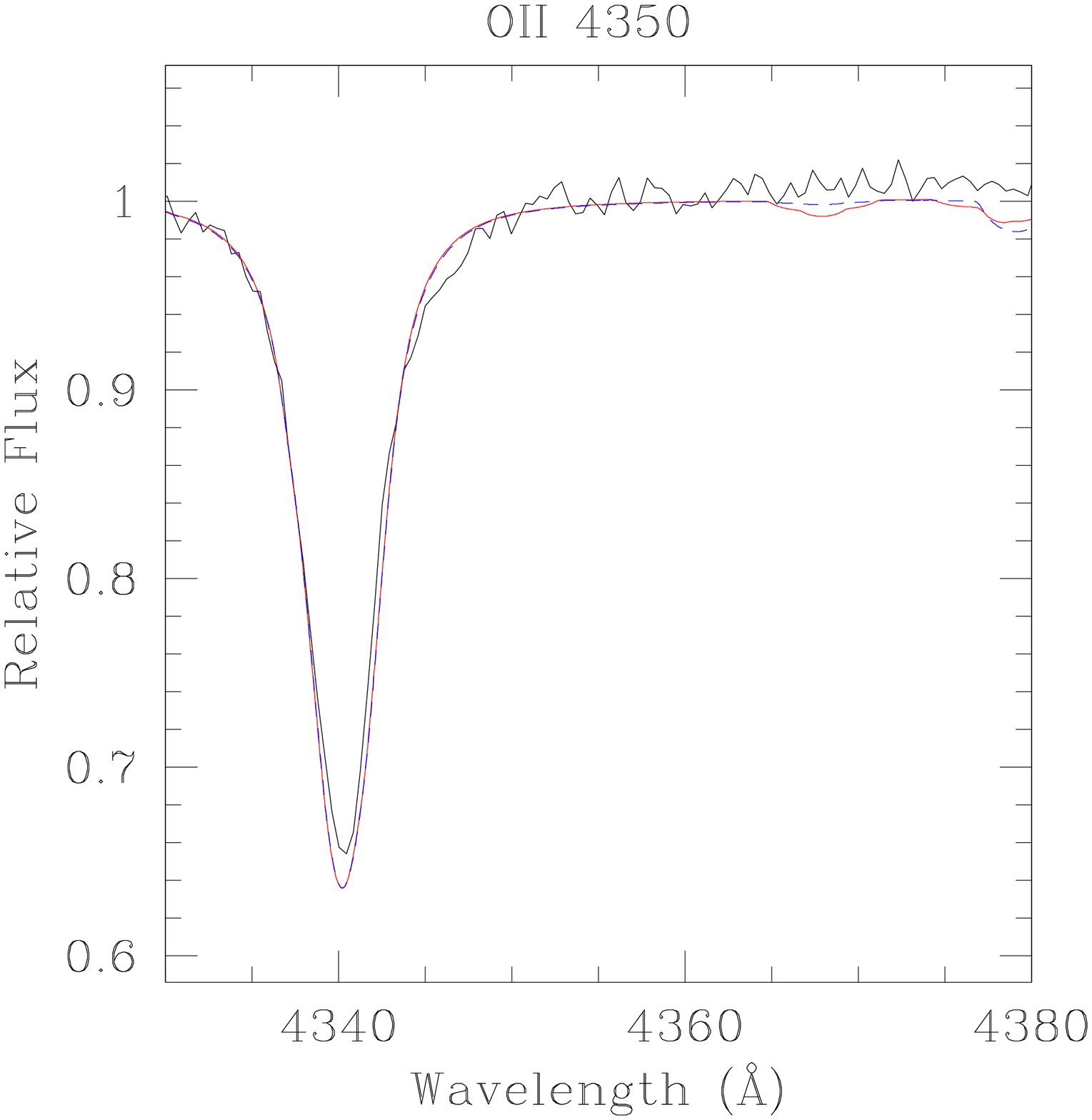}
\plotone{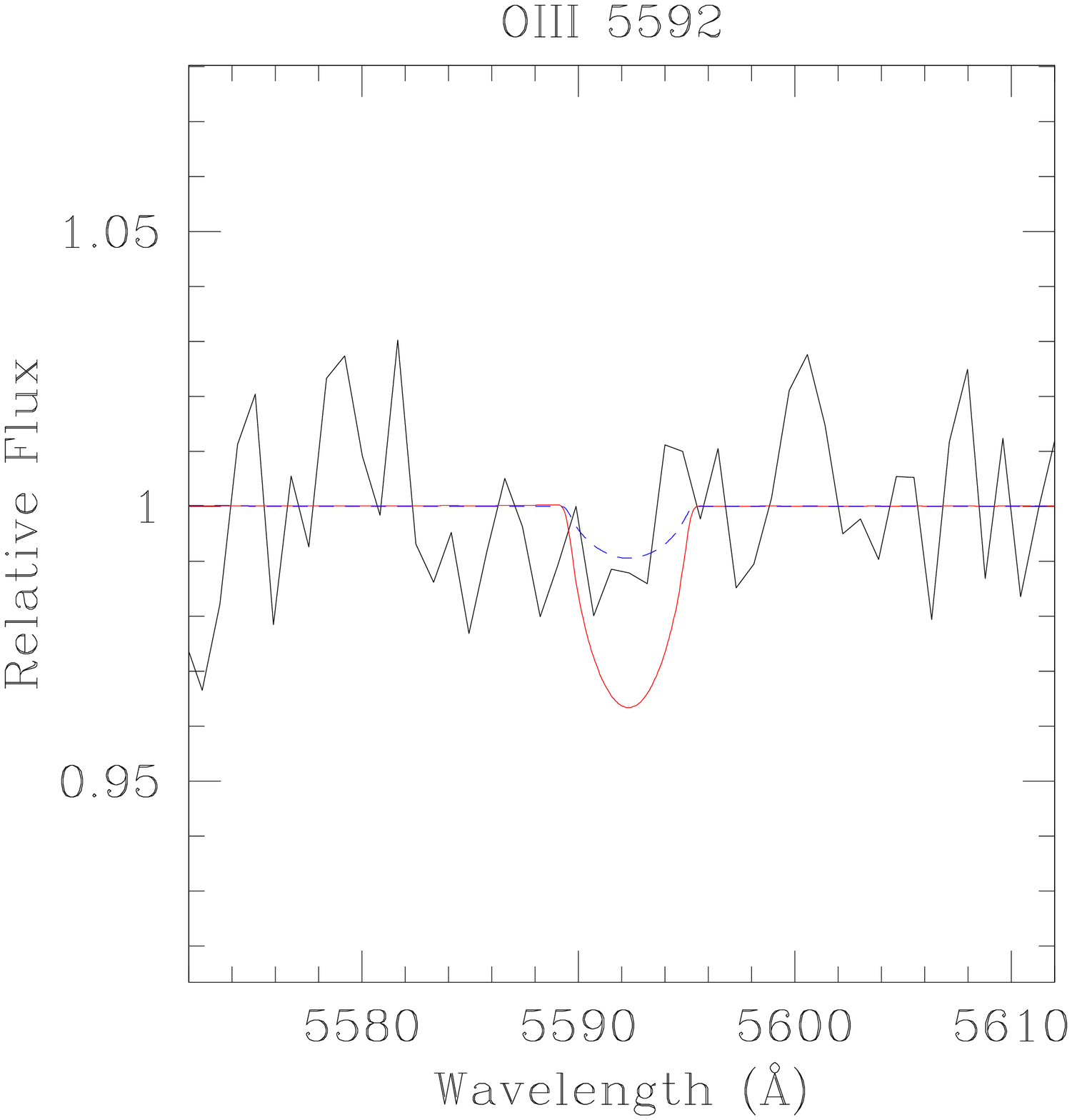}
\caption{\cmfgen\ fits for CNO lines for AzV 26, O6 I(f). Black denotes the observed spectrum, solid red indicates the \cmfgen\ model with SMC abundances ($Z/Z_\odot=0.2$), and the dashed blue shows the \cmfgen\ model with C and O decreased by a 4 and N increased by a factor of 3.}
\end{figure}
\clearpage
\begin{figure}
\epsscale{0.3}
\plotone{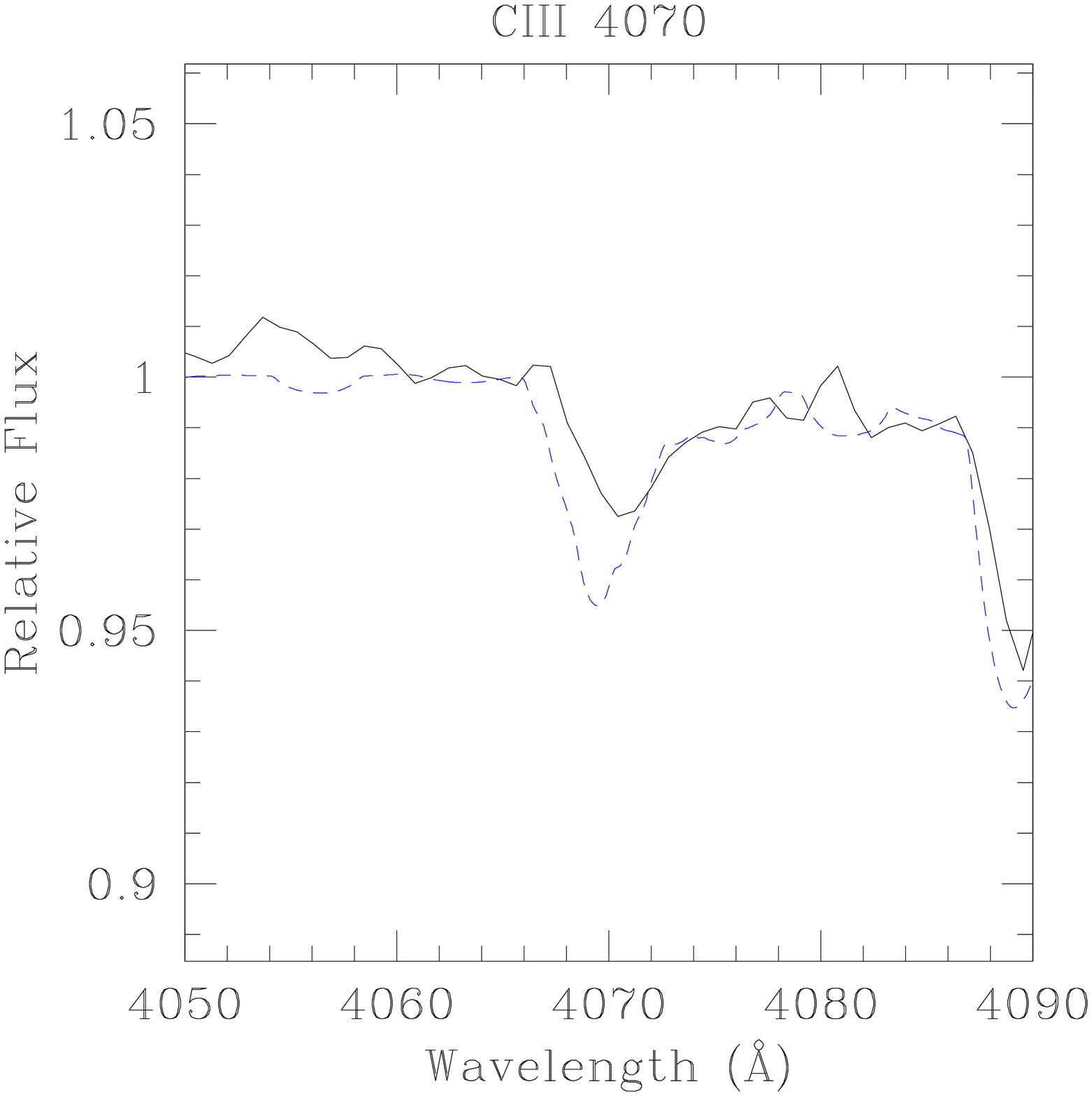}
\plotone{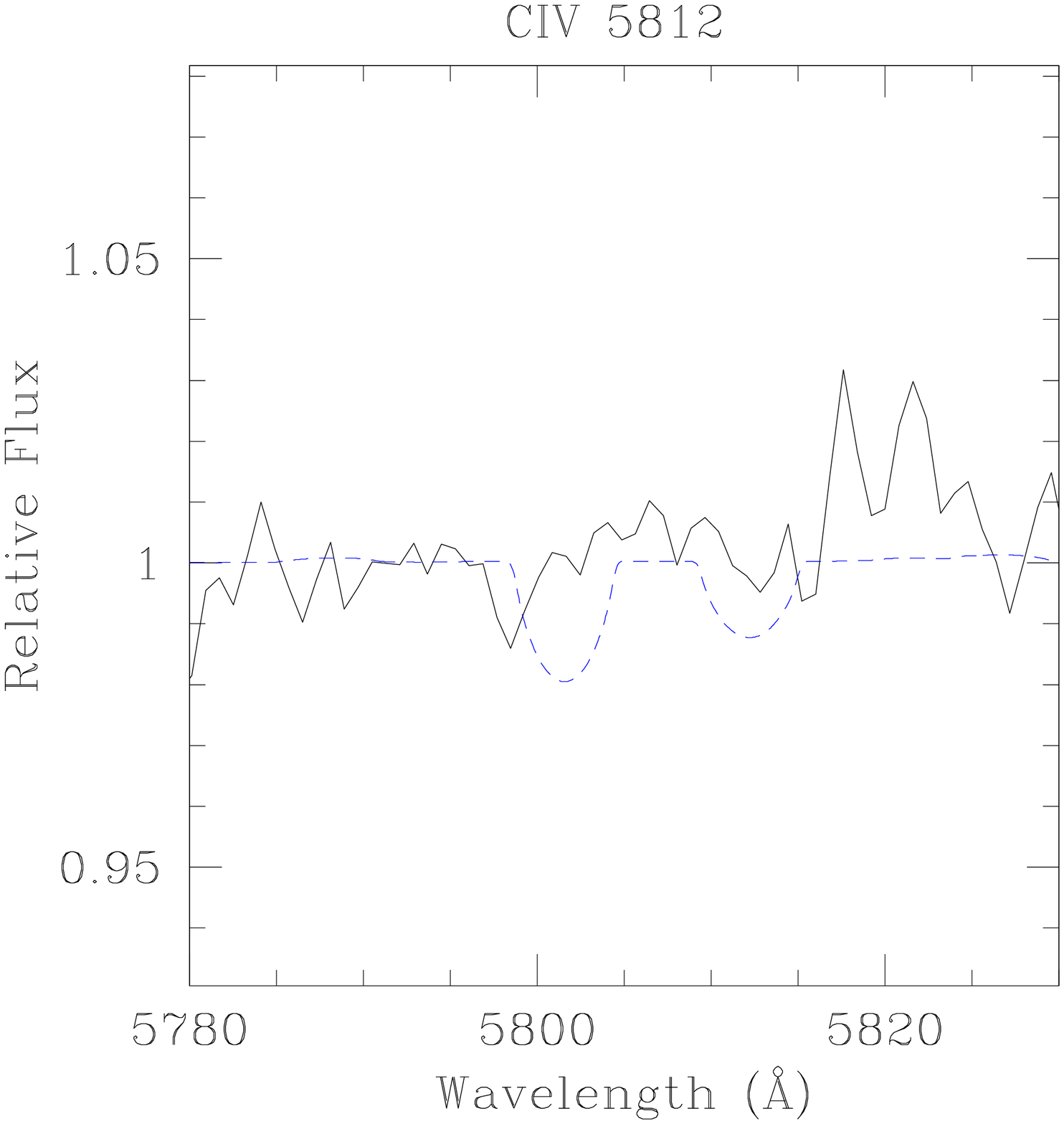}
\plotone{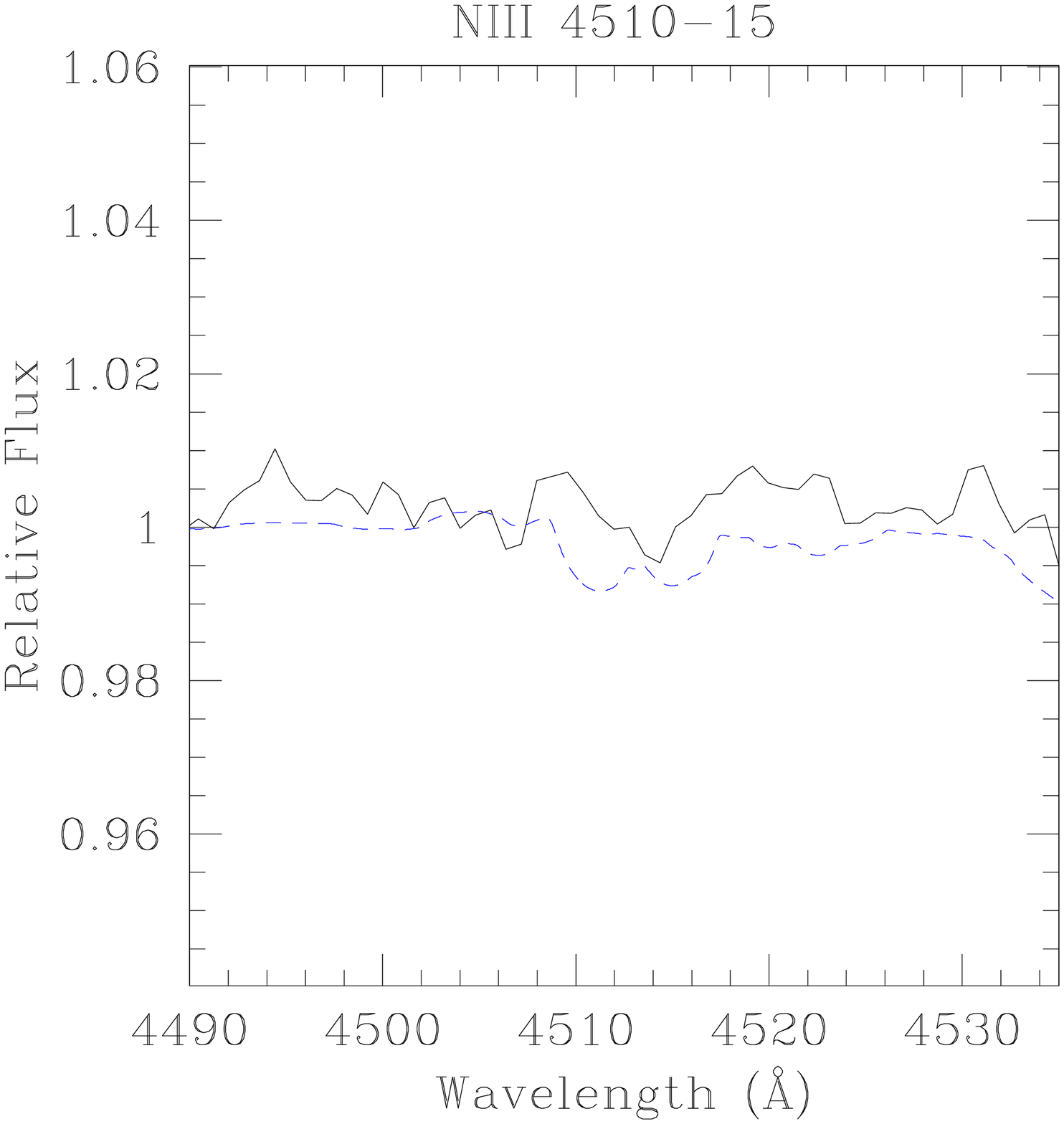}
\plotone{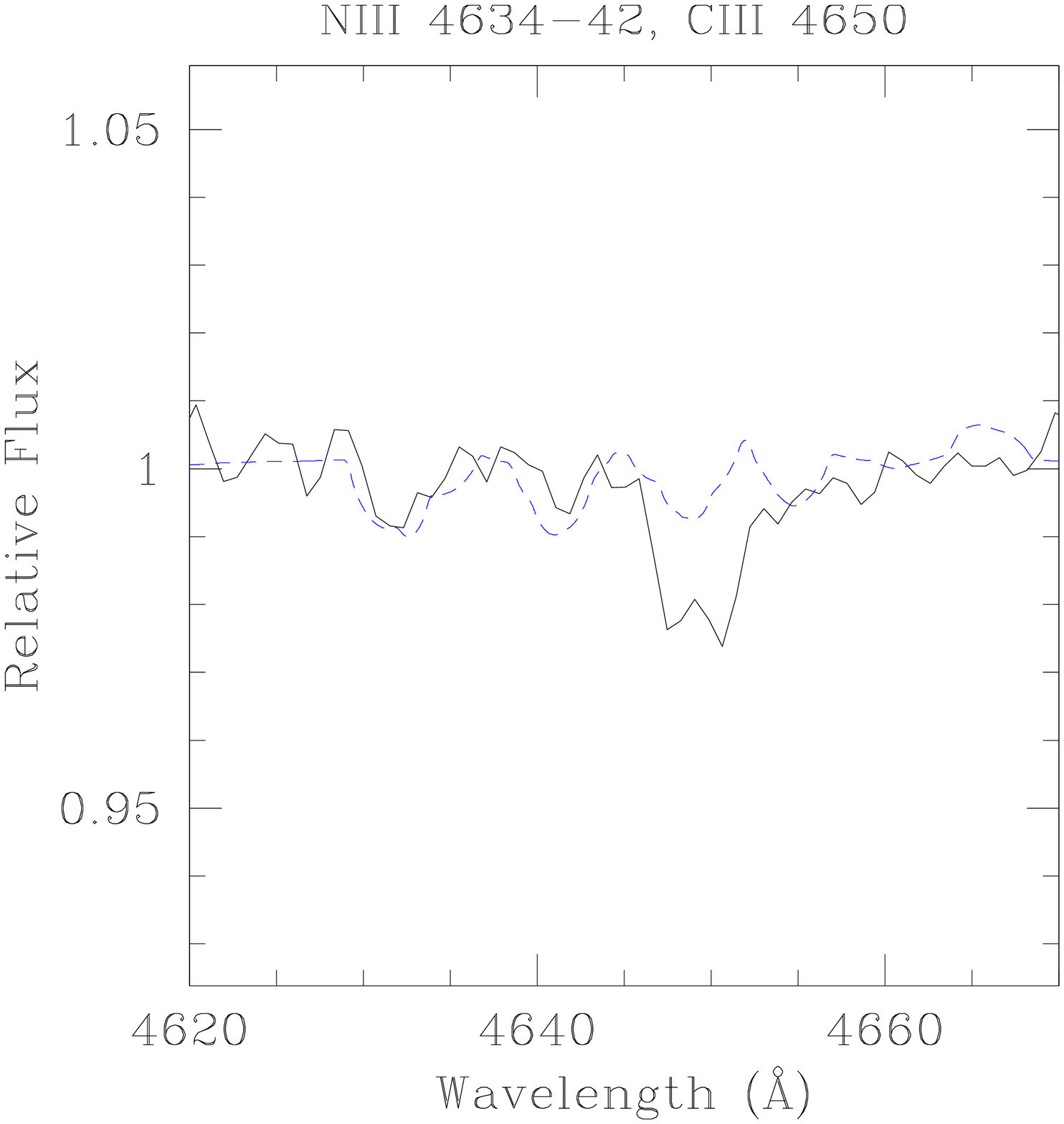}
\plotone{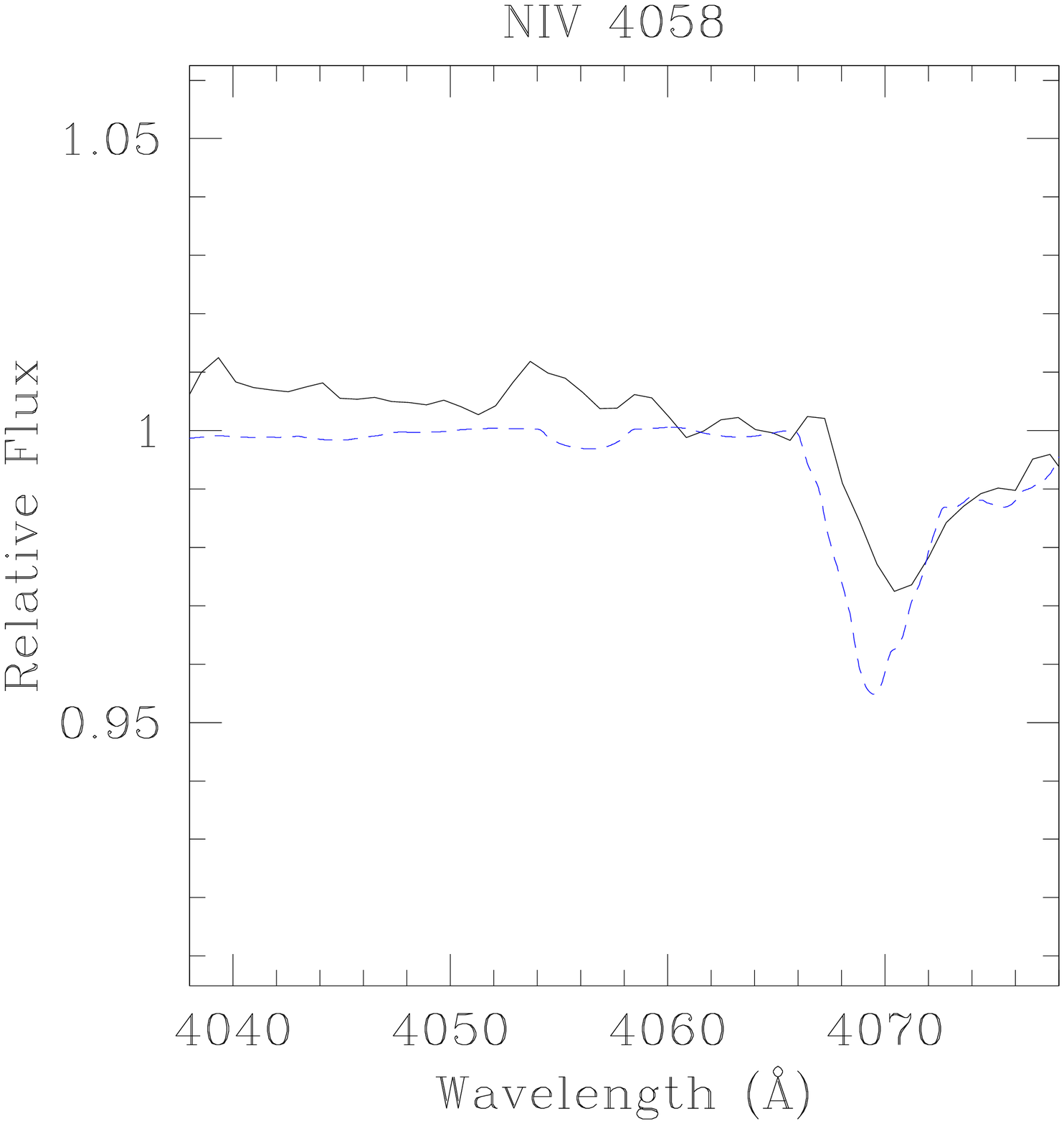}
\plotone{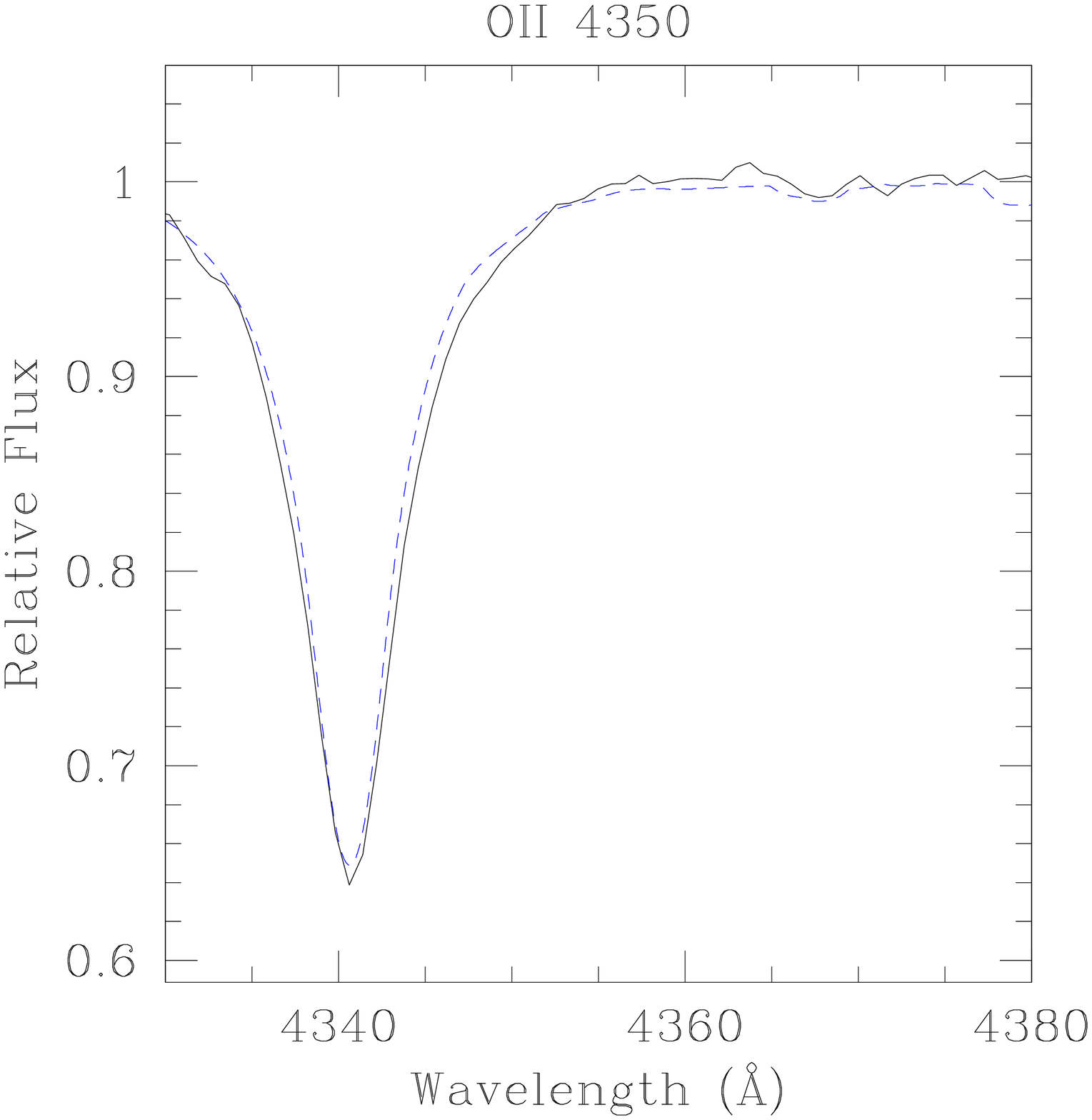}
\plotone{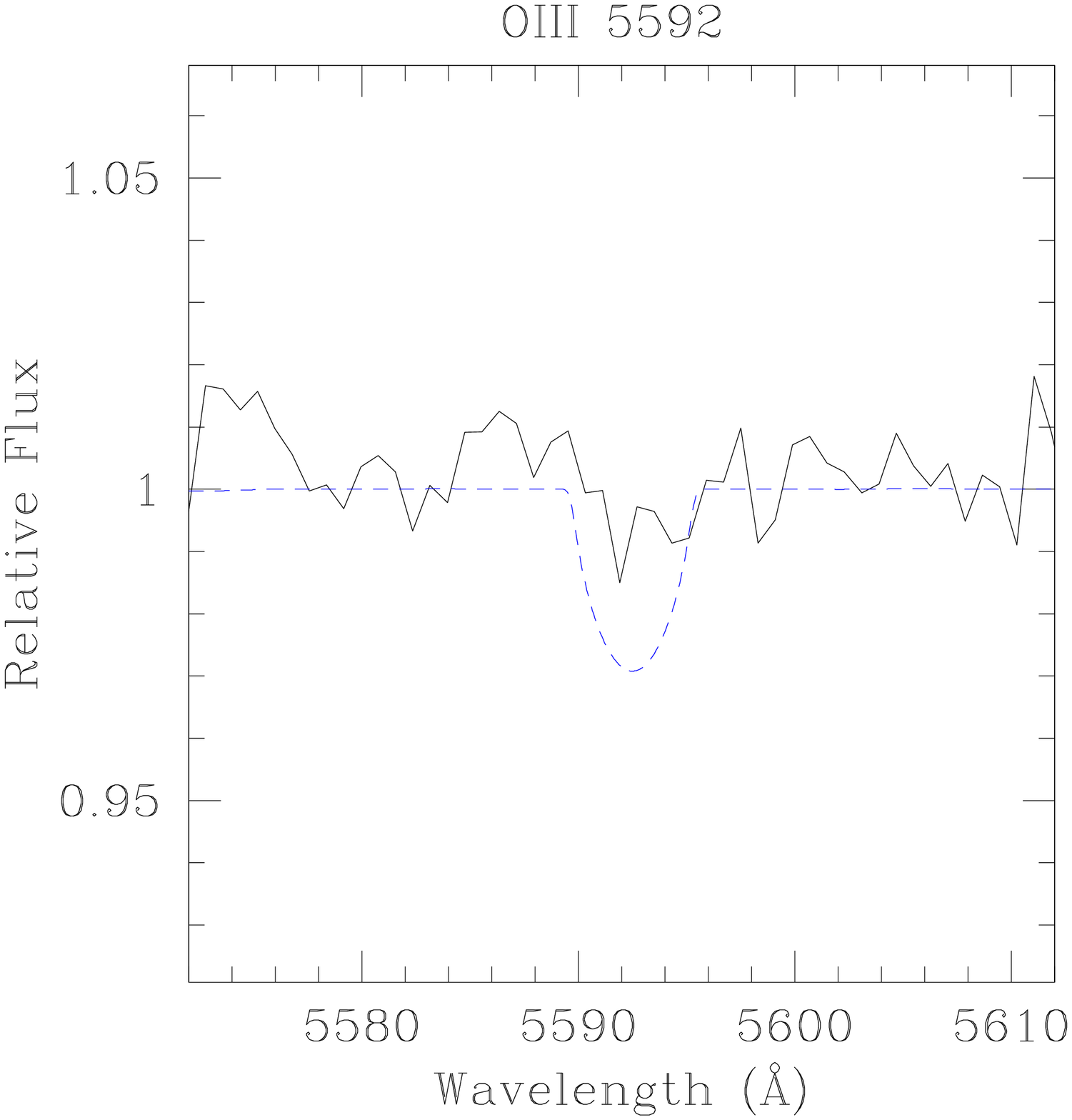}
\caption{\cmfgen\ fits for CNO lines for NGC346-682, O8 V. Black denotes the observed spectrum, and dashed blue shows the \cmfgen\ model with SMC abundances ($Z/Z_\odot=0.2$).}
\end{figure}
\clearpage
\begin{figure}
\epsscale{0.3}
\plotone{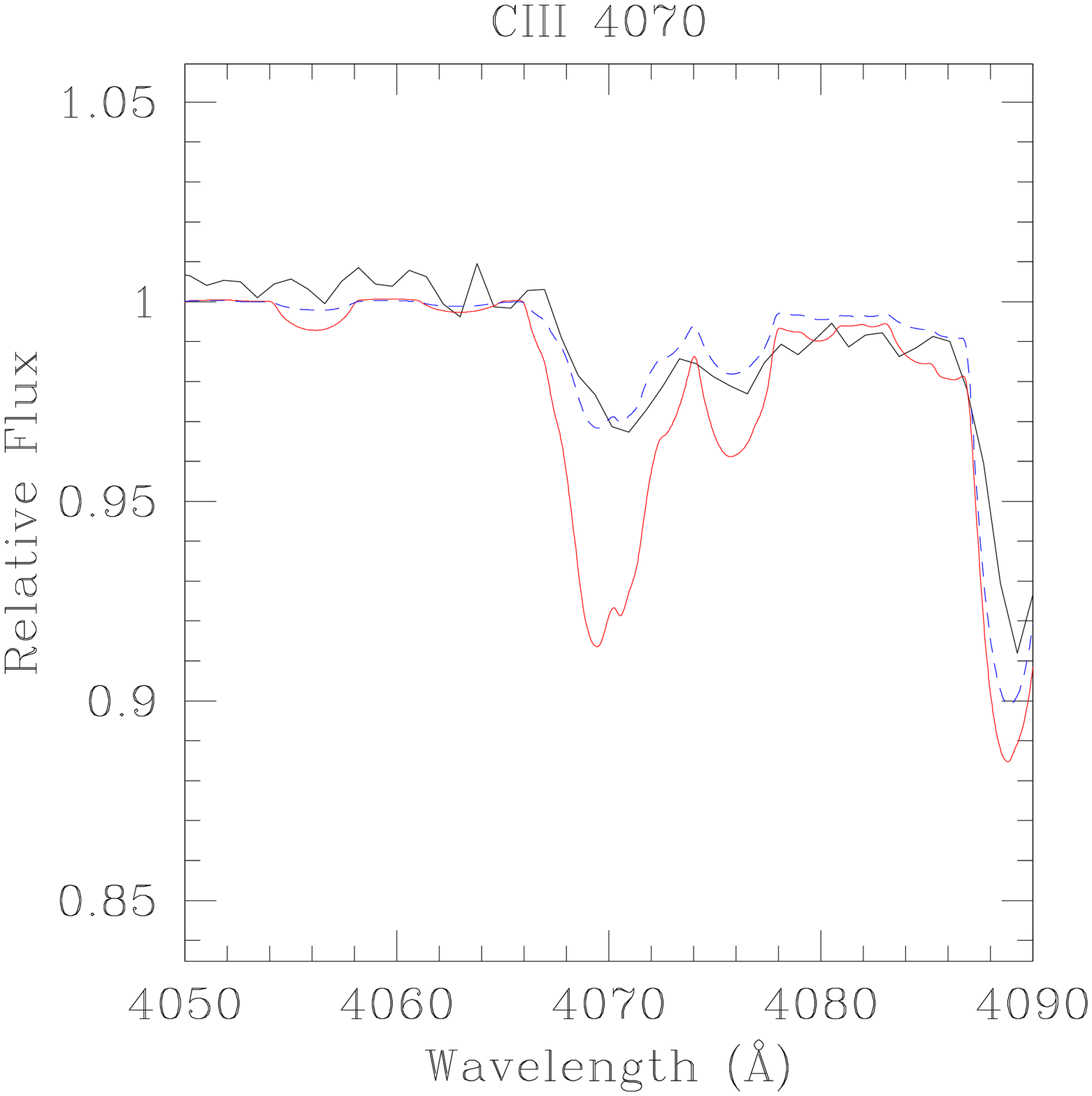}
\plotone{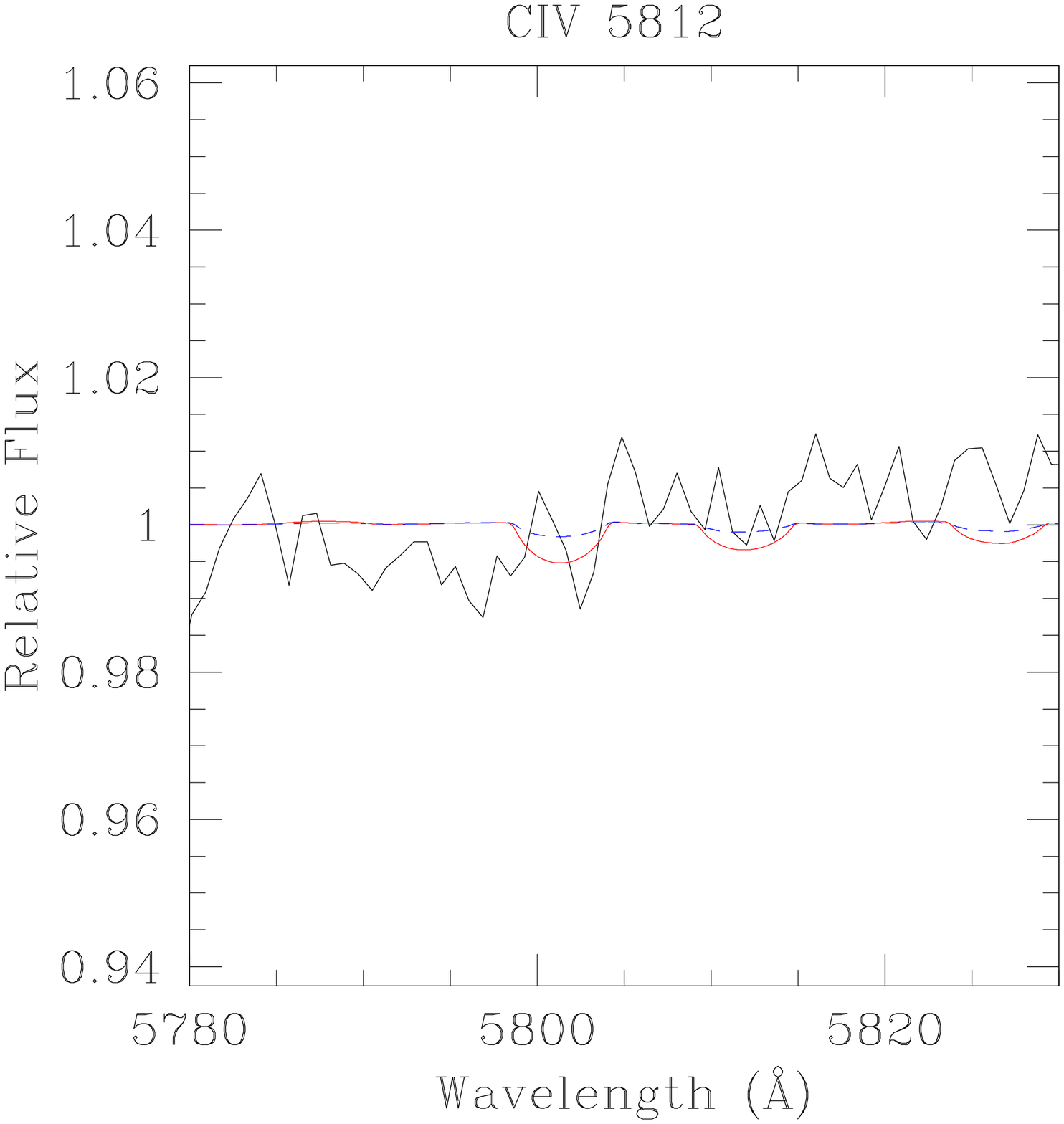}
\plotone{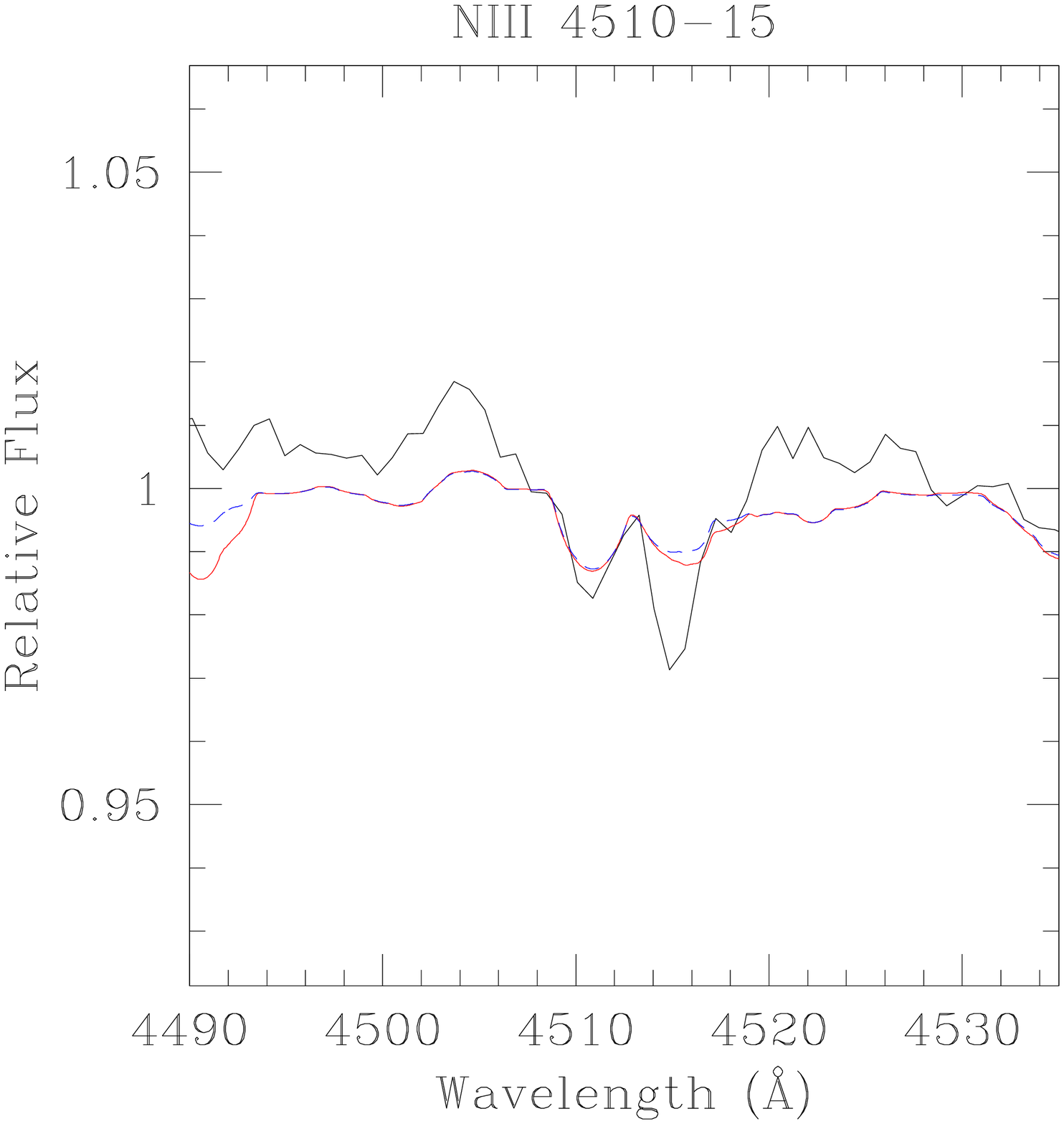}
\plotone{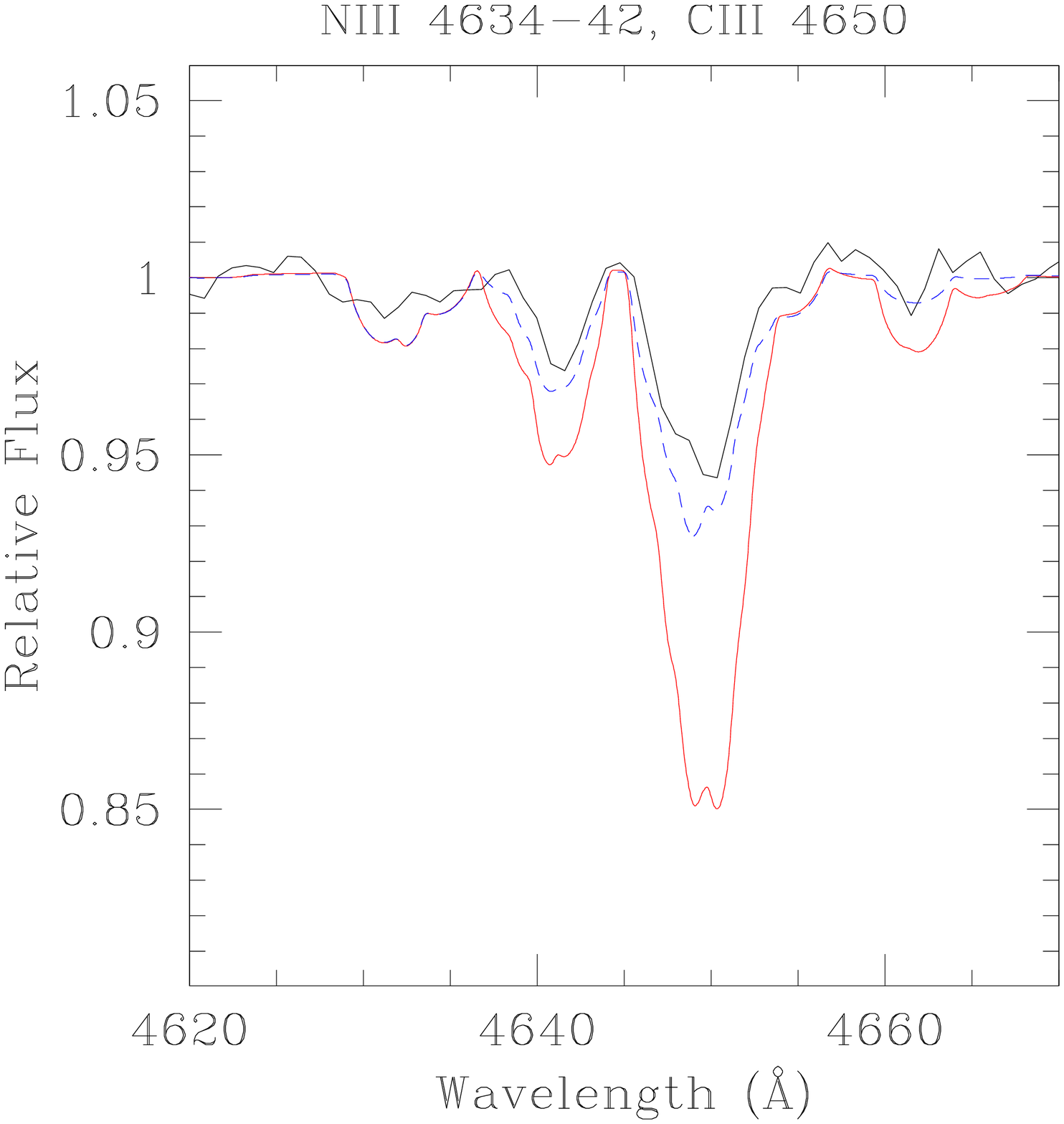}
\plotone{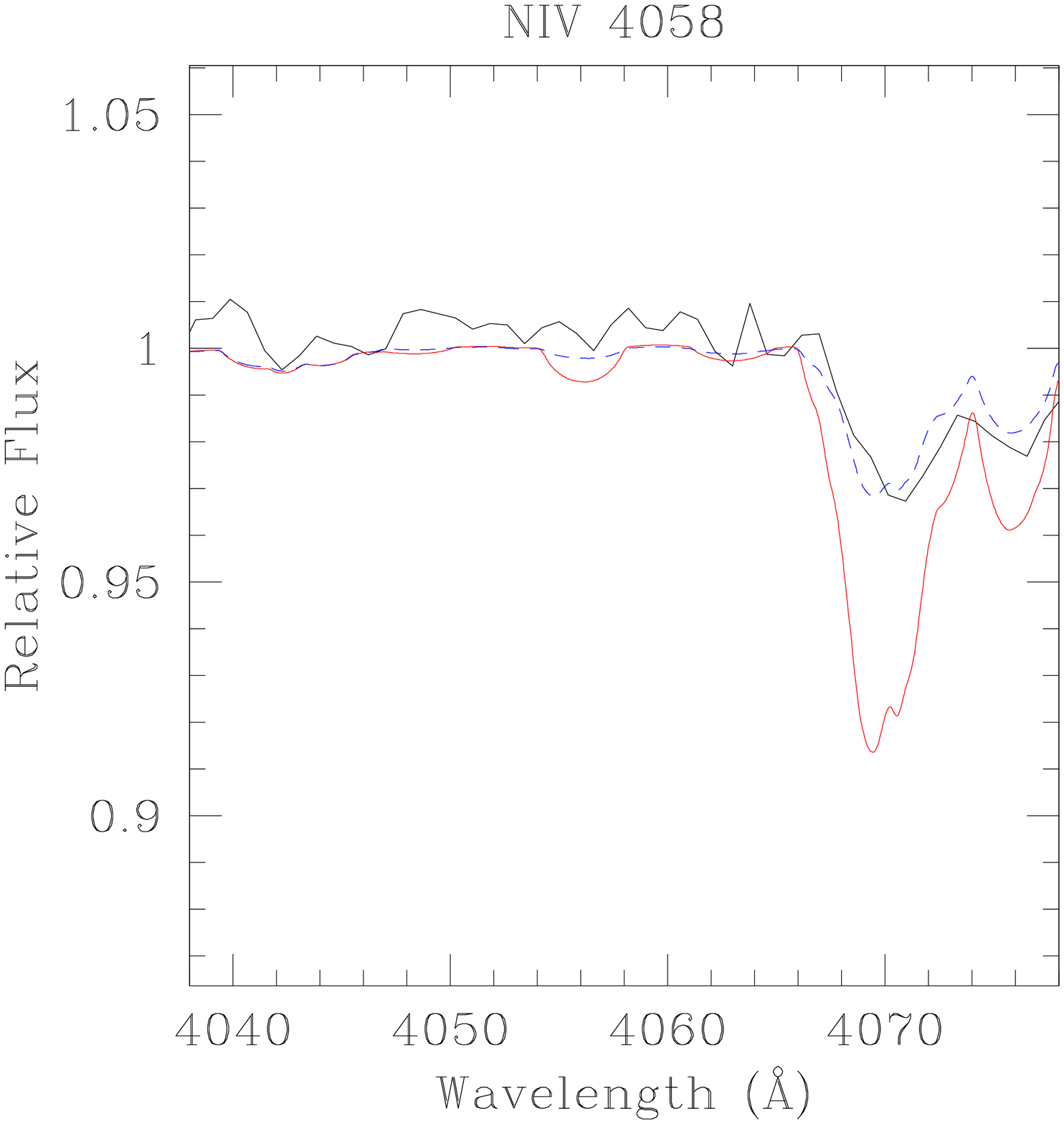}
\plotone{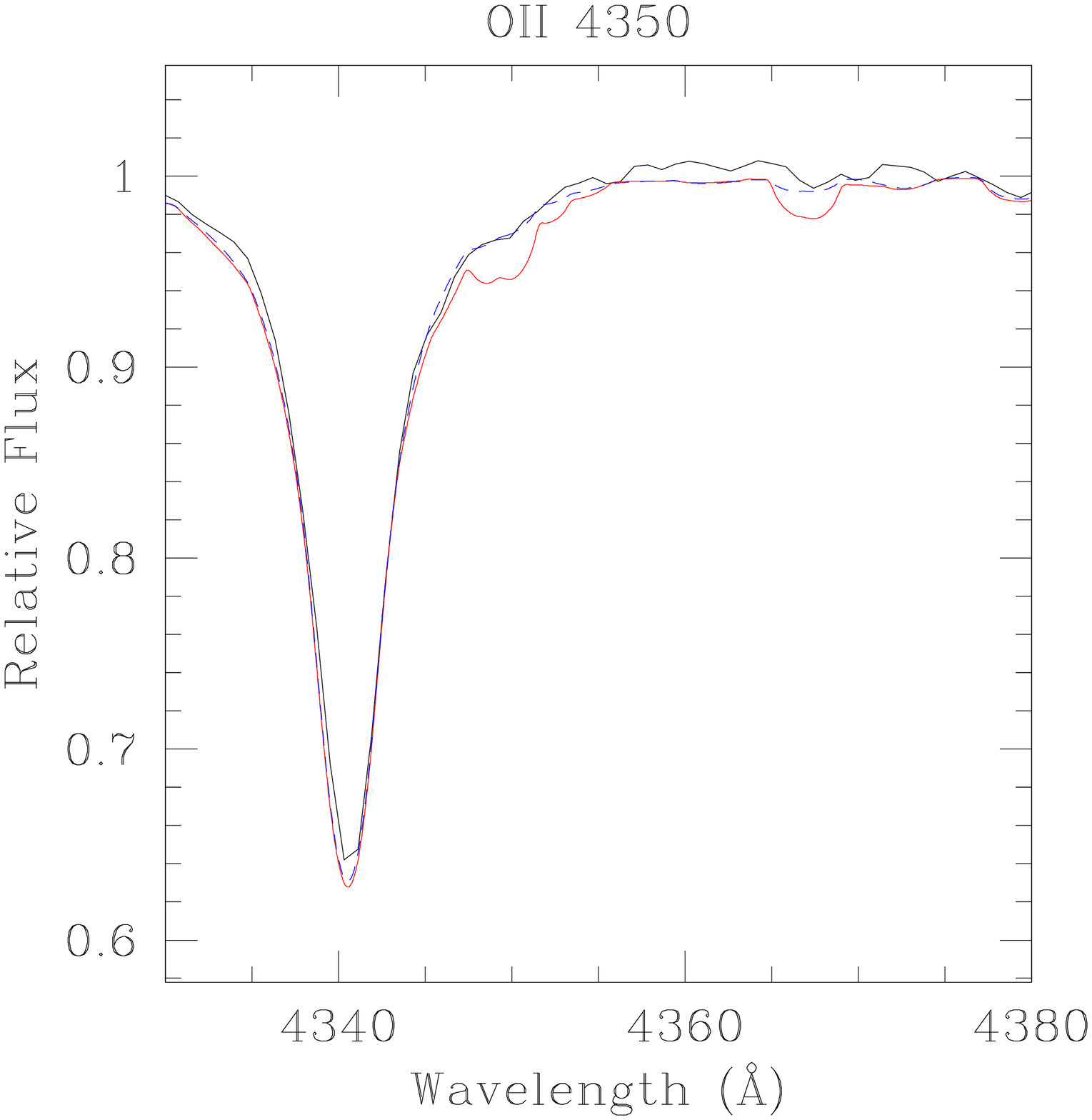}
\plotone{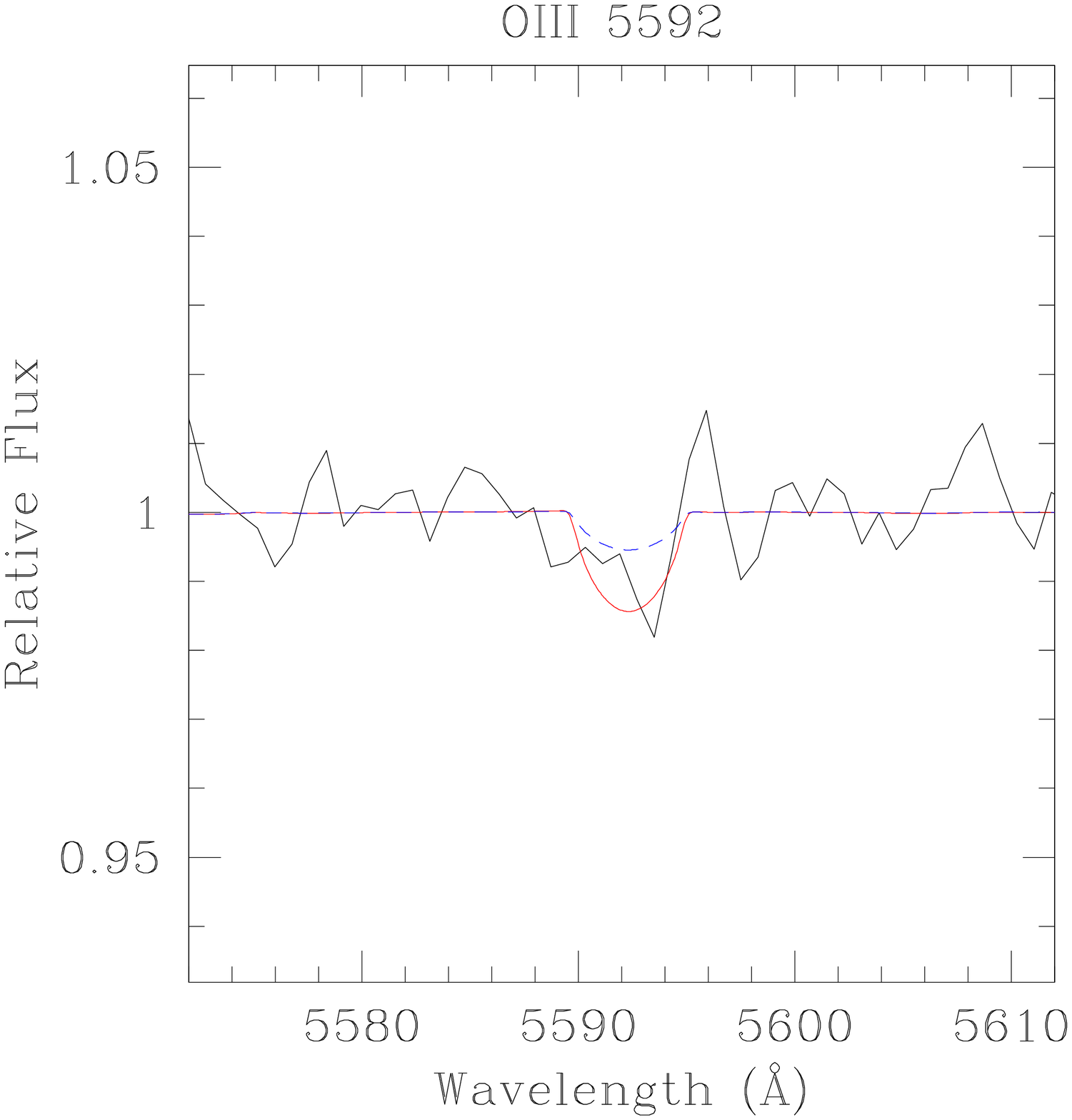}
\caption{\label{fig:CNOAzV233} \cmfgen\ fits for CNO lines for AzV 223, O9.5 II. Black denotes the observed spectrum, solid red indicates the \cmfgen\ model with SMC abundances ($Z/Z_\odot=0.2$), and the dashed blue shows the \cmfgen\ model with C decreased by a 4, N normal, and O decreased by a factor of increased by a factor of 2.8.}
\end{figure}
\clearpage
\begin{figure}
\epsscale{0.3}
\plotone{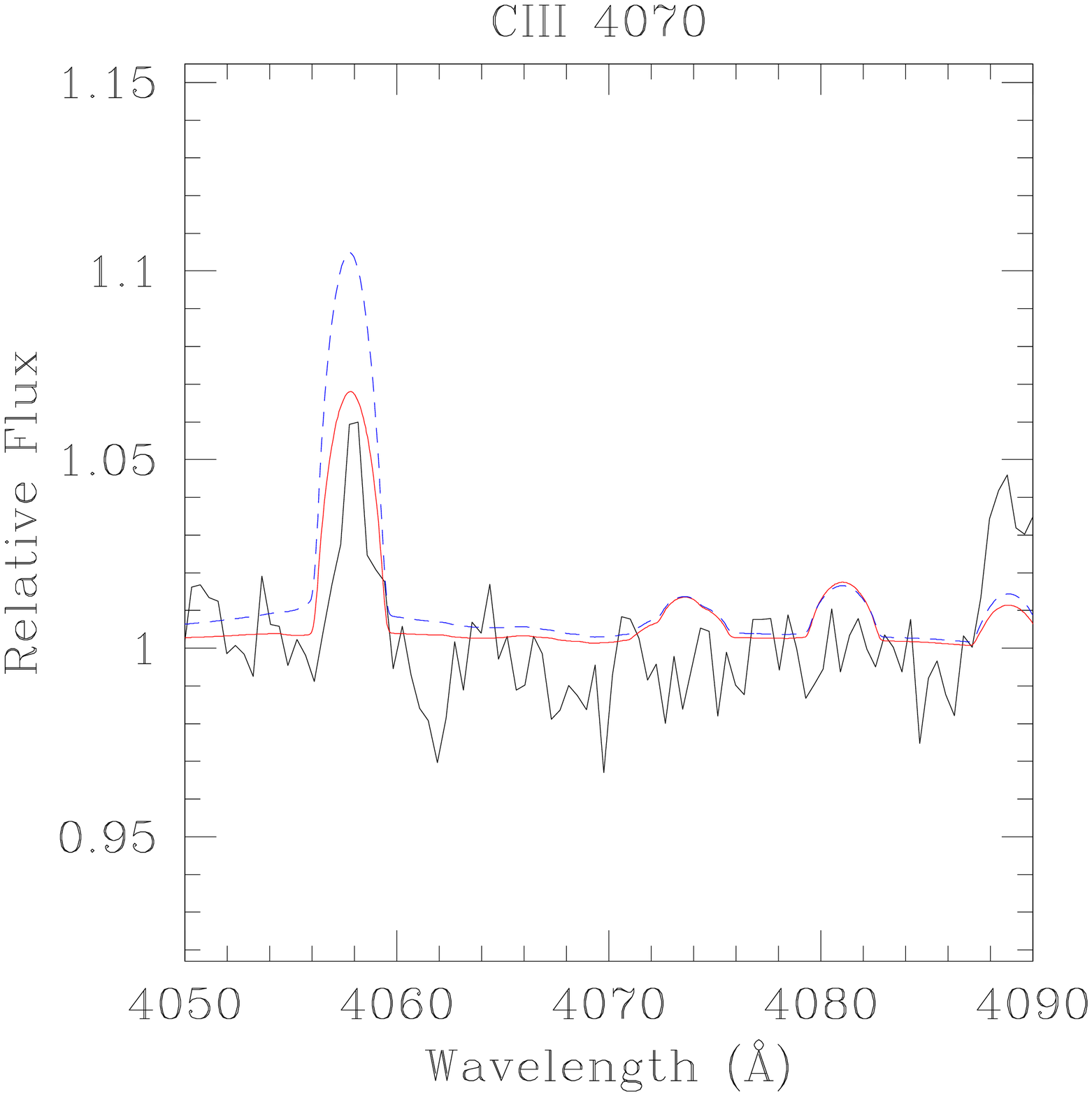}
\plotone{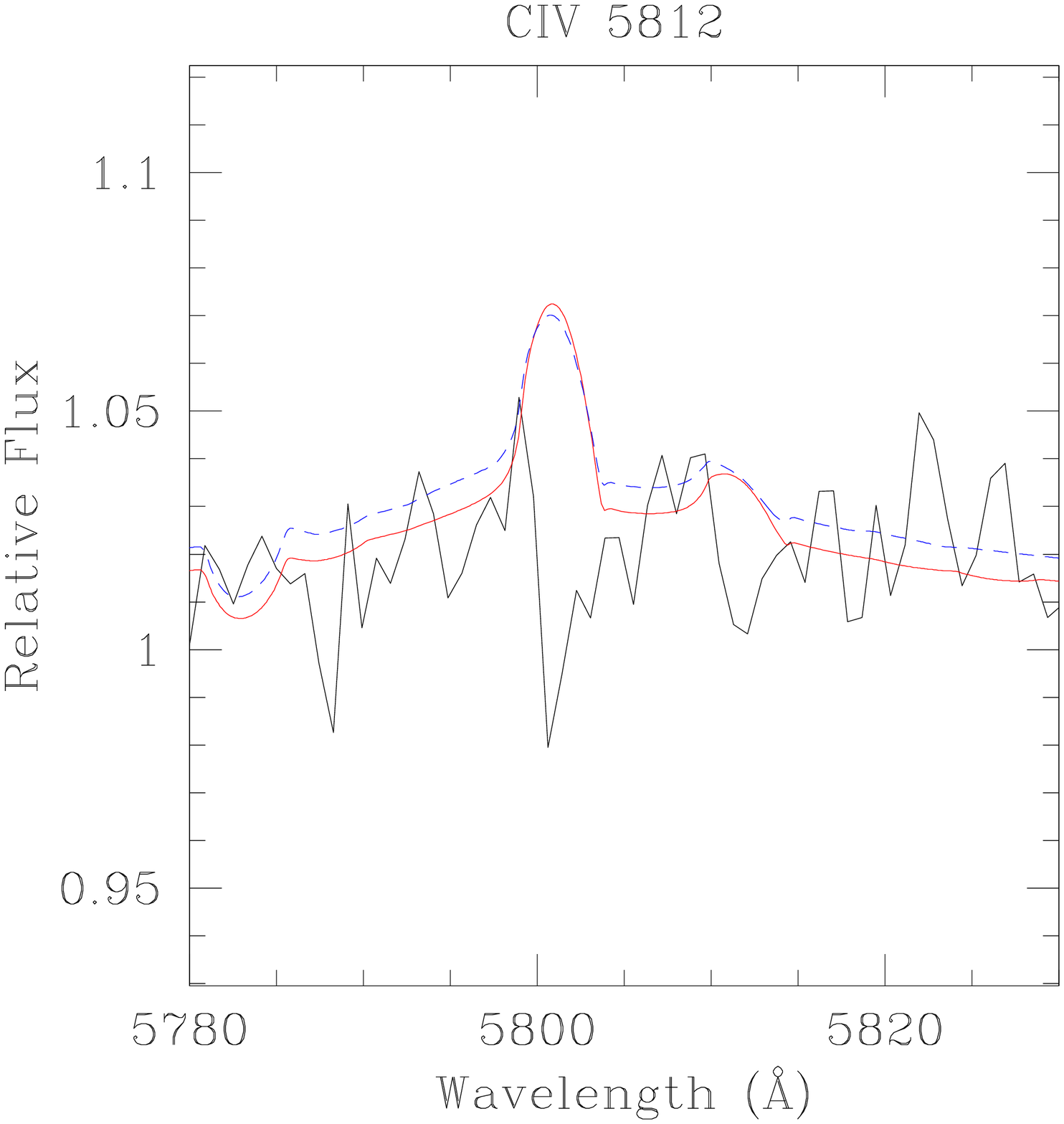}
\plotone{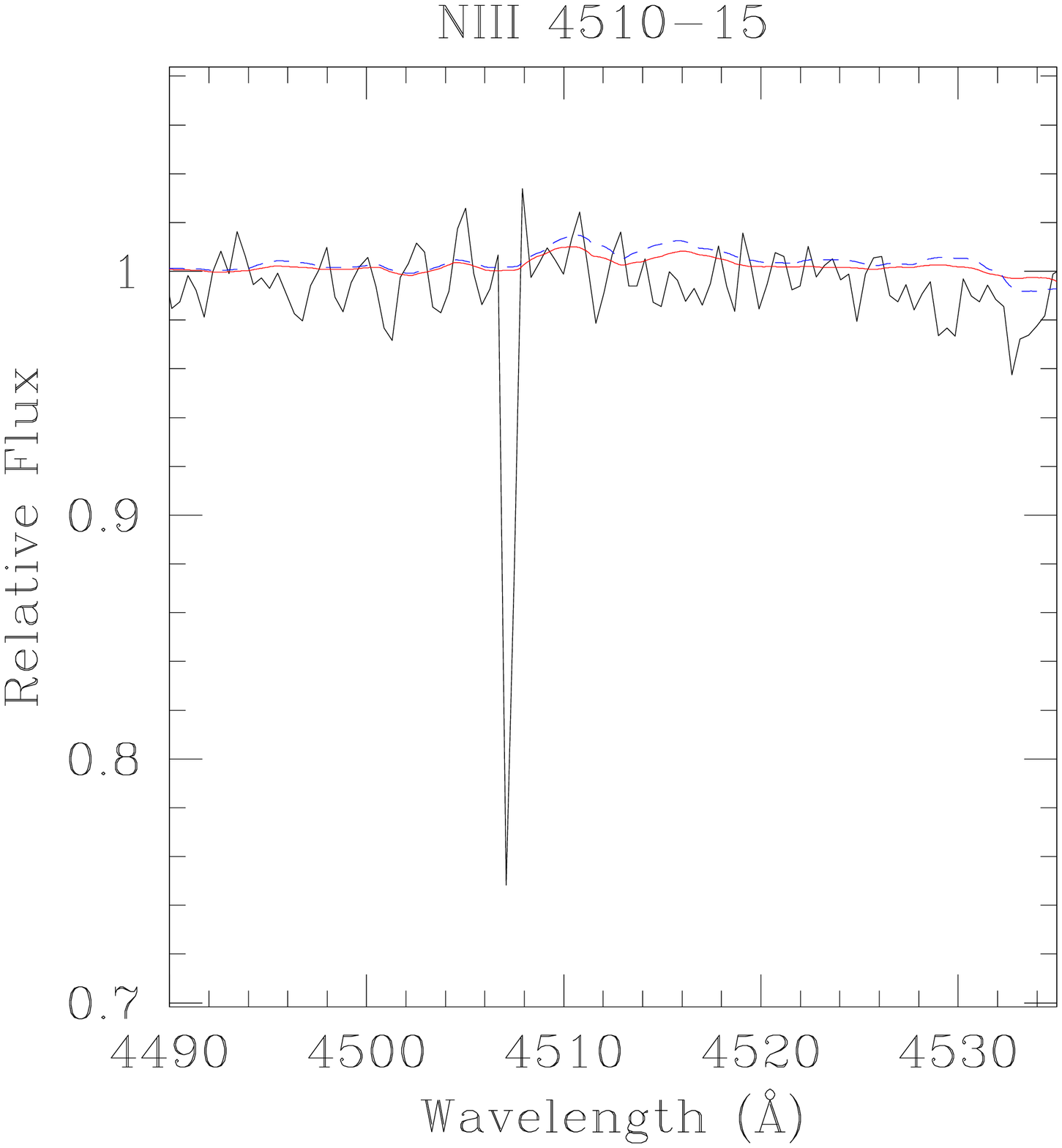}
\plotone{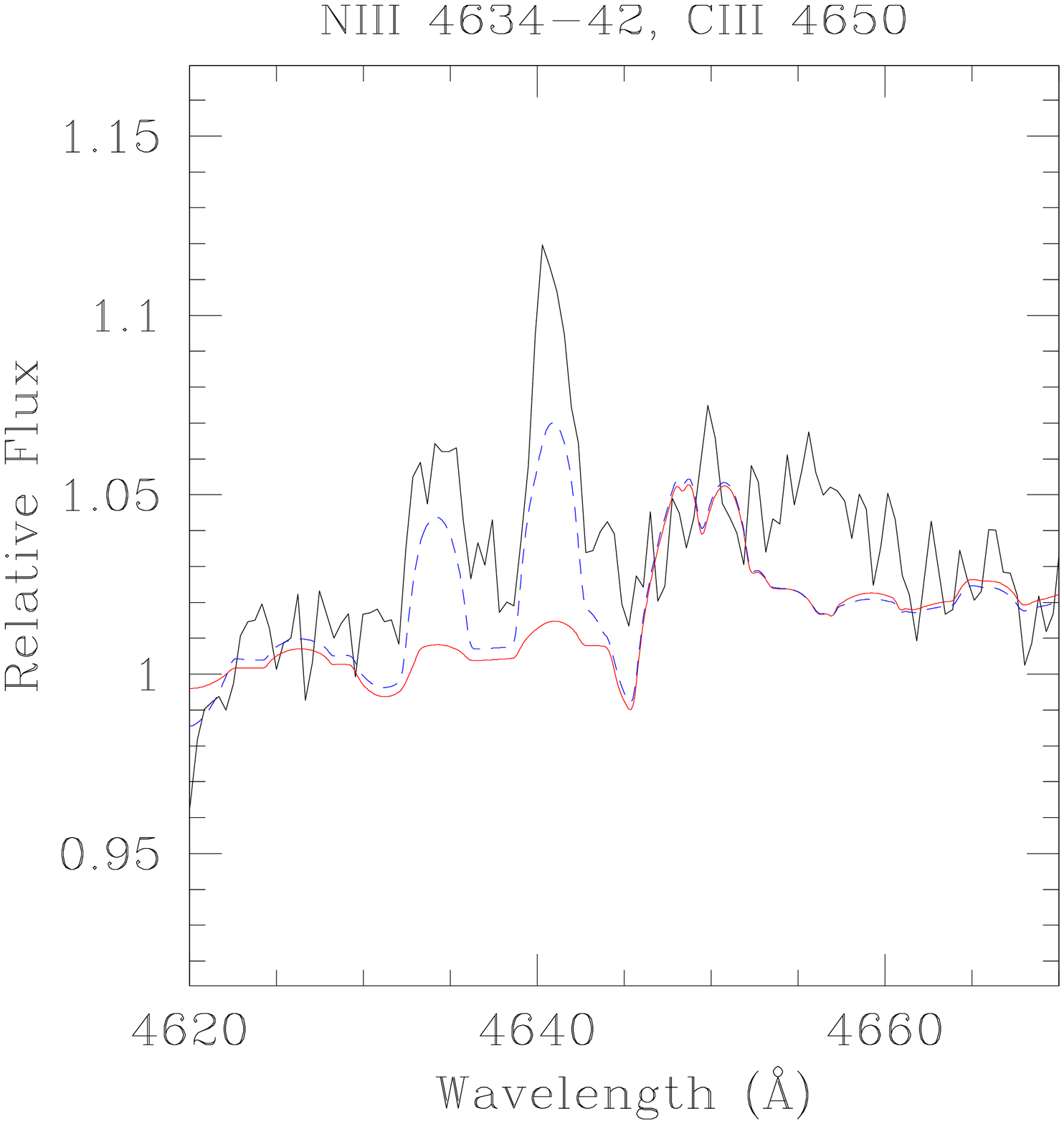}
\plotone{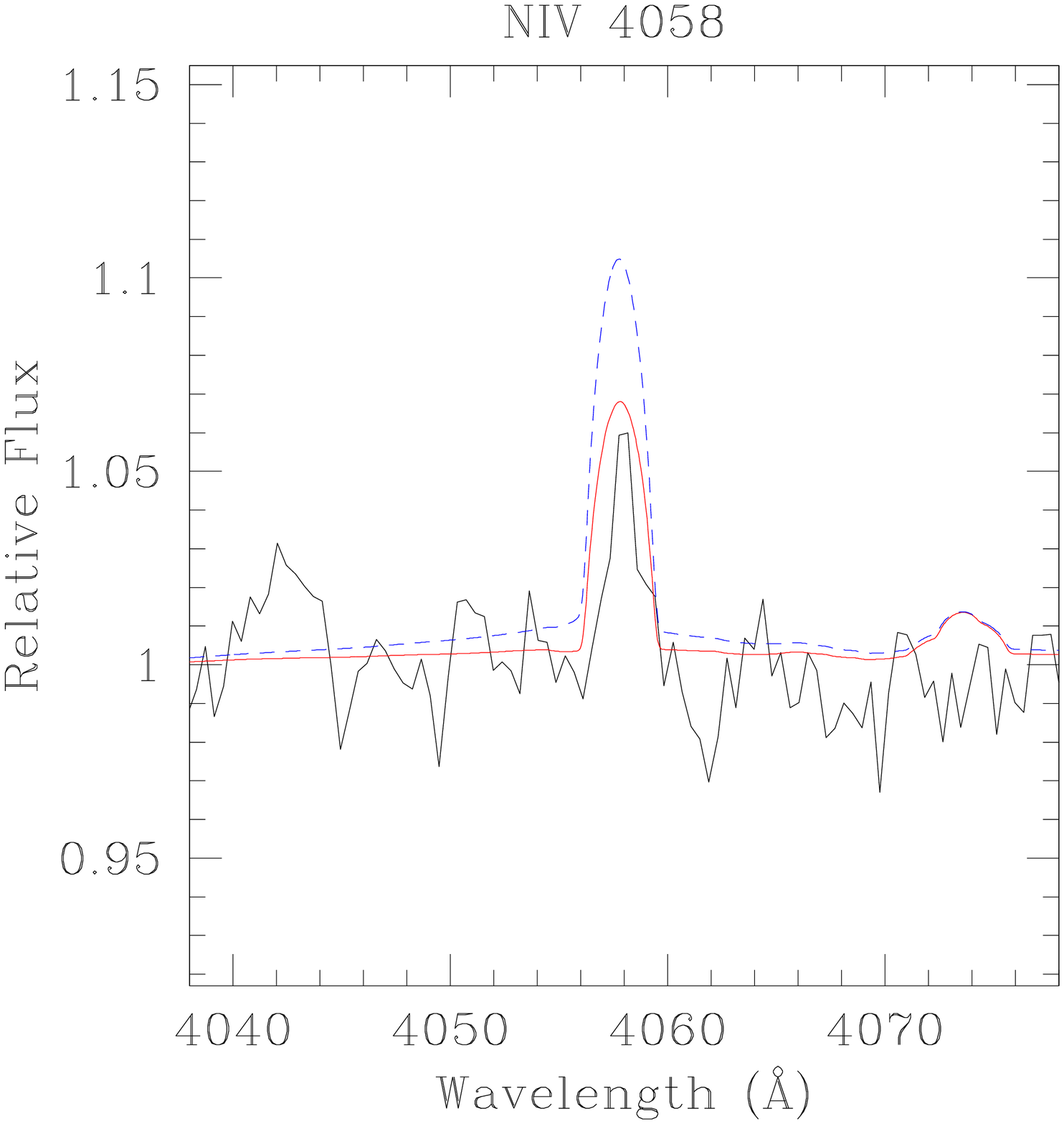}
\plotone{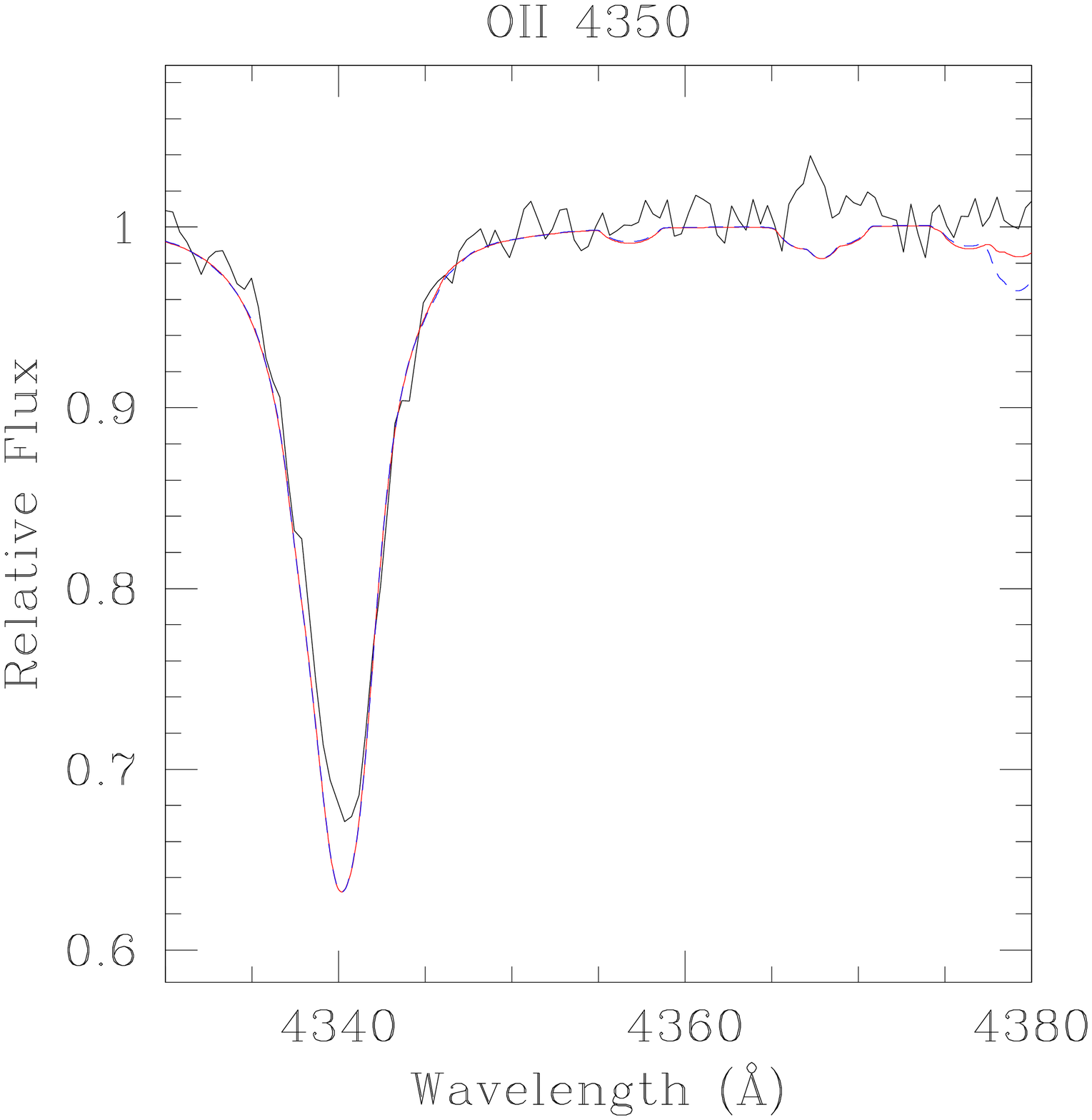}
\plotone{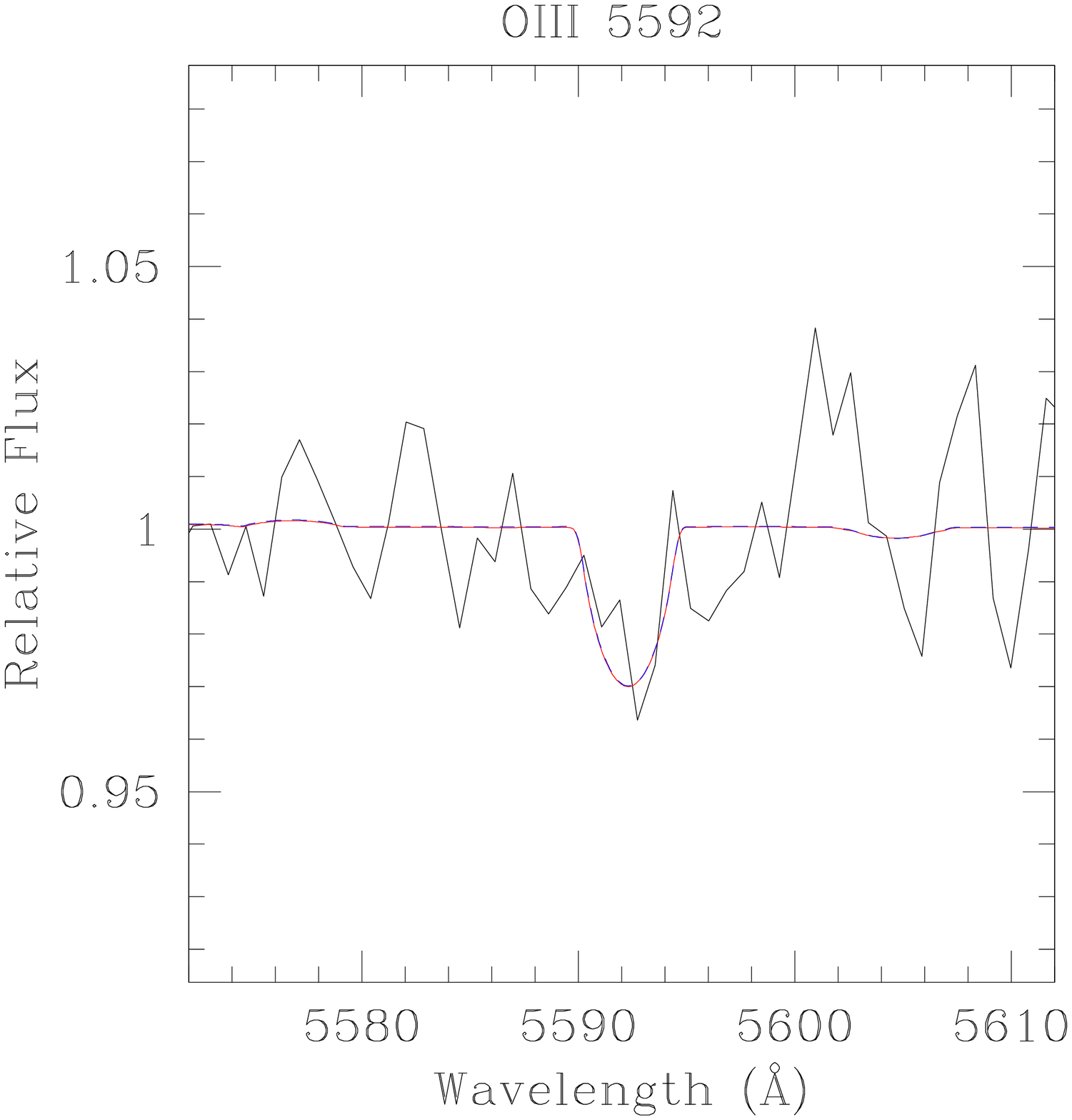}
\caption{\cmfgen\ fits for CNO lines for LH81:W28-23, O3.5 V((f+)). Black denotes the observed spectrum, solid red indicates the \cmfgen\ model with LMC abundances ($Z/Z_\odot=0.5$), and the dashed blue shows the \cmfgen\ model with decreased by a factor of 10, and N increased by a factor of 5, and O at LMC-metallicity.}
\end{figure}
\clearpage
\begin{figure}
\epsscale{0.3}
\plotone{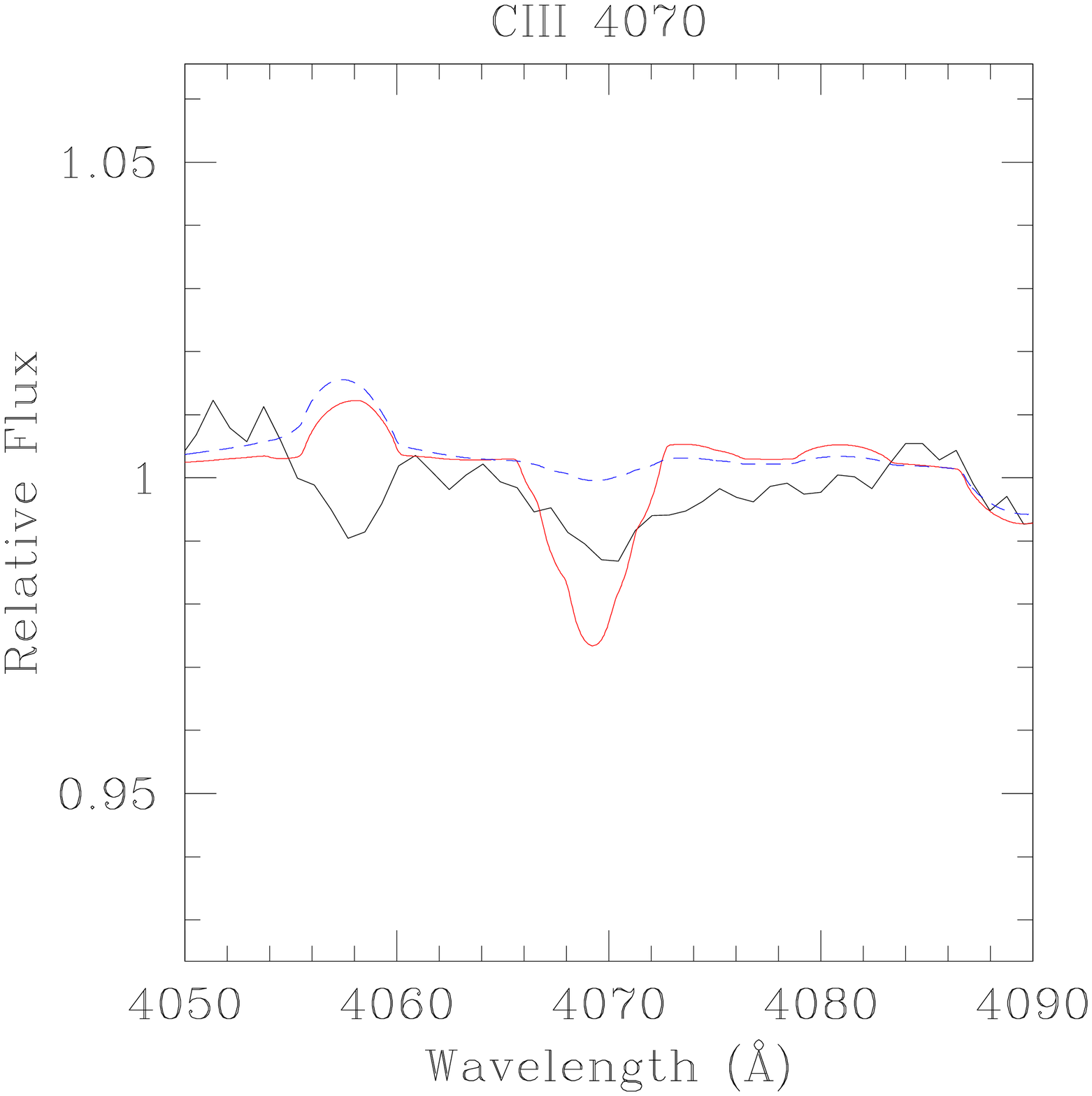}
\plotone{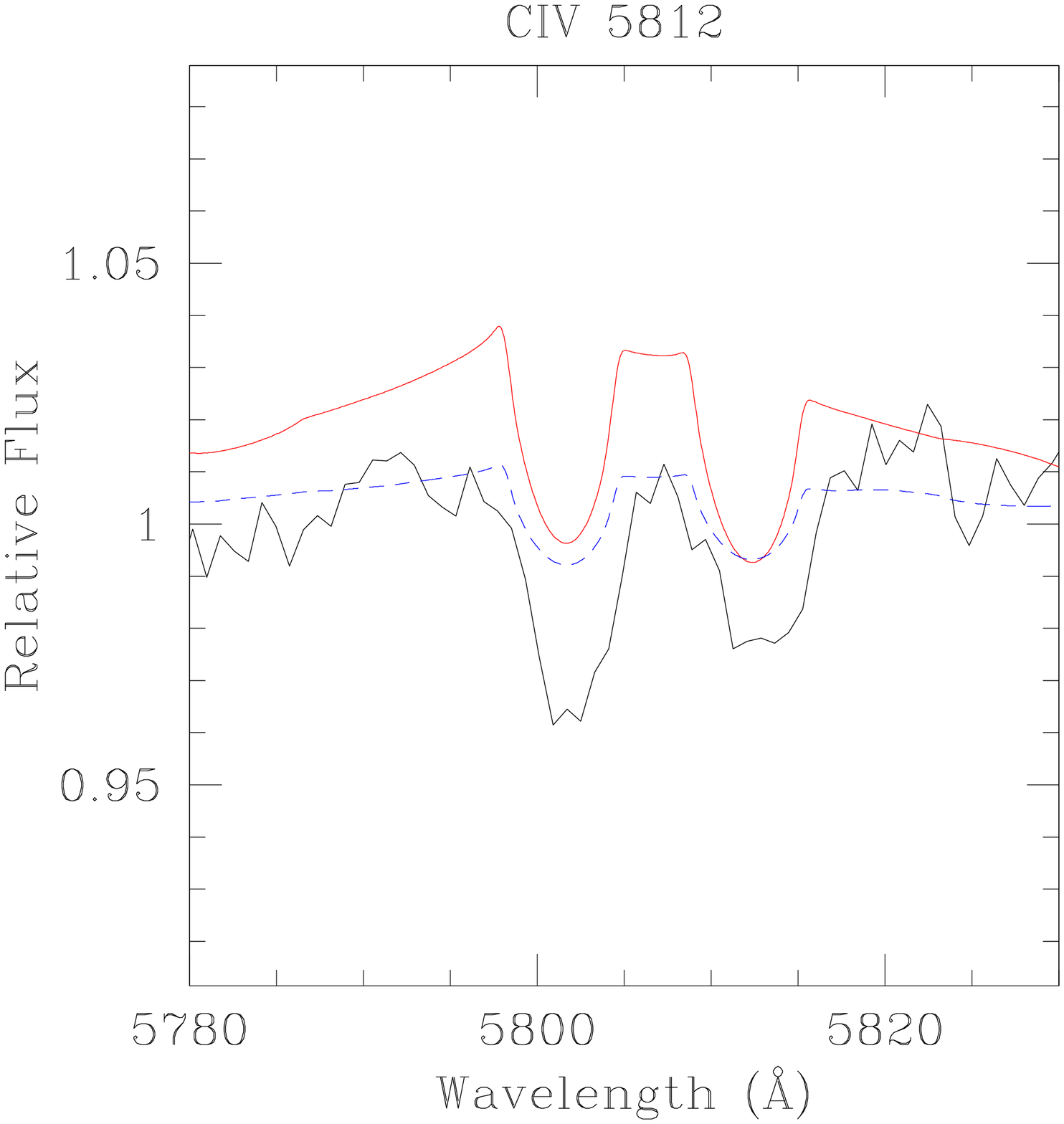}
\plotone{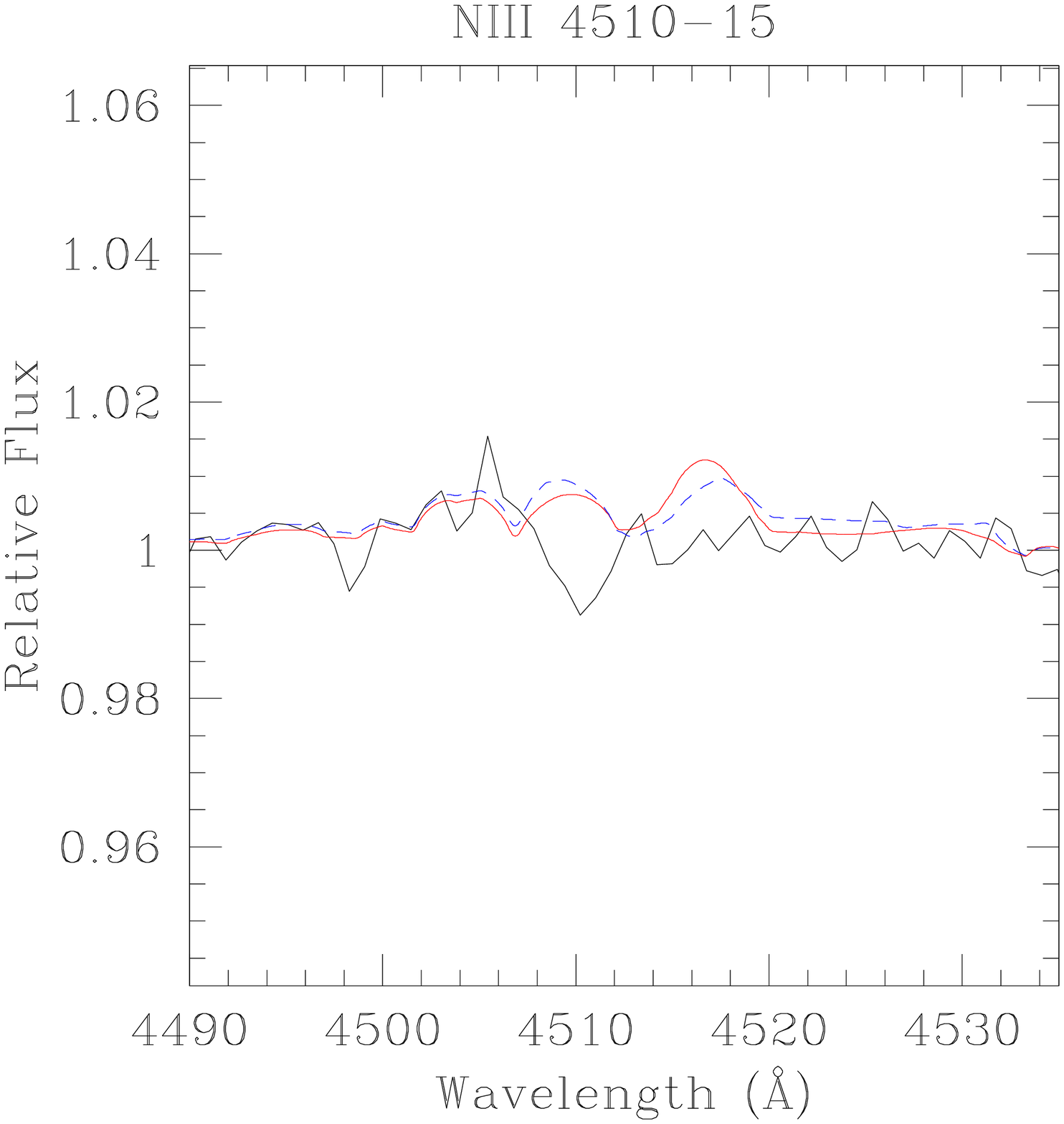}
\plotone{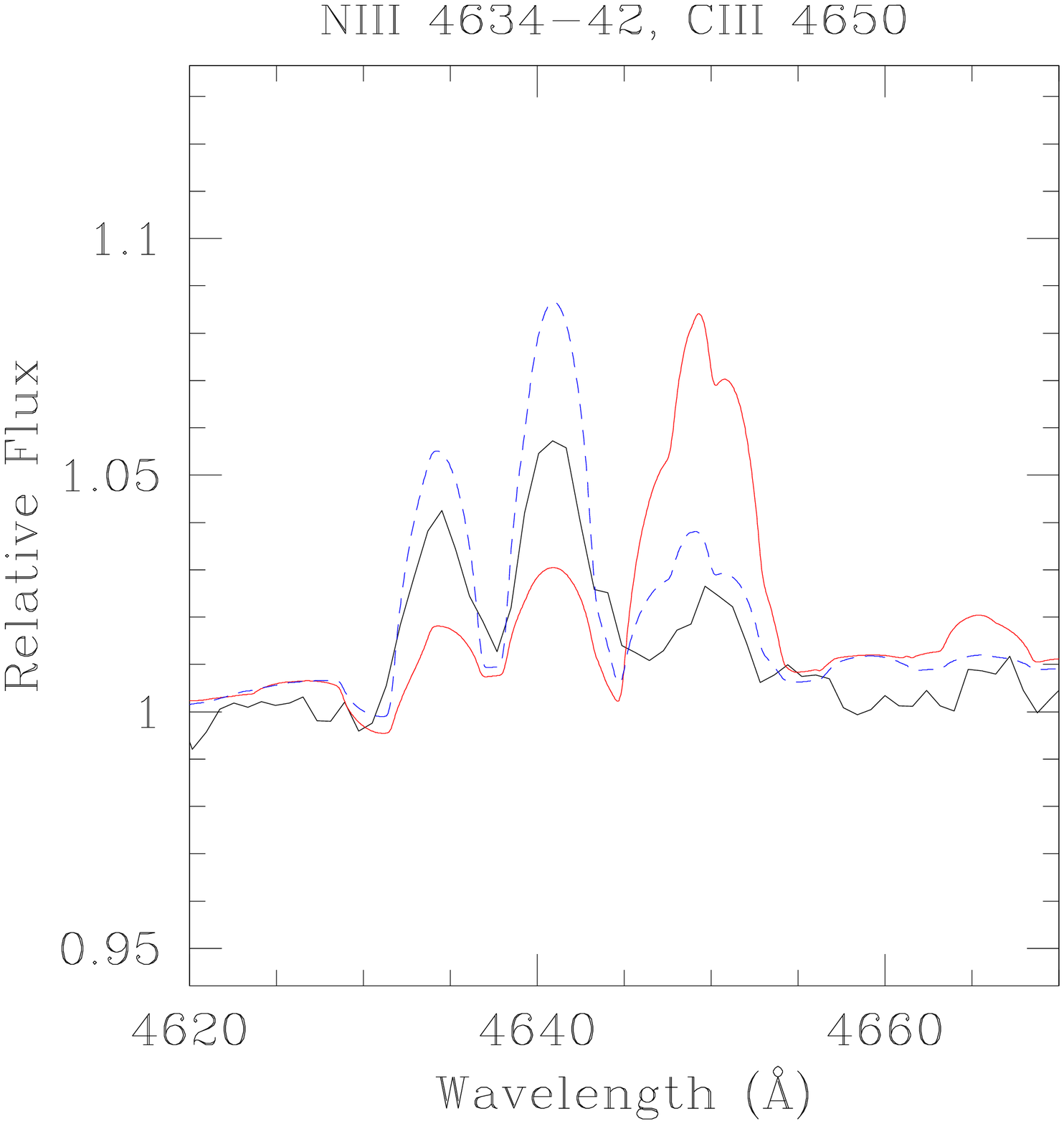}
\plotone{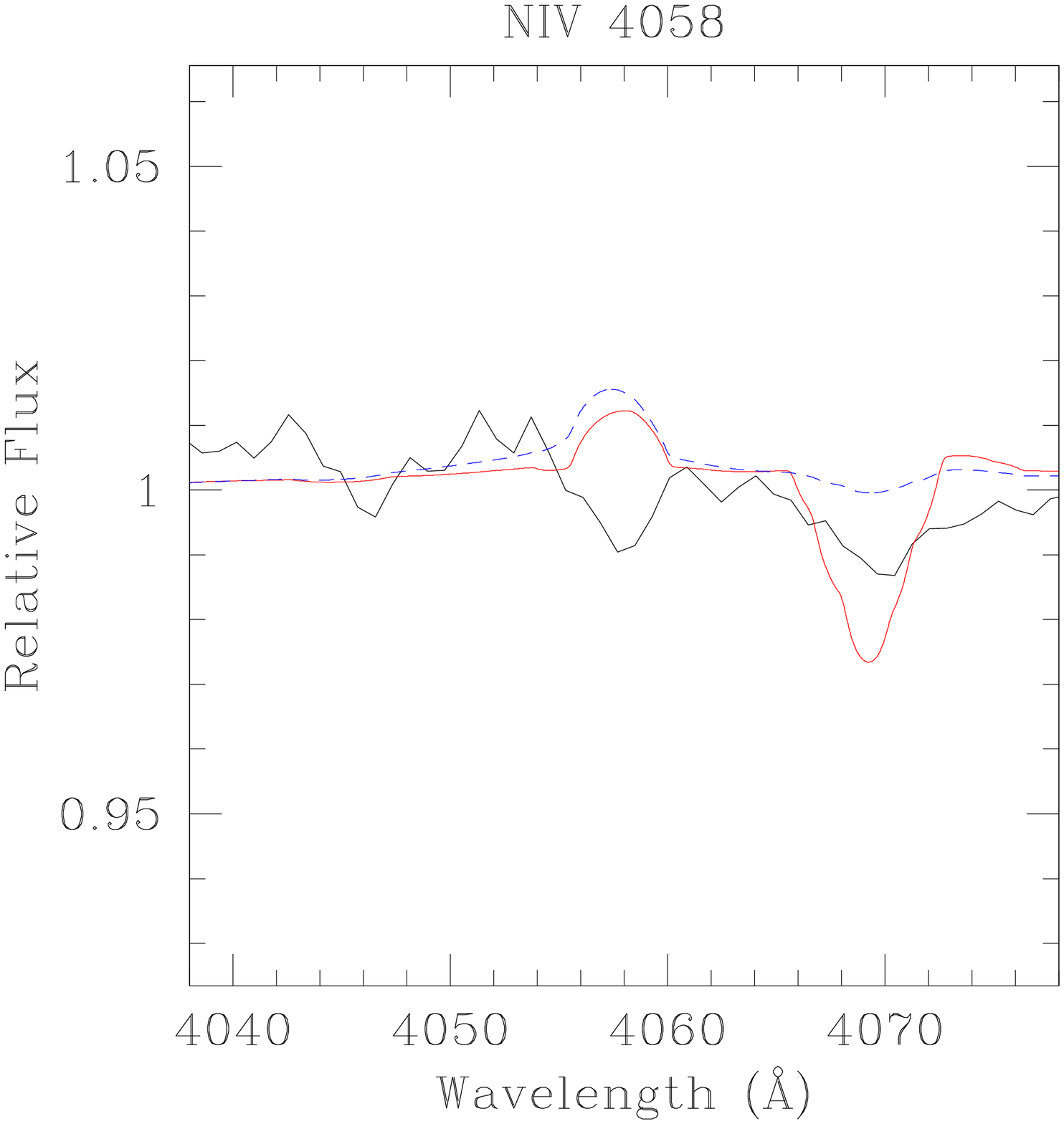}
\plotone{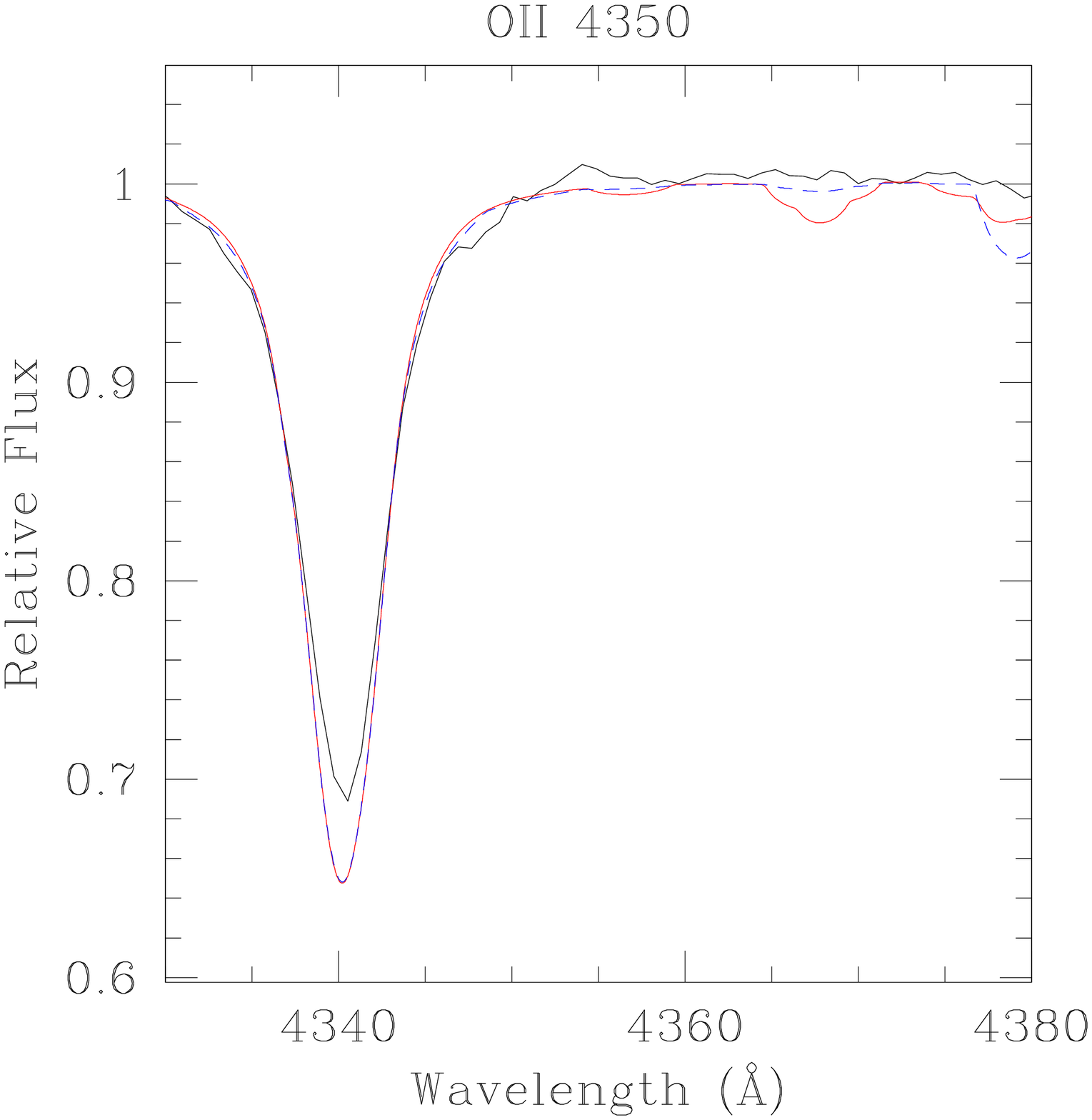}
\plotone{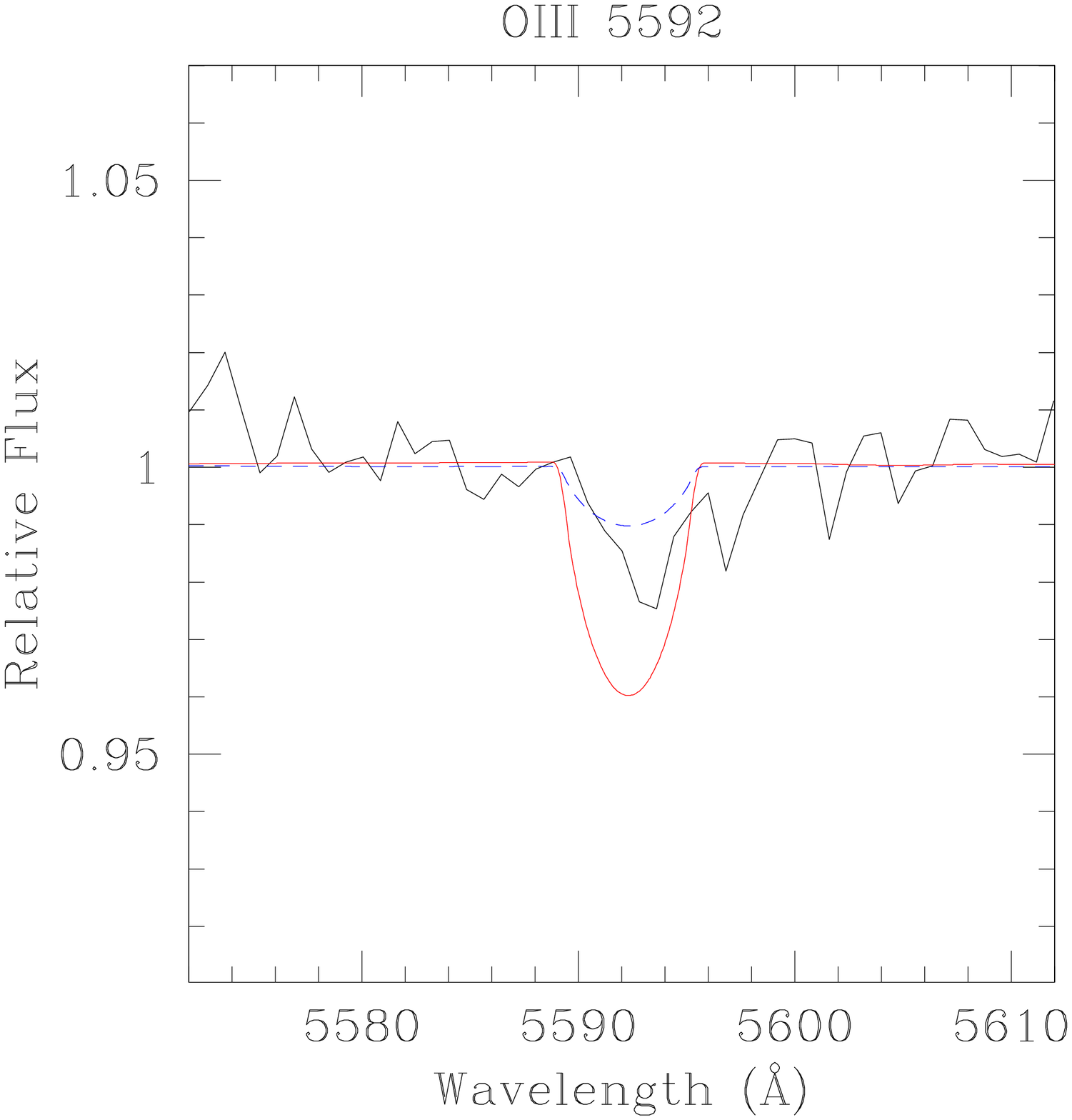}
\caption{\cmfgen\ fits for CNO lines for Sk $-70^\circ$69, O5.5 V((f)).  Black denotes the observed spectrum, solid red indicates the \cmfgen\ model with LMC abundances ($Z/Z_\odot=0.5$), and the dashed blue shows the \cmfgen\ model with C and O decreased by a factor of 5, and N increased by a factor of 5.}
\end{figure}
\clearpage
\begin{figure}
\epsscale{0.3}
\plotone{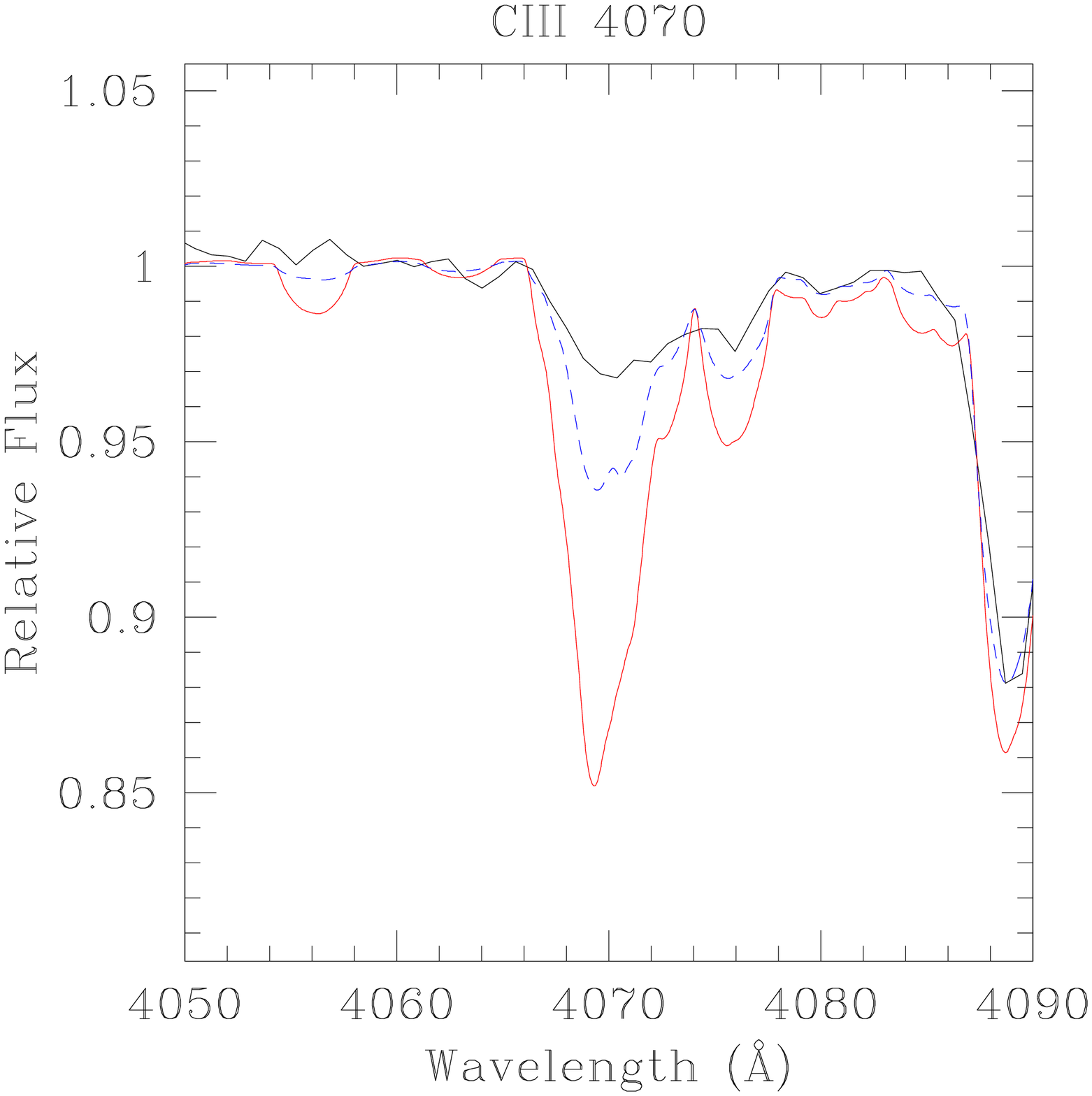}
\plotone{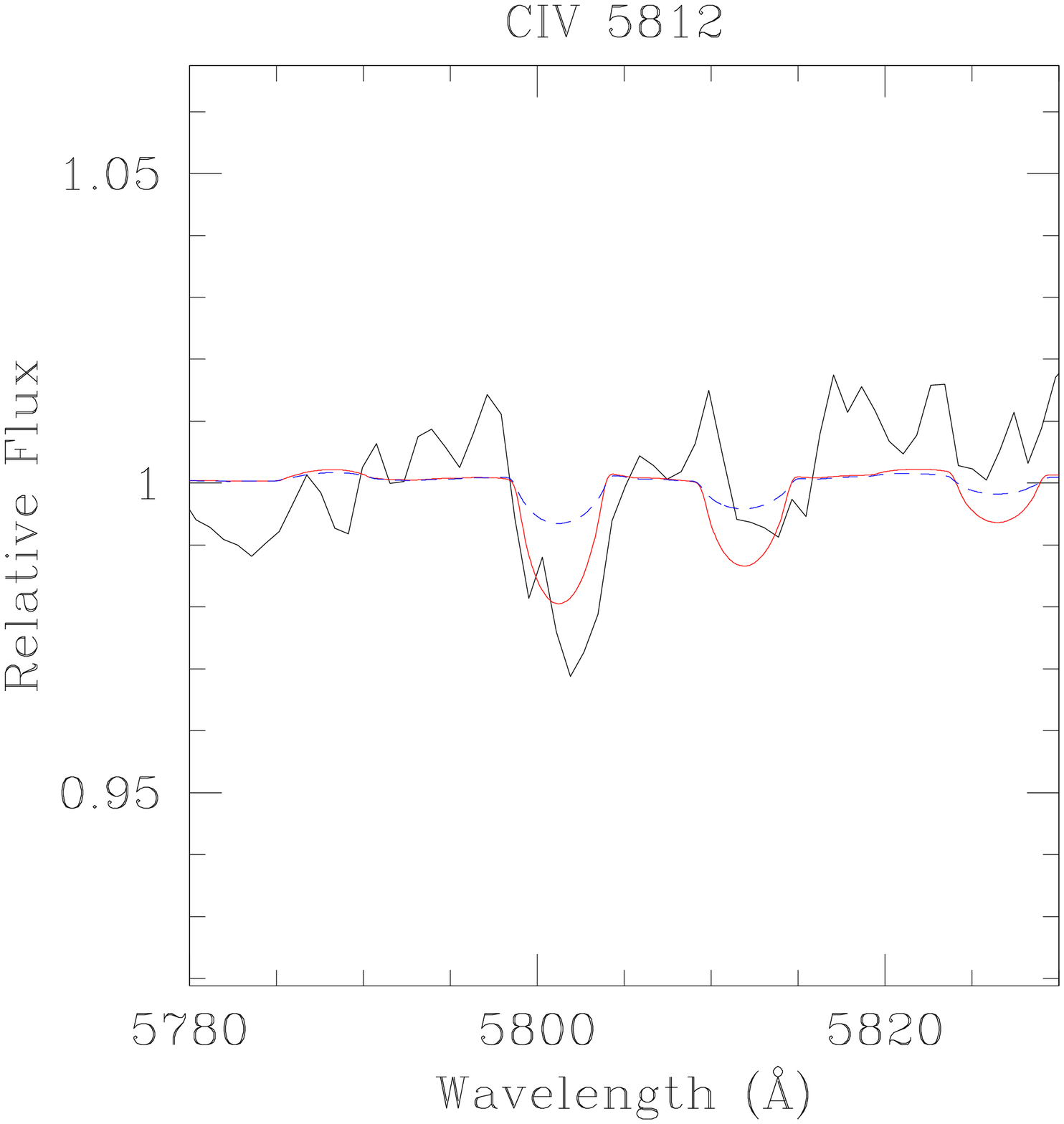}
\plotone{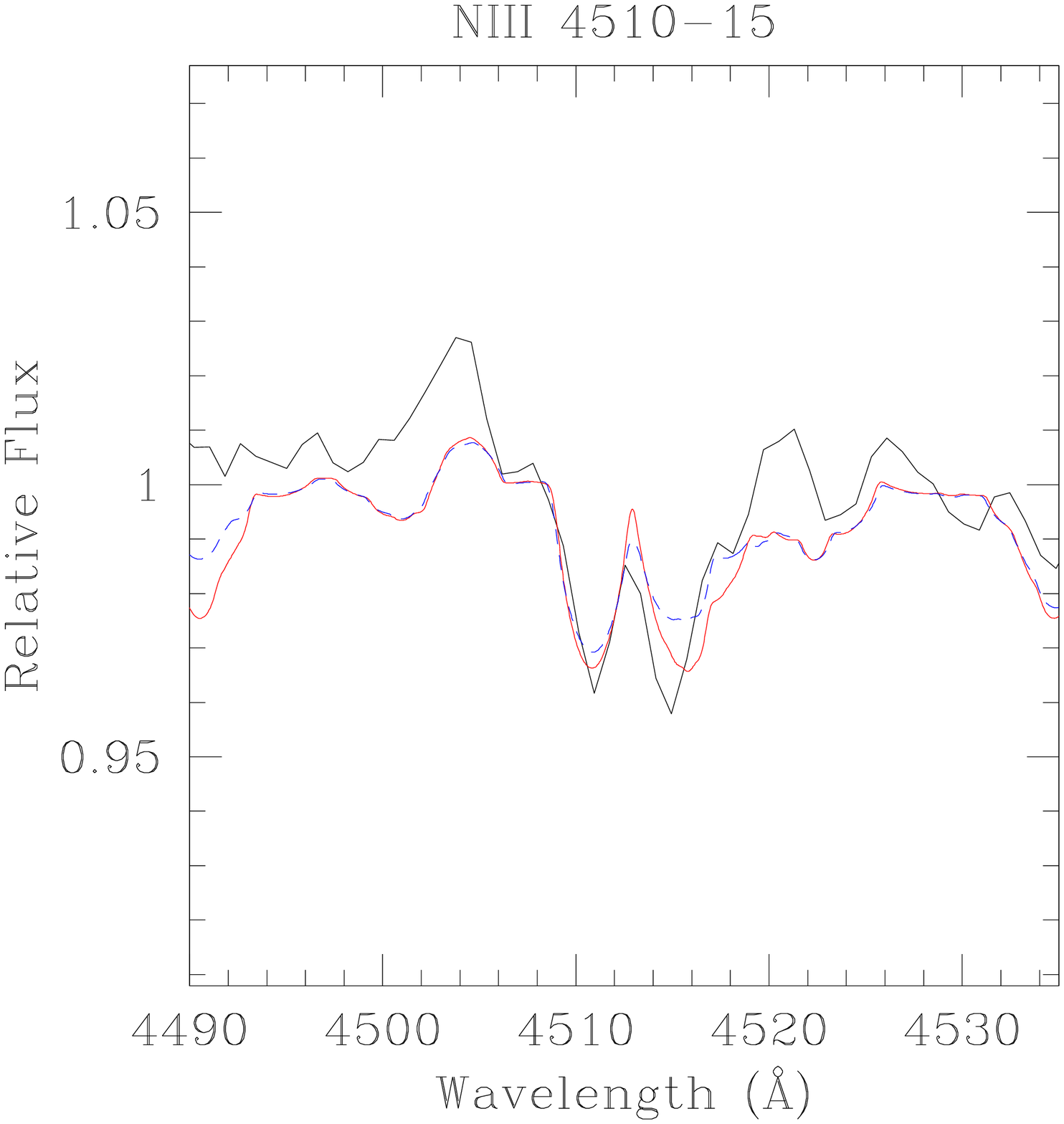}
\plotone{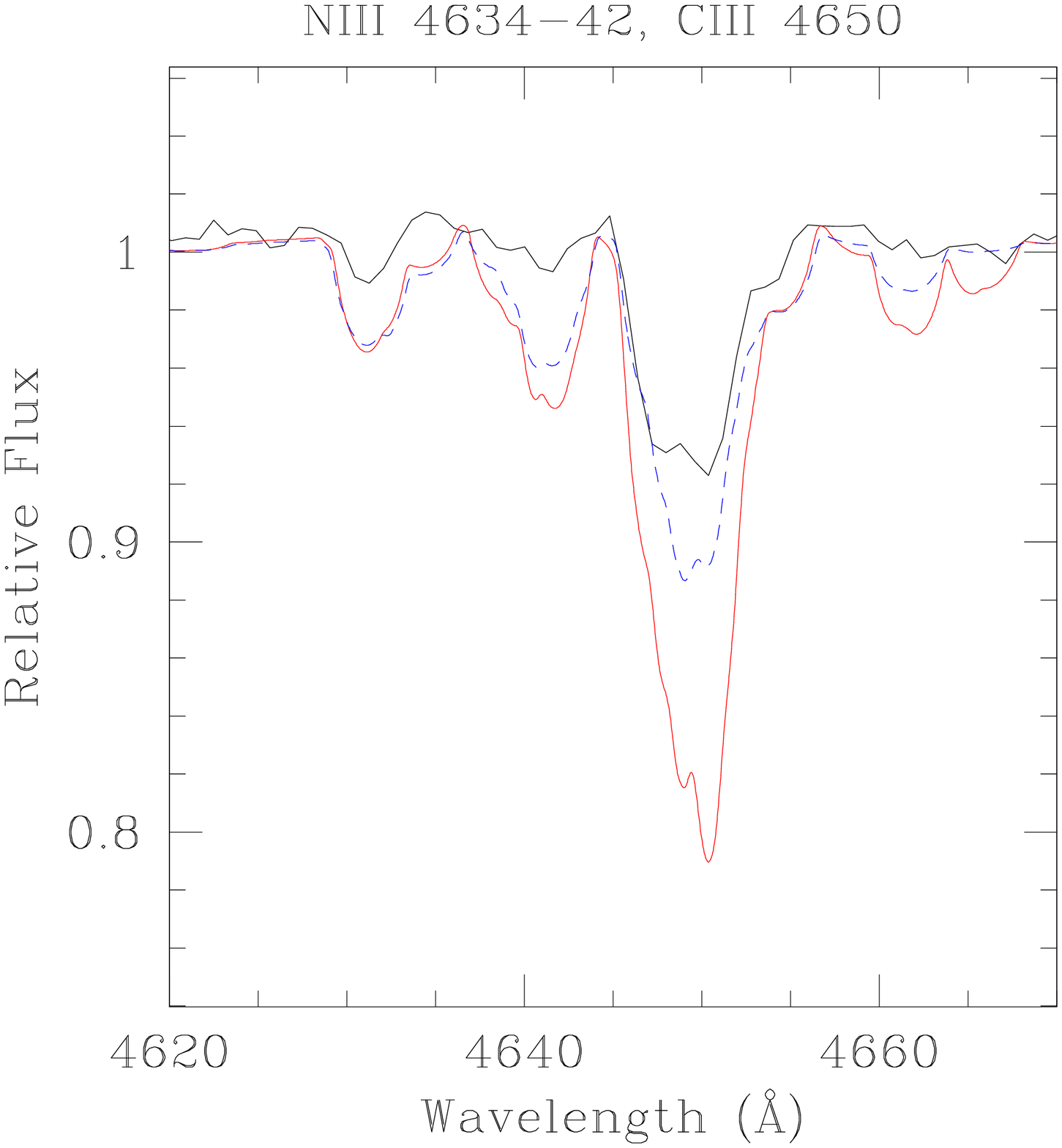}
\plotone{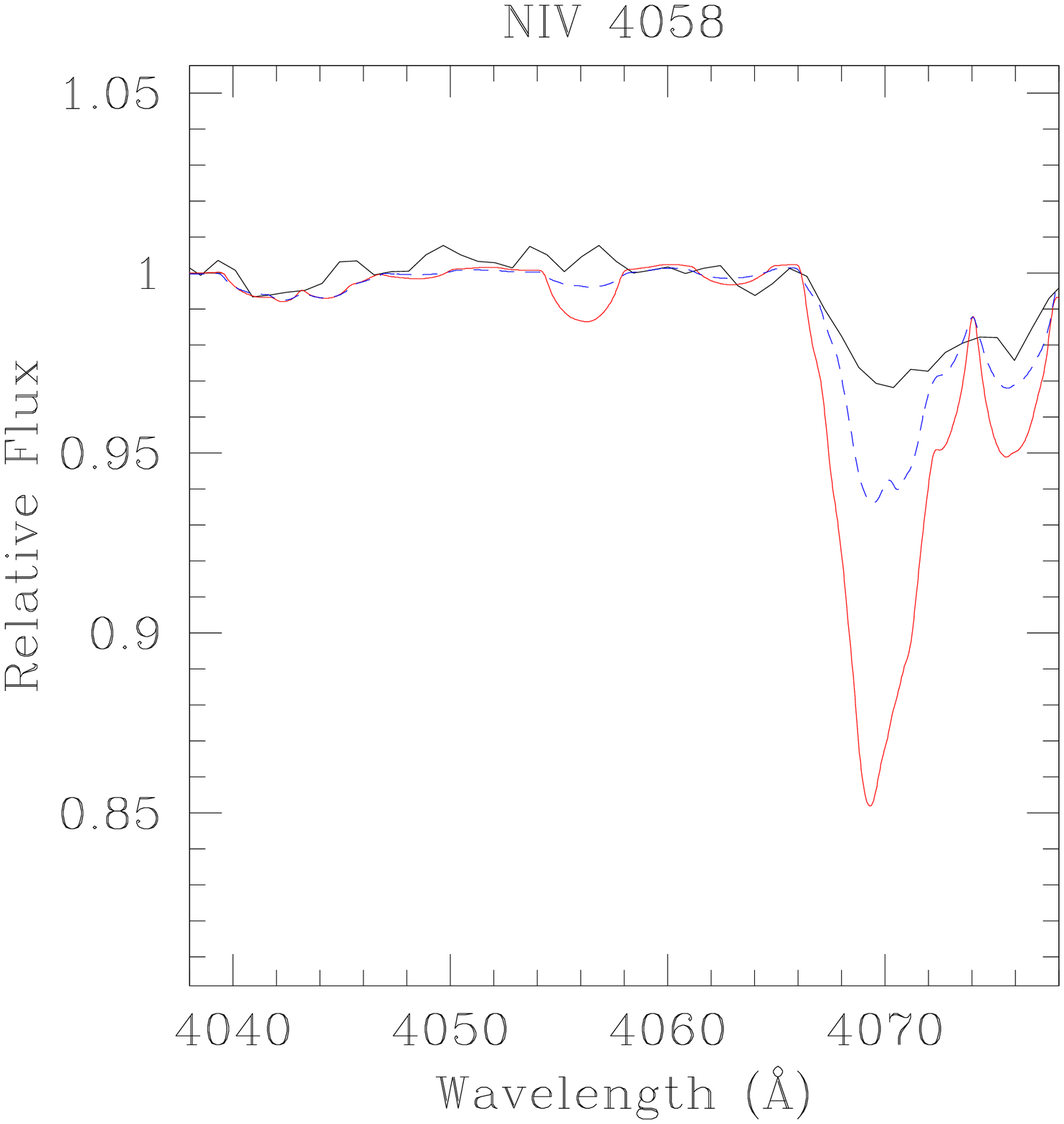}
\plotone{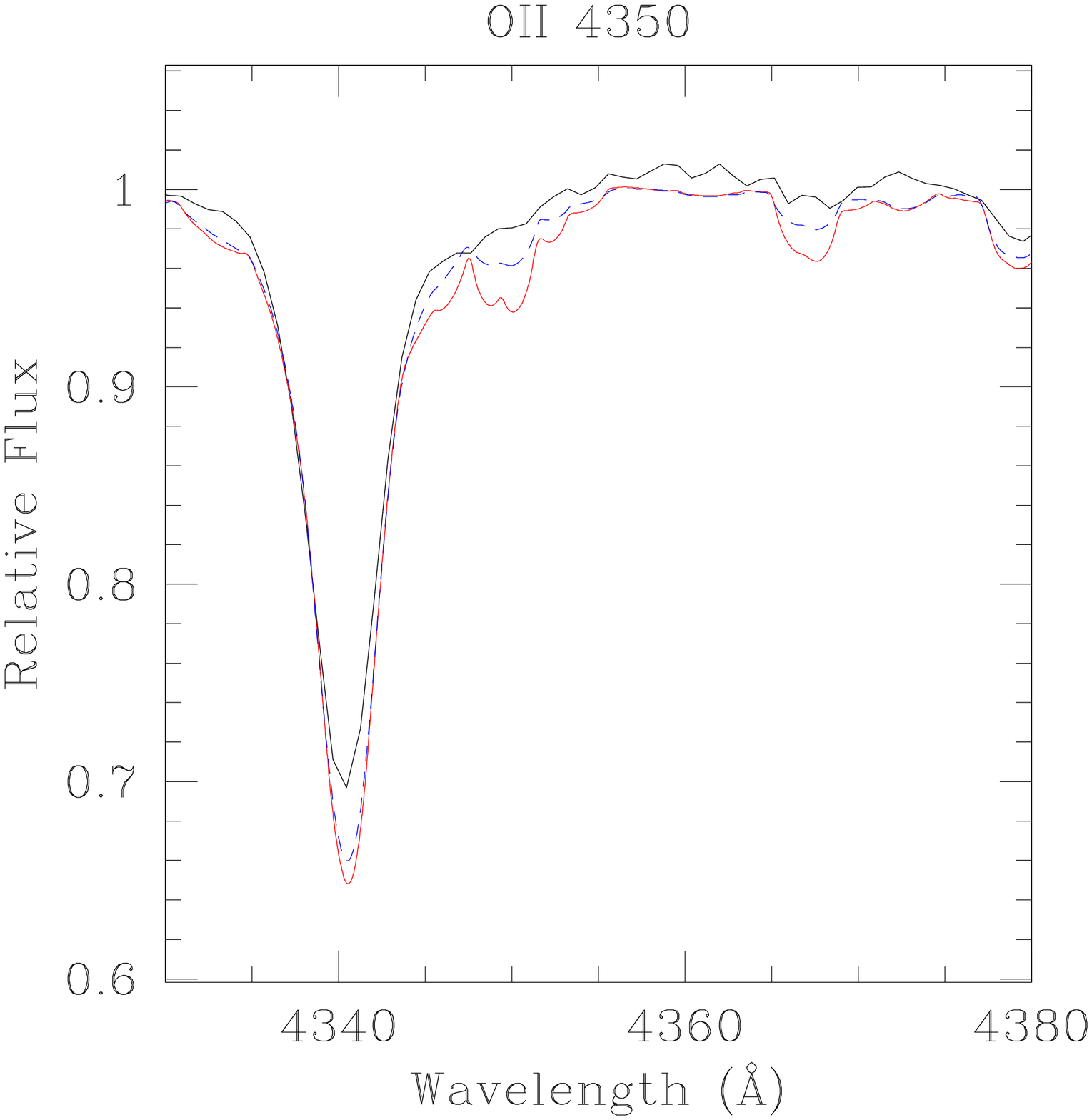}
\plotone{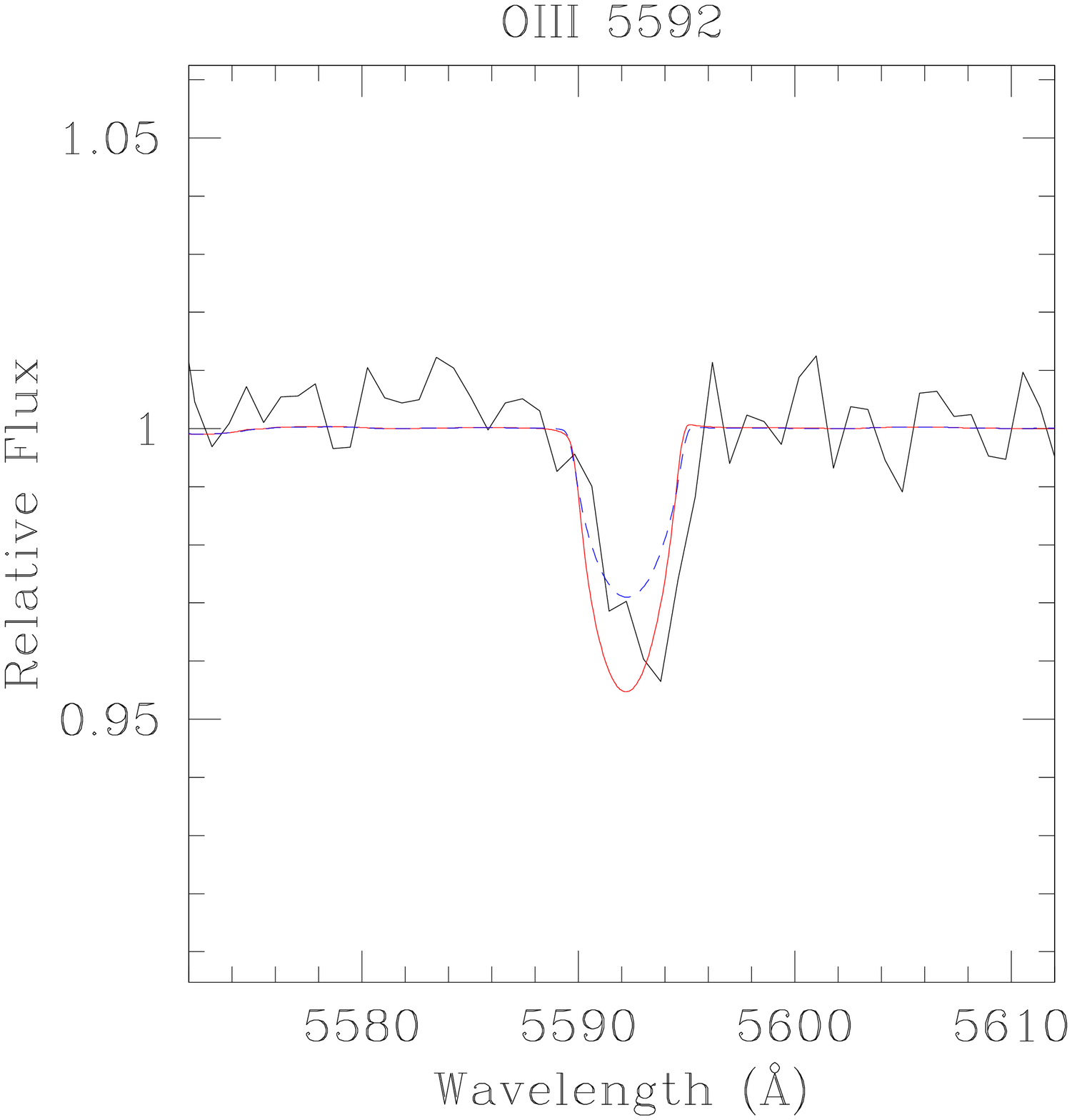}
\caption{\cmfgen\ fits for CNO lines for BI~170, O9.5~I. Black denotes the observed spectrum, solid red indicates the \cmfgen\ model with LMC abundances ($Z/Z_\odot=0.5$), and the dashed blue shows the \cmfgen\ model with C decreased by a factor of 5, N normal, and O decreased by a factor of 2.}
\end{figure}
\clearpage
\begin{figure}
\epsscale{0.3}
\plotone{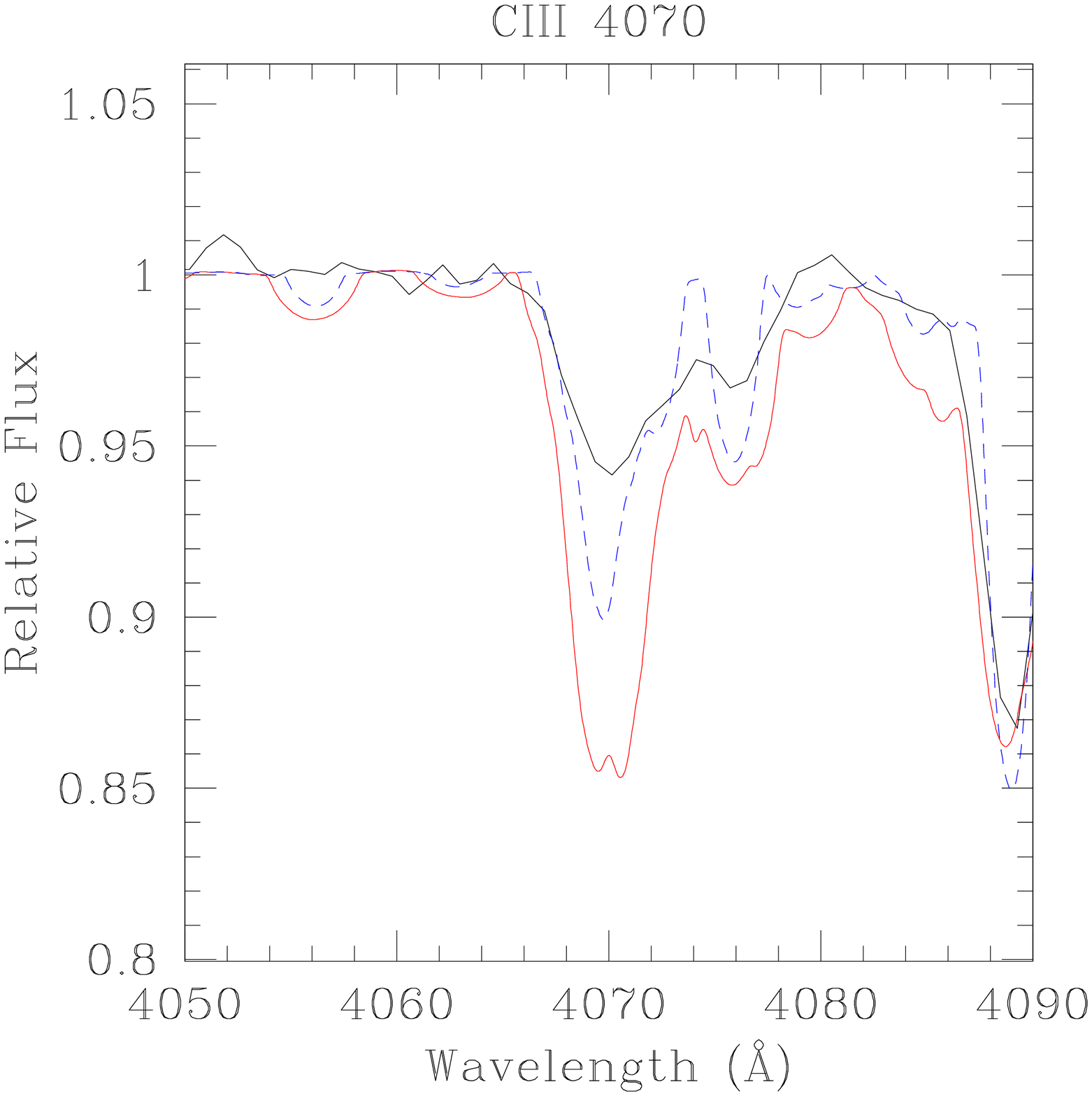}
\plotone{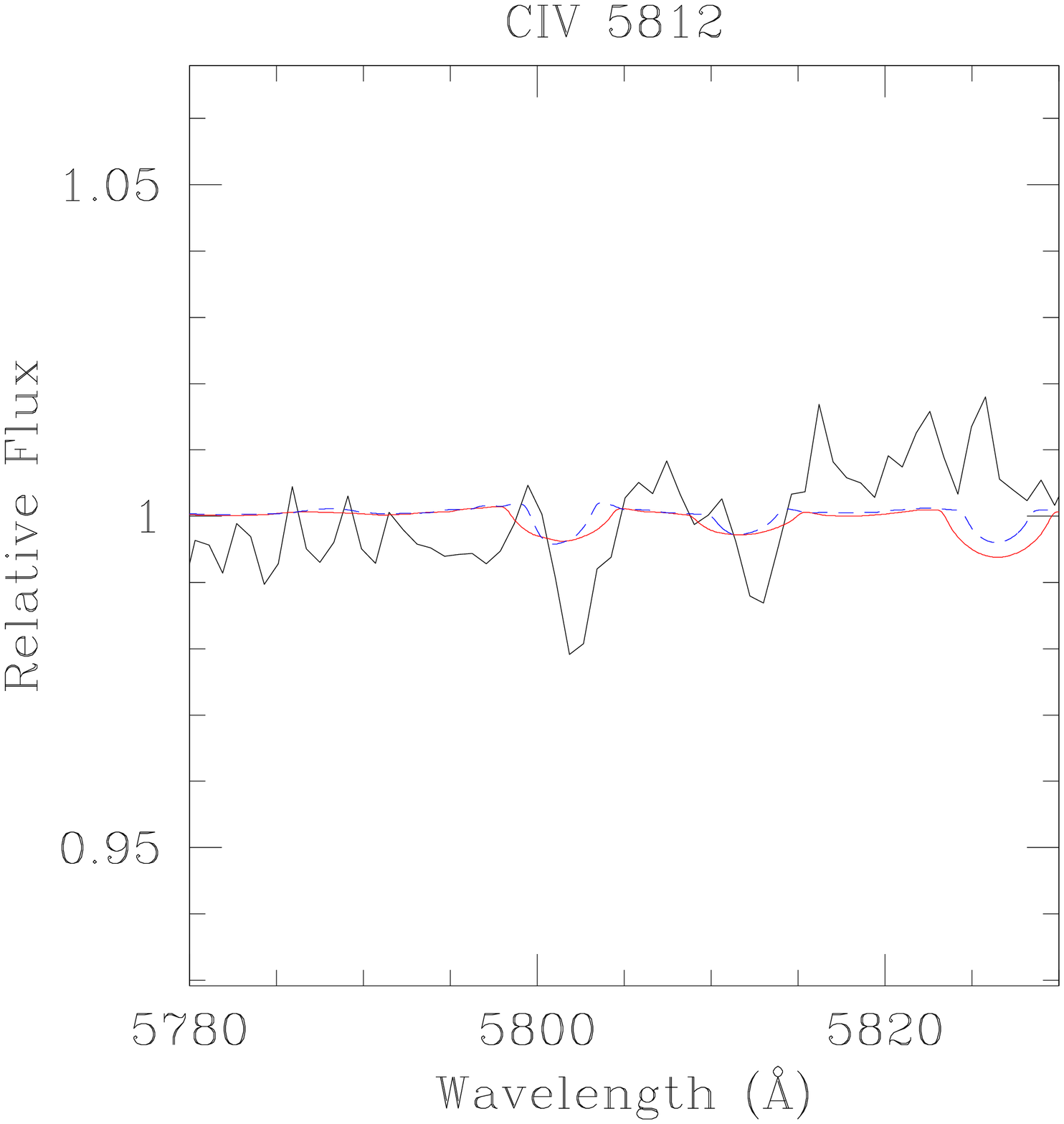}
\plotone{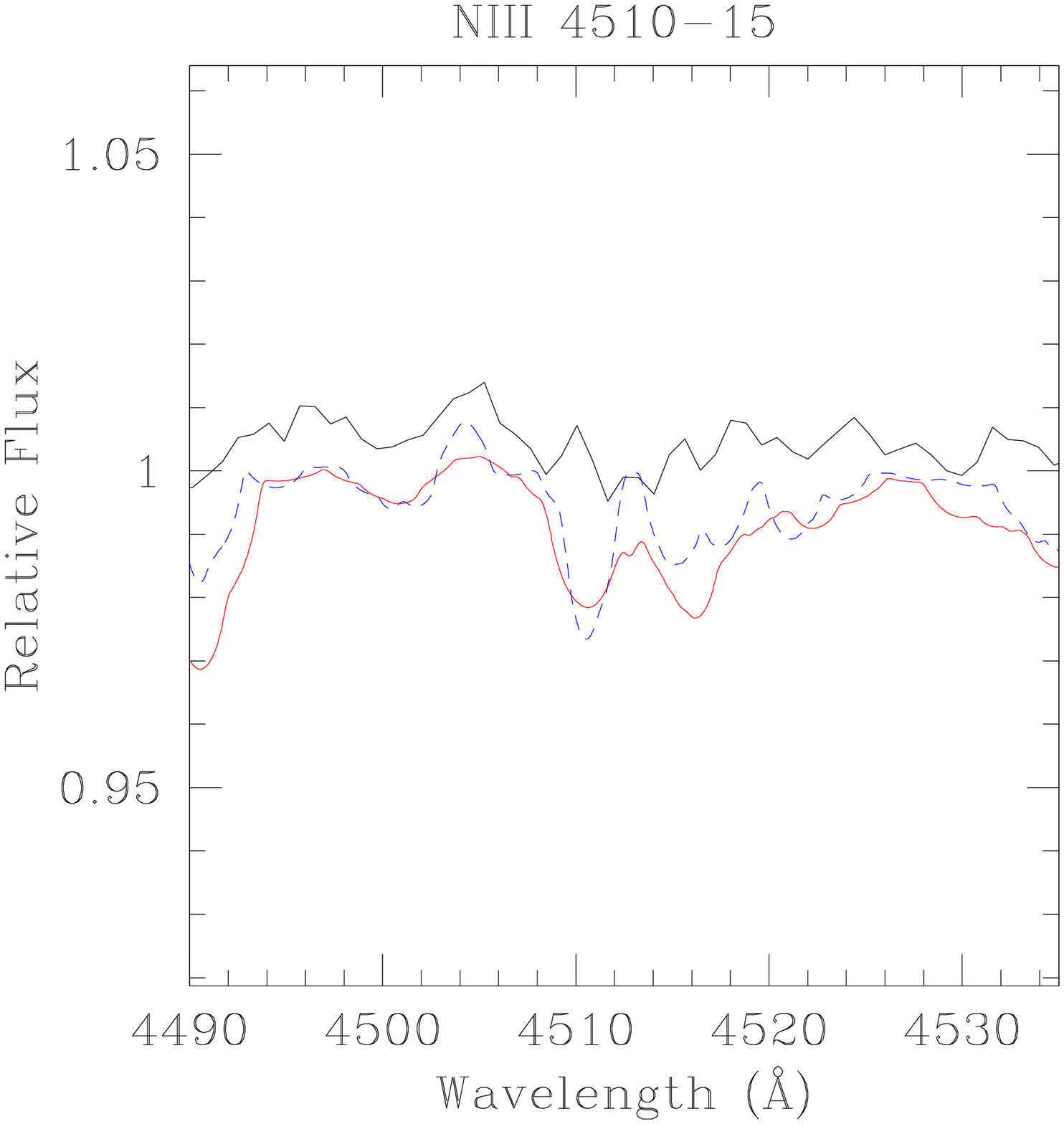}
\plotone{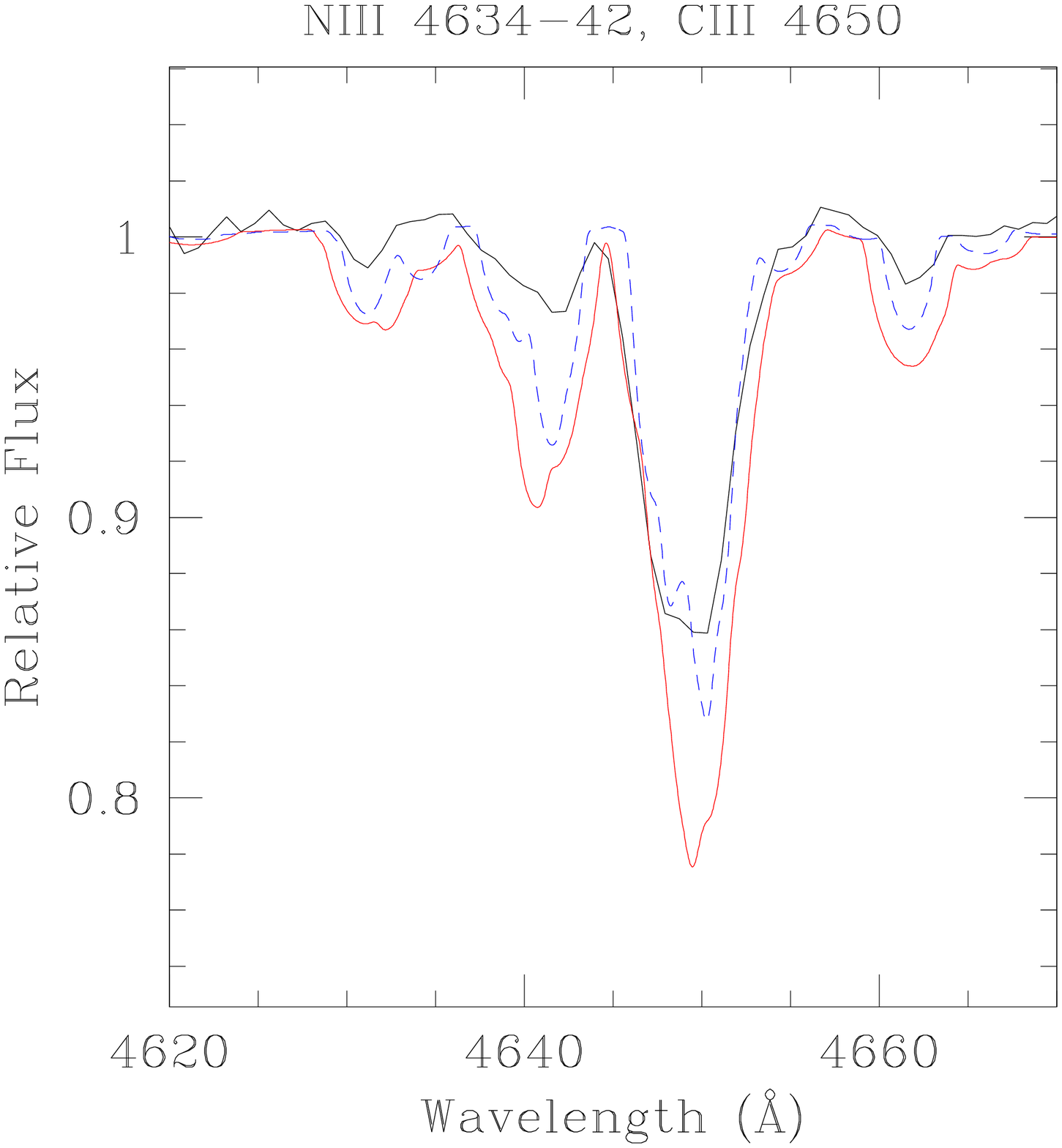}
\plotone{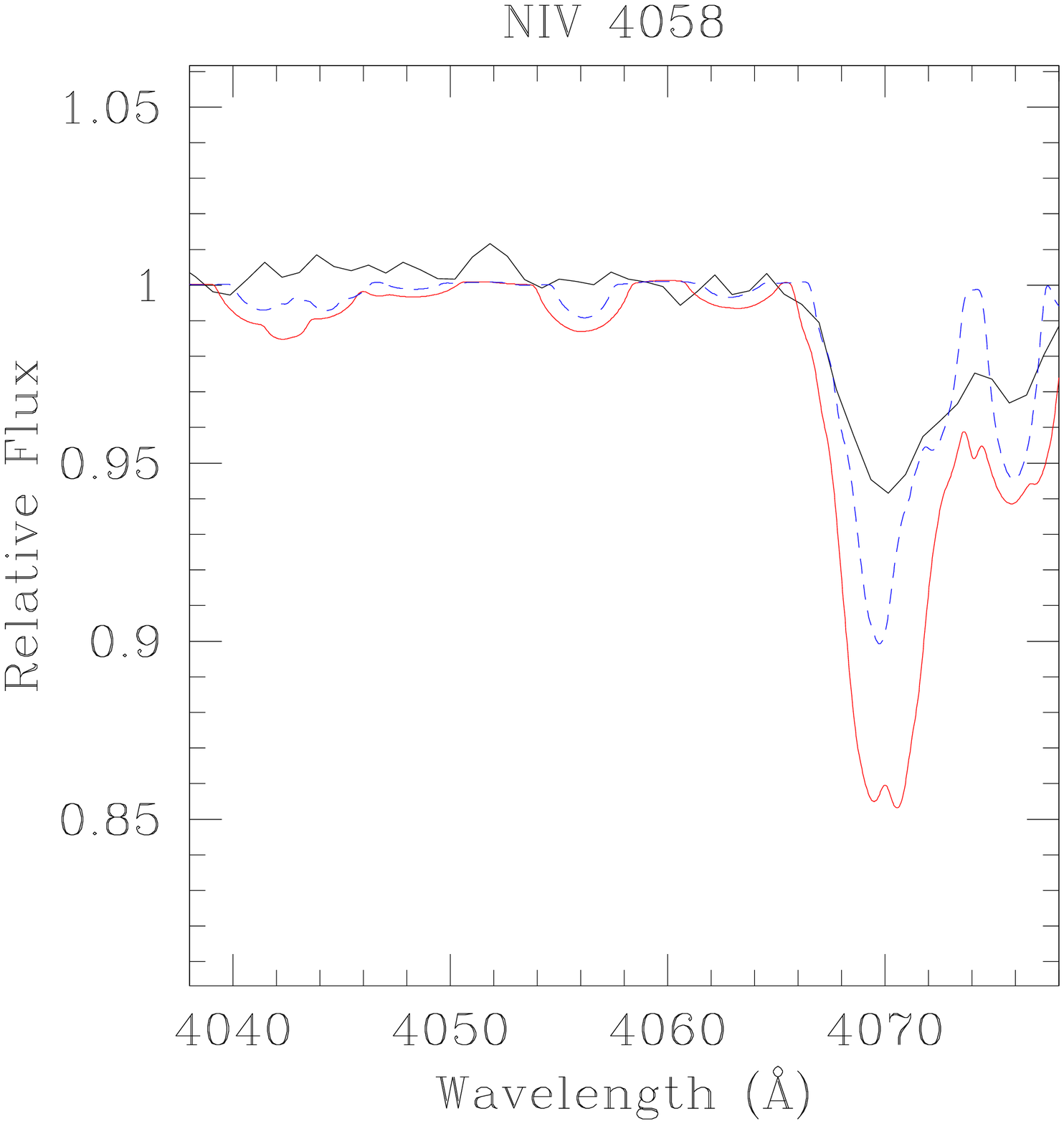}
\plotone{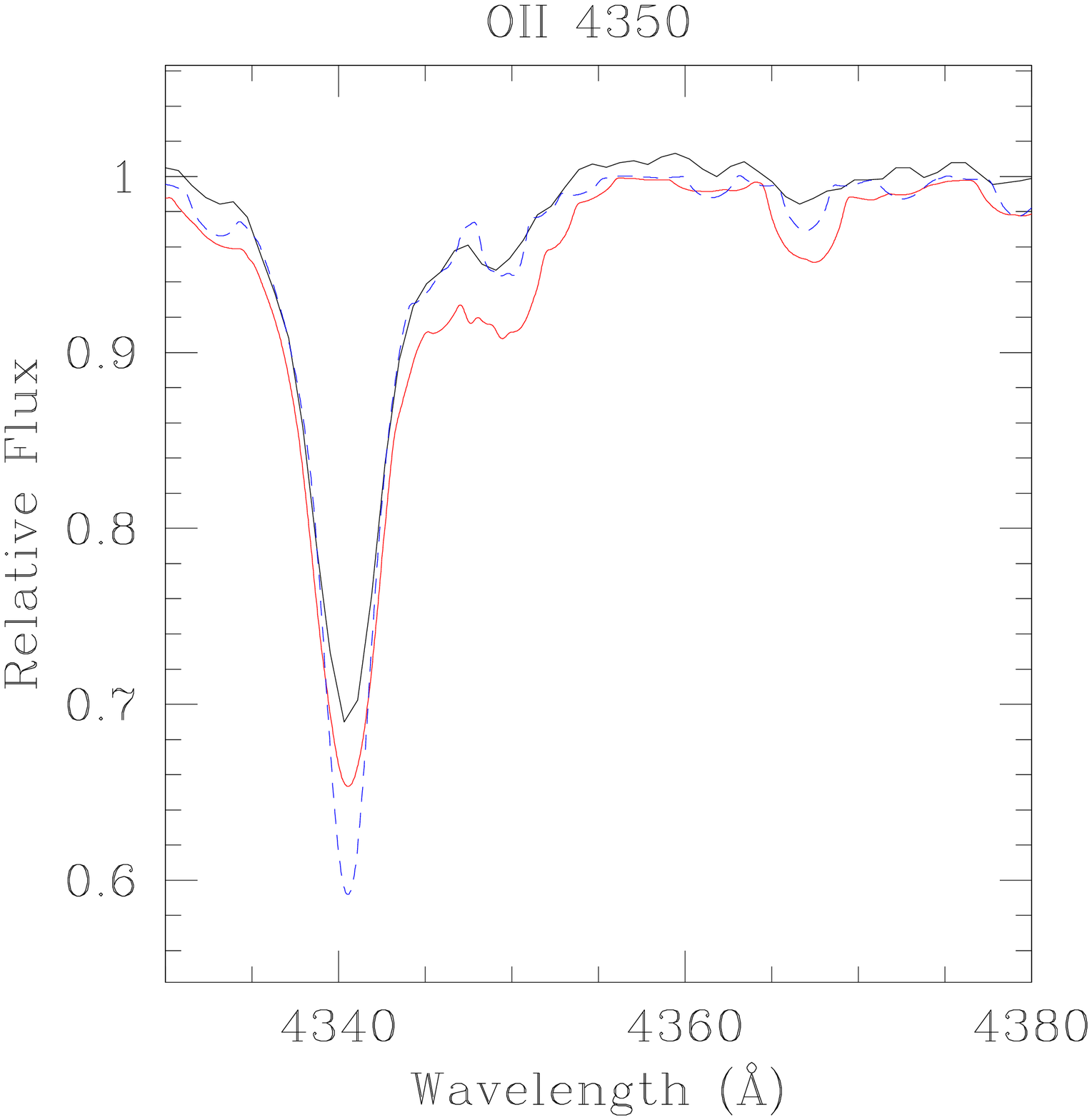}
\plotone{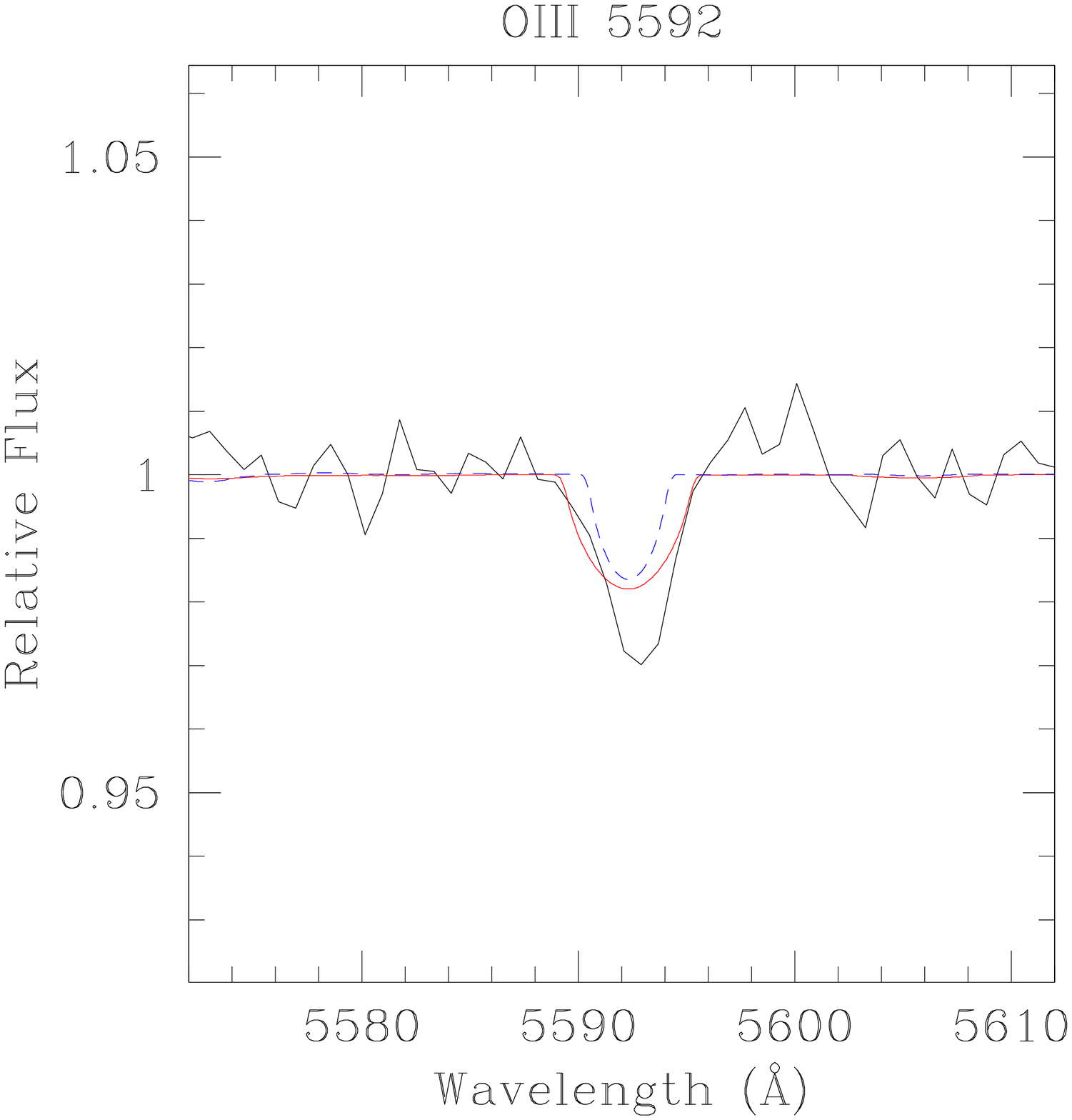}
\caption{\label{fig:sk69124metal} \cmfgen\ fits for CNO lines for Sk $-69^\circ$124, O9.7 I. Black denotes the observed spectrum while solid red indicates our original \cmfgen\ model with LMC abundances ($Z/Z_\odot=0.5$).  The dashed blue shows a second effort by D. J. H. with C decreased by a factor of 3, N decreased by a factor of 2.2, and O decreased by a factor of 2.9.}
\end{figure}
\clearpage
\begin{figure}
\epsscale{0.35}
\plotone{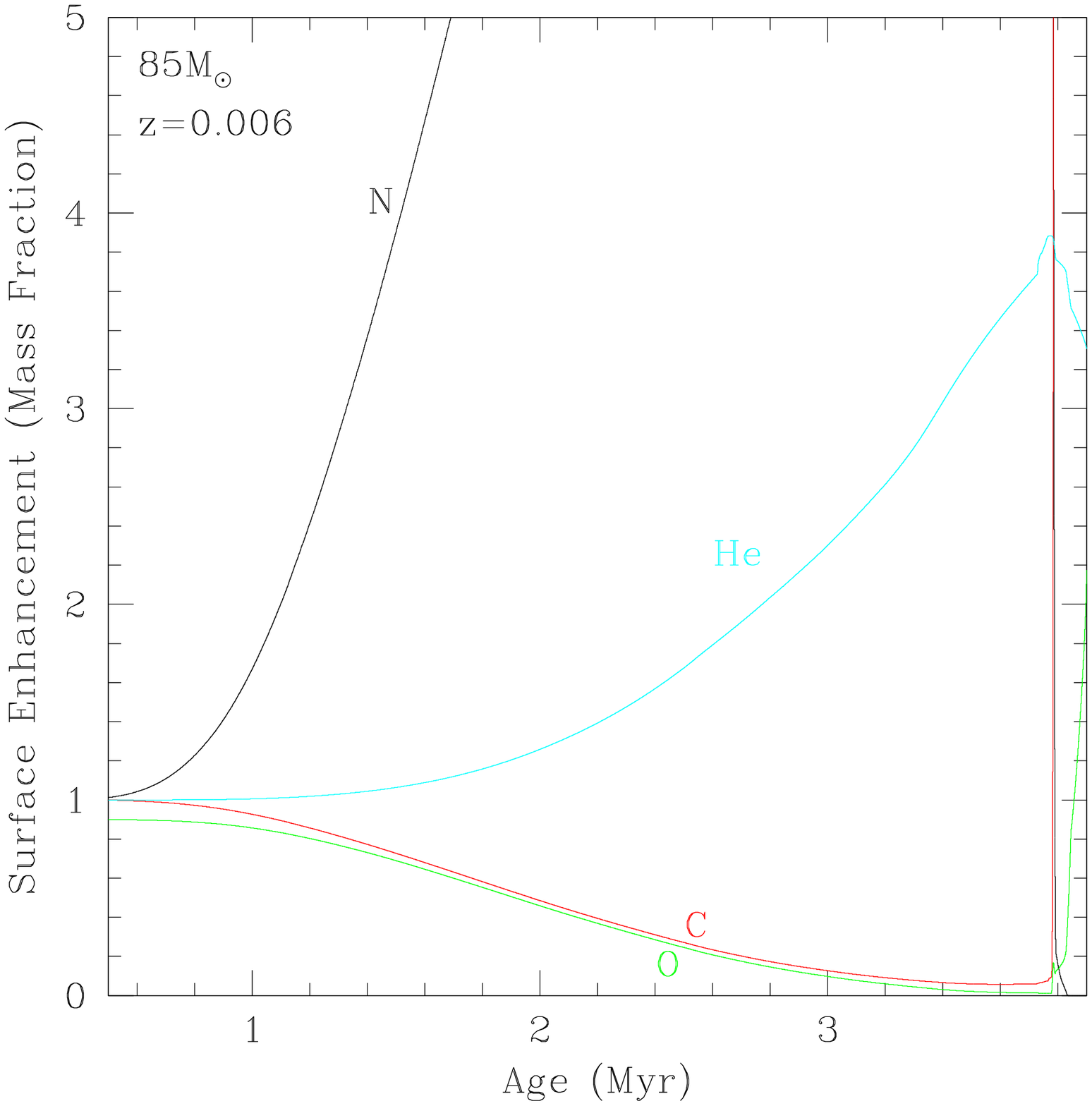}
\plotone{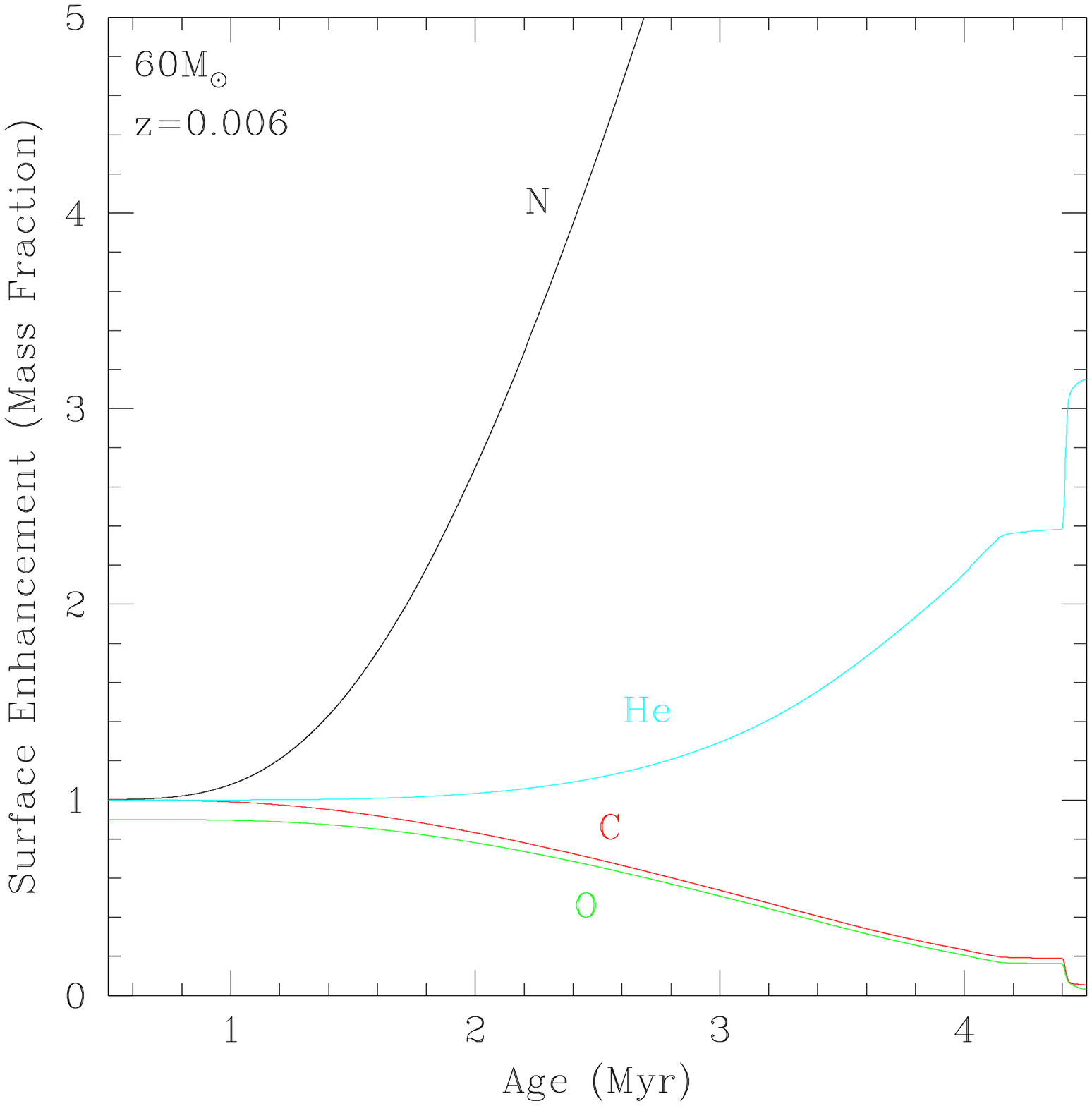}
\plotone{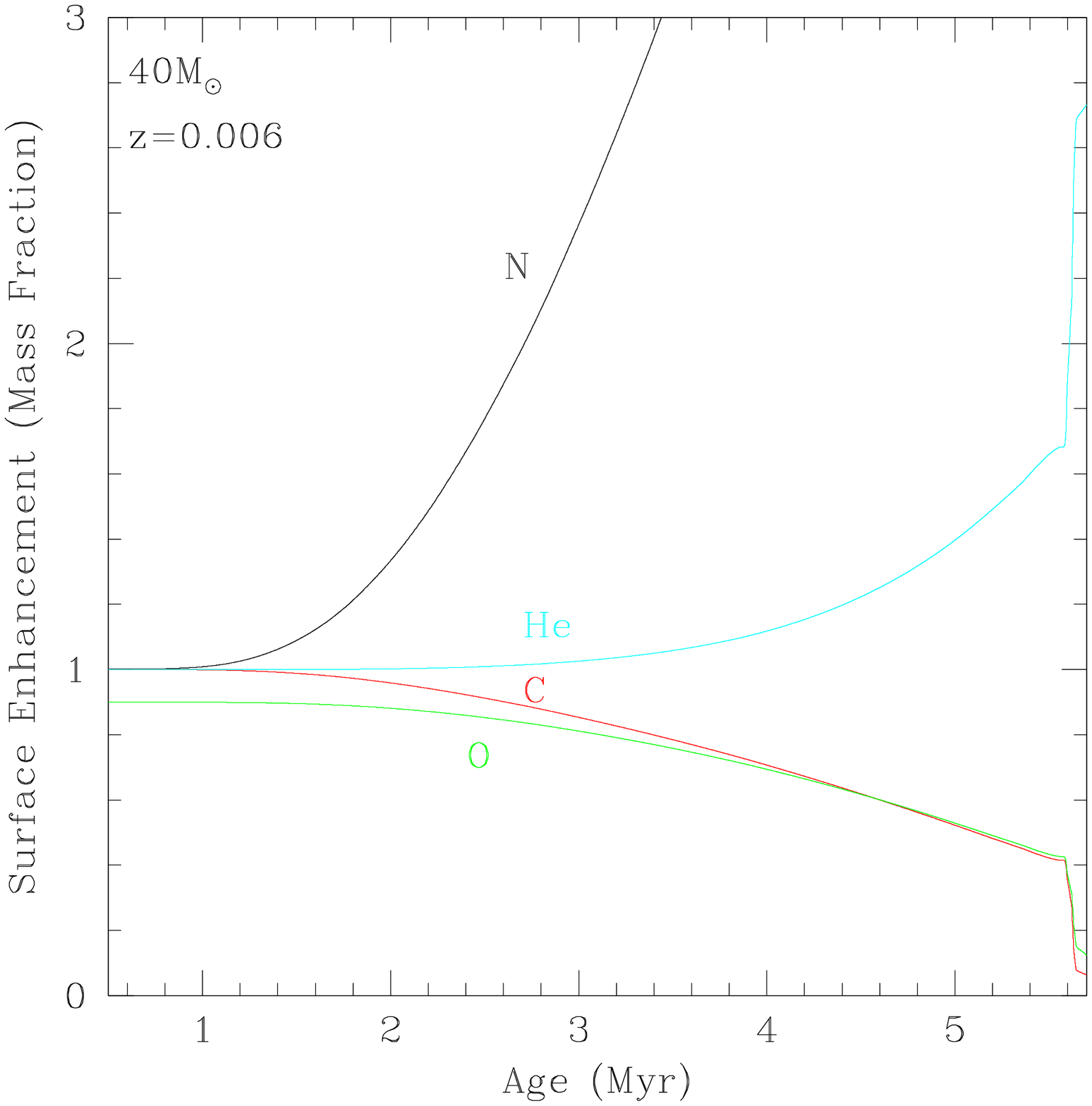}
\plotone{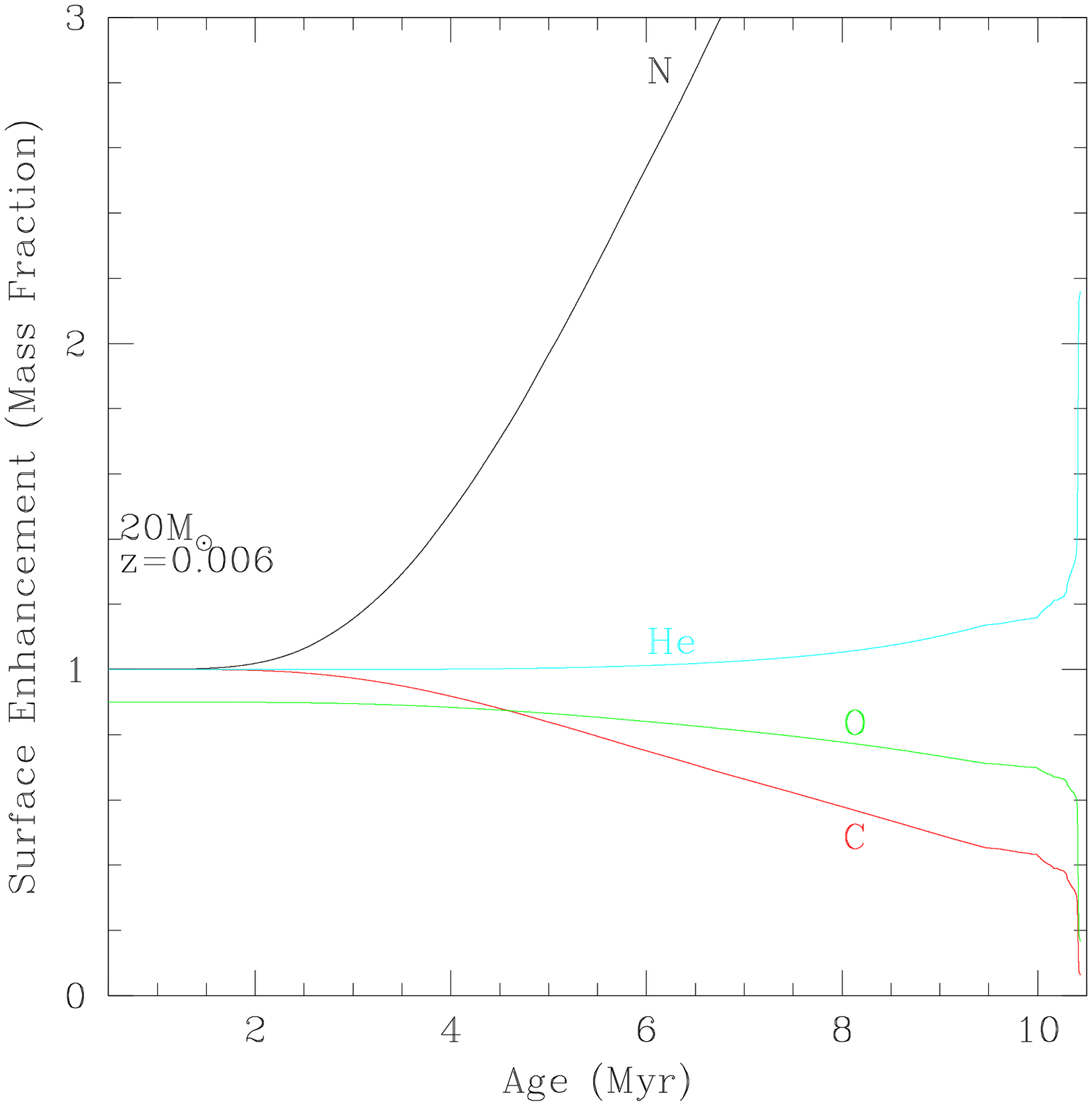}
\caption{\label{fig:evolabund} Expected surface enhancements in
terms of mass fractions.  The three curves show the predicted changes
of the surface enhancements of carbon, oxygen, nitrogen, and helium
during the main-sequence evolution of massive stars.  The models
used include rotation and are computed for $z=0.006$, roughly
corresponding to the metallicity of the LMC (V. Chomienne et al.\
2013, in preparation), with further details given by  Ekstr\"{o}m
et al.\ (2012).}
\end{figure}
\clearpage
\begin{deluxetable}{l l l l c}
\tablecaption{\label{tab:stars} The Sample of Stars Modeled}
\tablewidth{0pt}
\tablehead{
\colhead{Star}
&\colhead{Type}
&\colhead{Galaxy}
&\colhead{$V$}
&\colhead{Paper\tablenotemark{a}}
}
\startdata
AzV 177 & O4 V((f)) & SMC& 14.53 & II \\
AzV 388 & O5.5 V((f)) & SMC& 14.09 & III \\
AzV 75 & O5.5 I(f) & SMC&  12.70 & III \\
AzV 26 & O6 I(f) & SMC&  12.46 & I \\
NGC 346-682 & O8 V & SMC&  14.87 & III \\
AzV 223 & O9.5 II & SMC&  13.66 & III \\ \\
LH 81:W28-23 & O3.5 V((f+)) &LMC& 13.81 & II \\
Sk $-70^\circ$69 & O5.5 V((f)) & LMC& 13.95 & III \\
BI 170 & O9.5 I & LMC& 13.06 & III \\
Sk $-69^\circ$124 & O9.7 I &LMC& 12.77 & III \\
\enddata
\tablenotetext{a}{Paper I = Massey et al.\ 2004; Paper II = Massey et al.\ 2005; Paper III = Massey et al.\ 2009.}
\end{deluxetable}

\clearpage

\begin{deluxetable}{l l c c c c c c c c}
\tablecaption{\label{tab:fixed} Fixed Parameters}
\tablewidth{0pt}
\tablehead{
\colhead{Star}
&\colhead{Type}
&\colhead{$M_V$}
&\colhead{$v_\infty$}
&\colhead{$\beta$}
&\colhead{$Z/Z_\odot$}
&\colhead{$v \sin{i}$}
&&\multicolumn{2}{c}{$v_{\rm rad}$\tablenotemark{b} (km s$^{-1}$)} \\ \cline{9-10}
&
&
&\colhead{km s$^{-1}$} 
&
&
&\colhead{km s$^{-1}$}
&
&\colhead{Blue}
&\colhead{Red} 
}
\startdata
AzV 177 & O4 V((f)) & -4.78    & 2650 &0.8 & 0.2 & 220 && 150 & 100 \\
AzV 388 & O5.5 V((f)) & -5.15 & 1935 &0.8 & 0.2 & 190 && 190 & 170  \\
AzV 75 & O5.5 I(f) & -6.70       & 2120 &0.8 & 0.2 & 160 &&140 & 140   \\
AzV 26 & O6 I(f) & -7.00          & 2150 & 0.8 & 0.2 & 150 &&120 & 160   \\
NGC 346-682 & O8 V & -4.25 & 2900\tablenotemark{a} &0.8 & 0.2 & 150 && 170 & 170  \\
AzV 223 & O9.5 II & -5.64       & 1680 & 0.8 & 0.2 & 140 && 190 & 170  \\ \\

LH 81:W28-23 & O3.5 V((f+)) & -5.14 & 3050 & 0.8 & 0.5 & 120 && 360 & 360   \\
Sk $-70^\circ$ 69 & O5.5 V((f)) & -4.83           & 2300 & 0.8 & 0.5 & 165 && 285 & 250  \\
BI I70 & O9.5 I & -5.66                         & 1370 & 1.2 &0.5 & 140 && 290 & 240  \\
Sk $-69^\circ$124 & O9.7 I & -6.07                & 1430 & 1.2 & 0.5 & 160 && 190 & 160  \\
\enddata
\tablenotetext{a}{$v_\infty$ determined from 2.6$\times$ the escape speed as determined in Paper III.}
\tablenotetext{b}{Note that these are the {\it observed} radial velocities, i.e., what is needed to align the spectra
and the models.  To convert to heliocentric radial velocity for the stars from Paper III, add $-12$ km s$^{-1}$ for the SMC and $-2$ km s$^{-1}$ for the LMC.  For the stars from Papers I-II, add $-6$ km s$^{-1}$ for the SMC,
and $-1$ km s$^{-1}$ for the LMC.  Note that which star comes from which paper is identified in Table~\ref{tab:stars}.}
\end{deluxetable}

\clearpage

\begin{deluxetable}{l l c c r c c c r c c c r c c c r}
\tabletypesize{\tiny}
\rotate
\tablecaption{\label{tab:results} Model Fits: \fastwind\ vs.\ \cmfgen\tablenotemark{a}}
\tablewidth{0pt}
\tablehead{
&&\multicolumn{3}{c}{$T_{\rm eff}$ [K]}
&&\multicolumn{3}{c}{$\log g$ [cgs]}
&&\multicolumn{3}{c}{$\dot{M}$ ($10^{-6} M_\odot$ yr$^{-1}$)\tablenotemark{b}}
&&\multicolumn{3}{c}{He/H (by number)} \\ \cline{3-5} \cline{7-9} \cline{11-13} \cline{15-17}
\colhead{Star}
&\colhead{Type}
&\colhead{FW}
&\colhead{CM}
&\colhead{$\Delta$}
&
&\colhead{FW}
&\colhead{CM}
&\colhead{$\Delta$}
&
&\colhead{FW}
&\colhead{CM}
&\colhead{$\Delta$}
&
&\colhead{FW}
&\colhead{CM}
&\colhead{$\Delta$}
}
\startdata
AzV 177 & O4 V((f))                &  44,000 & 44,500 &   -500 && 3.80 & 3.95 & -0.15 && 0.2 & 0.2 &  0.0 && 0.15 & 0.10 & 0.05 \\
AzV 388 & O5.5 V((f))             &  41,500 & 44,000 &-2,500 && 3.85 & 4.10 & -0.25 && 0.1 & 0.1 &  0.0  && 0.10 & 0.10 & 0.00 \\
AzV 75 & O5.5 I(f)                   &  40,000 & 39,500 &   500 &&3.65 & 3.70 & -0.05 && 1.3 & 1.7 &  -0.4  && 0.10 & 0.10 &  0.00 \\
AzV 26 & O6 I(f)                   &  38,000  & 38,000 &         0 &&3.50 & 3.60 & -0.10 && 1.3 & 1.3 &  0.0  && 0.15 & 0.10 & 0.05 \\
NGC 346-682 & O8 V             &  35,500  & 35,800 &   -300 && 4.10 & 4.10 & 0.00  && 0.0 & 0.0 &  0.0  && 0.10 & 0.10 & 0.00  \\
AzV 223 & O9.5 II                   &  31,600  & 31,000 &   600 && 3.45& 3.60  & -0.15 && 0.2& 0.4 & -0.2  && 0.10 & 0.15 & -0.05 \\
LH 81:W28-23 & O3.5 V((f+)) & 48,000 & 46,500 &  1,500  && 3.75 & 3.90 & -0.15 && 0.9 & 1.2 & -0.3 && 0.25 & 0.25 & 0.00\\
Sk $-70^\circ$69 & O5.5 V((f))           & 39,500 & 41,000 &   -1,500  && 3.70 & 3.80 & -0.10 && 0.3 & 0.5 & -0.2 && 0.10 & 0.10 & 0.00 \\
BI I70 & O9.5 I                      & 31,500 & 30,000 &  1,500  && 3.18 & 3.30  & -0.12  && 0.3 &1.0 & -0.7  && 0.15 & 0.18 & -0.03\\
Sk $-69^\circ$124\tablenotemark{c} & O9.7 I               & 29,000 & 27,800 &  1,200  && 3.12 & 3.20 &  -0.08  &&  0.3 & 0.5 & -0.2  && 0.10 & 0.15 & -0.05 \\

Mean difference   &\nodata & \nodata  & \nodata   &    -80 &&  \nodata  & \nodata & -0.12 && \nodata & \nodata & -0.2&& \nodata & \nodata & 0.00 \\
Median difference &\nodata &  \nodata  & \nodata   &      0 && \nodata  & \nodata  & -0.12 && \nodata & \nodata &-0.2&& \nodata & \nodata &  0.00 \\
$\sigma$               &\nodata &  \nodata &  \nodata  &1300 &&  \nodata &  \nodata &  0.07 &&  \nodata & \nodata & 0.2 &&  \nodata & \nodata & 0.03  \\
Typical fit uncertainty&\nodata &  \nodata & \nodata &500 && \nodata & \nodata &  0.05 && \nodata & \nodata & 0.2 && \nodata & \nodata & 0.02 \\
\enddata
\tablenotetext{a}{FW=\fastwind, CM=\cmfgen, $\Delta$=\fastwind\ result minus \cmfgen\ result.}
\tablenotetext{b}{All mass loss-rates $\dot{M}$ were determined using $\beta=0.8$, except for BI 170 and Sk $-69^\circ$124, for which we adopted $\beta=1.2$.}  
\tablenotetext{c}{Sk $-69^\circ$124 is excluded from the averages as we conclude it is likely a binary. See text.}
\end{deluxetable}

\clearpage

\begin{deluxetable}{l l c c r c c c r c c c r c c c r}
\tabletypesize{\tiny}
\rotate
\tablecaption{\label{tab:FWresults} Model Fits: \fastwind\ New vs. \fastwind\ Old\tablenotemark{a}}
\tablewidth{0pt}
\tablehead{
&&\multicolumn{3}{c}{$T_{\rm eff}$ [K]}
&&\multicolumn{3}{c}{$\log g$ [cgs]}
&&\multicolumn{3}{c}{$\dot{M}$ ($10^{-6} M_\odot$ yr$^{-1}$)\tablenotemark{b}}
&&\multicolumn{3}{c}{He/H (by number)} \\ \cline{3-5} \cline{7-9} \cline{11-13} \cline{15-17}
\colhead{Star}
&\colhead{Type}
&\colhead{New}
&\colhead{Old}
&\colhead{$\Delta$}
&
&\colhead{New}
&\colhead{Old}
&\colhead{$\Delta$}
&
&\colhead{New}
&\colhead{Old}
&\colhead{$\Delta$}
&
&\colhead{New}
&\colhead{Old}
&\colhead{$\Delta$}
}
\startdata
AzV 177 & O4 V((f))                &  44,000 & 44,000\tablenotemark{c} &   0 && 3.80 & 3.85\tablenotemark{c} & -0.05 && 0.2 & 0.1\tablenotemark{c} & 0.1 && 0.15 & 0.15\tablenotemark{c} & 0.00 \\
AzV 388 & O5.5 V((f))             &  41,500 & 42,500 &-1,000 && 3.85 & 3.90 & -0.05 && 0.1 & 0.1 &  0.0  && 0.10 & 0.10 & 0.00 \\
AzV 75 & O5.5 I(f)                   &  40,000 & 39,500 &   500 &&3.65 & 3.50 &  0.15 && 1.3 & 0.9 &  0.4  && 0.10 & 0.10 &  0.00  \\
AzV 26 & O6 I(f)                   &  38,000  & 38,000\tablenotemark{d} &         0 &&3.50 & 3.50\tablenotemark{d} & 0.00 && 1.3 & 0.8\tablenotemark{d} & 0.4  && 0.15 & 0.15\tablenotemark{d} & 0.00 \\
NGC 346-682 & O8 V             &  35,500  & 36,000 &   -500 && 4.10 & 4.10 & 0.00  && 0.0 & 0.0 &  0.0  && 0.10 & 0.10 & 0.00  \\
AzV 223 & O9.5 II                   &  31,600  & 32,000 &  -400 && 3.45& 3.50  & -0.05 && 0.2& 0.2 & 0.0  && 0.10 & 0.10 &  0.00 \\
LH 81:W28-23 & O3.5 V((f+)) & 48,000 & 47,500\tablenotemark{c} &  500  && 3.75 & 3.80\tablenotemark{c} & -0.05 && 0.9 & 0.8\tablenotemark{c} & 0.1 && 0.25 & 0.25\tablenotemark{c} & 0.00\\
Sk $-70^\circ$69 & O5.5 V((f))           & 39,500 & 40,500 &   -1,000  && 3.70 & 3.70 & 0.00 && 0.3 & 0.2 & 0.1 && 0.10 & 0.10 & 0.00 \\
BI I70 & O9.5 I                      & 31,500 & 31,500 &  0  && 3.18 & 3.15  & 0.03  && 0.3 &0.2 & 0.1  && 0.15 & 0.15 & 0.00\\
Sk $-69^\circ$124\tablenotemark{e} & O9.7 I               & 29,000 & 29,250 &  -250  && 3.12 & 3.10 &  0.02  &&  0.3 & 0.2 & 0.1  && 0.10 & 0.10 & 0.00 \\

Mean difference   &\nodata & \nodata  & \nodata   &     -210 &&  \nodata  & \nodata & 0.00 && \nodata & \nodata & 0.1&& \nodata & \nodata & 0.00 \\
Median difference &\nodata &  \nodata  & \nodata   &        0 && \nodata  & \nodata  & 0.00 && \nodata & \nodata & 0.1&& \nodata & \nodata &  0.00 \\
$\sigma$               &\nodata &  \nodata &  \nodata  & 560 &&  \nodata &  \nodata &  0.06 &&  \nodata & \nodata & 0.2 &&  \nodata & \nodata & 0.00  \\
Typical uncertainty&\nodata &  \nodata & \nodata &500 && \nodata & \nodata &  0.05 && \nodata & \nodata & 0.2 && \nodata & \nodata & 0.02 \\
\enddata
\tablenotetext{a}{New=\fastwind\ version 10.1, Old=\fastwind\ version 9.2, unless otherwise noted.  $\Delta$=New result minus Old result.  Old values from Papers I, II, and III.}
\tablenotetext{b}{All mass loss-rates $\dot{M}$ were determined using $\beta=0.8$, except for BI 170 and Sk $-69^\circ$124, for which we adopted $\beta=1.2$.  The (unclumped)
$\dot{M}$ from old \fastwind\ fits have been divided by 3.3 to allow direct comparisons with the clumped $\dot{M}$ rates from the new \fastwind\ fits.} 
\tablenotetext{c}{Old=\fastwind\ version 8.3.1.}
\tablenotetext{d}{Old=\fastwind\ version 6.6.}
\tablenotetext{e}{Sk $-69^\circ$124 is excluded from the averages as we conclude it is likely a binary. See text.}
\end{deluxetable}

\clearpage

\begin{deluxetable}{l l r r r r l}
\tablecaption{\label{tab:abund} \cmfgen\ CNO Mass Fractions\tablenotemark{a}}
\tablewidth{0pt}
\tablehead{
\colhead{Star}
&\colhead{Type}
&\colhead{He/H\tablenotemark{b}}
&\colhead{C}
&\colhead{N}
&\colhead{O} 
&\colhead{Comment}
}
\startdata
\cutinhead{SMC}\\
AzV 177 & O4 V((f))    & 0.10 & 0.20 & 0.20 & 0.20 & Very good \\
AzV 388 & O5.5 V((f))  & 0.10 &0.20 & 0.20 & 0.20 & Good \\
AzV 75  & O5.5 I(f)    & 0.10 &0.08 & 0.48 & 0.08 & Good but NIII weak\\
AzV 26  & O6 I(f)      &  0.10 & 0.05 & 0.60 & 0.05 & Very good\\
NGC346-682 & O8 V&  0.10  & 0.20 & 0.20 & 0.20 & Good \\
AzV 223 & O9.5 II      & 0.15 &  0.05 & 0.20 & 0.07 & Very good\\
\cutinhead{LMC}\\
LH81:W28-23      & O3.5 V((f+)) & 0.25 & 0.05 & 2.5 & 0.50 & Good \\
Sk $-70^\circ$69  & O5.5 V((f)) & 0.10 & 0.10 & 2.5 & 0.10 & C~IV weak\\
BI~170           & O9.5 I & 0.18 &  0.10 & 0.50 & 0.25 & C~III strong \\
Sk $-69^\circ$124 & O9.7 I & 0.15 & 0.50 & 0.50 & 0.50 & C~III and N~III poor\\
Sk $-69^\circ$124 & O9.7 I & 0.15 & 0.17 & 0.23 & 0.17 & Alternative fit\\
\enddata
\tablenotetext{a}{Compare to 0.20 for the SMC stars, and 0.50 for the LMC stars.}  
\tablenotetext{b}{He/H number fraction, 0.10 being unenhanced.}
\end{deluxetable}

\clearpage

\begin{deluxetable}{l c r c c r r r c c  r c c r r r}
\tabletypesize{\tiny}
\rotate
\tablecaption{\label{tab:MD} Derived Quantities and the Mass Discrepancy}
\tablewidth{0pt}
\tablehead{
\colhead{Star} 
&\multicolumn{7}{c}{\fastwind\ } 
& &\multicolumn{7}{c}{\cmfgen}   \\   \cline{2-8} \cline{10-16}
&\colhead{$T_{\rm eff}$}
&\colhead{$R$}
&\colhead{$\log g_{\rm true}$}
&\colhead{$\log L$}
&\colhead{$m_{\rm spect}$}
&\colhead{$m_{\rm evol}$}
&\colhead{$\log \frac {m_{\rm spec}}{m_{\rm evol}}$}
&
&\colhead{$T_{\rm eff}$}
&\colhead{$R$}
&\colhead{$\log g_{\rm true}$}
&\colhead{$\log L$}
&\colhead{$m_{\rm spect}$}
&\colhead{$m_{\rm evol}$}
&\colhead{$\log \frac {m_{\rm spec}}{m_{\rm evol}}$} \\
& \colhead{(K)} &\colhead{($R_\odot$)} & \colhead{(cgs)} &\colhead{($L_\odot$)} & \colhead{($M_\odot$)} & \colhead{($M_\odot$)} & \colhead{(dex)}  
& & {(K)} &\colhead{($R_\odot$)}  & \colhead{(cgs)} &\colhead{($L_\odot$)} & \colhead{($M_\odot$)} & \colhead{($M_\odot$)} & \colhead{(dex)}
}
\startdata
AzV 177         & 44,000   & 8.9  &3.85  &5.43   &$20.7^{+4.8}_{-3.9}$  &$41.6^{+5.1}_{-4.5}$ & $-0.30\pm0.12$   &&  44,500 & 9.0 & 3.99 & 5.45  & $28.6^{+6.6}_{-5.3}$ & $42.7^{+4.1}_{-3.8}$&$-0.18\pm0.12$ \\
AzV 388         & 41,500  &11.0  &3.88  & 5.51  & $33.6^{+7.7}_{-6.3}$  & $40.8^{+4.0}_{-3.6}$ & $-0.08\pm0.12$ &&  44,000 & 10.6 & 4.12 & 5.58  & $54.4^{+12.5}_{-10.2}$  & $44.3^{+4.3}_{-3.9}$ &$+0.09\pm0.12$ \\
AzV 75           & 40,000  &23.1  &3.67  & 6.09  & $91.4^{+21.0}_{-17.1}$  &$71.3^{+5.3}_{-4.9}$ &$+0.11\pm0.12$  &&  39,500 & 23.2 & 3.71 & 6.07  & $100.6^{+23.2}_{-18.8}$ &$70.0^{+5.2}_{-4.8}$  &$+0.16\pm0.12$ \\
AzV 26           & 38,000  &27.2  & 3.52 & 6.14  & $89.1^{+20.5}_{-16.7}$  &$75.8^{+6.2}_{-5.7}$ & $+0.07\pm0.04$  &&  38,000 & 27.2 & 3.61 & 6.14 & $109.6^{+25.2}_{-20.5}$ & $75.8^{+6.2}_{-5.7}$ & $+0.16\pm0.12$\\
NGC 346-682 & 35,500  & 8.0   & 4.11 & 4.96  & $30.1^{+6.9}_{-5.6}$  & $23.2^{+1.6}_{-1.4}$ &$+0.11\pm0.12$  &&   35,800 & 8.0 & 4.11 & 4.97 & $29.8^{+6.9}_{-5.6}$  & $23.6^{+1.3}_{-1.4}$&$+0.10\pm0.12$ \\
AzV 223          & 31,600 & 16.4  & 3.48 & 5.38 & $29.5^{+6.8}_{-5.5}$   &  $28.9^{+2.2}_{-2.1}$& $+0.01\pm0.12$ &&   31,000 & 16.6 & 3.62 &  5.36 & $42.0^{+9.7}_{-7.9}$ &$28.4^{+2.0}_{-1.9}$ & $+0.17\pm0.12$\\
LH 81:W28-23 & 48,000 & 10.0 & 3.77 & 5.68 & $21.6^{+5.0}_{-4.0}$    & $57.5^{+5.3}_{-4.8}$ &$-0.43\pm0.12$ &&   46,500 & 10.2 & 3.91 & 5.64 & $30.9^{+7.1}_{-5.8}$  & $52.8^{+4.9}_{-4.4}$&$-0.23\pm0.12$ \\
Sk $-70^\circ$69& 39,500 &  9.8 & 3.73 & 5.32 & 18.7$^{+4.3}_{-3.5}$    &$32.2^{+2.2}_{-2.0}$ &$-0.24\pm0.12$ &&   41,000 & 9.6 & 3.83 &  5.37 & $22.8^{+5.3}_{-4.3}$ & $35.2^{+2.5}_{-2.4}$ & $-0.19\pm0.12$ \\
BI 170              & 31,500  &  16.5 & 3.23 & 5.38 & $16.8^{+3.9}_{-3.1}$  &\nodata &\nodata &&  30,000 & 17.0& 3.33 & 5.32 & $^{+5.2}_{-4.2}$  &\nodata & \nodata \\
Sk $-69^\circ$124\tablenotemark{a} & 29,000 & 21.1 & 3.17 & 5.45 & $24.0^{+5.5}_{-4.5}$ & $29.4^{+1.9}_{-1.7}$ &$-0.09\pm0.12$ &&  27,800 & 21.7 & 3.24 & 5.40 & $29.7^{+6.8}_{-5.6}$  & $25.3^{+1.8}_{-1.7}$&$+0.07\pm0.12$ \\

Mean difference& \nodata &\nodata           &\nodata          & \nodata       &\nodata & \nodata   & $-0.09\pm0.04$  &&\nodata    & \nodata   &   \nodata &\nodata&\nodata&\nodata &$+0.01\pm0.04$ \\
Median difference & \nodata &\nodata           &\nodata          & \nodata     &\nodata & \nodata     & $-0.04$  &&\nodata    & \nodata   &    \nodata &\nodata&\nodata&\nodata &+0.09 \\
\enddata
\tablenotetext{a}{Sk $-69^\circ$124 is excluded from the averages as we conclude it is likely a binary. See text.}
\end{deluxetable}

\end{document}